\documentclass{SciPost}

\hypersetup{
    colorlinks,
    linkcolor={red!50!black},
    citecolor={blue!50!black},
    urlcolor={blue!80!black}
}

\usepackage[bitstream-charter]{mathdesign}
\DeclareSymbolFont{usualmathcal}{OMS}{cmsy}{m}{n}
\DeclareSymbolFontAlphabet{\mathcal}{usualmathcal}

\fancypagestyle{SPstyle}{
\fancyhf{}
\lhead{\colorbox{scipostblue}{\bf \color{white} ~SciPost Physics Lecture Notes }}
\rhead{{\bf \color{scipostdeepblue} ~Submission }}

\fancyfoot[C]{\textbf{\thepage}}
}

\usepackage{setspace}
\usepackage{soul}
\usepackage{amsmath, physics}
\usepackage{bbm}
\usepackage{slashed}
\usepackage{graphicx}
\usepackage{verbatim}
\usepackage[T1]{fontenc}
\usepackage[utf8]{inputenc}
\usepackage[normalem]{ulem}
\usepackage[shortlabels]{enumitem}
\usepackage[printonlyused]{acronym}
\def\acroswithtooltips{1}
\if\acroswithtooltips1
\usepackage{pdfcomment}
\newcommand{\tooltipacro}[2]{\acro{#1}[\pdftooltip{#1}{#2}]{#2}}
\fi
\usepackage[nameinlink]{cleveref}

\makeatletter
\crefformat{tcb@cnt@examplebox}{example~#2#1#3}
\Crefformat{tcb@cnt@examplebox}{Example~#2#1#3}
\crefformat{tcb@cnt@definitionbox}{definition~#2#1#3}
\Crefformat{tcb@cnt@definitionbox}{Definition~#2#1#3}
\crefformat{tcb@cnt@exercisebox}{exercise~#2#1#3}
\Crefformat{tcb@cnt@exercisebox}{Exercise~#2#1#3}
\crefformat{tcb@cnt@resultbox}{result~#2#1#3}
\Crefformat{tcb@cnt@resultbox}{Result~#2#1#3}
\makeatother

\usepackage{tcolorbox}
\tcbuselibrary{theorems}
\newtcbtheorem[number within=section]{definitionbox}{Definition}%
{colback=scipostblue!15!white,colframe=scipostblue,fonttitle=\bfseries}{def}
\newtcbtheorem[number within=section]{examplebox}{Example}%
{colback=scipostblue!15!white,colframe=scipostblue,fonttitle=\bfseries}{example}
\newtcbtheorem[number within=section]{exercisebox}{Exercise}%
{colback=scipostblue!15!white,colframe=scipostblue,fonttitle=\bfseries}{exercise}
\newtcbtheorem[number within=section]{resultbox}{Result}%
{colback=scipostblue!15!white,colframe=scipostblue,fonttitle=\bfseries}{result}

\newcommand{\eg}{e.g.}
\newcommand{\ie}{i.e.}
\newcommand{\rmd}{{\rm{d}}}

\newcommand{\GN}{\ensuremath{G_N}}
\newcommand{\MPl}{\ensuremath{M_\text{Pl}}}
\newcommand{\CC}{\ensuremath{\Lambda}}

\newcommand{\UVcutoff}{\ensuremath{\Lambda_\text{UV}}}

\newcommand{\covD}{\ensuremath{\nabla}}
\newcommand{\gaugecovD}{\ensuremath{\mathcal D}}

\newcommand{\GFalpha}{\ensuremath{\alpha}}
\newcommand{\GFbeta}{\ensuremath{\beta}}
\newcommand{\GFoperator}{\ensuremath{\mathfrak F}}
\newcommand{\GFcondition}{\ensuremath{\mathcal F}}

\newcommand{\propG}{\ensuremath{\mathcal G}}

\newcommand{\partitionfunction}{\ensuremath{\mathcal Z}}

\newcommand{\scatteringamplitude}{\ensuremath{\mathcal A}}
\newcommand{\polarizationtensor}{\ensuremath{\epsilon}}
\newcommand{\polarizationtensorbasis}{\ensuremath{e}}

\newcommand{\gaussbonnetterm}{\ensuremath{\mathfrak E}}

\newcommand{\lagrangian}{\ensuremath{\mathcal L}}

\newcommand{\orderneglected}{\ensuremath{\mathcal O}}

\allowdisplaybreaks

\usepackage{titlesec}

\titleclass{\subsubsubsection}{straight}[\subsection]

\newcounter{subsubsubsection}[subsubsection]
\renewcommand\thesubsubsubsection{\thesubsubsection.\arabic{subsubsubsection}}

\titleformat{\subsubsubsection}
  {\normalfont\normalsize\bfseries}{\thesubsubsubsection}{1em}{}
\titlespacing*{\subsubsubsection}
{0pt}{3.25ex plus 1ex minus .2ex}{1.5ex plus .2ex}

\makeatletter
\renewcommand\paragraph{\@startsection{paragraph}{5}{\z@}%
  {3.25ex \@plus1ex \@minus.2ex}%
  {-1em}%
  {\normalfont\normalsize\bfseries}}
\renewcommand\subparagraph{\@startsection{subparagraph}{6}{\parindent}%
  {3.25ex \@plus1ex \@minus .2ex}%
  {-1em}%
  {\normalfont\normalsize\bfseries}}
\def\toclevel@subsubsubsection{4}
\def\toclevel@paragraph{5}
\def\toclevel@subparagraph{6}
\def\l@subsubsubsection{\@dottedtocline{4}{7em}{4em}}
\def\l@paragraph{\@dottedtocline{5}{10em}{5em}}
\def\l@subparagraph{\@dottedtocline{6}{14em}{6em}}
\makeatother

\makeatletter
\renewcommand\@dotsep{1000}
\makeatother

\numberwithin{equation}{section}
\numberwithin{figure}{section}

\usepackage{pgf}
\usepackage{tabularx}
\usepackage{array}
\usepackage{colortbl}
\tcbuselibrary{skins}

\newcolumntype{Y}{>{\centering\arraybackslash}X}
\tcbset{tab2/.style={enhanced,fonttitle=\bfseries,fontupper=\normalsize\sffamily,
colback=yellow!10!white,colframe=red!50!black,colbacktitle=red!30!white,
coltitle=black,center title}}
\newenvironment{eqaed}
    {\begin{equation}
    \begin{aligned}
    }
    { 
    \end{aligned}
    \end{equation}
    \ignorespacesafterend
    }

\usepackage{dsfont}
\usepackage{appendix}
\usepackage{caption, subcaption}

\newcommand{\GNk}{\ensuremath{G_{k}}}
\newcommand{\CCk}{\ensuremath{\Lambda_k}}

\newcommand{\emailIvano}{ivano.basile@lmu.de}
\newcommand{\fullaffiliationIvano}{Arnold-Sommerfeld Center for Theoretical Physics, Ludwig Maximilians Universität M\"unchen, Theresienstraße 37, 80333 M\"unchen, Germany}
\newcommand{\briefaffiliationIvano}{LMU M\"unchen}
\newcommand{\orcidIvano}{0000-0001-6183-594X}
\newcommand{\titleIvano}{Introduction to string theory}
\newcommand{\emailLuca}{luca.buoninfante@ru.nl}
\newcommand{\fullaffiliationLuca}{High Energy Physics Department, Institute for Mathematics, Astrophysics, and Particle Physics, Radboud University, Nijmegen, The Netherlands}
\newcommand{\briefaffiliationLuca}{Radboud University Nijmegen}
\newcommand{\orcidLuca}{0000-0002-1875-8333}
\newcommand{\titleLuca}{Introduction to perturbative quantum gravity}
\newcommand{\emailFrancesco}{francesco.difilippo@mff.cuni.cz}
\newcommand{\fullaffiliationFrancesco}{Institute of Theoretical Physics, Faculty of Mathematics and Physics, Charles University, V Holešovičkách 2, 180 00 Prague 8, Czech Republic}
\newcommand{\briefaffiliationFrancesco}{Charles University Prague}
\newcommand{\orcidFrancesco}{0000-0002-4727-8953}
\newcommand{\titleFrancesco}{Quantum effects in black hole spacetimes}
\newcommand{\emailBenjamin}{knorr@thphys.uni-heidelberg.de}
\newcommand{\fullaffiliationBenjaminone}{Institute for Theoretical Physics, Heidelberg University, Philosophenweg 12, 69120 Heidelberg, Germany}
\newcommand{\briefaffiliationBenjamin}{Heidelberg University}
\newcommand{\orcidBenjamin}{0000-0001-6700-6501}

\newcommand{\emailAlessia}{alessia.platania@nbi.ku.dk}
\newcommand{\fullaffiliationAlessia}{Niels Bohr International Academy, The Niels Bohr Institute, Blegdamsvej 17, DK-2100 Copenhagen \O, Denmark}
\newcommand{\briefaffiliationAlessia}{NBI}
\newcommand{\orcidAlessia}{0000-0001-7789-344X}
\newcommand{\titleAlessiaBenjamin}{Non-perturbative renormalization group and asymptotic safety}
\newcommand{\titleAlessiaBenjaminnewline}{Non-perturbative renormalization group and \\ asymptotic safety}
\newcommand{\emailAnna}{tokareva@ucas.ac.cn}
\newcommand{\fullaffiliationAnnaone}{School of Fundamental Physics and Mathematical Sciences, Hangzhou Institute for Advanced Study, UCAS, Hangzhou 310024, China}
\newcommand{\fullaffiliationAnnatwo}{International Centre for Theoretical Physics Asia-Pacific, Beijing/Hangzhou, China}
\newcommand{\fullaffiliationAnnathree}{Theoretical Physics, Blackett Laboratory, Imperial College London, SW7 2AZ London, UK}
\newcommand{\briefaffiliationAnna}{HIAS, ICTP-AP Beijing/Hangzhou, Imperial College}
\newcommand{\orcidAnna}{0000-0003-3540-9028}
\newcommand{\titleAnna}{Gravitational effective field theory and positivity bounds}

\begin{document}

\pagestyle{SPstyle}
\begin{center}{\Large \textbf{\color{purple}{
Lectures in Quantum Gravity}}}\end{center}

\begin{center}\textbf{
Ivano Basile\,\href{https://orcid.org/\orcidIvano}{\protect \includegraphics[scale=.07]{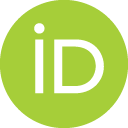}}\,\textsuperscript{1$\star$},
Luca Buoninfante\,\href{https://orcid.org/\orcidLuca}{\protect \includegraphics[scale=.07]{ORCIDiD_icon128x128.png}}\,\textsuperscript{2$\dagger$},
Francesco Di Filippo\,\href{https://orcid.org/\orcidFrancesco}{\protect \includegraphics[scale=.07]{ORCIDiD_icon128x128.png}}\,\textsuperscript{3$\ddagger$}, \\
Benjamin Knorr\,\href{https://orcid.org/\orcidBenjamin}{\protect \includegraphics[scale=.07]{ORCIDiD_icon128x128.png}}\,\textsuperscript{4$\circ$},
Alessia Platania\,\href{https://orcid.org/\orcidAlessia}{\protect \includegraphics[scale=.07]{ORCIDiD_icon128x128.png}}\,\textsuperscript{5$\P$} and
Anna Tokareva\,\href{https://orcid.org/\orcidAnna}{\protect \includegraphics[scale=.07]{ORCIDiD_icon128x128.png}}\,\textsuperscript{6$\parallel$}
}\end{center}

\begin{center}
$\star$ \href{mailto:\emailIvano}{\small \emailIvano}\,,\quad
$\dagger$ \href{mailto:\emailLuca}{\small \emailLuca}\,,\quad
$\ddagger$ \href{mailto:\emailFrancesco}{\small \emailFrancesco}\,,\quad \\
$\circ$ \href{mailto:\emailBenjamin}{\small \emailBenjamin}\,,\quad
$\P$ \href{mailto:\emailAlessia}{\small \emailAlessia}\,,\quad 
$\parallel$ \href{mailto:\emailAnna}{\small \emailAnna}
\end{center}

\section*{\color{scipostdeepblue}{Abstract}}
\textbf{\boldmath{%
Formulating a quantum theory of gravity lies at the heart of fundamental theoretical physics. This collection of lecture notes encompasses a selection of topics that were covered in six mini-courses at the Nordita PhD school \textit{``Towards Quantum Gravity''}. The scope was to provide a coherent picture, from its foundation to forefront research, emphasizing connections between different areas. The lectures begin with perturbative quantum gravity and effective field theory. Subsequently, two ultraviolet-complete approaches are presented: asymptotically safe gravity and string theory. Finally, elements of quantum effects in black hole spacetimes are discussed.
}} 
\vspace{\baselineskip}

\noindent\textcolor{white!90!black}{%
\fbox{\parbox{0.975\linewidth}{%
\textcolor{white!40!black}{\begin{tabular}{lr}%
  \begin{minipage}{0.6\textwidth}%
    {\small Copyright attribution to authors. \newline
    This work is a submission to SciPost Physics Lecture Notes. \newline
    License information to appear upon publication. \newline
    Publication information to appear upon publication.}
  \end{minipage} & \begin{minipage}{0.319\textwidth}
    {\small Received Date \newline Accepted Date \newline Published Date}%
  \end{minipage}
\end{tabular}}
}}
}

\begin{center}
    \textcolor{scipostdeepblue}{\rule{13cm}{1pt}}
\end{center}

\begin{center}
{\bf 1} \fullaffiliationIvano
\\
{\bf 2} \fullaffiliationLuca
\\
{\bf 3} \fullaffiliationFrancesco
\\
{\bf 4} \fullaffiliationBenjaminone 
\\
{\bf 5} \fullaffiliationAlessia
\\
{\bf 6} \fullaffiliationAnnaone \\ \fullaffiliationAnnatwo \\\fullaffiliationAnnathree 
\end{center}


\clearpage

{\setstretch{0.96}
\vspace{10pt}
\noindent\rule{\textwidth}{1pt}
\tableofcontents
\noindent\rule{\textwidth}{1pt}
\vspace{10pt}
}



\section{Introduction}
\label{sec:intro}

The formulation of a quantum theory of gravity is one of the most challenging and fascinating questions in fundamental physics. It has attracted increasing interest since the middle of the previous century. Especially in the last decades, new theoretical progress has been made in developing different \ac{QG} approaches and gaining new insights into quantum aspects of gravity. In addition, the new trinity of gravitational observations --- precision cosmology, \ac{GW} astronomy, and \ac{BH} shadows --- has opened up a unique possibility for testing new physics beyond classical \ac{GR}, thus offering concrete hopes of detecting quantum-gravitational signatures with future observations. 

In a broad and diverging research field such as \ac{QG}, it can be hard to keep up. On the one hand, working on different approaches and following orthogonal directions hinders constructive communication across communities. Indeed, experts disagree not only on the answers, but even on the questions that one should ask. On the other hand, researchers who are not yet familiar with the topic may find it difficult to grasp the big picture, the main essence underpinning specific approaches, and the reasons behind apparently contradicting ideas. In such a state of affairs, it becomes essential to debate, learn from the developments and milestones of other approaches, and find common grounds.

The Nordita Scientific Program \href{https://indico.fysik.su.se/event/8133/}{\textit{``Quantum Gravity: from gravitational effective field theories to ultraviolet complete approaches''}} was a one-of-a-kind event in the field of \ac{QG}. It included not only an intensive three-week workshop with talks and \textit{extensive} discussion sessions~\cite{Buoninfante:2024yth}, but also a one-week PhD school titled \textit{``Towards Quantum Gravity''} that kicked off the program. The school was structured into six mini-courses consisting of six hours of lectures each, for a total of 36 hours. The aim was to provide students and early-career researchers with a broad (yet partial) overview of the basics of \ac{QG}, enabling them to follow more advanced and specialized talks during the subsequent three-week workshop.

These \textit{``Lectures in Quantum Gravity''} collect and unify the content of five of the mini-courses taught at the PhD school, including some extra material. An important aspect is that the various sections are not disconnected from each other: a great effort has been made to provide a coherent picture, connecting topics that often appear disconnected in forefront research. Special care has been taken to use the same conventions and notations across sections. Additionally, where useful, references to other sections have been made in the hope of highlighting how different topics are connected or even build on one another. We hope that these arrangements contributed creating a pedagogical and coherent set of lectures, thus facilitating the reader in studying the material and grasping the subject \textit{as a whole}, rather than in disconnected patches. All mini-courses were recorded and are available on the YouTube channel \href{https://www.youtube.com/@Quantumgravity.nordita}{@Quantumgravity.nordita}. The links to the individual lectures are given at the beginning of each section.

The lectures start with an introduction to perturbative \ac{QG}, where \ac{GR} is quantized in the framework of perturbative \ac{QFT}. The degrees of freedom, the graviton propagator, and the failure of perturbative renormalizability are analyzed in detail. The last part of this first mini-course exploits these basics to discuss a first approach to \ac{QG}: quadratic gravity as a perturbatively renormalizable \ac{QFT}. Subsequently, in the second mini-course,  gravitational \ac{EFT} is introduced, presenting both applications and limitations. Consistency constraints from the requirements of unitarity and causality are derived. More advanced topics related to scattering amplitudes are then discussed as necessary tools to study the implications of \ac{EFT} in \ac{QG}. Perturbative \ac{QG} and \ac{EFT} are the building blocks that different \ac{QG} approaches must recover at low energies.

The third and fourth mini-courses focus on two examples of \ac{UV}-complete approaches to \ac{QG}. The third mini-course introduces the general notion of non-per\-tur\-ba\-tive renormalization and its application to \ac{QG}, resulting in a theory known as \ac{ASQG}. Advanced computational methods to study non-per\-tur\-ba\-tive \ac{RG} flows and the existence of interacting fixed points are presented, such as the heat kernel technique and the \ac{FRG}. In particular, an explicit \ac{FRG} calculation and fixed-point analysis are performed for the Einstein-Hilbert truncation. Some more advanced topics and physical consequences of \ac{ASQG} are then discussed. The fourth mini-course presents a \ac{QG}-oriented introduction to \ac{ST}. After motivating \ac{ST} as a proposal to address \ac{QG} problems, the lectures focus on weakly interacting closed strings, introduce the worldsheet formulation and study the implications at both low and high energies. In the low-energy regime, the connection with gravitational \acp{EFT} is outlined. Furthermore, high-energy scattering between closed strings is studied, including the derivation of the Virasoro-Shapiro amplitude for graviton scattering and a discussion of the \ac{BH} transition.

The last mini-course is devoted to the study of quantum effects in \ac{BH} spacetimes. The phenomenon of particle creation in a gravitational collapse is presented. After introducing elements of \ac{QFT} in curved spacetime, the lectures concentrate on the derivation of the Hawking radiation in the case of a collapsing null shell. Different choices of vacuum states are analyzed, and the distinction between the physical understanding of Hawking radiation in static and collapsing \acp{BH} is explained. In the final part, the information loss problem is also discussed.

The sections reflect the structure of the PhD school, and are organized as follows.
\begin{description}
    
    \item[Sec.~\ref{sec:LUCA}:] \textit{``\titleLuca''} by Luca Buoninfante.

    \item[Sec.~\ref{sec:ANNA}:] \textit{``\titleAnna{}''} by Anna Tokareva.

    \item[Sec.~\ref{sec:ALESSIABENJAMIN}:] \textit{``\titleAlessiaBenjamin''} by Benjamin Knorr and Alessia Platania.

    \item[Sec.~\ref{sec:IVANO}:] \textit{``\titleIvano''} by Ivano Basile.

    \item[Sec.~\ref{sec:FRANCESCO}:] \textit{``\titleFrancesco''} by Francesco Di Filippo.

    \item[Sec.~\ref{sec:FAQ}:] Several FAQ on aspects of \ac{QG} are answered, especially in relation to the topics that were covered in the lectures.

    \item[Sec.~\ref{sec:conclusions}:] Overall thoughts about the PhD school are jointly shared by all the lecturers and concluding remarks are drawn.

\end{description}
We hope that these lecture notes will become a useful reference on \ac{QG} for experts who might use them as a manual to refresh their minds on some specific topics when necessary, for lecturers who need a pedagogical and modern exposition of the subject to complement other textbooks, and for researchers who are less familiar with the basics or want to learn more about \ac{QG}. Having said that, it is now time to wish the reader a great journey into the \ac{QG} universe!

\clearpage

\section*{Conventions and notation}\label{conventions}

\paragraph{Units.} We work in \textit{natural units} (unless otherwise stated) that are defined by setting the reduced Planck constant and the speed of light equal to one:  
\begin{equation}
	\hbar=1=c\,.
\end{equation}
In this system of units the reduced Planck mass is related to Newton's constant by the formula
\begin{equation}
	\MPl{}\equiv \frac{1}{\sqrt{8\pi \GN{}}} \, .
\end{equation}
To avoid carrying factors of $8\pi$, we will find it useful to work in terms of \MPl{} instead of \GN{}.

\paragraph{Metric signature.} In these lectures we adopt the mostly plus convention for the metric signature. This means that the flat line element in Cartesian coordinates in a $d$-dimensional spacetime is given by
\begin{equation}
	{\rmd}s^2=- ({\rmd}x^0)^2 + ({\rmd}x^1)^2+ ({\rmd}x^2)^2+ ({\rmd}x^3)^2 + \dots \equiv \eta_{\mu\nu}{\rmd}x^\mu {\rmd}x^\nu\,,
\end{equation}
where the Minkowski metric reads
\begin{equation}
	(\eta_{\mu\nu})=\text{diag}(-1,+1,+1,+1,\dots)\,.
\end{equation}
In the mostly plus convention, timelike separations are negative definite and spacelike separations are positive definite.

\paragraph{Fourier transform.} The function $f(x)$ and its Fourier transform $\tilde{f}(p)$ are related by 
\begin{equation}
	f(x)=\int \frac{{\rmd}^dx}{(2\pi)^d} \tilde{f}(p) e^{i p\cdot x}\,,\qquad  \tilde{f}(p)=\int {\rmd}^dx f(x) e^{-ip\cdot x}\,,
	\label{fourier-transf-def}
\end{equation}
where $p\cdot x = p_\mu x^\mu=\eta_{\mu\nu} p^\mu x^\mu$.

Given the definitions in \eqref{fourier-transf-def}, the Fourier transform of the partial derivative is 
\begin{equation}
	\partial_\mu f(x) = \int \frac{{\rmd}^dp}{(2\pi)^d} (ip_\mu)\tilde{f}(p) e^{i p\cdot x}\qquad \Rightarrow\qquad \partial_\mu \to i p_\mu\,.
\end{equation}
This implies that the Fourier transform of the d'Alembertian in Minkowski spacetime, \ie{} $\Box=\eta^{\mu\nu}\partial_\mu\partial_\nu$, is given by $\Box\to -p^2$.

\paragraph{Curvature tensors.}  The Christoffel symbol is defined as
\begin{equation*}
	\Gamma^\rho_{\phantom{\rho}\mu\nu}=\frac{1}{2}g^{\rho\sigma}\left(\partial_{\mu}g_{\sigma\nu}+\partial_{\nu}g_{\mu\sigma}-\partial_{\sigma}g_{\mu\nu}\right)\,.
\end{equation*}
The covariant derivatives for contravariant and covariant vectors are defined as
\begin{equation*}
	\covD_\mu V^\nu = \partial_\mu V^\nu + \Gamma^{\nu}_{\phantom{\nu}\mu \rho} V^\rho\qquad \text{and}\qquad \covD_\mu V_\nu = \partial_\mu V_\nu - \Gamma^{\rho}_{\phantom{\rho}\mu \nu} V_\rho\,,
\end{equation*}
respectively, and with these formulas the generalization to tensors with a generic number of lower and upper indices can be easily obtained. Here we always assume that metric compatibility holds true, \ie{} $\covD_\rho g_{\mu\nu}=0$, and that torsion is zero. Therefore, we always work with the Levi-Civita connection, \ie{} the Christoffel symbol. 

The commutation relations for two covariant derivatives acting on contravariant and covariant vectors are 
\begin{equation}
	[\covD_\nu,\covD_\rho]V^\sigma=V^\mu R^\sigma_{\phantom{\sigma}\mu\nu\rho}\qquad \text{and}\qquad [\covD_\nu,\covD_\rho]V_\mu=-V_\sigma R^\sigma_{\phantom{\sigma}\mu\nu\rho}\,,
\end{equation}
respectively, where the Riemann tensor with one index up is defined as
\begin{equation*}
	R^{\sigma}_{\phantom{\sigma}\mu\nu\rho}=\partial_\nu\Gamma^\sigma_{\phantom{\sigma}\mu\rho}-\partial_\rho\Gamma^\sigma_{\phantom{\sigma}\mu\nu}+\Gamma^\sigma_{\phantom{\sigma}\alpha\nu}\Gamma^\alpha_{\phantom{\alpha}\mu\rho}-\Gamma^\sigma_{\phantom{\sigma}\alpha\rho}\Gamma_{\phantom{\alpha}\mu\nu}^\alpha\,.
\end{equation*}
Lowering the upper index with the metric tensor we obtain the
completely covariant Riemann tensor:
\begin{equation}
	R_{\mu\nu\rho\sigma}=\frac{1}{2}\left(\partial_{\nu}\partial_{\rho}g_{\mu\sigma}+\mathcal{\partial_{\mu}\partial_{\sigma}}g_{\nu\rho}-\mathcal{\partial_{\sigma}\partial_{\nu}}g_{\mu\rho}-\partial_{\mu}\partial_{\rho}g_{\nu\sigma}\right) +g_{\alpha\beta}\left(\Gamma_{\phantom{\alpha}\nu\rho}^{\alpha}\Gamma_{\phantom{\beta}\mu\sigma}^{\beta}-\Gamma_{\phantom{\alpha}\sigma\nu}^{\alpha}\Gamma_{\phantom{\beta}\mu\rho}^{\beta}\right).
\end{equation}
Finally, the Ricci tensor is defined by
\begin{equation}
	R_{\nu\sigma}=R^\rho_{\phantom{\rho}\nu\rho\sigma}=\delta^{\phantom{\mu}\rho}_\mu R^{\mu}_{\phantom{\mu}\nu\rho\sigma}=g^{\mu\rho}R_{\mu\nu\rho\sigma}
\end{equation}
and the Ricci scalar by
\begin{equation}
R=R^\nu_{\phantom{\nu}\nu}= g^{\nu\sigma}R_{\nu\sigma}\,.
\end{equation}

\clearpage


\section{\titleLuca}
\label{sec:LUCA}

\begin{tcolorbox}[colback=white,colframe=scipostblue]
{\bf Lecturer:} Luca Buoninfante, \briefaffiliationLuca

{\bf Email address:} \href{mailto:\emailLuca}{\emailLuca}
\tcblower
{\bf Lecture recordings:}
\begin{enumerate}[label= Lecture \arabic*:, leftmargin = 3.5cm, labelsep = 0.5cm, parsep = 0.0cm]
    \item \href{https://youtu.be/hVrlbQLAbck}{https://youtu.be/hVrlbQLAbck}
    \item \href{https://youtu.be/6aiJPDf-y4g}{https://youtu.be/6aiJPDf-y4g}
    \item \href{https://youtu.be/ySYiNvuiLsI}{https://youtu.be/ySYiNvuiLsI}
    \item \href{https://youtu.be/uY5vnB19BtQ}{https://youtu.be/uY5vnB19BtQ}
\end{enumerate}

{\bf Abstract:}

In this set of lectures we will challenge the framework of perturbative quantum field theory by applying it to the study of gravitational interaction. First, we will analyze quantum aspects of general relativity: we will identify on-shell and off-shell degrees of freedom, derive the graviton propagator and show the failure of perturbative renormalizability. In particular, we will determine the form of the propagator in covariant and non-covariant gauges, and provide a detailed study of one-, two- and higher-loop divergences. Second, we will demonstrate that by adding quadratic curvatures to the Lagrangian it is possible to achieve strict renormalizability. We will discuss several features of quadratic gravity, including uniqueness and predictivity. At the same time, we will highlight the open questions. These lectures are also intended for anyone interested in other approaches to quantum gravity, since a good understanding of perturbative quantum gravity is always a desirable starting point for doing something else.
\end{tcolorbox}

\subsection*{Preface}

The aim of this course is to study quantum aspects of gravity by applying the same tools and methods that we usually use for other fundamental interactions such as the electromagnetic, weak and strong ones. In other words, we want to formulate a \ac{QFT} of the gravitational interaction to describe phenomena in which both gravitational and quantum effects are relevant.

In these lectures what we really mean by the expression ``\ac{QFT}'' is ``\textit{perturbative} \ac{QFT}''. This means that we assume quantum field fluctuations to interact weakly and make an expansion in powers of the interaction couplings. You may be worried that this is not a satisfactory way to handle quantum aspects of gravity, but it is! Or, to put it more humbly, it is the best we can do to start analyzing quantum features of the gravitational interaction. In the same way that we quantize electromagnetic waves, we can ask whether a similar quantization prescription can be used to quantize gravitational waves in regimes where the interactions are weak. 

One of the successes of the perturbative \ac{QFT} framework when applied to the \ac{SM} is that it is very restrictive in terms of selecting physical theories. In fact, by assuming certain principles we can almost uniquely fix the kinetic and interaction terms in a Lagrangian. This feature makes the \ac{QFT} framework very predictive. At the same time, these Lagrangians are the same ones that are inserted into a path integral to perform non-perturbative analyses, such as studies of instanton configurations. In other words, the perturbative \ac{QFT} framework also provides a good starting point for non-perturbative studies that may be needed in regimes where the perturbative approach may fail.

For these reasons I strongly believe that a good understanding of perturbative \ac{QG} (\ie{} gravitational interaction quantized in the framework of perturbative \ac{QFT}) is fundamental to deal with quantum-gravitational physics. One of the important messages of this course will be that the expression ``perturbative \ac{QG}'' does \textit{not} just correspond to quantum \ac{GR}, but it refers to any possible consistent perturbative \ac{QFT} of gravitational interaction. In particular, we will show that in four spacetime dimensions there exists a unique gravitational Lagrangian that is compatible with the symmetries (\ie{} invariance under diffeomorphisms and parity) and geometric structure (\ie{} metric compatibility and zero torsion) of \ac{GR}, and that at the same time extends the Einstein-Hilbert Lagrangian with additional quadratic-curvature terms, giving rise to a strictly renormalizable \ac{QFT} of gravity.

The lecture notes are organized as follows.
\begin{description}

\item[Sec.~\ref{sec:reminders}:] We introduce elements of classical \ac{GR} by working in the Lagrangian formalism. 

\item[Sec.~\ref{sec:quant-I}:] We start analyzing quantum aspects of \ac{GR} in the framework of perturbative \ac{QFT}. We consider metric fluctuations around the Minkowski background and focus on the free theory (with no self-interactions). We determine the physical degrees of freedom, derive the graviton propagator in different ways, and discuss the canonical quantization. We will work in both cases of covariant and non-covariant gauges.

\item[Sec.~\ref{sec:quant-II}:] We introduce self-interactions for the graviton field, explain the need to introduce Faddeev-Popov fields, and discuss unitarity. Furthermore, we make a detailed analysis of one-loop, two-loop and higher-loop divergences without going into complicated technicalities. In particular, we show the failure of perturbative renormalizability in \ac{GR}.

\item[Sec.~\ref{sec:lecture4}:] We introduce operators of mass dimension equal to four in the action and show that the resulting gravitational theory --- known as quadratic gravity --- is strictly renormalizable in four spacetime dimensions. 
We discuss various features of quadratic gravity such as degrees of freedom, propagator, power counting renormalizability, and make a comparison with the \ac{EFT} of \ac{GR}. We explain the success of this gravitational \ac{QFT} in terms of uniqueness and predictivity and, at the same time, highlight the open questions.

\item[Sec.~\ref{sec:conclus-luca}:] We draw conclusions and share future perspectives for perturbative \ac{QG} and beyond.

\item[App.~\ref{sec:elemQFT}:] We present a concise review of the fundamental fundamental principles on which the standard perturbative \ac{QFT} framework is based, in particular the notions of locality, unitarity, and perturbative renormalizability.

\item[App.~\ref{app:spin-proj}:] We provide additional details about the spin-projector formalism that will be used for the computation of the graviton propagator in both \ac{GR} and quadratic gravity.

\end{description}

I will not follow a single reference such as a review article or a book, but I will use various sources scattered throughout the literature combined with a more personal (sometimes emotional!) way of presenting the topic. However, review articles, lecture notes and textbooks that I found particularly well-written and useful, and from which I have learned a lot about perturbative \ac{QG}, are
\begin{itemize}
    
    \item M.~J.~G.~Veltman, \emph{\href{https://ncatlab.org/nlab/files/VeltmanQuantumnGravity76.pdf}{Quantum Theory of Gravitation}}, Conf. Proc. C \textbf{7507281} (1975)~\cite{Veltman:1975vx} 
	
    \item G.~'t Hooft, \emph{\href{https://webspace.science.uu.nl/~hooft101/lectures/erice02.pdf}{Perturbative Quantum Gravity}}, World Scientific (2003)~\cite{tHooft:2002umw}
	
    \item R.~Percacci, \emph{\href{http://www.percacci.it/roberto/physics/book2/index.html}{An Introduction to Covariant Quantum Gravity and Asymptotic Safety}}, World Scientific (2017)~\cite{Percacci:2017fkn}
	
    \item J.~F.~Donoghue, M.~M.~Ivanov and A~Shkerin, \emph{EPFL Lectures on General Relativity as a Quantum Field Theory}, \href{https://arxiv.org/abs/1702.00319}{arXiv:1702.00319 [hep-th]}~\cite{Donoghue:2017pgk}

    \item I.~L.~Buchbinder and I.~Shapiro, \emph{Introduction to Quantum Field Theory with Applications to Quantum Gravity}, Oxford University Press (2023)~\cite{Buchbinder:2021wzv}
    
\end{itemize}
%


\subsection{Elements of GR}\label{sec:reminders}

\subsubsection{Action and field equations}

\ac{GR} has provided a fantastic description of classical aspects of the gravitational interaction over a wide range of length scales. In fact, by introducing minimal couplings between gravity and matter and assuming the existence of a small cosmological constant, \ac{GR} predictions have been tested from length scales of order $10^{-5}$m (through torsion balance tests of Newton's law~\cite{Lee:2020zjt}) to distances of order $10^{26}$m (through late-time cosmological observations~\cite{SupernovaSearchTeam:1998fmf}).

The starting point for the Lagrangian formulation of \ac{GR} is the Einstein-Hilbert action
\begin{equation}
	S_{\rm EH}[g]=\frac{1}{2\kappa^2}\int {\rmd}^4x \sqrt{-g} \, \left(R-2\CC\right)\,,\label{action-EH-Luca}
\end{equation}
where $\kappa^2\equiv 8\pi \GN$, \GN{} being Newton's constant, while \CC{} is the cosmological constant whose measured value is of order $10^{-52}\text{m}^{-2}$~\cite{SupernovaSearchTeam:1998fmf,Planck:2018vyg}. The coupling between gravity and matter can be described by introducing the matter action $S_m$ which a functional of the metric and any type of matter field such as scalars, fermions, and gauge bosons.

Using the relations
\begin{align}
	\delta(\sqrt{-g})&=-\frac{1}{2}\sqrt{-g} \,g_{\mu\nu} \,\delta{g}^{\mu\nu}\,, \\
	\delta R_{\mu\nu}&=\frac{1}{2}g^{\sigma \rho}\left[\covD_\sigma \covD_\mu \delta g_{\rho \nu}+ \covD_\sigma \covD_\nu \delta g_{\mu \rho}-\covD_\sigma \covD_\rho \delta g_{\mu \nu}-\covD_\nu \covD_\mu \delta g_{\rho \sigma} \right]\,,
\end{align}
we can vary the total action with respect to $g^{\mu\nu}$ and obtain the Einstein's field equations\footnote{Rigorously speaking, to have a well-defined variational problem we have to add the well-known Gibbons-Hawking-York boundary term to cancel total derivative terms containing covariant derivatives of metric variations, \ie{} $\covD_{\mu}\delta g_{\nu\rho}$, which do not vanish on the boundary. For simplicity, we do not explicitly consider this term in the action.}
\begin{align}
	0=\delta \left(S_{\rm EH}+S_m\right)&=\int {\rmd}^4x\sqrt{-g}\left[\frac{1}{2\kappa^2}\left(R_{\mu\nu}-\frac{1}{2}g_{\mu\nu}R+\CC{} g_{\mu\nu}\right)-\frac{1}{2}T_{\mu\nu}\right]\delta g^{\mu\nu}\nonumber\\
	\Rightarrow  G_{\mu\nu}+\CC{} g_{\mu\nu}&=\kappa^2 T_{\mu\nu}\,, 
\end{align}
where we have introduced the Einstein tensor, $G_{\mu\nu}=R_{\mu\nu}-\frac{1}{2}g_{\mu\nu}R,$ and the stress-energy tensor 
\begin{equation}
	T_{\mu\nu}=\frac{-2}{\sqrt{-g}}\frac{\delta S_m}{\delta g^{\mu\nu}}\,,
	\label{def-stess}
\end{equation}
which is conserved, $\covD_\mu T^{\mu\nu}=0$, consistently with the Bianchi identity,  $\covD_\mu G^{\mu\nu}=0$.

\subsubsection{Diffeomorphism invariance}

The total action $S_{\rm EH}+S_m$ is invariant under \textit{diffeomorphisms}, namely under the transformation
\begin{equation}
	x^{\mu}\to x^{\prime\mu} (x) \label{diffeo}
\end{equation}
such that both $x^{\mu}(x')$ and $x^{\prime\mu}(x)$ are invertible smooth functions. 

An infinitesimal diffeomorphism reads
\begin{equation}
	x^{\mu}\to x^{\prime\mu} (x)=x^\mu + \zeta^\mu(x)\,,\label{inf-diffeo}
\end{equation}
where $\zeta^\mu(x)$ is an infinitesimal vector that depends on the spacetime point $x^{\mu}$. We can easily find how the spacetime metric transforms under an infinitesimal diffeomorphism by recalling that $g_{\mu\nu}(x)$  is a $(0,2)$ tensor that under~\eqref{diffeo} transforms as
\begin{equation}\label{metric-diff}
\begin{aligned}
	g'_{\mu\nu}(x')&=\frac{\partial x^\rho}{\partial x^{\prime\mu}}\frac{\partial x^\sigma}{\partial x^{\prime\nu}}g_{\rho\sigma}(x)\\
	&= \left(\delta_\mu^{\phantom{\mu}\rho}-\frac{\partial \zeta^\rho}{\partial x'^\mu} \right)\left(\delta_\nu^{\phantom{\nu}\sigma}-\frac{\partial \zeta^\sigma}{\partial x'^\nu} \right) g_{\rho\sigma}(x) \\
	&= g_{\mu\nu}(x)-g_{\mu\rho}(x)\partial_\nu\zeta^\rho(x)-g_{\nu\rho}(x)\partial_\mu\zeta^\rho(x)+\orderneglected(\zeta^2)\,,
\end{aligned}
\end{equation}
where we used the fact that $\frac{\partial}{\partial x^{\prime\mu}}=\frac{\partial}{\partial x^\mu}+\orderneglected(\zeta)$ and introduced the notation $\partial_\mu\equiv \frac{\partial}{\partial x^\mu}$.

On the other hand, if we Taylor expand the metric, we get
\begin{equation}\label{taylor}
	g'_{\mu\nu}(x')=g'_{\mu\nu}(x)+\partial_\rho g_{\mu\nu}(x)\zeta^\rho+\orderneglected(\zeta^2)\,,
\end{equation}
where we used the fact that $\partial_\rho g'_{\mu\nu}(x)\zeta^\rho=\partial_\rho g_{\mu\nu}(x')\zeta^\rho+\orderneglected(\zeta^2)=\partial_\rho g_{\mu\nu}(x)\zeta^\rho+\orderneglected(\zeta^2)$.

Combining \eqref{metric-diff} and~\eqref{taylor} we can obtain the metric variation defined at the same spacetime coordinate $x$, which tells us how the metric field changes under an active infinitesimal diffeomorphism:\footnote{It is an active diffeomorphism because the metrics $g_{\mu\nu}^\prime(x)$ and $g_{\mu\nu}(x)$ are evaluated at two different spacetime points $P^\prime$ and $P$ that are described by the same value of the coordinate $x$ in their respective coordinate systems.}
\begin{equation}\label{metric-diff-2}
\begin{aligned}
	\delta_\zeta g_{\mu\nu}(x)&\equiv 	g'_{\mu\nu}(x)-g_{\mu\nu}(x) \\
	&= -\zeta^\rho(x)\partial_\rho  g_{\mu\nu}(x)-g_{\mu\rho}(x)\partial_\nu\zeta^\rho(x)-g_{\nu\rho}(x)\partial_\mu\zeta^\rho(x)\\
	&=-\covD_\mu \zeta_\nu(x) -\covD_\nu \zeta_\mu(x)\,.
\end{aligned}
\end{equation}

Using the last equation we can derive the following Noether identity:
\begin{equation}
	0=\delta_\zeta S_{\rm EH}=\frac{1}{2\kappa^2}\int {\rmd}^4x \frac{\delta\left(\sqrt{-g}(R-2\CC{})\right)}{\delta g_{\mu\nu}}\delta_\zeta g_{\mu\nu}=-\frac{1}{\kappa^2} \int {\rmd}^4x \sqrt{-g}\left(\covD_\mu G^{\mu\nu}\right)\zeta_{\nu} \,,
\end{equation}
which must be true for any arbitrary $\zeta_\nu$, thus we get the Bianchi identity for the Einstein tensor, $\covD_\mu G^{\mu\nu}=0$, as a consequence of diffeomorphism invariance of the action. If we demand the invariance of the matter action under diffeomorphism, we consistently obtain that the stress-energy tensor is covariantly conserved, \ie{} $\covD_\mu T^{\mu\nu}=0$.

\subsubsection{Degrees of freedom}\label{sec:dof-gr}

The spacetime metric $g_{\mu\nu}(x)$ is a rank-two tensor, thus in four spacetime dimensions it has $16$ components. We now want to determine the number of physically independent components.

First of all, since the metric tensor is symmetric in its two indices, we go from $16$ to $10$ components. Then, we can use diffeomorphism invariance to further reduce this number. Indeed, the invariance of the action under~\eqref{diffeo} or~\eqref{metric-diff-2} tells us that we can gauge away four metric components by making some suitable choice of the arbitrary vector $\zeta_\mu$ with $\mu=0,1,2,3$.  This allows us to kill four unphysical degrees of freedom off-shell, \ie{} without using the field equations: $10-4=6$.

Since gauge invariance hits twice, we should be able to kill four additional unphysical metric components on-shell, \ie{} using the field equations. In fact, some components of Einstein's equations are not dynamical because they do not contain second-order time derivatives. This can be understood by analyzing Bianchi's identity more closely:
\begin{equation}
	0=\covD_\mu G^{\mu\nu}= \partial_0 G^{0 \nu}+\partial_i G^{i \nu}+\Gamma_{\phantom{\mu}\mu\rho}^\mu G^{\rho\nu}+\Gamma_{\phantom{\nu}\mu\rho}^\nu G^{\mu\rho}\,.
\end{equation}
The right-hand side can vanish only if $G^{0\nu}$ is of first order in time derivatives. This implies that the $(\mu=0,\nu)$ components of the Einstein equations, \ie{} $G^{0\nu}=\kappa^2 T^{0\nu}$, are not dynamical, but they are four constraints on the metric. Although this is not a rigorous proof, it suggests that in the end we get $6-4=2$ independent metric components, \ie{} two physical dynamical degrees of freedom.

In the next section, we will show that in the language of \ac{QFT} these physical degrees of freedom correspond to the $\pm 2$ helicities of the graviton.

\subsubsection{Metric fluctuations and action expansion}\label{sec:metric-fluct}

To formulate \ac{GR} as a \ac{QFT}, the first thing to do is to identify the classical field fluctuation that needs to be quantized and promoted to an operator. We separate the spacetime metric into two parts:
\begin{equation}
	g_{\mu\nu}(x)=\bar{g}_{\mu\nu}(x)+2\kappa h_{\mu\nu}(x)\,,
	\label{metric-fluct}
\end{equation}
where $\bar{g}_{\mu\nu}(x)$ is treated as a \textit{background} that in general can be position-dependent, whereas $h_{\mu\nu}(x)$ is a metric \textit{fluctuation} such that $\kappa |h_{\mu\nu}|\ll 1$ in some coordinate system. The latter is the field fluctuation which will then be quantized, and whose excitations give rise to quantum states populated by particles called \textit{gravitons}. The constant factor $2\kappa$ has been chosen to have a canonically normalized field as we show below. By convention, all the quantities computed in terms of the background metric are indicated with a ``bar'', \eg{} the covariant derivative $\bar{\covD}_\mu$, the Riemann tensor $\bar{R}_{\mu\nu\rho\sigma}$ and all its contractions.  Consistency of the approach requires that the indices of $\bar{\covD}_\mu$, $\bar{R}_{\mu\nu\rho\sigma}$, $h_{\mu\nu}$ and all other fields (except $g_{\mu\nu}$) are raised and lowered with $\bar{g}_{\mu\nu}$, \eg{} the trace of the graviton field is $h\equiv \bar{g}^{\mu\nu}h_{\mu\nu}$. By contrast, the indices of $g_{\mu\nu}$, the full covariant derivative $\covD_\mu$ and the full Riemann tensor $R_{\mu\nu\rho\sigma}$ are raised and lowered with the full metric $g_{\mu\nu}$.

Our aim is to expand the Einstein-Hilbert action in powers of the fluctuations so that we can identify kinetic and interaction terms, \ie{}
\begin{equation}
	S_{\rm EH}[\bar{g}+2\kappa h]=S^{(0)}_{\rm EH}[\bar{g}]+S^{(1)}_{\rm EH}[\bar{g},h]+S^{(2)}_{\rm EH}[\bar{g},h]+ \dots +S^{(n)}_{\rm EH}[\bar{g},h]+\dots\,,
\end{equation}
where
%
\begin{align}
	S_{\rm EH}^{(0)}[\bar{g}]&= \frac{1}{2\kappa^2} \int {\rmd}^4x \sqrt{-\bar{g}} \, (\bar{R}-2\CC{}) \,,\nonumber\\
	S_{\rm EH}^{(1)}[\bar{g},h]&= \frac{1}{1!} 2\kappa \int {\rmd}^4y \left. \frac{\delta S_{\rm EH}}{\delta g_{\mu\nu}(y)}\right|_{g=\bar{g}}h_{\mu\nu}(y) \,,\nonumber\\
	S^{(2)}_{\rm EH}[\bar{g},h]&=  \frac{1}{2!}(2\kappa)^2 \int {\rmd}^4y_1 {\rmd}^4y_2 \left. \frac{\delta S_{\rm EH}}{\delta g_{\mu_1\nu_1}(y_1)\delta g_{\mu_2\nu_2}(y_2)}\right|_{g=\bar{g}}h_{\mu_1\nu_1}(y_1)h_{\mu_2\nu_2}(y_2)  \,,\nonumber\\
	&\vdots\nonumber\\
	S^{(n)}_{\rm EH}[\bar{g},h]&=  \frac{1}{n!}(2\kappa)^n \int {\rmd}^4y_1\cdots {\rmd}^4y_n \left. \frac{\delta S_{\rm EH}}{\delta g_{\mu_1\nu_1}(y_1)\cdots\delta g_{\mu_n\nu_n}(y_n)}\right|_{g=\bar{g}} \!\! h_{\mu_1\nu_1}(y_1)\cdots h_{\mu_n\nu_n}(y_n) \,, \nonumber\\
	&\vdots\,. 
\end{align}
%
The zeroth order term $S_{\rm EH}^{(0)}$ is a constant with respect to $h_{\mu\nu}$, therefore does not contribute to any dynamics involving graviton fluctuations and can be neglected. 
The first order term $S_{\rm EH}^{(1)}$ is proportional to the field equations evaluated on the background $g_{\mu\nu}=\bar{g}_{\mu\nu}$, which we assume to be a solution of the Einstein equations, therefore it gives a vanishing contribution:
\begin{equation}
\begin{aligned}
	S_{\rm EH}^{(1)}[\bar{g},h]&=\frac{1}{\kappa}\int {\rmd}^4y\left.\left[\frac{\delta}{\delta g_{\mu\nu}(y)}\int{\rmd}^4x \sqrt{-g(x)}\big(R(x)-2\CC{}\big) \right]\right|_{g=\bar{g}}h_{\mu\nu}(y) \\
	&= -\frac{1}{\kappa}\int {\rmd}^4y\sqrt{-g(y)}\left[\bar{R}^{\mu\nu}(y)-\frac{1}{2}\bar{g}^{\mu\nu}(y)\bar{R}(y)+\bar{g}^{\mu\nu}(y)\CC{}\right]h_{\mu\nu}(y) \\
	&=0\,.
\end{aligned}
\end{equation}
The higher-order terms are the relevant ones. $S^{(2)}_{\rm EH}$ is quadratic in $h_{\mu\nu}$ and corresponds to the kinetic part of the action from which we can derive the propagator, while $S^{(n)}_{\rm EH}$ with $n\geq 3$ are the interaction terms from which we can derive $n$-point vertices, \ie{} cubic, quartic, and so on.

To calculate $S^{(2)}_{\rm EH}$ it is simpler to directly consider the second order variation of the action, ignoring the fact that we have four-dimensional Dirac deltas to take into account when taking functional derivatives. Thus, we have
\begin{equation}\label{S-quad-expans}
\begin{aligned}
	S^{(2)}_{\rm EH}[\bar{g},h]&=\frac{1}{2!}\delta^{(2)}S_{\rm EH}[\bar{g},h] \\
	&= \frac{1}{4\kappa^2}\int {\rmd}^4x \left[\delta \left(\delta\left(\sqrt{-g}\right)R+\sqrt{-g}\delta R \right)\right]_{g=\bar{g}} \\
	&= \frac{1}{4\kappa^2} \int {\rmd}^4x \left[\delta^{(2)}\left(\sqrt{-g}\right)R+2\delta\left(\sqrt{-g}\right)\delta R+\sqrt{-g}\delta^{(2)} R \right]_{g=\bar{g}}\,.
\end{aligned}
\end{equation}
Using the formulas for the expansions of the inverse metric and the Christoffel symbol
\begin{align}
g^{\mu\nu}&=\bar{g}^{\mu\nu}-2\kappa h^{\mu\nu}+4\kappa^2 h^{\mu}_{\phantom{\mu}\rho} h^{\nu\rho}+\dots\,, \\
\Gamma^\mu_{\phantom{\mu}\rho\sigma}&= \bar{\Gamma}^\mu_{\phantom{\mu}\rho\sigma}+\kappa g^{\mu\nu}\left(\bar{\covD}_\rho h_{\nu\sigma}+\bar{\covD}_\sigma h_{\nu\rho}-\bar{\covD}_\nu h_{\rho\sigma}\right)\,,
\end{align}
we can derive the expansion of the curvature tensors, in particular that for the Ricci scalar (see also refs.~\cite{Percacci:2017fkn,Buchbinder:2021wzv}):
\begin{equation}
\begin{aligned}
	R&=\bar{R}+\delta R+\frac{1}{2}\delta^{(2)}R+\dots\,, \\
	\delta R&= 2\kappa\left(\bar{\covD}_\mu \bar{\covD}_{\nu} h^{\mu\nu}-\bar{\covD}^2h-\bar{R}^{\mu\nu}h_{\mu\nu}\right) \,, \\
	\delta^{(2)} R&= 4\kappa^2\left( \frac{3}{2}\bar{\covD}_\rho h_{\mu\nu}\bar{\covD}^\rho h^{\mu\nu}+2h_{\mu\nu}\bar{\covD}^2h^{\mu\nu}- 2\bar{\covD}_\rho h^\rho_{\phantom{\rho}\mu}\bar{\covD}_\sigma h^{\sigma\mu} \right. \\
	&\qquad\,\,\, +2\bar{\covD}_\rho h^\rho_{\phantom{\rho}\mu} \bar{\covD}^\mu h-4h_{\mu\nu}\bar{\covD}^\mu\bar{\covD}_\rho h^{\rho\nu}+2h_{\mu\nu}\bar{\covD}^\mu\bar{\covD}^\nu h \\
	&\qquad\,\,\, \left.-\bar{\covD}_\mu h_{\nu\rho}\bar{\covD}^\rho h^{\mu\nu}-\frac{1}{2}\bar{\covD}_\mu h \bar{\covD}^\mu h+2\bar{R}_{\mu\nu\rho\sigma}h^{\mu\rho}h^{\nu\sigma}\right)\,.
\end{aligned}
\end{equation}
Moreover, the expansion of the metric determinant up to second order is given by
\begin{equation}
\begin{aligned}
	&\qquad\sqrt{-g}=\sqrt{-\bar{g}}+\delta\left(\sqrt{-g}\right)+\frac{1}{2}\delta^{(2)}\left(\sqrt{-g}\right)+\dots \\
	&\delta\left(\sqrt{-g}\right)=\sqrt{-\bar{g}} \, \kappa h\,,\quad \delta^{(2)}\left(\sqrt{-g}\right)=\sqrt{-\bar{g}} \, \kappa^2\left(h^2-2h_{\mu\nu}h^{\mu\nu}\right)\,.
\end{aligned}
\end{equation}

We now have all the ingredients to explicitly evaluate \eqref{S-quad-expans}. Indeed, substituting the above expansions for the Ricci scalar and the metric determinant into \eqref{S-quad-expans}, integrating by parts and using commutation relations for the covariant derivatives, we get the following expression for the second-order contribution to the action:
\begin{equation}\label{quadratic-act-generic}
\begin{aligned}
	S^{(2)}_{\rm EH}[\bar{g},h] &= \int {\rmd}^4x\sqrt{-\bar{g}}\left[ -\frac{1}{2}\bar{\covD}_\rho h_{\mu\nu}\bar{\covD}^\rho h^{\mu\nu} +\bar{\covD}_\rho h^\rho_{\phantom{\rho}\mu} \bar{\covD}_\sigma h^{\sigma\mu}-\bar{\covD}_\mu h\bar{\covD}_\nu h^{\mu\nu}+\frac{1}{2}\bar{\covD}_\rho h\bar{\covD}^\rho h \right. \\
	&\hspace{1cm} \left.-\frac{1}{2}\left(\bar{R}-2\CC{}\right)\left(h_{\mu\nu}h^{\mu\nu}-\frac{1}{2}h^2\right)+\left(h^{\mu\rho}h_\rho^{\phantom{\rho}\nu}-hh^{\mu\nu}\right)\bar{R}_{\mu\nu}+h^{\mu\rho}h^{\nu\sigma}\bar{R}_{\mu\nu\rho\sigma}\right]\,.
\end{aligned}
\end{equation}
From the last equation we can now clearly understand the reason why we inserted a factor of $2\kappa$ in the metric perturbation~\eqref{metric-fluct}, so that the kinetic term is in canonical form, \ie{} the field $h_{\mu\nu}$ is canonically normalized.

The interaction terms $S_{\rm EH}^{(n)}[\bar{g},h]$ with $n\geq 3$ can be derived in a similar manner, by considering the higher-order expansions of the inverse metric, determinant and curvature tensors. The cubic order contribution is already too complicated, and its explicit form is not needed for the purpose of these lectures. What we must observe is that an $n$-th order interaction term has the following dependence on $\kappa$ and $h_{\mu\nu}$:
\begin{equation}\label{interaction-terms}
	S^{(n)}[\bar{g},h]\sim \orderneglected\left(\kappa^{n-2}h^n\right)\,,\qquad n\geq 3\,,
\end{equation}
namely the coupling of an $n$th-order interaction term is $\kappa^{n-2}=1/\MPl^{n-2}$.

\subsection{GR as a QFT: free theory}\label{sec:quant-I}

The main goal of this section is to formulate a perturbative \ac{QFT} of the gravitational interaction, \ie{} to quantize the metric fluctuation $h_{\mu\nu}$ in the framework of perturbative \ac{QFT}, compatible with the standard principles of locality, symmetries, unitarity and strict renormalizability. For readers unfamiliar with these concepts, especially that of \textit{strict} renormalizability, we recommend reading \cref{sec:elemQFT} before starting to study the quantization of \ac{GR}.

In this and subsequent sections we only focus on metric fluctuations around the Minkowski background, and analyze the free theory (no self-interaction) up to possible linear couplings to matter. We will discuss graviton polarization and helicity, determine the off-shell and on-shell degrees of freedom, and derive the propagator. It will be instructive to perform the analysis in both covariant and non-covariant gauges. In particular, we will compute the propagator in the Feynman gauge using the covariant de Donder gauge fixing, and in the Prentki gauge using a non-covariant gauge fixing. Furthermore, we will derive the graviton propagator for a generic de Donder gauge fixing using the spin-projector formalism, which will allow us to identify both off-shell and on-shell degrees of freedom.

\subsubsection{Linearization around Minkowski spacetime}

We set $\CC{}=0$ and $\bar{g}_{\mu\nu}=\eta_{\mu\nu}$.\footnote{From a physical point of view, we are assuming that we are in a region of spacetime where the cosmological constant is negligible and the background metric can be approximated by Minkowski.} This also means that the background covariant derivative becomes the ordinary partial derivative\footnote{To be more precise, this is true in Cartesian coordinates that are the ones we use here.}, $\bar{\covD}_\mu=\partial_\mu, $ and all the curvature tensors vanish when evaluated on the background, \ie{} $\bar{R}_{\mu\nu\rho\sigma}=R_{\mu\nu\rho\sigma}(\eta)=0$. Now the trace reads $h=\eta^{\mu\nu}h_{\mu\nu}$ and we simply use the box symbol for the flat d'Alembertian, $\Box= \eta^{\mu\nu}\partial_{\mu}\partial_\nu$.

\paragraph{Kinetic action.} The quadratic action~\eqref{quadratic-act-generic} around the Minkowski background reduces to
\begin{equation}\label{quadratic-act-mink}
	S^{(2)}_{\rm EH}[\eta,h]= \int {\rmd}^4x\left[ -\frac{1}{2}\partial_\rho h_{\mu\nu}\partial^\rho h^{\mu\nu} +\partial_\rho h^\rho_{\phantom{\rho}\mu} \partial_\sigma h^{\sigma\mu}-\partial_\mu h\partial_\nu h^{\mu\nu}+\frac{1}{2}\partial_\rho h\partial^\rho h \right] \,.
\end{equation}

It is convenient to recast the action~\eqref{quadratic-act-mink} into an equivalent form up to total derivatives. Integrating by parts and symmetrizing, we can write
\begin{equation}\label{quadratic-act-mink-2}
	S^{(2)}_{\rm EH}[\eta,h]= \int {\rmd}^4x \frac{1}{2}h_{\mu\nu}\mathbb{K}^{\mu\nu\rho\sigma}h_{\rho\sigma}\,,
\end{equation}
where the kinetic operator is defined as
\begin{tcolorbox}
\begin{equation}\label{kinetic-oper}
\begin{aligned}
	\mathbb{K}^{\mu\nu\rho\sigma}\equiv &\frac{1}{2}\left(\eta^{\mu\rho}\eta^{\nu\sigma}+\eta^{\mu\sigma}\eta^{\nu\rho} \right)\Box-\eta^{\mu\nu}\eta^{\rho\sigma}\Box +\eta^{\mu\nu}\partial^{\rho}\partial^\sigma +\eta^{\rho\sigma}\partial^{\mu}\partial^\nu \\
	& -\frac{1}{2}\left(\eta^{\mu\rho}\partial^{\nu}\partial^\sigma+\eta^{\mu\sigma}\partial^{\nu}\partial^\rho +\eta^{\nu\rho}\partial^{\mu}\partial^\sigma+\eta^{\nu\sigma}\partial^{\mu}\partial^\rho \right)\,,
\end{aligned}
\end{equation}
\end{tcolorbox}
\noindent and satisfies the following symmetry properties:
\begin{equation}\label{kinetic-op-GR-symm}
\mathbb{K}^{\mu\nu\rho\sigma}=\mathbb{K}^{\nu\mu\rho\sigma}=\mathbb{K}^{\mu\nu\sigma\rho}=\mathbb{K}^{\rho\sigma\mu\nu}\,.
\end{equation}

\paragraph{Matter coupling.} We can also add a matter contribution $S_m$ to the action and expand in metric fluctuations up to linear order in $h_{\mu\nu}$:
\begin{equation}
\begin{aligned}
	S_m[\eta+2\kappa h]&=S_m[\eta]+ 2\kappa \int {\rmd}^4x \, \frac{\delta S_m}{\delta g_{\mu\nu}} h_{\mu\nu}+\orderneglected(\kappa^2 h^2) \\
	&=S_m[\eta]+\kappa \int{\rmd}^4x \, T^{\mu\nu}h_{\mu\nu}+\orderneglected(\kappa^2 h^2)\,,
\end{aligned}
\end{equation}
where we have used the definition in~\eqref{def-stess} for the stress-energy tensor.

\paragraph{Linearized diffeomorphisms.}

The metric transformation under diffeomorphism in \eqref{metric-diff-2} can be written in terms of $\kappa h_{\mu\nu}$ as
\begin{equation}\label{diff-h}
\begin{aligned}
\delta_\zeta h_{\mu\nu}&= \covD_\mu \zeta_\nu + \covD_\nu \zeta_\mu \\
&= \partial_\mu \zeta_\nu +\partial_\nu \zeta_\mu + 2\kappa \left( h_{\mu\rho}\partial_\nu \zeta^\rho+h_{\nu\rho}\partial_\mu \zeta^\rho+\zeta^\rho\partial_\rho h_{\mu\nu} \right)\,,
\end{aligned}
\end{equation}
where we have replaced $\zeta_\mu\to -2\kappa \zeta_\mu$ so that $\zeta_\nu$ has mass dimension zero, consistent with a canonically normalized field $h_{\mu\nu}$ of mass dimension one.

It is easy to show that the action $S^{(2)}_{\rm EH}[\eta,h]$ is invariant under the zeroth order of the field transformation~\eqref{diff-h}, \ie{}
\begin{equation}\label{gauge-transf}
\delta_\zeta h_{\mu\nu}= \partial_\mu \zeta_\nu+  \partial_\nu \zeta_\mu\qquad \Rightarrow\qquad \delta_\zeta S^{(2)}_{\rm EH}[\eta,h]=0\,.
\end{equation}
This means that there is a gauge redundancy in the theory: most of the components of the symmetric tensor $h_{\mu\nu}$ are unphysical, as we will explain in more detail below.

Furthermore, the invariance of the matter action at the linear level requires that the stress-energy tensor satisfies the conservation law $\partial_\mu T^{\mu\nu}=0$.

\paragraph{Linearized field equations.} The linearized field equations are given by
\begin{equation}\label{EOM-mink}
\begin{aligned}
&\mathbb{K}_{\mu\nu}^{\phantom{\mu\nu} \rho\sigma}h_{\rho\sigma}=-\kappa T_{\mu\nu} \\
\Leftrightarrow \qquad &\Box h_{\mu\nu}-\frac{1}{2}\eta_{\mu\nu}\Box h + \eta_{\mu\nu}\partial_\rho \left(\partial_\sigma h^{\rho\sigma}-\frac{1}{2}\partial^\rho h \right) \\
\qquad &-\partial_\mu \left( \partial_\rho h^\rho_{\phantom{\rho}\nu}-\frac{1}{2}\partial_\nu h\right)-\partial_\nu \left( \partial_\rho h^\rho_{\phantom{\rho}\mu}-\frac{1}{2}\partial_\mu h\right)=-\kappa T_{\mu\nu}\, ,
\end{aligned}
\end{equation}
and the trace reads
\begin{equation}\label{EOM-mink-trace}
-2\Box h+2\partial_\rho\partial_\sigma h^{\rho\sigma}=-\kappa T\,,
\end{equation}
where $T=\eta^{\mu\nu}T_{\mu\nu}$. The solution to~\eqref{EOM-mink} is not uniquely determined because if $h_{\mu\nu}$ is a solution, then $h_{\mu\nu}+\partial_\mu\zeta_\nu+\partial_\nu\zeta_\mu$ will be a solution as well. In fact, we need to impose a gauge condition to eliminate this redundancy. In what follows we determine the independent physical solutions of the field equations by performing the analysis in two different equivalent ways: first, we impose the covariant de Donder gauge condition; second, we impose the non-covariant radiation (or Coulomb) gauge.

\subsubsection{Graviton polarizations: covariant gauge}

A convenient covariant gauge is the de Donder one and is defined by the condition
\begin{equation}\label{de-donder}
\partial_\rho h^\rho_{\phantom{\rho}\nu} -\frac{1}{2}\partial_\nu h=0\,,
\end{equation}
which corresponds to the linearized version of the harmonic gauge $g_{\mu\nu}g^{\rho\sigma} \Gamma^\mu_{\phantom{\mu}\rho\sigma}=0$.

If we impose~\eqref{de-donder}, the linearized field equations simplify enormously,
\begin{equation}\label{EOM-mink-dedonder}
\Box h_{\mu\nu}-\frac{1}{2}\eta_{\mu\nu}\Box h =-\kappa T_{\mu\nu}\,.
\end{equation}

Let us now work in vacuum, \ie{} in the spacetime region where $T_{\mu\nu}(x)=0$, and solve the linearized field equations there. The vacuum trace equation is $\Box h=0$, which gives the following wave equation:
\begin{equation}\label{wave-eq-h}
\Box h_{\mu\nu}=0\,.
\end{equation}
The solution is given by
\begin{equation}\label{wave-eq-h-solution}
h_{\mu\nu}(x)=\polarizationtensor_{\mu\nu}(p) e^{ip\cdot x}+\polarizationtensor_{\mu\nu}^\ast(p) e^{-ip\cdot x}\,,\qquad p^2=-p_0^2+\vec{p}^{\,2}=0\,,
\end{equation}
where $\polarizationtensor_{\mu\nu}$ is called polarization tensor, and it can in general depend on the momentum, and the condition $p^2=0$ means that the field $h_{\mu\nu}$ will be associated with massless particles when quantized.

We now want to determine the independent components of the polarization tensor. To do so, we can exploit the de Donder gauge condition expressed in terms of $\polarizationtensor_{\mu\nu} e^{ip\cdot x}$,
\begin{equation}\label{de-donder-polariz}
p^\mu \polarizationtensor_{\mu\nu}-\frac{1}{2}p_\nu \polarizationtensor =0\,,\qquad \polarizationtensor\equiv \eta^{\mu\nu}\polarizationtensor_{\mu\nu}\,,
\end{equation}
that can be used to eliminate four components of the polarization tensor: $10-4=6$. Moreover, we can still make a gauge transformation $h_{\mu\nu}'=h_{\mu\nu}+\partial_\mu \zeta_\nu+\partial_\nu\zeta_\mu$ as long as the condition~\eqref{de-donder} is preserved: 
\begin{equation}
0=\delta_\zeta \left(\partial^\rho h_{\rho\nu} -\frac{1}{2}\partial_\nu h\right)=\partial^\rho \left(\partial_\rho \zeta_\nu+\partial_\nu \zeta_\rho\right)-\frac{1}{2}\partial_\nu \left(2\partial_\rho \zeta^\rho\right)=\Box \zeta_\nu\,,
\end{equation}
which is the so-called residual gauge condition and whose solution reads
\begin{equation}
\zeta_{\nu}(x)=r_{\nu}(p) e^{ip\cdot x}+r_{\nu}^\ast(p) e^{-ip\cdot x}\,,\qquad p^2=-p_0^2+\vec{p}^{\,2}=0\,.
\end{equation}
We can choose the vector $r_\nu$ (\ie{} $\zeta_\nu$) with $\nu=0,1,2,3$ to eliminate four additional components of the polarization tensor. Therefore, we get $10-4-4=2$ independent on-shell degrees of freedom.

Let us explicitly find the independent physical on-shell components. To simplify the analysis, we can rotate the spatial vector $\vec{p}$ in such a way that it is parallel to the $\hat{z}$-axis, \ie{} we choose
\begin{equation}\label{momentum-along-z}
p^\mu=(p^0,0,0,p^3)\,, \qquad p^0=p^3\,,
\end{equation}
for both $h_{\mu\nu}$ and $\zeta_\nu$ since they both satisfy a homogeneous wave equation.

The de Donder gauge condition~\eqref{de-donder-polariz} gives the following four equations:
\begin{equation}\label{de-dond-compon}
\begin{aligned}
\nu&=0:&\polarizationtensor_{00}+\polarizationtensor_{30}+\frac{1}{2}\polarizationtensor&=0\,, \\
\nu&=1:&\polarizationtensor_{01}+\polarizationtensor_{31}&=0\,, \\
\nu&=2:&\polarizationtensor_{02}+\polarizationtensor_{32}&=0\,, \\
\nu&=3:&\polarizationtensor_{03}+\polarizationtensor_{33}-\frac{1}{2}\polarizationtensor&=0\,.
\end{aligned}
\end{equation}
Then, we can make the gauge transformations $\polarizationtensor'_{\mu\nu}=\polarizationtensor_{\mu\nu}+ip_\mu r_\nu+ip_\nu r_\mu$ and choose $r_{\nu}$ to set some of the polarization tensor components to zero:\footnote{For example, we can choose $r_0=-\polarizationtensor_{00}/(2ip_0)$ to get $\polarizationtensor^\prime_{00}=0;$ the analog procedure can be applied to the other three components. Note that with an abuse of notation we continue to denote the gauge-transformed polarization by the symbol $\polarizationtensor_{\mu\nu}$ and not with $\polarizationtensor^{\prime}_{\mu\nu}$.}
\begin{equation}\label{residual-cond}
	r_0:\quad \polarizationtensor_{00}=0\,,\qquad 	r_1:\quad \polarizationtensor_{01}=0\,,\qquad r_2:\quad \polarizationtensor_{02}=0\,,\qquad r_3:\quad \polarizationtensor_{33}=0\,. 
\end{equation}
The set of equations~\eqref{de-dond-compon} and~\eqref{residual-cond} give eight conditions which allow to express all components of the polarization tensor in terms of two independent ones. Combining~\eqref{de-dond-compon} and~\eqref{residual-cond} we get that the only non-vanishing components are $\polarizationtensor_{12}=\polarizationtensor_{21}$ and $\polarizationtensor_{11}=-\polarizationtensor_{22}$:
\begin{tcolorbox}
\begin{equation}\label{polarization-tensor-dedonder}
\polarizationtensor_{\mu\nu}=\left(\begin{array}{cccc}
0&0&0&0 \\
0&\polarizationtensor_{11}&\polarizationtensor_{12}&0 \\
0&\polarizationtensor_{12}&-\polarizationtensor_{11}&0 \\
0&0&0&0
\end{array}\right) = \polarizationtensor_{11}\, \polarizationtensorbasis^{(+)}_{\mu\nu} +\polarizationtensor_{12}\,\polarizationtensorbasis^{(\times)}_{\mu\nu}\,, 
\end{equation}
\end{tcolorbox}
\noindent where we have defined the two independent polarizations
\begin{equation}\label{2-polarizations}
	\polarizationtensorbasis^{(+)}_{\mu\nu}=\left(\begin{array}{cccc}
		0&0&0&0 \\
		0&1&0&0 \\
		0&0&-1&0 \\
		0&0&0&0
	\end{array}\right)\,,\qquad 	\polarizationtensorbasis^{(\times)}_{\mu\nu}=\left(\begin{array}{cccc}
	0&0&0&0 \\
	0&0&1&0 \\
	0&1&0&0 \\
	0&0&0&0
\end{array}\right)\,.
\end{equation}

In summary, working in the de Donder gauge, we found that the graviton field satisfies a wave equation, is transverse and traceless, and propagates only two physical degrees of freedom on-shell. In coordinate space, the transverse and traceless conditions are
\begin{equation}\label{de-donder-graviton}
\partial_\mu h^{\mu}_{\phantom{\mu}\nu}=0\,,\qquad h=\eta^{\mu\nu}h_{\mu\nu}=0\,.
\end{equation}

\paragraph{Helicity.} We are interested in finding the helicity of the two propagating degrees of freedom. This can be done by finding the eigenstates of the rotation matrix around the $\hat{z}$-axis,
\begin{equation}\label{rotation-matrix}
R^{\nu}_{\phantom{\nu}\mu}(\theta)=\left(\begin{array}{cccc}
		1&0&0&0 \\
		0&\cos\theta&\sin\theta&0 \\
		0&-\sin\theta&\cos\theta&0 \\
		0&0&0&1
	\end{array}\right)\,,
\end{equation}
whose eigenvalues are $e^{i\lambda \theta}$,  $\lambda\equiv j_z$ being the helicity.

We can easily check that $\polarizationtensorbasis^{(+)}_{\mu\nu}$ and $\polarizationtensorbasis^{(\times)}_{\mu\nu}$ are not eigenvectors of~\eqref{rotation-matrix}. However, we can make the change of basis
 \begin{equation}\label{basis-change}
\polarizationtensorbasis^{(+2)}_{\mu\nu}=\frac{1}{\sqrt{2}}\left(\polarizationtensorbasis^{(+)}_{\mu\nu}+i \polarizationtensorbasis^{(\times)}_{\mu\nu}\right)\,,\qquad \polarizationtensorbasis^{(-2)}_{\mu\nu}=\frac{1}{\sqrt{2}}\left(\polarizationtensorbasis^{(+)}_{\mu\nu}-i \polarizationtensorbasis^{(\times)}_{\mu\nu}\right)\,,
 \end{equation}
and show that 
\begin{equation}\label{helicity-eigenstates}
	R^\rho_{\phantom{\rho}\mu}(\theta)\polarizationtensorbasis^{(\pm2)}_{\rho\sigma} R_{\nu}^{\phantom{\nu}\sigma}(\theta)=e^{\pm 2i\theta} \polarizationtensorbasis^{(\pm2)}_{\mu\nu}\,,\qquad \lambda=\pm 2\,.
\end{equation}
This means that the graviton field propagates two independent massless degrees of freedom with helicity $+2$ and $-2$, respectively, \ie{} the graviton polarization tensor can be expressed as a linear combination of $\polarizationtensorbasis^{(+2)}_{\mu\nu}$ and $\polarizationtensorbasis^{(-2)}_{\mu\nu}$:
\begin{equation}\label{polarization-helocit}
\polarizationtensor_{\mu\nu}=\frac{1}{\sqrt{2}}\left(\polarizationtensor_{11}-i\polarizationtensor_{12}\right) \polarizationtensorbasis^{(+2)}_{\mu\nu}+\frac{1}{\sqrt{2}}\left(\polarizationtensor_{11}+i\polarizationtensor_{12}\right) \polarizationtensorbasis^{(-2)}_{\mu\nu}\equiv \polarizationtensor_{\mu\nu}^{(+2)}+\polarizationtensor_{\mu\nu}^{(-2)}\,, 
\end{equation}
where we have defined
\begin{equation}\label{helicity-polariz}
 \polarizationtensor_{\mu\nu}^{(+2)}\equiv \frac{1}{\sqrt{2}}\left(\polarizationtensor_{11}-i\polarizationtensor_{12}\right) \polarizationtensorbasis^{(+2)}_{\mu\nu}\,,\qquad   \polarizationtensor_{\mu\nu}^{(-2)}\equiv \frac{1}{\sqrt{2}}\left(\polarizationtensor_{11}+i\polarizationtensor_{12}\right) \polarizationtensorbasis^{(-2)}_{\mu\nu}\,.
\end{equation}

\subsubsection{Graviton polarizations: non-covariant gauge}

It is instructive to determine the independent physical components of the graviton field considering a different gauge and to show that the result is the same as in the previous subsection, as expected. In particular, we impose the following non-covariant gauge condition, also known as radiation gauge,
\begin{equation}\label{radiation-gauge}
\partial_i h^i_{\phantom{i}\mu}=0\,,\qquad \mu=0,1,2,3\,,
\end{equation}
where the Latin index only runs over the spatial coordinates $i=1,2,3$. In this gauge, the vacuum field equations become
\begin{equation}\label{radiation-gauge-EOM}
\Box h_{\mu\nu}-\eta_{\mu\nu}\Box h+\eta_{\mu\nu}\ddot{h}_{00}+\partial_\mu\partial_\nu h+\partial_\mu \dot{h}_{0\nu}+\partial_\nu \dot{h}_{0\mu}=0\,,
\end{equation}
where the dot stands for the derivative with respect to time, \ie{} $\dot{}\equiv \partial_0$. Unlike the covariant de Donder gauge, in the radiation gauge, some of the components of the field equations are not wave equations, but constraints that are important to determine the number of independent physical components of the graviton field. This also means that we cannot yet use the dispersion relation $p^2=0$, but we will derive it below after imposing various constraints.

To simplify our analysis, we choose again a frame in which $p_1=0=p_2$ as in \eqref{momentum-along-z}. For the time being, we Fourier transform only in space, \ie{} $\partial_j\to i p_j$, and with an abuse of notation we denote the Fourier-transformed graviton field by the same symbol, but now it is a function of the time coordinate and the spatial momentum, \ie{} $h_{\mu\nu}=h_{\mu\nu}(t,\vec{p})$. 

The radiation gauge assumes a very simple form in the spatial-momentum Fourier space,
\begin{equation}\label{radiation-gauge-momentum}
	0=ip_j h^j_{\mu}=ip_3 h_{3\mu} \qquad \Rightarrow\qquad h_{30}=h_{31}=h_{32}=h_{33}=0\,,
\end{equation}
thus four components are already killed: $10-4=6$. We now have to inspect the field equations and find the four constraints that will eliminate four additional components. 

The $(0,0)$ component gives
\begin{equation}
-p_3^2(h_{00}+h)=0 \qquad \Rightarrow\qquad h_{00}=-h\,.
\end{equation}
Then, since $h=-h_{00}+h_{11}+h_{22}=-h_{00}$ (with $h_{33}=0$) we also get $h_{11}=-h_{22}$. 

Using $h_{00}=-h$, the components $(0,1)$ and $(0,2)$ of the field equations give $h_{01}=h_{02}=0$, while the $(0,3)$ is identically satisfied.

The $(1,1)$ and $(2,2)$ components are
\begin{equation}
-\left(\ddot{h}_{11}+p_3^2 h_{11}\right)+p_3^2 h=0\,,\qquad -\left(\ddot{h}_{22}+p_3^2 h_{22}\right)+p_3^2 h=0\,,
\end{equation}
respectively. Using $h_{11}=-h_{22}$ we get the constraint $h=0$, which also implies $h_{00}=0$, thus it follows that $h_{11}$ and $h_{22}$ solve the same harmonic oscillator equation with frequency $p_3$ which corresponds to a wave equation if we Fourier transform back to space. If we choose $h_{11}$, we have
\begin{equation}
\ddot{h}_{11}+p_3^2 h_{11}=0\,.
\end{equation}

The $(1,2)$ component is already in the harmonic oscillator form, \ie{}
\begin{equation}
\ddot{h}_{12}+p_3^2 h_{12}=0\,.
\end{equation}
The remaining components are $(1,3)$, $(2,3)$ and $(3,3)$ that are identically satisfied after imposing the other constraints that we have derived.

As expected, we have found that the graviton field has only two independent physical components, \ie{} $h_{11}$ and $h_{12}$. If we Fourier transform also in the time coordinate and call the fully Fourier transformed field $\polarizationtensor_{\mu\nu}$, we get
\begin{equation}
	\left(-p_0^2+p_3^2\right) \polarizationtensor_{11}=0\,,\qquad 	\left(-p_0^2+p_3^2\right) \polarizationtensor_{12}=0\,,
\end{equation}
that are satisfied if and only the massless dispersion relation holds, \ie{} $p^2=-p_0^2+p_3^2=0$.

As done in the previous section, we can introduce the polarizations $\polarizationtensorbasis^{(+)}_{\mu\nu}$ and $\polarizationtensorbasis^{(\times)}_{\mu\nu}$ or the helicity eigenvectors $\polarizationtensorbasis^{(+2)}_{\mu\nu}$ and $\polarizationtensorbasis^{(-2)}_{\mu\nu}$, and reach the same conclusions as in the de Donder gauge. Therefore, we have shown that imposing two completely different gauges, we get the same result. This confirms that (on-shell) physics does not depend on the gauge choice.

Before concluding this part, it is worth mentioning that in the radiation gauge, it is not necessary to explicitly impose a residual gauge condition because the radiation gauge together with the on-shell constraints completely determine the physical polarizations. Indeed, if we consider gauge transformations that leave the radiation gauge-invariant, we get
\begin{equation}
    \covD^2 \zeta_\nu+\partial_\nu (\partial^j \zeta_j)=0 \,,
\end{equation}
which in Fourier space and in the frame $p_1=0=p_2$ reads $-p_3^2 r_\nu - p_\nu p_3 r_3=0$. It is easy to show that, if $p_3\neq 0$, the last equation is satisfied if and only if $r_\mu=0$ for $\mu=0,1,2,3$. The gauge redundancy always hits twice, but how the ``twice'' acts depends on the type of gauge condition.

\subsubsection{Graviton propagator: covariant gauge}

The propagator is defined as the inverse of the kinetic operator $\left(\mathbb{K}^{-1}\right)_{\mu\nu\rho\sigma}$. However, the kinetic operator in \eqref{kinetic-oper} or~\eqref{kinetic-oper-momentum} is \textit{not} invertible: this can be shown by noticing that there exists a non-zero tensor $V_{\rho\sigma}$ such that $\mathbb{K}^{\mu\nu\rho\sigma}V_{\rho\sigma}=0$, which implies that the kernel of the kinetic operator is not empty. It is easy to find such a tensor because we know that the theory is invariant under the gauge transformation~\eqref{gauge-transf}. Indeed, we have
\begin{equation}
V_{\rho\sigma}=\partial_\rho \zeta_\sigma+\partial_\sigma\zeta_\rho\,,\qquad \mathbb{K}^{\mu\nu\rho\sigma}V_{\rho\sigma}=0\,.
\end{equation}

\paragraph{Covariant gauge fixing.} To make the kinetic operator invertible, we have to add a \textit{gauge-fixing} term to the action. We now choose the de Donder gauge fixing defined as
\begin{equation}\label{de-donder-gauge-fix}
S_{\rm gf}[\eta,h]=-\frac{1}{\GFalpha}\int {\rmd}^4x \, \GFcondition_\mu \GFcondition^\mu\,,\qquad \GFcondition_{\mu}\equiv \partial_\nu h^{\nu}_{\phantom{\nu}\mu}-\frac{1}{2}\partial_\mu h\,,
\end{equation}
where $\GFalpha$ is a gauge-fixing parameter.

Integrating by parts, the gauge-fixing contribution can be written as 
\begin{equation}
	S_{\rm gf}[\eta,h]=\int {\rmd}^4x \, \frac{1}{2}h_{\mu\nu} \mathbb{K}_{\rm gf}^{\mu\nu\rho\sigma}h_{\rho\sigma}\,,
\end{equation}
where
\begin{equation}
\begin{aligned}
\mathbb{K}^{\mu\nu\rho\sigma}_{\rm gf}\equiv &\frac{1}{2\GFalpha}\eta^{\mu\nu}\eta^{\rho\sigma}\Box-\frac{1}{\GFalpha}\left(\eta^{\mu\nu}\partial^\rho\partial^\sigma + \eta^{\rho\sigma}\partial^\mu\partial^\nu \right) \\
&+\frac{1}{2\GFalpha}\left(\eta^{\mu\rho}\partial^{\nu}\partial^\sigma+\eta^{\mu\sigma}\partial^{\nu}\partial^\rho +\eta^{\nu\rho}\partial^{\mu}\partial^\sigma+\eta^{\nu\sigma}\partial^{\mu}\partial^\rho \right)\,.
\end{aligned}
\end{equation}
Therefore, the total quadratic action now reads
\begin{equation}
	\tilde{S}^{(2)}[\eta,h]\equiv S^{(2)}_{\rm EH}[\eta,h]+S_{\rm gf}[\eta,h]= \int {\rmd}^4x \, \frac{1}{2}h_{\mu\nu} \tilde{\mathbb{K}}^{\mu\nu\rho\sigma}h_{\rho\sigma}\,,
\end{equation}
where 
\begin{equation}\label{kinetic-oper-gf}
\begin{aligned}
\tilde{\mathbb{K}}^{\mu\nu\rho\sigma}\equiv \mathbb{K}^{\mu\nu\rho\sigma}+\mathbb{K}_{\rm gf}^{\mu\nu\rho\sigma} &= \frac{1}{2}\left(\eta^{\mu\rho}\eta^{\nu\sigma}+\eta^{\mu\sigma}\eta^{\nu\rho} \right)\Box-\left(1-\frac{1}{2\GFalpha}\right)\eta^{\mu\nu}\eta^{\rho\sigma}\Box \\
	&\qquad+\left(1-\frac{1}{\GFalpha}\right)\left(\eta^{\mu\nu}\partial^{\rho}\partial^\sigma +\eta^{\rho\sigma}\partial^{\mu}\partial^\nu\right) \\
	&\qquad -\frac{1}{2}\left(1-\frac{1}{\GFalpha}\right)\left(\eta^{\mu\rho}\partial^{\nu}\partial^\sigma+\eta^{\mu\sigma}\partial^{\nu}\partial^\rho +\eta^{\nu\rho}\partial^{\mu}\partial^\sigma+\eta^{\nu\sigma}\partial^{\mu}\partial^\rho \right)\,.
\end{aligned}
\end{equation}
The new kinetic operator including the gauge-fixing term is invertible,  in particular
\begin{equation}
    \tilde{\mathbb{K}}^{\mu\nu\rho\sigma}V_{\rho\sigma}\neq 0 \, .
\end{equation}

In momentum space ($\partial_\mu\to i p_\mu$) the kinetic operator reads\footnote{With an abuse of notation we call the momentum-space quantities with the same symbol of their position-space counterparts, but we explicitly write the momentum dependence.}
\begin{equation}\label{kinetic-oper-momentum}
\begin{aligned}
	\tilde{\mathbb{K}}^{\mu\nu\rho\sigma}(p) =& -\frac{1}{2}\left(\eta^{\mu\rho}\eta^{\nu\sigma}+\eta^{\mu\sigma}\eta^{\nu\rho} \right)p^2+\left(1-\frac{1}{2\GFalpha}\right)\eta^{\mu\nu}\eta^{\rho\sigma}p^2 \\
	&-\left(1-\frac{1}{\GFalpha}\right)\left(\eta^{\mu\nu}p^{\rho}p^\sigma +\eta^{\rho\sigma}p^{\mu}p^\nu \right) \\
	& +\frac{1}{2}\left(1-\frac{1}{\GFalpha}\right)\left(\eta^{\mu\rho}p^{\nu}p^\sigma+\eta^{\mu\sigma}p^{\nu}p^\rho +\eta^{\nu\rho}p^{\mu}p^\sigma+\eta^{\nu\sigma}p^{\mu}p^\rho \right)\,.
\end{aligned}
\end{equation}

\paragraph{Kinetic operator inversion.} The propagator in momentum space $\propG_{\mu\nu\rho\sigma}(p)$ is defined as the inverse of the kinetic operator \eqref{kinetic-oper-momentum} through the following relation:
\begin{equation}
\propG_{\mu\nu}^{\phantom{\mu\nu}\alpha\beta}(p)\,\tilde{\mathbb{K}}_{\alpha\beta}^{\phantom{\alpha\beta}\rho\sigma}(p)=i \mathbbm{1}_{\mu\nu}^{\phantom{\mu\nu}\rho\sigma}\,,
\end{equation}
or, equivalently,
\begin{equation}\label{propag-inversion}
	\propG_{\mu\nu\alpha\beta}(p)\,\tilde{\mathbb{K}}^{\alpha\beta}_{\phantom{\alpha\beta}\rho\sigma}(p)=i\mathbbm{1}_{\mu\nu\rho\sigma}\,,
\end{equation}
where
\begin{equation}\label{eq:LUCA_SYM2_ID}
    \mathbbm{1}_{\mu\nu}^{\phantom{\mu\nu}\rho\sigma}=\frac{1}{2}(\delta_\mu^{\phantom{\mu}\rho}\delta_\nu^{\phantom{\nu}\sigma}+\delta_\nu^{\phantom{\nu}\rho}\delta_\mu^{\phantom{\mu}\sigma}) \, , \qquad \mathbbm{1}_{\mu\nu\rho\sigma}=\frac{1}{2}(\eta_{\mu\rho}\eta_{\nu\sigma}+\eta_{\nu\rho}\eta_{\mu\sigma})
\end{equation}
is the identity in the space of rank-four symmetric tensors, and the factor $i$ is inserted accordingly to our convention for the Feynman rules.

The propagator $\propG_{\mu\nu\rho\sigma}(p)$ can be written as a linear combination of the elements of a basis in the space of rank-four symmetric tensors. From Lorentz invariance, we can easily find a basis given by the following five independent elements:
\begin{align}
B_{\phantom{(1)} \mu\nu\rho\sigma}^{(1)}(p)&= \eta_{\mu\rho}\eta_{\nu\sigma}+\eta_{\mu\sigma}\eta_{\nu\rho}\,, \\
B_{\phantom{(2)} \mu\nu\rho\sigma}^{(2)}(p)&=  \eta_{\mu\nu}\eta_{\rho\sigma}\,,  \\
B_{\phantom{(3)} \mu\nu\rho\sigma}^{(3)}(p)&= \frac{1}{p^2}\left(\eta_{\mu\nu}p_\rho p_\sigma + \eta_{\rho\sigma}p_\mu p_\nu \right) \, , \label{element-basis-3}\\
B_{\phantom{(4)} \mu\nu\rho\sigma}^{(4)}(p)&= \frac{1}{p^2}\left(\eta_{\mu\rho}p_\nu p_\sigma + \eta_{\mu\sigma}p_\nu p_\rho +\eta_{\nu\rho}p_\mu p_\sigma + \eta_{\nu\sigma}p_\mu p_\rho  \right) \, , \\
B_{\phantom{(5)} \mu\nu\rho\sigma}^{(5)}(p)&=\frac{1}{(p^2)^2}p_\mu p_\nu p_\rho p_\sigma \,.
\end{align}
Thus, the propagator can be written as
\begin{equation}\label{propag-basis}
\propG_{\mu\nu\rho\sigma}(p)=\sum_{j=1}^5 c_j(p) B_{\phantom{(j)} \mu\nu\rho\sigma}^{(j)}(p)\,,
\end{equation}
where $c_j(p)$ are momentum-dependent coefficients. To completely derive the propagator, we have to substitute \eqref{propag-basis} into \eqref{propag-inversion} and find the coefficients $c_j(p)$ that solve the tensor equation. 

Since the brute-force calculation can be lengthy, in this subsection we compute the propagator in the so-called Feynman gauge, in which the kinetic operator simplifies, thus rendering its inversion easier. Then, in \cref{sec:propag-III} we will use a more efficient method to derive the propagator for a generic de Donder gauge-fixing parameter by using the spin-projectors formalism.

\paragraph{Feynman gauge.} The Feynman gauge corresponds to the choice $\GFalpha=1$ of the gauge-fixing parameter. From \eqref{kinetic-oper-momentum} we can notice that in this gauge, all the terms containing non-contracted momenta (\ie{} non-contracted derivatives) are set to zero, and only the contribution proportional to $p^2$ (\ie{} to $\Box$) survive:
\begin{equation}\label{kinetic-oper-momentum-feynman}
\begin{aligned}
\tilde{\mathbb{K}}^{(\GFalpha=1)\,\mu\nu\rho\sigma}(p)&= -\frac{1}{2}\left(\eta^{\mu\rho}\eta^{\nu\sigma}+\eta^{\mu\sigma}\eta^{\nu\rho}-\eta^{\mu\nu}\eta^{\rho\sigma}\right)p^2 \\
&=a(p)\left(\eta^{\mu\rho}\eta^{\nu\sigma}+\eta^{\mu\sigma}\eta^{\nu\rho}\right)+b(p)\eta^{\mu\nu}\eta^{\rho\sigma}\,,
\end{aligned}
\end{equation}
where we have defined $a(p)\equiv -p^2/2$ and $b(p)\equiv p^2/2$.

We now make the following ansatz for the propagator in the Feynman gauge:
\begin{equation}\label{propag-ansatz-feynman}
\propG^{(\GFalpha=1)}_{\mu\nu\rho\sigma}(p)=A(p)\left(\eta_{\mu\rho}\eta_{\nu\sigma}+\eta_{\mu\sigma}\eta_{\nu\rho}\right)+B(p)\eta_{\mu\nu}\eta_{\rho\sigma}\,,
\end{equation}
where the momentum-dependent coefficients $A(p)$ and $B(p)$ are two unknowns to be determined. Substituting~\eqref{propag-ansatz-feynman} into~\eqref{propag-inversion} with the kinetic operator given by~\eqref{kinetic-oper-momentum-feynman}, we can find the expressions for $A(p)$ and $B(p)$ that solve the equation. We have
\begin{equation}
\begin{aligned}
	\propG^{(\GFalpha=1)}_{\mu\nu\alpha\beta}(p)\tilde{\mathbb{K}}^{(\GFalpha=1)\,\alpha\beta}_{\phantom{(\GFalpha=1)\,\alpha\beta}\rho\sigma}(p)&= 2a A\left[\eta_{\mu\rho}\eta_{\nu\sigma}+\eta_{\mu\sigma}\eta_{\nu\rho}\right]+\left[2bA+2aB+4bB\right]\eta_{\mu\nu}\eta_{\rho\sigma} \\
	&=\frac{i}{2}\left(\eta_{\mu\rho}\eta_{\nu\sigma}+\eta_{\nu\rho}\eta_{\mu\sigma}\right)\,,
\end{aligned}
\end{equation}
which is satisfied if and only if 
\begin{equation}
\left\lbrace 
\begin{array}{ll}
\displaystyle 2Aa=\frac{i}{2}\\
2bA+2aB+4bB=0
\end{array} \right.\quad \Leftrightarrow \quad A(p)=-B(p)=-\frac{i}{2p^2}\,.
\end{equation}

Therefore, the graviton propagator in the Feynman gauge reads

\begin{tcolorbox}
\begin{equation}\label{propag-feynman-gauge}
\propG^{(\GFalpha=1)}_{\mu\nu\rho\sigma}(p)=\frac{1}{2}\frac{-i}{p^2-i\epsilon}\left(\eta_{\mu\rho}\eta_{\nu\sigma}+\eta_{\mu\sigma}\eta_{\nu\rho}-\eta_{\mu\nu}\eta_{\rho\sigma}\right)\,,
\end{equation}
\end{tcolorbox}
\noindent where we also introduced the Feynman prescription for how to shift the poles, \ie{} $p^2\to p^2-i\epsilon$ with $\epsilon\to 0^+$.

\paragraph{Remark.} A nice feature of the propagator in the Feynman gauge is that it is manifestly Lorentz covariant and does not depend on uncontracted momenta, which makes computations more efficient in most of the cases. At the same time, one caveat of the covariant gauge is that before contracting with some conserved stress-energy tensor and going on-shell, it is not clear how many components of $\propG_{\mu\nu\rho\sigma}$ are the physical ones. Furthermore, from \eqref{propag-feynman-gauge} it is not clear what is the spin structure of the propagator, \ie{} what are the spin components of the off-shell degrees of freedom. In the next two subsections we address these two points.

\subsubsection{Graviton propagator: non-covariant gauge}\label{sec:grav-propag-non-cov}

The propagator in \eqref{propag-feynman-gauge} has many components with poles at $p^2=-(p_0)^2+\vec{p}^{\,2}=0$ and it is not clear how many independent ones there are and whether any of them have negative residues. A similar situation occurs in the case of Yang-Mills theory when using the covariant Lorenz gauge fixing. In that case, it is possible to use a non-covariant gauge fixing and to work in the Coulomb gauge to make the number of independent poles manifest. A similar procedure can be implemented in Einstein's gravity.

\paragraph{Non-covariant gauge fixing.} We now consider the following non-covariant gauge-fixing term:
\begin{equation}\label{non-covariant-gauge-fix}
	S_{\rm gf}[\eta,h]=-\frac{1}{\GFalpha}\int {\rmd}^4x \GFcondition_\mu \GFcondition^\mu\,,\qquad \GFcondition_{\mu}\equiv \partial_i h^i_\mu\,,
\end{equation}
where the Latin index only runs over the spatial coordinates $i=1,2,3$. Integrating by parts we can write $S_{\rm gf}=\int {\rmd}^4 x\frac{1}{2}h_{\mu\nu}\mathbb{K}_{\rm gf}^{\mu\nu\rho\sigma}h_{\rho\sigma}$, where
\begin{equation}
\mathbb{K}_{\rm gf}^{\mu\nu\rho\sigma}=\frac{1}{2\GFalpha} \left(\eta^{\mu\rho}\eta^{\nu i}\eta^{\sigma j}+\eta^{\mu\sigma}\eta^{\nu i}\eta^{\rho j}+\eta^{\nu\rho}\eta^{\mu i}\eta^{\sigma j}+\eta^{\nu\sigma}\eta^{\mu i}\eta^{\rho j}\right)\partial_i\partial_j\,.
\end{equation}

The full kinetic operator $\tilde{\mathbb{K}}^{\mu\nu\rho\sigma}=\mathbb{K}^{\mu\nu\rho\sigma}+\mathbb{K}_{\rm gf}^{\mu\nu\rho\sigma}$ in momentum space now reads
\begin{equation}\label{kinetic-oper-momentum-non-cov}
\begin{aligned}
\tilde{\mathbb{K}}^{\mu\nu\rho\sigma}(p,\bar{p})&= -\frac{1}{2}\left(\eta^{\mu\rho}\eta^{\nu\sigma}+\eta^{\mu\sigma}\eta^{\nu\rho} \right)p^2+\eta^{\mu\nu}\eta^{\rho\sigma}p^2-\left(\eta^{\mu\nu}p^{\rho}p^\sigma +\eta^{\rho\sigma}p^{\mu}p^\nu \right)   \\
	&\qquad +\frac{1}{2}\left[\eta^{\mu\rho}\left(p^{\nu}p^\sigma-\frac{1}{\GFalpha}\bar{p}^{\,\nu}\bar{p}^{\,\sigma}\right)+ \eta^{\mu\sigma}\left(p^{\nu}p^\rho-\frac{1}{\GFalpha}\bar{p}^{\,\nu}\bar{p}^{\,\rho}\right) \right. \\
	&\hspace{2cm}\left. + \eta^{\nu\rho}\left(p^{\mu}p^\sigma-\frac{1}{\GFalpha}\bar{p}^{\,\mu}\bar{p}^{\,\sigma}\right) + \eta^{\nu\sigma}\left(p^{\mu}p^\rho-\frac{1}{\GFalpha}\bar{p}^{\,\mu}\bar{p}^{\,\rho}\right) \right]\,,
\end{aligned}
\end{equation}
where we have defined $\bar{p}^{\,\mu}\equiv \left(0,p_1,p_2,p_3\right)=(0,\vec{p})$.

To simplify our analysis, we choose again the reference frame in which $p_2=0=p_3$, \ie{} we rotate the spatial part of the four-momentum $\vec{p}$ along the $\hat{z}$-axis, so that we have $p^\mu=(-p_0,0,0,p_3)$ and $\bar{p}^{\,\mu}=(0,0,0,p_3)$, and their squares are $p^2=-p_0^2+p_3^3$ and $\bar{p}^2=p_3^2$, respectively. In this frame, the independent non-zero components of the kinetic operator~\eqref{kinetic-oper-momentum-non-cov} are
\begin{equation}\label{indep-comp-kinetic}
\begin{aligned}
	&\tilde{\mathbb{K}}^{0101}= \tilde{\mathbb{K}}^{0202}=\frac{1}{2}p_3^2\,,\qquad \tilde{\mathbb{K}}^{0303}=\frac{p_3^2}{2\GFalpha}\,,\qquad \tilde{\mathbb{K}}^{1313}=\tilde{\mathbb{K}}^{2323}=-\frac{p^2}{2}+\frac{1}{2}\left(1-\frac{1}{\GFalpha}\right)p_3^2\,,\\
	&\tilde{\mathbb{K}}^{0011}=\tilde{\mathbb{K}}^{0022}=-p_3^2\,,\qquad \tilde{\mathbb{K}}^{0113}=\tilde{\mathbb{K}}^{0223}=-\frac{p_0p_3}{2}\,,\qquad \tilde{\mathbb{K}}^{1103}=\tilde{\mathbb{K}}^{2203}=p_0p_3\,, \\
	& \qquad \qquad\tilde{\mathbb{K}}^{1122}=p^2\,,\qquad \tilde{\mathbb{K}}^{1212}=-\frac{1}{2}p^2\,,\qquad \tilde{\mathbb{K}}^{1133}=\tilde{\mathbb{K}}^{2233}=p^2-p_3^2\,.
\end{aligned}
\end{equation}

\paragraph{Kinetic operator inversion.} It is convenient to recast the kinetic operator and the propagator into matrix form in order to make the inversion more efficient. We define the vector
\begin{equation}\label{vector-h}
	\hat{h}\equiv \left( h_{00},h_{01},h_{02},h_{03},h_{11},h_{12},h_{13},h_{22},h_{23},h_{33}\right)\,,
\end{equation}
and the $10\times 10$ symmetric matrix $\hat{\mathbb{K}}$ whose elements are
\begin{equation}\label{kinetic-elements}
\hat{\mathbb{K}}^{\hat{k}\hat{\ell}}\equiv s^{k\ell}\tilde{\mathbb{K}}^{k\ell}\,,\qquad k,\ell\in \left\lbrace 00,01,02,03,11,12,13,22,23,33\right\rbrace\,,\qquad \hat{k},\hat{\ell}\in \left\lbrace 1,2,\dots,10\right\rbrace\,,
\end{equation}
where each value of the hatted indices correspond to unique values of the unhatted ones, \ie{} $\hat{k}\leftrightarrow k$ and $\hat{\ell}\leftrightarrow \ell$, for example $1\leftrightarrow 00$, $2\leftrightarrow 01$, and so on.\footnote{Do not get confused by the notation in \eqref{kinetic-elements}: $\tilde{\mathbb{K}}^{kl}$ are the components of a rank-four tensor, while $\hat{\mathbb{K}}^{\hat{k}\hat{l}}$ is a $10\times 10$ matrix. For example, $\hat{\mathbb{K}}^{11}=s^{0000}\tilde{\mathbb{K}}^{0000}=\tilde{\mathbb{K}}^{0000}$, $\hat{\mathbb{K}}^{12}=s^{0001}\tilde{\mathbb{K}}^{0001}=2\tilde{\mathbb{K}}^{0001}$, and so on.}
The coefficients $s^{k\ell}$ are symmetry factors defined as $s^{\mu\mu\nu\nu}=1$, $s^{\mu\nu\rho\rho}=2$ with $\mu\neq \nu$, and $s^{\mu\nu\rho\sigma}=4$ with $\mu\neq \nu$,  $\rho\neq\sigma$. Thus, the quadratic action can be recast as
\begin{equation}\label{quadratic-matrix-form}
\tilde{S}^{(2)}= \frac{1}{2}\int {\rmd}^4x\, \sum_{\hat{k},\hat{\ell}=1}^{10} \hat{h}_{\hat{k}}\,\hat{\mathbb{K}}^{\hat{k}\hat{\ell}}\,\hat{h}_{\hat{\ell}} = \frac{1}{2}\int {\rmd}^4x\, \hat{h}\cdot\hat{\mathbb{K}}\cdot\hat{h}^{\rm \scriptscriptstyle T}\,,
\end{equation}
where $\hat{h}^{\rm \scriptscriptstyle T}$ is the transpose of $\hat{h}$ and the kinetic matrix reads
\begin{equation}
 \hat{\mathbb{K}}=
\left(\begin{smallmatrix}
    0 & 0 & 0 & 0 & -2p_3^2 & 0 & 0 & -2p_3^2 & 0 & 0 \\
	0 & \frac{p_3^2}{2} & 0 & 0 & 0 & 0 & -2p_0 p_3 & 0 & 0 & 0 \\
	0 & 0 & \frac{p_3^2}{2} & 0 & 0 & 0 & 0 & 0 & -2p_0 p_3 & 0 \\
	0 & 0 & 0 & \frac{p_3^2}{2\GFalpha} & 2p_0 p_3 & 0 & 0 & 2p_0 p_3 & 0 & 0 \\
	-2p_3^2 & 0 & 0 & 2p_0 p_3 & 0 & 0 & 0 & p^2 & 0 & p^2-p_3^2 \\
	0 & 0 & 0 & 0 & 0 & -2p^2 & 0 & 0 & 0 & 0 \\
	0 & -2p_0 p_3 & 0 & 0 & 0 & 0 & -2p^2+2 \big(1-\frac{1}{\GFalpha}\big) p_3^2 & 0 & 0 & 0 \\
	-2p_3^2 & 0 & 0 & 2p_0 p_3 & p^2 & 0 & 0 & 0 & 0 & p^2-p_3^2 \\
	0 & 0 & -2p_0 p_3 & 0 & 0 & 0 & 0 & 0 & -2p^2+2 \big(1-\frac{1}{\GFalpha}\big) p_3^2 & 0 \\
	0 & 0 & 0 & 0 & p^2-p_3^2 & 0 & 0 & p^2-p_3^2 & 0 & \frac{-2 p_3^2}{\GFalpha}   
\end{smallmatrix}\right) \, .
\label{K-matrix-simpler}
\end{equation}

Inverting~\eqref{K-matrix-simpler} we can find the propagator for a generic gauge-fixing parameter $\GFalpha$. In matrix form we have
\begin{equation}
 \hat{\propG}=
 i \left( \begin{smallmatrix}
     \frac{\GFalpha  p_0^2 \left(p^2+15p_3^2\right)-p_3^2 p^2}{8 p_3^6} & 0 & 0 & \frac{2\GFalpha  p_0}{p_3^3} & \frac{-1}{4 p_3^2} & 0 & 0 & \frac{-1}{4 p_3^2} & 0 & \frac{\GFalpha  p_0^2}{4 p_3^4} \\
	0 & \frac{2 \left(p_3^2-\GFalpha  p_0^2\right)}{p_3^2(3\GFalpha p_0^2+p_3^2)} & 0 & 0 & 0 & 0 & \frac{-2 \GFalpha  p_0}{p_3(3\GFalpha p_0^2+p_3^2)} & 0 & 0 & 0 \\
	0 & 0 & \frac{2 \left(p_3^2-\GFalpha  p_0^2\right)}{p_3^2(3\GFalpha p_0^2+p_3^2)}  & 0 & 0 & 0 & 0 & 0 & \frac{-2 \GFalpha  p_0}{p_3(3\GFalpha p_0^2+p_3^2)} & 0 \\
	\frac{2\GFalpha  p_0}{p_3^3} & 0 & 0 & \frac{2\GFalpha }{p_3^2} & 0 & 0 & 0 & 0 & 0 & 0 \\
	\frac{-1}{4 p_3^2} & 0 & 0 & 0 & \frac{-1}{2 p^2} & 0 & 0 & \frac{1}{2 p^2} & 0 & 0 \\
	0 & 0 & 0 & 0 & 0 & -\frac{1}{2p^2} & 0 & 0 & 0 & 0 \\
	0 & \frac{-2 \GFalpha  p_0}{p_3(3\GFalpha p_0^2+p_3^2)} & 0 & 0 & 0 & 0 & \frac{-\GFalpha}{6\GFalpha p_0^2+2p_3^2} & 0 & 0 & 0 \\
	\frac{-1}{4p_3^2} & 0 & 0 & 0 & \frac{1}{2 p^2} & 0 & 0 & \frac{-1}{2 p^2} & 0 & 0 \\
	0 & 0 & \frac{-2 \GFalpha  p_0}{p_3(3\GFalpha p_0^2+p_3^2)} & 0 & 0 & 0 & 0 & 0 & \frac{-\GFalpha}{6\GFalpha p_0^2+2p_3^2} & 0 \\
	\frac{\GFalpha  p_0^2}{4 p_3^4} & 0 & 0 & 0 & 0 & 0 & 0 & 0 & 0 & \frac{-\GFalpha}{2 p_3^2} 
 \end{smallmatrix} \right) \, ,
\end{equation}
where the imaginary unit was again inserted according to our convention for the Feynman rules, \ie{} $\hat{\propG}\, \hat{\mathbb{K}}=i\, \mathbbm{1}$.

\paragraph{Prentki gauge.} As done in the case of the covariant gauge fixing, we can further simplify the propagator by choosing a suitable value of the gauge-fixing parameter. We take $\GFalpha\to  0$, which corresponds to the so-called Prentki gauge~\cite{tHooft:1974toh,Veltman:1975vx}, and obtain the following expression for the graviton propagator in matrix form:
\begin{equation}
\hat{\propG}^{(\GFalpha=0)}= i
\left(
\begin{array}{cccccccccc}
	\frac{-p^2}{8 p_3^4} & 0 & 0 & 0 & \frac{-1}{4 p_3^2} & 0 & 0 & \frac{-1}{4 p_3^2} & 0 & 0 \\
	0 & \frac{2}{p_3^2} & 0 & 0 & 0 & 0 & 0 & 0 & 0 & 0 \\
	0 & 0 & \frac{2}{p_3^2} & 0 & 0 & 0 & 0 & 0 & 0 & 0 \\
	0 & 0 & 0 & 0 & 0 & 0 & 0 & 0 & 0 & 0 \\
	-\frac{1}{2 p_3^2} & 0 & 0 & 0 & \frac{-1}{2 p^2} & 0 & 0 & \frac{1}{2 p^2} & 0 & 0 \\
	0 & 0 & 0 & 0 & 0 & \frac{-1}{2p^2} & 0 & 0 & 0 & 0 \\
	0 & 0 & 0 & 0 & 0 & 0 & 0 & 0 & 0 & 0 \\
	\frac{-1}{4p_3^2} & 0 & 0 & 0 & \frac{1}{2 p^2} & 0 & 0 & \frac{-1}{2 p^2} & 0 & 0 \\
	0 & 0 & 0 & 0 & 0 & 0 & 0 & 0 & 0 & 0 \\
	0 & 0 & 0 & 0 & 0 & 0 & 0 & 0 & 0 & 0 
\end{array}
\right)\,.
\label{propag-matrix-prentki}
\end{equation}

To determine the number of independent physical components in the propagator we can compute its eigenvalues and check how many of them have a pole at $p^2=0$. The ten eigenvalues $\lambda_i$ can be found by solving the characteristic equation
\begin{equation}
{\det}\left(\hat{\propG}^{(\GFalpha=0)}-\lambda\mathbbm{1}\right)=0\quad \Leftrightarrow\quad \frac{\lambda ^4 \left(1+\lambda  p^2\right) \left(1+2\lambda  p^2\right) \left(2-\lambda  p_3^2\right)^2 \left(1-\lambda  p^2-8 \lambda ^2 p_3^4\right)}{2 (p^2)^2 p_3^8}=0\,,
\end{equation}
whose solutions are
\begin{equation}
\begin{aligned}
&\lambda_1=\lambda_2=\lambda_3=\lambda_4=0\,,\qquad  \lambda_5=\lambda_6= \frac{2}{p_3^2}\,,\qquad \lambda_7=-\frac{p^2+\sqrt{(p^2)^2+32 p_3^4}}{4 p_3^4}\,, \\
&\qquad  \lambda_8=\frac{-p^2+\sqrt{(p^2)^2+32 p_3^4}}{4 p_3^4}\,,\qquad \lambda_9=-\frac{2}{p^2}\,,\qquad \lambda_{10}=-\frac{1}{p^2}\,.
\end{aligned}
\end{equation}
We can now explicitly see that only two eigenvalues, $\lambda_9$ and $\lambda_{10}$, have a physical pole at $p^2=-p_0^2+p_3^2=0$. This means that the graviton propagates only two degrees of freedom, which is consistent with the counting of on-shell degrees of freedom performed above.

For the sake of completeness, we also show the tensor form of the physical part of the propagator in the Prentki gauge and in the reference frame $p_2=0=p_3$. This corresponds to the $3\times 3$ subspace $\left\lbrace 11,12,22\right\rbrace $ of the matrix~\eqref{propag-matrix-prentki}:
\begin{tcolorbox}
\begin{equation}\label{tensor-form-prentki}
\propG^{(\GFalpha=0)}_{\mu\nu\rho\sigma}(p,\bar{p})=\frac{1}{2}\frac{-i}{p^2-i\epsilon}\left(\bar{\delta}_{\mu\rho}\bar{\delta}_{\nu\sigma}+\bar{\delta}_{\mu\sigma}\bar{\delta}_{\nu\rho}-\bar{\delta}_{\mu\nu}\bar{\delta}_{\rho\sigma} \right)+\dots\,,
\end{equation}
\end{tcolorbox}
\noindent where we have defined 
\begin{equation}
\bar{\delta}_{\mu\nu}\equiv {\rm diag}(0,1,1,0)\,,
\end{equation}
and introduced the Feynman prescription. The dots stand for additional terms that do not have poles at $p^2=0$ and are proportional to $1/p_3^2$ and $1/p_3^4$: these contributions are associated to the remaining elements of the matrix~\eqref{propag-matrix-prentki}. 

The graviton field projected on the physical subspace $\left\lbrace 11,12,22\right\rbrace $ is transverse and traceless, and propagates only two helicities. Indeed, we can obtain the on-shell graviton field by acting with the residue of the propagator~\eqref{tensor-form-prentki} at $p^2=0$ on $h^{\rho\sigma}$:
\begin{equation}
\lim_{p^2\to 0} \left[i p^2	\propG^{(\GFalpha=0)}_{\mu\nu\rho\sigma}(p,\bar{p})\right]h^{\rho\sigma}= \frac{1}{2}\left(\polarizationtensorbasis_{\mu\nu}^{(+)} \polarizationtensorbasis_{\rho\sigma}^{(+)} + \polarizationtensorbasis_{\mu\nu}^{(\times)} \polarizationtensorbasis_{\rho\sigma}^{(\times)}\right)h^{\rho\sigma}\,,
\end{equation}
where we have used the relation
\begin{equation}
\bar{\delta}_{\mu\rho}\bar{\delta}_{\nu\sigma}+\bar{\delta}_{\mu\sigma}\bar{\delta}_{\nu\rho}-\bar{\delta}_{\mu\nu}\bar{\delta}_{\rho\sigma} = \polarizationtensorbasis_{\mu\nu}^{(+)} \polarizationtensorbasis_{\rho\sigma}^{(+)} + \polarizationtensorbasis_{\mu\nu}^{(\times)} \polarizationtensorbasis_{\rho\sigma}^{(\times)}\,.
\end{equation}

It is worth noting that the structure of the propagator in the Prentki gauge which contains the pole contributions is the same as in the Feynman gauge~\eqref{propag-feynman-gauge} up to the replacement $\eta_{\mu\nu} \to \bar{\delta}_{\mu\nu}$. This confirms what we alluded to above, that is, not all components of the propagator in the Feynman gauge correspond to physical propagating degrees of freedom, and that by switching to a non-covariant gauge we can manifestly reveal the physical ones.

\subsubsection{Graviton propagator: spin projectors}\label{sec:propag-III}

We now derive the propagator for a generic de Donder gauge-fixing parameter by using the spin-projector formalism through which we can easily identify the spin structure of the off-shell components. We introduce the spin projectors without giving too many details, but the reader is encouraged to check \cref{app:spin-proj} for an expanded discussion. To warm up, we first apply the formalism to the case of the photon propagator in \ac{QED} and then move to Einstein's gravity.

\subsubsubsection*{Warm up: photon propagator}  

Consider the free action for a photon with the Lorenz gauge fixing:
\begin{align}
S_A&=\int{\rmd}^4x \left[-\frac{1}{4}F_{\mu\nu}F^{\mu\nu}-\frac{1}{2\xi}\left(\partial_\mu A^\mu\right)^2\right]=\frac{1}{2}\int{\rmd}^4x A_{\mu}\tilde{\mathbb{K}}^{\mu\nu}A_\nu\,, \\
\tilde{\mathbb{K}}^{\mu\nu}&\equiv \eta^{\mu\nu} \Box-\left(1-\frac{1}{\xi}\right)\partial^\mu\partial^\nu\,.
\end{align}
Here, $\xi$ is a gauge parameter that plays the same role as \GFalpha{} in gravity. We want to rewrite the kinetic operator in terms of its spin components and then invert it to find the propagator whose spin structure will then be manifest.

Under the rotation group $SO(3)$ the four-vector $A_\mu$ can be decomposed into two irreducible representations: a scalar ($A_0$) and a three-vector ($A_i$), that is
\begin{equation}
	A_\mu \in \boldsymbol{0}\oplus \boldsymbol{1}\,.
\end{equation}
Working in momentum space, we can define the following projector operators 
\begin{equation}
\theta_{\mu\nu}=  \eta_{\mu\nu}-\frac{p_\mu p_\nu}{p^2}\,,\qquad \omega_{\mu\nu}=\frac{p_\mu p_\nu}{p^2}\,,
\end{equation}
that are idempotent and orthogonal,
\begin{equation}
	\theta_{\mu\rho}\theta^{\rho}_{\phantom{\rho}\nu}=\theta_{\mu\nu}\,,\qquad \omega_{\mu\rho}\omega^{\rho}_{\phantom{\rho}\nu}=\omega_{\mu\nu}\,,\qquad \theta_{\mu\rho}\omega^{\rho}_{\phantom{\rho}\nu}=0\,,
\end{equation}
and form a complete set
\begin{equation}
\theta_\mu^{\phantom{\mu}\nu}+\omega_\mu^{\phantom{\mu}\nu}=\delta_\mu^{\phantom{\mu}\nu}\,\qquad \Leftrightarrow \qquad \theta_{\mu\nu}+\omega_{\mu\nu}=\eta_{\mu\nu}\,.
\end{equation}
Since the projectors are idempotent, their trace equals their rank. This means that the trace is equal to the dimension of the corresponding irreducible representation (\ie{}  $2j+1$):
\begin{equation}
\eta^{\mu\nu}\theta_{\mu\nu}=3=2(1)+1\,,\qquad \eta^{\mu\nu}\omega_{\mu\nu}=1=2(0)+1\,,
\end{equation}
which means that $\theta_{\mu\nu}$ projects along the spin-one component and $\omega_{\mu\nu}$ along the spin-zero.

Besides forming a complete set of projectors, $\left\lbrace \theta,\omega \right\rbrace $ also form a basis in the space of symmetric rank-two tensors. Therefore, we can express the photon kinetic operator as a linear combination of the spin projectors, and in momentum space we have
\begin{equation}
\tilde{\mathbb{K}}^{\mu\nu}=-p^2\left[\theta^{\mu\nu}+\frac{1}{\xi}\omega^{\mu\nu} \right]\,.
\end{equation}
The propagator can be found by solving the tensor equation $\propG_{\mu\rho}\tilde{\mathbb{K}}^{\rho}_{\phantom{\rho}\nu}=i\eta_{\mu\nu}$. Making the ansatz $\propG_{\mu\nu}(p)=A(p) \theta_{\mu\nu}+B(p)\omega_{\mu\nu}$ and using the idempotency and orthogonality properties of the projectors, we can easily determine the two unknowns, \ie{} $A(p)=-i/p^2$ and $B(p)=-i\xi/p^2$. Thus we obtain
\begin{equation}
\propG_{\mu\nu}(p)=-\frac{i}{p^2}\left( \theta_{\mu\nu}+\xi \omega_{\mu\nu}  \right)\,.
\end{equation}
Note that the propagator contains all four (physical and unphysical) degrees of freedom. However, $\omega_{\mu\nu}$ is proportional to the four-momentum $p_\mu$, therefore it does not contribute to an amplitude when we contract the propagator with some external conserved current. By contrast, $\theta_{\mu\nu}$ does contribute and carries the off-shell propagating degrees of freedom of the photon field. Indeed, the photon has \textit{three} degrees of freedom off-shell. The longitudinal component of $\theta_{\mu\nu}$ disappears on-shell due to gauge invariance. This counting of off-shell and on-shell degrees of freedom shows that one component is killed purely off-shell, then another one is eliminated on-shell: gauge invariance hits twice.

\subsubsubsection*{Graviton propagator} 

We now apply the same procedure to the gravitational case where the graviton field is a symmetric rank-two tensor and the kinetic operator and the propagator are symmetric rank-four tensors. 

Under the rotation group $SO(3)$ a symmetric rank-two tensor $h_{\mu\nu}$ can be decomposed into two scalars (the spatial trace $\delta_{ij}h^{ij}$ and $h^{00}$), a spin-one ($h^{0i}$) and a traceless spin-two ($h^{ij}$):
\begin{equation}
h^{\mu\nu} \in \boldsymbol{0}\oplus \boldsymbol{0}\oplus\boldsymbol{1}\oplus\boldsymbol{2}\,.
\end{equation}
For these four irreducible blocks we can introduce the following four spin projectors
\begin{equation}
\begin{aligned}
\mathcal{P}^{(2)}_{\phantom{(2)}\mu\nu\rho\sigma}&=  \frac{1}{2}\left(\theta_{\mu\rho}\theta_{\nu\sigma}+\theta_{\mu\sigma}\theta_{\nu\rho}\right)-\frac{1}{3}\theta_{\mu\nu}\theta_{\rho\sigma}\,, \\
\mathcal{P}^{(1)}_{\phantom{(1)}\mu\nu\rho\sigma}&=  \frac{1}{2}\left(\theta_{\mu\rho}\omega_{\nu\sigma}+\theta_{\mu\sigma}\omega_{\nu\rho}+\theta_{\nu\rho}\omega_{\mu\sigma}+\theta_{\nu\sigma}\omega_{\mu\rho}\right)\,, \\
\mathcal{P}^{(0,s)}_{\phantom{(0,s)}\mu\nu\rho\sigma}&=  \frac{1}{3}\theta_{\mu\nu}\theta_{\rho\sigma}\,, \\
\mathcal{P}^{(0,w)}_{\phantom{(0,w)}\mu\nu\rho\sigma}&=\omega_{\mu\nu}\omega_{\rho\sigma}\,.
\end{aligned}
\end{equation}
They are idempotent and orthogonal, that is
\begin{equation}\label{orthog-1}
	\mathcal{P}^{(i,a)\phantom{\mu\nu}\alpha\beta}_{\phantom{(i,a)}\mu\nu} \mathcal{P}^{(j,b)\phantom{\alpha\beta}\rho\sigma}_{\phantom{(j,b)}\alpha\beta} =	\delta^{ij}\delta^{ab} \mathcal{P}^{(i,a)\phantom{\mu\nu}\rho\sigma}_{\phantom{(i,a)}\mu\nu}\,,
\end{equation}
and form a complete set
\begin{equation}\label{completeness-rel}
	\mathcal{P}^{(2)}_{\phantom{(2)}\mu\nu\rho\sigma} + \mathcal{P}^{(1)}_{\phantom{(1)}\mu\nu\rho\sigma} + \mathcal{P}^{(0,s)}_{\phantom{(0,s)}\mu\nu\rho\sigma} + \mathcal{P}^{(0,w)}_{\phantom{(0,w)}\mu\nu\rho\sigma}=\mathbbm{1}_{\mu\nu\rho\sigma} \,.
\end{equation}
Since the projectors are idempotent, their trace equals their rank. This means that the trace is equal to the dimension of the corresponding irreducible representation (\ie{}  $2j+1$):
\begin{equation}
\begin{aligned}
\mathbbm{1}^{\mu\nu\rho\sigma} \mathcal{P}^{(2)}_{\phantom{(2)}\mu\nu\rho\sigma}&=5=2(2)+1 \, , \\
\mathbbm{1}^{\mu\nu\rho\sigma} \mathcal{P}^{(1)}_{\phantom{(1)}\mu\nu\rho\sigma}&=3=2(1)+1 \, , \\
\mathbbm{1}^{\mu\nu\rho\sigma} \mathcal{P}^{(0,s)}_{\phantom{(0,s)}\mu\nu\rho\sigma}&=1=2(0)+1 \, , \\
\mathbbm{1}^{\mu\nu\rho\sigma}  \mathcal{P}^{(0,w)}_{\phantom{(0,w)}\mu\nu\rho\sigma}&=1=2(0)+1 \, ,
\end{aligned}
\end{equation}
which means that $\mathcal{P}^{(2)}$ projects along the spin-two component (the traceless $h_{ij}$), $\mathcal{P}^{(1)}$  along the spin-one ($h_{0i}$), $\mathcal{P}^{(0,s)}$ along one of the spin-zero (spatial trace) and  $\mathcal{P}^{(0,w)}$ along the other spin-zero ($h_{00}$).

Unlike the case of the photon, the complete set of projectors is not sufficient to form a basis in the space of symmetric rank-four tensors. Indeed, we can notice that the basis element $B^{(3)}_{\phantom{(3)}\mu\nu\rho\sigma}$ in \eqref{element-basis-3} cannot be obtained from the four projectors introduced above. This means that we need to add an additional element to close the basis, and we choose it as follows
\begin{equation}
\mathcal{P}^{(0,\times)}_{\phantom{(0,\times)}\mu\nu\rho\sigma}= \mathcal{P}^{(0,sw)}_{\phantom{(0,sw)}\mu\nu\rho\sigma} + \mathcal{P}^{(0,ws)}_{\phantom{(0,ws)}\mu\nu\rho\sigma}\,,
\end{equation}
where 
\begin{equation}
\mathcal{P}^{(0,sw)}_{\phantom{(0,sw)}\mu\nu\rho\sigma}=\frac{1}{\sqrt{3}}\theta_{\mu\nu}\omega_{\rho\sigma} \,,\qquad \mathcal{P}^{(0,ws)}_{\phantom{(0,ws)}\mu\nu\rho\sigma}=\frac{1}{\sqrt{3}}\omega_{\mu\nu}\theta_{\rho\sigma}\,,
\end{equation}
from which we can reconstruct the missing element $B^{(3)}_{\phantom{(3)}\mu\nu\rho\sigma}$ to close the basis.

Let us emphasize that the operators $\mathcal{P}^{(0,sw)}$ and $\mathcal{P}^{(0,ws)}$ are not projectors, indeed they are not idempotent, do not contribute to any completeness relation, and are not orthogonal to the spin projectors. However, they satisfy some relations which together with those in \eqref{orthog-1} can be written in the following compact form:
\begin{equation}\label{orthog-2}
	\mathcal{P}^{(i,ab)\phantom{\mu\nu}\alpha\beta}_{\phantom{(i,ab)}\mu\nu} 	\mathcal{P}^{(j,cd)\phantom{\alpha\beta}\rho\sigma}_{\phantom{(j,cd)}\alpha\beta} = \delta^{ij}\delta^{bc}\mathcal{P}^{(i,ad)\phantom{\mu\nu}\rho\sigma}_{\phantom{(i,ad)}\mu\nu} \, ,
\end{equation}
where the notation and conventions for the labels is explained in the appendix below \eqref{orthog-2-app}.

Using the completeness relation~\eqref{commut-relat} and the identities
\begin{equation}
\begin{aligned}
\eta_{\mu\nu}\eta_{\rho\sigma}&=\left(3\mathcal{P}^{(0,s)}+\mathcal{P}^{(0,w)}+\sqrt{3}\mathcal{P}^{(0,\times)}\right)_{\mu\nu\rho\sigma} \, , \\
\eta_{\mu\nu}\omega_{\rho\sigma}+\eta_{\rho\sigma}\omega_{\mu\nu}&=\left(\sqrt{3}\mathcal{P}^{(0,\times)}+2\mathcal{P}^{(0,w)}\right)_{\mu\nu\rho\sigma}\,,\\
\frac{1}{2}\left(\eta_{\mu\rho}\omega_{\nu\sigma}+\eta_{\mu\sigma}\omega_{\nu\rho}+\eta_{\nu\sigma}\omega_{\mu\rho}+\eta_{\nu\rho}\omega_{\mu\sigma}\right)&=\left(\mathcal{P}^{(1)}+2\mathcal{P}^{(0,w)}\right)_{\mu\nu\rho\sigma}\,,
\end{aligned}
\end{equation}
we can rewrite the momentum-space kinetic operator with the de Donder gauge fixing~\eqref{kinetic-oper-gf} as 
\begin{equation}\label{kinetic-spin-proj}
\begin{aligned}
\tilde{\mathbb{K}}^{\mu\nu\rho\sigma}(p)=-p^2\Bigg[ &\mathcal{P}^{(2)\, \mu\nu\rho\sigma} +\frac{1}{\GFalpha} \mathcal{P}^{(1)\,\mu\nu\rho\sigma}+\left(\frac{3}{2\GFalpha}-2\right) \mathcal{P}^{(0,s)\,\mu\nu\rho\sigma}\\
&+\frac{1}{2\GFalpha} \mathcal{P}^{(0,w)\, \mu\nu\rho\sigma}-\frac{\sqrt{3}}{2\GFalpha} \mathcal{P}^{(0,\times)\, \mu\nu\rho\sigma}\Bigg] \, .
\end{aligned}
\end{equation}

The propagator can be found by first expressing $\propG_{\mu\nu\rho\sigma}(p)$ as a linear combination of the basis elements with with some unknown coefficients, \ie{},
\begin{equation}\label{proapag-basis-pojec}
\begin{aligned}
    \propG_{\mu\nu\rho\sigma} (p) &= \,A(p) \mathcal{P}^{(2)}_{\phantom{(2)}\mu\nu\rho\sigma} + B(p) \mathcal{P}^{(1)}_{\phantom{(1)}\mu\nu\rho\sigma} + C(p) \mathcal{P}^{(0,s)}_{\phantom{(0,s)}\mu\nu\rho\sigma} \\
    &\qquad+ D(p) \mathcal{P}^{(0,w)}_{\phantom{(0,w)}\mu\nu\rho\sigma} + E(p) \mathcal{P}^{(0,\times)}_{\phantom{(0,\times)}\mu\nu\rho\sigma}\,,
    \end{aligned}
\end{equation}
substituting the latter into~\eqref{propag-inversion} and solving the tensor equation for the unknown coefficients. The relations~\eqref{orthog-2} make this computation very straightforward, and it can be shown that 
\begin{equation}		
A(p)=-\frac{i}{p^2}\,,\quad B(p)=-\frac{i}{p^2}\GFalpha\,,\quad C(p)=\frac{i}{2p^2}\,,\quad D(p)=-\frac{i}{p^2}\left(\frac{4\GFalpha-3}{2}\right)\,,\quad E(p)=\frac{i}{p^2}\frac{\sqrt{3}}{2}\,,  
\end{equation}
which give the following expression for the graviton propagator in a generic de Donder gauge fixing:
\begin{tcolorbox}
\begin{equation}\label{propag-spin-proj}
\begin{aligned}
	\propG_{\mu\nu\rho\sigma}(p)=-\frac{i}{p^2}\Bigg[ & \mathcal{P}^{(2)}_{\phantom{(2)}\mu\nu\rho\sigma}-\frac{1}{2} \mathcal{P}^{(0,s)}_{\phantom{(0,s)}\mu\nu\rho\sigma} +\GFalpha \mathcal{P}^{(1)}_{\phantom{(1)}\mu\nu\rho\sigma} \\
&+ \frac{4\GFalpha-3}{2} \mathcal{P}^{(0,w)}_{\phantom{(0,w)}\mu\nu\rho\sigma}-\frac{\sqrt{3}}{2} \mathcal{P}^{(0,\times)}_{\phantom{(0,\times)}\mu\nu\rho\sigma}\Bigg]\,.
\end{aligned}
\end{equation}
\end{tcolorbox}

Note that the propagator contains all ten (physical and unphysical) degrees of freedom. However, $\mathcal{P}^{(1)}$, $\mathcal{P}^{(0,w)}$, and also $\mathcal{P}^{(0,\times)}$, are proportional to the four-momentum $p_\mu$, therefore they do not contribute to an amplitude when we contract the propagator with some external conserved stress-energy tensor. By contrast, $\mathcal{P}^{(2)}$ and $\mathcal{P}^{(0,s)}$ do contribute and carry the off-shell degrees of freedom of the graviton field. Indeed, the graviton has \textit{six} degrees of freedom off-shell: five coming from the spin-two  ($j_{z}=+2,+1,0,-1,-2$) and one from the spin-zero. What happens when going on-shell is that, due to gauge invariance, the contribution of the $j_z=\pm 1$ helicities vanish and the longitudinal component $j_z=0$ is canceled by an equal term coming from the spin-zero projector, thus only the two helicities $\pm 2$ contribute on-shell. Similarly to the case of the photon propagator, this counting of the off-shell and on-shell degrees of freedom is consistent with the fact that gauge invariance hits twice: four degrees of freedom are killed purely off-shell (that is why we have six of them off-shell) and additional four degrees of freedom are eliminated on-shell.

Therefore, the gauge-independent spin structure of  the graviton propagator is given by
\begin{equation}\label{propag-spin-proj-2-and-0}
	\propG^{(\text{gauge-ind})}_{\mu\nu\rho\sigma}(p)= -\frac{i}{p^2}\left[ \mathcal{P}^{(2)}_{\phantom{(2)}\mu\nu\rho\sigma}-\frac{1}{2}\mathcal{P}^{(0,s)}_{\phantom{(0,s)}\mu\nu\rho\sigma}\right]\,.
\end{equation}

Let us also observe that the spin-zero component comes with a minus sign unlike the spin-two that comes with the usual positive sign (up to our convention for the Feynman rule that requires the factor $-i$). In general, such type of opposite signs in the propagator may create troubles. However, in this case the minus sign is harmless because the spin-zero is not a propagating degree of freedom on-shell and, moreover, its opposite sign is actually necessary to cancel the longitudinal component of the spin-two projector~\cite{Dicus:2004rt}, and to consistently obtain the correct counting of degrees of freedom on-shell,\footnote{It is worth to mention that in theories of massive gravity, the flat propagator is given by $(-i)\mathcal{P}^{(2)}_{\phantom{(2)}\mu\nu\rho\sigma}/(p^2+m^2)$, where $m$ is the mass of the massive graviton. In this case, the number of off-shell and on-shell degrees of freedom coincides and is equal to five. We can notice that the naive massless limit $m\to 0$ does not recover the graviton propagator in \ac{GR} because there is no spin-zero projector to start with. This is the well-known Van Dam-Veltman-Zakharov discontinuity~\cite{vanDam:1970vg,Zakharov:1970cc}.} as explained before. In~\cref{sec:lecture4} we will encounter propagators with additional components carrying opposite signs that can propagate on-shell. In this case, a more careful analysis is needed in order to understand whether such type of propagator can be physically viable.

\subsubsection{Canonical quantization}

We can implement the canonical quantization for the free graviton field $h_{\mu\nu}$ by following similar steps as in the case of the photon field. Since we have already identified the independent field components, for example those in \eqref{polarization-helocit} if we work in the helicity basis, we can promote $h_{\mu\nu}$ to an operator and decompose it in terms of creation and annihilation operators by summing over the two physical helicity eigenvalues $\pm2$.

\subsubsubsection*{Commutation relations} 

The quantum graviton field can be written as an infinite superposition of plane waves weighted by the polarization tensors in the helicity basis and the annihilation/creation operators:
\begin{equation}\label{quantum-field}
h_{\mu\nu}(x)=\sum_{\lambda=+2,-2} \int \frac{{\rmd}^3p}{(2\pi)^3}\frac{1}{2\omega_{\vec{p}}}\left(a_{\vec{p},\lambda} \polarizationtensor_{\mu\nu}^{(\lambda)} e^{ip\cdot x}+ a_{\vec{p},\lambda}^{\dagger} \polarizationtensor_{\mu\nu}^{(\lambda)\ast} e^{-ip\cdot x}  \right)\,,
\end{equation}
where $\omega_{\vec{p}}=\sqrt{\vec{p}^{\,2}}=|\vec{p}|$ and $\polarizationtensor_{\mu\nu}^{(\pm 2)}$ were defined in \eqref{helicity-polariz}. 
The annihilation and creation operators $a_{\vec{p},\lambda}$ and $a_{\vec{p},\lambda}^{\dagger}$, respectively, satisfy the following commutation relations:
\begin{equation}\label{commut-relat}
\big[a_{\vec{p},\lambda},a_{\vec{p}\,',\lambda'}\big]=0=\big[a_{\vec{p},\lambda}^{\dagger},a_{\vec{p}\,',\lambda'}^{\dagger}\big]\,,\qquad \big[a_{\vec{p},\lambda},a_{\vec{p}\,',\lambda'}^{\dagger}\big]=2\omega_{\vec{p}}(2\pi)^3 \delta_{\lambda\lambda'}\delta^{(3)}\big(\vec{p}-\vec{p}\,'\big)\,.
\end{equation}

We can define a unique Poincar\'e-invariant (non-interacting) vacuum $\left| 0 \right\rangle $ as
\begin{equation}
a_{\vec{p},\lambda}\left| 0 \right\rangle =0\,.
\end{equation}
Furthermore, we can construct states populated by free particles called gravitons by acting with the creation operator on the vacuum. The first state on the top of the vacuum is the one containing a single graviton with momentum $\vec{p}$ and helicity $\lambda$, and it is given by
\begin{equation}
\left| \vec{p},\lambda \right\rangle = a^\dagger_{\vec{p},\lambda}\left| 0 \right\rangle \,.
\end{equation}

\subsubsubsection*{Issues with self-interactions} 

The above quantization procedure works very well in the case of a free theory, but things might get very complicated and unclear when self-interactions $\orderneglected(\kappa^{n-2}h^3)$ are included. Complications are due to both non-linearities and gauge symmetry. It is true that we have managed to identify the physical on-shell degrees of freedom, \ie{} the states that would be attached to external legs in a Feynman diagram. However, degrees of freedom that do not propagate on-shell can still contribute to virtual processes and appear off-shell in an internal line propagator. Furthermore, to prove the unitarity of the $S$-matrix, generally speaking, we have to show that imaginary parts of loop diagrams (left-hand side of the optical theorem) are equal to the sum over cut diagrams of lower order (right-hand side of the optical theorem); see \cref{sec:app-unitarity}. In the cutting procedure, internal lines become external, and unwanted degrees of freedom could appear on-shell. This situation would be catastrophic for the consistency of the theory and its viability.

This type of difficulties were first noticed by Feynman in both Yang-Mills and \ac{GR}~\cite{Feynman:1963ax}. He realized that unitarity could be restored by manually adding new diagrams containing loops of spin-zero particles carrying $-1$ factors as if they obeyed fermionic statistics. Soon after, de Witt~\cite{DeWitt:1967ub}, Faddeev and Popov~\cite{Faddeev:1967fc} realized that these fields with the opposite statistics arise naturally from a path integral construction in gauge theories. They are now named \textit{Faddeev-Popov ghost fields}: they were probably called ghosts because they only appear as internal lines and never contribute as external on-shell degrees of freedom.\footnote{In \cref{sec:lecture4} we will use the word ``ghost'' with a different meaning, \ie{} for fields whose kinetic term (or their propagator) has the wrong sign. So, please, do not get confused by the two different meanings!}

It is worth mentioning that it is possible to find a gauge in which the Faddeev-Popov fields fully decouple, and the canonical quantization becomes more doable. This gauge is sometime called axial gauge. While this could be useful in Yang-Mills theory, working with the axial gauge in the gravitational case is horrible because the renormalization procedure becomes more obscure due to the appearance of non-covariant counterterms that one has to carefully take care of~\cite{Capper:1981rc, Capper:1982vk}.

In this section, we do not really need the Faddeev-Popov fields. However, since in the next section we will start introducing gravitational self-interactions, for completeness we will briefly introduce them by mentioning some of the relevant features and explicitly show that they never appear as propagating on-shell degrees of freedom.

\subsection{GR as a QFT: interacting theory}\label{sec:quant-II}

So far we have mainly focused on the quantum structure of the kinetic term and degrees of freedom in \ac{GR}. However, to do physics we need to take into account interactions, and that is what we are going to do in this section. In \cref{sec:faddeev} we introduce the Faddeev-Popov ghost fields in both cases of covariant and non-covariant gauge fixing. In \cref{sec:unitarity} we discuss some aspects of unitarity. Finally, in \cref{sec:failure} we prove that \ac{GR} is perturbatively non-renormalizable, and analyze the structure of \ac{UV} divergences.

\subsubsection{Faddeev-Popov fields}\label{sec:faddeev}

The action for the Faddeev-Popov fields can be found by determining how the gauge-fixing term transforms under a gauge transformation, \ie{} under the diffeomorphism~\eqref{diff-h} in the case of Einstein's gravity. The full gravitational action including gauge fixing $S_{\rm gf}$ and Faddeev-Popov ghost term $S_{\rm gh}$ is given by
\begin{equation}
S[g,\eta,h]=\frac{1}{2\kappa^2}\int {\rmd}^4x \sqrt{-g} \, R -\frac{1}{\GFalpha} \int {\rmd}^4x \, \GFcondition_\mu \eta^{\mu\nu} \GFcondition_{\nu} + S_{\rm gh}[g,\eta,h]\,,	
\end{equation}
where\footnote{To be more precise, we should write~\eqref{FP-action} as a double integral $\int{\rmd}^4x {\rmd}^4y$ because the functional derivative depends on two spacetime points $x$ and $y$, \ie{}  $ \frac{\delta \GFcondition_\mu(x)}{\delta \zeta_\nu(y)}$, and Dirac deltas $\delta^{(4)}(x-y)$ need to be taken into account to kill one of the integrations. However, as already mentioned when we expanded the gravitational action in metric fluctuations, for simplicity we neglect the four-dimensional Dirac deltas coming from the functional derivatives and assume that the additional integrations are carried out.}
\begin{equation}\label{FP-action}
S_{\rm gh}[g,\eta,h]=\int {\rmd}^4x \, \bar{c}^\mu \frac{\delta \GFcondition_\mu}{\delta \zeta_\nu}c_\nu\,,	
\end{equation}
$c_\mu$ is called ghost field, while $\bar{c}_\mu$ is the anti-ghost field, and they are anti-commuting. Here we only consider the Minkowski background, but the same construction can be easily generalized to a generic background metric $\bar{g}_{\mu\nu}$ (see \cref{sect:ALESSIABENJAMIN_EH}).

We can rewrite the functional derivative of the gauge fixing as
\begin{equation}
\frac{\delta \GFcondition_\mu}{\delta \zeta_\nu}=\frac{\delta \GFcondition_\mu}{\delta h_{\rho\sigma}}\frac{\delta h_{\rho\sigma}}{\delta \zeta_\nu}=-\frac{\delta \GFcondition_\mu}{\delta h_{\rho\sigma}} \left(\delta_\sigma^{\phantom{\sigma}\nu} \covD_\rho + \delta_\rho^{\phantom{\rho}\nu} \covD_\sigma \right)\,.
\end{equation}
Substituting this into the action~\eqref{FP-action}, we obtain
\begin{equation}\label{FP-action-2}
	S_{\rm gh}[g,\eta,h]=-\int {\rmd}^4x \, \bar{c}^\mu \frac{\delta \GFcondition_\mu}{\delta h_{\rho\sigma}}\left(\covD_\rho c_\sigma + \covD_\sigma c_\rho \right)\,.
\end{equation}

To evaluate $\frac{\delta \GFcondition_\mu}{\delta h_{\rho\sigma}}$ we need to first tell which gauge fixing we are working with. Let us now consider both cases analyzed above when we derived the propagator: first, the covariant gauge fixing in \eqref{de-donder-gauge-fix} and second, the non-covariant one in \eqref{non-covariant-gauge-fix}.

\paragraph{Covariant gauge fixing.}

To take the functional derivative it is convenient to first symmetrize $\GFcondition_\mu$ in \eqref{de-donder-gauge-fix} as follows
\begin{equation}
\GFcondition_\mu= \left[\frac{1}{2}\left(\delta_\mu^{\phantom{\mu}\alpha} \partial^\beta + \delta_\mu^{\phantom{\mu}\beta} \partial^\alpha \right)-\frac{1}{2}\eta^{\alpha\beta}\partial_\mu  \right] h_{\alpha\beta}\,.
\end{equation}
Then, the derivative with respect to $h_{\rho\sigma}$ is
\begin{equation}
	\frac{\delta \GFcondition_\mu}{\delta h_{\rho\sigma}}= -\left[\frac{1}{2}\left(\delta_\mu^{\phantom{\mu}\sigma} \partial^\rho + \delta_\mu^{\phantom{\mu}\rho} \partial^\sigma \right)-\frac{1}{2}\eta^{\rho\sigma}\partial_\mu  \right] \, .
\end{equation}
Substituting this into the action~\eqref{FP-action} we get the following expression for the Faddeev-Popov action:
\begin{equation}
S_{\rm gh}=\int {\rmd}^4x \, \bar{c}^\mu \left(\partial^\rho \covD_\rho c_\mu +\partial^\rho \covD_\mu c_\rho -\partial_\mu \covD^\rho c_\rho  \right)\,.
\end{equation}

If we expand in metric fluctuations, we can identify the zeroth order kinetic term of the Faddeev-Popov fields around the Minkowski background:
\begin{equation}\label{kinetic-FP-cov}
	S_{\rm gh}=\int {\rmd}^4x \, \bar{c}^\mu \Box c_\mu +\orderneglected(\kappa h) \,.
\end{equation}
The interaction contribution $\orderneglected(\kappa h)$ contains only terms that are linear in $h_{\mu\nu}$, $c_\mu$ and $\bar{c}^\mu$, that is, the graviton, the Faddeev-Popov ghost and anti-ghost fields couple only via a cubic vertex that is linear in the coupling $\kappa$. This fact becomes manifest if we use \eqref{diff-h} for the expression of $\covD_\rho c_\sigma + \covD_\sigma c_\rho$ in \eqref{FP-action-2}.

\paragraph{Non-covariant gauge fixing.} In the case of the non-covariant gauge fixing in \eqref{non-covariant-gauge-fix}, symmetrizing we get
\begin{equation}
	\GFcondition_\mu= \frac{1}{2}\left(\delta_\mu^{\phantom{\mu}\alpha} \delta_i^{\phantom{i}\beta} + \delta_\mu^{\phantom{\mu}\beta} \delta_i^{\phantom{i}\alpha} \right)\delta_\nu^{\phantom{\nu}i} \partial^\nu h_{\alpha\beta}\,,
\end{equation}
and the derivative with respect to $h_{\rho\sigma}$ gives
\begin{equation}
	\frac{\delta \GFcondition_\mu}{\delta h_{\rho\sigma}}= -\frac{1}{2}\left(\delta_\mu^{\phantom{\mu}\rho} \delta_i^{\phantom{i}\sigma} + \delta_\mu^{\phantom{\mu}\sigma} \delta_i^{\phantom{i}\rho} \right)\delta_\nu^{\phantom{\nu}i} \partial^\nu\,.
\end{equation}
Substituting into the action~\eqref{FP-action} we get
\begin{equation}
	S_{\rm gh}=\int {\rmd}^4x \, \bar{c}^\mu \left(\partial^i\covD_\mu c_i + \partial^i\covD_i c_\mu \right)\,,
\end{equation}
and if we expand in metric fluctuations we can identify the zeroth order kinetic term:
\begin{equation}\label{kinetic-FP-non-cov}
	S_{\rm gh}=\int {\rmd}^4x \, \bar{c}_\mu \left(\delta_i^{\phantom{i}\nu}\partial^i\partial^\mu + \eta^{\mu\nu} \partial^i\partial_i \right) c_\nu +\orderneglected(\kappa h) \,.
\end{equation}

\subsubsection{Unitarity}\label{sec:unitarity}

In our gravitational \ac{QFT}, we expect quantum probabilities to be conserved, for example the $S$-matrix operator should be unitary. One can actually show that the optical theorem is satisfied order by order in perturbation theory~\cite{Anselmi:2016fid}. Unitarity also requires that no gauge and Faddeev-Popov degrees of freedom propagate on-shell; instead, depending on the selected gauge fixing, they can propagate off-shell as internal lines in loop diagrams.

In what follows we show some tree-level computations to convince ourselves that unitarity is indeed preserved.

\subsubsubsection*{Tree-level optical theorem}

From \cref{sec:unitarity} we know that one of the implications of the optical theorem is that the imaginary part of an elastic amplitude must be positive. Let us verify this property for the simplest type of elastic amplitude we can imagine, that is, a graviton going into itself:
\begin{equation}
\scatteringamplitude_{1\to 1}(p^2)=(-i) (-i)^2 \polarizationtensor^{\ast\mu\nu}\propG_{\mu\nu\rho\sigma}(p^2)\polarizationtensor^{\rho\sigma}\,,
\end{equation}
where $(-i)^2$ comes from the two vertices and the third $-i$ comes from the overall multiplicative factor of the amplitude according to our convention for the Feynman diagrams.

We can choose any gauge we like for the propagator. For example, let us choose the expression in the Feynman gauge obtained in \eqref{propag-feynman-gauge}. Then, using the Sokhotski–Plemelj formula
\begin{equation}
    \frac{1}{p^2-i\epsilon}={\rm P.V.}\left(\frac{1}{p^2}\right)+i\pi\delta(p^2) \,,
\end{equation}
where ${\rm P.V.}$ stands for the Cauchy principal value, we obtain
\begin{equation}
	{\rm Im}\left[\scatteringamplitude_{1\to 1}(p^2)\right]= \pi \delta(p^2)\left[ \polarizationtensor^{\ast\mu\nu}\polarizationtensor_{\mu\nu}-\frac{1}{2}|\eta^{\mu\nu}\polarizationtensor_{\mu\nu}|^2\right]=\pi \delta(p^2) \polarizationtensor^{\ast\mu\nu}\polarizationtensor_{\mu\nu}\geq 0\,.
\end{equation}
In the last step we have used the fact that the Dirac delta $\delta(p^2)$ makes the graviton on-shell, and this implies that the trace of the polarization tensor is zero because $\polarizationtensor_{22}=-\polarizationtensor_{11}$. Therefore, we have shown that the imaginary part of the amplitude is positive, in agreement with the optical theorem.

\subsubsubsection*{Faddeev-Popov fields are unphysical}

As mentioned above, Faddeev-Popov fields are very important for the consistency of the theory, in particular their off-shell contribution is needed to cancel unwanted intermediate states on the right-hand side of the optical theorem, and thus to preserve perturbative unitarity. In fact, the contributions coming from the Faddeev-Popov ghost and anti-ghost fields always compensate the ones of the (non-residual and residual) gauge components of the graviton field~\cite{Feynman:1963ax, DeWitt:1967ub}. This also means that the ghost and anti-ghost states, together with the non-physical polarization states of the graviton, can be projected out of the physical Hilbert space in a way compatible with gauge invariance and unitarity.

Looking at the expression of the kinetic term in the covariant gauge~\eqref{kinetic-FP-cov}, it seems that a $1/\Box$ pole could arise after the inversion of the kinetic operator. However, this pole is not physical but a gauge artifact, similarly to the poles of the unphysical gauge components of the graviton propagator in the de Donder gauge~\eqref{propag-spin-proj}. The crucial point is that Faddeev-Popov ghosts can propagate off-shell through loops, but they never appear as on-shell degrees of freedom, otherwise unitarity would be violated.

As done for the graviton propagator in \cref{sec:grav-propag-non-cov}, we can work in the non-covariant gauge where the Faddeev-Popov propagator can be shown to have no pole at $p^2=0$~\cite{tHooft:1974toh, Veltman:1975vx}.
Choosing the frame $p^{\mu}=(p^0,0,0,p^3)$, the kinetic term of the Faddeev-Popov fields in the non-covariant gauge~\eqref{kinetic-FP-non-cov} can be written in the following form
\begin{equation}
S_{\rm gh}=\int {\rmd}^4x \, \bar{c}_\mu\mathbb{K}^{\mu\nu}_{\rm FP}c_\nu+\orderneglected(\kappa h)\,,
\end{equation}
where
\begin{equation}\label{kin-mat-FP}
	\mathbb{K}^{\mu\nu}_{\rm FP}=\left(\begin{array}{cccc}
		p^2_3&0&0&-\frac{1}{2}p^2_3 \\
		0&-p^2_3& 0&0 \\
		0&0&-p^2_3&0 \\
		-\frac{1}{2}p^2_3&0&0&-2p^2_3
	\end{array}\right)\,.
\end{equation}
Since the kinetic operator of the Faddeev-Popov fields in the non-covariant gauge does not depend on $p_0$, its inverse cannot depend on $p^2=-p_0^2+p^2_3$, therefore no pole at $p^2$ can appear in the propagator. Since this is an on-shell result, it will be valid in any other gauge. Thus, no Faddeev-Popov particle can ever appear on-shell.

\subsubsection{Failure of perturbative renormalizability}\label{sec:failure}

To determine whether a local \ac{QFT} is perturbatively renormalizable, we just need to look at the mass dimension of the interaction couplings (see \cref{sec:elemQFT}). From \eqref{interaction-terms} we know that the gravitational couplings in \ac{GR} are proportional to inverse powers of the Planck mass, \ie{} $\kappa^{n-2}=1/\MPl^{n-2}$ with $n\geq 3$, therefore the mass dimensions are always negative and the theory is perturbatively non-renormalizable:
\begin{equation}\label{mass-dim-grav-coupl}
\Delta_n\equiv \big[\kappa^{n-2}\big]=2-n<0\,,\qquad n\geq 3\,.
\end{equation}
Given a loop diagram $G$ containing $E$ external legs and $V_n$ vertices with $n\geq 3$ legs, the corresponding superficial degree of divergence is given by (see~\eqref{total-delta})
\begin{equation}\label{sup-degree-einstein}
\delta(G)=4-E-\sum_n V_n \Delta_n\,,
\end{equation}
and increases with the number of vertices since $\Delta_n<0$.

To check the failure of perturbative renormalizability even more explicitly, we can consider a generic $L$-loop integral and compute how it scales with the internal momenta at high energies. If we do that, we have
\begin{equation}\label{2L+L}
	{\int \underbrace{{\rmd}^4k\cdots {\rmd}^4k}_{L\text{-loops}}}\, \times {\underbrace{\frac{1}{k^2}\cdots \frac{1}{k^2}}_{I\text{-internal propagators}}} \times\, {\underbrace{k^{2}\cdots k^{2}}_{V\text{-vertices}}}\,\sim\,k^{2L+2(L-I+V)}=k^{2L+2}\,,
\end{equation}
where we have used the fact that the graviton propagator goes like $1/k^2$, that the vertex contains always two powers of momenta, and we chose the most divergent case for which all vertex momenta are internal. Thus, from \eqref{2L+L} we see that the \ac{UV} divergence of the integrals become worse if the number of loops increase, implying that the theory is perturbatively non-renormalizable.

We now analyze the structure of one-loop, two-loop and higher-loop divergences.

\subsubsection{One-loop divergences}\label{sec:one-loop-div-luca}

Loop computations in \ac{QG} are very hard, and an explicit derivation cannot be shown easily. However, thanks to symmetry and power counting arguments, we can easily guess the tensor structure of loop divergences without performing tedious computations which, instead, are needed to determine the exact numerical coefficients.

\begin{figure}[t!]
			\centering
                \begin{subfigure}[b]{0.45\textwidth}
                    \centering
                    \includegraphics[scale=0.4]{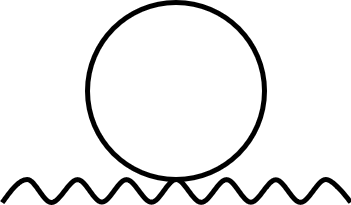}
                    \caption{}
                    \label{fig-tad-luca}
                \end{subfigure}
                \hfill
                \begin{subfigure}[b]{0.45\textwidth}
                    \centering
                    \includegraphics[scale=0.4]{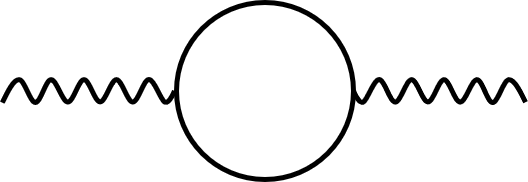}
                    \caption{}
                    \label{fig-bubb-luca}
                \end{subfigure}
                \hfill
                \begin{subfigure}[b]{0.45\textwidth}
                    \centering
                    \includegraphics[scale=0.4]{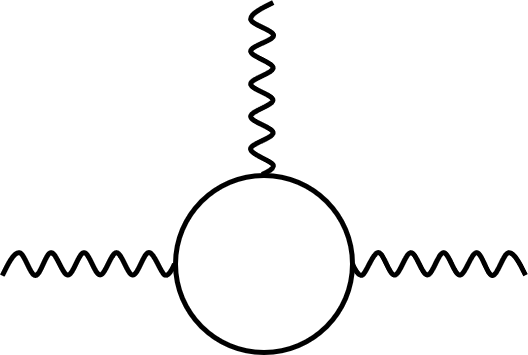}
                    \caption{}
                    \label{fig-one-loop-vert}
                \end{subfigure}
                \hfill
                \begin{subfigure}[b]{0.45\textwidth}
                    \centering
                    \includegraphics[scale=0.4]{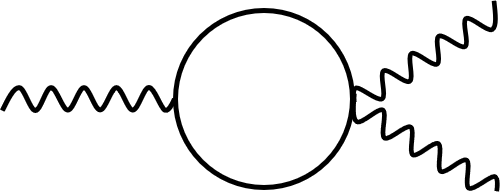}
                    \caption{}
                    \label{fig-one-loop-vert-2}
                \end{subfigure}
                \caption{(a) Tadpole diagram; (b) bubble diagram; (c) one-loop three-vertex diagram with external lines attached to three-vertices; (d) one-loop three-vertex diagram with one external line attached to a three-vertex and two external lines attached to a four-vertex. Wavy external lines correspond to gravitons, while the solid internal lines correspond to either graviton or Faddeev-Popov propagators.}
            \label{fig:three graphs}
		\end{figure}

Let us start with one-loop corrections to the graviton propagator in pure \ac{GR}, \ie{} with no coupling to matter, for the time being. In particular, we are interested in one-loop diagrams with two external graviton legs and either graviton or Faddeev-Popov propagators as internal lines (they both behave as $1/p^2$ in de Donder gauge). The loop will contain either one or two internal propagators, namely we can have either a \textit{tadpole} (\cref{fig-tad-luca}) or a \textit{bubble} (\cref{fig-bubb-luca}) diagram. Depending on the number of vertex momenta that are associated to internal lines, we can have different types of \ac{UV} divergences. 
\begin{itemize}
	
	\item The tadpole only contributes with power-law divergences. If all momenta in the vertex are internal, we have a quartic divergence; if only one vertex momentum is internal, we have a cubic divergence; if no vertex momentum is internal, we have a quadratic divergence.
	
	\item The bubble contributes with both power-law and logarithmic divergences. If all momenta in the two vertices are internal, we have a quartic divergence; if three vertex momenta are internal, we have a cubic divergence; if only two vertex momenta are internal, we have a quadratic divergence; if only one vertex momentum is internal, we have a linear divergence; if all vertex momenta are external, we have a logarithmic divergence.
	
\end{itemize}
%

		
The power-law divergences are set to zero if we work in dimensional regularization,\footnote{With other types of regularizations (such as a cutoff regularization), some of the power-law divergences contribute to the renormalization of Newton's constant and the cosmological constant, while others do not appear because of symmetry reason (\eg{} the cubic and linear divergences would violate parity invariance).} therefore we only need to focus on the logarithmic divergences coming from the bubble diagram which have the following schematic form: 
\begin{equation}\label{log-div}
pppp \int {\rmd}^4k \frac{1}{k^2}\frac{1}{k^2}k^2k^2\sim pppp \int \frac{{\rmd}k}{k}\,.
\end{equation}
This means that the counterterm needed to cancel the divergence has to contain four powers of the momentum, \ie{} in coordinate space fourth order derivatives acting on the metric fluctuation. Since the Einstein-Hilbert Lagrangian only contains second-order derivatives, the structure of the counterterm associated to~\eqref{log-div} is \textit{not} contained in the original bare Lagrangian. 

Furthermore, we can consider one-loop $n$-graviton vertex corrections with $n$ external gravitons. For example, the one-loop contributions to the three-graviton vertex are shown in \cref{fig-one-loop-vert} and~\cref{fig-one-loop-vert-2}. Also in this case, we can make a power counting analysis, and find the logarithmic divergences whose functional structure contains fourth order derivatives acting on the metric field.

The question we need to ask now is: what local terms containing fourth order derivatives and respecting diffeomorphism invariance can we add to the action? Answering this question is easy, and the possible fourth order terms contributing to the one-loop divergent part of the effective action are
\begin{equation}\label{fourth-order div}
\begin{aligned}
\Gamma^{(1)}_{\rm div}[g]=\frac{1}{\varepsilon}\int {\rmd}^4x \sqrt{-g}\Bigg[ &c_1 R^2+c_2 R_{\mu\nu}R^{\mu\nu}+c_3 R_{\mu\nu\rho\sigma}R^{\mu\nu\rho\sigma} \\
&+c_4 R_{\mu\nu\rho\sigma}R^{\mu\rho\nu\sigma}+c_5\covD_\mu\covD_\nu R^{\mu\nu}+c_6 \Box R\Bigg]\,,
\end{aligned}
\end{equation}
where $c_1,\dots, c_6$ are constant coefficients and $1/\varepsilon=1/(4-d)$ is the simple pole divergence in dimensional regularization. We now show that only two out of the six terms may contribute.

The sixth term is a total derivative and can be neglected up to surface terms. The fifth is also a total derivative and, actually, is equal to the sixth due to the Bianchi identity, \ie{} $\covD_\mu\covD_\nu R^{\mu\nu}=\Box R/2$. The fourth term can be written in terms of the third by using the cyclic identity $R_{\mu[\rho\nu\sigma]}=0$ which implies $R_{\mu\nu\rho\sigma}R^{\mu\rho\nu\sigma}=\frac{1}{2}R_{\mu\nu\rho\sigma}R^{\mu\nu\rho\sigma}$. Furthermore, using the fact that in $d=4$, the Gauss-Bonnet combination $\sqrt{-g}(R_{\mu\nu\rho\sigma}R^{\mu\nu\rho\sigma}-4R_{\mu\nu}R^{\mu\nu}+R^2)$ is locally a total derivative,\footnote{See~\cite{Percacci:2017fkn} for a pedagogical proof that the Gauss-Bonnet term is locally a total derivative.} we can express the square of the Riemann tensor in terms of $R^2$ and $R_{\mu\nu}R^{\mu\nu}$ plus boundary terms. Thus, in end we can recast the divergent part of the one-loop effective action into
\begin{tcolorbox}
\begin{equation}\label{fourth-order div-2}
	\Gamma^{(1)}_{\rm div}[g]=\frac{1}{\varepsilon}\int {\rmd}^4x \sqrt{-g}\left[ a R^2+ b R_{\mu\nu}R^{\mu\nu}\right]\,,
\end{equation}
\end{tcolorbox}
\noindent where $a\equiv c_1-c_3-c_4/2$ and $b\equiv c_2+4(c_3+c_4/2)$ are two redefined coefficients.

The exact numerical factors were computed for the first time by 't Hooft and Veltman in the de Donder-Feynman gauge~\cite{tHooft:1974toh}, and they found
\begin{equation}\label{thooft-1-loop}
a=-\frac{1}{(4\pi)^2} \frac{1}{120}\,,\qquad b= -\frac{1}{(4\pi)^2} \frac{7}{20}\,.
\end{equation}

Although we are neglecting the boundary terms in~\eqref{fourth-order div-2}, it is worth to mention that they are important for the renormalization of the theory, as they do contribute to the one-loop effective action~\cite{tHooft:1974toh,Goroff:1985sz,Goroff:1985th}. For example, the Gauss-Bonnet invariant appears with coefficient $-\frac{1}{(4\pi)^2\varepsilon}\frac{53}{90}$ and can be computed from the one-loop three-graviton vertex diagrams in \cref{fig-one-loop-vert} and \cref{fig-one-loop-vert-2}~\cite{Goroff:1985th}. In fact, the Gauss-Bonnet and $\Box R$ divergences also have important physical implications due to their contribution to the conformal anomaly when gravity is coupled to conformal matter~\cite{Percacci:2017fkn,Buchbinder:2021wzv}.

\subsubsubsection*{One-loop finiteness in pure gravity: field redefinition} 

It can be shown that in pure gravity, the one-loop divergent terms in \eqref{fourth-order div-2} can be absorbed by a ``renormalization'' of the metric field due to the fact that the integrand in \eqref{thooft-1-loop} is proportional to the vacuum field equations of the classical theory. More precisely, given a metric field $g_{\mu\nu}$ and the action
\begin{equation}
S'[g]=S_{\rm EH}[g]+\Gamma^{(1)}_{\rm div}[g]\,,
\end{equation}
we can find a field redefinition 
\begin{equation}\label{metric-transf}
g_{\mu\nu}\to g'_{\mu\nu}(g,a,b)=g_{\mu\nu}+\Delta g_{\mu\nu}(a,b)
\end{equation}
such that, perturbatively in $a$ and $b$, we have 
\begin{equation}
S_{\rm EH}[g]+\Gamma^{(1)}_{\rm div}[g]=S_{\rm EH}[g+\Delta g]+\orderneglected(a^2,b^2,ab)=S_{\rm EH}[g']+\orderneglected(a^2,b^2,ab)\,,
\end{equation}
where the terms $\orderneglected(a^2,b^2,ab)$ are expected to contribute to higher-loop and higher-curvature orders. While the bare metric $g_{\mu\nu}$ and the counterterm $\Delta g_{\mu\nu}$ are infinite,  the one-loop renormalized metric field $g'_{\mu\nu}$ is finite.

The expression of $\Delta g_{\mu\nu}$ can be found by imposing that the first-order variation of the Einstein-Hilbert action under the metric transformation~\eqref{metric-transf} is equal to $\Gamma^{(1)}_{\rm div}[g]$, \ie{} 
\begin{equation}
\Delta S_{\rm EH}[g]\equiv \int {\rmd}^4x \sqrt{-g} \, \frac{\delta S_{\rm EH}}{\delta g_{\mu\nu}}\Delta g_{\mu\nu} = \Gamma_{\rm div}^{(1)}[g]\,.
\end{equation}
Making the ansatz $\Delta g_{\mu\nu}=A g_{\mu\nu} R+B R_{\mu\nu}$, where $A$ and $B$ are two coefficients to be determined, we find
\begin{equation}
\begin{aligned}
	\Delta S_{\rm EH}[g]&= - \int {\rmd}^4x \sqrt{-g} \, \frac{1}{2\kappa^2}\left(R^{\mu\nu}-\frac{1}{2}g^{\mu\nu}R \right)\left( A g_{\mu\nu}R+B R_{\mu\nu} \right) \\
	&=-\int {\rmd}^4 x\sqrt{-g} \, \frac{1}{2\kappa ^2}\left[-\left( A+\frac{1}{2}B\right)R^2+B R_{\mu\nu}R^{\mu\nu}\right]\,,
\end{aligned}
\end{equation}
which equals $\Gamma^{(1)}_{\rm div}[g]$ if 
\begin{equation}\label{A-B-expressions}
\left\lbrace \begin{array}{l}
\displaystyle \frac{a}{\varepsilon}=\frac{1}{2\kappa^2}\left(A+\frac{B}{2}\right) \\[2.5mm]
\displaystyle \frac{b}{\varepsilon}=-\frac{B}{2\kappa^2}
\end{array}	
\right. \quad \Leftrightarrow \quad  \left\lbrace \begin{array}{l}
	\displaystyle A=\frac{\kappa^2}{\varepsilon} (2a+b) \\[2.5mm]
	\displaystyle B= -\frac{\kappa^2}{\varepsilon} 2b
\end{array}	
\right.\,.
\end{equation}
Therefore, the field redefinition needed to absorb the one-loop divergence in pure gravity reads
\begin{tcolorbox}
\begin{equation}\label{metric-transf-2}
\begin{aligned}
g'_{\mu\nu}&=g_{\mu\nu}+\frac{\kappa^2}{\varepsilon}\left[(2a+b) \, g_{\mu\nu} \, R -2b \, R_{\mu\nu}\right] \\
&= g_{\mu\nu}+\frac{\kappa^2}{10(4\pi)^2\varepsilon}\left(\frac{11}{3}g_{\mu\nu} \, R-7 \, R_{\mu\nu} \right)\,.
\end{aligned}
\end{equation}
\end{tcolorbox}

Note that, strictly speaking, we never imposed the field equations, but only used the fact that $\Gamma^{(1)}_{\rm div}$ is proportional to them, \ie{} the field redefinition was performed off-shell. Another way to rephrase the one-loop finiteness is saying that the one-loop divergent contribution vanishes on-shell, \ie{} $\Gamma^{(1)}_{\rm div}\big|_{\text{on-shell}}=0$ when $G_{\mu\nu}=0\Rightarrow R_{\mu\nu}=0$.

In the presence of a non-vanishing cosmological constant $\CC{}\neq 0$, pure Einstein's gravity is still one-loop renormalizable because the one-loop divergence can be absorbed by a logarithmic renormalization of the metric field and the Newton's constant off-shell. In this case, we would have
\begin{equation}
S_{\rm EH}^{(\CC{}\neq 0)}[g]+\Gamma^{(1)}_{\rm div}[g]=S_{\rm EH}^{(\CC{}\neq 0)}[g']-\frac{23}{(4\pi)^230\varepsilon}\int \rmd^4x \sqrt{-g'} \CC{} R(g')\,,
\end{equation}
where the last term can be absorbed into a renormalization of the Newton's constant. Another way to rephrase this argument is to say that on-shell (\ie{} $R_{\mu\nu}=g_{\mu\nu} \CC{}$) the one-loop divergence becomes $\Gamma^{(1)}_{\rm div}\big|_{\text{on-shell}}=-\frac{1}{(4\pi)^2\varepsilon}\frac{169}{25}\int\rmd^4x\sqrt{-g}\CC^2$, which can be absorbed by a logarithmic renormalization of the cosmological constant.

\paragraph{Remark.} Although the field redefinition~\eqref{metric-transf-2} does not change the physics at the level of the action, one may still wonder whether it changes the quantum properties of the theory, for example by introducing new terms in the path-integral through the generation of a non-trivial Jacobian from the transformation of the path-integral measure. The Jacobian associated to the field redefinition~\eqref{metric-transf-2} has the schematic form ${\det}(1+X)$ where $X$ takes into account derivatives of $\Delta g_{\mu\nu}$ with respect to $g_{\mu\nu}$. Introducing anti-commuting fields $\bar{q}$ and $q$, the Jacobian can be written as
\begin{equation}
{\det}(1+X)=\int {\mathcal D}\bar{q}{\mathcal D}q \, e^{i\int {\rmd}^4x\,\bar{q}(1+X)q}\,,
\end{equation}
from which it follows that the propagators of $\bar{q}$ and $q$ are constant. Since the field redefinition is local, $X$ is also local and the interactions involving $\bar{q}$ and $q$ are local as well. Therefore, all loops made with $\bar{q}$ and $q$ contain integrals of polynomials which vanish in dimensional regularization. It follows that the Jacobian is equal to one, which means that perturbatively the local field redefinition~\eqref{metric-transf-2} does not affect the path integral.

\subsubsubsection*{One-loop finiteness in pure gravity: gauge-fixing dependence}

Remember that the one-loop result in \eqref{thooft-1-loop} was obtained in the de Donder-Feynman gauge. In fact, the divergent part of the one-loop effective action depends on the gauge fixing, and this is the key feature behind the one-loop finiteness of pure Einstein's gravity, because there exists a gauge choice such that $\Gamma^{(1)}_{\rm div}\big|_{\text{off-shell}}=0$. In other words, if we generalize the de Donder gauge fixing~\eqref{de-donder-gauge-fix} to
\begin{equation}\label{de-donder-gauge-fix-generaliz}
	S_{\rm gf}[\eta,h]=-\frac{1}{\GFalpha}\int {\rmd}^4x \GFcondition_\mu \GFcondition^\mu\,,\qquad \GFcondition_{\mu}\equiv \partial_\nu h^{\phantom{\mu}\nu}_\mu - \frac{1+\GFbeta}{4}\partial_\mu h\,,
\end{equation}
the quadratic-curvature coefficients in \eqref{thooft-1-loop} turn out to be gauge-fixing dependent, \ie{} $a=a(\GFalpha,\GFbeta)$ and $b=b(\GFalpha,\GFbeta)$. Then, there exist specific values of the gauge parameters $\GFalpha_\ast$ and $\GFbeta_\ast$ such that~\cite{Kallosh:1978wt}
\begin{equation}
\Gamma^{(1)}_{\rm div}\left(\GFalpha_\ast,\GFbeta_\ast\right)\big|_{\text{off-shell}}=0\,.
\end{equation}

This fact related to the one-loop finiteness via the field redefinition discussed above. Since on-shell physics is independent of the gauge-fixing parameters, the difference of the one-loop divergent effective action evaluated for two different choices of gauge parameters must be proportional to the field equations, \ie{} it must vanish on-shell. Indeed, it can be shown that~\cite{Kallosh:1978wt,Buchbinder:2021wzv}
\begin{equation}
\Gamma^{(1)}_{\rm div}\left(\GFalpha_1,\GFbeta_1\right)-\Gamma^{(1)}_{\rm div}\left(\GFalpha_2,\GFbeta_2\right)=\frac{1}{\varepsilon} \int {\rmd}^4x f_{\mu\nu}\frac{\delta S_{\rm EH}}{\delta g_{\mu\nu}}\,,
\end{equation}
where $f_{\mu\nu}$ is some symmetric tensor function that depends on $\GFalpha_i$, $\GFbeta_i$.

If we choose $\GFalpha_2=\GFalpha_\ast$ and $\GFbeta_2=\GFbeta_\ast$ such that $\Gamma^{(1)}_{\rm div}\left(\GFalpha_2,\GFbeta_2\right)=0$, we get
\begin{equation}
	\Gamma^{(1)}_{\rm div}\left(\GFalpha_1,\GFbeta_1\right)=\frac{1}{\varepsilon} \int {\rmd}^4x f_{\mu\nu}\frac{\delta S_{\rm EH}}{\delta g_{\mu\nu}}\,,
\end{equation}
namely the divergent part of the one-loop effective action is proportional to the field equations. In particular, if we work in the de Donder-Feynman gauge $\GFalpha_1=1$, $\GFbeta_1=1$ we get the result in \eqref{thooft-1-loop} with $f_{\mu\nu}=Ag_{\mu\nu}R+B R_{\mu\nu}$, where $A$ and $B$ are those in \eqref{A-B-expressions}.

In summary, since there is a choice of gauge parameters such that the divergent part of the one-loop effective action vanishes, it follows that $\Gamma^{(1)}_{\rm div}$ is zero or proportional to the field equations of the classical theory. However, it should be remarked that the coefficient of the one-loop divergent contribution proportional to the Gauss-Bonnet invariant is gauge-independent and never vanishes.

\subsubsubsection*{One-loop divergences and coupling to matter}

The previous discussion strongly relies on the fact that matter couplings were absent. However, if the interactions with matter are switched on, then Einstein's gravity turns out to be one-loop non-renormalizable. In fact, the one-loop divergence \eqref{thooft-1-loop} is no longer proportional to the field equations since
\begin{equation}
\frac{\delta (S_{\rm EH}+S_{m})}{\delta g_{\mu\nu}}=-\frac{1}{2\kappa^2}\sqrt{-g}\left(R^{\mu\nu}-\frac{1}{2}g^{\mu\nu}R-\kappa^2T^{\mu\nu}\right)\,.
\end{equation}
In this case the field redefinition of the metric would generate terms like $\frac{\kappa^2}{\varepsilon} R_{\mu\nu}T^{\mu\nu}$ and $\frac{\kappa^2}{\varepsilon} R T$ that cannot be absorbed in the initial bare action $S_{\rm EH}+S_{m}$, and that on-shell would be proportional to squares of the stress-energy tensor, \ie{} $T_{\mu\nu}T^{\mu\nu}$ and $T^2$. This also means that there is no choice of the gauge-fixing parameters that would make the one-loop divergence vanish. 

One may hope that some special combinations of matter fields make the one-loop divergence disappear, but this generically does not happen, especially if we consider the physical scenario of \ac{SM} Lagrangians coupled to \ac{GR}.

In summary, pure Einstein's gravity is one-loop renormalizable but Einstein's gravity coupled to matter is not.

\subsubsection{Two-loop divergences}\label{sec:LUCA_twoloopdivs}

One might hope that the same cancellations could also occur at higher loops, but this is not the case. In fact, pure \ac{GR} is perturbatively non-renormalizable starting from two loops, as we now explain.

Making again a power-counting analysis, we can understand that the two-loop divergences can appear in different ways: sextic divergences that renormalize the cosmological constant, quartic divergences that are proportional to $R$ and renormalize Newton's constant, quadratic divergences proportional to four-derivative terms, \ie{} $R^2$ and $R_{\mu\nu}R^{\mu\nu}$, and logarithmic divergences multiplying terms with six derivatives acting on the metric. 
The latter is the only type of two-loop divergence that would show up in dimensional regularization. Since the vertices contain only two powers of momenta, the logarithmic two-loop divergences will come from two-loop diagrams containing three external gravitons on-shell.

The tensorial structure of the two-loop divergence as a function of the metric perturbation $h_{\mu\nu}$ and the derivatives $\partial_\mu$, with the three external gravitons taken on-shell, can be fully determined up to a numerical coefficient. Indeed, if we work in de Donder gauge, the graviton field is traceless and transverse (see \eqref{de-donder-graviton}), thus the only functional structure which contains six derivatives acting on the metric perturbation and does not vanish on-shell is~\cite{Goroff:1985sz,Goroff:1985th}
\begin{tcolorbox}
\begin{equation}\label{only-two-loop-struc}
\kappa^5 h^{\alpha\beta}\partial_\alpha \partial_\mu\partial_\nu h^{\rho\sigma}\partial_\beta \partial_\rho\partial_\sigma h^{\mu\nu}\,.
\end{equation}
\end{tcolorbox}
Note that other functional structures depending on the d'Alembertian $\Box=\partial_\mu \partial^\mu$ do not contribute because they vanish on-shell. The five powers of $\kappa$ in~\eqref{only-two-loop-struc} can be understood as follows: three powers are needed for the canonical normalization of the graviton field (\ie{} $\kappa h_{\mu\nu}$ is dimensionless), while the remaining two powers are needed to give the right dimension to the divergent part of the two-loop effective action. Therefore, we expect the covariant form of~\eqref{only-two-loop-struc} to be proportional to curvature invariants of mass dimension equal to six multiplied by $\kappa^2=1/\MPl^2$.

\begin{figure}[t!]
			\centering
                \begin{subfigure}[b]{0.3\textwidth}
                    \centering
                    \includegraphics[scale=0.4]{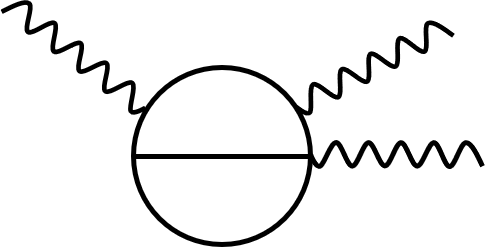}
                    \caption{}
                    \label{two-loop-1-luca}
                \end{subfigure}
                \hfill
                \begin{subfigure}[b]{0.3\textwidth}
                    \centering
                    \includegraphics[scale=0.4]{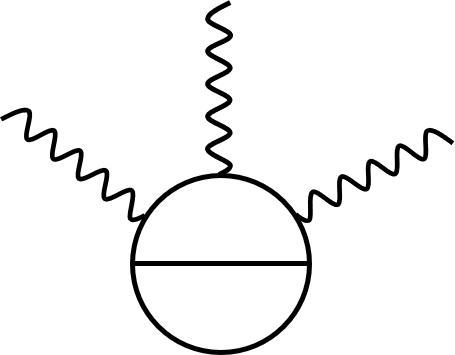}
                    \caption{}
                    \label{two-loop-2-luca}
                \end{subfigure}
                \hfill
                \begin{subfigure}[b]{0.3\textwidth}
                    \centering
                    \includegraphics[scale=0.4]{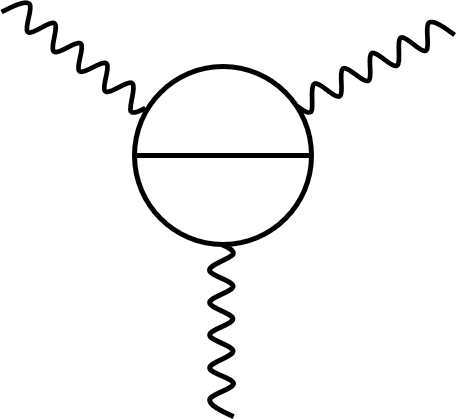}
                    \caption{}
                    \label{two-loop-3-luca}
                \end{subfigure}
                \hfill
                \begin{subfigure}[b]{0.3\textwidth}
                    \centering
                    \includegraphics[scale=0.4]{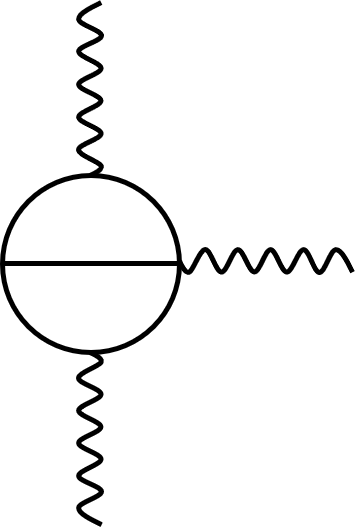}
                    \caption{}
                    \label{two-loop-4-luca}
                \end{subfigure}
                  \hfill
                \begin{subfigure}[b]{0.3\textwidth}
                    \centering
                    \includegraphics[scale=0.4]{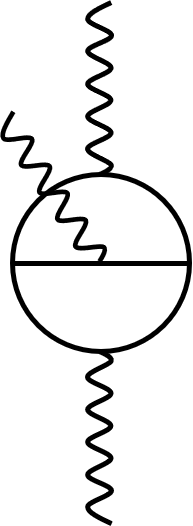}
                    \caption{}
                    \label{two-loop-5-luca}
                \end{subfigure}
                  \hfill
                \begin{subfigure}[b]{0.3\textwidth}
                    \centering
                    \includegraphics[scale=0.4]{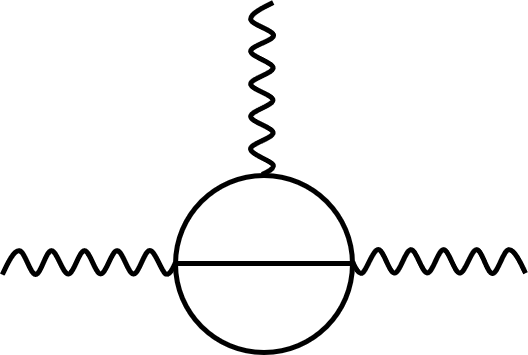}
                    \caption{}
                    \label{two-loop-6-luca}
                \end{subfigure}
                \hfill
                \begin{subfigure}[b]{0.45\textwidth}
                    \centering
                    \includegraphics[scale=0.4]{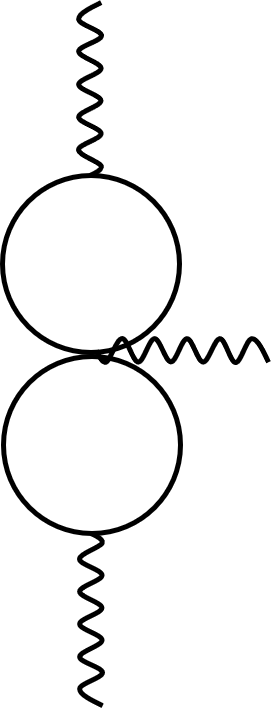}
                    \caption{}
                    \label{two-loop-7-luca}
                \end{subfigure}
                \hfill
                \begin{subfigure}[b]{0.45\textwidth}
                    \centering
                    \includegraphics[scale=0.4]{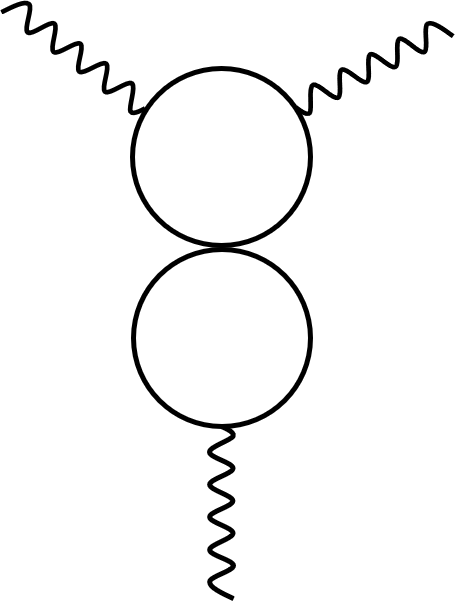}
                    \caption{}
                    \label{two-loop-8-luca}
                \end{subfigure}
                \caption{Set of diagrams that contribute with logarithmic divergences to the divergent part of the two-loop quantum effective action. Wavy external lines correspond to gravitons, while the solid internal lines correspond to either graviton or Faddeev-Popov propagators.}
            \label{fig:two-loop graphs}
		\end{figure}

All possible diagrams that could give a non-vanishing contribution proportional to~\eqref{only-two-loop-struc} are those with three external gravitons attached to three distinct vertices, and are shown in \cref{fig:two-loop graphs}. Two-loop diagrams with two external lines attached to the same vertex cannot generate two-loop divergences proportional to~\eqref{only-two-loop-struc}, thus we do not show them. Moreover, one can find that the divergent part of the two-loop diagram in \cref{two-loop-8-luca} vanishes, whereas the diagram in \cref{two-loop-7-luca} only contributes with a double pole divergence in dimensional regularization~\cite{Goroff:1985sz,Goroff:1985th}. However, we will show below that double pole divergences are absent on-shell.

		
As done in the one-loop case, we can find the covariant tensor structure of the divergent part of the two-loop effective action using the locality of counterterms and diffeomorphism invariance, whereas the exact numerical coefficient can only be determined by performing an explicit computation. Furthermore, without much effort, we can also determine the pole order of the two-loop divergence in dimensional regularization, \ie{} whether it is a simple pole $1/\varepsilon$, or a double pole $1/\varepsilon^2$. We first derive the tensor structure of the two-loop divergence, and then focus on the order of the pole.

\subsubsubsection*{Tensor structure of counterterms}

The two-loop counterterm must be proportional to something like
\begin{equation}
\Gamma_{\rm div}^{(2)}[g]\propto \kappa^2\int {\rmd}^4x\sqrt{-g}\, \mathcal{R}\,,
\end{equation}
where $\mathcal{R}$ is an invariant of mass dimension six that contains contractions of metric tensors, Riemann curvatures and covariant derivatives. The possible contractions, neglecting irrelevant boundary terms, are~\cite{Fulling:1992vm} 
\begin{equation}\label{six-dimens-counterterms}
\begin{aligned}
 \covD_\mu R\covD^\mu R\,,\quad \covD_\rho R_{\mu\nu}\covD^\rho R^{\mu\nu}\,,\quad R^3\,,\quad R R_{\mu\nu}R^{\mu\nu}\,,\quad R_{\mu\nu} R^{\mu\rho}R_\rho^{\phantom{\rho}\nu}\,,\quad R R_{\mu\nu\rho\sigma}R^{\mu\nu\rho\sigma} \,,  \\
 R_{\mu\nu}R_{\rho\sigma} R^{\mu\rho\nu\sigma}\,,\quad   R^{\phantom{\alpha}\beta}_\alpha R^{\alpha \mu\nu\rho}R_{\beta\mu\nu\rho} \,,\quad R_{\mu\nu\rho\sigma} R^{\mu\nu}_{\phantom{\mu\nu} \alpha\beta}R^{\rho\sigma\alpha\beta}\,,\quad R_{\mu\nu\rho\sigma} R^{\mu\phantom{\alpha}\rho}_{\phantom{\mu}\alpha\phantom{\rho}\beta} R^{\nu\alpha\sigma\beta} \,.
\end{aligned}
\end{equation}
First, we note that all terms except the last two are proportional to the classical vacuum field equations. This means that they can be absorbed by a redefinition of the metric field that generalizes~\eqref{metric-transf-2} with additional terms of mass dimension four, \eg{} $g_{\mu\nu}R^2$, $g_{\mu\nu} R_{\rho\sigma}R^{\rho\sigma}$, $R R_{\mu\nu}$, $ \covD_\mu\covD_\nu R$, and so on. Moreover, it can be shown that the last two cubic invariants in~\eqref{six-dimens-counterterms} are linearly dependent. This follows from the identity
\begin{equation}
R^{[\mu\nu}_{\phantom{[\mu\nu}\rho\sigma}R^{\rho\sigma}_{\phantom{\rho\sigma} \alpha\beta} R^{\alpha]\beta}_{\phantom{\alpha]\beta} \mu\nu}=0\,,
\end{equation}
that in four spacetime dimensions is always true due to antisymmetrization over five indices. More explicitly, it reads~\cite{Percacci:2017fkn}
\begin{equation}
0=4 R_{\mu\nu\rho\sigma} R^{\rho\sigma\alpha\beta} R^{\mu\nu}_{\phantom{\mu\nu} \alpha\beta}-8R_{\mu\nu\rho\sigma} R^{\mu\phantom{\alpha}\rho}_{\phantom{\mu}\alpha\phantom{\rho}\beta} R^{\nu\alpha\sigma\beta}+ (\text{terms proportional to $R_{\mu\nu}$ and $R$})\,.
\end{equation}
This means that up to contributions proportional to the vacuum field equations the last invariant in \eqref{six-dimens-counterterms} can be expressed in terms of the next-to-last one. Therefore, the tensor structure of the two-loop counterterm is given by
\begin{equation}
\Gamma_{\rm div}^{(2)}[g]\propto\kappa^2 \int{\rmd}^4x \sqrt{-g} \, R_{\mu\nu\rho\sigma} R^{\mu\nu}_{\phantom{\mu\nu} \alpha\beta}R^{\rho\sigma\alpha\beta} \,.
\end{equation}
What still remains to be determined is the pole order of the two-loop divergence and, of course, the exact numerical coefficient.

\subsubsubsection*{Absence of double pole divergence}

In general, in the framework of perturbative \ac{QFT}, a two-loop divergence can appear with both simple ($1/\varepsilon$) and double ($1/\varepsilon^2$) poles in dimensional regularization; the latter is determined by one-loop subtractions that must be implemented correctly to avoid the appearance of inconsistent non-local \ac{UV} divergences. It can actually be shown that if a \ac{QFT} is finite up to $N$ loops, then the $N+1$ loop divergence will contain only a simple pole~\cite{Chase:1982sf,Marcus:1984ei}. This means that since Einstein's gravity (without matter) is finite at one loop, the two-loop divergence must contain only a simple pole.

This result can be proven as follows. Since pure gravity is finite at one loop, if we implement the field redefinition we can then calculate the two-loop contribution to the effective action without using counterterms and one-loop subtractions. This means that the two-loop divergence will have the following schematic form:
\begin{equation}\label{two-loop-1st-expr}
\mu^{2\varepsilon}\left(\frac{C_1}{\varepsilon}+\frac{C_2}{\varepsilon^2}\right) \,,
\end{equation}
where $\mu$ is the renormalization scale whose power $2\varepsilon$ is needed to compensate the $2(4-\varepsilon)$ dimension of two-loop integrals in dimensional regularization, while $C_1$ and $C_2$ are local functions.

The two-loop calculation can also be performed in an alternative way, using counterterms and one-loop subtractions, \ie{} without making any field redefinition. In this case, the expression of the two-loop divergence will not contain any $\mu^{\varepsilon}$ and schematically reads
\begin{equation}\label{two-loop-end-expr}
\frac{C_1^\prime}{\varepsilon}+\frac{C_2^\prime}{\varepsilon^2} \,,
\end{equation}
where $C_1^\prime$ and $C_2^\prime$ are local functions.

Since the two computations must give the same result on-shell, we have
\begin{equation}
\left(1+2\varepsilon\ln\mu+\dots\right)\left(\frac{C_1}{\varepsilon}+\frac{C_2}{\varepsilon^2}\right)=\frac{C_1^\prime}{\varepsilon}+\frac{C_2^\prime}{\varepsilon^2}\,.
\end{equation}
Equating equal powers of $1/\varepsilon$, it follows that $C_2$ must vanish because $C_2^\prime$ is a local function and cannot contain any logarithmic dependence. We get $C_1=C_1^\prime\neq 0$ and $C_2=C_2^\prime=0$, so double poles are absent in the two-loop divergence of pure Einstein's gravity.

This result was confirmed by an explicit calculation of the diagrams in~\cref{fig:two-loop graphs}, and the final expression of the gauge-independent divergent part of the two-loop effective action, including the exact numerical coefficient, is given by~\cite{Goroff:1985sz,Goroff:1985th,vandeVen:1991gw}
\begin{tcolorbox}
\begin{equation}\label{goroff-sagnotti-counter-luca}
\Gamma_{\rm div}^{(2)}[g]=\frac{209}{2880(4\pi)^4}\frac{\kappa^2}{\varepsilon} \int{\rmd}^4x \sqrt{-g} \, R_{\mu\nu\rho\sigma} R^{\mu\nu}_{\phantom{\mu\nu} \alpha\beta}R^{\rho\sigma\alpha\beta} \,,
\end{equation}
\end{tcolorbox}
\noindent and is also known as the Goroff-Sagnotti counterterm. It is worth to mention that the full off-shell expression of the divergent part of the two-loop effective action contains double poles, but their tensor structure is proportional to the classical vacuum equations and vanishes on-shell. In other words, a field redefinition can be performed to reabsorb these divergences in a renormalization of the metric tensor, thus the only true gauge-independent divergence at two loops is the one in~\eqref{goroff-sagnotti-counter-luca}.

\subsubsection{Higher-loop divergences}

No explicit computations of loop divergences higher than two have been performed in the perturbative \ac{QFT} approach to \ac{GR}. However, it is still possible to determine the covariant form of the leading divergences up to numerical coefficients. Indeed, it can be easily shown that the leading divergent contribution to the $L$-loop effective action in $d=4$ is given by~\cite{vanNieuwenhuizen:1976vb}
\begin{equation}\label{structure-L-loo-div}
\Gamma_{\rm div}^{(L)}[g]\propto \kappa^{2L-2}\int {\rmd}^4x\sqrt{-g}\, \mathcal{R}\,,
\end{equation}
where now $\mathcal{R}$ is a local function of mass dimension equal to $2L+2$, independent of $\kappa$, and constructed in terms of $L-k+1$ Riemann tensors and $2k$ covariant derivatives, where $0\leq k\leq L$. 

The power of $\kappa$ can be determined as follows. Since an $n_i$-leg vertex is proportional to $\kappa^{n_i-2}$, an $L$-loop diagram with $V$ vertices will be proportional to $\kappa$ to the power
\begin{equation}\label{power-kappa-L-loop}
\sum_{i=1}^V(n_i-2)=\sum_{i=1}^Vn_i-2V\,.
\end{equation}
Using the fact that an internal propagator is always connected to two vertices, we can write
\begin{equation}
\sum_{i=1}^Vn_i=2I+E\,,
\end{equation}
where $I$ is the number of internal propagators and $E$ the number of external graviton legs. Substituting the last equation into \eqref{power-kappa-L-loop}, we get
\begin{equation}\label{power-kappa-L-loop-2}
\sum_{i=1}^V(n_i-2)=2I+E-2V=2L-2+E\,,
\end{equation}
where in the last step we used the topological identity $I-V=L-1$. Since $E$ powers of $\kappa$ are needed for the canonical normalization of $E$ external graviton fields, we are only left with $2L-2$, which is exactly the power of $\kappa$ shown in \eqref{structure-L-loo-div}.

Let us emphasize that the structure \eqref{structure-L-loo-div} is only valid for the leading $L$-loop divergences, while power law divergences are not captured. For example, in the case of two loops we can have quadratic power-law divergences that are proportional to $R^2$ and $R_{\mu\nu}R^{\mu\nu}$. Furthermore, it is also worth to mention that, while at two loops the double pole divergence is absent due to the one-loop finiteness of pure \ac{GR}, at loops higher than two similar cancellations do not happen, and all possible pole orders are expected to be present.   

\subsubsection{What to do next?}

At this point, several questions could be raised: what does the failure of perturbative renormalizability of \ac{GR} mean? Are we implementing the quantization of gravity in the right way? Is the perturbative approach insufficient for the quantization of \ac{GR}? Is the \ac{QFT} framework not suitable to describe quantum aspects of gravity? Are we working with the wrong set of degrees of freedom? Are we missing some enhanced or broken symmetry? Can the quantum and gravitational worlds be compatible at all?

\subsubsubsection*{\ac{EFT} of \ac{GR}} 

First of all, let us clarify that \ac{QFT} and \ac{GR} are compatible, at least in the low-energy regime: the failure of perturbative renormalizability does not imply that \ac{GR} cannot be treated as a \ac{QFT}. In fact, one can formulate a consistent \ac{EFT} of \ac{GR} that is valid up to some cutoff energy scale (\eg{}, the Planck mass $\MPl\sim 10^{18}$ GeV in pure gravity), in such a way that \ac{QG} predictions can be trusted up to finite errors proportional to inverse powers of the cutoff~\cite{Donoghue:1993eb,Donoghue:1994dn}. In this case, the local part of the gravitational Lagrangian will contain all possible infinite counterterms that are compatible with the symmetries of \ac{GR}:
\begin{equation}\label{EFT-action-Luca}
\begin{aligned}
S_{\rm EFT}=\int {\rmd}^4x\sqrt{-g}&\left[\frac{1}{2\kappa^2}\left(R-2\CC{}\right) +  a_1 R^2 + a_2 R_{\mu\nu}R^{\mu\nu}\right. \\
&\left.+a_3 \kappa^2 R^3+a_4\kappa^2 R_{\mu\nu\rho\sigma}R^{\rho\sigma}_{\phantom{\rho\sigma}\alpha\beta}R^{\alpha\beta\mu\nu}+\dots\right]\,,
\end{aligned}
\end{equation}
where the dots stand for all possible local contractions of metric tensors, Riemann tensors and covariant derivatives. If matter is absent, we know that all terms proportional to the classical field equations can be removed by a metric field redefinition, while if matter is present, the \ac{EFT} of \ac{GR} can also capture corrections coming from terms like $R^2$, $R_{\mu\nu}R^{\mu\nu}$, $R^3$, etc. The coefficients $a_1,a_2,a_3,a_4,\dots$ are dimensionless, and their renormalized value can be determined up to errors proportional to powers of $E/\MPl$, where $E$ is the characteristic energy scale of the physical process. When the coupling with matter is switched on, it can be shown that the \ac{UV} cutoff of the \ac{EFT} becomes lower than the Planck scale~\cite{Han:2004wt}; see also~\cref{sec:part_i}.

It is important to observe that in the \ac{EFT} of \ac{GR}, the additional operators in \eqref{EFT-action-Luca} do not introduce new degrees of freedom, that is, the massless graviton remains the \textit{only} physical particle in the gravitational spectrum.

One approach to \ac{QG} is to be (temporarily) happy with a low-energy \ac{EFT} description and to forget about any specific form of \ac{UV} completions and/or non-trivial departures from \ac{GR}. See \cref{sec:ANNA} for applications of the \ac{EFT} of \ac{GR}.

\subsubsubsection*{Non-perturbative renormalizability?}

Although \ac{GR} is perturbatively non-renormalizable, it might happen that the quantum theory possesses an interacting \ac{UV}  fixed point. This would imply that the Einstein-Hilbert action supplemented with all possible infinite tower of higher-curvature contractions that are compatible with diffeomorphism invariance is non-perturbatively renormalizable, a property that the perturbative \ac{QFT} approach would fail to capture. In this case, quantum aspects of gravity would be better described in the non-perturbative \ac{QFT} framework, and the corresponding \ac{QG} approach is known as \ac{ASQG}~\cite{Weinberg:1980gg}. Both continuum \ac{FRG} techniques~\cite{Percacci:2017fkn, Reuter:2019byg} and lattice methods~\cite{Loll:2019rdj} have been proposed as computational tools to prove the existence of an interacting fixed point. One such lattice method is known as \ac{CDT}, and it is sometimes also seen as a different \ac{QG} approach. 

In \cref{sec:ALESSIABENJAMIN}, the general notion of non-perturbative renormalization and the \ac{FRG} method are introduced, and their application to \ac{QG} is discussed.

\subsubsubsection*{Beyond-\ac{QFT} approaches?}

Another possibility could be that the framework of \ac{QFT} is not suitable for capturing quantum aspects of gravity at the fundamental level. While the \ac{EFT} treatment, which is \ac{QFT}-based, could be trusted to study quantum-gravitational physics at low energies, it might happen that the high-energy behavior of the gravitational interaction cannot be consistently described with the tools of \ac{QFT}. 
Along this direction, there are several types of proposals. On one side, there are approaches that try to formulate a theory of quantum spacetime at a full non-perturbative level: these include loop quantum gravity~\cite{Ashtekar:2021kfp}, causal set theory~\cite{Surya:2019ndm}, and non-commutative geometry~\cite{Chamseddine:2022rnn}, and in some of them spacetime is fundamentally discrete. On the other side, there is the framework of \ac{ST}~\cite{Green:2012oqa, Green:2012pqa} where the concept of a spacetime metric is somehow emergent from more fundamental \textit{stringy} degrees of freedom in terms of which one aims at formulating a unified \ac{UV}-complete quantum theory of all interactions.

An introduction to \ac{ST} is given in \cref{sec:IVANO}.

\subsubsubsection*{Insist a little more on the perturbative \ac{QFT} framework?}

While we might be strongly tempted to immediately abandon the perturbative \ac{QFT} framework after learning that \ac{GR} is perturbatively non-renormalizable, there is actually an obvious question we should still ask ourselves before starting to think about other approaches to \ac{QG}:
is the perturbative \ac{QFT} framework applied to gravity only useful for describing \ac{GR} as an \ac{EFT} in the low-energy regime? In the remainder of this section, we will show that the answer is negative.

In fact, a similar story happened for the other forces of nature, for example the weak interaction. In this case, Fermi theory turned out to be perturbatively non-renormalizable. Was this a problem for a consistent quantization of the weak interaction in the framework of \ac{QFT}? Obviously not! Indeed, the real story is that Fermi theory is a non-renormalizable \ac{EFT} valid at energy scales below $100$ GeV, and its perturbative completion is the electroweak theory which is a ``strictly'' renormalizable \ac{QFT}~\cite{Schwartz:2014sze}. Let us recall that in this context the word ``strictly'' means that the interaction couplings that govern the \ac{UV} behavior of the theory are dimensionless  (see~\cref{sec:app-renormaliz} for a detailed discussion of power-counting renormalizability and the definition of strict renormalizability).

It should be noted that in the last century, a very important theoretical achievement in fundamental physics has been the realization that ``strict'' renormalizability can be a powerful and predictive criterion for selecting fundamental Lagrangians in the framework of perturbative \ac{QFT}. In fact, all \ac{SM} interactions (electromagnetic, weak and strong) are described as ``strictly'' renormalizable \acp{QFT}. Therefore, the most logical and conservative question to ask is:
\begin{tcolorbox}
\centering Is there any ``strictly'' renormalizable \ac{QFT} of gravitational interaction? 
\end{tcolorbox}
The answer is positive: in $d=4$ there exists a unique ``strictly'' renormalizable \ac{QFT} of gravity that is metric compatible, has zero torsion and preserves the symmetries (diffeomorphism and parity invariance) of \ac{GR}. Its Lagrangian contains operators of mass dimensions up to four, in particular these include curvature invariants up to quadratic order. This theory is known as \textit{quadratic gravity}~\cite{Stelle:1976gc, Salvio:2018crh, Anselmi:2017ygm, Donoghue:2021cza, Holdom:2021hlo, Piva:2023bcf}.

Hence, it is important to understand what perturbative \ac{QFT} and strict renormalizability can still tell us about gravity at a more fundamental level and, in particular, whether they can still be considered a powerful framework and a predictive criterion to describe new physics beyond \ac{GR}. These are the questions we are going to critically assess in the last part of this section.

\subsection{Quadratic gravity}\label{sec:lecture4}

In this section, we aim at exploring several features of quadratic gravity. We will first introduce the classical action and the field equations. We will then move to quantum aspects, derive the propagator, determine the degrees of freedom, and show power-counting renormalizability. We will make a comparison between quadratic gravity and the \ac{EFT} of \ac{GR} in order to clarify some points that are sometimes misunderstood. We will also highlight the uniqueness and predictive power of the theory as consequences of strict renormalizability. Finally, we will discuss the main open questions in relation to the massive spin-two ghost and the high-energy behavior of the theory, around which debates are still ongoing. Due to the lack of ``spacetime'', in this last section we will provide less computational details than in previous ones. However, we will touch on all the relevant conceptual points and also make contact with experiments, while trying to keep the explanations pedagogical.

\subsubsection{Action and field equations}

We want to consider the general gravitational action that contains invariants of mass dimension up to four. From the analysis of one-loop divergences in \ac{GR} performed in \cref{sec:one-loop-div-luca}, we already know that the general expression in \eqref{fourth-order div} can be reduced to two independent curvature invariants up to boundary terms, these are $R^2$ and $R_{\mu\nu}R^{\mu\nu}$. However, instead of working with the square of the Ricci tensor, it is now convenient to introduce the squared Weyl tensor, as we will explain below. 

The Weyl tensor is defined as the trace-free part of the Riemann tensor, and in $d=4$ it reads
\begin{equation}\label{weyl-tensor-Luca}
C_{\mu\nu\rho\sigma}=R_{\mu\nu\rho\sigma}+\frac{1}{2}\left(g_{\mu\sigma}R_{\nu\rho}-g_{\mu\rho}R_{\nu\sigma}+g_{\nu\rho}R_{\mu\sigma}-g_{\nu\sigma}R_{\mu\rho} \right) +\frac{1}{6}\left(g_{\mu\rho}g_{\nu\sigma}-g_{\mu\sigma}g_{\nu\rho}\right) R\,.
\end{equation}
Using the relation
\begin{equation}\label{weyl-tensor-square-Luca}
C_{\mu\nu\rho\sigma}C^{\mu\nu\rho\sigma}=R_{\mu\nu\rho\sigma}R^{\mu\nu\rho\sigma}-2 R_{\mu\nu}R^{\mu\nu}+\frac{1}{3}R^2\,,
\end{equation}
and the fact that the Gauss-Bonnet term 
\begin{equation}\label{gauss-bonnet-luca}
\sqrt{-g} \, \gaussbonnetterm \equiv \sqrt{-g} \, \left(R_{\mu\nu\rho\sigma}R^{\mu\nu\rho\sigma}-4 R_{\mu\nu}R^{\mu\nu}+R^2\right) 
\end{equation}
is locally a total derivative, we can replace $R_{\mu\nu}R^{\mu\nu}$ in terms of $R^2$ and $C_{\mu\nu\rho\sigma}C^{\mu\nu\rho\sigma}$ plus boundary contributions:
\begin{equation}
R_{\mu\nu}R^{\mu\nu}=\frac{1}{2}C_{\mu\nu\rho\sigma}C^{\mu\nu\rho\sigma}+\frac{1}{3}R^2-\frac{1}{2}\gaussbonnetterm\,.
\end{equation}
Therefore, the classical action of quadratic gravity can be written as\footnote{The subscript ``qg'' stands for ``quadratic gravity'' and should not be confused with ``\ac{QG}'' which is used as the acronym for ``quantum gravity'' in these lecture notes.}
\begin{tcolorbox}
\begin{equation}
S_{\rm qg}=\frac{1}{2}\int {\rmd}^4x\sqrt{-g}\left[\frac{1}{\kappa^2}\left(R-2\CC{}\right) + \frac{c_0}{6} R^2 -\frac{c_2}{2}C_{\mu\nu\rho\sigma}C^{\mu\nu\rho\sigma}\right]\,,
\label{quad-gravity-action}
\end{equation}
\end{tcolorbox}
\noindent where $c_0$, $c_2$ are dimensionless constants; the numerical coefficients and signs have been chosen for convenience, as will become clear when the propagator is derived and the masses of the degrees of freedom are determined. 

The quadratic gravity action is still invariant under diffeomorphisms: if we perform the metric transformation $\delta_\zeta g_{\mu\nu}$ in \eqref{metric-diff-2}, we get $\delta_\zeta S_{\rm qg}=0$.

The classical field equations are given by~\cite{Stelle:1977ry}
\begin{equation}\label{lhs-eom}
\frac{1}{\kappa^2}\left(R_{\mu\nu}-\frac{1}{2}g_{\mu\nu}R + \CC \, g_{\mu\nu} \right)+\frac{c_0}{3}\left(g_{\mu\nu}\Box R-\covD_\mu\covD_\nu R+R R_{\mu\nu}-\frac{1}{4}g_{\mu\nu} R^2\right)-2c_2 B_{\mu\nu}= T_{\mu\nu} \,,
\end{equation}
where $B_{\mu\nu}\equiv (\covD^\rho\covD^\sigma+\frac{1}{2}R^{\rho\sigma})C_{\mu\rho\nu\sigma}$ is the so-called Bach tensor, which is traceless. Since in four spacetime dimensions, the Bach tensor can be written solely in terms of Ricci scalar and Ricci tensor,
\begin{equation}
\begin{aligned}
B_{\mu\nu}&=\frac{1}{2}\Box R_{\mu\nu}+\frac{1}{6}\left(2\covD_\mu\covD_\nu-\frac{1}{2}g_{\mu\nu}\Box\right)R-\frac{1}{2}\covD_{\rho}\covD_\mu R_{\phantom{\rho}\nu}^\rho\\
&\qquad-\frac{1}{2}\covD_{\rho}\covD_\nu R_{\phantom{\rho}\mu}^\rho-\frac{1}{3}RR_{\mu\nu}+R_{\mu\rho}R^\rho_{\phantom{\rho}\nu}-\frac{1}{4}(R^{\rho\sigma}R_{\rho\sigma}-\frac{1}{3}R^2)g_{\mu\nu}\,,
\end{aligned}
\end{equation}
then we can also rewrite the field equations solely in terms of the Ricci scalar and the Ricci tensor, without Riemann or Weyl tensors, \ie{}
\begin{equation}\label{quad-grav-eom}
\begin{aligned}
&\frac{1}{\kappa^2}\left(R_{\mu\nu}-\frac{1}{2}g_{\mu\nu}R + \CC \, g_{\mu\nu} \right)+\frac{2}{3}\left(\frac{c_0}{2}+c_2\right)\left(g_{\mu\nu}\Box R-\covD_\mu\covD_\nu R+R R_{\mu\nu}-\frac{1}{4}g_{\mu\nu} R^2\right) \\
&\qquad-c_2\left(\Box R_{\mu\nu} +\frac{1}{2}g_{\mu\nu}\Box R-\covD_\rho \covD_{\mu}R^\rho_{\phantom{\rho}\nu}-\covD_\rho \covD_{\nu}R^\rho_{\phantom{\rho}\mu} +2R_\mu^{\phantom{\mu}\rho} R_{\rho\nu}-\frac{1}{2}g_{\mu\nu}R_{\rho\sigma}R^{\rho\sigma} \right)= T_{\mu\nu} \, .
\end{aligned}
\end{equation}
The trace equation does not depend on $c_2$ because the Weyl tensor, and thus also the Bach tensor, is traceless. Indeed, it reads
\begin{eqnarray}
-\frac{1}{\kappa^2}(R-4\CC)+c_0 \Box R = T\,.
\label{quad-grav-trace-eom}
\end{eqnarray}
The presence of operators with mass dimension equal to four implies that the dynamics of the metric in quadratic gravity is determined by fourth-order derivative field equations. This means that more boundary conditions are needed to find the solutions, and we may also expect the appearance of new degrees of freedom in addition to those of \ac{GR}.

\subsubsection{Propagator and degrees of freedom}\label{sec:LUCA_quadgrav_prop_dofs}

As done in the case of \ac{GR}, we can formulate quadratic gravity as a perturbative \ac{QFT} by quantizing metric fluctuations around some background and in a regime in which interactions are sufficiently weak. Then, we can identify kinetic and interaction terms, derive the propagator, determine the physical degrees of freedom, and investigate the \ac{UV} behavior of the theory to understand whether the criterion of (strict) renormalizability is satisfied.

Since the action contains fourth order derivatives acting on the metric, we expect two types of contributions to the kinetic term: a standard two-derivative term $\partial h \partial h$, and an additional one with four derivatives $\partial^2h \partial^2 h$. This means that we have \textit{two} options to normalize the mass dimension of the field and define the expansion in fluctuations: we can choose a mass dimension equal to one as done in \ac{GR}, or we can work with a dimensionless field. It turns out that the former option is more suitable for low-energy physics, while the latter captures the essence of the high-energy behavior of the theory. Having a dimensionless field in the \ac{UV} regime drastically distinguishes quadratic gravity from the \ac{EFT} of \ac{GR}, as we will explain in \cref{sec:quad-grav-vs-EFT}.

\paragraph{Expansion in metric fluctuations.} Assuming that the cosmological constant is negligible, let us expand around the Minkowski background and, for the time being, choose the following form of the metric perturbation: 
\begin{equation}
g_{\mu\nu}=\eta_{\mu\nu}+2 h_{\mu\nu}\,, 
\end{equation}
where $h_{\mu\nu}$ is dimensionless.

The action can be expanded as
\begin{eqnarray}
S_{\rm qg}[\eta+2 h]= S_{\rm qg}^{(2)}[\eta,h]+S_{\rm qg}^{(n\geq 3)}[\eta,h]\,,
\end{eqnarray}
where the kinetic term is given by
\begin{equation}\label{quad-h-QG}
\begin{aligned}
S_{\rm qg}^{(2)}[\eta,h]= \int {\rmd}^4 x &\left[ \frac{1}{2}h_{\mu\nu}\Box \left(\frac{1}{\kappa^2}-c_2\Box\right)h^{\mu\nu}-h_\mu^\rho \left(\frac{1}{\kappa^2}-c_2 \Box\right)\partial_\rho \partial_\nu h^{\mu\nu} \right. \\
& +h \left( \frac{1}{\kappa^2}-\frac{1}{3}(2c_0+c_2)\Box \right)\partial_\mu\partial_\nu h^{\mu\nu}-\frac{1}{2}h \left(\frac{1}{\kappa^2}-\frac{1}{3}(2c_0+c_2)\Box \right)\Box h\\
& \left.  + \frac{1}{3}(c_0-c_2)h_{\mu\nu}\partial^\mu\partial^\nu\partial^\rho\partial^\sigma h_{\rho\sigma}  \right]\,,
\end{aligned}
\end{equation}
while the interaction part $S_{\rm qg}^{(n\geq 3)}[\eta,h]$ contains terms of different type. Schematically, we have
\begin{eqnarray}
\frac{1}{\kappa^2}\partial^2 h^n\,, \qquad c_0 \partial^4 h^{n}\,,\qquad  c_2 \partial^4 h^{n}\,.
\end{eqnarray}
Note that the first two lines in the kinetic term~\eqref{quad-h-QG} contain the same tensor structures as in \ac{GR}, while the third line contains a new type of structure with four uncontracted derivatives. If we take the limits $c_0\to 0$ and $c_2\to 0$, while keeping  all the other parameters and the field fixed, we recover the expression for the kinetic term in \ac{GR} derived in \eqref{quadratic-act-mink} up to a canonical normalization factor for the graviton field.

\paragraph{Kinetic operator.} Since we are interested in deriving the propagator, it is convenient to rewrite the kinetic term as a bilinear form, so that we can identify the kinetic operator. After symmetrization of the various terms in \eqref{quad-h-QG}, we obtain
\begin{equation}
	S^{(2)}_{\rm qg}[\eta,h]= \int {\rmd}^4x \, \frac{1}{2}h_{\mu\nu}\mathbb{K}_{\rm qg}^{\mu\nu\rho\sigma}h_{\rho\sigma}\,,
	\label{quad-h-QG-2}
\end{equation}
where the kinetic operator is defined as
\begin{equation}\label{kinetic-oper-QG}
\begin{aligned}	
    \mathbb{K}_{\rm qg}^{\mu\nu\rho\sigma}\equiv& \frac{1}{2}\left(\eta^{\mu\rho}\eta^{\nu\sigma}+\eta^{\mu\sigma}\eta^{\nu\rho} \right)\left(\frac{1}{\kappa^2}-c_2 \Box\right)\Box-\eta^{\mu\nu}\eta^{\rho\sigma}\left(\frac{1}{\kappa^2}-\frac{1}{3}(2c_0+c_2) \Box\right)\Box \\
	&+\left(\eta^{\mu\nu}\partial^{\rho}\partial^\sigma +\eta^{\rho\sigma}\partial^{\mu}\partial^\nu  \right)\left(\frac{1}{\kappa^2}-\frac{1}{3}(2c_0+c_2) \Box\right)\\
    &-\frac{1}{2}\left(\eta^{\mu\rho}\partial^{\nu}\partial^\sigma+\eta^{\mu\sigma}\partial^{\nu}\partial^\rho +\eta^{\nu\rho}\partial^{\mu}\partial^\sigma+\eta^{\nu\sigma}\partial^{\mu}\partial^\rho \right)\left(\frac{1}{\kappa^2}-c_2 \Box\right) \\
	& +\frac{2}{3} (c_0-c_2)\partial^{\mu}\partial^\nu \partial^\rho\partial^\sigma \, ,
\end{aligned}	
\end{equation}
and satisfies the following symmetry properties:
\begin{equation}\label{kinetic-op-quad-grav-symm}
	\mathbb{K}_{\rm qg}^{\mu\nu\rho\sigma}=\mathbb{K}_{\rm qg}^{\nu\mu\rho\sigma}=\mathbb{K}_{\rm qg}^{\mu\nu\sigma\rho}=\mathbb{K}_{\rm qg}^{\rho\sigma\mu\nu}\,.
\end{equation}
In terms of the spin-projector operators introduced in \cref{sec:propag-III} (see also \cref{app:spin-proj}), the kinetic operator in momentum space can be written as
\begin{equation}\label{kinetic-quad-spin-proj}
\mathbb{K}_{\rm qg}^{\mu\nu\rho\sigma}=-\frac{p^2}{\kappa^2}\left[\mathcal{P}^{(2)\,\mu\nu\rho\sigma}\left(1+\kappa^2c_2 p^2\right) -2\mathcal{P}^{(0,s)\,\mu\nu\rho\sigma}\left(1+\kappa^2c_0p^2\right)\right]\,.
\end{equation}

\paragraph{Propagator.} Similarly to the \ac{GR} case, also here the kinetic operator is not invertible. This means that we need to introduce a gauge fixing. The same de Donder gauge used in \ac{GR} would already be enough to invert the kinetic operator but, since we now have derivatives up to fourth order, other types of covariant gauge-fixing terms could also be chosen. A generic local form for the gauge fixing is given by~\cite{Percacci:2017fkn, Buchbinder:2021wzv}
\begin{equation}\label{gauge-fix-quad-grav}
S_{\rm gf}[\eta,h]=-\frac{1}{\GFalpha \kappa^2}\int {\rmd}^4x \, \GFcondition_\mu \mathcal{Y}^{\mu\nu} \GFcondition_\nu\,,
\end{equation}
where 
\begin{equation}
\GFcondition_{\mu}\equiv \partial_\nu h^{\nu}_{\phantom{\nu}\mu}-\frac{1+\GFbeta}{4}\partial_\mu h\,,\qquad \mathcal{Y}^{\mu\nu}\equiv \eta^{\mu\nu}(1+\gamma \Box)+\omega \partial^{\mu}\partial^\nu\,,
\end{equation}
$\GFalpha$, $\GFbeta$, $\gamma$, and $\omega$ being gauge-fixing parameters. For example, if one is interested in studying the \ac{UV} behavior of the theory, then a gauge fixing containing fourth-order derivatives may be computationally more suitable.

Rewriting \eqref{gauge-fix-quad-grav} in terms of the spin projectors and adding it to \eqref{kinetic-quad-spin-proj}, we obtain an invertible operator which can be easily inverted by using the orthogonality properties of the spin projectors. Since the procedure is similar to that performed in the case of \ac{GR}, we will skip the details of the calculation here. In particular, we are mainly interested in the gauge-independent part of the propagator which contains also information about the particle content of the theory. Up to a normalization factor of $\kappa^2$, the propagator is given by~\cite{Stelle:1976gc}
\begin{tcolorbox}
\begin{equation}\label{propag-spin-proj-quad-gra}
\begin{aligned}
\propG_{{\rm qg}\,\mu\nu\rho\sigma}(p)&= -i\left[\frac{m_2^2 \mathcal{P}^{(2)}_{\phantom{(2)}\mu\nu\rho\sigma}}{p^2(p^2+m_2^2)}-\frac{m_0^2 \mathcal{P}^{(0,s)}_{\phantom{(0,s)}\mu\nu\rho\sigma}}{2p^2(p^2+m_0^2)}\right]+\dots \\
&=-\frac{i}{p^2}\left[ \mathcal{P}^{(2)}_{\phantom{(2)}\mu\nu\rho\sigma}-\frac{1}{2}\mathcal{P}^{(0,s)}_{\phantom{(0,s)}\mu\nu\rho\sigma}\right]-\frac{i}{2}\frac{\mathcal{P}^{(0,s)}_{\phantom{(0,s)}\mu\nu\rho\sigma}}{p^2+m_0^2}+i\frac{\mathcal{P}^{(2)}_{\phantom{(2)}\mu\nu\rho\sigma}}{p^2+m^2_2}+\dots\,,
\end{aligned}
\end{equation}
\end{tcolorbox}
\noindent where the dots represent gauge-dependent terms whose spin structure would not contribute if we multiply the propagator by a conserved stress-energy tensor.

Let us now make some comments.
\begin{itemize}

\item Because of fourth-order derivatives coming from the quadratic-curvature terms in the action~\eqref{quad-gravity-action}, the \ac{UV} behavior of the propagator in quadratic gravity is more suppressed than the \ac{GR} one, that is, it falls off like $\sim 1/p^4$ instead of $\sim 1/p^2$. This aspect is crucial for the renormalizability of the theory, as we will explain in \cref{sec:renorm-quad-grav}.

\item Using a partial-fraction decomposition, in the second line of \eqref{propag-spin-proj-quad-gra} we split the propagator into three terms: the \ac{GR} one (with massless poles) plus two additional contributions. The second term corresponds to a massive spin-zero degree of freedom, while the third term to a massive spin-two. Their squared masses are respectively given by
\begin{equation}\label{masses-spin-0-2}
m_0^2\equiv \frac{1}{\kappa^2c_0}=\frac{\MPl^2}{c_0}\,,\qquad m_2^2\equiv \frac{1}{\kappa^2c_2}=\frac{\MPl^2}{c_2}\,. 
\end{equation}
To avoid tachyons we must have $c_0>0$ and $c_2>0$, which explains the chosen convention for the signs in front of the quadratic-curvature invariants in the action~\eqref{quad-gravity-action}.

\item The additional degrees of freedom are not gauge artifacts and, in general, all contribute to the gravitational particle spectrum. In total, quadratic gravity has $2+1+5=8$ on-shell degrees of freedom.

\item We could have guessed the gauge-independent part of the propagator just by looking at \eqref{propag-spin-proj-quad-gra}. Indeed, since the additional modes are massive, their individual contributions in~\eqref{kinetic-quad-spin-proj} can be (formally) inverted without requiring any gauge fixing. Therefore, knowing the gauge-independent part of the massless graviton propagator in \ac{GR}, then the two additional massive components can be found by inverting the terms proportional to $c_0$ and $c_2$ in~\eqref{kinetic-quad-spin-proj}.

\item While the additional spin-zero component has the standard sign in front, the additional spin-two component has the opposite sign. This also means that the individual kinetic term corresponding to this massive spin-two field will have a sign opposite to standard two-derivative fields.\footnote{It is possible to introduce auxiliary fields and rewrite the action~\eqref{quad-gravity-action} in terms of an Einstein-Hilbert term plus contributions containing kinetic terms for the additional spin-zero and spin-two fields plus interactions. In particular, it can be shown that the kinetic term of the massive spin-two is equal to a Fierz-Pauli Lagrangian with a minus sign in front~\cite{Anselmi:2018tmf,Buoninfante:2023ryt}.} Fields characterized by an opposite sign in front of their kinetic term and propagator are called \textit{ghosts}. In \cref{sec:ghost-puzzle}, we will discuss the significance and implications of the presence of such a massive spin-two ghost, and in \cref{sec:open-quest-quad-grav} we will highlight the open questions in relation to unitarity and stability.

\item The \ac{GR} propagator is recovered in the limits $c_0\to 0$ and $c_2\to 0$, that in terms of the masses translate into $m_0^2/ \MPl^2\to \infty$ and $m_2^2/\MPl^2\to \infty$, respectively. From a more physical point of view, the \ac{GR} limit can be understood as a low-energy regime where we can expand in $|p^2|/m_0^2\ll 1$ and $|p^2|/m_2^2\ll 1$, so that the additional massive degrees of freedom are integrated out.\footnote{In general, in theories with ghosts, the vacuum may be unstable, and the notion of integrating out a ghost may not be justified. For example, even if the ghost is very heavy, but has negative energy, it could be easily produced at very low energies, thus destabilizing the vacuum. However, by implementing alternative quantizations that are compatible with unitarity and renormalizability (see \cref{sec:ghost-puzzle}) one can avoid this problem and prove the stability of the vacuum (at least perturbatively).}

\end{itemize}

\subsubsection{Renormalizability}\label{sec:renorm-quad-grav}

We now want to show that quadratic gravity is strictly renormalizable by power counting. To do so, we already know that we have to look at the mass dimension of the interaction couplings that are relevant in the \ac{UV} regime (see \cref{sec:app-renormaliz}). 
At high energies, the terms with four powers of momenta dominate in both the propagator and the vertices in quadratic gravity. This means that the \ac{UV} behavior of quadratic gravity is governed by a dimensionless field $h_{\mu\nu}$ and dimensionless couplings $c_0$ and $c_2$. 

\subsubsubsection*{Power counting} 

Given a loop diagram $G$ containing $E$ external legs, the corresponding superficial degree of divergence in quadratic gravity is given by (see \eqref{total-delta})
\begin{equation}\label{sup-degree-quad-grav}
\delta(G)=4-E\,,
\end{equation}
which is independent of the number of vertices because the relevant couplings are dimensionless, \ie{} $\Delta_n=0$ in the general formula~\eqref{total-delta} derived in the appendix. Therefore, only Green's functions with $E\leq 4$ external legs need to be renormalized, and the renormalization of these has to be implemented at all loop orders. This implies that quadratic gravity is strictly renormalizable (not super-renormalizable).

More explicitly, let us consider a generic $L$-loop integral and calculate how it scales with the internal momentum at high energies. If we do that, we have
\begin{equation}\label{loop-integ-scale-qg}
	{\int \underbrace{{\rmd}^4k\cdots {\rmd}^4k}_{L\text{-loops}}}\, \times {\underbrace{\frac{1}{k^4}\cdots \frac{1}{k^4}}_{I\text{-internal propagators}}} \times\, {\underbrace{k^{4}\cdots k^{4}}_{V\text{-vertices}}}\,\sim\,k^{4(L-I+V)}=k^{4}\,,
\end{equation}
where we have used the fact that the propagator goes like $1/k^4$, the dominant vertices contain four powers of momenta $k^4$, and we chose the most divergent case for which all vertex momenta are internal. From \eqref{loop-integ-scale-qg} we see that the degree of \ac{UV} divergence of the loop integrals do not become worse if the number of loops increases, it actually remains the same. This implies that the same counterterms can be used to renormalize the theory at any loop order. Therefore, the theory is perturbatively (strictly) renormalizable.

\subsubsubsection*{Tensor structure of counterterms} 

Some skeptics might still wonder whether loop divergences require counterterms not contained in the initial bare action. For example, since \ac{UV} divergences are local, then someone could still ask why no local higher-curvature counterterm, such as $R^3$, $R_{\mu\nu\rho\sigma}R^{\rho\sigma}_{\phantom{\rho\sigma}\alpha\beta}R^{\alpha\beta\mu\nu},\,\dots$, is generated by renormalization. Although the answer to this question is already contained in \eqref{sup-degree-quad-grav} and~\eqref{loop-integ-scale-qg}, we can explicitly show that no operator of mass dimension greater than four appears at any loop order.

In $d=4$ spacetime dimensions, operators of mass dimension higher than four need to be multiplied by couplings of negative mass dimension. Since the quadratic-curvature coefficients are dimensionless, $[c_0]=0=[c_2]$, the only dimensionful scales that could possibly appear as coefficients of higher-dimensional operators generated at some loop order are $\kappa=1/\MPl$ and $\CC{}$. However, we know that in the \ac{UV} regime the most dominant part of the propagator contains four powers of momenta. Schematically, the \ac{UV} behavior of the propagator, including the cosmological constant, can be written as 
\begin{equation}\label{schematic-propag-quad-grav}
\frac{1}{\CC{}\MPl^2+\MPl^2p^2+c_0p^4+c_2p^4}\stackrel{\text{UV}}{\sim} \frac{1}{c_0p^4+c_2p^4}\,.
\end{equation}
Moreover, the possible vertices are schematically proportional to $\CC{}\MPl^2$, $\MPl^2p^2$, $c_0 p^4$ and $c_2 p^4$. Therefore, the \ac{UV}-divergent contribution of any loop diagram can only be proportional to combinations of the couplings $\CC{}$, $\MPl$, $c_0$ and $c_2$ that have mass dimension equal or greater than zero. In particular, no combination which has an overall negative mass dimension can appear.

Therefore, the possible counterterms have the same form as the terms already contained in the initial bare action~\eqref{quad-gravity-action}, \ie{} they are given by
\begin{equation}\label{counterterms-quad-grav}
\lambda \mathbb{1}\,,\qquad R\,,\qquad R^2\,, \qquad C_{\mu\nu\rho\sigma}C^{\mu\nu\rho\sigma}\,,
\end{equation}
where $\lambda$ is a constant of dimension four which renormalizes the cosmological constant $\CC \MPl^2$, the second term renormalizes the Planck mass, and the other two renormalize the dimensionless couplings $c_0$ and $c_2$. It is important to mention that the renormalization procedure also generates boundary terms, \ie{} the Gauss-Bonnet invariant $\gaussbonnetterm$ in \eqref{gauss-bonnet-luca} and $\Box R$~\cite{Julve:1978xn, Fradkin:1981iu, Avramidi:1985ki, Percacci:2017fkn, Buchbinder:2021wzv}, but since these in turn do not feed back into the renormalization of the other terms, they can be safely neglected.

We have thus confirmed that quadratic gravity does not require additional counterterms to be renormalized, and is therefore perturbatively renormalizable. For completeness, we should note that the full proof of renormalizability also requires showing that the renormalization procedure does not break the gauge symmetry of the theory, and this can be done by proving the \ac{BRST} invariance of the effective action at any loop order. This was proven for the first time in~\cite{Stelle:1976gc}; see also~\cite{Piva:2023bcf} for a recent review.

\subsubsubsection*{Loop expansion}

While in \ac{GR}, the loop expansion is controlled by inverse powers of the Planck mass at both low and high energies, \ie{} $(1/\MPl^2)^{L-1}$ where $L$ is the number of loops, in quadratic gravity it is controlled by dimensionless couplings in the high-energy regime. Indeed, if we consider an $L$-loop diagram whose integrand contains $I_0$ spin-zero and $I_2$ spin-two components of the four-derivative propagator, $V_0$ vertices proportional to $c_0$ and $V_2$ vertices proportional to $c_2$, we can check that the loop expansion in the \ac{UV} is controlled by the following dimensionless combination of parameters:
\begin{equation}\label{dimensionless-combin-c}
\left(\frac{c_2}{c_0}\right)^{I_0-V_0}\left(\frac{1}{c_2}\right)^{L-1}=\left(\frac{c_0}{c_2}\right)^{I_2-V_2}\left(\frac{1}{c_0}\right)^{L-1}\,,
\end{equation}
where we have used the relations $I=I_0+I_2$, $V=V_0+V_2$, and the topological identity $L-1=I-V$. The smaller this dimensionless quantity, the better the behavior of the perturbative expansion.

If we define the couplings $g_0$ and $g_2$ via the relations
\begin{equation}\label{couplings-g0g2}
g_0\equiv \frac{1}{c_0}\,,\qquad  g_2\equiv \frac{1}{c_2}\,,
\end{equation}
the quantity in \eqref{dimensionless-combin-c} can be written as
\begin{equation}\label{dimensionless-combin-g}
\left(\frac{g_0}{g_2}\right)^{I_0-V_0}g_2^{L-1}=\left(\frac{g_2}{g_0}\right)^{I_2-V_2}g_0^{L-1}\,.
\end{equation}
If $g_0$ and $g_2$ (\ie{} $c_0$ and $c_2$) are not too different in order of magnitude, then the loop expansion is controlled by $g_2^{L-1}$ and $g_0^{L-1}$. Therefore, the larger the quadratic-curvature coefficients, the better the behavior of the perturbative expansion.

Because of this kind of high-energy behavior, in quadratic gravity the following canonical normalization for the metric perturbation is often used:
\begin{equation}\label{canonic-normal-dimens}
h_{\mu\nu}\to \frac{1}{\sqrt{c_2}}h_{\mu\nu}\,, 
\end{equation}
which would give the following schematic expansion for the action:
\begin{equation}\label{action-expanded-dimens-normal}
\begin{aligned}
S_{\rm qg}\sim \int \rmd^4 x &\left[ h\partial^4 h+ \frac{c_0}{c_2}h\partial^4 h+\frac{\MPl^2}{c_2}h\partial^2 h+\frac{1}{\sqrt{c_2}}\left(h\partial^4 h^2+\frac{c_0}{c_2}h\partial^4 h^2+\frac{\MPl^2}{c_2}h\partial^2 h^{2}\right)\right.\\
&\left. +\dots+ \left(\frac{1}{\sqrt{c_2}}\right)^{n-2}\left(h\partial^4 h^{n-1}+\frac{c_0}{c_2}h\partial^4 h^{n-1}+\frac{\MPl^2}{c_2}h\partial^2 h^{n-1}\right)+\dots \right]\,.
\end{aligned}
\end{equation}

\subsubsubsection*{Running couplings}

One-loop computations of beta functions in quadratic gravity have been performed by various authors~\cite{Julve:1978xn, Fradkin:1981iu, Avramidi:1985ki, Buccio:2024hys}. The Planck mass (\ie{} Newton's coupling) and the cosmological constant do not exhibit a physical running since no unique momentum dependence in physical observables can be identified and no momentum-dependent form factors can be derived for the Einstein-Hilbert part of the action~\cite{Anber:2011ut,Donoghue:2024uay}. On the other hand, the coefficients $c_0$ and $c_2$ (or, equivalently, $g_0$ and $g_2$) do run as a function of the physical momentum.\footnote{Beta functions can also be computed for the boundary terms $\gaussbonnetterm$ and $\Box R$~\cite{Fradkin:1981iu, Avramidi:1985ki, Percacci:2010af,  Buchbinder:2021wzv}. Even though they are usually neglected when boundary effects are not important, they still contribute to the conformal anomaly~\cite{Buchbinder:2021wzv}. While the beta function for $\gaussbonnetterm$ has been fully computed, for the surface term $\Box R$ only partial results exist in the literature~\cite{Fradkin:1981iu}.}

The beta functions of the quadratic-curvature couplings were initially computed in~\cite{Julve:1978xn, Fradkin:1981iu, Avramidi:1985ki}. In particular,~\cite{Avramidi:1985ki} concluded that quadratic gravity can be asymptotically free at the price of having a negative $R^2$ coefficient. This means that the couplings $g_0$ and $g_2$ would flow to zero in the \ac{UV} regime if and only if the additional massive spin-zero is a tachyon $(c_0=\MPl^2/m_0^2<0)$. However, a recent calculation~\cite{Buccio:2024hys} has argued that the old results do not capture the physical running, and cannot be relied upon to study the momentum dependence of $g_0$ and $g_2$. This criticism is based on the fact that previous calculations included contributions from tadpoles that have no dependence on the physical momentum, and left out finite infrared terms due to the four-derivative nature of quadratic gravity. The main outcome of the new computation in~\cite{Buccio:2024hys} was that asymptotic freedom could be achieved without tachyons.

It is believed that further investigations are needed to definitely answer the question of what the physical and gauge-invariant running couplings are in quadratic gravity.

\subsubsubsection*{Coupling to matter}

So far we have not considered the matter contribution and its coupling to gravity. In fact, we must ask whether the property of renormalizability is preserved also when matter is coupled to gravity. It can be easily understood that if we couple the \ac{SM} to quadratic gravity, we can still obtain a strictly renormalizable \ac{QFT} of gravity and matter if we also introduce the following non-minimal coupling between the Higgs field $H$ and the metric field~\cite{Stelle:1976gc,Salvio:2018crh}:
\begin{equation}\label{non-min-coup-quad}
\int{\rmd}^4x\sqrt{-g} \, \xi \, |H|^2R\,,
\end{equation}
where $\xi$ is a dimensionless interaction coupling. Here we assume that the \ac{SM} also includes renormalizable mass terms and Yukawa interactions to take into account the non-zero masses of neutrinos. In this strictly renormalizable \ac{QFT} of quadratic gravity coupled to the \ac{SM} with the inclusion of the non-minimal coupling~\eqref{non-min-coup-quad}, the free parameters that have to be renormalized and fixed by experiments are finite in number: they are given by those of \ac{SM} plus $\GN$, $\CC{}$, $c_0$, $c_2$, and $\xi$.

\subsubsection{Quadratic gravity vs. EFT of GR}\label{sec:quad-grav-vs-EFT}

We now want to address a simple question regarding the difference between the \ac{EFT} of \ac{GR} and quadratic gravity, which sometimes causes confusion and misunderstanding.

Let us consider a characteristic energy $E\ll \MPl$ such that we can reliably truncate the \ac{EFT} expansion in \eqref{EFT-action-Luca} up to quadratic order in the curvatures by committing errors proportional to $(E/\MPl)^4$:
\begin{equation}\label{EFT-action-Luca-quad}
S_{\rm EFT}\simeq \int {\rmd}^4x\sqrt{-g}\left[\frac{1}{2\kappa^2}R+  a_1 R^2 + a_2 R_{\mu\nu}R^{\mu\nu}\right] \,,
\end{equation}
where we have again neglected the cosmological constant as it is not important for our discussion. In addition, couplings to matter can also be present. Up to boundary terms, the expression~\eqref{EFT-action-Luca-quad} has the same structure as the action of quadratic gravity in \eqref{quad-gravity-action}. So, the question we want to ask is: since these two actions have the same structure, do they describe the same physics at energies of order $E\ll \MPl$? The short answer is: No!

Let us make some remarks and explain the crucial differences between the \ac{EFT} of \ac{GR} and quadratic gravity.
\begin{itemize}
    
    \item \textbf{Perturbative renormalizability:} First of all, we emphasize again that while the \ac{EFT} of \ac{GR} is perturbatively non-renormalizable, quadratic gravity is strictly renormalizable. This crucial distinction between the two theories can be understood in several ways and brings completely different physical implications.

    \item \textbf{Perturbation theory:} The perturbative expansion in the two theories is defined in two drastically different ways. In the \ac{EFT}, the kinetic term is still given by that of \ac{GR}, \ie{} it only contains second-order derivatives. On the other hand, in quadratic gravity the kinetic term contains both second- and fourth-order derivatives. This difference also implies that only in quadratic gravity it makes sense to describe the \ac{UV} regime in terms of a metric fluctuation with mass dimension equal to zero.
    
    \item \textbf{Quadratic-curvature coefficients:} In the \ac{EFT} of \ac{GR}, the quadratic-curvature terms can be generated by loop corrections or could also have a microscopic origin (\eg{} see \cref{sec:scattering_amplitudes}). If we are in the regime of validity of the perturbative \ac{QFT} framework, we expect the renormalized value of the dimensionless coefficients $a_1$ and $a_2$ to be of order one or smaller. On the other hand, the coefficients $c_0$ and $c_2$ in quadratic gravity are free parameters that in principle could be renormalized to any value, even very large ones. In fact, the larger their value, the better the behavior of the loop expansion.
    
    \item \textbf{Weak vs. strong coupling:} While $a_1$ and $a_2$ appear only in the vertices in the \ac{EFT}, the coefficients $c_0$ and $c_2$ appear also in the propagator due to the four-derivative structure of the kinetic term. This means that the loop expansion in the \ac{UV} regime in the two theories is defined in terms of two different expansion parameters. Indeed, as partly discussed in \cref{sec:renorm-quad-grav}, the loop diagrams in the \ac{EFT} of \ac{GR} are proportional to powers of $a_1$, $a_2$ and to $(1/\MPl^2)^{L-1}$, where $L$ is the number of loops. By contrast, loop diagrams in quadratic gravity are proportional to $(1/c_0)^{L-1}$, $(1/c_2)^{L-1}$, and other types of dimensionless combinations; see \eqref{dimensionless-combin-c}. This difference makes the meaning of weak and strong coupling depend on whether one considers one theory or the other. In particular, what is perturbative and weakly coupled for quadratic gravity (large $c_0$ and $c_2$) would be non-perturbative and strongly coupled for the \ac{EFT} (large $a_1$ and $a_2$).\footnote{It is worth emphasizing that here, large values of $a_i$ can be seen as a sign of strong coupling only if the \ac{EFT} cutoff is $\MPl$ (or slightly lower due to matter), since the energy $E$ at which the couplings would become large is below the cutoff. However, if the cutoff is not actually given by $\MPl$, the large values of $a_i$ could indicate that in fact, the cutoff is much lower and no strong coupling would arise. For instance, if quadratic gravity is considered as a perturbative completion, the cutoff of the \ac{EFT} of \ac{GR} would be the lower of the two masses $m_0$ and $m_2$. In this case, the theory would still be weakly coupled for energies of order $m_i$, and even above, despite the fact that $a_i\sim \MPl^2/m_i^2\gg 1$. Another example is perturbative \ac{ST} (see \cref{sec:IVANO}) where the gravitational \ac{EFT} is still weakly coupled, because the Wilson coefficients $a_i$, when expressed in terms of the correct cutoff $M_s\ll \MPl$, are order-one.} Vice versa, what is non-perturbative and strongly coupled for quadratic gravity (small $c_0$ and $c_2$) would be perturbative and weakly coupled for the \ac{EFT} of \ac{GR} (small $a_1$ and $a_2$).

    \item \textbf{Field redefinition:} From the last items, it becomes clear that the same field redefinition used in \cref{sec:one-loop-div-luca} to remove the quadratic-curvature terms in pure \ac{GR} at one loop, does \textit{not} work in the case of quadratic gravity. This latter theory, in fact, is characterized by classical field equations different from Einstein's, and contains additional degrees of freedom that can be understood as non-perturbative modifications with respect to the propagator in \ac{GR}: new physical poles appear.

    \item \textbf{Degrees of freedom:} In the \ac{EFT} of \ac{GR}, the massless graviton is the only active degree of freedom at energies $E\ll \MPl$. By contrast, in quadratic gravity, additional degrees of freedom can become active at energy scales $E$ of the order of or larger than the masses $m_0=\MPl/\sqrt{c_0}$ and $m_2=\MPl/\sqrt{c_2}$ defined in \eqref{masses-spin-0-2}. 
    
\end{itemize}

All these differences can also have important physical implications. Indeed, besides higher-curvature corrections, the \ac{EFT} of \ac{GR} in \eqref{EFT-action-Luca-quad} with its coupling to matter does not really introduce new physics at energy scales $E\ll \MPl$. On the other hand, quadratic gravity could in principle describe new physical phenomena beyond \ac{GR} thanks to the presence of the additional degrees of freedom that could give rise to a new dynamics. In particular, if $c_0\gg 1$ and $c_2\gg 1$, which is indeed necessary for a well-behaved perturbative expansion, we get $m_0,m_2\ll \MPl$ and hence new physics may arise in the sub-Planckian regime; see \cref{sec:uniqueness} for more details on physical implications of quadratic gravity.

\subsubsection{Ghost puzzle}\label{sec:ghost-puzzle}

So far so good. However, it is time to discuss some important open questions in quadratic gravity for which debates are still ongoing. Let us start with the ``ghost puzzle'' and its implications for unitarity.

The improved \ac{UV} behavior of the propagator in \eqref{propag-spin-proj-quad-gra} is obtained at the price of introducing an additional massive spin-two ghost in the spectrum. Typically ghosts are considered pathological because they could give rise to classical Hamiltonian instabilities and break unitarity at the quantum level~\cite{Woodard:2015zca}. We will now show why unitarity can be violated when ghosts are present in the theory, and discuss recent proposals to address the problem. Our philosophy will be to first try to solve the ghost problem in the quantum theory, and then discuss the classical limit and its stability. Different proposals correspond to different types of quantization of quadratic gravity.

\subsubsubsection*{Ghosts and unitarity violation}

To explain why unitarity can be violated in the presence of ghosts, we can neglect the tensor structure of the propagator and work with a simpler scalar version that has the same pole structure as that of the spin-two component in \eqref{propag-spin-proj-quad-gra}. This is given by
\begin{equation}
\propG(p^2)=\frac{-i m^2}{p^2(p^2+m^2)}=-i\left[\frac{1}{p^2}-\frac{1}{p^2+m^2}\right]\,,
\label{scalar-propag-main}
\end{equation}
where $1/p^2$ and $-1/(p^2+m^2)$ can mimic the massless spin-two graviton and the massive spin-two ghost, respectively.

From \cref{sec:app-unitarity}, we know that the unitarity condition on the S-matrix, \ie{} $S^\dagger S=\mathbb{1}$, can be expressed in terms of the transfer matrix $T$ defined through the relation $S=\mathbb{1}+iT$, so that one obtains the well known optical theorem $i(T^\dagger-T)=T^\dagger T$. Introducing in and out states $\left| a\right\rangle $ and $\left| b\right\rangle$ belonging to the Hilbert space $\mathcal{H}$, and using the completeness relation $\mathbb{1}=\sum_{\left| n \right\rangle\in \mathcal{H}}\left| n\right\rangle \left\langle n\right|$, where all physical states have positive norms, we can write\footnote{To be more precise we should introduce the Feynman amplitude $\scatteringamplitude$ as done in \cref{sec:app-unitarity}. However, to make the formula simpler, we now intentionally work with $T$ and neglect phase-space integrals as well as momentum-conserving Dirac deltas.}
\begin{equation}
i\left[\left\langle b\big|T^{\dagger}\big|a \right\rangle - \big\langle b\big|T\big|a \big\rangle \right]=\sum\limits_{\left| n \right\rangle\in \mathcal{H}} \left\langle b\big|T^{\dagger}\big|n \right\rangle\big\langle n\big|T\big|a \big\rangle\,.
\label{optical theorem-components-main}
\end{equation}
For an elastic process, \ie{} $\left| b\right\rangle=\left| a\right\rangle$, we have $2{\rm Im} \, [\left\langle a|T|a \right\rangle ]\geq 0$. This positivity constraint must also be respected by the propagator, as the latter can be seen as a $1\to 1$ tree-level amplitude. In particular, we should have ${\rm Im} \, [-i(-i)^2\propG(p^2)]\geq 0$, where the $(-i)^2$ comes from the two vertices, and $-i$ from the overall multiplicative factor in the definition of the transfer matrix. 

If we prescribe both components of the propagator in \eqref{scalar-propag-main} with the standard causal Feynman shift, \ie{} $p^2\to p^2-i\epsilon$ with  $\epsilon\to 0^+$, the imaginary part of the propagator multiplied by $i$ reads
\begin{equation}
{\rm Im}\left[i\propG_{\rm F}(p^2)\right]={\rm Im}\left[\frac{1}{p^2-i\epsilon}-\frac{1}{p^2+m^2-i\epsilon} \right]=\pi[\delta(p^2)-\delta(p^2+m^2)]\,.\label{imag-part-main}
\end{equation}
Because of the relative minus sign caused by the ghost component, the imaginary part of the amplitude can also be negative, and therefore the unitarity condition is not respected.

In such a  situation, we can introduce a \textit{pseudo-unitarity} equation given by
\begin{equation}
i\left[\left\langle b\big|T^{\dagger}\big|a \right\rangle - \big\langle b\big|T\big|a \big\rangle \right]=\sum\limits_{\left| n \right\rangle\in \mathcal{H}} \sigma_n\left\langle b\big|T^{\dagger}\big|n \right\rangle\big\langle n\big|T\big|a \big\rangle\,,
\label{pseudo-unit}
\end{equation}
where $\sigma_n=1$ for normal (non-ghost) states, while $\sigma_n=-1$ for ghost states. In general, the pseudo-unitarity equation~\eqref{pseudo-unit} is not equivalent to $S^\dagger S=\mathbb{1}$. However, the two conditions can be made equivalent by implementing an alternative quantization, as we now explain.

\subsubsubsection*{Three approaches to the ghost puzzle}

It is important to note that the above argument leading to the violation of unitarity relies heavily on two assumptions: the causal Feynman prescription for the propagator, and positive norms for physical states. We now want to show that by modifying one or both assumptions, the unitarity condition can be recovered. In the following, we will only consider a tree-level analysis, and briefly comment on what happens at loop orders, referring to the original works.

\paragraph{Feynman prescription and negative norms.} By looking at the expression for the pseudo-unitarity equation~\eqref{pseudo-unit}, the first thing that might come to mind is that, by admitting some negative coefficients $\sigma_n$ from the beginning, we could recover the unitarity condition. In fact, this can be done by assuming that some physical states populated by ghost particles have negative norms~\cite{Stelle:1976gc, Salvio:2018crh, Holdom:2021hlo}. In this case, the standard concept of a Hilbert space is replaced by a vector space with an indefinite metric~\cite{Lee:1969fy}, whose completeness relation reads
\begin{equation}
	\mathbbm{1}=\sum_{\left| n \right\rangle\in \mathcal{H}} \sigma_n \left| n \right\rangle \left\langle n\right|  \,,\label{complet-negative norms}
\end{equation}
where the coefficients $\sigma_n$ are defined via  $\langle n |m \rangle=\sigma_n\delta_{nm}$. If the (squared) norms of states containing an odd number of ghost particles are negative, \ie{} $\sigma_{2n+1}<0$, then unitarity would be preserved. This means that the minus sign appearing in~\eqref{imag-part-main} would correspond to another minus sign on the right-hand side of the optical theorem due to the negative norm of a one-ghost state. We refer to the ghost quantized in this way as a \textit{Feynman ghost}.

In this type of quantization, probabilities are conserved and sum up to one, but their sign can be negative. While from a standard perspective, the presence of negative probabilities (\eg{} negative cross sections) is a pathology, some recent work investigated whether a new quantum interpretation can solve the ghost puzzle in this type of quantization~\cite{Salvio:2015gsi, Strumia:2017dvt, Holdom:2024onr, Woodard:2023tgb, Kubo:2023lpz}. On the other hand, since the poles of the ordinary and ghost propagators are shifted according to the Feynman prescription, the standard rules for computing loop integrals in perturbation theory still apply, including the Wick rotation.

\paragraph{Anti-Feynman prescription and positive norms.} Another possibility to satisfy the unitarity condition in the presence of ghosts is to keep all norms positive, but to shift the poles in the ghost propagator according to the anti-Feynman prescription, \ie{} $p^2\to p^2+i\epsilon$ with  $\epsilon\to 0^+$. The propagator now reads
\begin{equation}
\propG_{\rm anti\text{-}F}(p^2)=-i\left[\frac{1}{p^2-i\epsilon}-\frac{1}{p^2+m^2+i\epsilon}\right]\,.
\label{scalar-propag-anti-f}
\end{equation}
In this case, the imaginary part of the amplitude $i\propG_{\rm anti\text{-}F}(p^2)$ will be positive, because the sign of the anti-Feynman shift compensates the minus sign of the ghost propagator:
\begin{equation}
{\rm Im} \, \left[i\propG_{\rm anti\text{-}F}(p^2)\right]={\rm Im}\left[\frac{1}{p^2-i\epsilon}-\frac{1}{p^2+m^2+i\epsilon} \right]=\pi[\delta(p^2)+\delta(p^2+m^2)]\,.\label{imag-part-main-anti}
\end{equation}
Thus, the optical theorem is satisfied with $\sigma_n>0$ for all $n$, but at the price of admitting both Feynman (for ordinary particles) and anti-Feynman (for ghost particles) shifts in the same theory. We refer to the ghost quantized in this way as an \textit{anti-Feynman ghost}~\cite{Donoghue:2019fcb,Donoghue:2021cza}.

Recall that, as also reviewed in \cref{sec:app-unitarity}, the Feynman shift physically means that positive (negative) energies propagate forward (backward) in time. By contrast, the anti-Feynman prescription describes positive (negative) energies propagating backward (forward) in time. This means that, if both types of prescriptions are introduced in the same theory, then two arrows of time are simultaneously present, and some form of causality is violated~\cite{Donoghue:2019ecz}. Furthermore, the presence of both shifts introduces non-analyticity in Feynman diagrams and loop integrals, complicating the proof of renormalizability: for example, the standard Wick rotation cannot be applied in this case. It is not yet clear whether alternative deformations of the integration contour that are consistent with both unitarity and renormalizability can be implemented at any loop order. Some attempts have been made up to one loop~\cite{Donoghue:2019fcb}.

\paragraph{Fakeon prescription.} The third approach is to implement a quantization such that the ghost can be converted into a purely off-shell degree of freedom, which does not appear as an on-shell particle but propagates only through internal lines in Feynman diagrams. In this case, the ghost becomes a \textit{fake} particle~\cite{Anselmi:2017ygm,Anselmi:2018bra,Piva:2023bcf}. According to the fakeon prescription, the ghost propagator is defined as an average of Feynman and anti-Feynman propagators, \ie{}
\begin{equation}
\frac{i}{p^2+m^2}\qquad\to\qquad i\frac{p^2+m^2}{(p^2+m^2)^2+\mathcal{\epsilon}^2}=\frac{i}{2}\left[\frac{1}{p^2+m^2+i\epsilon} + \frac{1}{p^2+m^2-i\epsilon} \right]\,.
\label{fake-ghost}
\end{equation}
Thus, the full propagator reads
\begin{equation}
\propG_{\rm fake}(p^2)=-i\left[\frac{1}{p^2-i\epsilon}-\frac{1}{2}\left(\frac{1}{p^2+m^2+i\epsilon} + \frac{1}{p^2+m^2-i\epsilon} \right)\right]\,.
\label{fake-propag}
\end{equation}
If we now compute the imaginary part of the amplitude, \ie{} the left-hand side of the optical theorem, the ghost does not contribute and we get
\begin{equation}
{\rm Im}\left[i\propG_{\rm fake}(p^2)\right]=\pi\delta(p^2)\,.\label{imag-part-main-fake}
\end{equation}
The fact that there is no Dirac delta associated with the ghost component means that indeed no ghost particle can appear on-shell. To satisfy unitarity, we should also ensure that the right-hand side of the optical theorem gives the same result. This can be done by projecting the ghost-like states out of the ``physical'' Hilbert space. In practice, the projection is made by imposing that only normal (non-ghost) states appear in the completeness relation, so that the sum on the right-hand side of the optical theorem runs over the states belonging to the ``physical'' subspace $\mathcal{H}_{\rm ph}\subset \mathcal{H}$, which does not contain any ghost state. In a formula, we have~\cite{Anselmi:2022qor}
\begin{equation}
\mathbb{1}=\sum\limits_{\left|n \right\rangle \in \mathcal{H}}\sigma_n\left| n \right\rangle \left\langle n \right| \qquad \to \qquad \mathbb{1}_{\rm ph}=\sum\limits_{\left|n \right\rangle \in \mathcal{H}_{\rm ph}}\left| n \right\rangle \left\langle n \right|\,,
\end{equation}
where $\mathbb{1}_{\rm ph}$ is the identity on the Hilbert space $\mathcal{H}_{\rm ph}$, and all normal states have been chosen with unit norm. We refer to the ghost quantized in this way as a \textit{fakeon ghost}.

Since both types of $\pm i\epsilon$ shifts appear in the same theory, analyticity is lost, in particular the standard Wick rotation cannot be performed. Unlike the anti-Feynman ghost case, however, for quadratic gravity with the spin-two ghost quantized as a fakeon, renormalizability and unitarity have been proven at all orders in perturbation theory by implementing alternative rules for the definition and deformation of the integration contour in loop integrals~\cite{Anselmi:2018kgz, Anselmi:2021hab}. In particular, this quantization defines the theory in the Euclidean signature, and then an appropriate non-analytic continuation to Lorentzian signature is made. Also in this approach causality is violated, but in a different form~\cite{Anselmi:2018kgz}.

\subsubsubsection*{Open questions}

Each of these unitary quantizations of quadratic gravity leave several questions open. It should be clear by now that reconciling perturbative renormalizability with unitarity in the presence of the spin-two ghost requires giving up something. For example, we have already mentioned that some form of causality is violated by the anti-Feynman and fakeon ghosts. In fact, causality is also violated by the Feynman ghost because the resummed propagator contains a pair of complex-conjugate poles on the first Riemann sheet. This feature modifies the analyticity property of the ghost propagator and the amplitudes in a non-trivial way~\cite{Grinstein:2008bg, Salvio:2018crh, Anselmi:2020lfx, Donoghue:2021cza, Kubo:2024ysu}. It is important to ensure that no macroscopic violation of causality occurs in these quantizations. On the other hand, acausal effects on microscopic scales (\ie{} on energy scales of the order of the ghost mass) could be in principle allowed and it is interesting to explore this further.

Unlike the case of the Feynman ghost, the anti-Feynman and fakeon prescriptions can preserve unitarity without introducing negative norms. However, it is still unclear whether they are compatible with the operator formalism of \ac{QFT}~\cite{Kubo:2023lpz}. The presence of both $\pm i \epsilon$ shifts in the same theory makes it unclear how to compute the amplitudes in terms of expectation values of ``ordered'' products of field operators. In particular, it is unclear what ``ordered'' means, since in principle both time-ordering and anti-time-ordering would be available as options. It is also worth to mention that a definition of non-time-ordered products was recently proposed at least in the case of fakeons~\cite{Anselmi:2022qor}.

Even if these approaches manage to solve the ghost puzzle of quadratic gravity at the quantum level, it is still important to ask what the classical limit of the theory is. Lagrangians containing ghost fields are typically considered pathological because of classical Hamiltonian instabilities and potential runaway solutions. It has been argued that both Feynman and anti-Feynman spin-two ghosts can decay due to a non-vanishing width, so that they could disappear from the set of asymptotic states~\cite{Grinstein:2008bg, Donoghue:2019fcb}. If true, this would imply that the spin-two ghost will not propagate on macroscopic scales, ensuring the classical stability of the theory. However, these claims have been criticized because the would-be unstable pole of the spin-two ghost propagator does not appear on the second Riemann sheet as in the case of ordinary unstable particles, but it shows up on the first Riemann sheet~\cite{Kubo:2024ysu, Buoninfante:2025klm}. This seems to suggest that the spin-two ghost does not decay, but still belongs to the set of asymptotic states. In this regard, it is interesting to further investigate the properties of a (stable) ghost-like resonance in contrast to that of an (unstable) ordinary resonance~\cite{Buoninfante:2025klm}. Furthermore, if the spin-two ghost is quantized as a fakeon, then it is not an on-shell degree of freedom from the start. The classical limit is taken consistently with the fakeon prescription by solving the field equations for the spin-two ghost field and integrating the latter out with appropriate boundary conditions~\cite{Anselmi:2018bra}. In this case, the classical dynamics becomes non-local, and its stability has not yet been proven.

\subsubsection{More open questions}\label{sec:open-quest-quad-grav}

In addition to the ghost puzzle, there are also important open questions related to the high-energy behavior of quadratic gravity. 

First of all, despite the property of renormalizability, it can actually be shown that the tree-level amplitude of the $2\to 2$ graviton scattering is still the same as in \ac{GR}~\cite{Dona:2015tra, Holdom:2021hlo}, \ie{} it grows with the square of the center-of-mass energy as $\sim E^2/\MPl^2$. This situation is very different from what happens in standard renormalizable \acp{QFT}. For example, in the \ac{SM} the introduction of the Higgs boson is not only important to ensure renormalizability, but also to suppress the cross section of the $W$-$W$ scattering. However, in the case of quadratic gravity, the additional spin-zero and spin-two degrees of freedom make the gravitational theory renormalizable, but they do not improve the behavior of the tree-level scattering amplitudes.

In \cite{Holdom:2021hlo, Holdom:2023usn} it has been argued that the only relevant physical quantities that can be measured in quadratic gravity are the totally-inclusive cross sections.\footnote{In a totally-inclusive cross section, not only are all possible final states summed over, but also all possible initial states. For example, in the case of quadratic gravity, one would add an amplitude with two ingoing gravitons to an amplitude with two ingoing spin-two ghosts, and so on.} These can be shown to be suppressed in the \ac{UV} regime due to special cancellations among contributions coming from the massless graviton and the massive spin-two ghost. This result applies only to the Feynman ghost case because negative norms are needed to have negative exclusive cross sections that can allow for cancellations in the totally-inclusive one. While this may shed new light on the high-energy behavior of quadratic gravity as a \ac{UV} completion and help distinguish between different approaches to the ghost puzzle, there are still ongoing debates about whether the only physical cross sections should be totally inclusive in this context. 

Furthermore, even if the tree-level exclusive cross section grows, one may hope that loop effects could improve the high-energy behavior if the interaction couplings are asymptotically free, \ie{} if $g_0$ and $g_2$ tend to zero in the \ac{UV} limit. However, it is not at all clear whether this would then imply that the amplitudes get suppressed at high energies. For instance, at least in a higher-derivative toy model~\cite{Buccio:2023lzo}, it has been shown that the scattering amplitudes at one loop continue to grow despite the fact that the interaction couplings tend to zero. This feature is due to the fourth-order derivative nature of the model: higher positive powers of momentum appear in the vertices and always overcome the logarithmic suppression of the couplings. It is still important to perform an analog calculation in the case of quadratic gravity and understand how loop effects influence the graviton-graviton scattering.

So far, we have mainly considered pure gravity without matter. In fact, another question that could be asked is whether the coupling of quadratic gravity to matter can help solve the Landau pole problems in the \ac{SM}. The answer appears to be negative. Indeed, despite the property of renormalizability of the gravity-matter system and the possibility to have asymptotic freedom in the gravitational sector, it can be shown that in this theory the \ac{SM} gauge couplings do not receive any gravitational correction~\cite{Salvio:2014soa, Anselmi:2018ibi}. Thus, the hypercharge gauge coupling will still hit a Landau pole in the trans-Planckian regime, leaving the question of quantum triviality open. However, it should also be noted that it is generally not known whether this is a problem that should be solved by \ac{QG} or by physics beyond the \ac{SM}.

These open questions still make it unclear whether quadratic gravity can be considered a valid \ac{UV} completion of the gravitational interaction. An alternative point of view might be that these facts are merely indications of the breakdown of the local perturbative description. In such a case, some new non-perturbative and/or non-local mechanism should then play a role. For instance, we know that gravity is more special than other interactions due to the existence of \acp{BH}, which are also expected to be present in quadratic gravity~\cite{Lu:2015cqa, Buoninfante:2024oyi}. The growth of the tree-level amplitude might suggest that a non-perturbative phenomenon, such as the formation of \acp{BH} by the scattering of energetic particles, should be taken into account as we approach the Planckian regime, and might play an important role for the \ac{UV} completion of the theory. However, these last words are not supported by any calculation. But perhaps some readers might find them interesting, or at least think about them and then conclude that they make no sense.

\subsubsection{Uniqueness and falsifiability}\label{sec:uniqueness}

Despite our critical assessment of the open questions about quadratic gravity, before concluding this section, let us mention something of which the theory could be proud of.

It is important to emphasize that in $d=4$ spacetime dimensions, quadratic gravity is the \textit{only} strictly renormalizable \ac{QFT} of gravity that is metric compatible, torsion-free and preserves the symmetries of \ac{GR}. This uniqueness property makes the theory predictive and falsifiable. Indeed, if an experiment will prove that the theory is wrong, then no other Lagrangian based on the same principles can be chosen, but one or more of the starting assumptions, for example that of strict renormalizability, must be changed.\footnote{It is worth mentioning that this is not the case for super-renormalizable \acp{QFT} of gravity, which can be constructed in terms of Lagrangians containing sixth- or higher-order derivatives of the metric field~\cite{Asorey:1996hz,Anselmi:2017ygm}. In this case, we have an infinite number of super-renormalizable \acp{QFT}. This means that even if an experiment proves one of them wrong, then we can still play the same game by choosing another super-renormalizable Lagrangian and still keep the starting principles unchanged. In this respect, the super-renormalizability criterion is less falsifiable.}   

The natural questions to ask now are: How can we in principle falsify quadratic gravity? Is there any new physical phenomenon beyond classical \ac{GR} that could be explained by quadratic gravity? Does the theory make predictions that could be tested with future observations? We can provide positive answers to these questions.

The most interesting application of quadratic gravity is to the evolution of the early universe. Indeed, the $R+R^2$ part of the action corresponds to the Starobinsky Lagrangian~\cite{Starobinsky:1980te,Starobinsky:1983zz,Kofman:1985aw}, where the additional massive spin-zero can represent a natural candidate for the inflaton field, and is sometimes called \textit{scalaron}. The observation of \ac{CMB} anisotropies can be explained by fitting the mass of the scalaron to a value of the order of $m_0\sim 10^{13}$ GeV or, equivalently, $c_0\sim 10^{10}$~\cite{Starobinsky:1983zz,Planck:2018jri}. This large value of the $R^2$-coefficient is consistent with the validity of the perturbative expansion.

The next step would be to study the effects of the Weyl-squared term. The latter can influence the dynamics of the primordial \acp{GW} and thus the tensor power spectrum. An important observable quantity that experimentalists hope to measure in the near future is the so-called tensor-to-scalar ratio, which is defined as the ratio of the tensor power spectrum to the scalar one. In quadratic gravity, its leading contribution in the slow roll expansion has been computed in various works and reads~\cite{Deruelle:2012xv, Salvio:2017xul, Anselmi:2020lpp, Salvio:2022mld}
\begin{equation}\label{tensor-to-scalar-luca} 
r=\frac{24}{N_{e}^2}\frac{m_2^2}{m_0^2+2m_2^2}=\frac{24}{N_{e}^2}\frac{c_0}{c_2+2c_0}\,,
\end{equation}
where $N_{e}\sim 55\text{-}65$ is the number of e-folds. The Starobinsky prediction is recovered in the limit $m_2/\MPl\to \infty$ or, equivalently, $c_2\to 0$, and is given by $r\to 12/N_{e}^2\simeq 0.003$.

Since $c_0$ is already fixed by experiments, then a future measurement of $r$ could be used to fix the value of $c_2$. If the value of $c_2$ turns out to be too large, meaning that the mass of the spin-two ghost $m_2$ is too small, then we may get contradictions with other well-established low-energy experiments, such as solar system observations. In this respect, quadratic gravity is falsifiable. Furthermore, once $c_2$ is fixed through the measurement of the tensor-to-scalar ratio, then all the other cosmological parameters (such as the tensor tilt and running of the scalar spectral index) can be predicted by the theory.

From a high-energy point of view, since no additional effect has been observed yet, it is natural to expect that the mass $m_2$ is at least of the same order or larger than $m_0\sim 10^{13}$ GeV. The hierarchy $m_2\gtrsim m_0$ (or, equivalently, $c_0\gtrsim c_2$) would also be consistent with the fakeon prescription whose consistency in the inflationary background requires $m_2\geq m_0/4$~\cite{Anselmi:2020lpp}. If we impose $m_2\gtrsim m_0/4$ in \eqref{tensor-to-scalar-luca} and take $N_{e}\sim 55\text{-}65$, we would obtain $r\sim \orderneglected(10^{-4}\text{-}10^{-3})$. Therefore, the expectation in quadratic gravity for the tensor-to-scalar ratio would still be of the same order or very close to Starobinsky's. 

Current experimental constraints give $r\lesssim 10^{-2}$~\cite{Planck:2018jri}, while future satellite missions, such as LiteBird~\cite{Paoletti:2022kij}, aim to look for values of the tensor-to-scalar ratio of the order of $\mathcal{O}(10^{-3}).$ This means that in ten years or so we may already be able to test Starobinsky's prediction and quadratic gravity as a \ac{UV} completion of the Starobinsky model.\footnote{Among the many exciting and fun things that happened during the Nordita Scientific Program, there was a bet between me and my dear friend, co-author and co-organizer Benjamin Knorr. If in the future no tensor-to-scalar ratio is measured in the range $r\sim \orderneglected(10^{-4}\text{-}10^{-3})$, then you will know that I have lost the bet. We thank Ivano Basile and Alessia Platania for acting as bet commissioners.}

\subsection{Conclusions}\label{sec:conclus-luca}

It is now time to conclude this ``perturbative'' journey in  \ac{QG}. Before we say goodbye, let us briefly summarize the main encounters made during this trip and what we learned. After introducing key elements of classical \ac{GR}, we quantized Einstein's theory in the framework of perturbative \ac{QFT}. First, we analyzed the free theory (\ie{} without interactions). Off-shell and on-shell degrees of freedom were identified and counted by exploiting the gauge freedom of \ac{GR} due to diffeomorphism invariance. The graviton propagator was derived in three different ways: in the covariant Feynman gauge, in the non-covariant Prentki gauge, and in the covariant de Donder gauge (keeping the gauge-fixing parameter generic) by using the spin-projector formalism. Second, we introduced (self-)interactions and showed that \ac{GR} as a \ac{QFT} is perturbatively non-renormalizable. Without going into sophisticated technicalities, we have managed to determine in detail, up to numerical coefficients, the structure of one-loop and two-loop divergences, and we have also briefly commented on the higher loops. In particular, we showed that pure Einstein's gravity is one-loop finite.

One of the messages of this section was that the expression ``perturbative \ac{QG}'' does not just refer to the perturbative \ac{QFT} of \ac{GR}, but to any possible consistent perturbative \ac{QFT} of the gravitational interaction. Indeed, we presented quadratic gravity, which in $d=4$ spacetime dimensions is a unique strictly renormalizable \ac{QFT} of gravity that is metric compatible, torsion-free and respects the symmetries of \ac{GR}. We studied its main features, such as degrees of freedom, propagator and power-counting renormalizability, discussed the falsifiability and predictive power of the theory, and highlighted the open questions. In particular, we explained that a future measurement of the tensor-to-scalar ratio of primordial inflationary fluctuations can provide a way to test or rule out the theory. Therefore, the interplay between theory and experiments could play a very important role for the field of \ac{QG} in the coming years and give rise to new surprises. 

Despite the end of this journey, the \ac{QG} adventure is not over yet. In fact, this first set of lectures also provides the basics for the \ac{EFT} treatment of \ac{GR} whose formalism and applications will be studied in \cref{sec:ANNA} and, at the same time, establishes a common ground on which other \ac{QG} approaches can build. Indeed, \ac{ASQG} and hence the idea that gravity could be perturbatively non-renormalizable but non-perturbatively renormalizable will be discussed in \cref{sec:ALESSIABENJAMIN}. Furthermore, the possibility that \ac{ST} might be needed beyond the \ac{QFT} framework to consistently describe quantum aspects of gravity in the high-energy regime will be presented in \cref{sec:IVANO}.

We hope you have enjoyed this \ac{QG} experience so far and that you will continue to have fun with the next sets of lectures. Ciao!



\begin{subappendices}

\subsection{Elements of perturbative QFT}\label{sec:elemQFT}

In this appendix, we take a quick tour through some of the fundamental principles on which the perturbative \ac{QFT} framework is based. Our aim is not to be very detailed, but to transmit the main features and physical understanding behind these principles. Examples using toy models and known theories such as \ac{QED} will be provided in order to illustrate the basic aspects.

\subsubsection{Locality}\label{sec:Luca locality}

The principle of locality states that all bare Lagrangians must depend polynomially on the derivatives acting on the fields. This means that we can only have differential operators of finite order, \ie{} given a generic tensorial field $\phi(x)$ we can only have
\begin{tcolorbox}
\begin{equation}
	\lagrangian{}=\lagrangian{}\left(\phi,\partial \phi,\partial^2\phi,\dots,\partial^{(n)}\phi
	\right)\,,\qquad n<\infty\,.\label{local-lagran}
\end{equation}
\end{tcolorbox}
Therefore, the bare Lagrangians in local \ac{QFT} cannot contain non-polynomial differential operators like $1/\Box$, $e^{\Box},\ln \Box$, etc. However, non-localities can arise in the quantum effective action due to loop corrections. For example, it is very natural to generate logarithmic non-localities at one loop.

The name ``locality'' comes from the fact that if we go to momentum space, the derivatives become momenta, and the Fourier transform of a polynomial of the momenta would generate terms involving a Dirac delta and its derivatives, which are distributions localized at a single point.  

All \ac{SM} interactions are described in terms of local bare Lagrangians. The kinetic terms of bosons contain second order derivatives (\eg{} $\partial_\mu\phi\partial^\mu\phi$ and $\partial_\mu A_{\nu} \partial^\mu A^\nu$) and those of fermions contain first order derivatives (\eg{} $\bar{\psi}i\gamma^\mu\partial_\mu \psi$). The interaction terms contain up to first order derivatives. In this section, we studied two local gravitational \acp{QFT}: \ac{GR} and quadratic gravity. The latter contains derivatives of order up to four.

\subsubsection{Symmetries}

Strong constraints on the type of Lagrangians we can write down arise from symmetry requirements. Actions can be characterized by spacetime symmetries as well as global and local internal symmetries.

\paragraph{Poincar\'e symmetry.} Standard local \acp{QFT} in Minkowski spacetime are invariant under the action of the global Poincar\'e group (also called inhomogeneous Lorentz group) that includes proper Lorentz transformations (spatial rotations and boosts) and spacetime translations: 
\begin{equation}
	x^\mu\to x^{\prime\mu}=\Lambda^\mu_{\phantom{\mu}\nu} x^\nu + a^\mu\,,\qquad {\det}\, \Lambda=1\,,
\end{equation}
where $\Lambda^\mu_{\phantom{\mu}\nu}$ is the matrix of proper Lorentz transformations with ${\det}\,\Lambda=1$, and $a^\mu$ is a constant four-vector. In four spacetime dimensions the Poincar\'e group is denoted by $ISO(3,1)$ where the ``$I$'' means ``inhomogeneous'' and $SO(3,1)$ is the proper Lorentz group: spacetime translations make the Lorentz transformations inhomogeneous. 

The group $SO(3,1)$ has two disconnected components: orthochronous ($\Lambda^0_{\phantom{0}0}=1$) and non-orthochronous ($\Lambda^0_{\phantom{0}0}=-1$). The latter can also be seen as the composition of the former with a discrete inversion of the time coordinate. Therefore, it is sufficient to focus on orthochronous proper Lorentz transformations.

The requirement of invariance under the action of the Poincar\'e group is very important for at least  \textit{two} reasons.
\begin{itemize}
    \item We can define tensorial fields as functions of coordinates with well-defined transformation properties under the Poincar\'e group. All fields are invariant under spacetime translations, but transform in different ways under $SO(3,1)$ depending on their spin value. Invariance of the action under the Lorentz group non-trivially constrains the form of kinetic and interaction terms in a Lagrangian, in particular time and space derivatives must appear with the same order in a manifestly Poincar\'e-invariant description. 
	
    \item Using the Hilbert space of one-particle states as representation basis of the Poincar\'e group, we can classify massive particles in terms of their mass $m$ and their spin $j$, and massless particles in terms of their helicity $\lambda$, which is defined as the projection of the angular momentum along the direction of the particle momentum, \ie{} $\lambda=\hat{p}\cdot \vec{J}$. In four spacetime dimensions, massive particles have $2j+1$ on-shell degrees of freedom (\eg{} a massive vector field has $3$ degrees of freedom), while massless particles have one on-shell degree of freedom associated to each helicity state. Since the helicity transforms as a pseudoscalar under parity, to each particle with helicity $\lambda$ we can associate another particle with opposite helicity $-\lambda$. In a theory that is invariant under parity transformations, these two helicity components must appear in a symmetric way. Therefore, in such situations it is convenient to assemble the two states with opposite helicities. For example, this is what is done with photons and gravitons, which are defined as particles with two values of helicity, $\lambda=\pm 1$ and $\lambda=\pm 2$, respectively.
\end{itemize}

The above considerations are valid for fields defined in Minkowski spacetime. Around less symmetric curved backgrounds, the above considerations do not generally follow. In other maximally symmetric spacetimes, such as \ac{dS}~\cite{Sun:2021thf}, we still have enough symmetry to define the concept of a particle and make a classification. Furthermore, in a generic spacetime the global Lorentz invariance is replaced by a local version.

In this section we have considered metric perturbations around the flat Minkowski spacetime. Therefore, the metric field fluctuation $h_{\mu\nu}$ lives on a Minkowski background, and the corresponding one-particle states can be classified according to the global Poincar\'e group.

\paragraph{Internal symmetries.} In addition to spacetime symmetries, actions in \ac{QFT} can also be invariant under symmetry groups that only act on the fields, while leaving the spacetime coordinates unchanged. We can have global and local internal symmetries. The latter are also known as \textit{gauge} symmetries. These are not true physical symmetries, but rather a redundancy regarding the number of physical degrees of freedom needed to describe the system in a local fashion. Gauge symmetries are also very important to constrain the structure of the interaction terms in a Lagrangian.

\medskip

\noindent\textit{Example.} \ac{QED} is a gauge theory describing interactions between photons $(A_\mu)$ and electrons $(\psi, \bar{\psi})$. The action is given by 
\begin{equation}
	S=\int {\rmd}^4x \left[-\frac{1}{4}F_{\mu\nu}F^{\mu\nu}-\bar{\psi}(i\gamma^\mu\partial_\mu+m)\psi-e\bar{\psi}\gamma^\mu\psi A_\mu\right]\,,
\end{equation}
where $F_{\mu\nu}=\partial_\mu A_\nu-\partial_\nu A_\mu$, and it is invariant under the gauge transformation
\begin{equation}
	\psi\to e^{i\alpha(x)}\psi\,,\qquad \bar{\psi}\to e^{-i\alpha(x)}\bar{\psi}\,,\qquad A_\mu\to A_{\mu}+\frac{1}{e}\partial_\mu \alpha(x)\,.
\end{equation}

\subsubsection{Unitarity}\label{sec:app-unitarity}

Unitarity in quantum mechanics is the statement that quantum probabilities are conserved. Given an isolated quantum mechanical system, we cannot suddenly lose information about some of its degrees of freedom, nor can we gain new information from nothing.

We now define the concept of unitarity with respect to the S-matrix operator. Given an initial state $\left| a\right\rangle$ and a final state $\left| b\right\rangle$ that belong to some physical Hilbert space, we define the S-matrix as an evolution operator that acts on $\left| a\right\rangle$ at $t=-\infty$ and evolves it to $\left| b\right\rangle$ at $t=+\infty$: $\left| b\right\rangle= S \left| a\right\rangle$. Probability is conserved if $\left\langle a|a \right\rangle  = \left\langle b|b \right\rangle = \left\langle a \big|S^{\dagger}S\big| a \right\rangle$, which is true if and only if $S$ is a unitary operator:
\begin{equation}
	S^\dagger S=  \mathbbm{1}\,.\label{S-unitarity-Luca}
\end{equation}

It is convenient to divide the S-matrix in two parts as $S=\mathbbm{1}+iT$, where $T$ is known as transfer matrix. In terms of $T$, the unitarity relation~\eqref{S-unitarity-Luca} becomes
\begin{equation}\label{optical-operat}
	-i(T-T^\dagger)=T^\dagger T\,,
\end{equation}
which is the \textit{optical theorem} in operator form. In practice, we work with matrix elements, therefore we would like to obtain the optical theorem in terms of the elements of $T$. This can be done by using the completeness relation
\begin{equation}
	\mathbbm{1}=\sum_{n} \sigma_n \left| n \right\rangle \left\langle n\right|  \,,\label{complet-n}
\end{equation}
where the coefficients $\sigma_n$  are defined via $\left\langle n|m \right\rangle = \delta_{nm}\sigma_n$, and sandwiching~\eqref{optical-operat} with an initial and a final state:
\begin{equation}
	-i\left[\big\langle b |T| a\big\rangle -\left\langle b |T^\dagger| a\right\rangle \right]=\sum_n \sigma_n \left\langle b |T^\dagger| n\right\rangle \big\langle n |T| a\big\rangle\,. 
	\label{optical-n}
\end{equation}
The optical theorem \eqref{optical-n} is still not written in an explicit form because the definition of $T$ hides Dirac deltas that take into account momentum conservation, and the completeness relation~\eqref{complet-n} hides the phase-space integration. 

If we introduce the Feynman amplitude $\scatteringamplitude$ as $\left\langle b |T| a\right\rangle=(2\pi)^4\delta^{(4)}(P_b-P_a)\left\langle b |\scatteringamplitude| a\right\rangle$, $P_b$ and $P_a$ being the total final and initial momenta, and use  the more explicit form of the completeness relation,
\begin{equation}
	\mathbbm{1}=\sum_{\left\lbrace n \right\rbrace}\prod_{l=1}^n\int \frac{{\rmd}^3k_l}{(2\pi)^3}\frac{1}{2\omega_l} \left|\left\lbrace k_l\right\rbrace  \right\rangle \left\langle \left\lbrace k_l\right\rbrace \right|\,,
	\label{complet-integ}
\end{equation}
where the summation is over all possible sets $\left\lbrace n \right\rbrace $ of intermediate states $\left|\left\lbrace k_l\right\rbrace\right\rangle $ containing $l$ momenta, and $\omega_l=\sqrt{\vec{k}_l^2+m_l^2}$ are the frequencies (energies) for each single momentum $k_l$, we can recast \eqref{optical-n} into the following form:
\begin{tcolorbox}
\begin{equation}\label{optical theorem-feynman}
\begin{aligned}
	i\left[\left\langle b|\scatteringamplitude^{\dagger}|a \right\rangle - \big\langle b\big|\scatteringamplitude\big|a \big\rangle \right]=\sum_{\left\lbrace n \right\rbrace}\sigma_n&\prod_{l=1}^n\int\frac{{\rmd}^3k_l}{(2\pi)^3}\frac{1}{2\omega_l} 
		(2\pi)^4\times \\
       & \delta^{(4)} \left(P_a-\sum_{l=1}^nk_l\right)\left\langle b\big|\scatteringamplitude^{\dagger}\big|\left\lbrace k_l\right\rbrace  \right\rangle \big\langle \left\lbrace k_l\right\rbrace\big|\scatteringamplitude\big|a \big\rangle\,,
        \end{aligned}
\end{equation}
\end{tcolorbox}
\noindent which must hold order by order in perturbation theory. 

The evaluation of the two sides of the optical theorem in \eqref{optical theorem-feynman} requires some more input about the sign of $\sigma_n$ and the prescription for shifting the singularities of the propagator, which is also related to the type of contour deformation needed to compute the matrix elements at a generic loop order. The sign of $\sigma_n$ is linked to the sign of probabilities, such as cross sections, while the prescription for the propagator is connected to the arrow of causality. These two choices have to be made compatibly with the condition of unitarity. Let us make some additional remarks.

\paragraph{Signs of the norms.} Unitarity (\ie{} the optical theorem) does not say anything about the sign of the probabilities. Indeed, some of the coefficients $\sigma_n$ can be negative and the unitarity relation can still hold: the total probability can be conserved even if some of the states have negative (squared) norm.\footnote{Strictly speaking, $\sigma_n=\left\langle n|n \right\rangle$ is a squared norm since the norm is defined as $\norm{|n\rangle} \equiv \sqrt{\langle n | n \rangle}$. However, in this section we refer to $\sigma_n$ simply as the norm, as is often the case in the literature.} In most of this section we considered physical states to have positive norms normalized to one, \ie{} $\sigma_n=1$. However, in \cref{sec:lecture4} we saw an explicit example of a situation where negative norms may be needed to preserve unitarity, indeed this may happen in the context of \acp{QFT} where the kinetic term contains higher-order derivatives, \ie{} theories with ghosts,\footnote{Remember that here a ghost is defined as a field whose kinetic term has a sign that is opposite to ordinary fields. Despite the same name, do not confuse this notion of ghost field with the Faddeev-Popov ghost fields introduced in \cref{sec:faddeev} .} such as quadratic gravity.

\paragraph{Positivity constraint.} Unitarity plus the assumption of positive-definite norms imply positivity constraints on the imaginary part of an amplitude. Indeed, if we consider an elastic process ($\left\lbrace b \right\rbrace = \left\lbrace a \right\rbrace $) and $\sigma_n=1$, \eqref{optical theorem-feynman} becomes

\begin{tcolorbox}
\begin{equation}
	2{\rm Im}\left[\left\langle a|\scatteringamplitude|a \right\rangle \right]=\sum_{\left\lbrace n \right\rbrace}\prod_{l=1}^n\int \frac{{\rmd}^3k_l}{(2\pi)^3}\frac{1}{2\omega_l} 
	(2\pi)^4 \delta^{(4)}\left(P_a-\sum_{l=1}^nk_l\right) \big|\big\langle \left\lbrace k_l\right\rbrace\big|\scatteringamplitude\big|a \big\rangle\big|^2\geq 0\,.
	\label{optical theorem-positivity}
\end{equation}
\end{tcolorbox}

\paragraph{Arrow of causality.} The notion of causality we adopted in this section is that according to which particles with positive energy propagate forward in time, while particles with negative energy travel backward in time. The second part of the statement means that anti-particles propagate positive energy forward in time. This notion is associated with the time-ordered structure of the propagator, which is also related to the \textit{Feynman prescription} that tells us how to shift the propagator poles in the complex energy plane to avoid singularities on the real axis.

Given a scalar field $\phi$ and a free vacuum $\left|0 \right\rangle$, the non-interacting causal Feynman propagator reads
\begin{equation} \label{space-feynm-propag}
\begin{aligned}
	\propG_\phi(x-y)&\equiv\left\langle 0|T\left\lbrace \phi(x)\phi(y)\right\rbrace |0\right\rangle \\
	&= \theta(x^0-y^0)\left\langle 0| \phi(x)\phi(y) |0\right\rangle + \theta(y^0-x^0)\left\langle 0| \phi(y)\phi(x) |0\right\rangle \\
	&=\int \frac{{\rmd}^3p}{(2\pi)^3}\frac{e^{i\vec{p}\cdot(\vec{x}-\vec{y})}}{2\omega_{\vec{p}}}\left[\theta(x^0-y^0)e^{-i\omega_{\vec{p}}(x^0-y^0)}+\theta(y^0-x^0)e^{i\omega_{\vec{p}}(x^0-y^0)} \right] \\
	&= \int \frac{{\rmd}^4p}{(2\pi)^4}\tilde{\propG}_\phi(p,\epsilon)e^{ip\cdot (x-y)}\,,
	\end{aligned}
\end{equation}
where $\omega_{\vec{p}}\equiv \sqrt{\vec{p}^2+m^2}$ and $\theta(x)$ is the Heaviside function defined as
\begin{equation}
	\theta(x)=\left\lbrace \begin{array}{cc}
		1\,,& x>0\,,\\
		1/2\,,& x=0\,,\\
		0\,,& x<0\,.
	\end{array}	\right.
	\label{theta-H}
\end{equation}
In the last line of \eqref{space-feynm-propag}, we have introduced the Feynman propagator in momentum space
\begin{tcolorbox}
\begin{equation}
\tilde{\propG}_\phi(p,\epsilon)\equiv \frac{-i}{p^2+m^2-i\epsilon}\,,\label{momentum-feyn-propag}
\end{equation}
\end{tcolorbox}
\noindent where the shift $p^2\to p^2-i\epsilon$ with $\epsilon\to 0^+$ is the Feynman prescription that displaces the poles in the complex energy plane as $p_0=\pm \omega_{\vec{p}}\to \pm(\omega_{\vec{p}}-i\epsilon)$. In the third line of \eqref{space-feynm-propag}, the term proportional to $e^{-i\omega_{\vec{p}}t}$ (where $t\equiv x^0-y^0\geq 0$) is associated to the propagation of positive energy as $i\frac{\partial}{\partial t} e^{-i\omega_{\vec{p}}t}=+\omega_{\vec{p}} \,e^{-i\omega_{\vec{p}}t}$, while the second term is associated to negative energy propagating backward in time $(t<0)$.

The Feynman prescription is accompanied by the Wick rotation when evaluating diagrams at higher-loop orders. If we use this prescription together with the assumption of positive norms for the physical states in the Hilbert space, then it can be shown that in ghost-free theories, the optical theorem can be satisfied. This is indeed the case for the \acp{QFT} of \ac{SM} interactions.

\medskip
\noindent \textit{Example.} To understand how the Feynman prescription and positive norms can coexist in ghost-free theories, let us show tree-level unitarity of $\lambda \phi^3$ theory as a toy example. The propagator with Feynman prescription is given by $-i/(p^2+m^2-i\epsilon)$ and the three-vertex by $-i\lambda$. We want to show that the following tree-level $2\to2$ scattering amplitude is unitary:
\begin{equation}
	\left\langle p_3,p_4 |\scatteringamplitude| p_1,p_2\right\rangle = (-i)(-i\lambda) \frac{-i}{p^2+m^2-i\epsilon}(-i\lambda)=\frac{\lambda^2}{p^2+m^2-i\epsilon}\equiv \left\langle p |\scatteringamplitude| p\right\rangle \,,
\end{equation}
where $p_1,p_2$ ($p_3,p_4$) are the initial (final) momenta, while $p=p_1+p_2=p_3+p_4$.
The left-hand side of the optical theorem \eqref{optical theorem-feynman} reads
\begin{equation}
	2{\rm Im}[\left\langle p |\scatteringamplitude| p\right\rangle]=2\pi\lambda^2 \delta(p^2+m^2)\,,\label{lhs-example}
\end{equation}
where we have used the Sokhotski–Plemelj formula 
\begin{equation}
\frac{1}{x\pm i\epsilon}={\rm P.V.}\left(\frac{1}{x}\right)\mp i\pi \delta(x).
\end{equation}
We now want to show that the right-hand side of \eqref{optical theorem-feynman} coincides with \eqref{lhs-example}. Since we have only one set of intermediate states, namely only one intermediate state (one internal leg) $n=1$, $\left| \left\lbrace k_1 \right\rbrace \right\rangle  \equiv  \left| k \right\rangle$, $\omega_1\equiv \omega= \sqrt{\vec{k}^2+m^2}$, we can write the right-hand side as
\begin{equation}
	\int \frac{{\rmd}^3k}{(2\pi)^3}\frac{1}{2\omega}(2\pi)^4\delta^{(4)}(p-k)\left\langle p_3,p_4\big|\scatteringamplitude^\dagger\big| k  \right\rangle \big\langle k\big|\scatteringamplitude\big| p_1,p_2  \big\rangle\,.
\end{equation}
Using  $\left\langle k|\scatteringamplitude| p_1,p_2  \right\rangle=(-i)(-i\lambda)=-\lambda$,  $\langle p_3,p_4|\scatteringamplitude^\dagger| k  \rangle=(\langle k|\scatteringamplitude| p_3,p_4  \rangle)^\ast=(-\lambda)^\ast=-\lambda$ and $\int \frac{{\rmd}^3k}{(2\pi)^3}\frac{1}{2\omega}=\int \frac{{\rmd}^4k}{(2\pi)^4}2\pi \delta(k^2+m^2)$, we obtain 
\begin{equation}
	\lambda^2 \int \frac{{\rmd}^4k}{(2\pi)^4}2\pi \delta^{(4)}(p-k)\delta(k^2+m^2) = 2\pi \lambda^2 \delta(p^2+m^2)\,,
\end{equation}
which is equal to~\eqref{lhs-example}, consistently with unitarity.

\paragraph{Remark.} It is important to emphasize that in standard \acp{QFT}, the definition of a causal propagator with positive (negative) energy propagating forward (backward) in time is a convention. In fact, we could flip the overall sign in front of the Lagrangian and choose the opposite arrow of causality. In this case, unitarity with positive norms would force us to implement the so-called anti-Feynman prescription with $p^2\to p^2+i\epsilon$, $\epsilon\to 0^+$, which would physically imply that positive/negative energy propagates backward/forward in time. However, this freedom in terms of conventions for the flow of energy and the overall sign in front of the Lagrangian is not available anymore when both normal and ghost fields are present in the theory. In such a case, two opposite arrows of time, one associated to ordinary (\ie{} non-ghost) particles, and the other to ghost particles, could be considered simultaneously, and would lead to a violation of causality at the energy scale of the ghost mass. This kind of situation can arise in higher-derivative theories: depending on the quantization prescription adopted, an example could be quadratic gravity as discussed in~\cref{sec:lecture4}.

\subsubsection{Renormalizability}\label{sec:app-renormaliz}

In perturbative \ac{QFT}, we are interested in computing various quantities such as correlators, form factors and S-matrix elements, from which we can extract physical observables. Quantum corrections computed in perturbation theory are usually divergent in the \ac{UV} regime because of divergent integrals. For instance, S-matrix elements $\left\langle b|\scatteringamplitude|a \right\rangle$ can diverge at some or any loop order. 

The standard way to deal with \ac{UV} divergences is to assume that all quantities in the starting Lagrangian are bare and divergent. Then, we can apply renormalization theory that consists in two main steps: (i) implement a regularization method to isolate the divergent part of a diagram/integral; (ii) absorb the divergences into a redefinition of the quantities in the Lagrangian (\ie{} masses, couplings and fields). Powerful theorems ensure that for local theories, the renormalization procedure can be carried out consistently at all orders in perturbation theory~\cite{Anselmi:2019pdm}. In particular, if the starting bare Lagrangian is local, then the divergences are also local in the momenta (\ie{} in the derivatives). In other words, the functional form of the counterterms needed to cancel the divergences is guaranteed to be local if the bare Lagrangian is also local~\cite{Anselmi:2019pdm, tHooft:1973wag, Weinberg:1995mt}.

This procedure is predictive for \textit{renormalizable} \acp{QFT} because only a finite number of terms in the Lagrangian is needed to completely renormalize all physical quantities and make predictions up to energy scales where the perturbative regime is valid. In what follows, we first give some examples of common \ac{UV}-divergent integrals in \ac{QFT}, and then define the concept of \textit{power-counting renormalizability}.

\paragraph{Superficial degree of divergence.} We can check whether a diagram diverges at a certain loop order by computing the superficial degree of divergence. Let us consider a diagram $G$ with the following associated loop integral 
\begin{equation}
	I^{(L)}(G)=\int {\rmd}^4k_1\cdots {\rmd}^4k_L \mathcal{I}(\{k_i\})\,,   \label{loop-integral}
\end{equation}
where $L$ denotes the number of loops, $k_i$ with $i=1,\dots,L$ the internal momenta, and $\mathcal{I}(\{k_i\})$ is the integrand that can also depend on the external momenta. We define the superficial degree of divergence $\delta(G)$ of the integral~\eqref{loop-integral} via the relation 
\begin{tcolorbox}
\begin{equation}
	\lim_{\lambda\to \infty} \lambda^{4L}\mathcal{I}(\{\lambda k_i\})\sim \lambda^{\delta(G)}\,.   \label{sup-degree-div}
\end{equation}
\end{tcolorbox}
\noindent If $\delta(G)<0$ the integral is convergent; if $\delta(G)=0$ we have a logarithmic divergence; while if $\delta(G)>0$ the integral has a power-law divergence.

\medskip
\noindent\textit{Examples.} The tadpole diagram
\begin{equation}
\int {\rmd}^4k \frac{1}{k^2+m^2}
\end{equation}
has $\delta(G)=2$ because $\lim\limits_{\lambda\to \infty}\lambda^4(\lambda^2k^2+m^2)^{-1}\sim \lambda^2$ (power-law divergence).

The bubble diagram
\begin{eqnarray}
	\int {\rmd}^4k \frac{1}{k^2+m^2}\frac{1}{(k-p)^2+m^2}
\end{eqnarray}
has $\delta(G)=0$ because $\lim\limits_{\lambda\to \infty}\lambda^4(\lambda^2k^2)^{-2}\sim \lambda^0$ (logarithmic divergence).

\paragraph{Power-counting renormalizability.} We now want to find a formula that expresses $\delta(G)$ in terms of the mass dimensions of couplings and fields in the Lagrangian. This will allow us to distinguish different types of \acp{QFT} based on their \ac{UV} properties.

Let us define the following parameters:
\begin{itemize}
    \item $n_V$: number of different types of vertices that are relevant in the \ac{UV},
    \item $d_i$: number of derivatives in the \textit{i}-th vertex,
    \item $f$: type of field,
    \item $n_{i,f}$: number of fields of type \textit{f} in the \textit{i}-th vertex.
\end{itemize}
%
Then, a generic diagram $G$ is characterized by the following quantities:
\begin{itemize}
    \item $L$: number of loops,
    \item $I_f$: number of internal propagators of type $f$,
    \item $E_f$: number of external legs of type $f$,
    \item $V_i$: number of \textit{i}-th vertices.
\end{itemize}
%
%
Consider a generic Lagrangian density $\lagrangian{}=\mathcal{K}-\mathcal{V}$, whose kinetic and interaction terms can be written in the following schematic form:
\begin{equation}
	\mathcal{K}\sim \sum_f \phi_f \left(\partial^{(2-2s_f)}\right)^{r_f}\phi_f\,,\qquad \mathcal{V}\sim \sum_i g_i \partial^{d_i}\prod_f \phi_f^{n_{i,f}}\,,
	\label{kinetic-vertex}
\end{equation}
where $s_f=0$ for bosons and $s_f=1/2$ for fermions, and $r_f$ takes into account the possibility to have kinetic operators that contain higher-order derivatives. The propagator for a field of type $f$ behaves as
\begin{equation}
	\tilde{\propG}_{\phi_f}(p)\sim p^{(2s_f-2)r_f}\,.
\end{equation}
Equipped with all the above definitions, a generic diagram $G$ can be written as
\begin{equation}
	{\int \underbrace{{\rmd}^4k\cdots {\rmd}^4k}_{L\text{-loops}}}\,\,\, \times\,\,\,{\underbrace{k^{(2s_f-2)r_f}\cdots k^{(2s_f-2)r_f}}_{I_f\text{-internal propagators}}}\,\,\,\times\, {\underbrace{k^{d_i}\cdots k^{d_i}}_{V_i\text{-vertices of type \textit{i}}}}\,.
	\label{generic-diagram}
\end{equation}
The \ac{UV} behavior is given by
\begin{eqnarray}
	k^{4L+I_f(2s_f-2)r_f+d_iV_i}=k^{4(I_f-V_i+1)+I_f(2s_f-2)r_f+d_iV_i} \,,\label{single-delta}
\end{eqnarray}
where we have used the topological identity $L=I_f-V_i+1$.\footnote{This can be easily proven as follows. The number of independent momenta flowing through the legs of a generic loop diagram is $(E-1)+L$, where the $-1$ is needed because the external momenta satisfy a conservation law. At the same time, the number of independent momenta can also be written as $E+I-V$ where $E+I$ is the total number of legs (external plus internal) and $V$ takes into account the fact that not all the momenta flowing through the legs are independent because of momentum-conserving Dirac deltas in the vertices. Thus, equating these two expressions we get the desired topological identity.} The exponent in~\eqref{single-delta} is the superficial degree of divergence for a single type of field $f$ and a single type of vertex $i$. To obtain the total superficial degree of divergence of the diagram $G$ we have to sum over $f$ and $i$, thus we get
\begin{equation}	\label{total-delta}
\begin{aligned}
	\delta(G)&=4\left(\sum_fI_f-\sum_iV_i+1\right)+\sum_fI_f\left(2s_f-2\right)r_f+\sum_id_iV_i \\
	&= 4 + \sum_f I_f \big[(2s_f-2)r_f+4\big]+\sum_i(d_i-4)V_i \, .
    \end{aligned}
\end{equation}

Note that each internal line is always attached to two vertices, while each external line is attached to one vertex. This means that we can write
\begin{equation}
	2I_f+E_f=\sum_i V_i n_{i,f}\qquad \Leftrightarrow \qquad I_f=-\frac{1}{2}E_{f}+\frac{1}{2}\sum_i V_i n_{i,f}\,,
\end{equation}
which implies
\begin{equation}
	\delta(G)=4 - \sum_f E_f \big[(s_f-1)r_f+2\big]-\sum_{i}V_i\Big[ 4-d_i-\sum_fn_{i,f}\big((s_f-1)r_f+2\big)  \Big]\,.
	\label{total-delta-2}
\end{equation}

We now want to show that the expression in the first square bracket corresponds to the mass dimension of the field of type $f$, and that the expression in the second square brackets is equal to the mass dimension of the \textit{i}-th coupling constant $g_i$. If we call $F_f\equiv [\phi_f]$ and $\Delta_i\equiv [g_i]$ the mass dimensions of $\phi_f$ and $g_i$, respectively, from \eqref{kinetic-vertex} we get
\begin{equation}
	F_f=(s_f-1)r_f+2\qquad \text{and}\qquad \Delta_i=4-d_i-\sum_f n_{i,f}\big[(s_f-1)r_f+2\big]\,.\label{fields-coupling-dimensions}
\end{equation}
Then, using~\eqref{fields-coupling-dimensions} we can write the superficial degree of divergence as
\begin{tcolorbox}	\label{total-delta-3}
    \begin{equation}
	\delta(G)=4 - \sum_f E_f F_f-\sum_{i}V_i\Delta_i\,.
\end{equation}
\end{tcolorbox}

It is worth emphasizing that the quantities $\Delta_i$ are the dimensions of the couplings $g_i$ that are relevant for the \ac{UV} behavior of the theory, and $n_V$ counts the number of these vertices. For example, mass parameters are usually not important and do not contribute to the superficial degree of divergence. In theories with vertices containing different powers of derivatives, the couplings that are relevant in the \ac{UV} are those associated with the highest derivative order.

\paragraph{\textit{Non}-renormalizability and~\textit{strict}/\textit{super}-renormalizability.} We can now distinguish two different cases. 
\begin{itemize}
	
	\item  If $\exists$ $i$ such that $\Delta_i<0$, then the superficial degree of divergence increases with the number of vertices. In this case, there is an infinite number of divergent Green's functions $G_{E_f}$ that need to be renormalized. These \acp{QFT} are called \textit{(perturbatively) non-renormalizable}.
	
	\item If $\Delta_i\geq 0$ for all $i=1,\dots,n_V$, then the superficial degree of divergence decreases with the number of vertices or remains constant. In this case there is a finite number of divergent Green's functions $G_{E_f}$ that need to be renormalized, \ie{} those for which $E_f$ satisfies the inequality $4-\sum_f E_f F_f\geq 0$. These \acp{QFT} are called \textit{renormalizable}.
	
\end{itemize}

Perturbatively non-renormalizable \acp{QFT} need an infinite number of counterterms to be fully renormalized, this means that infinite number of couplings need to be introduced. This implies that an infinite number of experiments need to be performed in order to measure this infinite number of couplings. Therefore, predictivity is lost at very high energies. However, perturbatively non-renormalizable \acp{QFT} can still be considered legitimate physical theories by treating them as \acp{EFT} in some low-energy regime. In this case, predictions can be made up to errors proportional to inverse powers of the cutoff at which the \ac{EFT} description is expected to break down. In contrast, renormalizable \acp{QFT} only need a finite number of counterterms to be renormalized, therefore only a finite number of couplings needs to be measured. This means that this type of \acp{QFT} have a higher predictive power. Note that these statements only hold in perturbation theory --- a theory can be perturbatively non-renormalizable but non-perturbatively renormalizable, see~\cref{sec:ALESSIABENJAMIN}. For the sake of readability, in this section by renormalizable we mean perturbatively renormalizable.

Renormalizable \acp{QFT} can be divided in two main subclasses:
\begin{itemize}
	
\item  If $\Delta_i> 0$ for all $i=1,\dots,n_V$, the Green's functions become superficially finite for sufficiently large values of $V_i$. These \acp{QFT} are called \textit{super}-renormalizable.

\item If $\Delta_i\geq 0$ for all $i=1,\dots,n_V$, and $\exists$ $i$ such that $\Delta_i=0$, the superficial degree of divergence does not decrease with the number of vertices, and divergences need to be renormalized at every loop order.
In particular, when all the interaction couplings that are relevant in the \ac{UV} are dimensionless, \ie{} $\Delta_i=0$ for all $i=1,\dots,n_V$, the \ac{QFT} is called \textit{strictly} renormalizable.

\end{itemize}

Therefore, we have learned that the superficial degree of divergence is a very important quantity that can be easily computed by looking at the mass dimensions of fields and interaction couplings. In particular, it captures the main features of the \ac{UV} behavior of a \ac{QFT} and tells us whether perturbative renormalizability can be achieved or not. However, we should also mention that the superficial degree of divergence does not always capture the true divergence of a loop diagram because at higher loops there could be sub-divergences. In the case of local \acp{QFT}, it can be shown that it is possible to handle sub-divergences in a consistent way by first renormalizing the lower-order divergences and substitute the renormalized result into the the higher loop orders. In the end, the superficial degree of divergence is still a good indicator to measure \ac{UV} properties of a perturbative \ac{QFT} at a generic loop order~\cite{Weinberg:1995mt, Anselmi:2019pdm}.

\medskip
\noindent\textit{Examples.} Consider the following Lagrangian in four spacetime dimensions:
\begin{equation}
	\lagrangian{}=-\frac{1}{2}\partial_\mu \phi \partial^\mu\phi-V(\phi)\,.
\end{equation}
An example of a super-renormalizable \ac{QFT} is given by the above Lagrangian with the potential equal to $V=\frac{\lambda_3}{3!} \phi^3$, since $[\lambda_3]=1$. A strictly renormalizable \ac{QFT} is given by the quartic potential $V=\frac{\lambda_4}{4!} \phi^4$, since $[\lambda_4]=0$. While perturbatively non-renormalizable \acp{QFT} are given by higher-order potentials $V=\frac{\lambda_n}{n!} \phi^n$  where the couplings have negative mass dimensions, \ie{} $[\lambda_n]=4-n<0$ for $n> 4$.

Fermi theory with a four-fermion interaction is an example of a perturbatively non-re\-nor\-ma\-lizable theory that works well up to energy scales of the order of $100$ GeV, but at higher energies it is completed by the Glashow-Salam-Weinberg theory of the weak interaction, which is strictly renormalizable. The important aspect to highlight is that all \ac{SM} interactions are described in terms of strictly renormalizable \acp{QFT}~\cite{Weinberg:1995mt}. In this section we have pushed the criterion of strict renormalizability to the last drop. In particular, in~\cref{sec:lecture4} we studied a gravitational \ac{QFT} that is strictly renormalizable and known as \textit{quadratic gravity}~\cite{Stelle:1976gc, Salvio:2018crh, Donoghue:2021cza, Holdom:2021hlo, Piva:2023bcf}.

\subsection{Spin-projector formalism}\label{app:spin-proj}

In this section we found that the spin-projectors formalism is very useful for writing the gravitational propagators in \ac{GR} and quadratic gravity in a form that makes the spin structure and the off-shell degrees of freedom explicit. However, we have not provided enough details on how the formalism is constructed. The aim of this appendix is to present a more detailed treatment of the spin projectors. Useful references are~\cite{Maggiore:2005qv} on tensor representations and~\cite{Rivers:1964nfl, Barnes, VanNieuwenhuizen:1973fi} on the spin projectors; see also the textbooks~\cite{Percacci:2017fkn, Buchbinder:2021wzv}.

\subsubsection{Lorentz tensor representation}

If we speak of tensors, what we usually mean are representations of $SO(3,1)$. For example, vectors transform as $V^\mu\to \Lambda^\mu_{\phantom{\mu}\nu}V^\nu$, contravariant rank-two tensors as $T^{\mu\nu}\to \Lambda^\mu_{\phantom{\mu}\rho}\Lambda^\mu_{\phantom{\mu}\sigma}T^{\rho\sigma}$, and so on. The vector representation is irreducible under $SO(3,1)$ because the action of $\Lambda^\mu_{\phantom{\mu}\nu}$ mixes all the four components. The rank-two tensor representation is reducible because its symmetric and antisymmetric parts do not mix, and the trace is an invariant. Thus, the $16$ components of $T^{\mu\nu}$ can be decomposed as a direct sum of a scalar, an antisymmetric tensor representation and an irreducible traceless symmetric representation ($16=6+10=6+(1+9))$. Let us show this explicitly.

Consider a rank-two tensor $\varphi^{\mu\nu}$ in Minkowski spacetime. Under Lorentz transformations, it transforms as
\begin{equation}\label{eq:b1}
\varphi'^{\mu\nu}=\Lambda_{\phantom{\mu}\rho}^{\mu}\Lambda_{\phantom{\nu}\sigma}^{\nu}\varphi^{\rho\sigma}\,.
\end{equation}
This $16$-dimensional representation of the Lorentz group is reducible. First, the symmetric and antisymmetric parts do not mix, \ie{} if $\varphi$ is symmetric (antisymmetric)
then also $\varphi'$ will be symmetric (antisymmetric), which implies that the $16$-dimensional representation can be decomposed as 
\begin{equation}	\varphi^{\mu\nu}=h^{\mu\nu}+\psi^{\mu\nu}\,,\qquad\begin{cases}
	\displaystyle 	h^{\mu\nu}\equiv\frac{1}{2}\left(\varphi^{\mu\nu}+\varphi^{\nu\mu}\right)\,,\\[2.5mm]
	\displaystyle	\psi^{\mu\nu}\equiv\frac{1}{2}\left(\varphi^{\mu\nu}-\varphi^{\nu\mu}\right)\,,
	\end{cases}\label{eq:b2}
\end{equation}
where $h^{\mu\nu}$ is a $10$-dimensional symmetric representation, and $\psi^{\mu\nu}$ a six-dimensional antisymmetric representation.

Furthermore, since trace and traceless components of the symmetric representation do not mix, \ie{}
\begin{equation}
h'=\eta_{\mu\nu}h'^{\mu\nu}=\eta_{\mu\nu}\Lambda_{\phantom{\mu}\rho}^{\mu}\Lambda_{\phantom{\nu}\sigma}^{\nu}h^{\rho\sigma}=\eta_{\rho\sigma}h^{\rho\sigma}=h\,,
\end{equation}
we can decompose the symmetric representation into a nine-dimensional irreducible symmetric traceless representation ($h^{T\mu\nu}$) and a one-dimensional scalar representation ($h=\eta_{\mu\nu}h^{\mu\nu}$):
\begin{equation}
	h^{T\mu\nu}\equiv h^{\mu\nu}-\frac{1}{4}\eta_{\mu\nu}h\,.\label{eq:b3}
\end{equation}
Therefore, a rank-two tensor can be decomposed into the direct sum of three irreducible representations. From \eqref{eq:b1} it follows that $\varphi^{\mu\nu}$ is a tensor product of two four-vector representations since the two indices transform separately as four-vector indices.

\subsubsection{Decomposition of Lorentz tensors under \texorpdfstring{$SO(3)$}{SO(3)}}

The decomposition of Lorentz tensors under the $SO(3)$ subgroup allows to reduce a Lorentz tensor representation into multiple irreducible representations of $SO(3)$ that are labeled by integer values of the angular momentum $j=0,1,2,\dots$. The dimension of each representation is $2j+1$ and the states are labeled by $j_z=-j,\dots,j$. If we name each representation as $\boldsymbol{j}$, a four-vector $V^\mu$ can be decomposed into a scalar $\boldsymbol{0}$ and a three-vector $\boldsymbol{1}$:
\begin{equation}
	V^{\mu}\in\mathbf{0}\oplus\mathbf{1}\,.\label{eq:b5}
\end{equation}
From \eqref{eq:b5} we can also obtain the decomposition of a rank-two tensor under $SO(3)$:
\begin{equation}\label{eq:b7}
\begin{aligned}
\varphi^{\mu\nu}\in(\mathbf{0}\oplus\mathbf{1})\otimes(\mathbf{0}\oplus\mathbf{1})&=  (\mathbf{0}\otimes\mathbf{0})\oplus(\mathbf{0}\otimes\mathbf{1})\oplus(\mathbf{1}\otimes\mathbf{0})\oplus(\mathbf{1}\otimes\mathbf{1}) \\
&=  \mathbf{0}\oplus\mathbf{1}\oplus\mathbf{1}\oplus(\mathbf{0}\oplus\mathbf{1}\oplus\mathbf{2})\,,
\end{aligned}
\end{equation}
where we have used the fact that the composition of two angular momenta $j_{1}$
and $j_{2}$ is given by all angular momenta between $|j_{1}-j_{2}|$
and $j_{1}+j_{2}$, \ie{}
\begin{equation}\label{eq:b8}
	\mathbf{0}\otimes\mathbf{0}=\mathbf{0}\,,\qquad \mathbf{0}\otimes\mathbf{1}=\mathbf{1}\otimes\mathbf{0}=\mathbf{1}\,\qquad \mathbf{1}\otimes\mathbf{1}=\mathbf{0}\oplus\mathbf{1}\oplus\mathbf{2}\,.
\end{equation}
Therefore, $\varphi^{\mu\nu}$ decomposes
into two spin-zero, three spin-one, and one spin-two representation under $SO(3)$.

Let us now explain how the various representations are spread over the symmetric and antisymmetric parts. Since the trace is a Lorentz scalar, it will transform as a scalar under $SO(3)$,
\begin{equation}
    h \in \mathbf{0} \, .
\end{equation}
The antisymmetric part $\psi^{\mu\nu}$ has six components, and it can be written as the direct sum of the two three-vectors\footnote{In the case of the field strength in \ac{QED}, \ie{} $F_{\mu\nu}=\partial_{\mu}A_{\nu}-\partial_{\nu}A_{\mu}$, the two three-vectors are the electric and magnetic fields, respectively given by $E^{i}=-F^{0i}$ and $B^{i}=-\frac{1}{2}\epsilon^{ijk}F^{jk}$.} $\psi^{0i}$ and $\frac{1}{2}\epsilon^{ijk}\psi^{jk}$:
\begin{equation}
	\psi^{\mu\nu}\in\mathbf{1}\oplus\mathbf{1}\,.\label{eq:b9}
\end{equation}
Since we have identified the trace $h$ with $\mathbf{0}$ and $\psi^{\mu\nu}$ with $\mathbf{1}\oplus\mathbf{1}$, from \eqref{eq:b7} we understand that $h^{T\mu\nu}$ decomposes as
\begin{equation}
h^{T\mu\nu}\in\mathbf{0}\oplus\mathbf{1}\oplus\mathbf{2}\,.\label{eq:b10b10}
\end{equation}

\subsubsection{Spin projector operators}

The next two tasks are: (i) introduce a complete set of projection operators through which we can project the tensor $\varphi^{\mu\nu}$ along its scalar, vector and tensor components; (ii) find a basis in the space of rank-four tensors in terms of which we can express the tensor $\mathcal{O}^{\mu\nu\rho\sigma}$ appearing in a parity-invariant kinetic term, \ie{}
\begin{equation}
	\lagrangian{}=\frac{1}{2}\varphi_{\mu\nu}\mathcal{O}^{\mu\nu\rho\sigma}\varphi_{\rho\sigma}\,.\label{eq:b361}
\end{equation}
The kinetic operator satisfies $\mathcal{O}^{\mu\nu\rho\sigma}=\mathcal{O}^{\rho\sigma\mu\nu}$ but, in general, it is not symmetric under the exchange of the indices in the individual pairs $(\mu\nu)$ and $(\rho\sigma)$.
In this section, we only worked with symmetric rank-four kinetic operators. However, in this appendix, we make the discussion more general and also allow for the presence of an antisymmetric part.

In~\cref{sec:propag-III} we showed that a generic four-vector $A^{\mu}$ can be projected along the two irreducible representations of the $SO(3)$ group ($A^{\mu}\in\mathbf{0}\oplus\mathbf{1}$) by using the set of projector operators $\left\{ \theta_{\mu\nu},\omega_{\mu\nu}\right\}$, that project along the spin-one and spin-zero components, respectively. The same projectors also form a basis in the space of rank-two symmetric tensors, and this property is very useful to decompose kinetic operator and propagator of a spin-one gauge boson into its spin components. We now want to show how to find spin projectors that can form a complete set for the decomposition of a rank-two tensor, and a basis for rank-four tensors.

Since a rank-two tensor behaves like the tensor product of two four-vectors, we can find the projection operators for $\varphi^{\mu\nu}$ by decomposing each of the two indices in terms of $\theta_{\mu\nu}$ and $\omega_{\mu\nu}$. Let us first focus on the symmetric part, and then move on to the antisymmetric one.

\subsubsubsection*{Symmetric decomposition}

The symmetric rank-two tensor $h_{\mu\nu}$ can be seen as the tensor product of two four-vectors and can be decomposed in its irreducible representations under $SO(3)$ as follows,
\begin{equation}		\label{eq:b350}
\begin{aligned}
		h_{\mu\nu}&=  \left(\theta_{\mu\rho}+\omega_{\mu\rho}\right)\left(\theta_{\nu\sigma}+\omega_{\nu\sigma}\right)h^{\rho\sigma}\\
		&= \left(\theta_{\mu\rho}\theta_{\nu\sigma}+\theta_{\mu\rho}\omega_{\nu\sigma}+\omega_{\mu\rho}\theta_{\nu\sigma}+\omega_{\mu\rho}\omega_{\nu\sigma}\right)h^{\rho\sigma}\\
	&= \frac{1}{2}\left(\theta_{\mu\rho}\theta_{\nu\sigma}+\theta_{\mu\sigma}\theta_{\nu\rho}\right)h^{\rho\sigma}-\frac{1}{3}\theta_{\mu\nu}\theta_{\rho\sigma}h^{\rho\sigma} +\frac{1}{3}\theta_{\mu\nu}\theta_{\rho\sigma}h^{\rho\sigma}+\omega_{\mu\nu}\omega_{\rho\sigma}h^{\rho\sigma}\\
		&\qquad +\frac{1}{2}\left(\theta_{\mu\rho}\omega_{\nu\sigma}+\theta_{\mu\sigma}\omega_{\nu\rho}+\theta_{\nu\rho}\omega_{\mu\sigma}+\theta_{\nu\sigma}\omega_{\mu\rho}\right)h^{\rho\sigma} \\
		&= \mathcal{P}^{(2)}_{\phantom{(2)}\mu\nu\rho\sigma} h^{\rho\sigma}+ \mathcal{P}^{(1,m)}_{\phantom{(1,m)}\mu\nu\rho\sigma} h^{\rho\sigma}+ \mathcal{P}^{(0,s)}_{\phantom{(0,s)}\mu\nu\rho\sigma} h^{\rho\sigma}+ \mathcal{P}^{(0,w)}_{\phantom{(0,w)}\mu\nu\rho\sigma} h^{\rho\sigma}\,,
        \end{aligned}
\end{equation}
where we have introduced four spin projectors that are defined as\footnote{We are labeling the spin-one projector operator with the additional letter $m$ in order to distinguish it from the other spin-one operators that will be introduced in the case of the antisymmetric decomposition. When only symmetric tensors are present, then we can simply call it ${\mathcal{P}^{(1)}}$ (without the $m$), as we have done in the main text of this section.}
\begin{tcolorbox}
\begin{equation}\label{eq-app-spin-projectors}
\begin{aligned}
    \mathcal{P}^{(2)}_{\phantom{(2)}\mu\nu\rho\sigma}&= \frac{1}{2}\left(\theta_{\mu\rho}\theta_{\nu\sigma}+\theta_{\mu\sigma}\theta_{\nu\rho}\right)-\frac{1}{3}\theta_{\mu\nu}\theta_{\rho\sigma}\,,\\	\mathcal{P}^{(1,m)}_{\phantom{(1,m)}\mu\nu\rho\sigma}&= \frac{1}{2}\left(\theta_{\mu\rho}\omega_{\nu\sigma}+\theta_{\mu\sigma}\omega_{\nu\rho}+\theta_{\nu\rho}\omega_{\mu\sigma}+\theta_{\nu\sigma}\omega_{\mu\rho}\right)\,,\\
    \mathcal{P}^{(0,s)}_{\phantom{(0,s)}\mu\nu\rho\sigma}&= \frac{1}{3}\theta_{\mu\nu}\theta_{\rho\sigma}\,,\\
\mathcal{P}^{(0,w)}_{\phantom{(0,w)}\mu\nu\rho\sigma}&= \omega_{\mu\nu}\omega_{\rho\sigma}\,.
    \end{aligned}
\end{equation}
\end{tcolorbox}
\noindent The label $m$ stands for ``momentum'' as the operator projects along the $h^{0i}$ components, the label $s$ stands for ``stress'' as the corresponding operator projects along the trace of the stress scalar $h^{i}_{\phantom{i}i}$, and the label $w$ stands for ``work'' (energy) as the projection is along the $h^{00}$ component~\cite{vanNieuwenhuizen:1976vb}.

The projectors are idempotent and orthogonal, that is
\begin{equation}
\mathcal{P}^{(i,a)\phantom{\mu\nu}\alpha\beta}_{\phantom{(i,a)}\mu\nu} 	\mathcal{P}^{(j,b)\phantom{\alpha\beta}\rho\sigma}_{\phantom{(j,b)}\alpha\beta}=	\delta^{ij}\delta^{ab}\mathcal{P}^{(i,a)\phantom{\mu\nu}\rho\sigma}_{\phantom{(i,a)}\mu\nu}\,,
	\label{orthog-1-app}
\end{equation}
and form a complete set,
\begin{equation}
\mathcal{P}^{(2)}_{\phantom{(2)}\mu\nu\rho\sigma}+\mathcal{P}^{(1,m)}_{\phantom{(1,m)}\mu\nu\rho\sigma}+\mathcal{P}^{(0,s)}_{\phantom{(0,s)}\mu\nu\rho\sigma}+\mathcal{P}^{(0,w)}_{\phantom{(0,w)}\mu\nu\rho\sigma}=\mathbb{1}_{\mu\nu\rho\sigma}\,,
\label{completeness-rel-app}
\end{equation}
where $ \mathbb{1}_{\mu\nu\rho\sigma}=
\frac{1}{2}(\eta_{\mu\rho}\eta_{\nu\sigma}+\eta_{\mu\sigma}\eta_{\nu\rho})$ is the identity operator in the space of symmetric rank-four tensors, that was already defined in the main text in~\eqref{eq:LUCA_SYM2_ID}.

Since the projectors are idempotent, their trace equals their rank. This means that the trace is equal to the dimension of the corresponding irreducible representation (\ie{}  $2j+1$):
\begin{equation}
\begin{aligned}
	\mathbbm{1}^{\mu\nu\rho\sigma}\mathcal{P}^{(2)}_{\phantom{(2)}\mu\nu\rho\sigma}&=5=2(2)+1\quad (\text{spin-two})\,, \\
	\mathbbm{1}^{\mu\nu\rho\sigma}\mathcal{P}^{(1,m)}_{\phantom{(1,m)}\mu\nu\rho\sigma}&=3=2(1)+1\quad (\text{spin-one})\,,\\
	\mathbbm{1}^{\mu\nu\rho\sigma}\mathcal{P}^{(0,s)}_{\phantom{(0,s)}\mu\nu\rho\sigma}&=1=2(0)+1\quad (\text{spin-zero})\,, \\
	\mathbbm{1}^{\mu\nu\rho\sigma} \mathcal{P}^{(0,w)}_{\phantom{(0,w)}\mu\nu\rho\sigma}&= 1=2(0)+1\quad (\text{spin-zero})\,,
\end{aligned}
\end{equation}
which means that $\mathcal{P}^{(2)}$ projects along the spin-two  component (the traceless $h_{ij}$), $\mathcal{P}^{(1,m)}$  along the spin-one ($h_{0i}$), $\mathcal{P}^{(0,s)}$ along one spin-zero (spatial trace) and  $\mathcal{P}^{(0,w)}$ along the other spin-zero ($h_{00}$). As an exercise, let us verify the first identity for the spin-two projector:
\begin{equation}
\begin{aligned}
	\mathbbm{1}^{\mu\nu\rho\sigma}\mathcal{P}^{(2)}_{\phantom{(2)}\mu\nu\rho\sigma}&= \mathbbm{1}^{\mu\nu\rho\sigma}\theta_{\mu\rho}\theta_{\nu\sigma}-\frac{1}{3}\mathbbm{1}^{\mu\nu\rho\sigma}\theta_{\mu\nu}\theta_{\rho\sigma} \\
    &= \frac{1}{2}\left(\theta_\mu^{\phantom{\mu}\mu}\theta_\nu^{\phantom{\nu}\nu}+\theta_\mu^{\phantom{\mu}\mu}\right)-\frac{1}{6}\left(2\theta_\mu^{\phantom{\mu}\mu}\right)\\
    &=\frac{1}{2}\left(3\times 3+3\right)-\frac{3}{3}\\
    &=5\,.
\end{aligned}
\end{equation}

The next question to ask is whether the complete set of projectors also forms a basis in the space of symmetric rank-four tensors. Using Lorentz invariance, a set of basis elements constructed in terms of $\eta_{\mu\nu}$ and $p_\mu$ can be easily found and was given in \eqref{propag-basis}. However, as already explained in \cref{sec:propag-III}, the projectors in \eqref{eq-app-spin-projectors} are not enough to form a basis because they cannot generate the terms $\eta_{\mu\nu}p_\rho p_\sigma$ and $\eta_{\rho\sigma}p_\mu p_\nu$. Indeed, we need to add an additional element to close the basis, and we choose it to be
\begin{tcolorbox}
\begin{equation}\label{eq-app-operator-cross}
\mathcal{P}^{(0,\times)}_{\phantom{(0,\times)}\mu\nu\rho\sigma}= \mathcal{P}^{(0,sw)}_{\phantom{(0,sw)}\mu\nu\rho\sigma}+\mathcal{P}^{(0,ws)}_{\phantom{(0,ws)}\mu\nu\rho\sigma}\,,
\end{equation}
where 
\begin{equation}\label{eq-app-operators-sw-ws}
\mathcal{P}^{(0,sw)}_{\phantom{(0,sw)}\mu\nu\rho\sigma}=\frac{1}{\sqrt{3}}\theta_{\mu\nu}\omega_{\rho\sigma} \,,\qquad \mathcal{P}^{(0,ws)}_{\phantom{(0,ws)}\mu\nu\rho\sigma}=\frac{1}{\sqrt{3}}\omega_{\mu\nu}\theta_{\rho\sigma}\,.
\end{equation}
\end{tcolorbox}

Note that the operators $\mathcal{P}^{(0,sw)}$ and $\mathcal{P}^{(0,ws)}$ are not projectors, indeed they are not idempotent, do not contribute to any completeness relation, and are not orthogonal to the spin projectors. However, they satisfy some relations that together with those in \eqref{orthog-1} can be written in the following compact form:
\begin{equation}
\mathcal{P}^{(i,ab)\phantom{\mu\nu}\alpha\beta}_{\phantom{(i,ab)}\mu\nu} 	\mathcal{P}^{(j,cd)\phantom{\alpha\beta}\rho\sigma}_{\phantom{(j,cd)}\alpha\beta}=	\delta^{ij}\delta^{bc}\mathcal{P}^{(i,ad)\phantom{\mu\nu}\rho\sigma}_{\phantom{(i,ad)}\mu\nu}\,,
	\label{orthog-2-app}
\end{equation}
where the various indices can take the values $i,j\in\{2,1,0\}$ and $a,b,c,d\in\{m,s,w\}$. Note that when one of the indices $a,b,c,d$ is absent, \ie{} for the spin-two projector, the quantities $\mathcal{P}^{(2,a)}$ and $\mathcal{P}^{(2,ab)}$ would correspond to $\mathcal{P}^{(2)}$. Moreover, the notation $\mathcal{P}^{(i,aa)}$ means  $\mathcal{P}^{(i,a)}$. Therefore, the set $\{\mathcal{P}^{(2)},\mathcal{P}^{(1,m)},\mathcal{P}^{(0,s)},\mathcal{P}^{(0,w)},\mathcal{P}^{(0,\times)}\}$ is closed, \ie{} the product between any two operators does not generate a new one that is not already included in the set, and also forms a basis in the space of symmetric rank-four tensors. 

As an exercise, let us verify that $\mathcal{P}^{(2)}$ projects $h_{\mu\nu}$ along the spin-two component which is traceless and transverse:
\begin{equation}
\begin{aligned}
\eta^{\mu\nu}\left(\mathcal{P}^{(2)}_{\phantom{(2)}\mu\nu\rho\sigma}h^{\rho\sigma}\right)&= \left[\frac{1}{2}\left(\eta^{\mu\nu}\theta_{\mu\rho}\theta_{\nu\sigma}+\eta^{\mu\nu}\theta_{\mu\sigma}\theta_{\nu\rho}\right)-\frac{1}{3}\eta^{\mu\nu}\theta_{\mu\nu}\theta_{\rho\sigma}\right]h^{\rho\sigma}\\
&= \left[\frac{1}{2}\left(\theta_{\rho}^{\phantom{\rho}\nu}\theta_{\nu\sigma}+\theta_{\sigma}^{\phantom{\sigma}\nu}\theta_{\nu\rho}\right)-\frac{1}{3}\left(4-1\right)\theta_{\rho\sigma}\right]h^{\rho\sigma}\\
&= \left[\frac{1}{2}\left(\theta_{\rho\sigma}+\theta_{\rho\sigma}\right)-\theta_{\rho\sigma}\right]h^{\rho\sigma}=0\,,
\end{aligned}
\end{equation}
and
\begin{equation}	p^{\mu}\left(\mathcal{P}^{(2)}_{\phantom{(2)}\mu\nu\rho\sigma}h^{\rho\sigma}\right)=\left[\frac{1}{2}\left(p^{\mu}\theta_{\mu\rho}\theta_{\nu\sigma}+p^{\mu}\theta_{\mu\sigma}\theta_{\nu\rho}\right)-\frac{1}{3}p^{\mu}\theta_{\mu\nu}\theta_{\rho\sigma}\right]h^{\rho\sigma}=0\,,\label{eq:b2.13}
\end{equation}
where we used $p^{\mu}\theta_{\mu\rho}=0$.

\subsubsubsection*{Antisymmetric decomposition}

Let us now focus on the antisymmetric part $\psi^{\mu\nu}\in\mathbf{1}\oplus\mathbf{1}$. Following steps similar to those of the symmetric case, we can write
\begin{equation}
\begin{aligned}
\psi_{\mu\nu}&= \left(\theta_{\mu\rho}+\omega_{\mu\rho}\right)\left(\theta_{\nu\sigma}+\omega_{\nu\sigma}\right)\psi^{\rho\sigma}\\
&=
\left(\theta_{\mu\rho}\theta_{\nu\sigma}+\theta_{\mu\rho}\omega_{\nu\sigma}+\omega_{\mu\rho}\theta_{\nu\sigma}+\omega_{\mu\rho}\omega_{\nu\sigma}\right)\psi^{\rho\sigma} \\
&= \frac{1}{2}\left(\theta_{\mu\rho}\theta_{\nu\sigma}-\theta_{\mu\sigma}\theta_{\nu\rho}\right)\psi^{\rho\sigma}+\frac{1}{2}\left(\theta_{\mu\rho}\omega_{\nu\sigma}-\theta_{\mu\sigma}\omega_{\nu\rho}-\theta_{\nu\rho}\omega_{\mu\sigma}+\theta_{\nu\sigma}\omega_{\mu\rho}\right)\psi^{\rho\sigma}\\
&=\mathcal{P}^{(1,b)}_{\phantom{(1,b)}\mu\nu\rho\sigma}\psi^{\rho\sigma}+\mathcal{P}^{(1,e)}_{\phantom{(1,e)}\mu\nu\rho\sigma}\psi^{\rho\sigma}\,,
\end{aligned}
\end{equation}
where we have defined the two projector operators 
\begin{tcolorbox}
\begin{equation}
\begin{aligned}
\mathcal{P}^{(1,b)}_{\phantom{(1,b)}\mu\nu\rho\sigma}&= \frac{1}{2}\left(\theta_{\mu\rho}\theta_{\nu\sigma}-\theta_{\mu\sigma}\theta_{\nu\rho}\right),\\
\mathcal{P}^{(1,e)}_{\phantom{(1,e)}\mu\nu\rho\sigma}&= \frac{1}{2}\left(\theta_{\mu\rho}\omega_{\nu\sigma}-\theta_{\mu\sigma}\omega_{\nu\rho}-\theta_{\nu\rho}\omega_{\mu\sigma}+\theta_{\nu\sigma}\omega_{\mu\rho}\right)\,.
\end{aligned}
\end{equation}
\end{tcolorbox}
\noindent They are idempotent and orthogonal, that is
\begin{equation}	\label{orthog-1-anitsymm}
\mathcal{P}^{(1,c)\phantom{\mu\nu}\alpha\beta}_{\phantom{(1,c)}\mu\nu} 	\mathcal{P}^{(1,d)\phantom{\alpha\beta}\rho\sigma}_{\phantom{(1,d)}\alpha\beta}=\delta^{cd}\mathcal{P}^{(1,c)\phantom{\mu\nu}\rho\sigma}_{\phantom{(1,c)}\mu\nu}\,,
\end{equation}
where $c,d\in\{b,e\}$. Thus, they form a complete set
\begin{equation}
\mathcal{P}^{(1,b)}_{\phantom{(1,b)}\mu\nu\rho\sigma}+\mathcal{P}^{(1,e)}_{\phantom{(1,e)}\mu\nu\rho\sigma}=\frac{1}{2}\left(\eta_{\mu\rho}\eta_{\nu\sigma}-\eta_{\mu\sigma}\eta_{\nu\rho}\right)\,,
\label{completeness-rel-anisymm}
\end{equation}
where $\frac{1}{2}(\eta_{\mu\rho}\eta_{\nu\sigma}-\eta_{\mu\sigma}\eta_{\nu\rho})$ is the identity operator in the space of antisymmetric rank-four tensors.

As before, since the projectors are idempotent, their trace equals their rank. This means that the trace is equal to the dimension of the corresponding irreducible representation (\ie{}  $2j+1$):
\begin{equation}
\begin{aligned}
\frac{1}{2}\left(\eta_{\mu\rho}\eta_{\nu\sigma}-\eta_{\mu\sigma}\eta_{\nu\rho}\right)\mathcal{P}^{(1,b)}_{\phantom{(1,b)}\mu\nu\rho\sigma}&=1=2(1)+1\quad (\text{spin-one})\,, \\
\frac{1}{2}\left(\eta_{\mu\rho}\eta_{\nu\sigma}-\eta_{\mu\sigma}\eta_{\nu\rho}\right)\mathcal{P}^{(1,e)}_{\phantom{(1,e)}\mu\nu\rho\sigma}&=1=2(1)+1\quad (\text{spin-one})\,.
\end{aligned}
\end{equation}
Both operators project onto spin-one components: the so called magnetic ($\frac{1}{2}\epsilon^{ijk}\psi_{jk}$) and electric ($\psi^{0i}$) components of the antisymmetric tensor $\psi^{\mu\nu}$. The use of the letters $b$ and $e$ is inspired by the magnetic field $\vec{B}$ and electric field $\vec{E}$ in \ac{QED}.

\subsubsubsection*{Full decomposition}

We can now decompose any rank-two tensor $\varphi^{\mu\nu}$ into its irreducible $SO(3)$ representations by using the complete set of projectors $\{ \mathcal{P}^{(2)},\mathcal{P}^{(1,m)},\mathcal{P}^{(0,s)},\mathcal{P}^{(0,w)},\mathcal{P}^{(1,b)},\mathcal{P}^{(1,e)}\}$.
The full completeness relation is
\begin{equation}
\begin{aligned}
\left(\mathcal{P}^{(2)}+\mathcal{P}^{(1,m)}+\mathcal{P}^{(0,s)}+\mathcal{P}^{(0,w)}+\mathcal{P}^{(1,b)}+\mathcal{P}^{(1,e)}\right)_{\mu\nu\rho\sigma}&=  \frac{1}{2}\left(\eta_{\mu\rho}\eta_{\nu\sigma}+\eta_{\mu\sigma}\eta_{\nu\rho}\right)\\
&\qquad+\frac{1}{2}\left(\eta_{\mu\rho}\eta_{\nu\sigma}-\eta_{\mu\sigma}\eta_{\nu\rho}\right) \\
&=  \eta_{\mu\rho}\eta_{\nu\sigma}\,.
\end{aligned}
\end{equation}

Furthermore, we can also find a basis to decompose any rank-four tensor $\mathcal{O}^{\mu\nu\rho\sigma}$ that could play the role of a kinetic operator in a parity-invariant Lagrangian as that in \eqref{eq:b361}. However, all the operators introduced so far, including $\mathcal{P}^{(0,sw)}$ and $\mathcal{P}^{(0,ws)}$, are not enough to form a basis. Indeed, we need an additional basis element that can allow to write rank-four tensors that are symmetric in the first pair $(\mu\nu)$ and antisymmetric in the second pair $(\rho\sigma)$, and vice versa. This additional element can be chosen to be
\begin{tcolorbox}
\begin{equation}
\mathcal{P}^{(1,\times)}_{\phantom{(1,\times)}\mu\nu\rho\sigma}=	\mathcal{P}^{(1,me)}_{\phantom{(1,me)}\mu\nu\rho\sigma}+	\mathcal{P}^{(1,em)}_{\phantom{(1,em)}\mu\nu\rho\sigma}\,,
\end{equation}
where
\begin{equation}
\begin{aligned}
\mathcal{P}^{(1,me)}_{\phantom{(1,me)}\mu\nu\rho\sigma}&= \frac{1}{2}\left(\theta_{\mu\rho}\omega_{\nu\sigma}-\theta_{\mu\sigma}\omega_{\nu\rho}+\theta_{\nu\rho}\omega_{\mu\sigma}-\theta_{\nu\sigma}\omega_{\mu\rho}\right)\,,\\
\mathcal{P}^{(1,em)}_{\phantom{(1,em)}\mu\nu\rho\sigma}&= \frac{1}{2}\left(\theta_{\mu\rho}\omega_{\nu\sigma}+\theta_{\mu\sigma}\omega_{\nu\rho}-\theta_{\nu\rho}\omega_{\mu\sigma}-\theta_{\nu\sigma}\omega_{\mu\rho}\right)\,.
\end{aligned}
\end{equation}
\end{tcolorbox}
\noindent Note that, similarly to $\mathcal{P}^{(0,sw)}$ and $\mathcal{P}^{(0,ws)}$, also $\mathcal{P}^{(1,em)}$ and $\mathcal{P}^{(1,me)}$ are not projectors.

The full basis of rank-four tensors can be written in the following compact form:\footnote{Note that no operators that connect electric and magnetic spin-one spaces, \ie{} $\mathcal{P}^{(1,eb)}$ and $\mathcal{P}^{(1,be)}$, nor operators of the type $\mathcal{P}^{(1,bm)}$ and $\mathcal{P}^{(1,mb)}$, are present. Their absence is due to the fact that we are considering the case of parity-invariant Lagrangians for which these types of spin-one transitions are not allowed. On the other hand, if we admit parity-violating operators, we could write terms like $\epsilon_{\mu\nu\rho\sigma}\varphi^{\mu\nu}\varphi^{\rho\sigma}$ or $\epsilon_{\mu\nu\rho\sigma}\varphi^{\mu\nu}\partial^{\rho}\psi^{\sigma}$, and so the operators $\mathcal{P}^{(1,eb)}, \mathcal{P}^{(1,be)}, \mathcal{P}^{(1,mb)}$ and $\mathcal{P}^{(1,bm)}$ would appear. See~\cite{VanNieuwenhuizen:1973fi} for more details.}
\begin{equation}
	\{\mathcal{O}^{(i)}\}\equiv \left\{ \mathcal{P}^{(2)},\mathcal{P}^{(1,m)},\mathcal{P}^{(0,s)},\mathcal{P}^{(0,w)},\mathcal{P}^{(0,sw)},\mathcal{P}^{(0,ws)},\mathcal{P}^{(1,b)},\mathcal{P}^{(1,e)},\mathcal{P}^{(1,em)},\mathcal{P}^{(1,me)}\right\} \, ,
	\label{full-basis}
\end{equation}
where $i=1,\dots,10$. Let us remark that the operators $\mathcal{P}^{(0,sw)}$ and $\mathcal{P}^{(0,ws)}$ are not two independent elements because they always come in the combination $\mathcal{P}^{(0,\times)}$. However, working explicitly in terms of $\mathcal{P}^{(0,sw)}$ and $\mathcal{P}^{(0,ws)}$ simplifies the various product relations. The same reasoning is true for the operators $\mathcal{P}^{(1,me)}$, $\mathcal{P}^{(1,em)}$ and $\mathcal{P}^{(1,\times)}$.

Indeed, the operators in \eqref{full-basis} satisfy the following relations:
\begin{equation}	\label{orthog-full}	\mathcal{P}^{(i,AB)\phantom{\mu\nu}\alpha\beta}_{\phantom{(i,AB)}\mu\nu} 	\mathcal{P}^{(j,CD)\phantom{\alpha\beta}\rho\sigma}_{\phantom{(j,CD)}\alpha\beta}=\delta^{ij}\delta^{BC}\mathcal{P}^{(i,AD)\phantom{\mu\nu}\rho\sigma}_{\phantom{(i,AD)}\mu\nu}\,,
\end{equation}
where now $i,j\in\{2,1,0\}$, $A,B,C,D\in\{m,s,w,b,e\}$,  $\mathcal{P}^{(2,AB)}=\mathcal{P}^{(2)}$ and $\mathcal{P}^{(i,AA)}=\mathcal{P}^{(i,A)}$.

We can now finally express the parity-invariant kinetic Lagrangian~\eqref{eq:b361} in momentum space in terms of the full basis of rank-four tensors in \eqref{full-basis}:
\begin{equation}
\lagrangian{}=\frac{1}{2}\varphi_{\mu\nu}\mathcal{O}^{\mu\nu\rho\sigma}\varphi_{\rho\sigma}=
\frac{1}{2}\varphi_{\mu\nu} \left(\sum_{i=1}^{10}c_{i}(p)\mathcal{O}^{(i)\,\mu\nu\rho\sigma}\right)\varphi_{\rho\sigma}\,,	
\end{equation}
where $c_i(p)$ are some coefficients that can be either constant or depend on $p^2$.

In this section, we only worked with symmetric kinetic operators, in \ac{GR} and quadratic gravity; see \eqref{kinetic-op-GR-symm} and~\eqref{kinetic-op-quad-grav-symm}. Therefore, we only needed the four symmetric spin projectors in \eqref{eq-app-spin-projectors} plus the operator~\eqref{eq-app-operator-cross} to form a basis in the space of rank-four symmetric tensors. 

\end{subappendices}


\section{\titleAnna}
\label{sec:ANNA}

\begin{tcolorbox}[colback=white,colframe=scipostblue]
{\bf Lecturer:} Anna Tokareva, \briefaffiliationAnna

{\bf Email address:} \href{mailto:\emailAnna}{\emailAnna}
\tcblower
{\bf Lecture recordings:}
\begin{enumerate}[label= Lecture \arabic*:, leftmargin = 3.5cm, labelsep = 0.5cm, parsep = 0.0cm]
    \item \href{https://youtu.be/5Ns78W_Dhqk}{https://youtu.be/5Ns78W\_Dhqk}
    \item \href{https://youtu.be/PyND_7lxEt4}{https://youtu.be/PyND\_7lxEt4}
    \item \href{https://youtu.be/HYWqZMsz4Cs}{https://youtu.be/HYWqZMsz4Cs}
    \item \href{https://youtu.be/Mw1v9vXi2vo}{https://youtu.be/Mw1v9vXi2vo}
\end{enumerate}

{\bf Abstract:}

In these lectures I introduce an \ac{EFT} approach in field theory and its application to gravity. I show the relation between the \ac{EFT} and scattering amplitudes, and give several examples of the techniques based on the general properties of scattering amplitudes. I derive analytic results for positivity bounds and show that under certain assumptions they lead to a compact allowed space for the Wilson coefficients. I finish the discussion of \ac{EFT} and amplitude methods by eikonal resummation of \ac{GR} amplitudes and positivity bounds for the amplitudes involving graviton exchange.
\end{tcolorbox}

\subsection*{Preface}

\ac{GR} being a perfect description of classical gravity certainly requires a completion at higher energies than the Planck mass. Curiously, all efforts to build a field theory-based complete description of gravity show at least one conceptual problem among this list:
\begin{itemize}
    \item {\bf Lack of unitarity:} Quadratic gravity (described in detail in \cref{sec:lecture4}) is a renormalizable weakly coupled theory. However, it suffers from the inevitable presence of a ghost state in the tensor sector. Moreover, this ghost state is coupled to normal gravitons, which leads to either classical instabilities, or negative probabilities for certain processes. This makes it impossible to consider quadratic gravity unless some revisiting of quantum-mechanical laws is done.
    
    \item {\bf Strong coupling:} Gravity can be non-perturbatively renormalizable, as it is assumed in the program of \ac{ASQG} introduced in \cref{sec:ALESSIABENJAMIN}. However, it means that it is necessary to use non-perturbative approaches in the \ac{UV} which is a dramatic complication from the point of view of the computations and consistency checks. Moreover, in a complete theory, in general, an infinite number of terms can play a role in the \ac{RG} running. Some luck is needed, in order to get just a finite set of relevant operators which is enough for computations.
    
    \item {\bf Non-locality:} Locality of field theory means, in simple words, that the theory admits a description in terms of operators with a finite number of derivatives. It is not fully clear at this moment whether this property can be preserved at all in a unitary and ghost-free field theory description of gravity. An infinite-derivative version of quadratic Stelle gravity~\cite{Tomboulis:1997gg} represents an example of a theory where unitarity is restored by means of sacrificing locality.
    
    \item {\bf Violation of Lorentz invariance:} Projectable Ho\v{r}ava gravity represents an example of a Lorentz-violating four-dimensional theory of gravity flowing from an asymptotically free fixed point in the \ac{IR} to the asymptotically free fixed point in the \ac{UV} \cite{Barvinsky:2023uir}. Thus, sacrificing Lorentz invariance, one has a chance to get weakly coupled ghost-free \ac{QG}.
\end{itemize}
It seems that, in order to get a complete description of gravity, one always has to give up some desired property. Even in \ac{ST} (described in detail in \cref{sec:IVANO}), a lot of complications, uncertainties, and a whole landscape of possibilities emerge. In general, \acp{EFT} coupled to gravity have landscapes. In that case, it is much easier to pinpoint a good realistic sector and vacuum and control the whole landscape. In \ac{ST}, the complications arise because it is not a \ac{QFT} in spacetime and, thus, it requires a different framework (infinite towers of states, etc.), and controlling the landscape is much harder than in \ac{QFT}. Thus, it seems there is no simple solution here. So,
    \textit{what is wrong with gravity}?
This is a good question but, perhaps, it is more fruitful to address a different question instead:
\begin{tcolorbox}
\begin{center}
    \textit{What to do with gravity?}
\end{center}
\end{tcolorbox}
Fortunately for us, gravity admits a good description for almost all observed phenomena, such that low energy \ac{EFT} remains valid. The reason for that is related to the extremely high value of the Planck mass $\MPl=10^{18}$ GeV, compared to all energy scales of particle physics. The highest energies which can be directly probed in collider experiments (including the LHC at CERN) are at most $10^3$ GeV. Thus, we have a gap of 15 orders of magnitude in energy, allowing us to live safely and be well protected from any \ac{QG} effects in ordinary life and in laboratory experiments. For all applications, just \ac{GR} or the \ac{EFT} of \ac{GR} seems to be enough. Why should we care at all about the \ac{UV} completion of gravity?

This section has an answer to this question. The main observation done in recent decades is that 
\begin{tcolorbox}
\begin{center}
    \textit{Not all \acp{EFT} are consistent with \ac{QFT} principles.}
\end{center}
\end{tcolorbox}
\noindent Naively, one may expect that we are supposed to measure the Wilson coefficients in front of all counterterms that we have to add to the action, and there is no fundamental principle telling us which values are allowed. However, the \ac{QFT} rules are more restrictive. Not all \acp{EFT} can be \ac{UV}-completed by a good theory which satisfy the general principles of unitarity, locality, symmetries, and causality. These principles can be formulated even beyond the \ac{QFT} framework, for example, as the properties of non-perturbative S-matrix, when \ac{ST} is assumed to be a \ac{UV} completion for gravity.

How can we verify whether a given \ac{EFT} can be completed by something nice, say, \ac{ST}, or a consistent \ac{QFT}? The main tool which makes it possible to classify \acp{EFT} is related to the scattering amplitude, or S-matrix formalism. In this formalism, we study the scattering process of the asymptotic states. The above-mentioned \ac{QFT} principles are conveniently encoded in mathematical properties of the scattering amplitude, such as their fixed analyticity structure in the complex plane of momenta. Due to these properties, it is possible to use complex analysis theorems which would literally relate the \ac{UV} completion to the \ac{IR} theory where scattering amplitudes can be directly computed from low-energy \ac{EFT}. The \ac{UV}-\ac{IR} relations of this type lead to constraints on Wilson coefficients of \acp{EFT} which are often called \emph{positivity bounds}. Historically, the first constraints were formulated in~\cite{Adams:2006sv} as only positive signs for a set of Wilson coefficients, hence the name. These constraints were further generalized and optimized, thus, nowadays they form compact islands in the parameter space of Wilson coefficients after the mass scale is fixed. 

The other application of \ac{UV}-\ac{IR} relations imposed by analyticity of the scattering amplitude is a partial reconstruction of the \ac{UV} completion through the scattering amplitude beyond the regime of validity of the \ac{EFT}. Remarkably, this can be a meaningful procedure in graviton-mediated scattering because in this case, the saturation of the \ac{FU} condition is a good approximation (especially in higher dimensions) leading to the possibility of an eikonal-based unitarization of tree-level scattering.

I am deeply indebted to several colleagues of mine, discussions and collaboration with whom shaped my understanding of \ac{EFT}, amplitudes and positivity bounds: M. Carrillo Gon\-z\'a\-lez, C. de Rham, S. Jaitly, M. Herrero-Valea, A. Koshelev, K. Mktrchyan, A. Tolley, P. Tourkine, A. Tseytlin, A. Zhiboedov, S.-Y. Zhou. I have to remember here my great teacher Valery Rubakov who unfortunately passed away too early. He has never been working on amplitudes but somehow it happened that I am still trying to give more and more rigorous answers to his sharp questions on loop positivity bounds and issues with the graviton pole. I am grateful to the MSc students from my group in HIAS, GuangZhuo Peng and YongJun Xu\footnote{Order is alphabetical.} for motivating me to make a more systematic review on \ac{EFT} and positivity bounds, for correcting my mistakes and misprints and for guiding me through a variety of delicious Chinese food. I am delighted to appreciate recent very deep discussions with my PhD student Long-Qi Shao (with his insights from low energy \ac{QCD}), which, in fact, formed a significant part of the FAQ on \ac{EFT} (\cref{sec:FAQ}), as well as my current understanding of causality. I wish to see in the near future that, despite certain \ac{IR} divergences, his research career will come to a very good \ac{UV} completion.

These lectures are organized as follows:

\begin{description}

\item[Sec.~\ref{sec:EFTofGR}:] We go all the way from \ac{GR} to the \ac{EFT} of \ac{GR}, explaining some steps with the use of a shift-symmetric scalar field as a toy model. We show how to reduce the tensor structure in the \ac{EFT} action to a very limited number of contractions of the Riemann tensor. We will also discuss the relation between the \ac{EFT} action and the amplitude.

\item[Sec.~\ref{sec:amplitudes}:] We review several important properties of scattering amplitudes, such as \ac{PWU} and polynomial boundedness. We derive one of the crucial consequences of unitarity --- the Martin-Froissart bound --- which makes it possible at all to use dispersion relations to constrain \ac{EFT} Wilson coefficients. In addition, we briefly introduce the idea of the amplitude's bootstrap techniques, and show how it works in a very simple toy model where it allows us to get a loop correction in a two-line computation.

\item[Sec.~\ref{sec:positivity}:] We introduce the techniques of the dispersion relations in non-gravitational theories, and derive constraints on Wilson coefficients for generic \acp{EFT} following from unitarity, causality and the Martin-Froissart bound. We go from the simplest positivity bounds to more advanced techniques based on non-linear integral inequalities and partial wave expansions.

\item[Sec.~\ref{sec:regge}:] We concentrate on the applications of the dispersion relations for \ac{UV}-\ac{IR} relations in graviton-mediated scattering. In addition, we introduce the gravitational eikonal and show how to obtain an analog of the Martin-Froissart bound in dimensions higher than four for graviton-mediated scattering via the unitarization of the tree-level graviton scattering amplitude. We also discuss the fate of positivity bounds based on a twice-subtracted dispersion relation in the presence of graviton exchange.

\item[Sec.~\ref{sect:ANNA-conclusions}:] We describe the general view on relations between \ac{UV} and \ac{IR} theories, and prospects for future studies.

\item[Sec.~\ref{sect:ANNA-dictionary}:] We provide a dictionary of important concepts in the field of \ac{EFT} and amplitudes.

\end{description}

\subsection{EFT of gravity: vertices, amplitudes, field redefinitions}
\label{sec:EFTofGR}

In this section, we make a relation between the \ac{EFT} of gravity and scattering amplitudes. We start with the explicit computation of the graviton exchange amplitude in \ac{GR} with a minimally coupled scalar field, based on the result for the Feynman graviton propagator obtained in \cref{sec:quant-I}.

The Einstein-Hilbert action, as it is a non-linear functional of the metric, contains vertices with an arbitrary number of gravitons after a perturbative expansion around flat spacetime. In this section, we will mainly concentrate only on the contributions to $2\to 2$ scattering process. We will also go beyond \ac{GR}, extending it by a sequence of higher curvature terms inevitably emerging as counterterms required to cancel loop divergences in \ac{GR}~\cite{Burgess:2003jk, Donoghue:1994dn, Calmet:2013hfa}, see also the detailed discussion in \cref{sec:LUCA_twoloopdivs}.

\subsubsection{The concept of EFT}

\ac{EFT} is a powerful tool for the description of low energy physics. It is useful as a dramatic simplification in computations even if a complete theory is known. In the case when the ultimate theory is unknown or too complicated, it represents a solid ground and organizing tool for experimental measurements, predictions and computations. 

  \begin{figure}[ht]
        \centering
        \includegraphics[width=4in]{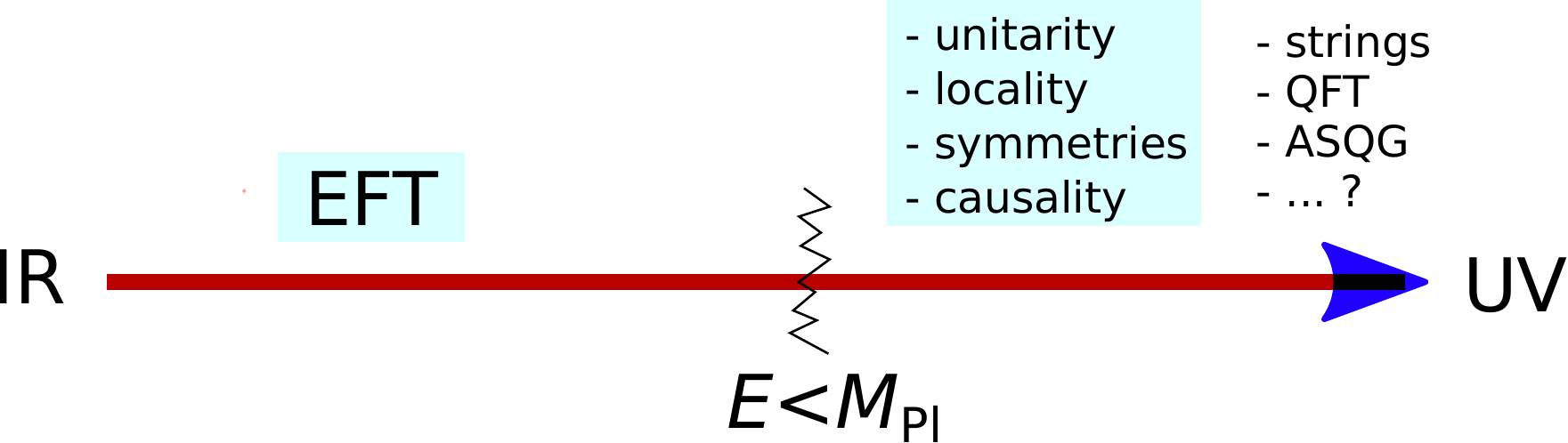}
        \caption{\ac{EFT} description, range of validity, properties and proposals for \ac{UV} completions.}
        \label{fig:EFT concept}
    \end{figure}

\begin{tcolorbox}
\begin{center}
    \textit{The first step in the construction of any \ac{EFT} is related to defining the low-energy degrees of freedom on top of flat spacetime and in the vacuum.}
\end{center}
\end{tcolorbox}

A very frequent confusion in quadratic gravity (or in the Starobinsky model) with the \ac{EFT} of gravity is related to the fact that the quadratic part of the \ac{EFT} action in four dimensions is trivialized, as we will see later. However, in quadratic gravity, these terms are supposed to improve or restore renormalizability! The origin and the resolution of this confusion is related to the proper definition of degrees of freedom. The \ac{EFT} of \ac{GR} is a theory of a single massless spin-two particle, while quadratic gravity has more degrees of freedom (a scalar, and a spin-two ghost, see \cref{sec:LUCA_quadgrav_prop_dofs}).

After determining the degrees of freedom from the linearized action, one can construct interaction terms in the Lagrangian which respect all the symmetries. In the case of gravity, general covariance must be preserved, so all interactions are different contractions of Riemann tensor. This way, the \ac{EFT} is organized as an infinite sum of all terms allowed by symmetries of the action, with Wilson coefficients which are supposed to be measured in experiments. Why is it still predictive?

The expansion can be ordered by mass dimension (or the power of the cutoff in the denominator) of the \ac{EFT} terms. Remarkably, only the first few terms are needed in most cases, if the relevant energies are far away from the scale of the \ac{EFT} breakdown --- the cutoff scale. Due to that, one compute scattering amplitudes and other observables just in an \ac{EFT} with a few terms, and it is expected to be a very good approximation in most cases.

We discussed how to construct an \ac{EFT} bottom-up. However, if we know a complete theory, the \ac{EFT} expansion can be obtained by means of integrating out heavy degrees of freedom and keeping the light ones. Let us show how this works in a simple example of integrating out a heavy scalar field. We start from the Lagrangian
\begin{equation}
    \lagrangian{}(\varphi,\psi)=-\frac{1}{2}\left( \partial _{\mu }\varphi \right)^{2}-\frac{1}{2}m^{2}\varphi^{2}+\varphi F\left( \psi \right) +G\left( \psi \right) \, ,
\end{equation}
where $\varphi$ is a heavy field, and $\psi$ represents light degrees of freedom. A partition function can be written in the form of a functional integral over the fields $\varphi$ and $\psi$,
\begin{equation}
\partitionfunction = \int \mathcal D\varphi \mathcal D\psi \, e^{i\int \rmd^{4}x \, \lagrangian{}\left( \varphi ,\psi\right)} \, .
\end{equation}
We can use an equivalent form for the Lagrangian:
\begin{equation}
    \lagrangian{}\left( \varphi ,\psi \right) =\frac{1}{2}\varphi \left( \Box-m^{2}\right) \varphi +\varphi F\left( \psi \right)+G\left( \psi \right) \, .
\end{equation}
Now we want to integrate out the heavy scalar $\varphi$. We define a shifted field $\overline{\varphi}$,
\begin{equation}
    \overline{\varphi }\left( x\right) =\varphi \left( x\right) +\int \rmd^{4}y \, D_{F}\left( x-y\right) F\left( \psi \left( y\right) \right) \, ,
\end{equation}
where $D_{F}\left( x-y\right)$ is the Green's function of the field $\varphi$ defined by the equation
\begin{equation}
    \left( \Box -m^{2}\right) D_{F}\left( x-y\right) =\delta^4 \left( x-y\right) \, .
\end{equation}
In terms of the field $\overline{\varphi }$, we obtain
\begin{equation}
    \frac{1}{2}\varphi \left( \Box-m^{2}\right) \varphi +\varphi F\left( \psi \right)=\frac{1}{2}\overline{\varphi }\left( \Box -m^{2}\right) \overline{\varphi }-\frac{1}{2}\int \rmd^4y \, F(\psi(x))D _{F}\left( x-y\right) F\left( \psi \left( y\right) \right) \, .
\end{equation}
This transformation is just a shift of the field, so it does not affect the measure in the functional integral, $\mathcal D \varphi = \mathcal D\overline{\varphi}$. Thus, we obtain the partition function in the form
\begin{equation}
\label{partition}
    \partitionfunction=\int \mathcal D{\psi} \, e^{i\int \rmd^{4}x \, G\left( \psi\right) }e^{-\frac{i}{2} \langle FDF \rangle} \, ,
\end{equation}
where we defined
\begin{equation}
\label{Non-local}
    \langle F D F\rangle =\int \rmd^{4}x \rmd^{4}y \, F\left( \psi \left( x\right) \right) D _{F}\left( x-y\right) F\left( \psi \left( y\right) \right) \, .
\end{equation}
The Green's function can be expanded in a series in momenta if the mass is large,
\begin{equation}
    D_{F}\left( x-y\right) =-\int \frac{\rmd^{4}q}{\left( 2\pi \right)^{4}}\frac{e^{-iq \cdot \left( x-y\right) }}{q^{2}+m^{2}}=-\int \frac{\rmd^{4}q}{\left( 2\pi\right)^{4}}e^{-i q \cdot \left( x-y\right) }\left( \frac{1}{m^{2}} - \frac{q^{2}}{m^{4}}+\dots \right) \, ,
\end{equation}
or
\begin{equation}
    D_{F}\left( x-y\right) =-\left( \frac{1}{m^{2}}+\frac{\Box }{m^{4}}+\frac{\Box^{2}}{m^{6}}+\dots \right) \delta^{4}\left( x-y\right) \, .
\end{equation}
Thus, in this limit, the effective low energy Lagrangian can be written as a local expansion in powers of the d'Alembertian operator $\Box$ acting on $F(\psi(x))$, instead of the non-local integral form \eqref{Non-local},
\begin{equation}
    \lagrangian{}_\text{eff}\left( \psi\right) =G\left( \psi \right) -\frac{1}{2}F\left( \psi \right) \frac{1}{m^{2}}F\left( \psi\right) -\frac{1}{2m^{4}}F\left( \psi \right) \Box F\left( \psi\right) + \dots \, .
\end{equation}
The partition function \eqref{partition} within this expansion is given by the functional integral of the local effective Lagrangian $\lagrangian{}_\text{eff}$,
\begin{equation}
    \partitionfunction=\int \mathcal D\psi \, e^{i\int \rmd^{4} x \, \lagrangian{}_\text{eff} \left( \psi\right) } \, .
\end{equation}
This Lagrangian represents a derivative expansion of all interactions, which can be ordered by the mass dimension of each term. Clearly, it looks like a non-renormalizable theory, although we obtained it by integrating out a heavy state in the theory which could be also renormalizable. 

We have just found that an \ac{EFT} can be obtained as a derivative expansion after integrating out heavy states. Does it mean that any \ac{EFT} corresponds to a \ac{UV} completion with a finite number of states? The answer is --- no! For example, the low energy expansion of the \ac{ST} effective action certainly contains an infinite number of states in the \ac{UV}. There are recent results arguing that the \ac{EFT} of \ac{GR} cannot be completed by a finite number of states~\cite{Caron-Huot:2024lbf}. Moreover, not all \acp{EFT} admit a good \ac{UV} completion, as we will see in the next sections. And, even more, not all \acp{EFT} are consistent with causality requirements formulated only in the low-energy domain of their validity~\cite{CarrilloGonzalez:2022fwg, CarrilloGonzalez:2023cbf}.

\subsubsection{Scattering amplitudes in GR with a minimally coupled scalar}

We start examining the structure and allowed space for \acp{EFT} from building a relation between the effective action and scattering amplitudes. In this section, we compute a graviton-mediated scattering amplitude, and present the result for graviton-graviton scattering in pure \ac{GR}.


We first recall the Feynman graviton propagator in Fourier space. In harmonic gauge (also called the de Donder gauge, \eqref{de-donder})
\begin{equation}
    \covD_{\mu}h^{\mu\nu}-\frac{1}{2}\eta^{\mu\nu}\covD_{\mu}h^{\alpha}_{\phantom{\alpha}\alpha}=0 \, .
\end{equation}
we have the graviton propagator (recall \eqref{propag-feynman-gauge})
\begin{tcolorbox}
    \begin{equation}
\propG_{\mu\nu\rho\sigma}(q)=\frac{1}{2}\frac{-i}{q^2-i\epsilon}\left(\eta_{\mu\rho}\eta_{\nu\sigma}+\eta_{\mu\sigma}\eta_{\nu\rho}-\eta_{\mu\nu}\eta_{\rho\sigma}\right) \, .
\end{equation}
\end{tcolorbox}
\noindent If we introduce the tensor
\begin{equation}
    P^{\alpha\beta\gamma\delta}=\frac{1}{2}\left(\eta^{\alpha\gamma}\eta^{\beta\delta}+\eta^{\alpha\delta}\eta^{\beta\gamma}-\eta^{\alpha\beta}\eta^{\gamma\delta}\right) \, ,
\end{equation}
we get
\begin{equation}\label{eq:ANNA_Feynman_grav_prop}
    \propG_{\mu\nu\rho\sigma}(q)=\frac{-i}{q^2-i\epsilon}P_{\alpha\beta\gamma\delta} \, .
\end{equation}
The vertices with three and four gravitons are very long and complicated expressions derived years ago in~\cite{DeWitt:1967yk, DeWitt:1967uc, DeWitt:1967ub, Sannan:1986tz}, and summarized, for example, in \cite{Donoghue:2017pgk}. They can also be easily reproduced with the use of computer algebra packages, such as \emph{xAct}~\cite{xActwebpage, Martin-Garcia:2007bqa, Brizuela:2008ra, Martin-Garcia:2008ysv, Nutma:2013zea}, dealing with tensors in covariant form. 

\begin{figure}[ht]
        \centering
        \includegraphics[width=0.8\textwidth]{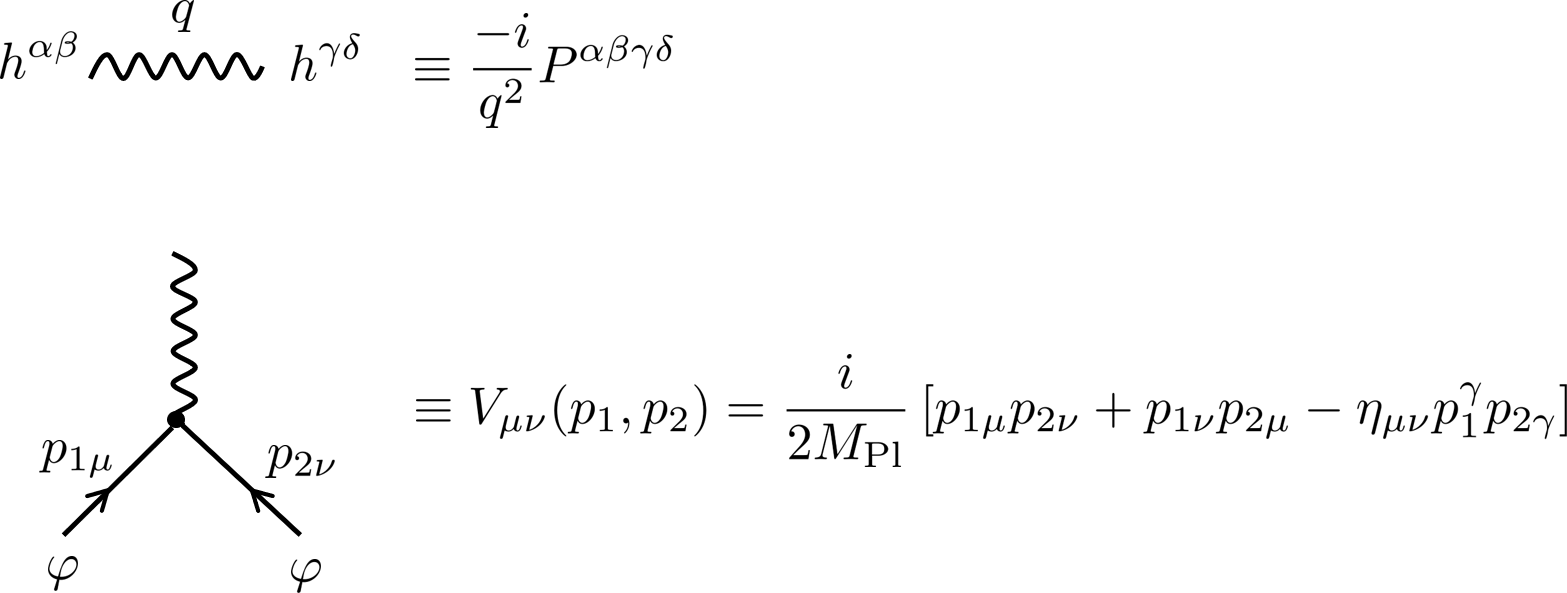}
        \caption{Feynman rules for minimally coupled scalar.}
        \label{feynrules1}
    \end{figure}

We will show the main properties of graviton-mediated scattering amplitudes using the example of a massless scalar field minimally coupled to gravity. Let us introduce
\begin{equation}
    S=\int \rmd^4 x\sqrt{-g}\left(\frac{\MPl^2\,R}{2}-\frac{1}{2}(\partial_{\mu}\varphi)^2\right) \, .
\end{equation}
The Feynman rules for this theory are given in \cref{feynrules1}~\cite{Weinberg:1964cn, Weinberg:1964ev}. The propagator was given above in \eqref{eq:ANNA_Feynman_grav_prop}, whereas the scalar-scalar-graviton vertex is
\begin{equation}
    V_{\mu\nu}(p_1,p_2)=\frac{i}{2 \MPl}\left(p_{1\mu}p_{2\nu}+p_{2\mu}p_{1\nu}-\eta_{\mu\nu}p_1^{\gamma}p_{2\gamma}\right) \, .
\end{equation}

   \begin{figure}[ht]
        \centering
        \includegraphics[width=0.15\textwidth]{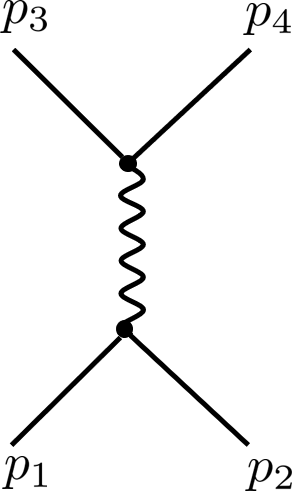}
        \caption{$t$-channel graviton exchange diagram contributing to \eqref{graviton-mediated amplitude}.}
        \label{feynrules-t}
    \end{figure}

Combining the Feynman rules from \cref{feynrules1}, we can find the matrix element corresponding to the graviton-mediated scalar scattering. The $t$-channel Feynman diagram is shown in \cref{feynrules-t}. For the $s$-channel, we have
\begin{equation}
\label{Gexch}
\scatteringamplitude_s=i \,V_{\mu \nu }\left( p_{1},p_{2}\right)\, \frac{i}{(p_1+p_2)^{2}}\,P^{\mu \nu \alpha \beta } \,V_{\alpha \beta }\left( p_{3},p_{4}\right) \, .
\end{equation}
By direct substitution, we find
\begin{equation}
    V_{\mu \nu }\left( p_{1},p_{2}\right) P^{\mu \nu \alpha \beta }=\frac{i}{2 \MPl}\left(p_{1}^{\alpha }p_{2}^{\beta }+p_{1}^{\beta }p_{2}^{\alpha }\right) \, .
\end{equation}
Contracting with the second vertex, we obtain
\begin{equation}
\begin{aligned}
    &2p_{1\alpha }p_{2_{\beta }}\left( p_{3}^{\alpha }p_{4}^{\beta }+p_{3}^{\beta }p_{4}^{\alpha }-\eta^{\alpha \beta }\left( p_{3} \cdot p_{4}\right) \right)\\
    = &2\left( p_{1} \cdot p_{3}\right) \left( p_{2} \cdot p_{4}\right) +2\left( p_{1} \cdot p_{4}\right) \left( p_{2} \cdot p_{3}\right) -2\left( p_{2} \cdot p_{1}\right) \left( p_{3} \cdot p_{4}\right) \, .
\end{aligned}
\end{equation}
It is convenient to express the amplitude in terms of Mandelstam variables defined as
\begin{equation}
\begin{aligned}
s&=-\left( p_{1}+p_{2}\right)^{2}=-\left( p_{3}+p_{4}\right)^{2} \, , \\
t&=-\left( p_{1}+p_{3}\right)^{2}=-\left( p_{2}+p_{4}\right)^{2} \, , \\
u&=-\left( p_{2}+p_{4}\right)^{2}=-\left( p_{2}+p_{3}\right)^{2} \, ,
\end{aligned}
\end{equation}
for all-ingoing momenta. For the scattering of the massless states, we have $s+t+u=0$, and
\begin{equation}
\begin{aligned}
s&=-2\left( p_{1} \cdot p_{2}\right) =-2\left( p_{3} \cdot p_{4}\right) \, , \\
t&=-2\left( p_{1} \cdot p_{3}\right) =-2\left( p_{2} \cdot p_{4}\right) \, , \\
u&=-2\left( p_{1} \cdot p_{4}\right) =-2\left( p_{2} \cdot p_{3}\right) \, .
\end{aligned}
\end{equation}
Here $s$ has the meaning of the total energy of the particles in the center-of-mass frame, and $t$ determines the momentum exchange. Also, we can find the scattering angle $\theta$ between ingoing and outgoing spatial momenta as
\begin{equation}
\cos\theta=1+\frac{2 t}{s} \, .
\end{equation}
Therefore, we get
\begin{equation}
    \scatteringamplitude_s=-\frac{1}{2 \MPl^2 s}\left(t^2+u^2-s^2\right)=\frac{1}{\MPl^2}\frac{t u}{s} \, .
\end{equation}
The two other crossed diagrams will contribute similarly (we need to change $s\to t$ and $s\to u$ for $t$ and $u$ channels, respectively). The final result for the total amplitude is
\begin{tcolorbox}
    \begin{equation}\label{graviton-mediated amplitude}
   \scatteringamplitude=\scatteringamplitude_s+\scatteringamplitude_t+\scatteringamplitude_u=\frac{1}{ \MPl^2}\left(\frac{t u}{s}+\frac{s u}{t}+\frac{t s}{u}\right) \, .
\end{equation}
\end{tcolorbox}
Notice that in the forward limit $t\to 0$, the amplitude has a pole $s^2/t$. This behavior is a typical feature of a graviton exchange. This type of singularity appears in all scattering amplitudes obtained from the exchange of a massive (in this case, a pole coming from the propagator is located at the value of the squared mass) or massless particles. However, graviton-mediated scattering is different from scalar, spinor or vector particles mediating the interaction. For example, in massless $\varphi^3$ theory, the pole contribution has a form
\begin{equation}
    \frac{1}{t}+\frac{1}{s}+\frac{1}{u} \, .
\end{equation}
There is no $s^2/t$ contribution. Even derivative couplings in a scalar theory would never combine into that structure. This behavior is a feature of the exchange of a spin-two particle. Another problem of the amplitude \eqref{graviton-mediated amplitude} is emergent in the limit of fixed scattering angle $\theta$ defined below in \eqref{momenta}.

\begin{figure}[ht]
        \centering
        \includegraphics[width=0.35\textwidth]{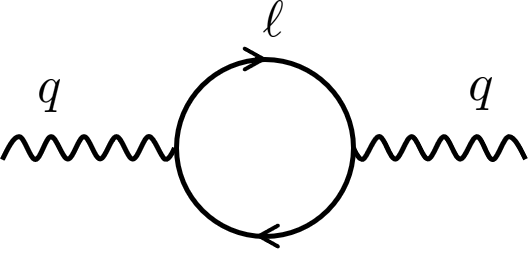}
        \caption{Matter loop correction to graviton propagator \eqref{loop}.}
        \label{feynrules}
\end{figure}

The one-loop diagram describing the contribution of matter particles to the graviton propagator, \cref{feynrules}, will be reduced to 
\begin{equation}
\label{loop}
\scatteringamplitude_{1l}=\int \frac{\rmd^{4} \ell}{\left( 2\pi \right)^{4}}\frac{i}{2\MPl}\left( \ell_{\alpha }\left( \ell+q\right) _{\beta }+ \ell_{\beta }\left( \ell+q\right) _{\alpha }\right) \frac{i}{\ell^{2}}\frac{i}{\left( \ell+q\right)^{2}} \frac{i}{2 \MPl}\left( \ell_{\delta }\left( \ell +q\right) _{\gamma }+ \ell_{\gamma}\left( \ell +q\right) _{\delta}\right) .
\end{equation}
The divergent part has the form
\begin{equation}
    \left( \ln q^{2}+\frac{1}{\varepsilon }\right) q_{\gamma}q_{\delta}q_{\alpha }q_{\beta }\propto q^4 \, ,
\end{equation}
which means that counterterms with four derivatives are required for renormalization~\cite{tHooft:1974toh, Bjerrum-Bohr:2002gqz, Akhundov:1996jd}, as discussed in \cref{sec:quant-II}. This result is a reflection of the well-known fact that \ac{GR} is a non-renormalizable theory \cite{tHooft:1974toh, Goroff:1985th}, and has to be embedded into some other \ac{UV}-complete theory at energies larger than the Planck mass.

\subsubsection{Graviton-graviton amplitudes in GR}

\begin{figure}[ht]
        \centering
        \includegraphics[width=0.9\textwidth]{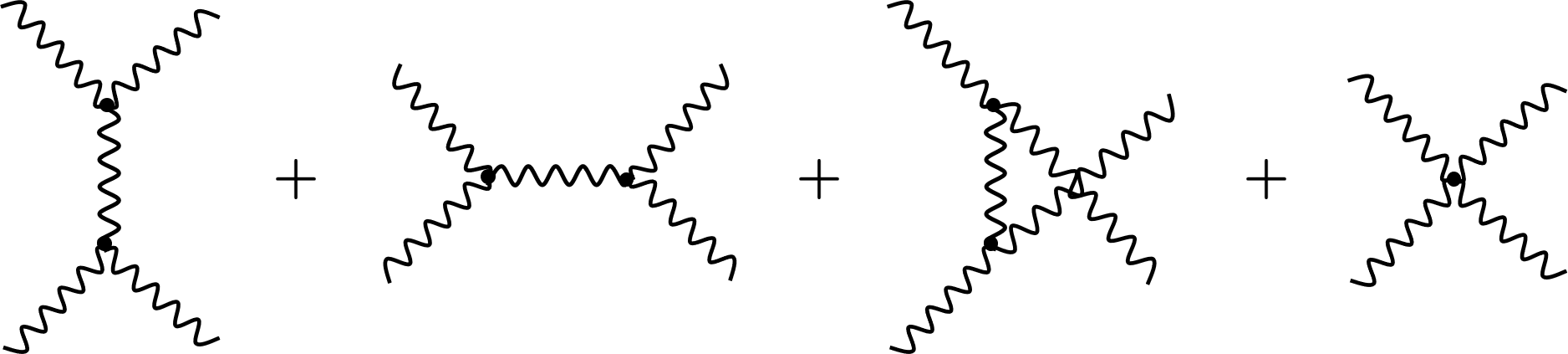}
        \caption{Diagrams contributing to $2\to 2$ graviton scattering.}
        \label{4diags}
    \end{figure}

The kinematics of graviton scattering can be described through the scattering angle $\theta$ and the total energy $p$. It is convenient to use all-ingoing momenta explicitly defined as
\begin{equation}\label{momenta}
\begin{aligned}
p_{1}^{\mu}&=\left( p,0,0,p\right) \, , \\
p_{2}^{\mu}&=\left( p,0,0,-p\right) \, , \\
p_{3}^{\mu}&=\left( -p,-p\sin \theta ,0,-p\cos \theta \right) \, , \\
p_{4}^{\mu}&=\left( -p, p\sin \theta ,0,p\cos \theta \right) \, .
\end{aligned}
\end{equation}
The momentum $p$ and the scattering angle $\theta$ are connected to the Mandelstam variables via
\begin{equation}
    p=\frac{s}{2}\, , \qquad\cos\theta=1+\frac{2 t}{s} \, .
\end{equation}
The graviton polarization tensor must be traceless and orthogonal to the corresponding momentum. Also, it should transform according to its helicity with respect to rotations around the momentum vector. Tensors with these properties can be composed from the four-vectors orthogonal to the corresponding momentum $p$:
\begin{align}
\polarizationtensorbasis^{\mu \pm }(p_1)&=\frac{1}{\sqrt{2}}\left( 0,1,\pm i,0\right) \, , \nonumber \\
\polarizationtensorbasis^{\mu \pm }(p_2)&=\frac{1}{\sqrt{2}}\left( 0,-1,\pm i,0\right) \, , \nonumber \\
\polarizationtensorbasis^{\mu \pm }(p_3)&=\frac{1}{\sqrt{2}}\left( 0,\cos \theta ,\pm i,-\sin \theta \right) \, , \nonumber \\
\polarizationtensorbasis^{\mu \pm }(p_4)&=\frac{1}{\sqrt{2}}\left( 0,-\cos \theta ,\pm i, \sin \theta \right) \, .
\end{align}
Polarization tensors can be obtained simply as a tensor product of the corresponding polarization vectors,
\begin{equation}
    \polarizationtensorbasis^{\mu\nu}_{\pm}(p_i) = \polarizationtensorbasis^{\mu\pm}(p_i) \polarizationtensorbasis^{\nu\pm}(p_i) \, .
\end{equation}
There are many different combinations of graviton polarizations, however, there are only three independent scattering amplitudes among all the combinations of polarization states, as the others can be related to them with the use of parity symmetry and crossing symmetry. For example, exchanging particles labeled '1' and '2', we get
\begin{equation}
    \scatteringamplitude_{+\,-\,+\,-}(s,t,u) = \scatteringamplitude_{+\,+\,-\,-}(t,s,u) \, .
\end{equation}
Parity symmetry implies the invariance of the amplitude with respect to flipping the helicity of all particles. Thus,
\begin{equation}
    \scatteringamplitude_{-\,-\,-\,+} = \scatteringamplitude_{+\,+\,+\,-} \, , \qquad \scatteringamplitude_{+\,+\,+\,+} = \scatteringamplitude_{-\,-\,-\,-}\, ,\dots \, .
\end{equation}
These transformations allow us to express all the amplitudes through three independent structures, represented for example by
\begin{equation}
    \scatteringamplitude_{+\,+\,+\,+}\, , \scatteringamplitude_{+\,+\,+\,-} \, , \scatteringamplitude_{+\,+\,-\,-} \, .
\end{equation}
The result of computing graviton scattering in pure \ac{GR} is
\begin{tcolorbox}
\begin{equation}
    \scatteringamplitude_{+\,+\,-\,-} = \frac{1}{\MPl^2}\frac{s^3}{t u} \, , \qquad \scatteringamplitude_{+\,+\,+\,+} = \scatteringamplitude_{+\,+\,+\,-} = 0 \, .
\end{equation}
\end{tcolorbox}
This computation can be done straightforwardly in Mathematica with the use of the \emph{xAct} package ~\cite{xActwebpage, Martin-Garcia:2007bqa, Brizuela:2008ra, Martin-Garcia:2008ysv, Nutma:2013zea}, see \eg{} \cite{xActAnna}. The idea of the computation is based on obtaining vertices from the perturbed Lagrangian, and further substituting explicit expressions for polarization tensors (realized in the \emph{xCoba} sub-package). The same computation, of course, can be easily extended to the \ac{EFT} of \ac{GR}.

\subsubsection{EFT of a shift-symmetric scalar: field redefinitions}

In this section, we show how an \ac{EFT} can be constructed and reduced to a lower number of terms with the use of perturbative field redefinitions. For simplicity we use a shift-symmetric scalar field theory.

What could be the structure of the low energy theory of massless shift-symmetric scalar? One can write down a derivative expansion containing plenty of different combinations of fields and derivatives:
\begin{equation}
S=\int \rmd^{4}x \sqrt{-g} \, \left( -\frac{1}{2}\left( \partial_{\mu }\varphi \right)^{2}+\alpha \left( \partial _{\mu }\varphi \right)^{2}\Box\varphi +\beta \left( \left( \partial _{\mu }\varphi \right)^{2}\right)^{2} +\gamma \left( \partial^{6}\left( \varphi^{4} \right) \right)+\dots \right) \, .  
\end{equation}
However, many of these combinations are not independent. Several couplings that are left can be eliminated using perturbative field redefinitions. Consider the example
\begin{equation}
    \varphi =\chi-a\left( \partial _{\rho}\chi\right)^2 \, .
\end{equation}
Substituting this, we obtain for the Lagrangian
\begin{equation}
\begin{aligned}
 \lagrangian{}&=-\frac{1}{2}\left( \partial _{\mu }\chi\right)^{2}+2a \partial_{\mu }\chi \partial_{\rho} \chi \partial^{\mu }\partial^{\rho}\chi+2a^{2}\left( \partial_{\rho}\chi\right)^{2}\left( \partial _{\mu }\partial _{\nu}\chi\right)^{2}+\alpha \Box \chi \left( \partial_{\mu} \chi\right)^{2} \\
&\qquad +a\alpha \Box \left( \partial_{\rho}\chi\right)^{2}\left( \partial_{\mu} \chi\right)^{2}+2a\alpha \Box \chi\partial _{\mu }\chi\partial^{\mu }\partial _{\rho}\chi \partial^{\rho}\chi+\orderneglected\left( \chi^{5}\right) +\dots \, .
\end{aligned}
\end{equation}
Adjusting the value of $a$, we can get rid of the cubic interaction. Indeed, under the integral sign, we can rewrite 
\begin{equation}
\begin{aligned}
\partial_{\mu} \chi\partial_{\rho} \chi\partial^{\mu }\partial^{\rho}\chi=-\Box \chi \left( \partial_{\rho} \chi\right)^{2}-\partial_{\mu} \chi\partial ^{\mu }\partial_{\rho}\chi \partial^{\rho}\chi=-\frac{1}{2}\Box \chi \left( \partial_{\rho}\chi\right)^{2} \,,~ \text{(up to a total derivative)}
\end{aligned}
\end{equation}
such that there is only one $\chi^3$ combination left. Demanding $a=-\alpha$, we eliminate all cubic interaction terms, moving them to higher point terms. Thus, the minimal set of operators contributing to the $2\to 2$ scattering amplitude reads
\begin{equation}
\label{scalarEFT}
\lagrangian{}=-\frac{1}{2}\left( \partial _{\mu }\varphi \right)^{2}+\frac{g_2}{2}\left( \left( \partial _{\mu }\varphi\right)^2 \right)^{2}+\frac{g_3}{3} \left( \partial _{\mu }\varphi \right)^{2}\left( \partial _{\rho}\partial _{\sigma }\varphi \right)^{2}+4 g_4 \left( \partial _{\rho}\partial _{\sigma }\varphi \partial^{\rho}\partial^{\sigma }\varphi \right)^{2}+\dots \, .
\end{equation}
Here we introduced the couplings $g_2$, $g_3$, $g_4$ in the same way as in~\cite{Caron-Huot:2020cmc}, which we will use further in the amplitude computation. Notice the most important general rule helping to dramatically reduce the number of terms in each order of the derivative expansion: 
\begin{tcolorbox}
\begin{center}
    \textit{The terms proportional to the free equations of motion can \\always be eliminated by field redefinitions.}
\end{center}
\end{tcolorbox}
\noindent For this reason, there are no terms containing $\Box \varphi$ in \eqref{scalarEFT}. We can generalize this observation noticing that if the Lagrangian contains a term proportional to the free equations of motion,
\begin{equation}
    \lagrangian{}=F\left[ \varphi \right] \hat{E}\varphi +\frac{1}{2}\varphi \hat{E}\varphi \, , \qquad \hat{E}\varphi =0 \, ,
\end{equation}
this term can be eliminated by the following perturbative field redefinition,
\begin{equation}
    \varphi=\chi-F\left[ \chi\right].
\end{equation}
Indeed,
\begin{equation}
   \lagrangian{}=F\left[ \chi\right] \hat{E} \chi+\frac{1}{2}\chi\hat{E}\chi+F\left[ \chi\right] \hat{E} \left( F\left[ \chi\right] \right) -F\left[ \chi\right] \hat{E}\chi+\dots \, ,
\end{equation}
one can see that the terms linear in the equation of motion get canceled. As the operator $F[\chi]$ contains at least two fields $\chi$, the residual terms contribute to higher point vertices. This way, terms proportional to the equations of motion can be moved to the vertices which do not contribute to $2\to 2$ scattering processes.

Let us stress here that the term does not disappear completely from the theory, as this field redefinition would also modify other terms in the action. Instead, it contributes to higher order and higher point operators. Even though the latter do not affect tree-level $2\to 2$ scattering, we should remember about their possible presence because they can affect the loop corrections to the amplitude.

In addition, these terms can be important outside the framework of the S-matrix, for example, if one is searching for the classical background solution in such a theory. If the background values of the fields are not too large compared to the cutoff scale, this problem is well-posed, and perturbations around the classical solution can be studied. This procedure is very common when cosmological backgrounds are considered. But one should remember that before the splitting into the background and perturbations, the degrees of freedom and canonical fields (with respect to which we integrate in a functional integral) should be uniquely determined.

How can we check whether the terms left in the \ac{EFT} action after field redefinitions indeed represent a minimal set of couplings which cannot be reduced further? In fact, physical observables which are invariant under field redefinitions are scattering amplitudes, that is why they play an important role in the context of the \ac{EFT} construction. For the Lagrangian \eqref{scalarEFT}, we obtain a tree level amplitude (recall that $s+t+u=4 m^2$)
\begin{equation}\label{eq:ANNA_scattamp_sss}
    \scatteringamplitude(s,t)=g_2(s^2+t^2+u^2)+g_3 s t u+g_4 (s^2+t^2+u^2)^2+\dots \, .
\end{equation}
It has a crossing-symmetric form, and a polynomial structure in Mandelstam variables, thus, the further terms can be constructed as crossing-symmetric polynomials of the proper power of energy. From this observation, we can find the number of independent \ac{EFT} operators in each power of energy. It is interesting that the number of these independent terms grows slowly. Thus, finding an irreducible set of \ac{EFT} couplings can be efficiently done through the construction of the amplitude and matching it to a proper number of operators in the \ac{EFT} where their form can be just guessed.

\subsubsection{Structure of the gravitational EFT and amplitudes}

The action of \ac{GR} is
\begin{equation}
S=\frac{\MPl^{2}}{2}\int \rmd^{4}x \sqrt{-g} \, R \, .
\end{equation}
It is known to be non-renormalizable at one loop in the presence of matter, and at two loops without matter, see \cref{sec:quant-II}. Thus, for the low energy description of gravity, only an \ac{EFT} expansion makes sense. What could be the most general action? In fact, it can have all possible contractions of Riemann tensors with derivatives. However, many of these terms are redundant. For example,
\begin{equation}
\Gamma = \int \rmd^4 x  \sqrt{-g} \, \left[ \frac{\MPl^{2}}{2}R+aR^{2}+ b R_{\mu \nu }R^{\mu \nu }+c R_{\mu \nu \rho \sigma}R^{\mu \nu \rho \sigma}+d\Box R+\frac{e}{\UVcutoff^{2}}\text{Riem}^{3}+\dots\right] \, ,
\end{equation}
where \UVcutoff{} is the \ac{EFT} cutoff. Recall that the equations of motion are 
\begin{equation}
R_{\mu \nu }-\frac{1}{2}R g_{\mu \nu }=\frac{1}{\MPl^{2}}T_{\mu \nu } \, .
\end{equation}
Thus, all terms having at least one $R_{\mu\nu}$ can be rewritten as matter couplings. If we consider only graviton couplings, we can set 
\begin{equation}
R_{\mu \nu }=0 \, , \qquad R=0
\end{equation}
in the \ac{EFT} Lagrangian. In this way, the couplings $a$ and $b$ can be set to zero. Moreover, the term with coupling $d$ is a total derivative. In four dimensions, the $\text{Riem}^2$-term can be written as a combination of $R^2$, $R_{\mu\nu}R^{\mu\nu}$, and a total derivative --- the Gauss-Bonnet invariant (cf. \eqref{gauss-bonnet-luca})
\begin{equation}
\gaussbonnetterm=R^{2}-4R_{\mu \nu }R^{\mu \nu }+R_{\mu\nu\rho \sigma }R^{\mu \nu \rho \sigma } \, .
\end{equation}
For these reasons, all quadratic terms can be eliminated in the \ac{EFT}, and rewritten as terms of higher order in curvature. Therefore, the first non-trivial contribution starts at $\text{Riem}^3$. The general rules allowing to construct non-redundant operators are related to Hilbert series \cite{Ruhdorfer:2019qmk}.

We also have the following properties:
\begin{align}
R_{\mu \nu \rho \sigma }+R_{\mu \rho \sigma \nu }+R_{\mu \sigma \nu \rho}&=0 \, , \\
\covD _{\alpha }R_{\mu \nu \rho \sigma }+\covD _{\rho}R_{\mu \nu \sigma \alpha }+\covD _{\sigma }R_{\mu \nu \alpha \rho }&=0 \, .
\end{align}
The last relation is known as Bianchi identity. One can define the traceless Weyl tensor (see also \eqref{weyl-tensor-Luca})
\begin{equation}
C_{\mu \nu \rho \sigma }=R_{\mu \nu \rho \sigma }-\left( g_{\mu[\rho }R_{\sigma ] \nu}-g_{\nu [ \rho }R_{\sigma ]\mu }\right) +\frac{1}{3}g_{\mu [\rho} g_{\sigma]\nu }R \, .
\end{equation}
Here, we used the antisymmetrization over indices defined via
\begin{equation}
A_{\left[ \mu \nu \right]} =\frac{1}{2}\left( A_{\mu \nu }-A_{\nu \mu }\right) \, .
\end{equation}
We can see that the Weyl tensor is the same as the Riemann tensor, up to terms that vanish on the equations of motion. In addition, up to the equations of motion we have (cf. \eqref{eq:ALESSIABENJAMIN_boxRiem})
\begin{equation}
\covD^{\mu }C _{\mu \nu \rho \sigma }=0 \, , \qquad \Box C _{\mu \nu \rho \sigma }\sim C_{\mu \nu\rho \sigma }^2 \, .
\end{equation}
Thus, indices of derivatives should be contracted only with each other, avoiding $\Box$ combinations. We get the following building blocks for the \ac{EFT} of gravity:
\begin{equation}
C _{\mu \nu \rho \sigma }\, , \quad \covD _{( \mu _{1}}C_{  \mu ) \nu \rho \sigma } \, , \quad \covD _{( \mu_1}\covD _{\mu _{2}}C_{  \mu ) \nu \rho \sigma} \, , \dots \, .
\end{equation}
The Weyl tensor can be further decomposed into 
\begin{equation}
C_{L/R}=\frac{1}{2}\left( C\pm i\tilde{C}\right) \, , \qquad \tilde{C}^{\mu \nu \rho\sigma }=\frac{1}{2}\epsilon^{\mu\nu\alpha\beta}{C _{\alpha \beta }}^{\rho \sigma } \, .
\end{equation}
Thus, the action can be constructed in terms of $C_L$ and $C_R$ and their symmetrized derivatives. In this form, it is also easy to see which terms are CP-even (they contain an even number of $\epsilon$ tensors) and CP-odd.

Including terms that are suppressed by the cutoff scale as up to $\UVcutoff^8$, the action generalized to $d$ dimensions can be found to be of the form~\cite{Ruhdorfer:2019qmk, deRham:2022gfe}
\begin{equation}
\begin{aligned}
    S=\MPl^{d-2}\int \rmd^{d}x\sqrt{-g} \, \Bigg\{ \frac{R}{2}+\frac{a_{2}}{\UVcutoff^{2}}\gaussbonnetterm +\frac{a_{3}}{\UVcutoff^{4}}C^{3} &+\frac{a_{4}}{\UVcutoff^{6}}{\mathcal C}^{2}+\frac{\tilde{a}_{4}}{\UVcutoff^{6}}{\tilde{\mathcal C}^{2}} \\
    &+\frac{a_{5}}{\UVcutoff^{8}}F_{\phantom{\alpha}\alpha }^{\alpha } {\mathcal C}  +\frac{\tilde{a}_5}{\UVcutoff^{8}}\tilde{F}_{\phantom{\alpha}\alpha }^{\alpha}{\tilde{\mathcal C}}+\dots \Bigg\} \, ,
\end{aligned}
\end{equation}
where $\mathcal C = C_{\mu\nu\rho\sigma} C^{\mu\nu\rho\sigma}$ and $\tilde{\mathcal C} = C_{\mu\nu\rho\sigma} \tilde C^{\mu\nu\rho\sigma}$.
We leave here the Gauss-Bonnet term $\gaussbonnetterm$, as it will remain a non-trivial operator in $d>4$. The tensors $F_{\alpha\beta}$ and $\tilde F_{\alpha\beta}$ are defined as
\begin{equation}
    F_{\alpha \beta }=\covD _{\alpha }C_{\mu \nu \rho \sigma }\covD _{\beta}C^{\mu \nu \rho\sigma}\, , \qquad \tilde{F}_{\alpha\beta }=\covD _{\alpha }C_{\mu \nu \rho \sigma }\covD _{\beta}\tilde{C}^{\mu \nu \rho\sigma} \, .
\end{equation}
The corresponding scattering amplitudes for gravitons are~\cite{deRham:2022gfe}
\begin{align}
\MPl^{d-2}\scatteringamplitude_{+\,+\,-\,-}&=\frac{s^3}{tu}-\frac{8( d-4)}{d-2}\frac{a_{2}}{\UVcutoff^{4}}s^{3}-18 \frac{a_{3}^{2}}{\UVcutoff^{8}}s^{3}\left( \frac{d-4}{d-2}s^{2}+2st+2t^{2}\right) \nonumber \\
&\hspace{6cm}+\frac{8 s^4}{\UVcutoff^{6}}a_{4+}+\frac{4}{\UVcutoff^{8}}a_{5+}s^{5} \, , \\
\MPl^{d-2}\scatteringamplitude_{+\,+\,+\,+}&=\frac{12}{\UVcutoff^{2}}\left( 5a_{3}-\frac{2\left( d-4\right) }{d-2}a_{2}^{2}\right) x \nonumber \\
&\hspace{2.2cm} -\frac{2}{\UVcutoff^{8}}\left( \frac{9\left( 12-d\right) }{d-2}a_{3}^{2}+10a_{5-}\right) xy +\frac{16}{\UVcutoff^{6}}a_{4-}x^{2} \, , \\
\MPl^{d-2}\scatteringamplitude_{+\,+\,+\,-}&=\frac{6}{\UVcutoff^{4}}a_{3}y+\frac{\gamma y^{2}}{\UVcutoff^{10}} \, .
\end{align}
Here we defined the triple crossing-symmetric variables
\begin{equation}
    y= s t u\,, \qquad x=s t +t u +u s \, .
\end{equation}
The term $\gamma$ above corresponds to an operator suppressed by $\UVcutoff^{10}$. We also use $a_{n\pm}=a_n\pm \tilde{a}_n$. From the form of the amplitude, we can see that the basis of the \ac{EFT} operators is not redundant, as all coefficients which we introduced lead to different structures in the amplitudes. But it is interesting to note that it is not always possible to write an \ac{EFT} operator providing the given structure in the amplitude. It is clearly seen that in $d=4$ no local operator that contributes to the $s^3$-term in $\scatteringamplitude_{+\,+\,-\,-}$ can be constructed. Thus, the relation between the \ac{EFT} action and the amplitude may be more subtle, and the structure of even tree-level amplitudes is more constrained than it is expected from crossing symmetry alone. Another source of constraints may be related to the locality requirement for \ac{EFT} Lagrangians which does not follow from crossing symmetry either. It is a separate constraint on the \ac{IR} action.

\subsection{Scattering amplitudes: analyticity, unitarity, Martin-Froissart bound, bootstrap}
\label{sec:amplitudes}

\subsubsection{Analyticity properties of the scattering amplitudes}\label{sec:ANNAanalyticity}

Scattering amplitudes of Lorentz-invariant theories are functions of two Mandelstam variables $s$ and $t$ on shell. Physical scattering corresponds to the values of $s>0$ and $t<0$. However, it is still important how the amplitudes are analytically continued to the whole complex plane of $s$ and $t$. As an implication of causality, the amplitude must be an analytic function away from real values for $s$ and $t$. We will use the following property in further derivations of dispersion relations:
\begin{tcolorbox}
\begin{center}
    \textit{For fixed $t<0$ and ${\rm Im}\,s\ne 0$, the amplitude is an analytic function of $s$.}
\end{center}
\end{tcolorbox}
This analyticity property was originally derived from the fact that the commutators of the fields must vanish outside the light-cone by the definition of S-matrix causality~\cite{Gell-Mann:1954ttj} (also referred to as Bogolyubov causality~\cite{Bogolyubov:1959bfo,Bogolyubov:1975ps}). It can also be derived from the requirement of causal propagation of the signal~\cite{Toll:1956cya} (see also~\cite{Camanho:2014apa} for a more pedagogical derivation of the dispersion relation). Thus, the strict condition of the absence of a time advance in signal propagation (also referred to as asymptotic causality, or as the Gao-Wald condition~\cite{Gao:2000ga} in more recent literature~\cite{Bellazzini:2021shn}) implies the analyticity of the amplitude.

 \begin{figure}[ht]
        \centering
        \includegraphics[width=0.7\textwidth]{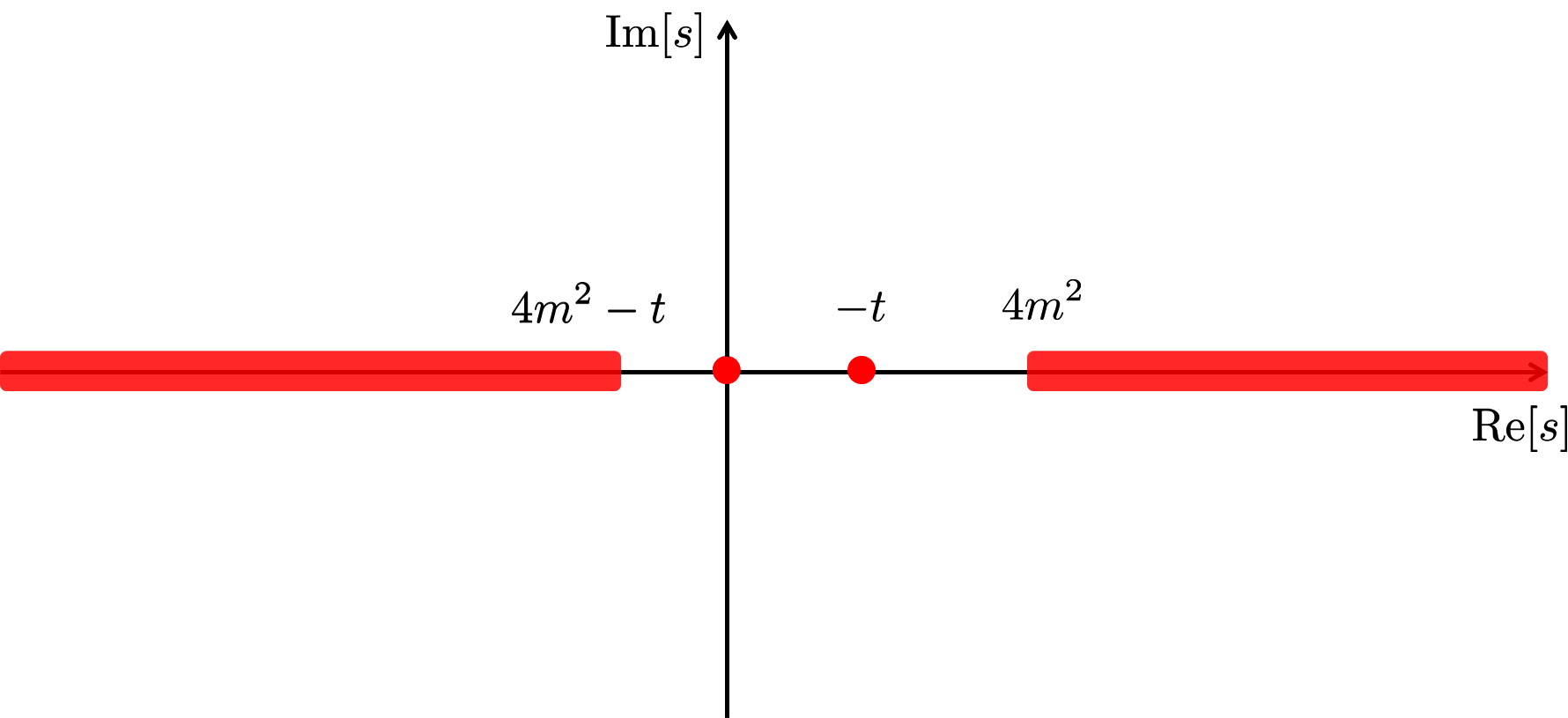}
        \caption{Analyticity structure of a scattering amplitude of equal mass particles at fixed $t<0$. Red areas represent branch cut singularities, red circles are poles. If there are interacting massless states in the theory, branch cuts are intersecting with each other, and the complex plane is split into two pieces.}
        \label{analyticity}
    \end{figure}

Even though the amplitude is physically defined only for real $s>0$, it can be analytically continued through the upper half of the complex plane to negative $s$. This will correspond to crossing relations and complex conjugation~\cite{Bros:1965kbd}. The analyticity structure of the amplitude at fixed $t<0$ is summarized in \cref{analyticity}. For theories with a mass gap, there is an analyticity region between $s=0$ and $s=4 m^2$ where the amplitude can have only pole singularities corresponding to an exchange of other massive or massless particles. For large $s$, amplitudes typically have a branch cut singularity along the real axis with a discontinuity given by its imaginary part. 

\subsubsection{Partial wave expansion}
\label{sec:AnnaPartialWaves}

The amplitude describing the scattering of plane waves with definite values of momenta can be decomposed into the scattering of angular momentum eigenstates --- partial waves. In the center-of-mass frame, this corresponds to the decomposition of its scattering angle dependence into the eigenfunctions of the angular momentum operator. If we consider scattering of scalars in four dimensions, these functions are Legendre polynomials $P_l(\cos\theta)$,
\begin{equation}
    \scatteringamplitude(s,\theta)=32\pi\sum_{l=0}^{\infty}\left(l+\frac{1}{2}\right)f_l(s) P_l(\cos\theta) \, .
\end{equation}
Legendre polynomials form an orthogonal system of functions,
\begin{equation}
    \int_{-1}^{1}\rmd\cos\theta \, P_j(\cos\theta)P_k(\cos\theta)=\frac{2}{2 j+1}\delta_{jk} \, .
\end{equation}

The \ac{PWU} requirement implies for each partial wave 
\begin{tcolorbox}
\begin{equation}
    |f_j(s)|<1 \, , \qquad {\rm Im}\,f_j(s)>0 \, .
\end{equation}
\end{tcolorbox}
All \ac{EFT} amplitudes have partial wave amplitudes growing with $s$, which means that the first condition must be violated when $s$ is large enough. The minimal value of $s$ when this condition is violated can be taken as the definition of the \ac{EFT} cutoff scale \UVcutoff{}.

There is a stronger non-linear unitarity condition which can be derived from the optical theorem. The total cross section can be found as a sum of partial wave amplitudes, 
\begin{equation}
    \sigma_\text{tot} = \frac{1}{32\pi s}\int_{-1}^1 \rmd\cos\theta \left| \scatteringamplitude\left( \theta \right) \right|^{2} \, .
\end{equation}
Substituting the partial wave decomposition, we obtain
\begin{equation}
\begin{aligned}
\sigma_\text{tot} &= \frac{1}{32\pi s}\sum _{j,k}\int_{-1}^1 \rmd\cos\theta \, \left( 32\pi \right)^{2}f_{j}\left( s\right) f_{k}^{\ast }\left( s\right) P_{j}\left( \cos \theta \right) P_{k}\left( \cos\theta \right) \left( j+\frac{1}{2}\right) \left( k+\frac{1}{2}\right) \\
&=\frac{32\pi }{s}\sum _{j}\left( j+\frac{1}{2}\right) \left| f_{j}\left( s\right) \right|^{2} \, .
\end{aligned}
\end{equation}
The optical theorem (see also \eqref{optical theorem-feynman} in \cref{sec:LUCA}) implies for the scattering of scalars with the same mass $m$
\begin{align}
    {\rm Im} \, \scatteringamplitude( s,\theta = 0) &\geq \frac{1}{2}\sqrt{s\left( s-4m^{2}\right) } \, \sigma_\text{tot}\left( s\right) \, , \\
    32\pi \sum _{j}{\rm Im}\, f_{j}\left( s\right) \left( j+\frac{1}{2}\right) &\geq \frac{16\pi }{s}\sqrt{s\left( s-4m^{2}\right) }\sum _{j}\left( j+\frac{1}{2}\right) \left| f_{j}\left( s\right) \right|^{2} \, .
\end{align}
Thus, for each partial wave in the sum one can obtain the \ac{FU} condition
\begin{tcolorbox}
\begin{equation}
    2\,{\rm Im}\,f_{j}\left( s\right) \geq \left| f_{j}\left( s\right) \right|^{2}\sqrt{\frac{s-4m^{2}}{s}} \, .
\end{equation}
\end{tcolorbox}
Let us mention here that the \ac{FU} condition tells us that unitary amplitudes must have an imaginary part! Thus, it is not possible to find a theory where the amplitudes have only real tree-level contributions, while all loops are zero due to some symmetry.

The partial wave decomposition is crucial for implementing non-perturbative unitarity conditions and the reconstruction of the amplitudes based on unitarity requirements. It allows to derive a plethora of non-trivial, model-independent results following from the basic properties of the \ac{QFT}. 

In $d$-dimensional theories, the eigenfunctions of the angular momentum are given by Gegenbauer polynomials, such that the amplitude can be decomposed as
\begin{equation}
    \scatteringamplitude\left( s,t\right) =\frac{1}{2}\sum n\left( l,d\right) f_{l}\left( s\right) P_{l}^{(d)}\left( 1+\frac{2t}{s-4m^{2}}\right) \, ,
\end{equation}
where $P_{l}^{(d)}$ are related to the Gegenbauer polynomials $C^{\alpha}_l$ by 
\begin{equation}
    P_{l}^{(d)}\left( z\right) =\frac{\Gamma \left( 1+l\right) \Gamma \left( d-3\right) }{\Gamma \left( l+d-3\right) }C_{l}^{\frac{d-3}{2}}\left( z\right) \, .
\end{equation}
They are normalized as
\begin{equation}
    \frac{1}{2}\int^{1}_{-1}\rmd z \, \left( 1-z^{2}\right)^{\frac{d-4}{2}}P_{l}^{\left( d\right)}\left( z\right)P_{l'}^{\left( d\right) }\left( z\right) =\frac{\delta_{ll'}}{N\left( d\right) n\left( l,d\right) } \, , 
\end{equation}
where
\begin{equation}
    N \left( d\right) =\frac{\left( 16\pi \right)^{\frac{2-d}{2}}}{\Gamma \left( \frac{d-2}{2}\right) } \, , \qquad n\left( l,d\right) =\frac{\left( 4\pi \right)^{\frac{d}{2}}\left( d+2l-3\right) \Gamma \left( d+l-3\right) }{\pi \Gamma \left( \frac{d-2}{2}\right) \Gamma \left( l+1\right) } \, .
\end{equation}
Another expression for $d$-dimensional Legendre polynomials relates them to the hypergeometric function:
\begin{equation}
    P_l^{(d)}(x)={}_{2}F_1\left(-l,l+d-3,\frac{d-2}{2},\frac{1-x}{2}\right) \, .
\end{equation}
The corresponding partial wave amplitudes can be found as
\begin{equation}
    f_{l}\left( s\right) =N\left( d\right) \int _{-1}^1 \rmd x \, \left( 1-x^{2}\right)^{\frac{d-4}{2}}P_{l}^{\left( d\right) }\left( x\right) \scatteringamplitude\left( s,t\left( x\right) \right).
\end{equation}
In $d$ dimensions, the \ac{PWU} conditions get modified,
\begin{tcolorbox}
\begin{equation}
    2\, {\rm Im} f_{l}\left( s\right) \geq \frac{\left( s-4m^{2}\right)^{\frac{d-3}{2}}}{\sqrt{s}}\left| f_{l}\left( s\right) \right|^{2} \, , \qquad \left| f_{l}(s)\right|  <\frac{\sqrt{2s}}{\left( s-4m^{2}\right)^{\frac{d-3}{2}}} \, .
\end{equation}
\end{tcolorbox}
The partial wave decomposition can also be generalized straightforwardly to the scattering of particles with spin. In this case, the eigenfunctions of the angular momentum are given by Wigner d-functions (\emph{WignerD} in Mathematica),
\begin{equation}
    \scatteringamplitude_{h_{1}h_{2}h_{3}h_{4}}=32\pi \sum _{l}\left( 2l+1\right) d_{\lambda \mu }^{l}\left( \cos \theta \right) f_{h_{1}h_{1}h_{3}h_{4}}^{l}\left( s\right) \, ,
\end{equation}
where
\begin{equation}
\lambda =h_{2}-h_{1} \, , \qquad \mu =h_{4}-h_{3} \, , \qquad d_{\lambda \mu }\left( -\theta \right) =\left( -1\right)^{\lambda -\mu }d_{\lambda \mu }\left( \theta \right) \, .
\end{equation}
For the scattering of gravitons, we always have
\begin{equation}
\left( -1\right)^{\lambda -\mu }=1 \, ,
\end{equation}
therefore the amplitudes are even functions of $\theta$, 
\begin{equation}
\scatteringamplitude_{h_{1}h_{2}h_{3}h_{4}}=16\pi \sum _{l}\left( 2l+1\right) \left( d_{\lambda \mu }^{l}\left( \theta \right) +d_{\lambda \mu }^{l}\left( -\theta \right) \right) f_{h_{1}h_{1}h_{3}h_{4}}^{l}(s) \, .
\end{equation}
The latter combination can be simplified,
\begin{equation}
    d_{\lambda \mu }^{l}\left( \theta \right) +d_{\lambda \mu }^{l}\left( -\theta \right) = 2e^{ i\frac{\pi }{2}( \lambda -\mu ) }\sum^{l}_{\nu =-l}d_{\lambda \nu}^{l} \left( \frac{\pi }{2}\right) d_{\mu \nu }^{l}\left( \frac{\pi }{2}\right) \cos \nu \theta.
\end{equation}
In this way, the Wigner d-functions are expressed through elementary functions.

\subsubsection{Martin-Froissart bound}
\label{MFbound}

One of the important consequences of unitarity and analyticity is a bound on the scattering amplitude at fixed $t$ and large $s$~\cite{Froissart:1961ux, Martin:1962rt,Jin:1964zza}. In this section, we present its detailed derivation illustrating the practical use of \ac{PWU} and analyticity based on the review in~\cite{Tokuda:2019nqb}. Consider the scattering of massive scalars with the same mass $m^2$ for simplicity. The amplitude can be decomposed into partial waves,
\begin{equation}
\scatteringamplitude\left( s, t\right) =16\pi \sum^{\infty }_{l=0}\left( 2l+1\right) f_{l}\left( s\right) P_{l}\left( 1+\frac{t}{2q^{2}}\right) \, ,
\end{equation}
where we defined
\begin{equation}
    q=\frac{1}{2}\sqrt{s-4 m^{2}} \, .
\end{equation}
Given the expression for the amplitude, one can compute partial wave amplitudes using the orthogonality of the Legendre polynomials. However, we can have a better estimate of the values of the partial wave amplitudes expressing them through Legendre functions $Q_l(z)$ because they decay exponentially for large $l$, unlike $P_l(z)$ which oscillate at large $l$. Recall that Legendre functions $Q_l(z)$ can be expressed as
\begin{equation}
    Q_{l}\left( z\right) =\frac{1}{2}\int^{1}_{-1} \rmd\mu \, \frac{P_{l}\left( \mu \right) }{z-\mu} \, .
\end{equation}
We can compute an integral of an amplitude multiplied with $Q_l(z)$ over a contour which we will specify a bit later, based on the convenience of further computations:
\begin{equation}
    \oint_{\gamma} \rmd z \, \scatteringamplitude\left( s,t\left( z\right) \right) Q_{n}\left( z\right)=\frac{1}{2}\sum _{l}\left( 2l+1\right) f_{l}\left( s\right) \int^{1}_{-1}\rmd\mu \oint _{\gamma}\rmd z \, P_{l}\left( 1+\frac{t(z)}{2q^{2}}\right) \frac{P_{n}\left( \mu \right) }{z-\mu } \, .
\end{equation}

 \begin{figure}[ht]
        \centering
        \includegraphics[width=0.7\textwidth]{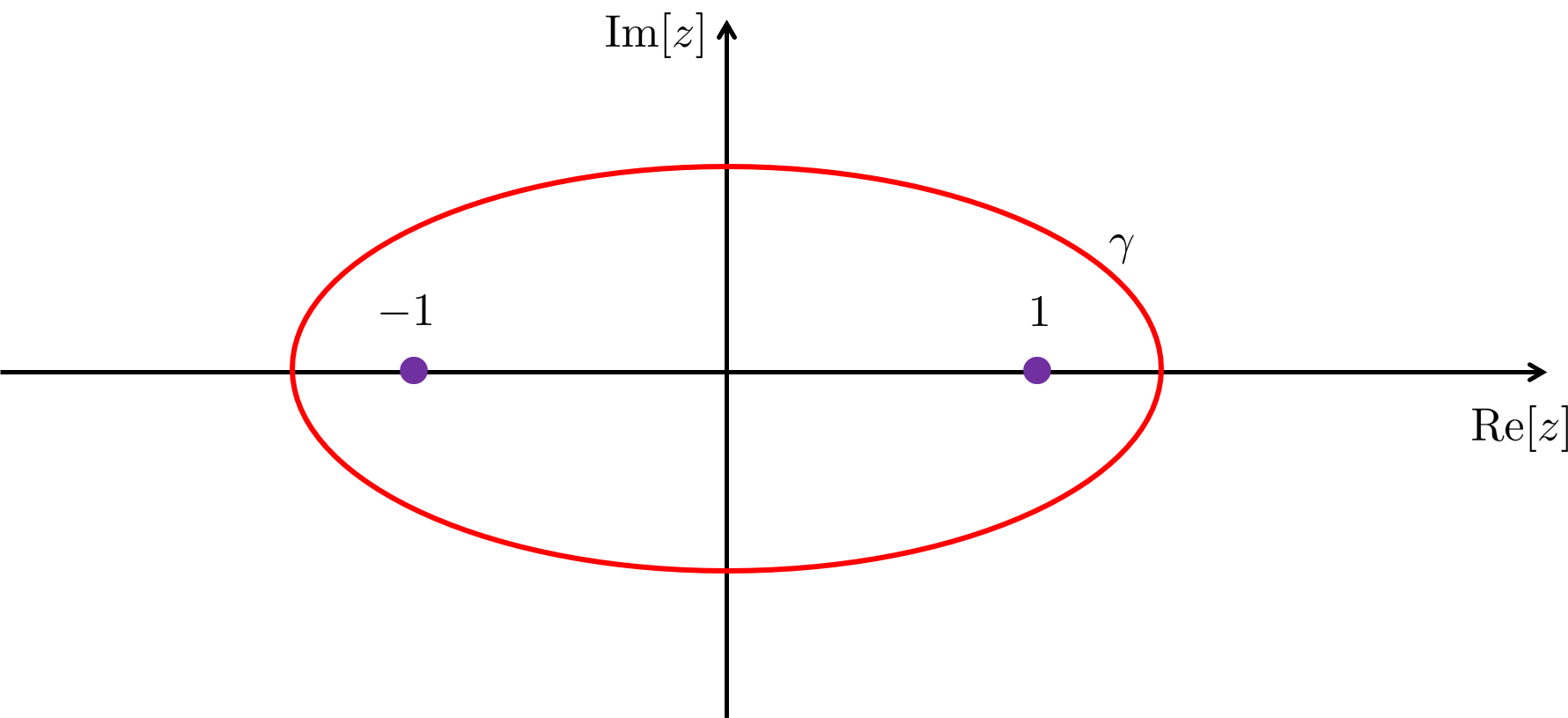}
        \caption{The integration contour $\gamma$ on the complex $z$-plane should include the points $(-1, 1)$, while its precise form is not important.}
        \label{MFcontour}
    \end{figure}

Given the analyticity of Legendre polynomials, we can compute the integral over $z$ as a residue at the point $z=\mu$ on the real axis placed in the interval $(-1,1)$, see \cref{MFcontour}. Thus, $\gamma$ should include this interval, and for further derivations it is essential to have analyticity properties of the amplitude inside this contour. We obtain
\begin{equation}
    \oint_{\gamma} \rmd z \, \scatteringamplitude\left( s,t\left( z\right) \right) Q_{n}\left( z\right) =(2\pi i) \, \frac{1}{2}\sum _{l}\left( 2l+1\right) f_{l}\left( s\right) \int^{1}_{-1}\rmd\mu \, P_{n}\left( \mu \right) P_{l}\left( 1+\frac{t\left( \mu \right) }{2q^{2}}\right) \, .
\end{equation}
We can get the desired expression for $f_l(s)$ based on the orthogonality properties of $P_l(\mu)$ if we choose
\begin{equation}
    1+\frac{t\left( \mu \right) }{2q^{2}}=\mu \, , \qquad t\left( \mu \right) =2q^{2}\left( \mu -1\right) \, .
\end{equation}
In this way, we obtain the so-called Neumann expression for the partial wave amplitudes,
\begin{equation}
    f_{l}\left( s\right) =\frac{1}{2\pi i}\oint_{\gamma } \rmd z \, \scatteringamplitude\left( s,2q^{2}\left( z-1\right) \right) Q_{l}\left( z\right) \, .
\end{equation}
For large $l$, we have
\begin{equation}
    Q_{l}\left( z\right) \approx K_{0}\left( l\sqrt{2\left( z-1\right) }\right) \approx e^{-l\sqrt{2\left( z-1\right) }}\sqrt{\frac{\pi }{2l\sqrt{2\left( z -1\right) }}} \, , \qquad l\to\infty \, ,
\end{equation}
where $K_0$ is the Macdonald function of index $0$ (or simply a specific Bessel function --- in Mathematica, it is \emph{BesselK[0,z]}). Recall that
\begin{equation}
    z-1= \frac{t}{2 q^2} \, .
\end{equation}
If the amplitude is polynomially bounded in all directions, we can have an estimate for fixed $t$ and large $l$:
\begin{equation}
\label{pwave}
    \left| f_{l}(s)\right| \propto\frac{q}{2\pi \sqrt{s}}e^{-\frac{\sqrt{t}}{q}\left( l+\frac{1}{2}\right) }\sqrt{\frac{\pi q}{2 l \sqrt{t}}}B_{0}\left( s\right) \, , \qquad B_0\left( s\right) \lesssim s^{N} \, .
\end{equation}
For small $l$, we can simply use the unitarity requirement,
\begin{equation}
    \left| f_{l}(s)\right|  <1 \, , \qquad l < L \, .
\end{equation}
Until which value of $L$ is it a stronger constraint, compared to \eqref{pwave}? We can find this via
\begin{align}
    e^{-\frac{\sqrt{t}}{q}\left( L+\frac{1}{2}\right) }\sqrt{\frac{q}{L\sqrt{t}}}B_{0}\left( s\right) &=1 \, , \\
    \frac{\sqrt{t}}{q}\left( L+\frac{1}{2}\right) &=\ln \frac{q}{\sqrt{t}}+\ln B_{0}\left( s\right) \, .
\end{align}
Solving for $L$ and using $q\sim \sqrt{s}$, we get
\begin{equation}
     L=\frac{q}{2\sqrt{t}}\ln \frac{q}{\sqrt{t}}+\frac{q}{\sqrt{t}}N\ln \frac{s}{s_{0}}\propto \sqrt{s}\ln s \, .
\end{equation}
Contributions of partial waves to the total amplitude can be split into $l<L$ and $l>L$,
\begin{equation}
    \scatteringamplitude(s,t\to0^+)=\sum^{L}_{l=0}P_{l}\left( 1\right) \left( 2l+1\right) +\sum^{\infty }_{l=L+1}P_{l}\left( 1\right) \frac{\left( 2l+1\right) }{\sqrt{l}}e^{ -\frac{\sqrt{t}}{q}(l+\frac{1}{2}) }s^N\sqrt{\frac{\pi q}{2\sqrt{t}}} \, .
\end{equation}
The large-$l$ contribution is exponentially suppressed at large $s$. We can see this from the estimate
\begin{equation}
    \sum^{\infty }_{l=L+1}\sqrt{l}e^{-\frac{\sqrt{t}}{q}l}\sim\int^{\infty }_{L} \rmd y \, \sqrt{y}e^{-\frac{\sqrt{t}}{q}y} \sim e^{-\frac{\sqrt{t}}{q}L}\sim e^{-\sqrt{t}\sqrt{s}\ln s} \, .
\end{equation}
Even the factor $s^N$ will be suppressed by this exponent,
\begin{equation}
    s^{N}e^{-\sqrt{t}\sqrt{s}\ln s}\to 0 \, , \qquad s\to \infty \, .
\end{equation}
Thus, large spins $l$ give zero contribution in the limit of large $s$. The contribution of low spins can be evaluated as
\begin{equation}
    \scatteringamplitude(s,t\to0^+)=\sum^{L}_{1=0}P_{l}\left( 1\right) \left( 2l+1\right)\sim (L+1)^2\propto s(\ln{s})^2 \, .
\end{equation}
The latter result represents the celebrated Martin-Froissart bound, limiting the growth of the amplitude as a function of $s$ at fixed small $t$ to
\begin{tcolorbox}
\begin{equation}
  \scatteringamplitude(s,t\to 0)<C |s|(\ln{|s|})^2 \, ,
\end{equation}
\end{tcolorbox}
\noindent where $C$ is a constant. This behavior can be extended to the whole complex $s$-plane, as the same discussion can be repeated for complex-valued $s$. Let us list here the assumptions required for the derivation of this result:
\begin{itemize}
    \item The amplitude must be an analytic function in the domain of integration $\gamma$. It is important to mention here that this assumption is violated in theories with loops of massless particles, as well as for the scattering of massless states. In this case, there is no analyticity domain at small $t$, as the branch cut starts at zero.
    \item \ac{PWU} has to be fulfilled.
    \item The amplitude must be polynomially bounded (this is usually thought to be related to the locality of the theory).
\end{itemize}

\subsubsection{Bootstrap in EFT: loops from trees}
\label{bootstrap}
In this section, we consider another example of the use of \ac{PWU} for the construction of loop corrections in the scalar field \ac{EFT} example. The tree-level amplitude for the shift-symmetric scalar field can be obtained as
\begin{equation}
    \scatteringamplitude(s,t)=g_2 (s^2+t^2+u^2)+g_3 s t u +g_4 (s^2+t^2+u^2)^2+ g_5 (s^2+t^2+u^2) s t u+ \dots \, .
\end{equation}
If we compute loops in this \ac{EFT}, we expect to get a running of the Wilson coefficients, leading to logarithmic corrections in the amplitude. But what is the exact form of these corrections? In certain cases, we can obtain them from unitarity requirements alone, avoiding rather complicated computations that use Feynman diagram techniques~\cite{Guerrieri:2018uew, Karateev:2019ymz, Guerrieri:2020bto}. The methods related to the amplitude reconstruction from unitarity and analyticity are usually referred to as a ``bootstrap program'', which can also be implemented numerically in the non-perturbative regime~\cite{Eden:1966dnq, Martin:1969ina, Paulos:2016fap, Kruczenski:2022lot,Karateev:2022jdb,Tourkine:2023xtu}. The generalization of the technique for spinning particles is described in a recent work~\cite{Haring:2022sdp}. Writing the amplitude as a function of $s,t,u$, we can write the partial wave amplitude as
\begin{equation}
    f_{l}\left( s\right) =\frac{1}{16\pi }\int^{1}_{-1}\rmd x \, P_{l}\left( x\right) \scatteringamplitude\left( s,-\frac{s}{2}\left( 1-x\right) ,-\frac{s}{2}\left( 1+x\right) \right) \, .
\end{equation}
The first partial waves for the mentioned amplitude are
\begin{equation}
f_{0}\left( s\right) =\frac{5g_2 s^{2}}{48\pi }+\frac{g_3 s^{3}}{96\pi }+\frac{7g_4 s^{5}}{40\pi } \, , \qquad
f_{2}\left( s\right) =\frac{g_2 s^{2}}{240\pi }-\frac{g_3 s^{3}}{480\pi }+\frac{g_4 s^{5}}{70\pi} \, .
\end{equation}
Obviously, they do not satisfy the \ac{FU} condition $2\, {\rm Im}\, f_{l}\geq|f_l|^2$ unless we add an imaginary part which is expected to appear from the logarithmic running in the initial amplitude. But we can deduce that, at leading order,
\begin{equation}
\label{Im}
    {\rm Im}\, f_0=\left( \frac{5 g_2}{48\pi }\right)^{2}s^{4} \, , \qquad  {\rm Im}\, f_{2}=\left( \frac{g_2 }{240\pi }\right)^{2}s^{4} \, .
\end{equation}
This pattern in partial waves can be provided by the amplitude with $s$-discontinuity (imaginary part)
\begin{equation}
     {\rm Im}\, \scatteringamplitude_{1l}=a_{1}s^{2}\left( s^{2}+a_{2} t u\right) \, , \qquad 2 \, {\rm Im}\, f_0=\frac{a_1 s^4}{8\pi} + \frac{a_1 a_2 s^4}{48\pi} \, , \qquad 2 \, {\rm Im}\, f_2=-\frac{a_1 a_2 s^4}{240\pi} \, .
\end{equation}
Matching with \eqref{Im} gives 
\begin{equation}
    a_1=\frac{7 g_2^2}{40 \pi} \, , \qquad a_2=-\frac{1}{21} \, .
\end{equation}
Thus, the first one-loop correction to the amplitude is (see also~\cite{Bellazzini:2021oaj})
\begin{equation}
\label{1loop}
    \scatteringamplitude_{1l}=-\frac{21 g_2^2}{240 \pi^2}s^2 \left(s^2-\frac{1}{21} t u \right)\ln{(-s)}+ (s\to t)+(s\to u) \, .
\end{equation}
Recall that the discontinuity coming from $\ln{(-s)}$ is $- i\pi$ (we define the branch cut of the logarithm as $\ln{s}-\ln{(-s)}=i\pi$).

Similarly, one can restore loop corrections at the order of the $s^5$ term. Moreover, for these terms, all loop corrections are reduced to one loop only, because higher loops will require counterterms with higher derivatives and higher powers of $s$ and $t$. However, at the order $s^6$, two-loop contribution will also be relevant. In addition, we can have contributions from operators like $(\partial_{\mu}\varphi)^6$ which do not contribute to the tree-level scattering amplitude. Their potential presence would spoil the uniqueness of the reconstruction of the full \ac{EFT} amplitude (bootstrap) without knowledge about higher point interactions. However, the described procedure can be useful for finding the boundaries of the allowed region in the \ac{EFT} parameters corresponding to the $2\to 2$ scattering domination over multi-particle processes.

\subsection{Positivity bounds in EFTs: selected analytic results, compactness of Wilson coefficients space}\label{sec:positivity}

\subsubsection{The simplest positivity bound}

Not all \acp{EFT} can have a \ac{UV} completion that is consistent with a set of expectations for a good \ac{QFT}. Here is a list of some desired properties:

\begin{itemize}
    \item {\bf Unitarity:} Unitarity is a fundamental requirement in quantum theory to have probabilities for all processes only between zero and one. Concerning scattering amplitudes, this implies \ac{PWU}, encoding the scattering of eigenstates of the angular momentum operator. At the level of the \ac{EFT} degrees of freedom, it also requires the absence of negative norm states (ghosts) interacting with positive norm states, as these processes would include negative probabilities. The condition of \ac{FU} also reflects the completeness of the Hilbert space of the \ac{QFT}. As it was shown in \cref{sec:AnnaPartialWaves}, it follows from the optical theorem where the complete set of intermediate states is used.
    
    \item {\bf Lorentz invariance:} This requirement implies limitations on the kinematics of the scattering processes, such that the amplitude is a function of certain scalar combinations of momenta. For $2\to 2$ scattering, it restricts the amplitude to be a function of two scalar variables on-shell.
    
    \item {\bf Causality:} As it was discussed in \cref{sec:ANNAanalyticity}, the propagation of a particle on top of any background, or scattering off another state can lead only to a time delay compared to the free propagation (asymptotic causality). This implies analyticity properties for the scattering amplitude at fixed $t<0$ in the complex $s$-plane away from the real axis.
   
    \item {\bf Locality:} This is perhaps the most subtle requirement imposed on \acp{QFT}. We can separate this requirement into its \ac{IR} part described also in \cref{sec:Luca locality}, meaning that amplitudes follow from \acp{EFT} which have a standard derivative expansion where at each power of the cutoff scale, there are a finite number of derivatives. An example of a non-local \ac{IR} operator could be (but is not limited to) the $1/\Box$ operator acting on any combination of fields. In our discussion, we assume there are no such kind of terms. \ac{UV} locality is an even more subtle property of the \ac{UV} completion. Perhaps the most concise definition of it can be given as the requirement of polynomial boundedness of the amplitudes everywhere in the complex $s$-plane at fixed $t<0$. Hereafter, we assume this definition, although it cannot be derived from unitarity and causality requirements, and has to be assumed additionally.
    
\end{itemize}

Having only an \ac{EFT} description at hand, how can we decide whether it can be \ac{UV}-com\-pleted at all? The listed set of assumptions can be transformed into precise mathematical properties of the scattering amplitude. Namely, we will be using the following set of \ac{QFT} properties applied both to the \ac{EFT} and to the unknown \ac{UV}-completion.

\begin{tcolorbox}
\ac{QFT} axioms in a language of the scattering amplitudes:
\begin{itemize}
    \item Lorentz invariance $~\Rightarrow \, \scatteringamplitude=\scatteringamplitude(s,t,u)$,
    \item Unitarity $~\Rightarrow~ 2\, {\rm Im}\, f_l(s)>\left| f_{j}\left( s\right) \right|^{2}\sqrt{\frac{s-4m^{2}}{s}}>0,~|f_l(s)|<1$,
    \item Causality $~\Rightarrow \, \scatteringamplitude(s,t)$ is an analytic function outside the real axes,
    \item Locality $~\Rightarrow \, \scatteringamplitude(s,t)<s^N t^N$ for finite $N$.
\end{itemize}
\end{tcolorbox}

\begin{figure}[ht]
        \centering
        \includegraphics[width=0.7\textwidth]{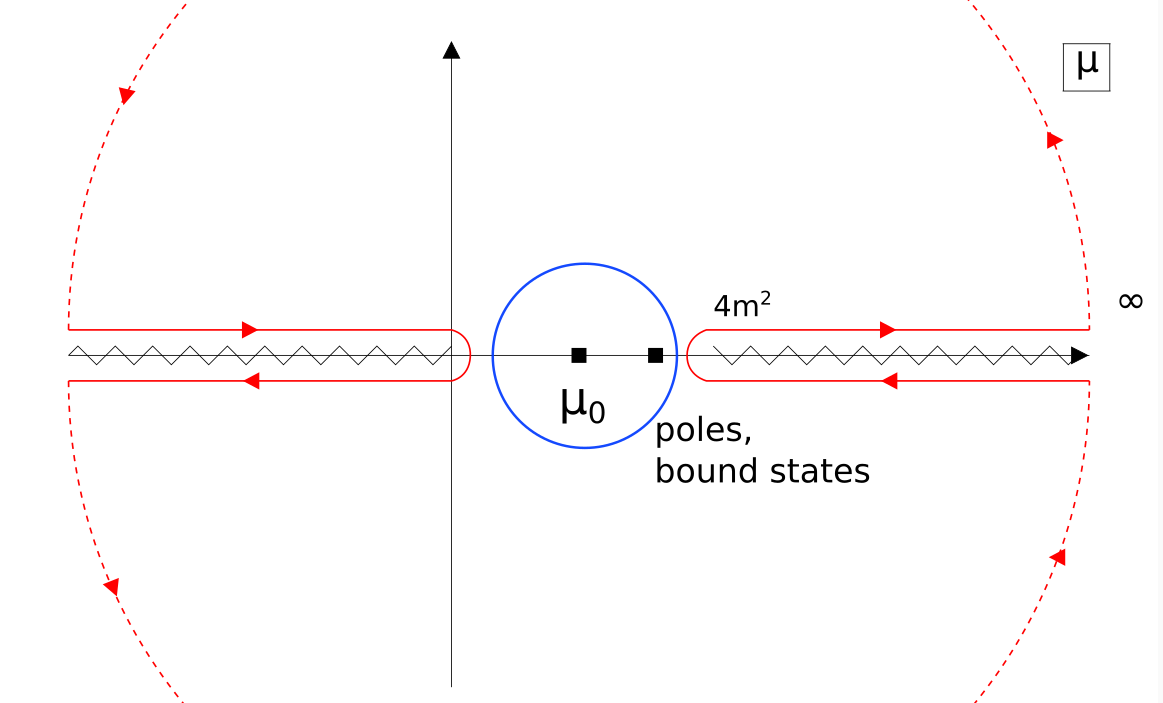}
        \caption{Integration contour $\Gamma$ used for the derivation of positivity bounds.}
        \label{c0}
    \end{figure}

Following the procedure first suggested in~\cite{Adams:2006sv}, we define a quantity that is computable in the \ac{IR} theory, 
    \begin{equation}
        \Sigma _\text{IR}=\frac{1}{2\pi i}\oint _{\Gamma}\rmd\mu \frac{\scatteringamplitude\left( \mu ,0\right) }{\left( \mu -\mu _{0}\right)^{3}} \, .
    \end{equation}
The contour is shown in \cref{c0}.

It can be also expressed through a sum over residues,
\begin{equation}
    \Sigma _\text{IR}=\sum \text{Res}\,\frac{\scatteringamplitude\left( \mu,0 \right) }{\left( \mu -\mu _{0}\right)^{3}}={\left.\frac{1}{2}\frac{\partial^{2}\scatteringamplitude\left( \mu ,0\right) }{\partial \mu^{2}}\right|}_{ \mu =\mu _{0}} \, .
\end{equation}
On the other side, using the analyticity of the amplitude, we can deform the integration contour and obtain
\begin{equation}
\begin{aligned}
    \Sigma _\text{IR}&=\frac{1}{2\pi i}\left( \int^{\infty }_{4m^{2}}\rmd\mu +\int^{0}_{-\infty }\rmd\mu \right) \frac{\scatteringamplitude\left( \mu +i\varepsilon ,0\right) -\scatteringamplitude\left( \mu -i\varepsilon ,0\right) }{\left( \mu -\mu _{0}\right)^{3}} \\
    &=\int^{\infty }_{4m}\frac{\rmd\mu }{\pi }\left( \frac{\text{Im}\, \scatteringamplitude\left( \mu,0 \right) }{\left( \mu -\mu _{0}\right)^{3}}+\frac{\text{Im}\, \scatteringamplitude^{\ast }\left( \mu ,0\right) }{\left( \mu -4m^{2}+\mu _{0}\right)^{3}}\right) \, .
\end{aligned}
\end{equation}
This way, $\Sigma_\text{IR}$ is expressed through integrals of the imaginary part of the amplitude. Using the optical theorem (see also \cref{sec:app-unitarity}) we can relate it to positive-valued physical observables, such as the cross section,
\begin{equation}
    \text{Im} \, \scatteringamplitude\left( s,0\right) =\sqrt{s-4m^{2}} \, \sigma _\text{tot}\left( s\right)  >0 \, .
\end{equation}
As the \ac{UV} integral is always positive in a unitary \ac{UV} completion, we obtain a bound on the low energy \ac{EFT},
\begin{equation}
\label{positivity0}
     {\left.\frac{\partial^{2}\scatteringamplitude\left( \mu ,0\right) }{\partial \mu^{2}}\right|}_{ \mu =\mu _{0}}>0 \, .
\end{equation}
This bound has crucial implications for theories with Galileon symmetry, making this symmetry incompatible with a good \ac{UV} completion~\cite{Adams:2006sv}. Indeed, the \ac{EFT} of a shift-symmetric scalar with Lagrangian
\begin{equation}
    \lagrangian{}=-\frac{1}{2}\left( \partial _{\mu }\varphi \right)^{2}+\alpha \left( \partial _{\mu }\varphi \right)^{2}\Box \varphi + g_2 \left( \left(\partial _{\mu }\varphi\right)^2 \right)^{2}
\end{equation}
possesses the Galileon symmetry
\begin{equation}
    \varphi \to q+a_{\rho}x^{\rho}\, , \qquad \partial _{\mu }\varphi \to \partial _{\mu }\varphi +a_{\mu } \, ,
\end{equation}
if $g_2=0$. Indeed,
\begin{equation}
\begin{aligned}
\lagrangian{} & =-\frac{1}{2}\left( \partial _{\mu }\varphi +a_{\mu }\right)^{2}+\alpha \Box \varphi \left( \partial _{\mu }\varphi +a_{\mu }\right)^{2} \\
&=-\frac{1}{2}\left( \partial _{\mu }\varphi \right)^{2}+\alpha  \varphi \left( \partial _{\mu }\varphi \right)^{2}+\alpha  \varphi a^{\mu }\partial _{\mu }\varphi +\alpha (a_{\mu })^{2} \varphi \\
&=-\frac{1}{2}\left( \partial _{\mu }\varphi \right)^{2}+\alpha \Box \varphi ( \partial _{\mu }\varphi )^{2}+\partial _{\mu }\left( \dots \right) \, .
\end{aligned}
\end{equation}
Here we used that up to a total derivative
\begin{equation}
(\Box \varphi) a^{\mu }\partial _{\mu }\varphi =-a^{\mu }(\Box \partial _{\mu }\varphi) \varphi =-a^{\mu }(\partial _{\mu }\varphi) \Box \varphi =0 \, .
\end{equation}
However, the term $((\partial_{\mu}\varphi)^2)^2$ is not invariant under the Galileon symmetry. 

Let us come back to the more general case. Recall that the amplitude of a general shift-symmetric scalar field, \eqref{eq:ANNA_scattamp_sss}, is given by
\begin{equation}
    \scatteringamplitude(s,t,u)=g_2 (s^2+t^2+u^2)+g_3 s t u+\dots \, .
\end{equation}
One can see that the bound \eqref{positivity0} implies
\begin{equation}
\label{beta}
    g_2 >0 \, ,
\end{equation}
and does not allow $g_2=0$ in an interacting theory. Thus, the exact Galileon symmetry (which strictly requires $g_2=0$) is in contradiction with the positivity bound. The importance of this result is related to the fact that a scalar with unbroken Galileon symmetry is a part of all massive gravity proposals~\cite{Nicolis:2008in,deRham:2010ik,deRham:2010kj}. For this reason, positivity bounds imply strong constraints on the phenomenological implications of massive gravity~\cite{Adams:2006sv,deRham:2017imi,Bellazzini:2017fep,Bellazzini:2023nqj}. In addition, the bound \eqref{beta} implies the a-theorem~\cite{Komargodski:2011vj}, which constrains the \ac{RG} running of the coefficient of the conformal anomaly \cite{Duff:1977ay,Duff:1993wm,Deser:1993yx,Shaposhnikov:2022dou,Shaposhnikov:2022zhj,Karateev:2023mrb,Schwimmer:2024vxw}.

\subsubsection{Linear bounds}

In the recent literature, the described technique found a lot of developments and applications, including infinite sets of positivity bounds beyond the forward limit~\cite{Vecchi:2007na,deRham:2017avq,Manohar:2008tc,Nicolis:2009qm}, for massive spinning particles (particularly spin two)~\cite{Bellazzini:2016xrt,deRham:2017zjm,Cheung:2016yqr, Bonifacio:2016wcb,Alberte:2019zhd,Alberte:2019xfh,Wang:2020xlt,Davighi:2021osh,Bellazzini:2019bzh}, gravity~\cite{Haring:2022cyf, Bern:2021ppb, Chowdhury:2021ynh, Chiang:2022jep, Alberte:2020jsk, Alberte:2020bdz, Alberte:2021dnj, deRham:2022gfe, Hamada:2018dde, Noumi:2022wwf, Tokuda:2020mlf,Herrero-Valea:2022lfd}, and non-linear bounds from crossing symmetry, Cauchy-Schwarz-type inequalities and properties of Gegenbauer polynomials~\cite{Chiang:2021ziz,Arkani-Hamed:2020blm, Bellazzini:2020cot, Tolley:2020gtv, Caron-Huot:2020cmc}, as well as numerous applications to the \ac{SM} \ac{EFT}~\cite{Zhang:2018shp, Bi:2019phv, Remmen:2019cyz,Zhang:2020jyn, Fuks:2020ujk,Remmen:2020vts, Yamashita:2020gtt}. In this section, we will show how to analytically derive an infinite set of linear inequalities providing compact bounds on the Wilson coefficients of the \ac{EFT}.

 \begin{figure}[ht]
        \centering
        \includegraphics[width=0.7\textwidth]{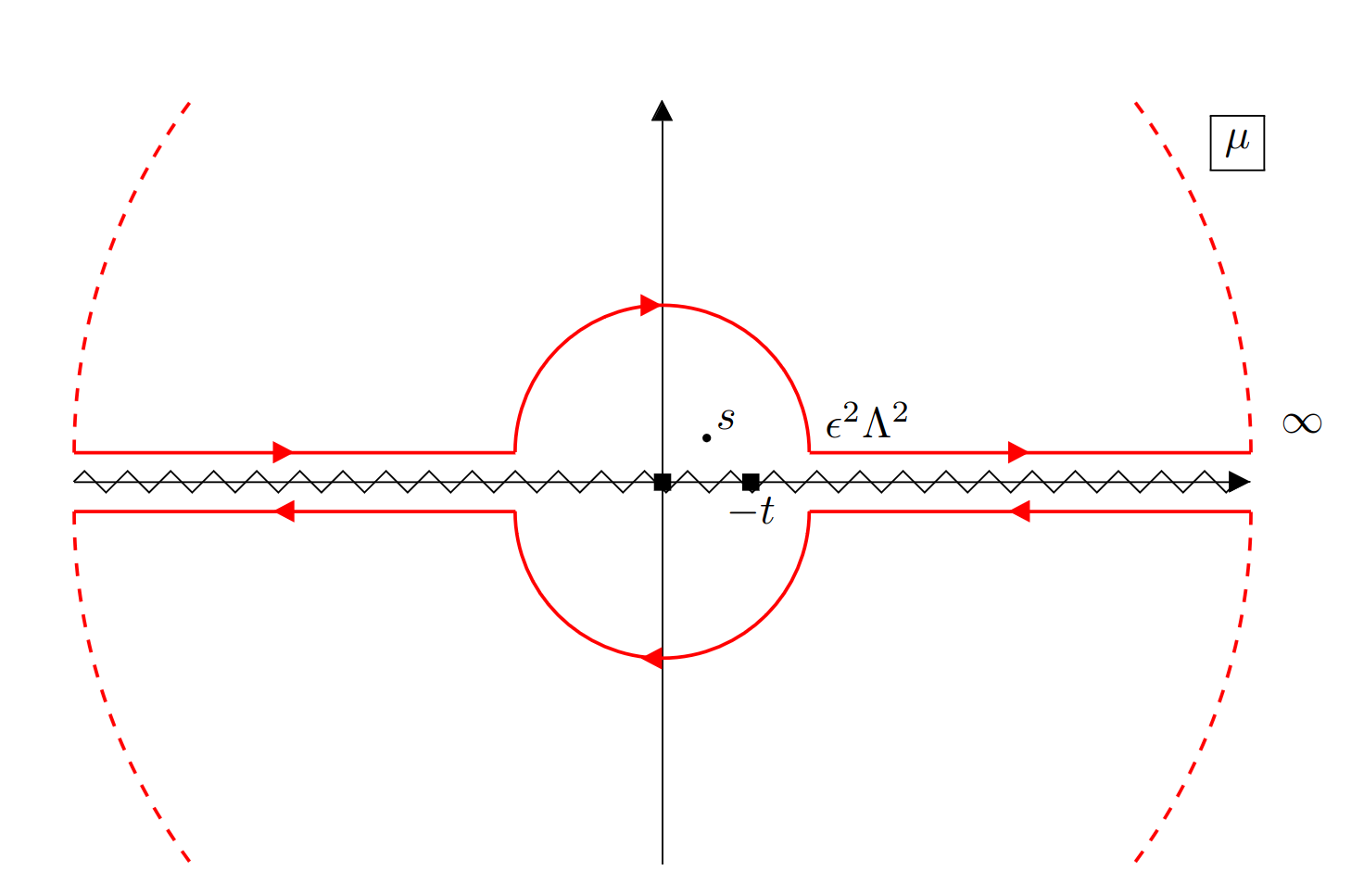}
        \caption{Integration contour used for the derivation of positivity bounds at non-zero $t$.}
        \label{cc}
    \end{figure}

We will also concentrate on the derivation of bounds that are applicable to massless particles. This case is very different from theories with a mass gap, because the two branch cuts in the massless limit at $t\to 0$ tend to merge, splitting the $s$-plane into two disconnected parts. For this reason, computing the residue at the point $\mu$, as we have done in the previous section, does not work. The other way proposed in~\cite{Bellazzini:2020cot} suggests dealing with the arc integrals instead of the single-point residues. The Cauchy theorem applied to the contours in the upper and lower half-plane (marked $+$ and $-$ respectively) in \cref{cc} leads to 
\begin{equation}
    \frac{1}{2\pi i}\left( \oint _{+}+\oint _{-}\right) \rmd\mu \frac{\scatteringamplitude\left( \mu ,t\right) }{\left( \mu -s\right)^{3}}=0 \, .
\end{equation}
If the amplitude is bounded by $s^2$, we can express the arc integrals through the discontinuities of the amplitude,
\begin{equation}
    \text{Disc}_{s} \, \scatteringamplitude\left( s,t\right) =\frac{1}{2i}\left( \scatteringamplitude\left( s+i\varepsilon,t \right) - \scatteringamplitude\left( s-i\varepsilon,t \right) \right) \, .
\end{equation}
Thus, the sum of the upper and lower arcs with the radius chosen to be $\epsilon^2 \UVcutoff^2$ ($\epsilon<1$, $\UVcutoff$ is the cutoff scale) would  be given by
\begin{equation}
\frac{1}{2\pi i}\int _\text{arc}\rmd\mu \frac{\scatteringamplitude\left( \mu ,t\right) }{\left( \mu -s\right)^{3}}=\int^{\infty }_{\varepsilon^{2}\UVcutoff^{2}}\frac{\rmd\mu }{\pi }\frac{\text{Disc}_{s} \, \scatteringamplitude_{s}\left( \mu ,t\right) }{\left( \mu -s\right)^{3}}+\int^{\infty }_{\varepsilon^{2}\UVcutoff^{2}-t}\frac{\rmd\mu }{\pi }\frac{\text{Disc}_{s} \, \scatteringamplitude_{u}\left( \mu ,t\right) }{\left( \mu -u\right)^{3}} \, .
\end{equation}
If we take a tree-level approximation for the amplitude, the arcs can be related to the derivatives of the amplitude at $s=0$:
\begin{equation}
\begin{aligned}
    C\left( n,m\right) &=\left. \frac{1}{2}\frac{n!}{2\pi i}\int _\text{arc}\rmd\mu \frac{\partial _{t}^{m} \scatteringamplitude\left( \mu ,t\right) }{\mu^{n+1}}\right| _{t=0} \\
     &={\left. \frac{1}{2}\partial _{t}^{m}\partial _{s}^{n} \scatteringamplitude\left( s,t\right) \right|} _{s=0,t=0} \, .
\end{aligned}
\end{equation}
We can also introduce the unknown but positive-definite quantities
\begin{equation}
    I_{s,u}\left( n,m\right) = {\left.\int^{\infty }_{\varepsilon^{2}\UVcutoff^{2}}\frac{\rmd\mu }{\pi }\frac{\partial _{t}^{m} \text{Disc}_{s} \, \scatteringamplitude_{s,u}\left( \mu, t\right) }{\mu^{n+1}}\right| }_{t=0} \, .
\end{equation}
The discontinuity of the amplitude, or its imaginary part, can be expanded in partial waves, such that the imaginary parts of all partial waves are positive. In addition, we can use the property of Legendre polynomials which tells us that all derivatives of $P_l(z)$ are positive at $z=1$ (corresponding to $t=0$).\footnote{Note that this is not true for finite negative $t$, hence we cannot trivially apply positive definiteness of discontinuities at finite $t$.} 

Now we need to express the $C(n,m)$, which are directly related to the Wilson coefficients in low energy theory, through positive $I(n,m)>0$. The following expression and its $s$- and $t$-derivatives allow us to derive several relations:
\begin{equation}
\scatteringamplitude\left( s,t\right) =s^{2}\int^{\infty }_{\varepsilon^{2}\UVcutoff^{2}} \rmd\mu \frac{\text{Disc}_s \, \scatteringamplitude\left( \mu ,t\right)}{\mu^{2}\left( \mu -s\right) }+\left( s+t\right)^{2}\int^{\infty }_{\varepsilon^2 \UVcutoff^2} \rmd\mu \frac{\text{Disc}_{u} \, \scatteringamplitude\left( \mu ,t\right)}{\mu^{2}\left( \mu +s+t\right) }+ \left.\frac{\partial \scatteringamplitude(s,t)}{\partial s}\right|_{s=0} s^2 \, .
\end{equation}

To keep the discussion more general, we will not assume full crossing symmetry here, as for spinning particles and for particles of different types, some amplitudes do not have this symmetry. However, they still can be bounded. If the amplitude is $(s - t - u)$-symmetric, $I_u=I_s$, an even number of $s$-derivatives is always zero. Here are several relations between $C$ and $I$:
\begin{align}
 C\left( 2,0\right) &= I_{s}\left( 2,0\right) +I_{u}\left( 2,0\right) \, , \\
 C\left( 3,0\right) &= 3I_{s}\left( 3,0\right) -3I_{u}\left( 3,0\right) \, , \\
 C\left( 2,1\right) &=\frac{\text{Disc}_{s} \, \scatteringamplitude_u\left( \varepsilon^{2}\UVcutoff^{2},0\right) }{\pi \left( \varepsilon^{2}\UVcutoff^{2}\right)^{3}}+I_{s}\left( 2,1\right) -3I_{u}\left( 3,0\right) +I_{u}\left( 2,1\right) \, .
\end{align}
From the first equation, it is easy to derive that $C(2,0)>0$, which is the same bound as obtained in the previous section. We can see that already for $C(3,0)$ or $C(2,1)$, we have negative terms on the right-hand side, so we cannot immediately conclude anything about their sign. However, we can compensate for the negative terms by making use of the trivial inequality
\begin{equation}
    \frac{1}{\mu } <\frac{1}{\varepsilon^{2}\UVcutoff^{2}}\, , \qquad \mu  >\varepsilon^{2}\UVcutoff^{2} \, ,
\end{equation}
implying
\begin{equation}
    \frac{\partial _{t}^{m}\text{Disc}_{s} \, \scatteringamplitude_{s,u}}{\mu^{n+1}} <\frac{\partial _{t}^{m}\text{Disc}_{s} \, \scatteringamplitude_{s,u}}{\varepsilon^{2}\UVcutoff^{2}\mu^{n}} \, , \qquad I_{s,u}\left( n+1,m\right)  <\frac{1}{\varepsilon^{2}\UVcutoff^{2}}I_{s,u}\left( n,m\right) \, .
\end{equation}
Now we are in a position to derive a chain of inequalities for the Wilson coefficients. For example,
\begin{align}
    C\left( 2,1\right) +\frac{3}{2\varepsilon^2\UVcutoff^2}C\left( 2,0\right) -\frac{1}{2}C\left( 3,0\right)  &>0 \, , \\
    \frac{12}{\left( \varepsilon^{2}\UVcutoff^{2}\right)^{2}}C\left( 2,0\right) > C\left( 4,0\right) &> 0 \, , \\
    C\left( 3,1\right) +\frac{3}{\varepsilon^{2}\UVcutoff^{2}}C\left( 2,1\right) -\frac{3}{2 \varepsilon^{2}\UVcutoff^{2}}C\left( 3,0\right) +\frac{9 C\left( 2,0\right) }{2\left( \varepsilon^{2}\UVcutoff^{2}\right)^{2}} &>0 \, .
\end{align}
The general form of the inequalities of this type can be found in~\cite{deRham:2017avq}. The most optimal bounds may require building linear combinations of several inequalities with coefficients being dependent on the Wilson coefficients.

\subsubsection{Non-linear bounds}

Making use of several types of integral inequalities, one can derive bounds on non-linear combinations of Wilson coefficients. For example, using the Cauchy-Schwarz-Buniakowsky integral inequality,
\begin{equation}
    \left( \int^{b}_{a} \rmd x \, f\left(  x\right)g\left( x\right) \right)^{2}\leq \left( \int^{b}_{a} \rmd x \, f^{2}\left( x\right) \right) \left( \int^{b}_{a} \rmd x \, g^{2}\left( x\right) \right) \, ,
\end{equation}
we can derive
\begin{equation}
    I_{s,u}\left( 3,0\right)^{2} <I_{s,u}\left( 2,0\right) I_{s,u}\left( 4,0\right) \, .
\end{equation}
These integrals can be directly expressed through Wilson coefficients,
\begin{equation}
    \frac{4}{3}C\left( 3,0\right)^{2} <C\left( 2,0\right) C\left( 4,0\right) \, .
\end{equation}
In general, the Wilson coefficient $C(n,0)$ can be squeezed between $C(n+1,0)$ and $C(n-1,0)$, or between $C(n+2,0)$ and $C(n-2,0)$. The latter is important for fully crossing-symmetric amplitudes when only the $C(2n,0)$ are non-zero.

A similar technique can be applied to the Wilson coefficients including $t$-derivatives. This way, we can derive a bound on $C(2,1)$,
\begin{equation}
    C\left( 2,1\right) -\frac{1}{2}C\left( 3,0\right) =I_{s}\left( 2,1\right) +I_{u}\left( 2,1\right) -\frac{3}{2}\left( I_{s}\left( 3,0\right) +I_{u}\left( 3,0\right) \right) \, .
\end{equation}
We can constrain the negative part using the Cauchy-Schwartz inequality,
\begin{equation}
    ( I_{s}\left( 3,0\right) +I_{u}\left( 3,0\right))^{2} < \left( I_{s}\left(2,0\right) +I_{u}\left( 2,0\right) \right) \left( I_{s}\left( 4,0\right) +I_{u}\left( 4,0\right) \right) \, .
\end{equation}
Thus, we finally get a bound on $C(2,1)$, reproducing the coefficient in front of the $s t u$-term in the amplitude,
\begin{equation}
    C\left( 2,1\right) -\frac{1}{2}C\left( 3,0\right) +\frac{\sqrt{3}}{4}\sqrt{C\left( 2,0\right) C\left( 4,0\right) } >0 \, .
\end{equation}
This can be straightforwardly generalized to obtaining bounds on $C(2n,0)$ in the form
\begin{equation}
    C(2n,0)^2<\alpha(n) C(2n+2,0)C(2n-2,0) \, .
\end{equation}
These bounds actually forbid a hierarchy between certain Wilson coefficients contributing to $C(2n,0)$, squeezing them between higher and lower order ones. 

Obtaining these bounds can be formalized as a math problem from the theory of moments. The state-of-the-art technique applying these mathematical results to positivity bounds for tree-level \acp{EFT} is described in~\cite{Chiang:2022ltp}.

Summarizing, at tree-level approximation in the \ac{EFT} without gravity (we postpone the discussion of the problems caused by the graviton exchange pole to \cref{sec:regge}), one can obtain a set of two-sided constraints on the parameter space of Wilson coefficients. One of these coefficients actually sets the cutoff scale of the \ac{EFT}, thus after fixing this scale, all parameters live in a compact area of a multidimensional space of the \ac{EFT} parameters.

\subsubsection{Loop corrections}

Naively, one could expect that loop corrections should be small in a weakly-coupled theory, thus they should not affect the procedure of obtaining positivity bounds too much. However, here we will show that this is not always the case. 

Let us take the amplitude computed in \cref{bootstrap} in the limit $t\to 0$,
\begin{equation}
    \scatteringamplitude=2 g_2 s^{2}+2 g_4s ^{4}+\beta s^{4}\left( \ln s+\ln \left( -s\right) \right) +2 g_6 s^{6}+\dots \, .
\end{equation}
We can compute the arc integral over a circle with radius $\epsilon^2\UVcutoff^2$:
\begin{equation}
    C\left( 4,0\right) =\int _\text{arc}\frac{\rmd\mu \scatteringamplitude\left( \mu ,0\right) }{\mu^{5}}=2g_2+4 g_4\ln \left( \varepsilon^{2}\UVcutoff^{2}\right) \, .
\end{equation}
The arc integral is no longer the same as the $s$-derivative of the amplitude, due to the logarithm. In addition, one cannot take the limit $\epsilon\to 0$. In fact, the arc integral $C(4,0)$ probes the Wilson coefficient in front of the $s^4$ term at the scale $\epsilon \UVcutoff$. The next arc integral,
\begin{equation}
    C\left( 6,0\right) =2 g_6-\frac{\beta }{\varepsilon^{4}}\UVcutoff^{4} \, ,
\end{equation}
would have a $1/\varepsilon^4$ contribution which is dominant for small $\varepsilon$. This way, one can get rigorous bounds on beta functions. 

The Wilson coefficient $c$ can be also bounded if $\epsilon$ is taken to be close to $1$. However, this is possible only under the assumption of weak coupling, such that the term $\beta$, as well as the other terms in the amplitude ($s^8$, $s^{10}$, $\dots$), are also suppressed. In this sense, using arcs with $\epsilon\approx 1$ reproduces the tree-level bounds under the extra assumption of weak coupling. More specifically, these bounds can only probe Wilson coefficients around the cutoff scale.

So far, we considered only bounds without $t$-derivatives. The addition of $t$-dependent terms leads to more complications at loop level in massless theories. For example, the amplitude \eqref{1loop} contains terms which make $t$-derivatives of certain arcs divergent. For example, for
\begin{equation}
    \frac{21g_2^{2}}{240\pi^{2}}t^{2}\left( t^{2}+\frac{1}{21}s\left( s+t\right) \right) \ln t \, ,
\end{equation}
we get
\begin{equation}
    C\left( 2,2\right) \sim \ln t \, .
\end{equation}
This value is infinite when $t\to 0$. For this reason, $t$-derivatives can lead to extra \ac{IR} divergences, preventing us from obtaining some bounds based on linear and non-linear inequalities. This problem can be avoided if one uses dispersion relations at finite $t$ and numerical optimization, as it was proposed in \cite{Bellazzini:2021oaj,Beadle:2024hqg,Bertucci:2024qzt}. The recent application of these methods for loop-level bounds with gravity can be found in \cite{Chang:2025cxc,Beadle:2025cdx}. Analytic results, based on the construction of those combinations of the arcs and their $t$-derivatives which are finite at $t\rightarrow -0$ were also obtained in \cite{Peng:2025klv}.

Will the conclusion about the compact nature of the bounds on \ac{EFT} coefficients hold at the loop level? The answer to this question seems to be ``yes'' because even with \ac{IR} divergences coming from loops, one can still constrain the Wilson coefficients through their beta functions. For this reason, even though the bounds at the loop level can significantly change compared to the tree level ones if weak coupling is not additionally assumed, they still lead to allowed regions that are compact in the parameter space.

\subsection{Regge bounds on graviton-mediated scattering}\label{sec:regge}

The Martin-Froissart bound was derived for the scattering of massive states. To what extent can it be valid for an \ac{EFT} of gravity? As we have seen, all positivity bound techniques crucially depend on these properties. In this section, we show the derivation of bounds on high-energy scattering through gravitons based on the unitarization of \ac{GR} scattering amplitude. This can be done as a resummation of ``improved'' partial waves saturating the elastic scattering bound --- eikonal resummation. This method works well in dimensions higher than five, where elastic scattering is a good approximation for large spin partial waves~\cite{Haring:2022cyf}. We will also discuss the issues that arise in four dimensions with this approach.

\subsubsection{Eikonal resummation in \texorpdfstring{$d$}{d} dimensions}

Recall that in $d$ dimensions, the amplitude can be expressed in partial waves as
\begin{equation}
    \scatteringamplitude\left( s,t\right) =\frac{1}{2}\sum_{l=0}^{\infty} n\left( l,d\right) f_{l}\left( s\right) P_{l}^{(d)}\left( 1+\frac{2t}{s-4m^{2}}\right) \, .
\end{equation}
Here, the partial wave amplitudes can be found as
\begin{equation}
    f_{l}\left( s\right) =N\left( d\right) \int _{-1}^1\rmd x \, \left( 1-x^{2}\right)^{\frac{d-4}{2}}P_{l}^{\left( d\right) }\left( x\right) \scatteringamplitude\left( s,t\left( x\right) \right) \, .
\end{equation}
Recall that the normalization constants are
\begin{equation}
    N \left( d\right) =\frac{\left( 16\pi \right)^{\frac{2-d}{2}}}{\Gamma \left( \frac{d-2}{2}\right) } \, , \qquad n\left( l,d\right) =\frac{\left( 4\pi \right)^{\frac{d}{2}}\left( d+2l-3\right) \Gamma \left( d+l-3\right) }{\pi \Gamma \left( \frac{d-2}{2}\right) \Gamma \left( l+1\right) } \, .
\end{equation}
The non-linear \ac{FU} condition in $d$ dimensions reads
\begin{equation}
    2\,{\rm Im}\,f_{l}\left( s\right) \geq \frac{\left( s-4m^{2}\right)^{\frac{d-3}{2}}}{\sqrt{s}}\left| f_{l}\left( s\right) \right|^{2} \, .
\end{equation}
This unitarity condition is saturated if 
\begin{equation}
\label{eikonal}
    f_{l}\left( s\right) =\frac{\sqrt{s}}{\left( s-4m^{2}\right)^{\frac{d-3}{2}}}i\left( 1-e^{2i\delta _{l}\left( s\right) }\right) \, ,
\end{equation}
where $\delta_l$ is a real-valued phase shift of the amplitude. 

If the amplitude is a sum of partial waves constructed as in \eqref{eikonal}, it automatically satisfies all \ac{PWU} conditions. This observation provides a hint to searching for non-perturbative amplitudes with good properties without any reference to the field theory that produces it. The possible procedure is as follows.
\begin{itemize}
    \item Take a tree-level \ac{EFT} amplitude which is unitary only at low energies.
    \item Decompose it into partial waves.
    \item Take the tree-level partial wave amplitudes and promote them to the phase shift of the partial waves of non-perturbative amplitudes which will be unitary at all energy scales.
    \item Perform a resummation of these new unitarized partial waves.
\end{itemize}

This method of the unitarization of the amplitude interplays with the eikonal limit in quantum mechanics. In this limit, quasi-classical approximation works well, and the full answer can be obtained from the exponentiation of partial waves. This works for graviton-mediated scattering because, at large distances between scattering particles (large impact parameter), gravity is almost classical, nevertheless the center-of-mass energy can be very high and can even exceed the Planck scale. 

The whole procedure of computing an infinite sum can be technically complicated. For this reason, if we expect the main contribution to the amplitude to be given by large values of $l$ rather than small $l$, we can try to approximate this sum with respect to discrete values of spins by the integral with respect to a proper continuous parameter.

 \begin{figure}[ht]
        \centering
        \includegraphics[width=2in]{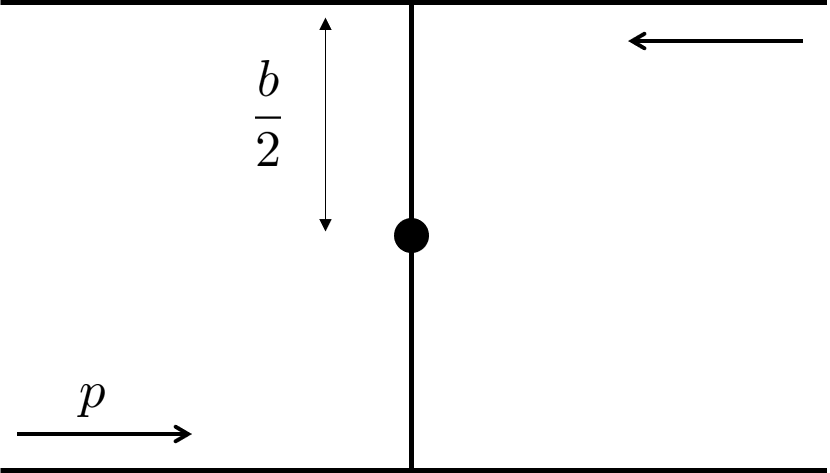}
        \caption{Meaning of the impact parameter in the center-of-mass frame.}
        \label{J}
    \end{figure}
    
In order to perform a resummation of large-$l$ partial waves, it is more convenient to use the impact parameter representation of the amplitude, see \cref{J}. The angular momentum vector is related to the impact parameter as
\begin{equation}
    |\vec{l}|=|\vec{b}\times \vec{p}|=\frac{b\sqrt{s-4m^{2}}}{2}=\frac{b\sqrt{s}}{2} \, , \qquad b=\frac{2 l}{\sqrt{s}} \, .
\end{equation}
At large $l$, the partial wave expansion of the amplitude can be directly transformed into the impact parameter representation. Indeed, the asymptotics of Legendre polynomials at large $l$ imply (here we define $t=-q^2<0$)
\begin{equation}
    \frac{1}{l^{d-3}}n\left( l,d\right) P_{l}^{ (d) }\left( 1-\frac{q^{2}b^{2}}{2l^{2}}\right) \approx 2^{d}\left( 2\pi \right)^{\frac{d-2}{2}}\left( bq\right)^{\frac{4-d}{2}}J_{\frac{d-4}{2}}\left( bq\right) \, .
\end{equation}
The sum over momenta $l$ can be substituted by an integral over the impact parameter $b$,
\begin{equation}
\sum_{l}\to \frac{\sqrt{s}}{2}\int \rmd b \, .
\end{equation}
For graviton-mediated scattering, one can find $\delta_\text{tree}(s,b)$ from the partial waves of the tree-level amplitude,
\begin{align}
   f_l(s)\to \delta \left( s,b\right) =\delta _\text{tree}\left( s,b\right) =\frac{\Gamma \left( \frac{d-4}{2}\right) \GN{}\,s}{\left( \pi \right)^{\frac{d-4}{2}}b^{d-4}}\, , \qquad b=\frac{2 l}{\sqrt{s}} \, .
\end{align}
The improved amplitude can be divided into two parts corresponding to low and high spins (or small and large impact parameters),
\begin{equation}
    \scatteringamplitude = \scatteringamplitude_\text{low} + \scatteringamplitude_\text{eik} \, .
\end{equation}
We will constrain the low-spin contribution from \ac{PWU}, and the high-spin part from eikonal resummation of the exponentiated partial waves. Using the integral instead of the sum, we get
\begin{equation}
    \scatteringamplitude_\text{eik}\left( s,t\right) =2is\left( 2\pi \right)^{\frac{d-2}{2}}\int^{\infty }_{b^{\ast }}\rmd b \, b^{d-3}\left( bq\right)^{\frac{4-d}{2}}J_{\frac{4-d}{2}}\left( bq\right) \left( 1-e^{2i\delta _\text{tree}\left( s,b\right) }\right) \, .
\end{equation}
This expression has been shown to be valid for $b>b^{\ast}\propto s^{\frac{1}{d-2}}$~\cite{Amati:1987uf}. At lower $b$, inelastic effects, such as the emission of several gravitons, become important. Despite our lack of knowledge about this regime, we can still bound the contribution of low spins $l<l^{\ast}=b^{\ast}\sqrt{s}/2$ from \ac{PWU},
\begin{equation}
    \left| f_{l}\left( s\right) \right|  <s^{2-\frac{d}{2}} \, .
\end{equation}
The Legendre polynomials have the following property which can be checked numerically for large $d$ (to the best of our knowledge, no analytic proof exists for the general case),
\begin{equation}
    \left| P_{l}^{ (d) }\left( 1-\frac{2q^{2}}{s}\right) \right|  <\left( \frac{q\,l}{\sqrt{s}}\right)^{\frac{3-d}{2}} \, .
\end{equation}
Thus, the low spin contribution can be estimated as
\begin{equation}
    \left| \sum^{l^{\ast }}_{l=0}n\left( l,d\right) f_{l}\left( s\right) P  _{l}^{\left(d\right)}\left( 1-\frac{2q^{2}}{s}\right) \right|  <\sum^{l^{\ast}}_{l=0}n\left( l,d\right) s^{2-\frac{d}{2}}\left( ql\right)^{\frac{3-d}{2}}s^{\frac{d-3}{4}} \, ,
\end{equation}
with
\begin{equation}
    b^{\ast }=s^{\frac{1}{d-2}},~l^{\ast }=\frac{b^{\ast }\sqrt{s}}{2}=s^{\frac{d}{2(d-2)}} \, .
\end{equation}
We get the low-spin contribution to be bounded by
\begin{equation}
    \left| \scatteringamplitude_\text{low}\right|  <s^{\frac{5-d}{4}}\left( l^{\ast }\right)^{\frac{d-1}{2}}=s^{2-\frac{d-3}{ 2(d-2) }}<s^2 \, .
\end{equation}
Thus, in higher dimensions, this bound still allows to use twice-subtracted dispersion relations.

\subsubsection{Derivation of Regge bounds in \texorpdfstring{$d>4$}{d larger 4}}

Here we will evaluate the large-$l$ contribution encoded in $\scatteringamplitude_\text{eik}$ assuming elastic scattering domination (which must be the case for \ac{GR} at large distances). Namely, we assume that $2\to 2$ processes are dominating over the other contributions to the imaginary parts of partial waves. In other words, we assume the saturation of the \ac{FU} condition. As we are dealing with an oscillatory function under the integral, we can use a stationary phase approximation,
\begin{equation}
    \int \rmd x \,  g\left( x\right) e^{if\left( x\right) }=g\left( x_{0}\right) e^{if\left( x_{0}\right) }e^{i\frac{\pi }{4}\text{sign}(f''\left( x_{0}\right)) }\sqrt{\frac{2\pi }{f''\left( x_{0}\right) }} \, .
\end{equation}
We also use that for large $b$,
\begin{equation}
    J_{\frac{d-4}{2}}\left( bq\right) =\sqrt{\frac{2}{\pi bq}}\cos \left( bq+\pi \frac{d-5}{4}\right) \, .
\end{equation}
Thus, the stationary point of the eikonal integral can be found as a solution to
\begin{equation}
    \frac{\partial }{\partial b}\left( \pm bq+\delta _\text{tree}\left( b\right) \right) =0 \, .
\end{equation}
Recall that here
\begin{equation}
    \delta _\text{tree }=\alpha s b^{4-d} \, , \qquad \alpha=\text{const.} \, ,
\end{equation}
so that
\begin{equation}
    \pm q-\left( d-4\right) \alpha s b^{3-d}=0 \, , \qquad b >0 \, .
\end{equation}
Thus, we can find the stationary point $b_0$ around which the integral receives the main contribution,
\begin{equation}
    b_{0}=\left( \frac{q}{\left(d-4\right) \alpha s}\right)^{\frac{1}{3-d}}\propto s^{\frac{1}{d-3}} \, .
\end{equation}
Taking an integral around this point, we get
\begin{equation}
    \scatteringamplitude_\text{eik}=-2is\left( 2\pi \right)^{\frac{d-2}{2}}b_{0}^{d-3}\left( b_{0}q\right)^{\frac{4-d}{2}}\sqrt{\frac{2}{\pi b_{0}q}}\frac{1}{2}e^{i\left( qb_0+\alpha sb_{0}^{4-d}\right) }\sqrt{ {\frac{2\pi }{\alpha b_{0}^{2-d}s}}}\frac{e^{i\pi\left( \frac{d}{4}-1\right) }}{\sqrt{(d-4)\left( d-3\right) }} \, ,
\end{equation}
meaning that
\begin{tcolorbox}
\begin{equation}
    \scatteringamplitude_\text{eik}\propto e^{i\zeta q^\frac{4-d}{3-d}s^\frac{1}{d-3}}s^{2-\frac{d-4}{2\left( d-3\right) }}q^{\frac{(d-2)^{2}}{2(3-d)}} \, .
\end{equation}
\end{tcolorbox}
From this result, we can see that if $d>4$, the property of being bounded by $s^2$ is justified. In addition, the power of $s$ in the exponent is less than $1/2$, which means that the exponent is bounded on the first (physical) sheet of the complex plane. For the formal limit $d\to 4$, none of these nice properties remain, which makes it a much more complicated case to study. In fact, the eikonal approximation may be invalid in $d=4$, as the emission of many soft gravitons is known to be a less suppressed process compared to $2\to 2$ scattering in four dimensions.

\subsubsection{Regge behavior from dispersion relations in \texorpdfstring{$d=4$}{4d}}

Even though it is not obvious that the Regge bound on the scattering with graviton exchange holds in $d=4$, one can still assume that $|\scatteringamplitude(s,0)|<s^2$, which naively seems to be enough to justify the use of positivity bounds based on twice-subtracted dispersion relations. In fact, this corresponds to the approximation that all scattering amplitudes (including those involving gravitons at large $s$) are dominated only by low spin contributions. It is often assumed in the literature, and it leads to the same positivity bounds as those based on the Martin-Froissart bound. However, it seems that it is not possible to derive such a behavior in four dimensions from first principles, so it can be treated as an extra assumption about the \ac{UV} theory.

We can start from the assumption 
\begin{equation}
    |\scatteringamplitude(s,t)|<s^2
\end{equation}
at fixed small $t$. The dispersion relation will lead to a certain connection between the \ac{IR} asymptotics of the amplitude and its \ac{UV} behavior. The amplitude with the one-loop graviton correction in four dimensions will have the form~\cite{Herrero-Valea:2020wxz}
\begin{equation}
    \scatteringamplitude\left( s,t\right) =A_{0}\frac{s^{2}}{t}+A_{1}s^{2}\ln \left( \frac{-t}{\mu^{2}}\right) \, .
\end{equation}
\begin{figure}[ht]
        \centering
        \includegraphics[width=4in]{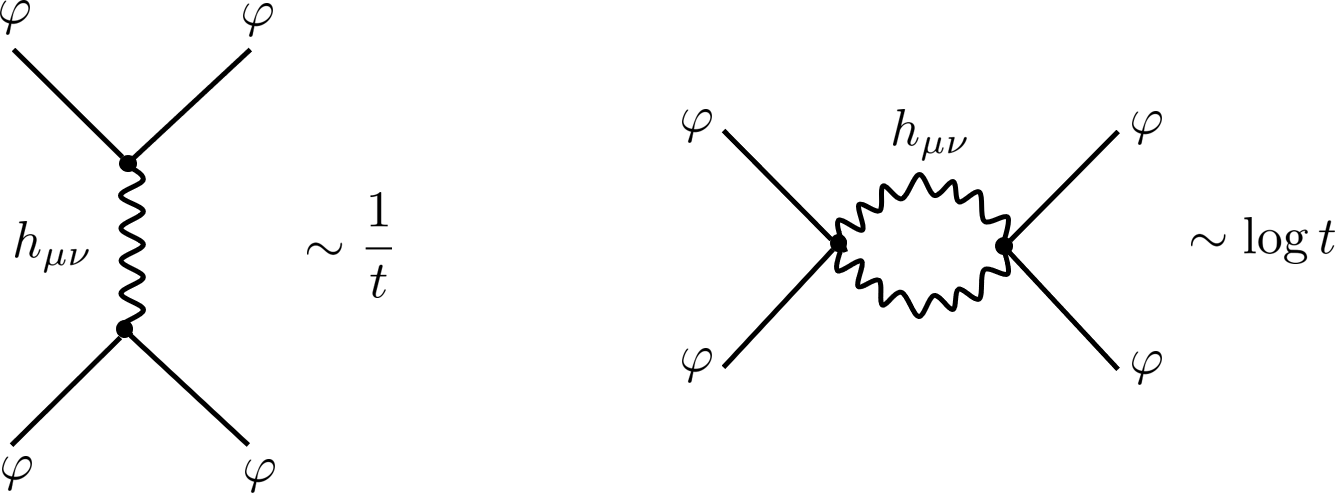}
        \caption{Diagrams for scalar scattering with tree-level (left) and one-loop (right) graviton exchange, respectively.}
        \label{g-loop}
    \end{figure}
Indeed, the presence of $s^2/t$- and $s^2\ln{t}$-terms in the \ac{IR} leads to a divergence in the twice-subtracted dispersion relation,
\begin{equation}
    C(2,0)=\frac{1}{2}\left(\frac{A_0}{t}+A_1\ln{t}+\dots\right)=\int_{\epsilon^2\UVcutoff^2}^{\infty} \rmd\mu \, \frac{{\rm Im}\, \scatteringamplitude(\mu, t)}{\mu^3} \, .
\end{equation}
Suppose we assume that the imaginary part of the amplitude is a regular function at $t=0$. In that case, the divergences on the left-hand side can be reproduced only by the infinite tail of the integral on the right-hand side. Namely,
\begin{equation}\label{inteq}
 \int^{\infty }_{M^{\ast }} \rmd\mu \, \frac{{\rm Im}\, \scatteringamplitude\left( s,t\right) }{\mu^{3}} =\frac{A_{0}}{t}+A_{1}\ln t+\text{(finite)} \, . 
\end{equation}
Here $M^{\ast2}>\UVcutoff^2$ corresponds to the scale after which the main contribution to \ac{IR} divergences is provided. This can be interpreted as an integral equation for the function $\text{Im} \, \scatteringamplitude(s,t)$. What is the general form of the solution to this equation? One of the possible solutions is given by a linear Regge trajectory~\cite{Tokuda:2020mlf}, namely,
\begin{equation}
\label{regge1}
    {\rm Im}\,\scatteringamplitude=s^{2+\alpha t} \, .
\end{equation}
It can be checked by direct substitution that this form reproduces the graviton pole. Is it the unique solution? We can search for other possibilities by generalizing it to
\begin{equation}
\label{regge}
    {\rm Im}\,\scatteringamplitude=s^{2+\alpha t}\varphi \left( s,t\right) \, .
\end{equation}
Here, we assume that $\varphi(s,t)$ is a regular function of $t$ around $t=0$,
 \begin{equation}
     \varphi \left( s,t\right) =\varphi \left( s,0\right) +\varphi _{t}\left( s,0\right) t+\frac{1}{2}\varphi _{tt}\left( s,0\right) t^2+\dots \, .
 \end{equation}
For further convenience, we introduce a new variable $\sigma$, such that
\begin{equation}
  s=M^{2}e^{\sigma } \, ,
\end{equation}
where $M$ is the scale at which the discontinuity of the amplitude is well described by \eqref{regge}, rather than by an \ac{EFT} expansion.
As the \ac{IR} divergences are determined only by the \ac{UV} part of the integral, we can set the integration limit in \eqref{inteq} to be zero instead of $M^{\ast}$. After that, we have
\begin{equation}
    \int^{\infty }_{0}\frac{\rmd\mu }{\mu } \,2\varphi \left( \mu ,0\right) \mu^{\alpha t}=\int^{\infty }_{0} \rmd\sigma \, 2M^{2\alpha t}\varphi \left( \sigma ,0\right) e^{\alpha \sigma t} \, ,
\end{equation}
or
\begin{equation}
  2M^{2\alpha t} L\left[ \varphi \left( \sigma ,0\right) \right] =f\left( t\right) +\orderneglected\left( t\right) \, ,
\end{equation}
where $L[\varphi \left( \sigma ,0\right)]$ stands for the Laplace transform with respect to the variable $\sigma$.

Thus, we see that we got a relation between the leading \ac{IR} singularity and the \ac{UV} asymptotics of the correction to the amplitude $\varphi(\sigma, t)$. Recall that we will get $\varphi(\sigma, 0)=1$ for the $1/t$-pole, hence the Regge form \eqref{regge} is a relevant solution to the equation \eqref{inteq}~\cite{Tokuda:2020mlf}. We can also find the solution when the logarithmic correction is included~\cite{Herrero-Valea:2020wxz}. In this case, we have 
 \begin{equation}
   {\rm Im}\,\scatteringamplitude=s^{2+\alpha t}\left( 1+\frac{1}{\ln s}\right) \, . 
 \end{equation}
The relation between \ac{IR} singularities and the \ac{UV} amplitude through the Laplace transform allows us to get a solution for the \ac{UV} amplitude as an inverse Laplace transform. In many cases, this can be done analytically with the use of known results. We can compute $\varphi$ and its derivatives as 
\begin{equation}
    \varphi \left( \sigma ,0\right) =a_{0}L^{-1}\left[ f\left( t\right) \right] \, , ~~\varphi _{t}\left( \sigma ,0\right) =a_{2}L^{-1}\left[ \frac{f\left( t\right) }{t}\right] \, ,~~ \varphi _{tt}\left( \sigma ,0\right) =a_{2}L^{-1}\left[ \frac{f\left( t\right) }{t^{2}}\right], \, \dots \, .
\end{equation}
Thus, the general solution to \eqref{inteq} for a $1/t$-divergence has the form
 \begin{equation}
     \varphi \left( \sigma ,t\right) =\sum _{n}a_{n}\sigma^{n}t^{n}=\varphi \left( t\ln s\right) \, , \qquad \varphi \left( 0\right) \neq 0 \, .
 \end{equation}
From this expansion, we can conclude that
\begin{tcolorbox}
\textit{The linear Regge trajectory \eqref{regge1} is a good approximation if $t\ln{s}\to 0$ while $t\to 0$ and $s\to \infty$.}
\end{tcolorbox}
Using analyticity, we can reconstruct the whole amplitude from its imaginary part~\cite{Herrero-Valea:2022lfd},
 \begin{equation}
     \scatteringamplitude\left( s,t\right) =-\frac{a_{0}e^{-i\pi \alpha t}}{\sin \left( \pi \alpha t\right) }\left( s^{2+\alpha t}+\left( -s-t\right)^{2+\alpha t}\right) + \orderneglected\left( t\ln s\right) \, .
 \end{equation}
This result is a good approximation for the whole reconstructed amplitude in the above-mentioned specific limit $t\to 0$, $s\to \infty$, with $t \ln s\to 0$. It is also important to state that this result was obtained under the assumption that the imaginary part of the amplitude can be expanded in a Taylor series in $t$. Remarkably, string amplitudes have exactly this structure in the Regge limit, being an expansion in a combination of $s\ln t$!

\subsubsection{No positivity bounds for \texorpdfstring{$C(2,0)$}{C(2,0)}?}

Consider the amplitude of the form
\begin{equation}
    \scatteringamplitude\left( s,t\right) =-\frac{s^{2}}{\MPl^{2}t}-\gamma s^{2} \, , \qquad \gamma  >0 \, ,
\end{equation}
with a negative coefficient in front of the $s^2$-term. We can find the form of the amplitude in the \ac{UV} such that it reproduces these structures in the \ac{IR}. Namely, we can write
\begin{equation}
  \scatteringamplitude_\text{UV}\left( s,t\right) =e^{-i\pi \alpha t}\left( -\gamma-\frac{\pi^{2}}{\MPl^{2}\sin \pi \alpha t}\right) \left( s^{2+\alpha t}+u^{2+\alpha t}\right) \, , 
\end{equation}
so that
\begin{equation}
    {\rm Im}\, \scatteringamplitude_\text{UV}=\gamma s^{2+\alpha t}\sin \pi \alpha t=\gamma \left( \pi \alpha t\right) s^{2+\alpha t}
    =\orderneglected\left( t\ln s\right) \, .
\end{equation}
Thus, an arbitrarily negative constant can be explained by the proper sub-leading correction to the linear Regge trajectory. However, the term reproducing $\gamma$ is sub-dominant, compared to the expansion in $t \ln s$ which we derived. For this reason, in the $t\to 0$ limit, we have a possibility to hide the arbitrary constant in the sub-leading correction to the Regge trajectory. Thus, it does not contradict any of the fundamental principles.

It is also known that negative contributions to the $s^2$-term appear if loops of light particles are taken into account~\cite{Alberte:2020bdz, Alberte:2021dnj, deRham:2022gfe}. For example, if one integrates out the electron in \ac{QED}, one gets a negative $s^2$-term. There was an intense discussion in the literature whether this contradicts positivity bounds, but this is just a reflection of the fact that the graviton
pole cannot be simply eliminated from the amplitude if light or massless states are present in the theory. In this case, the justification of twice-subtracted bounds requires extra assumptions, like low spin domination in the amplitude.

Even if we relax the bounds based on the twice-subtracted dispersion relation, we still have compact bounds on Wilson coefficients entering higher-order subtracted relations. In addition, the coefficient $\gamma$ will contribute to beta functions of higher-order operators, and thus one can still get two-sided constraints on its value.

\subsection{Conclusions}\label{sect:ANNA-conclusions}

This section briefly introduces and reviews selected topics in the \ac{EFT} approach to gravity and its relation to the scattering amplitudes. \ac{EFT} is an indispensable tool for the description of all phenomena emerging at low energies within a limited framework which is, however, much simpler than a complete theory even if it is known. In the case of \ac{QG}, the \ac{UV} completion is not even known, and can well be not a field theory at all. For this reason, the \ac{EFT} expansion of the action for gravity is a unique framework for addressing many questions at energies much smaller than the Planck scale. The examples of questions that can be resolved without getting into a complete theory include early and late-time cosmology and astrophysical \acp{BH} (see \cref{sec:FRANCESCO}). In this case, the results do not depend on a particular form of \ac{UV} completion. 

However, as it is pointed out in \cref{sec:IVANO}, several effects related to \ac{BH} entropy are mixing \ac{IR} theory and \ac{UV} completion. In this section, a systematic framework allowing to relate \ac{UV} and \ac{IR} theories is built, based on the properties of scattering amplitudes and \ac{QFT} axioms. Namely, the requirements of unitarity, Lorentz symmetry, causality, and locality of the complete fundamental theory of nature imply constraints on \ac{EFT} parameters, as well as the relations between \ac{UV} and \ac{IR} limits of the non-perturbative S-matrix. In this section, the concept of \ac{EFT} is introduced with a focus on its relation to the scattering amplitudes. Further results, such as positivity bounds, are derived with the use of the language of the S-matrix formalism, where desired and reasonable requirements for any healthy \ac{QFT} are conveniently formulated as mathematical properties of the amplitudes being functions of complexified momenta. This way, their analytic properties relate \ac{UV} and \ac{IR} theories in a very non-trivial way. 

Summarizing, not all \acp{EFT} are allowed by the fundamental principles of \ac{QFT}. There are many ways to obtain more constraints and relations between the \ac{IR} theory and the desired \ac{UV} completion. At this moment, S-matrix constraints are a quickly developing research field, and many new results will help to shed light on relations between low- and high-energy theories, as well as on the properties of the complete quantum theory of gravity. Perhaps, in the near future there will be more ways and methods to distinguish between different \ac{QG} proposals, prove (or disprove) the uniqueness of \ac{ST} along the lines of~\cite{Guerrieri:2021ivu, Guerrieri:2022sod} (cf. \cref{sec:IVANO}), as well as test the \ac{ASQG} framework (cf. \cref{sec:ALESSIABENJAMIN}) as a field theory-based formulation of \ac{QG}.

\begin{subappendices}

\subsection{Vocabulary of EFT and amplitudes: definitions of concepts}\label{sect:ANNA-dictionary}

\begin{itemize}

\item \hypertarget{DICT_ANNA_UNITARITY}{\textbf{Unitarity}}
    
Unitarity is a basic concept of quantum mechanics and \ac{QFT} encoding the requirement that all probabilities should be positive and sum up to unity for a complete system. This translates into the fact that the evolution of the quantum states is always described by a unitary operator. In \ac{QFT}, this requirement can be formulated for the S-matrix, stating that 
\begin{equation}
   S^{\dagger}S=1 \, . 
\end{equation}
In addition, in conventional approaches to \ac{QFT}, all states must be quantified with a positive norm. 
    
    \item \hypertarget{DICT_ANNA_PWU}{\textbf{Partial wave unitarity (\ac{PWU})}}

    Unitarity of the S-matrix applied to a $2\to2$ process can be formulated for the scattering of states with fixed angular momentum in the center-of-mass frame. This means that the scattering amplitude, being a function of the scattering angle, can be decomposed in terms of eigenfunctions of the angular momentum operator (Legendre polynomials in four dimensions for spin-zero particles),
    
    \begin{equation}
        \scatteringamplitude(s,\theta)=32\pi\sum_{l=0}^{\infty}\left(l+\frac{1}{2}\right)f_l(s) P_l(\cos\theta) \, .
\end{equation}

Unitarity of the S-matrix can then be written as the condition 
    \begin{equation}
         {\rm Im}\,f_{j}\geq 0 \,, \qquad\left| f_{j}\left( s\right) \right|\leq 1.
    \end{equation}
    
    \item \hypertarget{DICT_ANNA_FU}{\textbf{Full unitarity (\ac{FU})}}

In addition to the \ac{PWU} conditions, the optical theorem applied to the scattering of fixed angular momentum states implies a non-linear condition on the partial wave amplitudes,
\begin{equation}
      2\,{\rm Im}\,f_{j}\left( s\right) \geq \left| f_{j}\left( s\right) \right|^{2}\sqrt{\frac{s-4m^{2}}{s}} \, .
\end{equation}
This inequality would be saturated if the $2\to2$ process is the only one possible in the theory. This can be the case for energies below the threshold when multi-particle production is kinematically forbidden. Also, this can be a good approximation in the scattering of massless states (gravitons in more than four dimensions).

\item \hypertarget{DICT_ANNA_ASYMPTOTIC_CAUSALITY}{\textbf{Asymptotic causality}}

Asymptotic causality or microcausality is usually referred to as a condition forbidding any time advance in the signal propagation provided by interactions, compared to a free propagation. Thus, all interactions must only cause time delays. This condition is proven to imply analyticity of the scattering amplitude outside the real axis in~\cite{Toll:1956cya}. In~\cite{Camanho:2014apa}, it was shown that this condition is violated by adding quadratic terms to the \ac{EFT} of gravity. However, in subsequent work~\cite{DAppollonio:2015fly}, it was shown that in the \ac{EFT} corresponding to \ac{ST}, asymptotic causality is restored if all terms are taken into account. 

\item \hypertarget{DICT_ANNA_MACROCAUSALITY}{\textbf{Macrocausality}}

Macrocausality is the requirement of not having macroscopic time-like loops, or, in simple words, forbidding a construction of a time machine and being able to go into the past at a macroscopic level. Asymptotic causality implies macrocausality but, in fact, the latter is a weaker condition which still allows for having unresolvable time advances in the signal propagation on top of the background~\cite{Hollowood:2015elj}. The time advance is unresolvable if it lies within the quantum-mechanical uncertainty principle. In this case, a violation of causality has a probabilistic nature, and cannot be made large by adding up several subsequent time advances.

    \item \hypertarget{DICT_ANNA_ANALYTICITY}{\textbf{Analyticity of the amplitude}}
    
     The scattering amplitude satisfies a set of analyticity properties in the complex plane of Mandelstam variables. Namely, it is an analytic function everywhere in the complex plane except for the real axis. If there are no massless particles in the theory, amplitudes at $t=0$ have an analyticity domain $0<s<s_\text{th}$ on the real $s$-axis. Here $s_\text{th}=4 m^2$ in the case of equal masses. In general, $s_\text{th}$ stands for the threshold center-of-mass energy which makes the scattering process kinematically allowed. For theories with massless states, the branch cut singularity always disconnects the upper and the lower half-plane in $s$ at fixed $t<0$. 

    \item \hypertarget{DICT_ANNA_LOCALITY}{\textbf{Locality}}

Locality of the \ac{IR} theory can be defined as the possibility to construct an \ac{EFT} expansion ordered by the number of fields and derivatives in each term. There should be no non-analytic terms (like poles, logarithms and fractional powers) in the bare Lagrangian both in fields (after canonical normalization) and in the derivative operators. Locality in the \ac{UV} in this section is defined in the language of scattering amplitudes as their polynomial boundedness at fixed $t<0$. For theories without massless states, it implies the Martin-Froissart bound (see \cref{MFbound}). For graviton-mediated scattering, the implications of locality crucially depend on the spacetime dimension, making it very special in $d=4$ (see \cref{sec:regge}).
     
    \item \hypertarget{DICT_ANNA_BOOTSTRAP}{\textbf{Bootstrap}}
    
In the context of this section, this term stands for the (partial or complete) reconstruction of a consistent non-perturbative amplitude based on its desired properties, such as unitarity, analyticity, locality and crossing symmetry. Usually, this procedure starts from a tree-level \ac{EFT} amplitude, which is improved following the \ac{QFT} axioms. An example for the perturbative bootstrap in reconstructing loop corrections from the \ac{FU} condition is given in \cref{bootstrap}. Regge bounds in graviton-mediated scattering correspond to another implementation of the same idea, see \cref{sec:regge}. Under certain conditions, the properties of the scattering amplitude allow for a complete reconstruction of the non-perturbative amplitude with the use of numerical techniques (see, for example, \cite{Elvang:2020lue} for a review of the bootstrap program).

\end{itemize}

\end{subappendices}

\section[\titleAlessiaBenjamin]{\texorpdfstring{\titleAlessiaBenjaminnewline}{\titleAlessiaBenjamin}}
\label{sec:ALESSIABENJAMIN}

\begin{tcolorbox}[colback=white,colframe=scipostblue]
{\bf Lecturers:} Alessia Platania, \briefaffiliationAlessia, and Benjamin Knorr, \briefaffiliationBenjamin

{\bf Email addresses:} \href{mailto:\emailAlessia}{\emailAlessia} and \href{mailto:\emailBenjamin}{\emailBenjamin}
\tcblower
{\bf Lecture recordings:}
\begin{enumerate}[label= Lecture \arabic*:, leftmargin = 3.5cm, labelsep = 0.5cm, parsep = 0.0cm]
    \item \href{https://youtu.be/OcrCvlnpLg8}{https://youtu.be/OcrCvlnpLg8}
    \item \href{https://youtu.be/YsnlKYM9S10}{https://youtu.be/YsnlKYM9S10}
    \item \href{https://youtu.be/2NqgTKG2g2c}{https://youtu.be/2NqgTKG2g2c}
    \item \href{https://youtu.be/dJ-6JKVP3HQ}{https://youtu.be/dJ-6JKVP3HQ}
\end{enumerate}

{\bf Abstract:}

The concept of asymptotic safety first emerged within the Wilsonian studies on the renormalization group in the context of condensed matter physics, but the parallelism between statistical and quantum fluctuations in a path integral approach soon led to the application of the Wilsonian ideas to the Standard Model of particle physics and, following Weinberg's proposal, to gravity. These lectures present a pedagogical introduction to the concept of asymptotic safety and its possible realization in quantum gravity. We will show how the concept of asymptotic safety is related to those of quantum scale invariance and non-perturbative renormalizability, how it generalizes asymptotic freedom, classical scale invariance, and perturbative/power-counting renormalizability, and how it can be useful to formulate quantum gravity within the framework of quantum field theory. We will then explain the functional renormalization group --- a powerful machinery to compute non-perturbative renormalization group flows, that is typically used to explore asymptotic safety and its consequences. Its applicability extends beyond asymptotic safety, and can be used to explore non-perturbative aspects of other field theories or even other approaches to quantum gravity. We will close the lectures with a presentation of the milestones and main results, as well as the key open questions and challenges that remain in the field.
\end{tcolorbox}

\subsection*{Preface}

\ac{QG} seeks to describe the quantum aspects of gravity, which are believed to become relevant in the early universe or in the interior of \acp{BH}. While gravity is well-described by \ac{GR} at macroscopic scales, the other three fundamental forces (electromagnetic, weak nuclear, and strong nuclear) are successfully formulated within the framework of \ac{QFT} at microscopic scales. Yet, as seen in \cref{sec:LUCA}, applying perturbation theory to Einstein's gravity yields \textit{incurable divergences}, \ie{}, divergences that cannot be reabsorbed in a finite number of free parameters. In this section, we explore the question of whether such perturbative divergences are a feature or a bug of the theory, and whether there can be a way out.

Indeed, facing the problem of perturbative divergences, one is left with two options: going beyond the framework of \ac{QFT} (\ac{ST} --- see \cref{sec:IVANO}, loop quantum gravity, causal sets, \dots), or finding loopholes/fixes that allow us to describe gravity as a \ac{QFT}. What is the price to pay for the latter option? There are different possibilities, including the following:
\begin{itemize}
    \item Giving up perturbativity: \ac{ASQG}~\cite{Percacci:2017fkn, Reuter:2019byg};
    \item Giving up ``standard'' unitarity: quadratic gravity~\cite{Donoghue:2021cza}, see \cref{sec:lecture4};
    \item Giving up locality: non-local \ac{QG}~\cite{BasiBeneito:2022wux};
    \item Giving up Lorentz invariance: Ho\v{r}ava gravity~\cite{Herrero-Valea:2023zex}.
\end{itemize}
While the last three possibilities may appear as big swings, the first one is not so wild, at least at first sight: why should nature (and in particular gravity) be weakly interacting at all scales? We know several regimes where physics is strongly coupled, the \ac{IR} regime of \ac{QCD} being a paradigmatic example. Even without thinking about fundamental interactions, the concept of asymptotic safety is well-known and physically realized in certain condensed matter systems described by, \eg{}, scalar fields in three dimensions~\cite{Braun:2010tt}. 
Asymptotic safety thus appears to be a natural possibility to consider and explore. Such a possibility was first proposed by Steven Weinberg in 1976~\cite{Weinberg:1976xy, Weinberg:1980gg}: the asymptotic safety conjecture states that gravity could be a consistent \ac{QFT} whose \ac{UV} completion is provided by an interacting (or, ``safe'') theory. Technically, this is realized if the gravitational \ac{RG} flow attains an interacting (\ie{}, not free) fixed point at high energy. The first calculation in $d=4$ was performed in~\cite{Reuter:1996cp} and the fixed point was first found in~\cite{Souma:1999at}.

These lecture notes focus on reviewing key features of this possibility, as well as the techniques that are used to investigate it --- them being general enough to find application in a broad range of subjects~\cite{Dupuis:2020fhh}, from fundamental interactions to condensed matter systems and fluid dynamics. Such techniques are known under the broad name of exact or functional \ac{RG}. The material is meant for students or non-experts. We refer the interested reader to the book chapters~\cite{Knorr:2022dsx, Eichhorn:2022gku, Morris:2022btf, Martini:2022sll, Wetterich:2022ncl, Platania:2023srt, Saueressig:2023irs, Pawlowski:2023gym,Bonanno:2024xne} for a comprehensive and more technical exposition of the subject, and to the books~\cite{Percacci:2017fkn, Reuter:2019byg} for more extensive reviews of individual topics.

The lecture notes are organized as follows. 

\begin{description}

\item[Sec.~\ref{sec:ALEBENsec1}:] We start introducing the distinction between perturbative and non-perturbative re\-normalizability. We explain the Wilsonian \ac{RG} in a general manner, taking a dynamical system approach. This is a simple, yet powerful way to grasp concepts like \ac{UV} completions, relevant directions, non-perturbative beta functions, and the cases when the perturbative picture can fail. This motivates us to move on and study the non-perturbative \ac{RG}, aka, the \ac{FRG}.

\item[Sec.~\ref{sect:ALESSIABENJAMIN_FRG}:] We introduce the \ac{FRG}. We start by recalling general concepts in \ac{QFT}, particularly the formalism of effective actions and how this connects to the computation of $n$-point correlation functions seen in \cref{sec:ANNA}. We then generalize the concept to that of an \ac{RG}-scale-dependent effective action, and we derive the Wetterich equation governing its variation in theory space. Next, we discuss approximation schemes to solve the Wetterich equation, and the heat-kernel technique --- an essential tool for its resolution but also for the treatment of non-local operators. Finally, we provide two simple hands-on computational examples of \ac{FRG} flows: the anharmonic oscillator and the case of gravity in the Einstein-Hilbert truncation. The latter is the simplest system where a non-trivial fixed point for gravity is found. This fixed point lies at the foundation of \ac{ASQG}. 

\item[Sec.~\ref{sec:ALEBEN-actionstoamplitudes}:] We spell out the significance of the \ac{FRG} in more detail, discussing the difference between physical and \ac{RG}-scale running, the relationship between bare and fixed-point actions, between \ac{RG}-scale-dependent effective actions and ordinary effective actions, and between effective actions and scattering amplitudes.

\item[Sec.~\ref{sec:ALEBEN-forefront}:] We describe some of the most recent and important advances in the field, in particular key consequences of \ac{ASQG} in particle physics, cosmology, and \ac{BH} physics. We then outline milestones and open questions in the field.

\item[Sec.~\ref{sect:ALESSIABENJAMIN_conclusions}:] We conclude by providing a zoomed-out perspective on \ac{ASQG} and how it relates to other research areas introduced in this set of lecture notes.

\end{description}

\subsection{The concept of asymptotic safety}\label{sec:ALEBENsec1}

In this section, we introduce the concepts of non-perturbative renormalizability and asymptotic safety. We will use as few formulas and assumptions as possible, showing the generality of these ideas as results of considerations based on dynamical system theory.

\subsubsection{Perturbative vs. non-perturbative renormalizability and asymptotic safety}\label{sect:ALESSIABENJAMIN_reno}   

The concept of perturbative (non-)renormalizability is tied to those of mass dimension and power counting. To illustrate the idea, let us consider the following Lagrangian of a scalar field $\phi$:
\begin{equation}
    \lagrangian= -\frac{1}{2} \partial_\mu\phi \partial^\mu \phi - \frac{1}{2} m^2\phi^2 - \sum_n \frac{\tilde{\lambda}_n}{n!} \phi^n \, ,
\end{equation}
where the couplings $\lambda_n$ generally have a dependence on an energy/momentum scale, namely they are not constant but ``running'' couplings.
In natural units, any Lagrangian has mass dimension\footnote{The mass dimension of any quantity follows from the definitions that masses have a mass dimension of one, $[m]=1$, that we can only add terms with the same mass dimension, and that the mass dimension of products is the sum of the mass dimensions of the factors. From this, we see directly that derivatives also have mass dimension one.} four (in four spacetime dimensions). Hence, the scalar field $\phi$ should have the dimension of a mass. It follows that
\begin{equation}
    [\tilde{\lambda}_n]\cdot M^n=M^4\,,
\end{equation}
and thus the coupling constants $\tilde{\lambda}_n$ have mass dimensions $d_{\tilde{\lambda}_n}\equiv [\tilde{\lambda}_n]=4-n$. For $n>4$, the mass dimension of $\tilde{\lambda}_n$ is negative, and one can show that this implies (perturbative) non-renormalizability. This follows from counting the mass dimensions of generic loop diagrams that indicate their divergence behavior.

The divergence has to do with the behavior of the running couplings. 
To understand what is going on and generalize the concept, it is useful to translate the mass dimension argument into the leading-order variation of couplings with respect to a momentum scale $k$, 
\begin{equation}
    k \partial_k \big(\tilde{\lambda}_n(k) k^{-d_{\tilde{\lambda}_n}}\big)= k^{-d_{\tilde{\lambda}_n}} \big(k \partial_k \tilde{\lambda}_n(k)\big) - d_{\tilde{\lambda}_n} k^{-d_{\tilde{\lambda}_n}} \tilde{\lambda}_n(k) \,.
\end{equation}
We now re-write this equation with respect to the dimensionless coupling ${\lambda}_n\equiv \tilde{\lambda}_n(k) k^{-d_{\tilde{\lambda}_n}}$,
\begin{equation}\label{eq:ALEBEN-betawithanomalous}
    k \partial_k {\lambda}_n(k) = \tilde{\lambda}_n(k) k^{-d_{\tilde{\lambda}_n}} \left(\frac{k \partial_k \tilde{\lambda}_n(k)}{\tilde{\lambda}_n(k)}\right) - d_{\tilde{\lambda}_n} {\lambda}_n(k) =(\eta[{\lambda}_n] -d_{\tilde{\lambda}_n})\cdot {\lambda}_n\,,
\end{equation}
where we have defined the \emph{anomalous dimension}
\begin{equation}
    \eta[{\lambda}_n]\equiv \frac{\partial \log \tilde{\lambda}_n}{\partial \log k}\,.
\end{equation}
The latter must be at least linear in the coupling according to perturbation theory, since the leading-order scaling has to match that of the classical scaling dimension. Indeed note that while here we are not doing any calculation (yet), the structure of the beta functions above is general, as it hides all quantum corrections stemming from loop diagrams in the anomalous dimension. Thus, close to ${\lambda}_n=0$, the anomalous dimension is negligible compared to the other terms, and the coupling constant displays a classical running 
\begin{equation}\label{eq:ALESSIABENJAMIN_class-scaling}
    {\lambda}_n(k)\simeq {\lambda}_n(k_0) \ (k/k_0)^{-d_{\tilde{\lambda}_n}}\,,
\end{equation}
with $k_0$ being an arbitrary reference scale. What does this formula mean? It means that when the mass dimension $d_{\tilde{\lambda}_n}=4-n$ of a coupling constant is positive, the coupling decreases as the momentum scale goes to infinity, and eventually vanishes in the limit $k\to\infty$: the coupling is asymptotically free. 

Vice versa, if $d_{\tilde{\lambda}_n}<0$, what can we say? We can certainly say that, \emph{close to the origin ${\lambda}_n=0$}, the coupling has the opposite behavior compared to the previous case: it increases as $k$ increases. This is a crucial point: perturbation theory, since it works close to zero coupling, would indicate that couplings with negative-mass dimensions keep increasing with $k$ and hence diverge in the limit $k\to\infty$. This would lead to divergences, and hence such \emph{irrelevant} couplings are typically excluded from the bare action to ensure \emph{perturbative renormalizability}. Note a problem here: we are extracting the behavior of the coupling from a small-coupling expansion and find that it diverges, which lies clearly beyond the range of applicability!

Finally, for $d_{\tilde{\lambda}_n}=0$, the coupling is classically scale-invariant, \ie{}, it doesn't change with energy classically. To extract the leading-order scaling in this case, we have to compute the anomalous dimension.

The simple characterization above however fails if ``non-trivial stuff'' happens away from ${\lambda}_n=0$, in particular if non-perturbative effects yield an anomalous dimension $\eta$ that is big enough to counterbalance~$d_{\tilde{\lambda}_n}$ at finite values for the $\lambda_n$. 

To understand this, let us now introduce the important concept of ``\emph{theory space}''. The name is self-explanatory: the theory space is the space of all possible theories. In the context of a \ac{QFT}, the axes of this space are all the couplings compatible with the symmetries and the field content of the physical system considered. Each point is identified by a specific set of values for all coupling constants. Since each axis corresponds to a coupling, if the \ac{RG} flow makes a coupling increase/decrease, one technically says that the \ac{RG} flow moves along the coupling's ``direction''. An important role is played by the origin $\mathcal{O}\equiv \{\lambda_n=0\}$ of the theory space, corresponding to a free theory.\footnote{The attentive reader might notice a subtlety. In gauge theory, we often parameterize the action with the inverse gauge coupling multiplying the square of the field strength in the Lagrangian. The definition of a free theory still works in this case once one has canonically normalized all fields. This opens up different possibilities due to different choices in how to normalize fields, but we shall not discuss this further here.}

\begin{figure}[t!]
    \centering
    \includegraphics[width=0.7\textwidth]{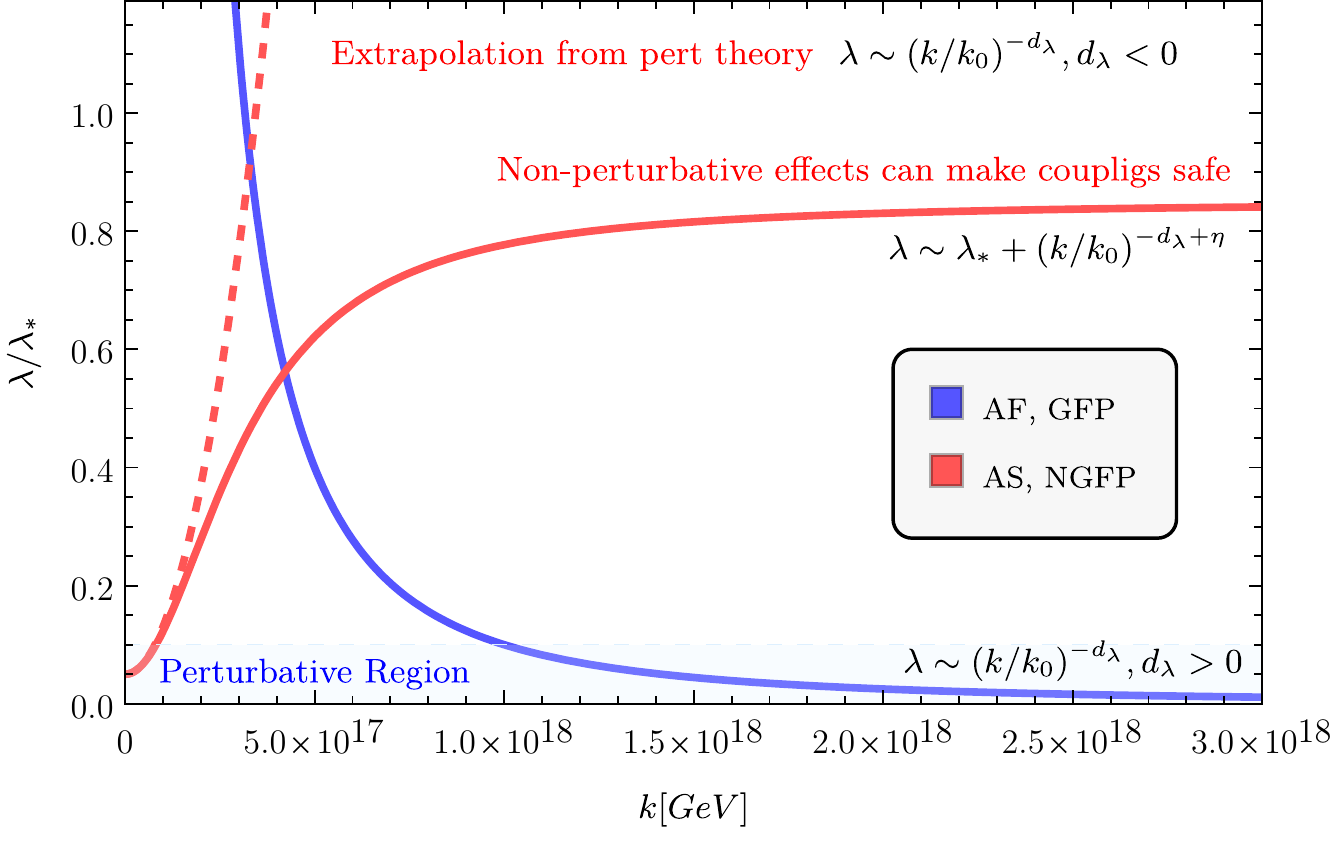}
    \caption{Intuitive generalization of asymptotic freedom to asymptotic safety: the divergences expected from perturbation theory may be tamed by non-perturbative effects at strong coupling, leading to a well-defined \ac{UV} completion characterized by non-vanishing interaction couplings.}
    \label{fig:ALESSIABENJAMIN_AS-AF}
\end{figure}

The perturbative argument and the classical scaling are linked to a special region of the theory space, namely the region close to the origin $\mathcal{O}$. This makes sense, as perturbation theory works when the interaction (measured by the couplings $\lambda_n$) is only a small perturbation of the free Lagrangian, corresponding to $\mathcal{O}$. From now on, we will refer to this origin as the \ac{GFP}. Why is this a fixed point of the system? Since a free theory is scale invariant,\footnote{This is only true if fields are canonically normalized.} \ie{} if it were sitting exactly at the \ac{GFP}, interactions would never be turned on by \ac{RG} transformations --- they would be zero at all energy scales. If instead, the initial conditions are such that interactions are non-zero at some energy scale, then couplings vary (under the \ac{RG} transformations) with the energy scale. In particular, they could approach the free theory as $k\to\infty$, realizing the well-known concept of \emph{asymptotic freedom}. When they do (as in the case of \ac{QCD}), the value of the couplings becomes zero asymptotically (blue line in \cref{fig:ALESSIABENJAMIN_AS-AF}). In this case, corresponding to positive mass dimension, the \ac{GFP}/free theory acts as an attractor in the \ac{UV}. 
If instead the mass dimension is negative, corresponding to the case of perturbative or power-counting non-renormalizability, the coupling increases with the energy (red dashed line in \cref{fig:ALESSIABENJAMIN_AS-AF}). However, the power-counting scaling leads the coupling to increase so much that it exits the region where perturbation theory holds (small/weak coupling), so that the power-counting approximation becomes unreliable. Indeed, away from the \ac{GFP} (\ie{}, when at least one coupling is large), the anomalous dimension $\eta$ is generally non-zero and can bend the dashed curved, preventing it from diverging (solid red curve in \cref{fig:ALESSIABENJAMIN_AS-AF}). If this happens, then the \ac{RG} trajectory is said to be ``\emph{asymptotically safe}'' and, as we shall see, this is the case if there exists an interacting generalization of the \ac{GFP}, called \ac{NGFP}, which is attractive in the \ac{UV} limit (at least with respect to some couplings). 

\subsubsection{RG evolution as a dynamical system: renormalizability and fixed points}\label{sect:ALESSIABENJAMIN_RG-concept}

We will now go deeper into understanding the relationship between re\-nor\-ma\-li\-za\-bi\-li\-ty and fixed points, and between relevant/irrelevant operators and critical exponents. These parallelisms, as well as the generalization of freedom$\to$safety and perturbative$\to$non-perturbative, stem from a deeper understanding of the \ac{RG} flow as a \emph{dynamical system} for the interaction couplings, whose evolution equation is the system of coupled beta functions,
\begin{equation}
    k\partial_k \lambda_n = \beta_{\lambda_n} \, .
\end{equation}
This set of equations is an autonomous dynamical system describing the evolution of the couplings $\lambda_n$ with respect to the \ac{RG} time $t_{RG}=\log k$. We thus arrive at the following consideration:
\begin{tcolorbox}
    The exact beta functions define a vector field on theory space, and their zeros correspond to points where the flow is frozen. These are the \emph{fixed points} of the \ac{RG} flow
\begin{equation}
    \text{Fixed Points:} \qquad k\partial_k \lambda_n^\ast = \beta_{\lambda_n^\ast}=0\,.
\end{equation}
\end{tcolorbox}
\noindent We will denote such points with an asterisk. As is clear from this equation, at a fixed point couplings do not run, they are constants equal to their fixed-point values. Specifically, we will define:
\begin{tcolorbox}
\begin{equation}
\begin{aligned}
    &\text{Gaussian Fixed Point:} \, & \forall \lambda_n : {\lambda_n^\ast}&=0 \, , \\
    &\text{Non-Gaussian Fixed Point:} \, & \exists n : {\lambda_n^\ast}&\neq 0\, .
\end{aligned}
\end{equation}
\end{tcolorbox}
\noindent The first one corresponds to the free theory/free Lagrangian and is the origin of the theory space, the second one instead corresponds to an interacting theory.

\begin{figure}[ht]
    \centering
    \includegraphics[width=0.57\textwidth]{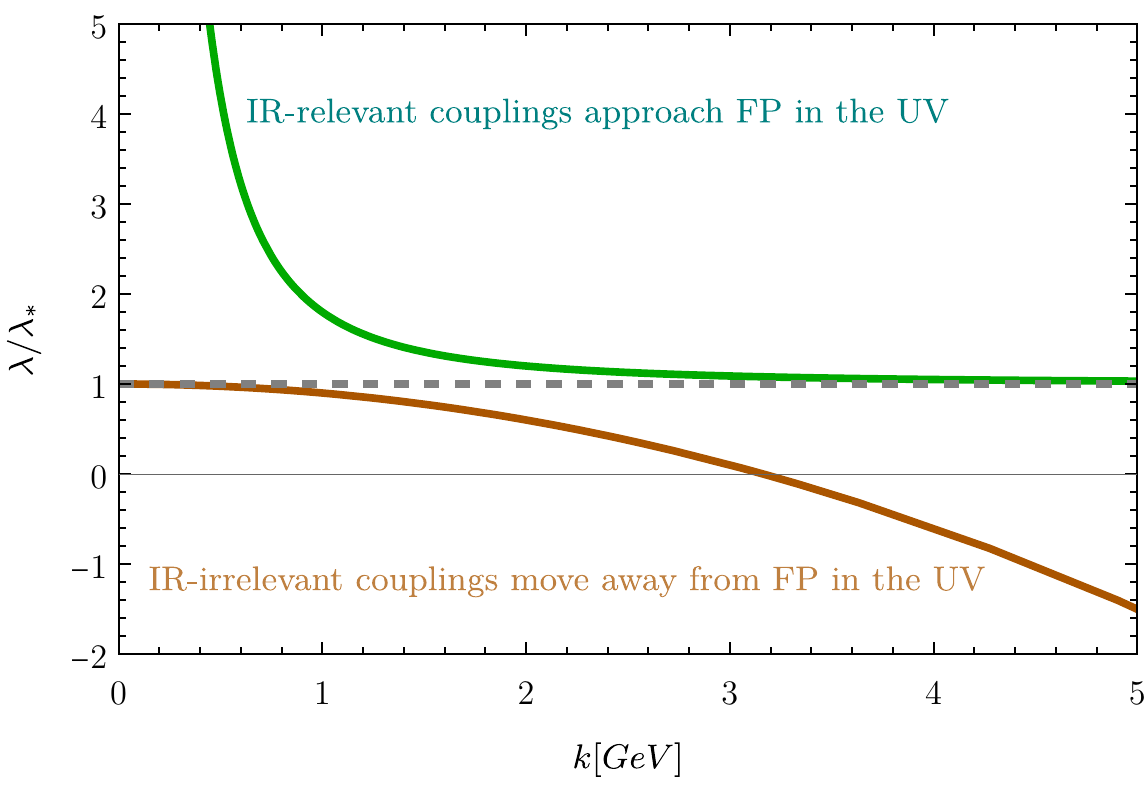}
    \caption{Relevant (green line) vs. irrelevant (orange line) couplings (case of a one-dimension theory space, spanned by the coupling $\lambda$ only). A fixed point at $\lambda=\lambda^\ast$ is depicted as a dashed gray line.}
    \label{fig:ALESSIABENJAMIN_it(relevant)}
\end{figure}

It is very important to understand the behavior of the flow in the proximity of a fixed point. In dynamical system theory, this tells us whether the evolution of the system drives trajectories towards or away from the given fixed point in theory space. To see this, let us expand the beta functions around a generic (free or interacting) fixed point. To leading order,
\begin{equation}\label{eq:ALESSIABENJAMIN_gen-scaling-multilambda}
    k\partial_k \lambda_n = \beta_{\lambda_n}\simeq \beta_{\lambda_n^\ast}+\sum_m \left.\frac{\partial \beta_{\lambda_n}}{\partial \lambda_m}\right|_{\lambda_m=\lambda_m^\ast}(\lambda_m-\lambda_m^\ast)\,,
\end{equation}
where $\beta_{\lambda_n^\ast}=0$ by the definition of a fixed point. In the case of a single coupling $\lambda_m\equiv \lambda$, the above equation has a simple solution
\begin{equation}\label{eq:ALESSIABENJAMIN_gen-eq-scaling}
    \lambda(k)=\lambda^\ast+(k/k_0)^{-\theta} \, ,
\end{equation}
where $k_0$ is an integration constant. Importantly, the constants $\theta$ assumes a special role.
\begin{tcolorbox}
    The \emph{critical exponent}
\begin{equation}
    \theta \equiv -\left.\frac{\partial \beta}{\partial \lambda}\right|_{\lambda=\lambda^\ast} \,
\end{equation}
defines the stability properties of the \ac{RG} flow in the proximity of a fixed point.
\end{tcolorbox}
\noindent Technically, we will call: 
\begin{itemize}
    \item $\theta>0 \quad \Rightarrow \lambda $ is an \ac{IR}-relevant coupling
    \item $\theta=0 \quad \Rightarrow \lambda $ is a marginal coupling
    \item $\theta<0 \quad \Rightarrow \lambda $ is an \ac{IR}-irrelevant coupling
\end{itemize}
The derivatives of the beta functions at the fixed point thus carry information about its attractiveness properties: if the exponent is negative, in the \ac{UV} limit the coupling will move away from the value $\lambda_\ast$, if it is positive, it will approach it asymptotically (see \cref{fig:ALESSIABENJAMIN_it(relevant)}).

\begin{figure}[ht]
    \centering
    \includegraphics[width=0.8\textwidth]{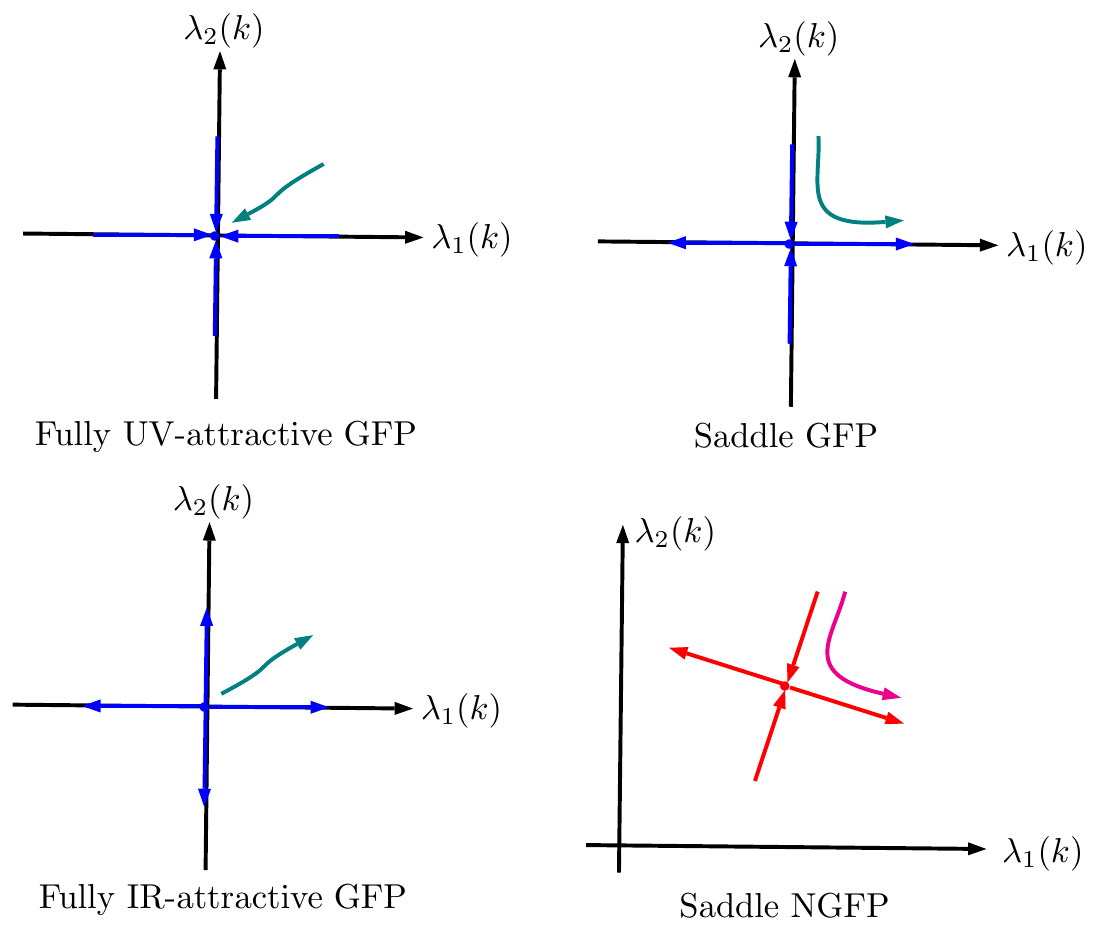}
    \caption{Examples of possible behaviors of \ac{RG} flows close to fixed points, with arrows pointing towards the \ac{UV}.}
    \label{fig:ALESSIABENJAMIN_various-beh}
\end{figure}

Does this remind you of something? Indeed, for the case of a \ac{GFP}, the equation above resembles \eqref{eq:ALESSIABENJAMIN_class-scaling}. You see now that \emph{the classical scaling with the mass dimension is a particular type of critical exponent, and the power-counting renormalizability dictated by the mass dimension of couplings depends on whether the \ac{GFP} is \ac{UV} attractive.} Yet, there can be other fixed points, and \eqref{eq:ALESSIABENJAMIN_gen-eq-scaling} is more general in this case, as it includes the possibility of non-trivial fixed points and non-trivial scalings. In the presence of multiple couplings, it is convenient to diagonalize \eqref{eq:ALESSIABENJAMIN_gen-scaling-multilambda}, arriving at the generalized formula for the scaling of couplings around a fixed point:
\begin{equation}\label{eq:ALESSIABENJAMIN_lin-stab-exp}
    \vec\beta \simeq \vec\beta^\ast + S_\text{stab} (\vec \lambda - \vec \lambda^\ast) \qquad \Rightarrow \qquad \vec \lambda = \vec \lambda^\ast + \sum_i \, c_i \, \vec{e}_i \, (k/k_0)^{-\theta_i} \, ,
\end{equation}
where we have defined the \emph{stability matrix}
\begin{equation}
    ( S_\text{stab} )_{nm} \equiv \left.\frac{\partial \beta_n}{\partial \lambda_m}\right|_{\lambda_i=\lambda_i^\ast}\,,
\end{equation}
$\vec{e}_i$ are its eigenvectors, $\theta_i$ are \emph{critical exponents} defined as minus its eigenvalues, and $c_i$ are integration constants. We now can see that terms with positive critical exponents will contribute to getting closer to the fixed point in the \ac{UV} ($k\to\infty$), while those associated with negative critical exponents will make the flow \dots flow away, in the same limit! Technically, we will call: 
\begin{itemize}
    \item $\theta_i>0\quad \Rightarrow \vec{e}_i$ is an \ac{IR}-relevant direction
    \item $\theta_i=0\quad \Rightarrow \vec{e}_i$ is a marginal direction
    \item $\theta_i<0\quad \Rightarrow \vec{e}_i$ is an \ac{IR}-irrelevant direction 
\end{itemize}
Note that critical exponents and eigendirections are generally different for each fixed point (and indeed this is key in gravity). Fixed points can be
\begin{itemize}
    \item Fully \ac{UV}-attractive if $\theta_i>0\quad \forall i$,
    \item Fully \ac{IR}-attractive if $\theta_i<0\quad \forall i$,
    \item Saddle points (generic case), if both relevant and irrelevant directions are present.
\end{itemize}
The \ac{UV} critical surface is then the space spanned by the (power-counting) renormalizable couplings. In systems with multiple fixed points, the different attractors are connected by \emph{separatrices}, \ie{}, \ac{RG} trajectories moving from an \ac{IR} source to a \ac{UV} sink. Examples with two couplings are shown in \cref{fig:ALESSIABENJAMIN_various-beh}.
As depicted, in terms of dynamical systems, you can see non-perturbative renormalizability in the space of couplings as a shift and potentially a rotation of what happens in the case of perturbative renormalizability, close to the \ac{GFP}. 

\begin{figure}[ht]
    \centering
    \includegraphics[width=0.45\textwidth]{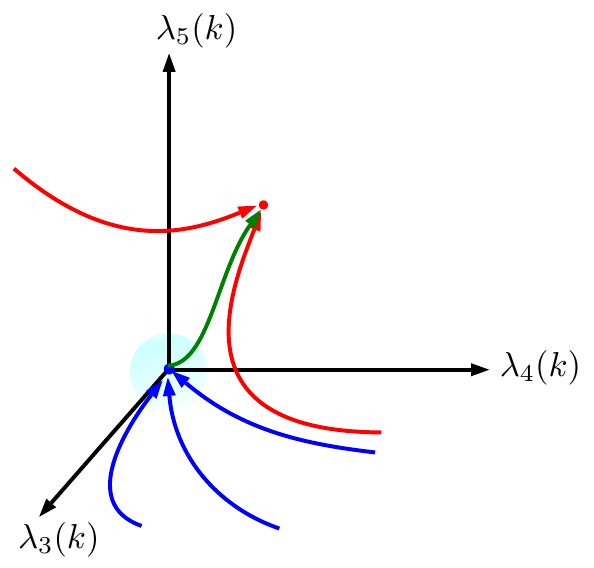}
    \caption{Example of an \ac{RG} flow driven by two fixed points in a three-dimensional theory space. Arrows point towards the \ac{UV}, and the light-blue region around the \ac{GFP} denotes the region where perturbation theory is valid. The \ac{GFP} with two \ac{IR}-relevant directions ($\lambda_3$ and $\lambda_4$) attracts the flow for all \ac{RG} trajectories whose initial conditions lie on the corresponding plane (blue lines). Such \ac{RG} trajectories are asymptotically free. The $\lambda_5$ direction is instead \ac{IR}-irrelevant for the \ac{GFP}. Hence, some \ac{RG} trajectories are repulsed by it. In the system however, there is also another interacting fixed point, \ie{}, an \ac{NGFP}, which has some attractive/\ac{IR}-relevant directions. Some \ac{RG} trajectories will be asymptotically safe (red lines). The \ac{RG} trajectory connecting the two (green line) is called \emph{separatrix}.}
    \label{fig:ALESSIABENJAMIN_theory-space}
\end{figure}

Fixed points and their stability matrix tell us a lot about the global properties of a dynamical system --- in our case, the \ac{RG} flow in theory space. To make a parallelism, if you want to draw the graph of a function, the first thing you study are the points where the derivatives of the functions vanish and where they diverge --- once this is done, one can approximately extrapolate and draw the function; with \ac{RG} flows and fixed points it is the same (see \cref{fig:ALESSIABENJAMIN_theory-space} for an example): once you know fixed points and their stability properties, you can have a good guess of how the \ac{RG} flow will look like! In particular, the parametric plot depicting the entire \ac{RG} flow and its properties is named the \emph{phase diagram} --- the name being inspired from similar flows in condensed matter theory, where fixed points have the physical interpretation of bulk phase and phase transitions, respectively, depending on whether the fixed point is trivial or not~\cite{Goldenfeld:1992qy}.

Now the important point comes that, based on the linearized formula~\eqref{eq:ALESSIABENJAMIN_lin-stab-exp}, \emph{\ac{UV}-complete trajectories are those with initial conditions $c_i=0\,\, \forall i\,:\theta_i<0$. Avoiding non-\ac{UV}-complete theories (or, equivalently, selecting renormalizable theories) fixes almost all of the $c_i$ (those corresponding to irrelevant directions), while the relevant directions yield free parameters that have to be fixed via experiments.} 

In this framework (built within the Wilsonian revolution~\cite{Wilson:1971bg, Wilson:1971dh, Wilson:1973jj} and constituting the modern idea of \ac{RG}) we thus reach a deeper understanding of renormalization, which is now phrased in new terms:
\begin{itemize}
    \item Fixed points with at least one relevant direction are the possible \ac{UV} completions of our theories. In other words, \emph{one cannot choose the \ac{UV} completion, rather, the \ac{RG} identifies all possible bare actions that can \ac{UV}-complete the flow as its fixed points}. The \ac{GFP} (free theory) is always there, even in perturbation theory. Once we take all-order effects into account, the perturbative beta functions generally acquire an anomalous dimension $\eta[\lambda_n]$, so that the fixed point equation
    
    \begin{equation}
        \beta_{\lambda_n}=(\eta[{\lambda}_n] -d_{\tilde{\lambda}_n})\cdot {\lambda}_n=0 \, ,
    \end{equation}
    
    can have, beyond the \ac{GFP} $\lambda_n^\ast=0$, also \acp{NGFP} solving the equation:
    
    \begin{equation}
        \eta[{\lambda}_n] -d_{\tilde{\lambda}_n}=0 \, .
    \end{equation}
    
    Note that the anomalous dimension $\eta[{\lambda}_n]$ encodes all-order (and beyond) effects and needs to be appropriately computed. This will be the subject of the next sections.
    
    \item Given one \ac{UV} completion, the number of free parameters equates to the number of relevant directions of the flow (modulo fixing the unit scale). We define a theory (or, \ac{RG} trajectory) to be \emph{non-perturbatively renormalizable} if it is \ac{UV}-completed at a (generally non-trivial) fixed point, with a finite number $N$ of relevant directions. Then the number of free parameters is $N-1$. This is because one of the integration constants $c_i$ can always be reabsorbed to redefine the transition scale $k_0$ to the fixed point regime. This can also be seen as redefining the speed of the flow. Finally, at the level of the effective action, it is equivalent to the statement that one of the couplings sets the unit scale of the system (\eg{}, the Newton coupling for Planck units). It is thus clear that the requirement of $N$ being finite ensures \emph{predictivity}. 
    
    \item Renormalizable theories correspond to \ac{RG} trajectories that are \ac{UV}-complete with respect to fixed points displaying a finite number of \ac{IR}-relevant directions.
    
    \item The \ac{UV} critical surface is the basin of attraction of the fixed point: all \ac{RG} trajectories with initial conditions on it will end up in the fixed point in the \ac{UV}.
    
    \item Whether an interaction is relevant or irrelevant depends on the \ac{UV} completion/fixed point.
    
    \item Different fixed points are connected via \emph{separatrices}: these are peculiar \ac{RG} trajectories departing from a fixed point in the \ac{IR} and ending up at another one in the \ac{UV}.
    
    \item A \ac{GFP}/free theory with at least one \ac{UV}-attractive direction is the simplest \ac{UV} completion, and corresponds to asymptotically free theories (like \ac{QCD}).
\end{itemize}
We can now understand the old \emph{perturbative statements} in the light of very general non-perturbative arguments:
\begin{itemize}
    \item We can now better understand the perturbative picture as an incomplete viewpoint, obtained by limiting ourselves to the perturbative regime close to the \ac{GFP}/free theory.
    
    \item Eigendirections at the \ac{GFP} are typically aligned with the axes, so couplings are typically either relevant or irrelevant.
    
    \item \ac{IR}-relevant/irrelevant couplings are attractive/repulsive directions of the \ac{GFP}, related to the (classical) scaling of couplings close to the free theory.
    
    \item Setting \ac{IR}-irrelevant couplings to zero to have renormalizable Lagrangians in the \ac{SM} corresponds to sitting in the sub-theory space of initial conditions such that all \ac{RG} trajectories are asymptotically free, \ie{}, in the basin of attraction of the \ac{GFP}; equivalently, the \ac{UV} critical surface is simply the space spanned by the (power-counting) renormalizable couplings.
    
    \item \ac{IR}-irrelevant directions are not necessarily dangerous, they only parameterize our ignorance about non-perturbative/all-order effects away from the \ac{GFP}.
\end{itemize}
With this better understanding of the \ac{RG}, we can now introduce \emph{Weinberg's asymptotic safety proposal}~\cite{Weinberg:1976xy, Weinberg:1980gg}: despite gravity being perturbatively non-renormalizable, \ie{}, not being renormalizable with respect to the free fixed point, it could still be non-perturbatively renormalizable if its \ac{RG} flow approaches a non-trivial fixed point in the \ac{UV}. Such a fixed point must have a finite number of relevant directions in order to ensure predictivity. Indeed, most of the Wilson coefficients in the \ac{EFT} expansion (see \cref{sec:ANNA}) ought to be predicted by \ac{QG}. This is indeed possible in \ac{ASQG}~\cite{Basile:2021krr, Knorr:2024yiu}.
We will see the simplest case of asymptotic safety in gravity in \cref{sect:ALESSIABENJAMIN_EH}, while some consequences of its predictive and falsifiability power will be listed in \cref{sect:ALESSIABENJAMIN_milestones}. The scope of the remaining part of this section is to show you ``how to get there'' (how to compute a gravitational flow in the simplest possible case) and summarize its most important physical consequences.

To summarize, in the Wilsonian picture, there is no assumption about the bare action: the theory tells you what the possible bare actions are in the space of all possible actions; and typically there are a few! One should appreciate how powerful this is in comparison to strict renormalizability (see \cref{sec:LUCA}), where one is restricted to the perturbative regime. The Wilsonian picture gives a deeper understanding of renormalizability and thus allows us to go beyond!

What we need to do then is to determine \ac{RG} fixed points in gravity. To this end, we should finally remark that \ac{RG} fixed points are associated with the more familiar concept of phase transitions in condensed matter physics. Based on this, there are two ways to look for asymptotic safety in a physical system and study its consequences:
\begin{itemize}
    \item \textbf{Find non-trivial \ac{RG} fixed points of the (non-perturbative) beta functions} (\eg{}, via the analytical methods of the \ac{FRG}~\cite{Dupuis:2020fhh}), or, equivalently,
    \item \textbf{Look for second-order phase transitions} (\eg{}, via the numerical simulations of dynamical triangulations~\cite{Loll:2019rdj} or Regge calculus).
\end{itemize}
Here we focus on the former approach, and we will discuss the \ac{FRG} next. In the following, we will only discuss \emph{Euclidean} signature.

\subsection{Computing non-perturbative beta functions: FRG}\label{sect:ALESSIABENJAMIN_FRG}

A central ingredient in discussing non-perturbative renormalization is of course a means to perform non-perturbative computations of beta functions. One such tool --- the one most widely used to investigate asymptotic safety --- is the {\bf \acf{FRG}}, that we will discuss now. For this, we will first make a short detour and derive the {\bf effective action~$\mathbf \Gamma$}. The latter is an extremely useful object in \ac{QFT}, for several reasons. It already includes all quantum corrections, so that the equations of motion derived from it are fully quantum. Moreover, to compute scattering amplitudes from it, only tree-level diagrams are needed to get the exact answer. In more technical words, $\Gamma$ is the generating functional of one-particle irreducible correlation functions. This subsection is inspired in part by~\cite{Gies:2006wv, Reichert:2020mja}.

\subsubsection{Prelude: the effective action}

The standard starting point to discuss \ac{QFT} is the {\bf path integral}, also called {\bf partition function}. Restricting for simplicity to a theory of a single scalar field $\varphi$,\footnote{This generalizes to other fields, with some care needed for gauge fields (due to gauge symmetry and the Gribov problem) and fermions (due to their Grassmann nature).} it reads\footnote{The attentive reader will notice that this is strictly speaking not a path integral, but a statistical partition function, since there is no factor of the imaginary unit $\mathbf{i}$. We do this because regularization in Lorentzian signature is difficult, and the subject of current research.}
\begin{tcolorbox}
\begin{equation}\label{eq:ALESSIABENJAMIN_path_integral}
    \partitionfunction[J] = \frac{1}{\mathcal N} \int \mathcal D \varphi \, e^{- S[\varphi] + \int \rmd^dx \, J(x) \, \varphi(x)} \, .
\end{equation}
\end{tcolorbox}
\noindent In this formula, $S$ is the microscopic (or ``bare'') action\footnote{Note that bare and classical actions are generally different: the bare action is the fundamental one that we want to quantize, which appears in the exponential within the path integral, whereas the classical action is the \ac{IR} limit of the effective action.} of the field $\varphi$, $J(x)$ is a source term, and $\mathcal N$ is a normalization constant which is formally infinite. This expression for $\partitionfunction$ contains divergences that need regularization and renormalization --- for now we assume that this has been done already, \eg{}, by a cutoff regularization with a cutoff $\UVcutoff$.

Let us briefly discuss the notation for more general fields. For this, we will use so-called super-indices. A general field $\Phi^A$ has an \emph{upper} super-index $A$, which can include both standard (spacetime or internal, like gauge or spinor) indices and any dependence on spacetime position (or momentum). For example, for a vector field, $A$ would be a single spacetime index, $\Phi^A \to v^\mu$. For a spinor, $A$ would be a Dirac index. Finally, for the metric, $A$ represents \emph{two lower} spacetime indices: $\Phi^A \to g_{\mu\nu}$. A contraction of super-indices means a sum over the discrete indices (spacetime, gauge), and an integral over the spacetime position.

A central object of study in \acp{QFT} that can be computed from $\partitionfunction[J]$ are {\bf $n$-point correlation functions} of the fundamental field.
\begin{tcolorbox}
Given a theory with bare action $S[\varphi]$, its {\bf $n$-point correlation functions} are defined as the normalized expectation value of the product of $n$ fields:
\begin{equation}
\begin{aligned}
    \langle \varphi(x_1) \cdots \varphi(x_n) \rangle_J &= \frac{\int \mathcal D \varphi \, \varphi(x_1) \cdots \varphi(x_n) \, e^{-S[\varphi] + \int \rmd^dx \, J(x) \, \varphi(x)}}{\int \mathcal D \varphi \, e^{-S[\varphi] + \int \rmd^dx \, J(x) \, \varphi(x)}} \\
    &= \frac{1}{\partitionfunction[J]} \frac{\delta^n}{\delta J(x_1) \cdots \delta J(x_n)} \partitionfunction[J] \, .
\end{aligned}
\end{equation}    
\end{tcolorbox}
\noindent In particular, we highlight the following important cases:
\begin{itemize}
    \item $n=0$: $\langle 1 \rangle_J = 1$ --- this is simply a statement of normalization.
    \item $n=1$: $\langle \varphi(x) \rangle_J \equiv \phi(x)$ --- this is the {\bf \ac{vev}} of the field.
    \item $n=2$: $\langle \varphi(x_1) \varphi(x_2) \rangle_J = \langle \varphi(x_1) \varphi(x_2) \rangle_{J,c} + \phi(x_1) \phi(x_2)$ --- this is the {\bf propagator}, that has a connected and a disconnected part.
    \item $n\geq 3$: much like the propagator, these higher order correlators (or vertices) split into connected and disconnected pieces.
\end{itemize}
From these examples, we find that $\partitionfunction[J]$ is not an efficient storage of information: for example for the propagator, only the connected part is ``new'' information, since the disconnected part is already completely determined in terms of the \ac{vev}. For this reason, in a first step we need to introduce a more useful entity.
\begin{tcolorbox}
We introduce the {\bf Schwinger functional $\mathbf{\mathcal W}$} as the logarithm of the partition function,
\begin{equation}
    \mathcal W[J] = \ln \partitionfunction[J] \, .
\end{equation}
\end{tcolorbox}
\noindent The Schwinger functional is a key object in \ac{QFT}, it being the generator of connected correlation functions:
\begin{equation}
    \langle \varphi(x_1) \cdots \varphi(x_n) \rangle_{J,c} = \frac{\delta^n}{\delta J(x_1) \cdots \delta J(x_n)} \mathcal W[J] \equiv \mathcal W^{(n)}[J] \, .
\end{equation}
Note that we will often use the notation $\propG(x_1,x_2) = \mathcal W^{(2)}[J]$ for the full quantum (\ie{}, connected) propagator. For a general field $\Phi^A$, the propagator $\propG^{AB}(x_1,x_2)$ is a \emph{bi-tensor}. This means that it is a tensor with super-index $A$ at the point $x_1$, and also a tensor with super-index $B$ at the point $x_2$, with all the ensuing transformation properties.

    Let us check that this is a reasonable definition. Consider the two-point function (\ie{} the case $n=2$ above). We compute
    \begin{equation}
    \begin{aligned}
        \mathcal W^{(2)}[J] &= \frac{\delta^2}{\delta J(x_1) \delta J(x_2)} \ln \partitionfunction[J] \\
        &= \frac{\delta}{\delta J(x_1)} \frac{1}{\partitionfunction[J]} \frac{\delta \partitionfunction[J]}{\delta J(x_2)} \\
        &= \left[ \frac{1}{\partitionfunction[J]} \frac{\delta^2 \partitionfunction[J]}{\delta J(x_1) \delta J(x_2)} \right] - \left[ \frac{1}{\partitionfunction[J]} \frac{\delta \partitionfunction[J]}{\delta J(x_1)} \right] \left[ \frac{1}{\partitionfunction[J]} \frac{\delta \partitionfunction[J]}{\delta J(x_2)} \right] \\
        &= \langle \varphi(x_1) \varphi(x_2) \rangle_J - \langle \varphi(x_1) \rangle_J \langle \varphi(x_2) \rangle_J = \langle \varphi(x_1) \varphi(x_2) \rangle_{J,c} \, .
    \end{aligned}
    \end{equation}
It turns out that we can store all information even more efficiently, via the generating functional of one-particle irreducible (1PI) $n$-point functions. The latter are related to Feynman diagrams that cannot be separated into two parts by cutting just a single internal line. This leads us to define an important quantity:
\begin{tcolorbox}
We define the \textbf{effective action},
\begin{equation}\label{eq:ALESSIABENJAMIN_effective_action_definition}
        \Gamma[\phi] = \sup_J \left\{ \int \rmd^dx \, J(x) \, \phi(x) \, - \mathcal W[J] \right\} \, ,
    \end{equation}
as the Legendre transform of the Schwinger functional with respect to the mean field.
\end{tcolorbox}
\noindent We will now discuss the 1PI property of $\Gamma$ in some detail. First, the conjugate variable of the source $J$ is indeed the \ac{vev} $\phi$. By definition, the conjugate variable is the derivative of the original function with respect to the original variable:
\begin{equation}
    \phi(x) = \frac{\delta \mathcal W[J_\text{sup}]}{\delta J_\text{sup}} = \frac{1}{\partitionfunction[J_\text{sup}]} \frac{\delta \partitionfunction[J_\text{sup}]}{\delta J_\text{sup}} = \langle \varphi(x) \rangle_{J_\text{sup}} \, .
\end{equation}
Here, $J_\text{sup}$ is the source for which the supremum is obtained in \eqref{eq:ALESSIABENJAMIN_effective_action_definition}. Note that it depends on the \ac{vev} $\phi$. Next, the quantum equation of motion for the mean field in the presence of a source is indeed given by the first variation of $\Gamma$. Using the definition,
\begin{equation}
    \Gamma^{(1)}[\phi] = J_\text{sup} + \frac{\delta J_\text{sup}}{\delta \phi} \underbrace{\left[ \phi - \frac{\delta \mathcal W[J_\text{sup}]}{\delta J_\text{sup}} \right]}_{=0} = J_\text{sup} \, .
\end{equation}
Finally, let us show that the inverse of the quantum propagator $\propG$ is $\Gamma^{(2)}$.
    Consider the following product and use the definitions above:
    \begin{equation}
    \begin{aligned}
        \int \rmd^dy \, \frac{\delta^2 \mathcal W}{\delta J(x_1) \delta J(y)} \frac{\delta^2 \Gamma}{\delta \phi(y) \delta \phi(x_2)} &= \int \rmd^dy \, \left[ \frac{\delta}{\delta J(x_1)} \frac{\delta \mathcal W}{\delta J(y)} \right] \left[ \frac{\delta}{\delta \phi(y)} \frac{\delta \Gamma}{\delta \phi(x_2)} \right] \\
        &= \int \rmd^dy \, \left[ \frac{\delta}{\delta J(x_1)} \phi(y) \right] \left[ \frac{\delta}{\delta \phi(y)} J(x_2) \right] \\
        &= \frac{\delta J(x_2)}{\delta J(x_1)} \equiv \delta(x_1-x_2) \, .
    \end{aligned}
    \end{equation}
    This shows that $\Gamma^{(2)}$ indeed is the inverse of $\propG$.

How is $\Gamma$ related to the original path integral? Let us use the definitions and the quantum equations of motion to find out. A short computation gives
\begin{equation}
\begin{aligned}
    e^{-\Gamma[\phi]} &= e^{-\int \rmd^dx \, J_\text{sup}(x) \, \phi(x) + \mathcal W[J_\text{sup}]} \\
    &= e^{-\int \rmd^dx \, \frac{\delta \Gamma[\phi]}{\delta \phi(x)} \, \phi(x)} e^{\mathcal W[J_\text{sup}]} \\
    &= e^{-\int \rmd^dx \, \frac{\delta \Gamma[\phi]}{\delta \phi(x)} \, \phi(x)} \int \mathcal D\varphi \, e^{-S[\varphi] + \int \rmd^dx \, J_\text{sup}(x) \, \varphi(x)} \, .
\end{aligned}
\end{equation}
A final shift of the integration variable then yields a complicated differential equation for the effective action in terms of the microscopic action:
\begin{equation}\label{eq:ALESSIABENJAMIN_effective_action_path_integral}
    e^{-\Gamma[\phi]} = \int \mathcal D\varphi' \, e^{-S[\phi+\varphi'] + \int \rmd^dx \, \frac{\delta \Gamma[\phi]}{\delta \phi(x)} \, \varphi'(x)} \, .
\end{equation}
Despite its complexity, the equation \eqref{eq:ALESSIABENJAMIN_effective_action_path_integral} is useful to discuss symmetries on the quantum level. 
\begin{tcolorbox}
In practice, using a systematic {\bf vertex expansion} of the effective action,
\begin{equation}\label{eq:ALESSIABENJAMIN_vertex_expansion}
    \Gamma[\phi] = \sum_{n\geq 0} \frac{1}{n!} \int \rmd^dx_1 \cdots \rmd^dx_n \, \Gamma^{(n)}[\phi=0](x_1,\dots,x_n) \, \phi(x_1) \cdots \phi(x_n) \, ,
\end{equation}
yields an infinite tower of integro-differential equations known as the Dyson-Schwinger equations~\cite{Dyson:1949ha, Schwinger:1951ex, Schwinger:1951hq}.
\end{tcolorbox}
\noindent They are particularly useful in asymptotically free theories like \ac{QCD}.

As mentioned before, $\Gamma$ is the quantum analog of the microscopic action $S$, it encodes the full quantum physics at tree level. As such, it is very useful for the computation of scattering amplitudes.

As an example, consider a two-to-two scattering of two different scalar fields $\phi$ and $\chi$, that is $\phi\phi\to\chi\chi$, mediated by gravitational interaction and a contact term. The full scattering amplitude $\scatteringamplitude_{\phi\phi\to\chi\chi}$ can directly be computed from correlation functions of the effective action. Schematically,
    \begin{equation}\label{eq:ALESSIABENJAMIN_scattering_amplitude_schematic}
        \scatteringamplitude_{\phi\phi\to\chi\chi} \simeq \frac{\delta^3 \Gamma}{\delta\phi\delta\phi\delta g_{\mu\nu}} \circ \frac{\delta^2 \Gamma}{\delta g^{\mu\nu} \delta g^{\rho\sigma}} \circ \frac{\delta^3 \Gamma}{\delta g_{\rho\sigma}\delta\chi\delta\chi} + \frac{\delta^4 \Gamma}{\delta\phi\delta\phi\delta\chi\delta\chi} \, .
    \end{equation}
    All quantum effects are included in this expression, there is no infinite tower of perturbative Feynman diagrams!
This example clearly shows the power residing in the effective action. However, so far we did not discuss a generic way to compute it. The \ac{FRG} provides such a way, and we will introduce it next.

\subsubsection{The Wetterich equation}

We will now introduce one systematic way to compute the effective action. The underlying physical idea is that of the Wilsonian \ac{RG}: instead of performing the path integral in one fell swoop, let us integrate out modes momentum-shell by momentum-shell. Concretely, we start integrating out the modes with large momenta (high energy), and continue with modes with successively lower momenta. Once we have integrated out all modes, the full path integral is performed.

\begin{figure}[h]
\centering
\includegraphics[width=0.5\textwidth]{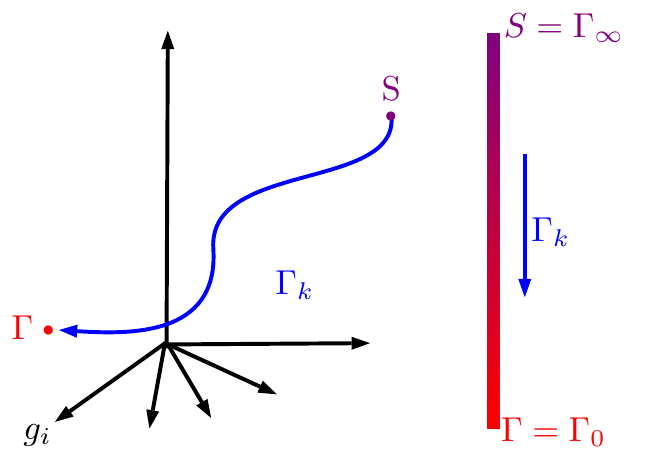}
\caption{\label{fig:ALESSIABENJAMIN_FRGidea} Central idea of the \ac{FRG}. The space of action functionals contains both the microscopic action $S$ and the effective action $\Gamma$. They are connected by the \ac{EAA} $\Gamma_k$. When $k\to\infty$, $\Gamma_k$ approaches $S$, whereas for $k\to0$, $\Gamma_k$ reproduces $\Gamma$. }
\end{figure}

With this procedure we can define the so-called {\bf \ac{EAA} $\mathbf{\Gamma_k}$}. It is the equivalent of the effective action where only modes with a momentum larger than the fiducial momentum scale $k$ have been integrated out. By construction, for $k\to\infty$ no modes have been integrated out, and we recover the microscopic action $S$. Likewise, for $k\to0$, all modes have been integrated out, and we obtain the standard effective action $\Gamma$. This means that $\Gamma_k$ acts as an interpolant between these two limits, as sketched in \cref{fig:ALESSIABENJAMIN_FRGidea}. The crucial advantage of this idea is that we can derive an \emph{exact} differential equation for $\Gamma_k$ in the fiducial scale $k$. This is highly advantageous: we know how to (approximately) solve differential equations systematically, whereas (path) integrals are generically very difficult to solve. The derivation of this differential equation, which usually goes under the name of \emph{Wetterich equation}, is our next goal.

\begin{figure}[ht]
\centering
\includegraphics[width=0.6\textwidth]{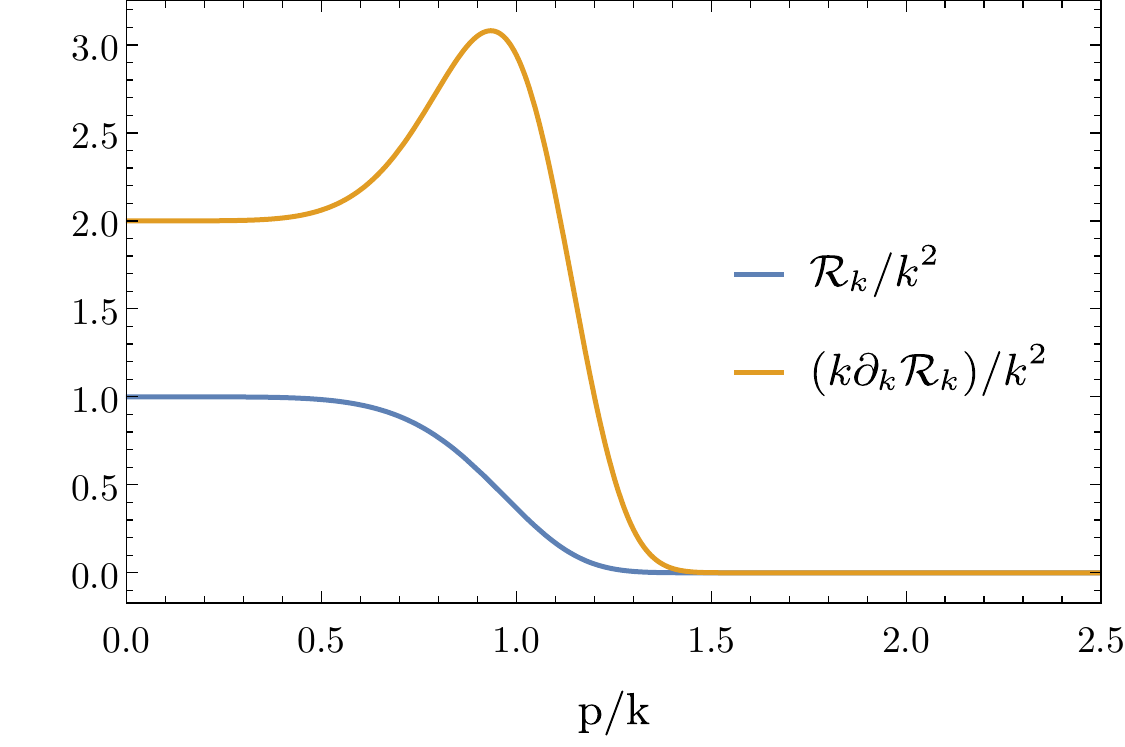}
\caption{\label{fig:ALESSIABENJAMIN_regulator} Illustration of a typical regulator function $\mathcal R_k$ and its scale derivative $k \partial_k \mathcal R_k$. The specific choice shown here is $\mathcal R_k = k^2 e^{-\left(p^2/k^2\right)^3}$. One can see how the regulator affects the regime of low momenta (compared to the scale $k$), whereas it leaves the regime of large momenta unaltered. The scale derivative peaks at $p\approx k$. This implements the idea of integrating out modes roughly at the scale $k$.}
\end{figure}

The derivation largely follows the derivation of the effective action as above, with small modifications that implement the Wilsonian idea. The first step is to modify the path integral by suppressing infrared modes, that is modes with momenta below the scale $k$. A strict implementation of this suppression leads to the Wegner-Houghton equation~\cite{Wegner:1972ih}, but it breaks the symmetries of many \acp{QFT}. We will implement this more smoothly. 
\begin{tcolorbox}
    We can define a \textbf{$k$-dependent partition function} by the \textit{ad hoc} addition of a smooth regulator functional to the bare action,
\begin{equation}\label{eq:ALESSIABENJAMIN_path_integral_with_regulator}
    \partitionfunction_k[J] = \frac{1}{\mathcal N} \int \mathcal D \varphi \, e^{-\Delta S_k[\varphi]} \, e^{- S[\varphi] + \int \rmd^dx \, J(x) \, \varphi(x)} \, .
\end{equation}
The regulator implements the Wilsonian shell-by-shell integration of fast-fluctuating modes.
\end{tcolorbox}
\noindent We now need to discuss the definition and properties of this Wilsonian regulator.
\begin{tcolorbox}
In momentum space, the \textbf{regulator functional} $\Delta S_k[\varphi]$ is defined as
    \begin{equation}
    \Delta S_k[\varphi] = \frac{1}{2} \int \frac{\rmd^dp}{(2\pi)^d} \, \varphi(-p) \, \mathcal R_k(p^2) \, \varphi(p) \, .
\end{equation}
\end{tcolorbox}
\noindent It is always quadratic in the \textit{fluctuation field}.\footnote{There are generalizations that allow for a more general term~\cite{Pawlowski:2005xe}. They have the disadvantage that the resulting differential equation is not of one-loop form, and thus in general more complicated.}
The {\bf regulator function $\mathbf{\mathcal R_k}$} appearing in this expression is essentially a momentum-dependent mass term that exactly implements the mode-by-mode integration: modes with momenta $p^2\gtrsim k^2$ are integrated out, the rest are suppressed and left unintegrated. To do so, the regulator has to fulfill three key properties:
\begin{itemize}
    \item It has to implement a mode suppression in the \ac{IR}:
        \begin{equation}
            \lim_{p^2\to0} \mathcal R_k(p^2) > 0 \, .
        \end{equation}
        Usually, we have $\mathcal R_k(0) \simeq k^2$.

    \item It has to vanish for $k\to0$ to ensure that we get back the effective action once all modes are integrated out:
        \begin{equation}
            \lim_{k\to0} \mathcal R_k(p^2) = 0 \, .
        \end{equation}

    \item It has to implement the limit $k\to\infty$ correctly:
        \begin{equation}
            \lim_{k\to\UVcutoff\to\infty} \mathcal R_k(p^2) \to \infty \, .
        \end{equation}
        Here, $\UVcutoff$ is the \ac{UV} cutoff related to the original regularization mentioned below \eqref{eq:ALESSIABENJAMIN_path_integral}. The explicit reconstruction of the microscopic action is somewhat involved~\cite{Manrique:2008zw, Fraaije:2022uhg}.
\end{itemize}
Note that the general shape of the regulator is largely arbitrary, reflecting the freedom of choosing a regularization scheme. A sketch of a typical regulator function is shown in \cref{fig:ALESSIABENJAMIN_regulator}.

    There are some commonly used regulator shapes. One choice that is very useful for analytical computations is the so-called linear, or Litim, regulator~\cite{Litim:2000ci, Litim:2001up}:
    \begin{equation}\label{eq:ALESSIABENJAMIN_litim-reg}
        \mathcal R_k(p^2) = (k^2 - p^2) \, \theta\left( 1 - \frac{p^2}{k^2} \right) \, .
    \end{equation}
    Here, $\theta$ is the Heaviside distribution. While generally very useful, its distributional character leads to problems in sophisticated computations. A smoother choice are regulators of exponential type, for example
    \begin{equation}
        \mathcal R_k(p^2) = k^2 \, e^{-\left(p^2/k^2\right)^n} \, ,
    \end{equation}
    or
    \begin{equation}
        \mathcal R_k(p^2) = \frac{p^2}{e^{p^2/k^2}-1} \, .
    \end{equation}

Coming back to the derivation of the Wetterich equation, the next step is to derive a differential equation for $\partitionfunction_k[J]$. For this, we take the $k$-derivative of \eqref{eq:ALESSIABENJAMIN_path_integral_with_regulator}. As we shall shortly see,
\begin{equation}
    k \partial_k \partitionfunction_k[J] = -\frac{1}{2} \int \frac{\rmd^dp}{(2\pi)^d} \left[ k \partial_k \mathcal R_k(p^2) \right] \, \frac{\delta^2 \partitionfunction_k[J]}{\delta J(-p) \delta J(p)} \, .
\end{equation}
This is already a functional (integro-)differential equation to compute the path integral, and in principle it can be used to obtain $\partitionfunction_{k=0}[J]$.

    Let us prove this equation. Using the definition \eqref{eq:ALESSIABENJAMIN_path_integral_with_regulator}, we compute
    \begin{equation}
    \begin{aligned}
        k \partial_k &\partitionfunction_k[J] = \frac{1}{\mathcal N} \int \! \mathcal D \varphi \, \left[ -k \partial_k \Delta S_k[\varphi] \right] \, e^{- S[\varphi] -\Delta S_k[\varphi] + \int \rmd^dx \, J(x) \, \varphi(x)} \\
        &= -\frac{1}{2\mathcal N} \int \! \mathcal D \varphi \int \! \frac{\rmd^dp}{(2\pi)^d} \, \varphi(-p) \left[ k \partial_k \mathcal R_k(p^2) \right] \varphi(p) \, e^{- S[\varphi] -\Delta S_k[\varphi] + \int \rmd^dx \, J(x) \, \varphi(x)} \\
        &= -\frac{1}{2\mathcal N} \int \! \mathcal D \varphi \int \! \frac{\rmd^dp}{(2\pi)^d} \left[ k \partial_k \mathcal R_k(p^2) \right] \frac{\delta^2}{\delta J(-p) \delta J(p)} \, e^{- S[\varphi] -\Delta S_k[\varphi] + \int \rmd^dx \, J(x) \, \varphi(x)} \\
        &= -\frac{1}{2} \int \frac{\rmd^dp}{(2\pi)^d} \left[ k \partial_k \mathcal R_k(p^2) \right] \, \frac{\delta^2 \partitionfunction_k[J]}{\delta J(-p) \delta J(p)} \, .
    \end{aligned}
    \end{equation}
As the next step of the derivation, we introduce the $k$-dependent Schwinger functional $\mathcal W_k[J]$,
\begin{equation}
    \mathcal W_k[J] = \ln \partitionfunction_k[J] \, .
\end{equation}
Just using this definition, we arrive at a similar functional integro-differential equation for $\mathcal W_k$:
\begin{equation}
    k \partial_k \mathcal W_k[J] = -\frac{1}{2} \int \frac{\rmd^dp}{(2\pi)^d} \left[ k \partial_k \mathcal R_k(p^2) \right] \, \left\{ \frac{\delta^2 \mathcal W_k[J]}{\delta J(-p) \delta J(p)} + \frac{\delta \mathcal W_k[J]}{\delta J(-p)} \frac{\delta \mathcal W_k[J]}{\delta J(p)} \right\} \, .
\end{equation}
The final step is to introduce $\Gamma_k$ by a slightly modified Legendre transform.
\begin{tcolorbox}
We define the \textbf{\acf{EAA}} as
\begin{equation}\label{eq:ALESSIABENJAMIN_EAA_definition}
        \Gamma_k[\phi] = \sup_J \left\{ \int \rmd^dx \, J(x) \, \phi(x) \, - \mathcal W_k[J] - \Delta S_k[\phi] \right\} \, .
    \end{equation}
\end{tcolorbox}
\noindent This is the central object of the \ac{FRG}: if a suitable fixed point exists, the \ac{RG} flow of $\Gamma_k[\phi]$ connects the bare action $S$ appearing in the path integral~\eqref{eq:ALESSIABENJAMIN_path_integral}\footnote{This is modulo a subtlety that we will discuss later: the reconstruction problem~\cite{Morris:2015oca,Fraaije:2022uhg}.} with the ordinary effective action $\Gamma$.
The addition of the regulator term on the right-hand side is for pure convenience --- the only thing that we have to ensure is that the limit $k\to0$ is not spoiled, but this is true since $\lim_{k\to0} \Delta S_k=0$. One way to read this equation is that it is actually the sum $\Gamma_k+\Delta S_k$ that is the Legendre transform of $\mathcal W_k$. In analogy to before, we find
\begin{align}
    J_\text{sup}[\phi] &= \frac{\delta \left( \Gamma_k[\phi] + \Delta S_k[\phi] \right)}{\delta\phi} \, , \\
    \propG_k[\phi] &\equiv \frac{\delta^2 \mathcal W_k[J_\text{sup}]}{\delta J_\text{sup}^2} = \left[ \frac{\delta^2}{\delta\phi^2} \left( \Gamma_k[\phi] + \Delta S_k[\phi] \right) \right]^{-1} = \left[ \Gamma_k^{(2)}[\phi] + \mathcal R_k \right]^{-1} \, .
\end{align}
Notably, the $k$-dependent propagator $\propG_k$ is the inverse of the regularized two-point function stemming from $\Gamma_k$. With this at hand, we can compute the differential equation for $\Gamma_k$.
Taking the $k$-derivative of \eqref{eq:ALESSIABENJAMIN_EAA_definition}, we find
    \begin{equation}
    \begin{aligned}
        k \partial_k \Gamma_k[\phi] &= -k \partial_k \mathcal W_k[J_\text{sup}[\phi]] - k \partial_k \Delta S_k[\phi] + \int \rmd^dx \, k \partial_k J_\text{sup}[\phi] \left[ \phi - \frac{\delta \mathcal W_k[J_\text{sup}[\phi]]}{\delta\phi} \right] \\
        &= \frac{1}{2} \int \frac{\rmd^dp}{(2\pi)^d} \left[ \propG_k(p,-p) + \phi(-p) \phi(p) \right] \, k \partial_k \mathcal R_k(p^2) \, - k \partial_k \Delta S_k[\phi] \\
        &= \frac{1}{2} \int \frac{\rmd^dp}{(2\pi)^d} \propG_k(p,-p) \, k \partial_k \mathcal R_k(p^2) \, ,
    \end{aligned}
    \end{equation}
where in the first line, the third term has a contribution both from the direct dependence of $\Gamma_k$ on $J_\text{sup}$ as well as its indirect dependence via $\mathcal W_k$.
This finally leads us to the main equation defining the \ac{FRG} flow of the \ac{EAA}.
\begin{tcolorbox}   
The \textbf{Wetterich equation}~\cite{Wetterich:1992yh, Morris:1993qb, Ellwanger:1993mw}
\begin{equation}\label{eq:ALESSIABENJAMIN_wetterich_equation}
        k \partial_k \Gamma_k[\phi] = \frac{1}{2} \text{STr} \left[ \left( \Gamma_k^{(2)}[\phi] + \mathcal R_k \right)^{-1} \, k \partial_k \mathcal R_k \right] \, 
    \end{equation}
is a formally exact functional integro-differential equation defining the \ac{RG} flow of the scale-dependent effective action $\Gamma_k$.
\end{tcolorbox}
\noindent In writing down this equation, we slightly generalized the computation above by the introduction of the {\bf super trace STr}. The super trace generally signifies:
\begin{itemize}
    \item a sum over discrete indices (spacetime, gauge),
    \item an integral over continuous variables (coordinates or momenta), and
    \item a minus sign for Grassmann-valued fields (fermions, Faddeev-Popov ghosts).
\end{itemize}
For a scalar field in flat space, the super trace simply represents the integral over the loop momentum. In the context of gravity, we have to be more careful in defining the super trace, since there is no momentum space available --- we will come back to this later. Note that \eqref{eq:ALESSIABENJAMIN_wetterich_equation} is a self-contained equation, \ie{}, it is a functional integro-differential equation that only makes reference to $\Gamma_k$ (and the regulator). In particular, no path integral has to be performed anymore.

Let us discuss some properties of the Wetterich equation:
\begin{itemize}
    \item The equation \eqref{eq:ALESSIABENJAMIN_wetterich_equation} is formally {\bf exact} and non-perturbative. No approximations have been involved in its derivation.
    
    \item The regulator $\mathcal R_k$ serves several purposes. First, it ensures \ac{IR} finiteness because
    \begin{equation}
        \Gamma_k^{(2)} + \mathcal R_k > 0 \, .
    \end{equation}
    Second, it ensures \ac{UV} finiteness due to
    \begin{equation}
        \lim_{p^2\to\infty} k \partial_k \mathcal R_k(p^2) = 0 \, .
    \end{equation}
    Finally, $k \partial_k \mathcal R_k$ is most sensitive to momenta $p \approx k$, implementing the Wilsonian idea of integrating out modes shell by shell.

    \item The equation is a first order differential equation in $k$. In this way, it defines a vector field on {\bf theory space}, as discussed previously.

    \item The explicit dependence on the regulator is a type of scheme dependence, any physical observable is independent of the specific choice. This however only holds at the exact level --- if approximations are introduced, one usually also finds a regulator dependence.
\end{itemize}

    Let us perform a sanity check, and compute the one-loop effective action with the help of the Wetterich equation. For this, we make the ansatz
    \begin{equation}
        \Gamma_k \simeq S + \Delta \Gamma_{k,1l} \, .
    \end{equation}
    By definition, $S$ is independent of $k$. Keeping only $S$ on the right-hand side to obtain the flow of the one-loop term, we find
    \begin{equation}
        k \partial_k \Delta \Gamma_{k,1l} = \frac{1}{2} \text{STr} \left[ \left( S^{(2)} + \mathcal R_k \right)^{-1} \, k \partial_k \mathcal R_k \right] = \frac{1}{2} k \partial_k \, \text{STr} \, \ln \left[ S^{(2)} + \mathcal R_k \right] \, .
    \end{equation}
    Integrating this equation, we find
    \begin{equation}
        \Gamma \simeq S + \frac{1}{2} \text{STr} \, \ln \, S^{(2)} \, ,
    \end{equation}
    which is the known expression for the one-loop effective action.

When it comes to scattering amplitudes, see \eqref{eq:ALESSIABENJAMIN_scattering_amplitude_schematic}, we need the correlation functions $\Gamma^{(n)}$. Their $k$-dependent counterparts also fulfill exact one-loop type flow equations, so that the correlators can be computed directly. Taking a field-derivative of the Wetterich equation, we find
\begin{equation}
    k \partial_k \Gamma_k^{(1)} = -\frac{1}{2} \text{STr} \left[ \propG_k \Gamma_k^{(3)} \propG_k \, k \partial_k \mathcal R_k \right] \, .
\end{equation}
Taking yet another derivative, we find the flow of the two-point function:
\begin{equation}
    k \partial_k \Gamma_k^{(2)} = -\frac{1}{2} \text{STr} \left[ \propG_k \Gamma_k^{(4)} \propG_k \, k \partial_k \mathcal R_k \right] + \text{STr} \left[ \propG_k \Gamma_k^{(3)} \propG_k \Gamma_k^{(3)} \propG_k \, k \partial_k \mathcal R_k \right] \, .
\end{equation}
The pattern continues for higher orders. Note the emerging hierarchy: the flow of the $n$-point function $\Gamma_k^{(n)}$ only depends on the correlation functions $\Gamma_k^{(2)}, \dots, \Gamma_k^{(n+2)}$. The aforementioned vertex expansion exploits this hierarchy to implement a systematic approximation scheme.

\subsubsection{Approximate resolution methods: truncation schemes}

The Wetterich equation is an exact, non-perturbative, functional integro-differential equation for the \ac{EAA} $\Gamma_k$. Yet, solving it requires some form of approximation: one needs an \emph{ansatz} to express $\Gamma_k$ and extract the flow projected onto a small part of the infinite-dimensional theory space spanned by all interaction couplings generated by the symmetry group of the theory.

To approximate the \ac{EAA}, one needs an \emph{ordering principle}, which boils down to choosing a way to expand the action, and a truncation order which defines the accuracy of the approximation within the given expansion scheme. Focusing on a theory with one real scalar field, two popular expansion schemes are the \emph{vertex expansion}
\begin{equation}\label{eq:ALESSIABENJAMIN_vertex-exp}
   \Gamma_k[\phi]=\sum_{n=0}^{\infty} \frac{1}{n!} \int \rmd^d x_1 \cdots \rmd^d x_n \Gamma_k^{(n)}\left(x_1, \dots, x_n\right) \phi\left(x_1\right) \cdots \phi\left(x_n\right) \,,
\end{equation}
and the \emph{derivative expansion}
\begin{equation}\label{eq:ALESSIABENJAMIN_der-exp}
    \Gamma_k=\int \rmd^d x\left[V_k(\phi)+\frac{1}{2} Z_k(\phi)\left(\partial_\mu \phi\right)^2+\orderneglected\left(\partial^4\right)\right]\,.
\end{equation}
The former accounts for the full momentum dependence and uses as the ordering principle the field (\ie{}, it is non-perturbative in the physical momentum $p$, but perturbative in the field, which acts as the expansion parameter), whereas the latter is a particular type of operator expansion where terms are ordered according to powers of the derivative (in other words, this expansion holds for small physical momenta $p$, but it is non-perturbative with respect to the field). In both cases, the expansions are \emph{non-perturbative in the couplings}.

Once an expansion scheme (like the vertex or derivative expansion) is fixed, a truncation order is chosen which depends on the scope and complexity of the calculation. This allows to perform computations, and to improve their accuracy systematically by increasing the truncation order in a step-by-step fashion.

\subsubsection{Symmetries and Ward identities}

Symmetries are a key element of \acp{QFT} that describe the real world. Important examples include the $SU(N)$ gauge symmetry of Yang-Mills theory, used to describe the forces in the \ac{SM}, the diffeomorphism symmetry of gravity, but also discrete symmetries like a $\mathbbm Z_2$-symmetry that plays a central role in the description of the Ising model. For our short discussion, we will now assume that the symmetry has no anomalies, that is, the path integral measure respects the symmetry.

The central point of our discussion is the interplay of the symmetry with the regularization. For example, dimensional regularization, which is widely used in perturbation theory, preserves gauge symmetry. On the contrary, a hard cutoff breaks it. This does not mean that one cannot use a hard cutoff to regularize a gauge theory --- it simply adds an extra layer of complexity as one has to ensure that the gauge symmetry is restored at the very end. This is exactly what happens with the \ac{FRG}: (gauge) symmetries are apparently broken by the regulator, but the breaking is encoded in symmetry identities that allow to restore the symmetry in the limit $k\to0$, \ie{}, when recovering the standard effective action.

Let us add some concreteness to this discussion. As a starting point, consider a theory where the microscopic action is invariant under a symmetry, like $O(N)$ or $SU(N)$, with an infinitesimal generator $\mathfrak G$.\footnote{This means that $\mathfrak G \phi$ is linear in $\phi$.} For example, a global $O(N)$ would have a generator
\begin{equation}
    \mathfrak G^a_{O(N)} = -f^{abc} \int \rmd^dx \, \phi^b(x) \, \frac{\delta}{\delta \phi^c(x)} \, ,
\end{equation}
whereas for an $SU(N)$ gauge symmetry,
\begin{equation}
    \mathfrak G^a_{SU(N)} = -\gaugecovD_\mu^{ab}(x) \, \frac{\delta}{\delta A_\mu^b(x)} = - \left( \partial_\mu \delta^{ab} - g \, f^{abc} A_\mu^c(x) \right) \, \frac{\delta}{\delta A_\mu^b(x)} \, .
\end{equation}
Let us first check what happens to the standard effective action. For this, we apply $\mathfrak G$ to the path integral (once again assuming that the measure not anomalous, \ie{}, invariant):
\begin{equation}
\begin{aligned}
    0 &= \frac{1}{\partitionfunction[J]} \int \mathcal D\varphi \, \mathfrak G \, e^{-S[\varphi] + \int \rmd^dx \, J(x) \, \varphi(x)} \\
    &= \frac{1}{\partitionfunction[J]} \int \mathcal D\varphi \, \left[ - \mathfrak G \, S[\varphi] + \int \rmd^dx \, J(x) \, \mathfrak G \varphi(x) \right] \, e^{-S[\varphi] + \int \rmd^dx \, J(x) \, \varphi(x)} \\
    &= - \langle \mathfrak G \, S \rangle_J + \langle \int \rmd^dx \, J(x) \, \mathfrak G \varphi(x) \rangle_J \, .
\end{aligned}
\end{equation}
To get to the second line, we used the linearity of $\mathfrak G$. Since we are interested in the effect of $\mathfrak G$ on $\Gamma$, let us evaluate this expression at the supremum source, $J = J_\text{sup}[\phi]$. The latter is independent of $\varphi$ by construction, since it only depends on the \ac{vev}. With this, we find
\begin{equation}
\begin{aligned}
    0 &= - \langle \mathfrak G \, S \rangle_{J_\text{sup}} + \int \rmd^dx \, J_\text{sup}(x) \, \langle \mathfrak G \varphi(x) \rangle_{J_\text{sup}} \\
    &= - \langle \mathfrak G \, S \rangle_{J_\text{sup}} + \int \rmd^dx \, \frac{\delta \Gamma}{\delta\phi(x)} \, \mathfrak G \phi(x) \\
    &= - \langle \mathfrak G \, S \rangle_{J_\text{sup}} + \mathfrak G \, \Gamma[\phi] \, .
\end{aligned}
\end{equation}
In this derivation, we used the quantum equation of motion and the definition of the \ac{vev} in the first step, and the linearity of $\mathfrak G$ in the second step.
\begin{tcolorbox}
    The effective action is invariant under a symmetry if the microscopic action is,
    \begin{equation}
        \mathfrak G \, \Gamma = \langle \mathfrak G \, S \rangle_{J_\text{sup}} \, .
    \end{equation}
\end{tcolorbox}
\noindent This simple and convenient result is complicated in the presence of gauge fixing and regularization.

Let us first discuss gauge symmetry. These symmetries require gauge fixing via a gauge-fixing action $S_\text{gf}$ and an accompanying Faddeev-Popov ghost action $S_\text{gh}$ to make the path integral well-defined. Both these extra terms break gauge symmetry, but they preserve what is called \ac{BRST} invariance~\cite{Becchi:1975nq, Tyutin:1975qk}. Gauge invariance of the effective action is then encoded in a generalized identity. 
\begin{tcolorbox}
Assuming that the microscopic action is gauge-invariant, the {\bf Ward-Takahashi identity}~\cite{Ward:1950xp, Takahashi:1957xn} holds
    \begin{equation}
        \mathcal W = \mathfrak G \, \Gamma - \langle \mathfrak G \, \left( S_\text{gf} + S_\text{gh} \right) \rangle_{J_\text{sup}} = 0 \, .
    \end{equation}
It dictates which breaking terms are, and are not, allowed in $\Gamma$. 
\end{tcolorbox}
\noindent For example, in Yang-Mills theory, it forbids a mass term for the gauge boson --- a term $m_A^2 A_\mu^a A^{a\mu}$ is incompatible with $\mathcal W$.

Second, we have to discuss symmetry breaking induced by the regularization. The best case scenario is of course if a regulator can be chosen that respects the symmetry. For linear global symmetries like $O(N)$, this is easily possible. Unfortunately, for non-linear symmetries like $SU(N)$ or diffeomorphism symmetry, the non-linear structure clashes with the fundamental requirement that the regulator term be quadratic in the field. Thus, the regulator generically represents a new source of symmetry breaking, and has to be accounted for as a modification to the Ward-Takahashi identity.
\begin{tcolorbox}
 A similar derivation as for the standard case then results in the \textbf{modified Ward-Takahashi identity}
    \begin{equation}
        \mathcal W_k = \mathfrak G \, \Gamma_k + \mathfrak G \, \Delta S_k - \langle \mathfrak G \, \left( S_\text{gf} + S_\text{gh} + \Delta S_k \right) \rangle_{J_\text{sup}} = 0 \, .
    \end{equation}
\end{tcolorbox}
\noindent By construction, when $k\to0$, the standard identity is recovered. One can furthermore show that the \ac{RG} flow of $\mathcal W_k$ is proportional to itself. At an exact level, this means that if it is fulfilled at an initial scale $k=\Lambda$, it is also fulfilled at all other scales $k$. This useful property however fails once approximations are used, and one has to control the corresponding error to restore the symmetry at $k=0$. How to do this best, in particular in \ac{QG}, is the topic of active research~\cite{Dupuis:2020fhh}.

To briefly illustrate this, let us come back to the mass term of an $SU(N)$ gauge boson. As stated above, $\mathcal W$ forbids such a term, so that $m_{A,k=0}^2=0$. However, at finite $k$, the regulator introduces such a term, and the modified Ward-Takahashi identity indeed does not forbid it. Solving $\mathcal W_k$, one indeed finds that the mass term runs approximately like $m_{A,k}^2 \approx g^2 k^2$. The naive choice $m_{A,k}^2=0$ for all $k$ would actually violate the identity, and would then not restore gauge symmetry at $k=0$~\cite{Cyrol:2016tym}.

\subsubsection{Making sense of the super trace in gravity: the heat kernel}\label{sec:ALESSIABENJAMIN_HK}

In this subsection, we will sketch how to perform the super trace when spacetime is curved. To simplify the discussion, we will discuss a free scalar field in a fixed curved spacetime, and explore which terms the super trace will generate. We will follow the exposition in~\cite{Groh:2011dw}, and more details can be found, \eg{}, in~\cite{Vassilevich:2003xt}.

As a starting point, we consider the action
\begin{equation}\label{eq:ALESSIABENJAMIN_free_scalar}
    \Gamma_k = \int \rmd^dx \, \sqrt{g} \, \frac{1}{2} (\covD_\mu \phi) (\covD^\mu \phi) \, ,
\end{equation}
and compute the quantum corrections produced by it. By diffeomorphism invariance, the resulting \ac{RG} flow has to have the form
\begin{equation}
    k \partial_k \Gamma_k = \int \rmd^dx \, \sqrt{g} \, \left[ c_0 + c_1 \, R + c_2 \, R^2 + c_3 R_{\mu\nu} R^{\mu\nu} + \dots \right] \, .
\end{equation}
The goal is to compute the numerical coefficients $c_i$. We will now work in position space (since a momentum space in general does not exist when the curvature is non-zero). To evaluate the flow, we first compute the second variation of the action \eqref{eq:ALESSIABENJAMIN_free_scalar},\footnote{Here and in the following, we are suppressing trivial factors of the Dirac delta distribution which (in momentum space) implement momentum conservation.}
\begin{equation}
    \Gamma_k^{(2)} = - g^{\mu\nu} \covD_\mu \covD_\nu \equiv -\covD^2 \equiv \Delta \, .
\end{equation}
Adding a general regulator, we thus have to evaluate
\begin{equation}
    \frac{1}{2} \text{STr} \left[ \left( \Delta + \mathcal R_k(\Delta) \right)^{-1} k \partial_k \mathcal R_k(\Delta) \right] \equiv \frac{1}{2} \text{STr} \, W(\Delta) \, .
\end{equation}
The task is thus to compute the trace of a general function of the Laplace operator $\Delta$. Before we deal with this, let us first compute a simpler super trace, and then relate the general trace to this simpler one. This idea goes under the name of the {\bf heat kernel method}. It turns out that we can compute
\begin{equation}\label{eq:ALESSIABENJAMIN_HK_def}
    \text{STr} \, e^{-s\Delta} \equiv \text{STr} \, H(s) \, ,
\end{equation}
in an expansion as indicated above. Defining the matrix elements of $H(s)$ in a position basis,
\begin{equation}
    H(x,y;s) = \langle y | H(s) | x \rangle \, ,
\end{equation}
the super trace would simply correspond to a sum over the eigenvalues. These matrix elements fulfill the heat equation\footnote{This is where the method derives its name from.}
    \begin{equation}
    \begin{aligned}
        \partial_s H(x,y;s) &= -\Delta_x \, H(x,y;s) \, , \\
        H(x,y;0) &= \delta(x-y) \, .
    \end{aligned}
    \end{equation}
The subscript $x$ on the Laplace operator indicates that it acts on the $x$-variable. In the following, we will not indicate this anymore; this is a standard omission in the literature. Now we can again use the fact that, as physicists, we are well-trained in solving differential equations. For example, for a flat spacetime where $\Delta=-\partial^2$, the fundamental solution is
\begin{equation}\label{eq:ALESSIABENJAMIN_HK_flat}
    H(x,y;s) = \left(\frac{1}{4\pi s} \right)^{d/2} \, e^{-\frac{(x-y)^2}{4s}} \, .
\end{equation}
With this, we can define the super trace as
\begin{tcolorbox}
\begin{equation}\label{eq:ALESSIABENJAMIN_HK_super_trace_def}
    \text{STr} \, e^{-s\Delta} = \text{tr} \int \rmd^dx \, \sqrt{g} \, \langle x | e^{-s\Delta} | x \rangle =  \text{tr} \int \rmd^dx \, \sqrt{g} \, H(x,x;s) \, .
\end{equation}
\end{tcolorbox}
\noindent Here, tr indicates the trace over discrete indices (\eg{} spacetime and gauge indices), and we call $H(x,x;s)$ the {\bf coincidence limit} of the heat kernel, since both points coincide. We will often denote this limit by an overbar, $H(x,x;s) \equiv \overline{H}(s)$.

    Let us make sure that this definition for the heat kernel makes sense, \ie{}, that it really corresponds to a simple integral over a loop momentum. For this, we take \eqref{eq:ALESSIABENJAMIN_HK_def} and go to momentum space to compute
    \begin{equation}
    \begin{aligned}
        \int \frac{\rmd^dp}{(2\pi)^d} e^{-s p^2} &= \frac{1}{(2\pi)^d} \int \rmd\Omega \int_0^\infty \rmd{}p \, p^{d-1} \, e^{-s p^2} \\
        &= \frac{1}{(2\pi)^d} \left[ \frac{2\pi^{d/2}}{\Gamma(d/2)} \right] \left[ \frac{\Gamma(d/2)}{2s^{d/2}} \right] = \left(\frac{1}{4\pi s} \right)^{d/2} \, .
    \end{aligned}
    \end{equation}
    For this, we used spherical coordinates. On the other hand, note that the coincidence limit of the flat heat kernel \eqref{eq:ALESSIABENJAMIN_HK_flat} indeed evaluates to the same expression,
    \begin{equation}
        H(x,x;s) = \left(\frac{1}{4\pi s} \right)^{d/2} \, .
    \end{equation}

The plan is now to compute $H(x,y;s)$ in a general spacetime. For this, we take inspiration from the flat heat kernel, and make the ansatz that
\begin{equation}
    H(x,y;s) = \left(\frac{1}{4\pi s} \right)^{d/2} \, e^{-\frac{\sigma(x,y)}{2s}} \, \Omega(x,y;s) \, .
\end{equation}
In this, $\sigma(x,y)$ is half of the squared geodesic distance between the points $x$ and $y$, sometimes also called the ``world function''. By this definition, its coincidence limit vanishes, 
\begin{equation}
    \sigma(x,x) \equiv \bar\sigma = 0 \, .
\end{equation}
It also satisfies the \textbf{fundamental equation for the world function}
    \begin{equation}\label{eq:ALESSIABENJAMIN_world_function}
        \frac{1}{2} (\covD_\mu \sigma) (\covD^\mu \sigma) = \sigma \, .
    \end{equation}
Moreover, $\Omega$ is the function that we have to solve for --- by consistency, it should be the identity plus curvature corrections. Note that since $s\Delta$ is dimensionless for the heat kernel to be well-defined, we expect an expansion of $\Omega$ where each factor of curvature is accompanied by a factor of $s$, and similarly, each factor of a covariant derivative comes with a factor of $\sqrt{s}$, for the overall dimension to be consistent.

We will now derive an equation that determines $\Omega$. For this, we insert our ansatz for $H(x,y;s)$ into the heat equation. Using \eqref{eq:ALESSIABENJAMIN_world_function}, after a short computation we find
\begin{equation}
    \left[ \left( -\frac{d}{2s} + \partial_s - \covD^2 + \frac{1}{2s} (\covD^2 \sigma(x,y)) \right) \Omega(x,y;s) + \frac{1}{s} (\covD_\mu \sigma(x,y)) (\covD^\mu \Omega(x,y;s)) \right] = 0 \, .
\end{equation}
All covariant derivatives here and in the following are with respect to the variable $x$. To solve this, we make a series expansion of $\Omega$ in $s$:\footnote{This is motivated by the consideration of the mass dimension above, that curvatures come with powers of $s$.}
\begin{equation}
    \Omega(x,y;s) \sim \sum_{n\geq 0} s^n \, A_n(x,y) \, , \qquad s \to 0 \, .
\end{equation}
with the boundary condition $A_0(x,x) \equiv \overline{A_0} = 1$. Inserting this ansatz into the above equation and solving individually for each power of $s$, we find an important recursive relation.
\begin{tcolorbox}
\textbf{Recursion relation for scalar heat kernel coefficients}:
    \begin{equation}\label{eq:ALESSIABENJAMIN_HK_recursion}
        \left( n - \frac{d}{2} + \frac{1}{2} \left( \covD^2 \sigma(x,y) \right) \right) A_n(x,y) + \left( \covD^\mu \sigma(x,y) \right) \left( \covD_\mu A_n(x,y) \right) - \covD^2 A_{n-1}(x,y) = 0 \, ,
    \end{equation}
    with $A_{-1}(x,y) = 0$ and $n \in \mathbbm N$.
\end{tcolorbox}
\noindent With this equation and \eqref{eq:ALESSIABENJAMIN_world_function}, one can recursively compute the coincidence limits of the heat kernel coefficients $A_n$. 

To illustrate the procedure, we will compute $\overline{A_1}$. For this, we first take the coincidence limit of \eqref{eq:ALESSIABENJAMIN_HK_recursion} for $n=1$ and get
\begin{equation}\label{eq:ALESSIABENJAMIN_A1}
    \left( 1 - \frac{d}{2} + \frac{1}{2} \overline{\covD^2\sigma} \right) \overline{A_1} + \overline{\covD^\mu \sigma} \, \overline{\covD_\mu A_1} - \overline{\covD^2 A_0} = 0 \, .
\end{equation}
From this we see that we have to compute several other coincidence limits to solve this equation for $\overline{A_1}$. We need $\overline{\covD^2\sigma}$, $\overline{\covD^\mu \sigma}$, $\overline{\covD^2 A_0}$ and potentially even $\overline{\covD_\mu A_1}$. The latter term looks problematic, as it would ruin the recursive solution strategy, but we will see that it is actually harmless. Let us also point out that a covariant derivative of a coincidence limit is not the same as the coincidence limit of a covariant derivative. For example, $\overline{\covD^2 A_0} \neq \covD^2\overline{A_0} = \covD^2 1 = 0$. We can however pull multiplicative operators out of the coincidence limit. This includes raising and lowering indices ``through'' the coincidence limit, \eg{}, $g^{\mu\nu} \overline{\covD_\nu A_0} = \overline{\covD^\mu A_0}$, but also pulling out curvature tensors, \eg{} $\overline{R_{\mu\nu} \, A_0} = R_{\mu\nu} \, \overline{A_0}$.

As a next step, let us try to derive an equation for $\overline{\covD^2 A_0}$. For this, we act with $\covD^2$ on \eqref{eq:ALESSIABENJAMIN_HK_recursion} with $n=0$ and take the coincidence limit. We find
\begin{equation}\label{eq:ALESSIABENJAMIN_D2A0}
\begin{aligned}
    \frac{1}{2} \overline{\covD^2 \covD^2 \sigma} \, \overline{A_0} &+ \overline{\covD^\mu \covD^2 \sigma} \, \overline{\covD_\mu A_0} + \left( -\frac{d}{2} + \frac{1}{2} \overline{\covD^2\sigma} \right) \overline{\covD^2 A_0} \\
    &+ \overline{\covD^2 \covD^\mu \sigma} \, \overline{\covD_\mu A_0} + \overline{\covD^\mu \sigma} \, \overline{\covD^2 \covD_\mu A_0} + 2 \overline{\covD^\mu \covD^\nu \sigma} \, \overline{\covD_\mu \covD_\nu A_0} = 0 \, .
\end{aligned}
\end{equation}
Once again, there are several terms that have the potential to break our recursive solution strategy. In any case, we will certainly need contractions of the coincidence limit of up to at least four covariant derivatives acting on the world function. Let us compute these and see if there are any simplifications.

The starting point for this computation is the property \eqref{eq:ALESSIABENJAMIN_world_function} of the world function. Taking the coincidence limit of it, we find
\begin{equation}
    \frac{1}{2} \overline{\covD^\mu \sigma} \, \overline{\covD_\mu \sigma} = \overline{\sigma} = 0 \qquad \Rightarrow \qquad \overline{\covD_\mu \sigma} = 0 \, .
\end{equation}
This relation is very helpful, as it eliminates problematic terms in both \eqref{eq:ALESSIABENJAMIN_A1} and \eqref{eq:ALESSIABENJAMIN_D2A0}. Next, we take a covariant derivative of \eqref{eq:ALESSIABENJAMIN_world_function} and once again take the coincidence limit. Here, we simply find a true relation:
\begin{equation}
    0 = \overline{\covD^\mu \sigma} \, \overline{\covD_\alpha \covD_\mu \sigma} = \overline{\covD_\alpha \sigma} = 0 \, .
\end{equation}
Following the same strategy, at the next step we find
\begin{equation}
    \left( \covD_\beta \covD_\alpha \covD_\mu \sigma(x,y) \right) \left( \covD^\mu \sigma(x,y) \right) + \left( \covD_\alpha \covD_\mu \sigma(x,y) \right) \left( \covD_\beta \covD^\mu \sigma(x,y) \right) = \covD_\beta \covD_\alpha \sigma(x,y) \, .
\end{equation}
Taking the coincidence limit, we have
\begin{equation}
    \overline{\covD_\alpha \covD_\mu \sigma} \, \overline{\covD_\beta \covD^\mu \sigma} = \overline{\covD_\beta \covD_\alpha \sigma} \, .
\end{equation}
This signifies that $\overline{\covD^\alpha \covD_\beta \sigma}$ is idempotent. Since it also has to be a geometric quantity, it has to be the identity/metric,
\begin{equation}
    \overline{\covD_\mu \covD_\nu \sigma} = g_{\mu\nu} \, .
\end{equation}
As a direct consequence,
\begin{equation}
    \overline{\covD^2\sigma} = d \, .
\end{equation}
Repeating the same procedure, at the next step we get as an intermediate result
\begin{equation}
    \overline{\covD_\beta \covD_\alpha \covD_\gamma \sigma} + \overline{\covD_\gamma \covD_\alpha \covD_\beta \sigma} = 0 \, .
\end{equation}
Let us quickly recall the commutator of covariant derivatives acting on an arbitrary tensor $T$. We have
    \begin{equation}\label{eq:ALESSIABENJAMIN_covDcommutator}
        \left[ \covD_\mu, \covD_\nu \right] T_{\alpha_1 \dots \alpha_n} = \sum_{i=1}^n \, R_{\mu\nu\alpha_i}^{\phantom{\mu\nu\alpha_i}\beta} T_{\alpha_1 \dots \alpha_{i-1} \beta \alpha_{i+1} \dots \alpha_n} \, .
    \end{equation}
Sorting covariant derivatives into the same order, we find
\begin{equation}
    2 \overline{\covD_\gamma \covD_\beta \covD_\alpha \sigma} + R_{\alpha\delta\beta\gamma} \overline{\covD^\delta \sigma} = 0 \, ,
\end{equation}
or
\begin{equation}
    \overline{\covD_\mu \covD_\nu \covD_\rho \sigma} = 0 \, .
\end{equation}
Finally, after a lengthier (but completely analogous) computation, one arrives at
\begin{equation}
    \overline{\covD_\mu \covD_\nu \covD_\rho \covD_\sigma \sigma} = -\frac{1}{3} \left( R_{\mu\rho\nu\sigma} + R_{\mu\sigma\nu\rho} \right) \, .
\end{equation}
Let us summarize these results:
    The first few coincidence limits of covariant derivatives of the world function $\sigma(x,y)$ read:
    \begin{align}
        \overline{\sigma} = \overline{\covD_\mu \sigma} = \overline{\covD_\mu \covD_\nu \covD_\rho \sigma} &= 0 \, , \\
        \overline{\covD_\mu \covD_\nu \sigma} &= g_{\mu\nu} \, , \\
        \overline{\covD_\mu \covD_\nu \covD_\rho \covD_\sigma \sigma} &= -\frac{1}{3} \left( R_{\mu\rho\nu\sigma} + R_{\mu\sigma\nu\rho} \right) \, .
    \end{align}
    As a direct consequence, we find
    \begin{equation}
        \overline{\covD^2 \sigma} = d \, , \qquad \overline{\covD^2 \covD^2 \sigma} = -\frac{2}{3} R \, .
    \end{equation}
Let us now use these results to first compute $\overline{\covD^2 A_0}$, and then finally $\overline{A_1}$. Using \eqref{eq:ALESSIABENJAMIN_D2A0} and the results just derived, we have
\begin{equation}
    \overline{\covD^2 A_0} = \frac{1}{6} R \, .
\end{equation}
Inserting this into \eqref{eq:ALESSIABENJAMIN_A1}, we find
\begin{equation}
    \overline{A_1} = \frac{1}{6} R \, .
\end{equation}
Note how all terms that would break the recursive solution strategy are harmless after all, as they multiply derivatives of the world function whose coincidence limits vanish. It is straightforward to check that this is the case for all heat kernel coefficients.

Summarizing, after a somewhat painful process, we have computed the first non-trivial heat kernel coefficient.
\begin{tcolorbox}
    The coincidence limit of the function $\Omega$ of the heat kernel has the asymptotic expansion
    \begin{equation}
        \overline{\Omega}(s) \sim 1 + \frac{1}{6} s \, R + \orderneglected(s^2) \, .
    \end{equation}
This defines the heat kernel coefficient $A_1$.
\end{tcolorbox}
\noindent While it is clear that this procedure can be extended order by order, the algebraic complexity becomes very large very quickly~\cite{Barvinsky:1993en}, and one should use computer tensor algebra to automatize the computation~\cite{xActwebpage, Nutma:2013zea}.

Before coming back to our original super trace, let us briefly comment on how the heat kernel can be generalized. First of all, we can consider a field with a super-index, $\Phi^A$. Then, the kinetic operator would have two such indices, $\Delta^A_{\phantom{A}B}$. The first change is that $\Omega(x,y;s)$ as well as the $A_n(x,y)$ become bi-tensors, with one super-index attached to each position. This for example also entails that the boundary condition for the recursion reads $\overline{A^{\phantom{0}A}_{0\phantom{A}B}} = \mathbbm 1^A_{\phantom{A}B}$, where $\mathbbm 1 = \delta \Phi/\delta\Phi$ refers to the identity in the corresponding space. The remaining trace that we indicated in the general formula \eqref{eq:ALESSIABENJAMIN_HK_super_trace_def} then corresponds to a contraction of the remaining open indices with this identity. The second change is that due to the additional indices, commutators of covariant derivatives acting on the $A_n$ generate generalized ``field strength'' terms, $\left[ \gaugecovD_\mu, \gaugecovD_\nu \right]^A_{\phantom{A}B} X^B = \mathcal F^{\phantom{\mu\nu}A}_{\mu\nu\phantom{A}B} X^B$, where $\gaugecovD$ is the gauge-covariant derivative attached to the corresponding bundle structure. Here it is important to remember the bi-tensor structure of all involved objects, and that covariant derivatives only act on $x$. The specific form of the generalized field strength tensor depends on the specific index content. Last but not least, one can add a non-derivative term to the original operator, that we usually call an endomorphism, without having to change the overall procedure. This is sometimes handy when one considers traces of operators shifted by a curvature, \eg{} $\Delta+R/6$.

Let us finally come back to our original task, namely to compute
\begin{equation}
    \frac{1}{2} \text{STr} \, W(\Delta) \, .
\end{equation}
How do we relate this to the heat kernel? To facilitate this, we can use a standard trick, and assume that $W$ has an inverse Laplace transform.
\begin{tcolorbox}
Suitable functions $W$ can be represented as
    \begin{equation}
        W(x) = \int_0^\infty \rmd{}s \, \tilde W(s) \, e^{-s \, x} \, .
    \end{equation}
    If the integral exists, the function $\tilde W$ defines the \textbf{inverse Laplace transform of $W$}.
\end{tcolorbox}
In practice, we can start by assuming that our function $W$ is the Laplace transform of some other function $\tilde W$. If we do this, we can then re-write
\begin{equation}
    \frac{1}{2} \text{STr} \, W(\Delta) = \frac{1}{2} \text{STr} \int_0^\infty \rmd{}s \, \tilde W(s) \, e^{-s\, \Delta} \, .
\end{equation}
Assuming that the above super trace and the integral commute, we can finally combine all the results of this subsection to get the relation
\begin{equation}
    \frac{1}{2} \text{STr} \, W(\Delta) \sim \frac{1}{2} \int \rmd^dx \, \sqrt{g} \, \int_0^\infty \rmd{}s \, \tilde W(s) \left(\frac{1}{4\pi s} \right)^{d/2} \, \left[ 1 + \frac{1}{6} s \, R + \orderneglected(s^2) \right] \, .
\end{equation}
This expression suggests that we have traded one evil for another --- we performed the super trace, but now it seems as if we have to compute the inverse Laplace transform. This is however not the case, as the inverse Laplace transform can be related back to the function $W$. Specifically, in the following we will prove that we have
    \begin{equation}
        \int_0^\infty \rmd{}s \, \tilde W(s) \, s^{-n} = \frac{1}{\Gamma(n)} \int_0^\infty \rmd{}z \, z^{n-1} \, W(z) \, , \qquad n > 0 \, ,
    \end{equation}
whereas we also have
    \begin{equation}
        \int_0^\infty \rmd{}s \, \tilde W(s) \, s^n = (-1)^n W^{(n)}(0) \, , \qquad n \geq 0 \, .
    \end{equation}
These two relations allow us to map integrals over the inverse Laplace transform to integrals over the original function.\footnote{Recall that this is indeed the best that we can hope for: in flat spacetime, the super trace is a loop integral.} Note that these cases cover all terms in the above expansion: for low orders, we can use the first case, whereas if the order of the expansion is high enough, we fall into the second case. Let us prove both formulas, starting with a ``backward'' proof of the first. Starting with the right-hand side, inserting the representation via the inverse Laplace transform, and commuting integrals, we find
\begin{equation}
\begin{aligned}
    \frac{1}{\Gamma(n)} \int_0^\infty \rmd{}z \, z^{n-1} \, W(z) &= \frac{1}{\Gamma(n)} \int_0^\infty \rmd{}z \, z^{n-1} \, \int_0^\infty \rmd{}s \, \tilde W(s) \, e^{-s \, z} \\
    &= \int_0^\infty \rmd{}s \, \tilde W(s) \, \frac{1}{\Gamma(n)} \, \int_0^\infty \rmd{}z \, z^{n-1} \, e^{-s \, z} \\
    &= \int_0^\infty \rmd{}s \, \tilde W(s) \, \frac{1}{\Gamma(n)} \, (-\partial_s)^{n-1} \, \int_0^\infty \rmd{}z \, e^{-s \, z} \\
    &= \int_0^\infty \rmd{}s \, \tilde W(s) \, \frac{1}{\Gamma(n)} \, (-\partial_s)^{n-1} \, \frac{1}{s} \\
    &= \int_0^\infty \rmd{}s \, \tilde W(s) \, s^{-n} \, .
\end{aligned}
\end{equation}
For the proof of the second formula, we compute
\begin{equation}
\begin{aligned}
    \int_0^\infty \rmd{}s \, \tilde W(s) \, s^n &= \left[ \int_0^\infty \rmd{}s \, \tilde W(s) \, s^n \, e^{-s \, z} \right] \Bigg|_{z=0} \\
    &= \left[ (-\partial_z)^n \int_0^\infty \rmd{}s \, \tilde W(s) \, e^{-s \, z} \right] \Bigg|_{z=0} \\
    &= \left[ (-\partial_z)^n W(z) \right] \Big|_{z=0} = (-1)^n W^{(n)}(0) \, .
\end{aligned}
\end{equation}
We thus find the final expression for the super trace.
\begin{tcolorbox}
The \textbf{super trace} of a function $W(\Delta)$ of the Laplace operator is given by
    \begin{equation}
    \begin{aligned}
        \frac{1}{2} \text{STr} \, W(\Delta) \sim \frac{1}{2} \frac{1}{(4\pi)^{d/2}} \int \rmd^dx \, \sqrt{g} \, \Bigg[ &\frac{1}{\Gamma\left(\frac{d}{2}\right)} \int_0^\infty \rmd{}z \, z^{\frac{d}{2}-1} \, W(z) \\
        &+ \frac{1}{6} \frac{1}{\Gamma\left(\frac{d}{2}-1\right)} \, R \,\int_0^\infty \rmd{}z \, z^{\frac{d}{2}-2} \, W(z) + \dots \Bigg] \, .
    \end{aligned}
    \end{equation}
\end{tcolorbox}
\noindent The remaining integrals are often called {\bf threshold integrals}. In our simple case, they are just regulator-dependent numbers multiplying the right power of $k$ --- more generally, they will depend on the couplings in the theory. For example, using the Litim regulator \eqref{eq:ALESSIABENJAMIN_litim-reg}, we find
\begin{equation}
    \int_0^\infty \rmd{}z \, z^{\frac{d}{2}-1} \, W(z) = \frac{4}{d} k^d \, , \qquad \int_0^\infty \rmd{}z \, z^{\frac{d}{2}-2} \, W(z) = \frac{4}{d-2} k^{d-2} \, .
\end{equation}
In the following, we will give two practical examples of the application of the \ac{FRG} to simple systems. We start with a simple anharmonic oscillator, and then discuss \ac{ASQG} in a simple approximation.

\subsubsection{Quantum-mechanical example: anharmonic oscillator}

The anharmonic oscillator is essentially a $(0+1)$-dimensional real scalar field theory, whose truncated \ac{EAA} can be written as
\begin{equation}
    \Gamma_k=\int \rmd\tau \left( \frac{1}{2}\dot{x}^2+ V_k(x) \right)\,.
\end{equation}
Technically, this can be seen as the lowest-order truncation of the derivative expansion~\eqref{eq:ALESSIABENJAMIN_der-exp} and goes under the name of \emph{local potential approximation}.

The first and simplest step to compute beta functions is always to evaluate the left-hand side of the flow equation, which in this case reads
\begin{equation}
    k\partial_k \Gamma_k\equiv \int \rmd\tau k \, \partial_k V_k\,.
\end{equation}
For the evaluation of the right-hand side, the general strategy is to always compute ``objects'' step-by-step, proceeding from the innermost bracket to the outermost one. The practical steps are thus~\cite{Benedetti:2010nr}:
\begin{enumerate}
    \item In the case of gauge theories, write down the gauge-fixing action (to be added to the physical starting action) and the Faddeev-Popov ghost action (which contributes separately to the right-hand side of the Wetterich equation),
    
    \item Determine the Hessian matrix $\Gamma_k^{(2)}$ with respect to all the fields in the system, and regularize it by adding the regulator terms $\mathcal{R}_k$ with appropriate tensorial structures,
    
    \item Invert the sum $(\Gamma_k^{(2)}+\mathcal{R}_k)$ to obtain the modified inverse propagator $\propG_k$,
    
    \item Perform the tensorial multiplication of the resulting modified inverse propagator and the $k$-derivative of the regulator, $k \partial_k \mathcal{R}_k$,
    
    \item Evaluate the functional traces, either by switching to momentum space and performing the momentum integrals (this is only possible when expanding about a flat spacetime), or by applying the heat kernel techniques and formulas, as detailed in the previous subsection,
    
    \item Adopt a projection scheme to extract the beta functions by comparing the left- and right-hand sides of the flow equation.
\end{enumerate}
The case of the anharmonic oscillator is among the simplest examples, and the steps above greatly simplify. Indeed, we have a single scalar field (that does not require gauge fixing, and thus no Faddeev-Popov ghosts). The Hessian of this system is the operator-valued $1\times1$ matrix
\begin{equation}
    \Gamma_k^{(2)}= (-\partial_\tau^2 + V_k''(x))\delta(\tau-\tau')\,.
\end{equation}
At this point, we need to choose and add a regulator term. We shall pick the Litim regulator~\eqref{eq:ALESSIABENJAMIN_litim-reg}, whose $k$-derivative reads
\begin{equation}
    k \partial_k \mathcal{R}_k^{\text{Litim}}=2k^2\theta(1-p^2/k^2) - 2 \frac{p^2}{k^2} \left( k^2 - p^2\right) \delta(1-p^2/k^2) \, .
\end{equation}
To project the flow equation, we now choose $x=\text{const.}$, so that we can apply the standard Fourier transform on the right-hand side, and the STr reduces to a momentum integral. Note that the second term in the $k$-derivative of the regulator integrates to zero, so that we find
\begin{equation}
    k\partial_k V_k=\frac{1}{2}\int_{-\infty}^{+\infty} \frac{\rmd{}p_\tau }{2\pi}\frac{2k^2\theta(1-p_\tau^2/k^2)}{k^2+V''_k} = \frac{1}{\pi} \frac{k^3}{k^2+V_k''} \, .
\end{equation}
This is the flow equation (or, beta functional) for the effective potential, and it can in principle be solved by only specifying initial conditions for $V_k$~\cite{Borchardt:2016pif}. Nonetheless, to illustrate one possible projection scheme to extract beta functions, we shall now focus on a \emph{truncated} potential with a finite number of terms, \eg{},
\begin{equation}\label{eq:ALESSIABENJAMIN_truncated-potential}
    V_k=E_k+\frac{1}{2!}\omega_kx^2+\frac{1}{4!}\lambda_k x^4\,.
\end{equation}
The truncated system has three coupling constants, $g\equiv\{E,\omega,\lambda\}$, for which we would like to compute the corresponding beta functions $\beta_{g_i}\equiv k\partial_k g_i$. To this scope, let us replace the potential above in its flow equation
\begin{equation}
    \beta_E+\frac{1}{2!}\beta_\omega x^2+\frac{1}{4!}\beta_\lambda x^4 = \frac{1}{\pi} \frac{k^3}{k^2+\omega_k+ \lambda_k x^2/2}\,.
\end{equation}
In principle, different projection schemes can be used at this point. A particularly straightforward one is to expand the right-hand side of the flow equation about $x=0$, so that it will be a polynomial with a similar structure as the one on the left-hand-side of the flow (modulo higher-order terms that, on the one hand, should be neglected to be consistent with the original truncation~\eqref{eq:ALESSIABENJAMIN_truncated-potential}, and, on the other hand, indicate that our original truncation is not closed under the \ac{RG} flow and needs to be systematically improved by adding higher-order terms in a step-by-step fashion). Notably, a similar projection strategy can be used in the case of $f(R)$ gravity, where one can use a constantly curved background like a sphere to evaluate the beta functional. If terms with other tensorial structure like $R_{\mu\nu}R^{\mu\nu}$ are included, this is not enough, and one needs a more general projection strategy.

Employing a Taylor expansion of the right-hand side, and comparing the coefficients of the corresponding powers of $x$ on the left- and right-hand sides finally yields the beta functions 
\begin{equation}
\begin{aligned}
\partial_k{E}_k&=\frac{1}{\pi} \frac{k^2}{k^2+\omega_k}\,,\\
\partial_k \omega_k & =-\frac{2}{\pi} \frac{k^2}{\left(k^2+\omega_k\right)^2} \frac{\lambda_k}{2}\,, \\
\partial_k\lambda_k & =\frac{24}{\pi} \frac{k^2}{\left(k^2+\omega_k\right)^3}\left(\frac{\lambda_k}{2}\right)^2\,.
\end{aligned}
\end{equation}
which can be used, \eg{}, to compute the energy levels of the anharmonic oscillator once the energy has been appropriately normalized. The latter is ensured by subtracting the vacuum energy from the flow, which would otherwise give rise to a flow of the ground state energy even in the absence of anharmonic terms. A more detailed pedagogical discussion of these beta functions can be found in~\cite{Reichert:2020mja}.

\subsubsection{Gravity in the Einstein-Hilbert truncation}\label{sect:ALESSIABENJAMIN_EH}

Let us now study the simplest approximation to investigate the mechanism of asymptotic safety in four dimensions. Concretely, we approximate the \ac{EAA} by just the Einstein-Hilbert action, with couplings now depending on $k$,
\begin{equation}
    \Gamma_k \simeq \frac{1}{16\pi \GNk} \int \rmd^4x \, \sqrt{g} \left[ 2\CCk - R \right] \, .
\end{equation}
The action possesses diffeomorphism invariance. The symmetry is generated by the Lie derivative along a vector field $v$,
\begin{equation}
    \mathfrak L_v g_{\mu\nu} = \covD_\mu v_\nu + \covD_\nu v_\mu \, .
\end{equation}
Consequently, $\Gamma_k$ is invariant under $g_{\mu\nu} \to g_{\mu\nu} + \mathfrak L_v g_{\mu\nu}$. To compute the \ac{RG} flow, we have to gauge-fix this symmetry. Beyond this, we also need a regularization quadratic in the field --- which clearly clashes with the non-linear diffeomorphism symmetry and the compatibility condition of the metric, $\covD_\mu g_{\alpha\beta}=0$. Both these problems can be solved by the background field method that we discuss next. Once we have introduced this, we discuss gauge fixing and Faddeev-Popov ghosts, the computation of the two-point function, regularization, performing the trace, and finally reading off and analyzing the beta functions. In spite of the simple approximation, the following is going to be \textbf{very technical}; the reader who is interested in the \textbf{core mechanism and physics of \ac{ASQG}} may jump to the final result of this section: the \hyperref[sect:ALE-BEN-betas-end]{beta functions}.

\subsubsubsection*{Background field method}

The {\bf background field method} is an ingenious way to perform \ac{RG} computations in gauge theories. The underlying idea is to split the gauge field --- in our case the metric --- into an \emph{arbitrary} but fixed background, plus fluctuations about the background. One then integrates over all admissible fluctuations. This split can be implemented in different ways. The most straightforward way is a {\bf linear parameterization},
    \begin{equation}\label{eq:ALESSIABENJAMIN_linear_parameterization}
    g_{\mu\nu} = \bar g_{\mu\nu} + h_{\mu\nu} \, .
\end{equation}
Here, $\bar g$ is the arbitrary (in general \emph{not} flat) background metric, and $h$ is the fluctuation. This is the parameterization that we will use here. An alternative with some nice conceptual properties is the {\bf exponential parameterization}~\cite{Nink:2014yya},
\begin{equation}
    g_{\mu\nu} = \bar g_{\mu\rho} \exp \left[ \bar g^{-1} h \right]^\rho_{\phantom{\rho}\nu} \, .
\end{equation}
Some aspects of different parameterizations are discussed in~\cite{Demmel:2015zfa, Knorr:2022mvn}. Let us emphasize two important points:
\begin{itemize}
    \item the background metric naturally induces geometric background quantities like background curvatures and covariant derivatives --- they are also indicated by an overbar, and
    \item indices of background quantities and fluctuations are raised and lowered with the \emph{background} metric.
\end{itemize}

What does the background field method do to the gauge symmetry? As a matter of fact, there are now two independent implementations of the symmetry that leave the action invariant:
\begin{tcolorbox}
    The \textbf{quantum diffeomorphism transformation} reads
    \begin{equation}
    \begin{aligned}
        \bar g_{\mu\nu} &\to \bar g_{\mu\nu} \, , \\
        h_{\mu\nu} &\to h_{\mu\nu} + \mathfrak L_v \left( \bar g_{\mu\nu} + h_{\mu\nu} \right) \, .
    \end{aligned}
    \end{equation}
    The \textbf{background diffeomorphism transformation} instead are
    \begin{equation}
    \begin{aligned}
        \bar g_{\mu\nu} &\to \bar g_{\mu\nu} + \mathfrak L_v \bar g_{\mu\nu} \, , \\
        h_{\mu\nu} &\to h_{\mu\nu} + \mathfrak L_v h_{\mu\nu} \, .
    \end{aligned}
    \end{equation}
\end{tcolorbox}
\noindent Gauge fixing will break the symmetry under quantum diffeomorphisms and turn it into a \ac{BRST} symmetry, which in turn is further broken by the regulator. As discussed before, the breaking is encoded by a symmetry identity, and in this context is called the modified Slavnov-Taylor identity. We will not discuss it further. On the other hand, the symmetry attached to background diffeomorphisms is kept intact.

Apart from the standard gauge symmetry, the background field method introduces a new symmetry --- {\bf split symmetry}. By construction, if the background metric is shifted by an arbitrary amount, and we adjust the corresponding shift of the fluctuation to be the negative amount, the total metric stays the same. The accompanying symmetry identity is called the {\bf Nielsen identity}~\cite{Nielsen:1975fs, Fukuda:1975di}. It simply encodes that in the end, we only have one physical metric and not two independent ones. In other words, it ensures background independence. The Nielsen identity is once again modified by the introduction of gauge-fixing and regulator terms, and reads
\begin{equation}
    \mathcal N_k = \frac{\delta \Gamma_k}{\delta \bar g_{\mu\nu}} - \frac{\delta \Gamma_k}{\delta h_{\mu\nu}} - \left\langle \left[ \frac{\delta}{\delta \bar g_{\mu\nu}} - \frac{\delta}{\delta h_{\mu\nu}} \right] \left( S_\text{gf} + S_\text{gh} \right) \right\rangle - \frac{1}{2} \text{STr} \left[ \frac{1}{\sqrt{\bar g}} \frac{\delta \sqrt{\bar g} \mathcal R_k}{\delta \bar g_{\mu\nu}} \propG_k \right] = 0 \, .
\end{equation}
The explicit form shows that the identity controls the difference between the background and fluctuation dependence of the \ac{EAA}, and that the difference is generated by gauge fixing and regularization. Like the other symmetry identities, the modified Nielsen identity is the subject of active research~\cite{Pawlowski:2023gym}.

\subsubsubsection*{Gauge fixing and Faddeev-Popov ghosts}

We can now discuss the gauge-fixing term in gravity together with the ensuing Faddeev-Popov ghosts. At a practical level, this is necessary because the graviton two-point function is not invertible on-shell, and we add (gauge fixing) and subtract (ghosts) terms to circumvent this issue, as also discussed in \cref{sec:LUCA}.
\begin{tcolorbox}
The \textbf{gauge-fixing action} reads
    \begin{equation}
        S_\text{gf} = \frac{1}{2} \int \rmd^4x \, \sqrt{\bar g} \, \GFcondition_\mu \, \bar g^{\mu\nu} \, \GFcondition_\nu \, .
    \end{equation}
\end{tcolorbox}
\noindent Here, $\GFcondition_\mu$ is a linear gauge-fixing condition:
    \begin{equation}
        \GFcondition_\mu \equiv \GFoperator_\mu^{\phantom{\mu}\alpha\beta} h_{\alpha\beta} = \frac{1}{\sqrt{16\pi \GNk \GFalpha_k}} \left[ \delta_\mu^{\phantom{\mu}(\alpha} \bar \covD^{\beta)} - \frac{1+\GFbeta_k}{4} \bar g^{\alpha\beta} \bar \covD_\mu \right] h_{\alpha\beta} \, .
    \end{equation}
    The two quantities $\GFalpha_k,\GFbeta_k$ are gauge-fixing parameters --- $\GFbeta_k$ dictates the type of gauge fixing, whereas $\GFalpha_k$ dictates how sharply the gauge fixing is implemented.
The round brackets around indices indicate symmetrization with unit strength. While this is not the most general way to gauge-fix, it will suffice for our purposes. One can show that the {\bf Landau limit}, $\GFalpha_k\to0$, is a fixed point for both gauge parameters for any $\GFbeta_k=\GFbeta < 3$~\cite{Litim:2002ce, Knorr:2017fus}. Let us briefly mention that we pulled the factor of $\GNk$ and $\GFalpha_k$ into the gauge-fixing condition, in contrast to the perturbative treatment (cf. \eg{} \eqref{de-donder-gauge-fix}). In \ac{FRG} flows, this makes a difference for the beta functions, and we have chosen this convention here for convenience. This way, the spin one gauge mode in the graviton sector is still exactly canceled by the Faddeev-Popov ghost.

\begin{tcolorbox}
    For the above gauge fixing, the corresponding \textbf{Faddeev-Popov ghost action} reads
    \begin{equation}
        S_\text{gh} = \int \rmd^4x \, \sqrt{\bar g} \, \bar c^{\mu} \GFoperator_\mu^{\phantom{\mu}\alpha\beta} \, \mathfrak L_c g_{\alpha\beta} = \int \rmd^4x \, \sqrt{\bar g} \, \bar c^{\mu} \GFoperator_\mu^{\phantom{\mu}\alpha\beta} \, \left( \covD_\alpha c_\beta + \covD_\beta c_\alpha \right) \, .
    \end{equation}
\end{tcolorbox}
\noindent Here, $c$ and $\bar c$ are the ghost and anti-ghost fields. Note the occurrence of the full covariant derivative in the Lie derivative. We will see below that surprisingly, there is only a graviton-ghost-ghost vertex, that is, the ghost action is linear in the fluctuation $h$ in the linear parameterization \eqref{eq:ALESSIABENJAMIN_linear_parameterization}.\footnote{This feature is however not stable: higher-order vertices are induced along the \ac{RG} flow.}

\subsubsubsection*{Evaluating the super trace}

Having specified our starting point, we can now set out and compute the resulting \ac{RG} flow. Our approximation for the gauge-fixed \ac{EAA} reads
\begin{equation}
    \Gamma_k \simeq \frac{1}{16\pi \GNk} \int \rmd^4x \, \sqrt{g} \left[ 2\CCk - R \right] + \frac{1}{2} \int \rmd^4x \, \sqrt{\bar g} \, \GFcondition_\mu \, \bar g^{\mu\nu} \, \GFcondition_\nu + \int \rmd^4x \, \sqrt{\bar g} \, \bar c^{\mu} \GFoperator_\mu^{\phantom{\mu}\alpha\beta} \, \mathfrak L_c g_{\alpha\beta} \, .
\end{equation}
Our approximations in this are as follows:
\begin{itemize}
    \item we only consider terms in the action with up to two derivatives,
    \item we employ the {\bf background field approximation} when it comes to the Nielsen identity, \ie{}, we identify fluctuation and background derivatives of the effective action,
    \item we will neglect the running of the gauge parameters $\GFalpha_k,\GFbeta_k$.
\end{itemize}
The regulating action now has tensorial structure:
\begin{equation}
    \Delta S_k = \frac{1}{2} \int \rmd^4x \, \sqrt{\bar g} \, h_{\mu\nu} \, \mathcal R_k^{h,\mu\nu\rho\sigma} \, h_{\rho\sigma} + \int \rmd^4x \, \sqrt{\bar g} \, \bar c_\mu \, \mathcal R_k^{c,\mu\nu} \, c_\nu \, .
\end{equation}
The tensor-valued regulator functions $\mathcal R_k^{h,c}$ depend only on background objects, \ie{}, the background metric and its associated covariant derivatives and curvatures. We will specify their exact form a little later.

With this setup, the Wetterich equation reads, in some more detail,
\begin{equation}
    k \partial_k \Gamma_k = \frac{1}{2} \text{STr} \left[ \begin{pmatrix}
        \frac{\delta^2 \Gamma_k}{\delta h^2} + \mathcal R_k^h & \frac{\delta^2 \Gamma_k}{\delta h \delta \bar c} & \frac{\delta^2 \Gamma_k}{\delta h \delta c} \\
        \frac{\delta^2 \Gamma_k}{\delta \bar c \delta h} & \frac{\delta^2 \Gamma_k}{\delta \bar c^2} & \frac{\delta^2 \Gamma_k}{\delta \bar c \delta c} + \mathcal R_k^c \\
        \frac{\delta^2 \Gamma_k}{\delta c \delta h} & \frac{\delta^2 \Gamma_k}{\delta c \delta \bar c} + \mathcal R_k^c & \frac{\delta^2 \Gamma_k}{\delta c^2}
    \end{pmatrix}^{-1} \, k \partial_k \begin{pmatrix}
        \mathcal R_k^h & 0 & 0 \\
        0 & 0 & \mathcal R_k^c \\
        0 & \mathcal R_k^c & 0
    \end{pmatrix} \right] \, .
\end{equation}
The background field approximation consists concretely of evaluating the above expression at $h=c=\bar c=0$ \emph{after} having performed the second variation. Note that now, the super trace also includes a trace over field space, \ie{}, it is also a matrix trace for the above three-by-three matrix. Taking both these things into account, we see that the graviton and the ghost sector decouple, since any expression containing both graviton fluctuation and ghost fields is at least cubic in the fields, and thus outside of our approximation. Thus, we can write the flow as
\begin{equation}
    k \partial_k \Gamma_k \simeq \frac{1}{2} \text{STr} \left[ \left( \frac{\delta^2 \Gamma_k}{\delta h^2} + \mathcal R_k^h \right)^{-1} \, k \partial_k \mathcal R_k^h \right] + \text{STr} \left[ \left( \frac{\delta^2 \Gamma_k}{\delta \bar c \delta c} + \mathcal R_k^c \right)^{-1} \, k \partial_k \mathcal R_k^c \right] \, \Bigg|_{h=\bar c=c=0} \, .
\end{equation}
Recall that there will be a minus sign for the ghost term, stemming from the fact that they are Grassmann-valued. The factor of two comes because $\bar c$ and $c$ are independent fields, and the contribution of the two terms above is the same. Note that in both these expressions, there is still tensor structure that we have suppressed.

\paragraph{Ghost contribution.}

Let us start with the contribution of the ghosts, since it is much easier to compute. For this, we take the ghost action, insert the gauge fixing, and simplify the expression by sorting covariant derivatives:
    To compute the two-point function, we can replace the full covariant derivative with the background covariant derivative, since the expression is already quadratic in the fields. This gives
    \begin{equation}
    \begin{aligned}
        S_\text{gh} &\simeq \frac{1}{\sqrt{16\pi \GNk \GFalpha_k}} \int \rmd^4x \, \sqrt{\bar g} \, \bar c^\mu \left[ \delta_\mu^{\phantom{\mu}(\alpha} \bar \covD^{\beta)} - \frac{1+\GFbeta_k}{4} \bar g^{\alpha\beta} \bar \covD_\mu \right] \left( \bar \covD_\alpha c_\beta + \bar \covD_\beta c_\alpha \right) \\
        &= \frac{1}{\sqrt{16\pi \GNk \GFalpha_k}} \int \rmd^4x \, \sqrt{\bar g} \, \bar c^\mu \left[ \bar \covD^2 \delta_\mu^{\phantom{\mu}\alpha} + \bar \covD^\alpha \bar \covD_\mu - \frac{1+\GFbeta_k}{2} \bar \covD_\mu \bar \covD^\alpha \right] c_\alpha \\
        &= \frac{1}{\sqrt{16\pi \GNk \GFalpha_k}} \int \rmd^4x \, \sqrt{\bar g} \, \bar c^\mu \left[ \bar \covD^2 \delta_\mu^{\phantom{\mu}\alpha} + \frac{1-\GFbeta_k}{2} \bar \covD_\mu \bar \covD^\alpha + \bar R_\mu^{\phantom{\mu}\alpha} \right] c_\alpha \, .
    \end{aligned}
    \end{equation}
    In the last step, we used \eqref{eq:ALESSIABENJAMIN_covDcommutator} to commute covariant derivatives.
We will now set $\GFbeta_k=1$ to simplify this two-point function. Doing so, note that the operator then is simply a Laplacian with an endomorphism,
\begin{equation}
    S_\text{gh} \simeq - \frac{1}{\sqrt{16\pi \GNk \GFalpha_k}} \int \rmd^4x \, \sqrt{\bar g} \, \bar c^\mu \left[ \bar \Delta \delta_\mu^{\phantom{\mu}\alpha} - \bar R_\mu^{\phantom{\mu}\alpha} \right] c_\alpha \, ,
\end{equation}
so its heat kernel is easy to compute. As expected from a ghost field, the overall sign of the kinetic term is negative. Let us now choose a regularization that mimics the exact same structure. In particular, we will choose
\begin{equation}
    \Delta S_k^\text{gh} = - \frac{1}{\sqrt{16\pi \GNk \GFalpha_k}} \int \rmd^4x \, \sqrt{\bar g} \, \bar c^\mu \, R_k^c\left( \bar \Delta \mathbbm 1 - \bar{\text{Ric}} \right)_\mu^{\phantom{\mu}\alpha} c_\alpha \, .
\end{equation}
Some comments are in order. First, by $\bar{\text{Ric}}$ we mean the background Ricci tensor. Second, we included the $k$-dependent prefactor also in the regulator action --- this will be ``seen'' by the $k \partial_k$ acting on the regulator in the trace.\footnote{Note how we used $R_k^c$ instead of $\mathcal R_k^c$ to indicate this --- in this way, $\mathcal R_k^c = - \frac{1}{\sqrt{16\pi \GNk \GFalpha_k}} R_k^c$. In the literature, this difference is sometimes not indicated.} Third, the tensor-valued regulator function can be defined via an inverse Laplace transform:
\begin{equation}
    R_k^c\left( \bar \Delta \mathbbm 1 - \bar{\text{Ric}} \right)_\mu^{\phantom{\mu}\alpha} = \int_0^\infty \rmd{}s \, \tilde R(s) \, \left( \exp \left[ -s \left( \bar\Delta \mathbbm 1 - \bar{\text{Ric}} \right) \right] \right)_\mu^{\phantom{\mu}\alpha} \, .
\end{equation}
This is well-defined since the exponential has a globally convergent Taylor series.

With this in place, the ghost contribution to the super trace has the structural form
\begin{equation}
    -\text{STr} \, W(\bar \Delta \mathbbm 1 - \bar{\text{Ric}}) = - \int_0^\infty \rmd{}s \, \tilde W(s) \, \text{STr} \, e^{-s(\bar \Delta \mathbbm 1 - \bar{\text{Ric}})} \, .
\end{equation}
As alluded to in \cref{sec:ALESSIABENJAMIN_HK}, there are some changes to be taken into account compared to the scalar heat kernel, since the ghosts are vector fields, see the corresponding literature~\cite{Vassilevich:2003xt, Groh:2011dw}. Truncating the heat kernel at second order in derivatives, we then find (after a similar calculation as for the scalar heat kernel)
\begin{tcolorbox}
    The \textbf{contribution from the Faddeev-Popov ghosts to the \ac{RG} flow} in our setup is
    \begin{equation}
    \begin{aligned}
        - \frac{1}{16\pi^2} \int \rmd^4x \, \sqrt{\bar g} \, \Bigg[ & 4 \int_0^\infty \rmd{}z \, z \, \frac{k \partial_k R_k^c(z) - \frac{1}{2} \frac{k \partial_k (\GNk \GFalpha_k)}{\GNk \GFalpha_k} R_k^c(z)}{z + R_k^c(z)} \\
        & + \frac{5}{3} \bar R \, \int_0^\infty \rmd{}z \, \frac{k \partial_k R_k^c(z) - \frac{1}{2} \frac{k \partial_k (\GNk \GFalpha_k)}{\GNk \GFalpha_k} R_k^c(z)}{z + R_k^c(z)} + \dots \Bigg] \, .
    \end{aligned}
    \end{equation}
\end{tcolorbox}
\noindent The derivation of the ghost contribution to the super trace was relatively straightforward. The graviton contribution, which we are going to discuss next, is going to be a bit more involved.

\paragraph{Graviton contribution.}

Let us now compute the graviton contribution to the \ac{RG} flow. For this, we have to expand the diffeomorphism-invariant part of the \ac{EAA} to second order in the fluctuation $h$. Recall that we use a linear parameterization. In the following, we will often omit the indices and use a matrix notation where indices of tensors are understood to be in their defining position, \eg{} two lower indices for both the metric and its fluctuation.\footnote{With this notation, we will indicate the determinant explicitly for clarity.} In such a notation, we can write
\begin{equation}
    g = \bar g + h = \bar g \left( \mathbbm 1 + \bar g^{-1} h \right) \, .
\end{equation}
Note now that the term in brackets is a tensor of rank $(1,1)$. For such tensors, matrix multiplication is covariant, \ie{}, the matrix product of two rank $(1,1)$ tensors is again a rank $(1,1)$ tensor. This observation will be extremely useful in the following.

To expand the action, we have to expand both $\sqrt{g}$ and $R$ to second order. Let us first discuss the determinant of the metric.
    Using matrix notation, we compute
    \begin{align}
        \det g = \det (\bar g + h) &= \det \left[ \bar g \left( \mathbbm 1 + \bar g^{-1} h \right) \right] \nonumber \\
        &= \left( \det \bar g \right) \, \det \left[ \mathbbm 1 + \bar g^{-1} h \right] \nonumber \\
        &= \left( \det \bar g \right) \, \exp \left[ \tr \ln \left( \mathbbm 1 + \bar g^{-1} h  \right) \right] \nonumber \\
        &= \left( \det \bar g \right) \, \exp \left[ \tr \sum_{n\geq 1} - \frac{(-1)^n}{n} \left( \bar g^{-1} h  \right)^n \right] \nonumber \\
        &= \left( \det \bar g \right) \, \exp \left[ - \sum_{n\geq 1} \frac{(-1)^n}{n} \tr \left\{ \left( \bar g^{-1} h  \right)^n \right\} \right] \nonumber \\
        &\simeq \left( \det \bar g \right) \, \left[ 1 + h^\alpha_{\phantom{\alpha}\alpha} + \frac{1}{2} h^\alpha_{\phantom{\alpha}\alpha} h^\beta_{\phantom{\beta}\beta} - \frac{1}{2} h^{\alpha\beta} h_{\alpha\beta} + \dots \right] \, .
    \end{align}
    Here, we used that the determinant of a product is the product of determinants. We also used a well-known formula to relate the determinant of a matrix to the exponential of the trace of its logarithm.\footnote{One can convince oneself of this formula by noting that for a finite-dimensional matrix $M$, $\det M$ is the product of its eigenvalues. Then writing $\lambda=e^{\ln \lambda}$ and combining the exponents, we find that $\det M$ can be written as the exponential of the sum of the logarithms of its eigenvalues.} We finally expanded the functions in a power series. Recall that indices are raised and lowered with the background metric, \eg{}, $h^\alpha_{\phantom{\alpha}\alpha} = \bar g^{\alpha\beta} h_{\alpha\beta}$.
Taking the square root of this expression, one can once more expand in powers of $h$, to wit
    \begin{equation}
        \sqrt{\det g} \simeq \sqrt{\det \bar g} \, \left[ 1 + \frac{1}{2} h^\alpha_{\phantom{\alpha}\alpha} + \frac{1}{8} h^\alpha_{\phantom{\alpha}\alpha} h^\beta_{\phantom{\beta}\beta} - \frac{1}{4} h^{\alpha\beta} h_{\alpha\beta} + \dots \right] \, .
    \end{equation}
To compute the Ricci scalar, we first note that
\begin{equation}
    R = g^{\mu\nu} R_{\mu\nu} \, ,
\end{equation}
and that the Ricci tensor can be expressed purely in terms of the connection and its partial derivatives. For the inverse metric above, using similar techniques as for the determinant, one can show that
    the inverse metric expanded in fluctuations reads
    \begin{equation}
        g^{-1} = \left( \mathbbm 1 + \bar g^{-1} h \right)^{-1} \, \bar g^{-1} \, .
    \end{equation}
    Restoring indices and truncating and second order, we have
    \begin{equation}
        g^{\mu\nu} \simeq \bar g^{\mu\nu} - h^{\mu\nu} + h^\mu_{\phantom{\mu}\alpha} h^{\alpha\nu} + \dots \, .
    \end{equation}
To expand the Christoffel symbol, let us first recall its definition:
\begin{equation}
    \Gamma^\mu_{\phantom{\mu}\alpha\beta} = \frac{1}{2} g^{\mu\nu} \left( \partial_\alpha g_{\nu\beta} + \partial_\beta g_{\nu\alpha} - \partial_\nu g_{\alpha\beta} \right) \, .
\end{equation}
It turns out that it is most convenient to first consider the Christoffel symbol with all lower indices, as it is linear in the metric.
    We expand the Christoffel symbol by inserting the split of the metric and then converting partial derivatives to background covariant derivatives:
    \begin{equation}
    \begin{aligned}
        \Gamma_{\nu\alpha\beta} &= \frac{1}{2} \left( \partial_\alpha g_{\nu\beta} + \partial_\beta g_{\nu\alpha} - \partial_\nu g_{\alpha\beta} \right) \\
        &= \frac{1}{2} \left( \partial_\alpha \bar g_{\nu\beta} + \partial_\beta \bar g_{\nu\alpha} - \partial_\nu \bar g_{\alpha\beta} \right) + \frac{1}{2} \left( \partial_\alpha h_{\nu\beta} + \partial_\beta h_{\nu\alpha} - \partial_\nu h_{\alpha\beta} \right) \\
        &= \bar \Gamma_{\nu\alpha\beta} + \frac{1}{2} \left( \bar \covD_\alpha h_{\nu\beta} + \bar \covD_\beta h_{\nu\alpha} - \bar \covD_\nu h_{\alpha\beta} \right) + \bar \Gamma^\mu_{\phantom{\mu}\alpha\beta} h_{\mu\nu} \\
        &= \bar \Gamma^\mu_{\phantom{\mu}\alpha\beta} \left( \bar g_{\mu\nu} + h_{\mu\nu} \right) + \frac{1}{2} \left( \bar \covD_\alpha h_{\nu\beta} + \bar \covD_\beta h_{\nu\alpha} - \bar \covD_\nu h_{\alpha\beta} \right) \\
        &= \bar \Gamma^\mu_{\phantom{\mu}\alpha\beta} g_{\mu\nu} + \frac{1}{2} \left( \bar \covD_\alpha h_{\nu\beta} + \bar \covD_\beta h_{\nu\alpha} - \bar \covD_\nu h_{\alpha\beta} \right) \, .
    \end{aligned}
    \end{equation}
After this short computation, we find
    \begin{equation}
        \Gamma^\mu_{\phantom{\mu}\alpha\beta} = \bar \Gamma^\mu_{\phantom{\mu}\alpha\beta} + \frac{1}{2} g^{\mu\nu} \left( \bar \covD_\alpha h_{\nu\beta} + \bar \covD_\beta h_{\nu\alpha} - \bar \covD_\nu h_{\alpha\beta} \right) \, .
    \end{equation}
    Note that we kept the inverse of the full metric to keep the notation compact.

With these computations done, all basic ingredients are ready, and everything in the action can be expanded. This is a somewhat lengthy process and we will just show the result. For our general gauge choice, and neglecting boundary terms, the part of the \ac{EAA} that is quadratic in $h$ reads
\begin{equation}\label{eq:ALESSIABENJAMIN_grav_two_point_function_gen}
\begin{aligned}
    \Gamma_k^{h^2} = \frac{1}{32\pi \GNk} \int \rmd^4x \, \sqrt{\bar g} \, h_{\mu\nu} \Bigg[ &\left( \bar\Delta - 2\CCk + \frac{2}{3} \bar R \right) \, \mathbbm 1^{\mu\nu\rho\sigma} - 2 \bar C^{\mu\rho\nu\sigma} \\
    &- \left( \left\{ 1 - \frac{(1+\GFbeta_k)^2}{8\GFalpha_k} \right\} \bar\Delta - \CCk + \frac{1}{6} \bar R \right) \bar g^{\mu\nu} \bar g^{\rho\sigma} \\
    & + \frac{1-2\GFalpha_k + \GFbeta_k}{\GFalpha_k} \bar g^{\mu\nu} \bar \covD^\rho \bar \covD^\sigma + 2 \left( 1 - \frac{1}{\GFalpha_k} \right) \bar g^{\mu\rho} \bar \covD^\nu \bar \covD^\sigma \Bigg] h_{\rho\sigma} \, .
\end{aligned}
\end{equation}
In this expression, $\bar C$ is the background Weyl tensor, that is the fully traceless version of the Riemann tensor (cf. \eqref{weyl-tensor-Luca}), and we introduced the identity on the space of symmetric rank two tensors (introduced in \eqref{eq:LUCA_SYM2_ID} for flat spacetime),
\begin{equation}
    \mathbbm 1^{\mu\nu\rho\sigma} = \frac{1}{2} \left( \bar g^{\mu\rho} \bar g^{\nu\sigma} + \bar g^{\mu\sigma} \bar g^{\nu\rho} \right) \, .
\end{equation}
Due to simplifications in the ghost sector, we have already picked $\GFbeta_k=1$. Inspecting the graviton sector, we see that setting $\GFalpha_k=1$ brings similar simplifications. This gauge choice is often called the harmonic (or de Donder) gauge. In general, the combined choice $\GFalpha_k=\GFbeta_k=1$ is however an approximation --- only $\GFalpha_k=0$ is known to be a fixed point~\cite{Litim:2002ce, Knorr:2017fus}. If we furthermore introduce trace and traceless projectors,
\begin{equation}
    \Pi^{\text{Tr}\mu\nu\rho\sigma} = \frac{1}{4} \bar g^{\mu\nu} \bar g^{\rho\sigma} \, , \qquad \Pi^{\text{TL}\mu\nu\rho\sigma} = \mathbbm 1^{\mu\nu\rho\sigma} - \Pi^{\text{Tr}\mu\nu\rho\sigma} \, ,
\end{equation}
the \emph{curved} two-point function neatly splits into the two sectors, without mixing. 
\begin{tcolorbox}
Fixing $\GFalpha_k=\GFbeta_k=1$, the \textbf{graviton two-point function} reads
    \begin{equation}
    \begin{aligned}
        \Gamma_k^{h^2} = \frac{1}{32\pi \GNk} \int \rmd^4x \, \sqrt{\bar g} \, h_{\mu\nu} \Bigg[ &\left\{ \bar\Delta + \frac{2}{3} \bar R - 2 \mathbbm C - 2\CCk \right\} \Pi^\text{TL} \\
        &\qquad\qquad\qquad - \left\{ \bar\Delta - 2\CCk \right\} \Pi^\text{Tr} \Bigg]^{\mu\nu\rho\sigma} h_{\rho\sigma} \, .
    \end{aligned}
    \end{equation}
\end{tcolorbox}
\noindent Here, $\mathbbm C$ stands for the Weyl tensor where indices are assigned as in \eqref{eq:ALESSIABENJAMIN_grav_two_point_function_gen}.
Note how we achieved the same structure as in the ghost sector: we have a pure Laplace-type kinetic term plus an endomorphism (plus the cosmological constant). In fact, it is useful to introduce the shorthand
\begin{equation}\label{eq:ALESSIABENJAMIN_Delta2}
    \bar\Delta_2 = \left\{ \bar\Delta + \frac{2}{3} \bar R - 2 \mathbbm C \right\} \Pi^\text{TL} \, ,
\end{equation}
that captures the kinetic term in the traceless sector.

A key take-away message here is to not do such computations by hand, but rather to use computer tensor algebra~\cite{xActwebpage, Nutma:2013zea}. This is much less error-prone and significantly faster --- for more advanced approximations, the use of computers is essentially unavoidable.

We can now discuss the regularization in the graviton sector. Mimicking our previous strategy, we use
\begin{equation}
    \Delta S_k^h = \frac{1}{32\pi \GNk} \int \rmd^4x \, \sqrt{\bar g} \, h_{\mu\nu} \left[ R_k^h(\bar\Delta_2) \Pi^\text{TL} - R_k^h(\bar\Delta) \Pi^\text{Tr} \right] h_{\rho\sigma} \, .
\end{equation}
The same remark as in the ghost sector applies: the $k$-dependent prefactor has to be taken into account when taking the $k$-derivative of the regulator. For simplicity, we have chosen the same shape function in the two sectors --- in general, different choices are possible. Note that we did not include the cosmological constant in the regulator, as it could spoil the $k\to0$ limit.

We can now discuss the graviton contribution to the super trace. Since the traceless and trace sectors do not mix, we get two individual contributions from the two sectors. Structurally, we have
\begin{equation}
    \frac{1}{2} \text{STr}_\text{TL} \, W(\bar\Delta_2) + \frac{1}{2} \text{STr}_\text{Tr} \, W(\bar\Delta) \, .
\end{equation}
Here, we indicated the different sectors as subscript on the super trace. Once again, a slight generalization for the heat kernel allows to evaluate these traces. The computation is cumbersome but straightforward, and gives the result
\begin{tcolorbox}
    The \textbf{contribution from the graviton to the \ac{RG} flow} in our setup is
    \begin{equation}
    \begin{aligned}
        \frac{1}{32\pi^2} \int \rmd^4x \, \sqrt{\bar g} \, \Bigg[ & 10 \int_0^\infty \rmd{}z \, z \, \frac{k \partial_k R_k^h(z) - \frac{k \partial_k \GNk}{\GNk} R_k^h(z)}{z + R_k^h(z) - 2\CCk} \\
        & - \frac{13}{3} \bar R \, \int_0^\infty \rmd{}z \, \frac{k \partial_k R_k^h(z) - \frac{k \partial_k \GNk}{\GNk} R_k^h(z)}{z + R_k^h(z) - 2\CCk} + \dots \Bigg] \, .
    \end{aligned}
    \end{equation}
\end{tcolorbox}
\noindent We are now ready to combine the different contributions to the super trace to determine the right-hand side of the Wetterich equation and compare it to the left-hand side to extract the beta functions in the approximation considered.

\subsubsubsection*{Beta functions}\label{sect:ALE-BEN-betas-end}

Now that we have evaluated the super trace, the last step on the road to the beta functions is to evaluate the left-hand side and to simply match coefficients. The left-hand side, evaluated at $h=\bar c=c=0$, reads
\begin{equation}
    k \partial_k \Gamma_k = \int \rmd^4x \, \sqrt{\bar g} \left[ \frac{k \partial_k \CCk - \frac{k \partial_k \GNk}{\GNk} \CCk}{8\pi \GNk} + \frac{k \partial_k \GNk}{16\pi \GNk^2} \, \bar R \right] \, .
\end{equation}
To extract the beta functions, we will now switch to dimensionless couplings,
\begin{equation}
    g = \GNk \, k^2 \, , \qquad \lambda = \CCk \, k^{-2} \, .
\end{equation}
We suppress the $k$-subscript on the dimensionless couplings for better readability, but both $g$ and $\lambda$ still depend on $k$. The beta functions are then given by
\begin{equation}
    \beta_g = k \partial_k g = \left( k \partial_k \GNk + 2 \GNk \right) \, k^2 \, , \qquad \beta_\lambda = k \partial_k \lambda = \left( k \partial_k \CCk - 2 \CCk \right) \, k^{-2} \, .
\end{equation}
To evaluate the threshold integrals, we will use the Litim regulator. Finally:
\begin{tcolorbox}
The \textbf{non-perturbative beta functions} in the Einstein-Hilbert truncation read
    \begin{align}\label{eq:betaskrunninggl}
        \beta_g &= 2g \, \left( 1 - g \, \frac{23-20\lambda}{6\pi(1-2\lambda)+g(-9+5\lambda)} \right) \, , \\
        \beta_\lambda &= \left( -4 + \frac{\beta_g}{g} \right) \lambda + \frac{5g}{12\pi} \frac{8 - \frac{\beta_g}{g}}{1-2\lambda} - \frac{7g}{3\pi} \left( 1 - \frac{1}{14} \frac{\beta_g}{g} \right) \, ,
    \end{align}
where we kept $\beta_g$ in $\beta_\lambda$ for compactness.
\end{tcolorbox}
\noindent Let us discuss this result in connection with the general arguments in \cref{sect:ALESSIABENJAMIN_reno} and \cref{sect:ALESSIABENJAMIN_RG-concept}. The function $\beta_g$ has the general structure of \eqref{eq:ALEBEN-betawithanomalous}. The term $2g$ comes from the canonical mass dimension of Newton's coupling, and is the term responsible for the perturbative non-renormalizability of gravity: the coupling would grow without bounds as $g_k\sim k^2$. Beyond linear order, however, this term may receive additional corrections. Such corrections are encoded in the anomalous dimension of the coupling, which in this case reads
\begin{equation}\label{eq:ALEBEN-anomalous}
    \eta_N\equiv - 2 g \, \frac{23-20\lambda}{6\pi(1-2\lambda)+g(-9+5\lambda)} \,.
\end{equation}
With this, we finally reached the key point: if there is any set of couplings where $\eta_N=-2$, then $\beta_g$ can vanish. In other words:
\begin{tcolorbox}
    \textit{Away from the origin of theory space, the non-perturbative corrections encoded in $\eta_N$ can become as important as the tree-level term $2g$, and possibly they can cancel it. This would make $\beta_g$ vanish, thus saving the theory from the divergences expected from perturbation theory. This is the mechanism at the core of \ac{ASQG}}.
\end{tcolorbox}
\noindent In the last part of this section, we are going to analyze the set of beta functions we obtained. We find out whether the anomalous dimension above can indeed cancel the canonical scaling term, leading to the appearance of an asymptotically safe fixed point which can make gravity well-defined at high energy.

\paragraph{Analysis.}

\begin{figure}[t!]
\centering
\includegraphics[width=0.7\textwidth]{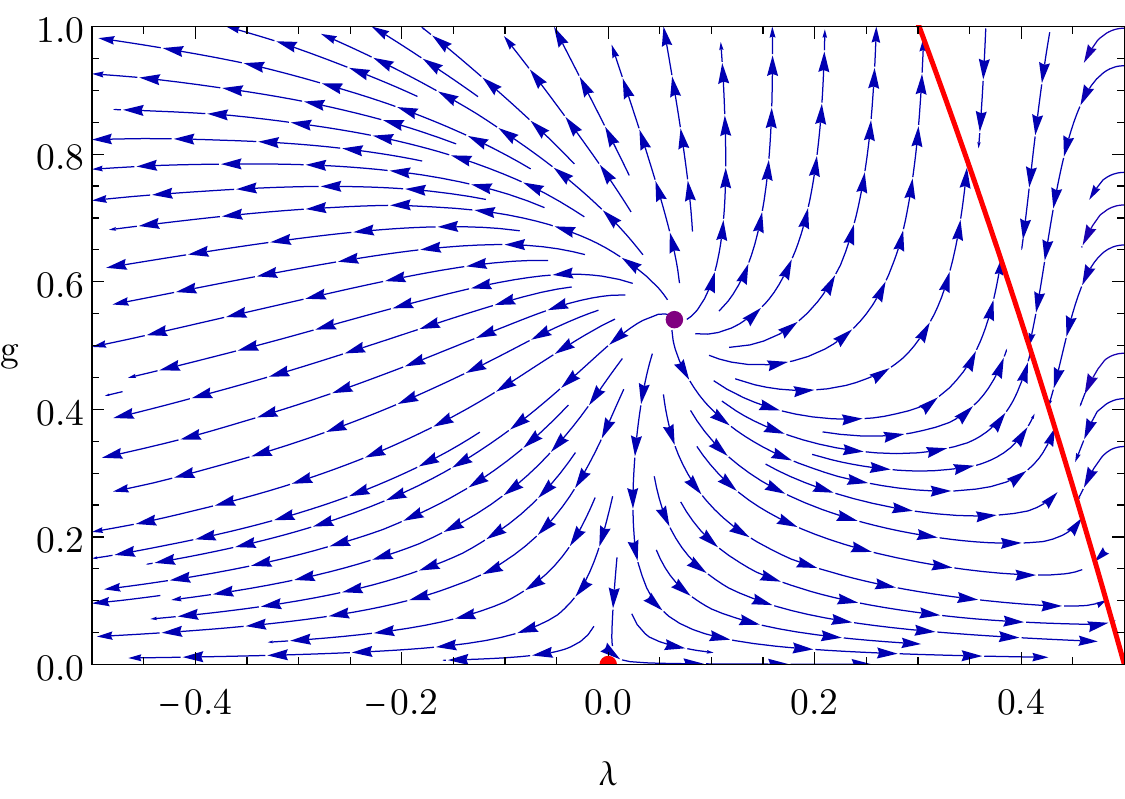}
\caption{\label{fig:ALESSIABENJAMIN_phase_diagram} Phase diagram obtained from the beta functions that we have computed. The red dot indicates the \ac{GFP}, the purple dot is the asymptotically safe \ac{NGFP}. The red line indicates a divergence where our approximation breaks down. The arrows point from large $k$ to small $k$.}
\end{figure}

Let us now analyze these beta functions. We first search for fixed points --- that is, coupling values where all beta functions vanish --- and then compute the corresponding critical exponents. In our system, there are three fixed points (or, the \emph{stars} of the game):
\begin{itemize}
    \item \textbf{The standard \ac{GFP} (free theory)} situated at
    \begin{equation}
        g_\ast=\lambda_\ast=0 \, ,
    \end{equation}
    with critical exponents
    \begin{equation}
        \theta_1 = 2 \, , \qquad \theta_2 = -2 \, ,
    \end{equation}
    The \ac{GFP} is a saddle point, with the irrelevant direction being the Newton coupling. This means that as soon as gravity is turned on (\ie{}, $\GN\neq0$), it cannot be renormalizable with respect to the free theory.

    \item \textbf{An \ac{NGFP} (interacting theory) with positive Newton coupling} which lies at
    \begin{equation}
        g_\ast = \frac{6\pi}{2615} \left( 39\sqrt{19} - 95 \right) \approx 0.541 \, , \qquad \lambda_\ast = \frac{5-\sqrt{19}}{10} \approx 0.064 \, ,
    \end{equation}
    with critical exponents
    \begin{equation}
        \theta_{1,2} \approx 2.667 \pm 0.958 i \, ,
    \end{equation}
    This \ac{NGFP} is fully attractive: the real part of both critical exponents is positive. Thus it can serve as a \textbf{\ac{UV} completion for gravity}. This fixed point (with a slightly different technical setup) has first been found in~\cite{Souma:1999at} and has later been named Reuter fixed point, recognizing the first non-perturbative \ac{RG} computation in gravity~\cite{Reuter:1996cp}. This fixed point is also found in a number of different truncation schemes beyond Einstein-Hilbert, and is at the core of \ac{ASQG}.

    \item \textbf{Another \ac{NGFP} (interacting theory) with negative gravitational coupling} sitting at
    \begin{equation}
        g_\ast = - \frac{6\pi}{2615} \left( 39\sqrt{19} + 95 \right) \approx -1.910 \, , \qquad \lambda_\ast = \frac{5+\sqrt{19}}{10} \approx 0.936 \, ,
    \end{equation}
    with critical exponents
    \begin{equation}
        \theta_1 \approx 7.451 \, , \qquad \theta_2 \approx -5.465 \, .
    \end{equation}
    This fixed point is of no interest, since gravity would be repulsive in this regime. Moreover, the line $g=0$ acts as a separatrix between sets of \ac{RG} trajectories with positive and negative $g$; hence, this fixed point has no impact on the physics of our universe.
\end{itemize}
The phase structure can be visualized by plotting integral curves of the beta functions. We show this in \cref{fig:ALESSIABENJAMIN_phase_diagram}, where the two relevant fixed points are also highlighted.

In the past two decades, numerous investigations employing the \ac{FRG} have extended these results beyond the simple case described here, \eg{}, including higher-derivative operators \cite{Falls:2014tra, Gies:2016con, Knorr:2021slg, Kluth:2022vnq, Baldazzi:2023pep} and matter fields~\cite{Dona:2013qba, Shaposhnikov:2009pv, Eichhorn:2017ylw, Eichhorn:2022gku}. Another essential step is to go beyond the background field approximation, see \eg{}~\cite{Christiansen:2015rva, Denz:2016qks, Christiansen:2017bsy, Pawlowski:2020qer, Knorr:2021niv, Pawlowski:2023gym}. These studies provide solid evidence for the existence of a non-trivial fixed point with two or three relevant directions. This lays the foundation for studies on the physical consequences of \ac{ASQG}, which we address next.

\subsection{From bare and fixed-point actions to amplitudes via the effective action}\label{sec:ALEBEN-actionstoamplitudes}

Grounded on the existence of the Reuter fixed point, whose simplest realization was derived in \cref{sect:ALESSIABENJAMIN_EH}, \ac{ASQG} has emerged as a conservative framework for \ac{QG}. As we have stressed, central to this (and other) approach(es) is the effective action $\Gamma$: it encodes the dynamics of gravity beyond \ac{GR} and serves as the foundation for understanding how the theory behaves across different scales. In the \ac{FRG} framework, the effective action is derived as the $k\to 0$-limit of the \ac{EAA} with a well-defined \ac{UV} limit, \ie{}, one described by a suitable \ac{RG} fixed point. We devote this section to the description of how fixed point actions are related to the bare actions, how to parameterize effective actions in a universal manner, and how this connects to the only observables we currently know how to define in \ac{QG}: scattering amplitudes. In between, we make an interlude on the significance of \ac{RG} running as compared to physical running.

\subsubsection{From fixed points to bare actions: reconstruction problem}

The central question of the reconstruction problem is how $\Gamma_\ast$, derived from the \ac{FRG}, relates to $S_\text{bare}$ used in the path integral formulation. In other words, given an \ac{FRG} flow and its fixed point action, how can one reconstruct the bare action of the theory? This relation is obscured by the fact that the regulator diverges as $k\to\infty$. An approximation to the relationship between the bare and the fixed-point action has been derived in~\cite{Manrique:2008zw}, and is encoded in the reconstruction map:
\begin{equation}
\Gamma_{\UVcutoff} = S_{\UVcutoff} + \frac{1}{2} \text{STr}_{\UVcutoff} \ln \left( S_{\UVcutoff}^{(2)} + R_{\UVcutoff} \right) \, ,
\end{equation}
where all quantities (including the super trace and the regulator) are evaluated at the \ac{UV} cutoff $\UVcutoff$ that, in principle, has to be taken to infinity at the end of the calculations. In this limit, one can read off the relationship between $\Gamma_\ast$ and $S_\text{bare}$.

To evaluate the reconstruction map, the background field method that we introduced in \cref{sect:ALESSIABENJAMIN_EH} is employed. In particular, it is convenient to use a linear split of the metric fluctuations.
The bare action $S_{\UVcutoff}[g]$ is then expanded in powers of $h_{\mu\nu}$. The trace in the reconstruction map is then evaluated using a non-local version~\cite{Codello:2012kq} of the heat kernel techniques we previously introduced.

A recent key result~\cite{Fraaije:2022uhg} is that non-local terms in $\Gamma_\ast$ arise only if a fundamental non-locality scale or \ac{UV} cutoff exists. Specifically~\cite{Fraaije:2022uhg}:
\begin{itemize}
    \item In discrete spacetime models, $\Gamma_\ast$ contains non-localities because the \ac{UV} limit corresponds to reaching the minimal length $\Lambda_\text{phys}^{-1}$.
    
    \item In non-local theories of \ac{QG}, the bare action by construction contains a finite non-locality scale $M$, and hence it contains functions of the ratio $\bar{\Delta}/M^2$. Non-locality is thus retained in the \ac{UV} limit. 
\end{itemize}
If no such physical cutoff scale exists, the locality of the reconstruction map is restored as $\UVcutoff \to \infty$,
\begin{equation}
\lim_{\UVcutoff \to \infty} \Gamma_\ast = S_\text{bare} + \text{local terms} \, .
\end{equation}
This result is important for \ac{ASQG}: scale invariance at the fixed point ensures (or, better, requires) the absence of fundamental mass scales. This prevents the introduction of logarithmic divergences in the reconstruction map~\cite{Fraaije:2022uhg}, supporting the consistency of the theory. In contrast, theories with explicit mass scales (\eg{}, $\Lambda_\text{phys}$) induce non-local corrections, complicating the relationship between $\Gamma_\ast$ and $S_\text{bare}$. A side effect is that in a \ac{QFT} context, the existence of a fixed point, where the theory is scale-invariant, requires the absence of fundamental scales. Vice versa, in theories featuring a fundamental mass or cutoff scale, a putative fixed point can only live at infinite distance in theory space.

\subsubsection{\texorpdfstring{Interlude: physical momentum dependence vs. \texorpdfstring{$k$}{k}-dependence \\ (or: ``\emph{prun\-ning}'' vs. ``\emph{krunning}'')}{Interlude: physical momentum dependence vs. \texorpdfstring{$k$}{k}-dependence (or: ``\emph{prun\-ning}'' vs. ``\emph{krunning}'')}}\label{sect:ALEBEN-kruprun}

At this point, we have to point out an important aspect of the \ac{FRG}. As emphasized in \cref{sect:ALESSIABENJAMIN_FRG}, the scale $k$ is a \emph{fiducial} momentum scale, not a physical one. This means that the beta functions that are computed with the \ac{FRG} have a somewhat different interpretation than standard beta function --- they parameterize the dependence on an \emph{\ac{IR} cutoff}. This is to be contrasted with the physical momentum dependence of correlation functions as in \eqref{eq:ALESSIABENJAMIN_vertex_expansion}.

The physical momentum dependence is instead obtained from momentum-dependent correlation functions in the limit $k\to0$. While the dependence on $k$ can sometimes reflect the dependence on a physical momentum scale (giving rise to the idea of ``\ac{RG} improvement''~\cite{Reuter:2003ca,Platania:2020lqb,Borissova:2022mgd, Platania:2023srt}), in general this is not the case.

What does that mean for gravity --- where is the physical momentum dependence stored? And what is momentum dependence in the first place? Recall that in a curved spacetime, no momentum space is available --- but (covariant) derivatives are available. We can thus clarify what physical momentum dependence generalizes to when gravity is considered: it is the dependence on the covariant derivative. For example, a general kinetic term for a scalar field is captured by a so-called form factor $f$ that depends on the (covariant) Laplacian,
\begin{equation}
    \frac{1}{2} \phi \, f(-\covD^2) \, \phi \, .
\end{equation}

In gravity, this has several immediate consequences~\cite{Knorr:2019atm}. First of all, both the cosmological constant and Newton's constant run with $k$, but do not display a physical running. This is because if we promoted them to form factors, they would either act on the identity (cosmological constant) or only generate boundary terms that we neglect (Newton's constant): 
\begin{equation}
\begin{aligned}
    \frac{1}{16\pi \GN{}} \int \rmd^4x \, \sqrt{g} \, (2\CC - R) \, &\mapsto \, \frac{1}{16\pi} \int \rmd^4x \, \sqrt{g} \, \frac{1}{\GN(\Delta)}(2\CC(\Delta) - R) \\
    &= \frac{1}{16\pi} \int \rmd^4x \, \sqrt{g} \, \frac{1}{\GN(0)}(2\CC(0) - R) \, ,
\end{aligned}
\end{equation}
with $\Delta = -\covD^2$, and we assumed that both $\GN(\Delta)$ and $\CC(\Delta)$ are analytic around zero. Second, any non-trivial physical momentum dependence of the graviton propagator is stored in form factors with two curvatures:
\begin{equation}
    R \, f_R(\Delta) \, R + R_{\mu\nu} \, f_{Ric}(\Delta) R^{\mu\nu} \, .
\end{equation}
A third form factor with two Riemann tensors is dependent, up to terms at least cubic in curvature. This follows from the relation
\begin{equation}\label{eq:ALESSIABENJAMIN_boxRiem}
\begin{aligned}
    \covD^2 R_{\mu\nu\rho\sigma} &= R_\nu^{\phantom{\nu}\alpha} R_{\mu\alpha\rho\sigma} - R_\mu^{\phantom{\mu}\alpha} R_{\nu\alpha\rho\sigma} + 2 R_{\mu\phantom{\alpha}\sigma}^{\phantom{\mu}\alpha\phantom{\sigma}\beta} R_{\nu\alpha\rho\beta} - 2 R_{\mu\phantom{\alpha}\rho}^{\phantom{\mu}\alpha\phantom{\rho}\beta} R_{\nu\alpha\sigma\beta} - 2 R_{\mu\phantom{\alpha}\nu}^{\phantom{\mu}\alpha\phantom{\nu}\beta} R_{\rho\sigma\alpha\beta} \\
    &\qquad + \covD_\mu \covD_\rho R_{\nu\sigma} - \covD_\mu \covD_\sigma R_{\nu\rho} - \covD_\nu \covD_\rho R_{\mu\sigma} + \covD_\nu \covD_\sigma R_{\mu\rho} \, .
\end{aligned}
\end{equation}
This can be proven using the Bianchi identity
\begin{equation}
    \covD_{[\alpha}R_{\mu\nu]\rho\sigma} = 0 \, ,
\end{equation}
and acting with $\covD^\alpha$ on it. Following the reasoning above, in the context of the \ac{FRG} it is useful to define:
\begin{itemize}
    \item \textit{krunning}: variation of the \ac{EAA} and related interaction couplings with respect to the \ac{RG} scale $k$.
    \item \textit{prunning}: physical running with respect to a physical momentum $p$; this corresponds, on curved spacetimes, to the couplings admitting form factors, and is the one appearing in scattering amplitudes.
\end{itemize}
In simple systems, particularly those involving a single scale, the krunning could give a good approximation to the prunning. A recent discussion of this has been given in~\cite{Buccio:2024hys}. 
We shall use these concepts in~\cref{sec:ALEBEN-forefront}.

\subsubsection{Effective actions and form factors}\label{sect:ALESSIABENJAMIN_formfactors}

In the previous subsection, we introduced the effective action, but we did not extensively discuss its properties. An effective action ought to resemble the key properties of \ac{EFT} (cf. \cref{sec:ANNA}) and in particular it should match the \ac{EFT} expansion of an action. In this context, it is useful to introduce the notion of form factors, which generalize the coupling constants of classical gravity to scale-dependent functions.

Form factors are non-local structures that arise in the effective action, reflecting the quantum corrections to classical gravitational interactions. Unlike local couplings, such as Newton's coupling or the cosmological constant, form factors depend explicitly on the momentum or curvature scales. 
\begin{tcolorbox}
In terms of form factors, the effective action reads~\cite{Knorr:2019atm,Knorr:2022dsx,Knorr:2021iwv}
\begin{equation}
\Gamma[g_{\mu\nu}] = \int \rmd^4x \sqrt{-g} \, \left[ \frac{-2\CC + R}{16 \pi \GN} + \frac{1}{6} R \, F_R(\Box) \, R - \frac{1}{2} C^{\mu\nu\rho\sigma} \, F_C(\Box) \, C_{\mu\nu\rho\sigma} + \dots \right],
\end{equation}
\end{tcolorbox}
\noindent where $F_R(\Box)$ and $F_C(\Box)$ are form factors capturing the scale-dependent corrections associated with scalar curvature $R^2$ and Weyl curvature $C^{\mu\nu\rho\sigma} C_{\mu\nu\rho\sigma}$, respectively, while ellipses contain higher-order operators.

Form factors provide a systematic way to describe how gravitational interactions are modified across different energy regimes in a diffeomorphism-invariant way, from \ac{IR} scales, where \ac{GR} ought to be recovered, to \ac{UV} scales, dominated by quantum effects. 
\begin{tcolorbox}
In momentum space, the form factors at quadratic order manifest as modifications of the \textit{graviton propagator}
\begin{equation}\label{eq:ALEBEN-propy}
\propG(p) \simeq\frac{1}{p^2(1 +p^2 F(p^2))}\,,
\end{equation}
where we have set $F\equiv F_C=F_R$ for simplicity.
\end{tcolorbox}
\noindent In the \ac{IR}, absent strong non-localities, this scales as in \ac{GR}, while in the \ac{UV} the graviton propagator and vertices are modified, and in \ac{ASQG} they should combine to yield asymptotically safe scattering amplitudes (see \cref{sect:ALESSIABENJAMIN_amplitudes}). 

In the context of \ac{ASQG} and similar \ac{QFT}-based approaches, form factors are important for several reasons. First, they allow for the explicit realization of scale invariance near the \ac{NGFP} by encoding how physical quantities, such as the graviton propagator or curvature invariants, scale under \ac{RG} transformations. Second, they provide a means to investigate the phenomenological consequences of \ac{QG}, such as modifications to \ac{BH} physics, cosmology, and \ac{GW} propagation (see \cref{sec:ALESSIABENJAMIN-cosmology} and \cref{{sec:ALESSIABENJAMIN-BH}}). Finally, form factors serve as a bridge between the microscopic, \ac{UV}-complete theory and the macroscopic, observable consequences of \ac{QG}, in principle enabling the comparison of \ac{ASQG} predictions with experimental data. Form factors are also useful in \ac{QG} theories going beyond the framework of \ac{QFT}, in which case their use is limited to the regime where \ac{QFT} is applicable.

The study of form factors in \ac{ASQG} often involves analyzing the non-local terms in the effective action, such as those appearing in curvature-squared or higher-derivative operators. These terms, controlled by the \ac{RG} flow, are crucial for understanding the emergence of physical phenomena like asymptotically safe inflation, scale-invariant cosmological perturbations, and potentially observable deviations from classical \ac{GR}. Notably, they are amenable to a definition of \ac{ASQG} in terms of scattering amplitudes, which we will discuss next.

\subsubsection{Asymptotic safety in amplitudes}\label{sect:ALESSIABENJAMIN_amplitudes}

In \cref{sect:ALESSIABENJAMIN_reno}, we introduced the notion of asymptotic safety as a condition on the \ac{RG} flow --- namely, that a suitable interacting fixed point exists. This formulation is very useful, as it directly connects to beta functions, which are central in computations. Nevertheless, there is a more refined formulation, that is actually the one employed by Weinberg in his original proposal~\cite{Weinberg:1980gg}. Specifically, the condition of asymptotic safety is that observables like scattering amplitudes asymptote to a generally non-vanishing constant at high energies. This also means that amplitudes are bounded when the energy goes to infinity, which is necessary to fulfill bounds connected to \eg{} unitarity, see \cref{sec:ANNA}. This notion introduces some complexity. For example, while beta functions are off-shell quantities and thus depend on the choice of gauge or the parameterization of the quantum fluctuations (see \eg{}~\cite{Gies:2015tca} for a comprehensive study within the Einstein-Hilbert truncation), on-shell quantities like amplitudes do not show such pathologies.  Even more importantly, couplings that can be absorbed by field redefinitions, and thus do not contribute to observables, would not need to approach a fixed point.\footnote{This issue has received a lot of interest recently, see \eg{}~\cite{Baldazzi:2021ydj, Baldazzi:2021orb, Knorr:2022ilz, Ihssen:2023nqd, Baldazzi:2023pep, Wetterich:2024uub, Ihssen:2024ihp, Falls:2024noj}.} Looking for fixed points for the physical running couplings (the prunning we defined in \cref{sect:ALEBEN-kruprun}, which is also in correspondence with scattering amplitudes) is thus essential. Nonetheless, computations to date support the picture that the krunning and prunning are approximately equal (see, \eg{},~\cite{Fehre:2021eob}), giving hope that the fixed point found in \ac{FRG} computations --- which is anyway essential to obtain a well-defined effective action --- is in (some non-trivial) correspondence with asymptotically safe scattering amplitudes.

A computation of scattering amplitudes in \ac{ASQG} from first principles is still out of reach, even though a lot of relevant progress has been made in the past years~\cite{Draper:2020knh, Knorr:2022lzn}. On the one hand, computations in Lorentzian signature are absolutely essential to formulate amplitudes~\cite{Bonanno:2021squ, Fehre:2021eob}. On the other hand, having a fixed point by itself is obviously not enough to ensure properties like unitarity and causality. What seems to be needed for unitary and safe amplitudes is a peculiar cancellation between different diagrams~\cite{Draper:2020bop}. There is some evidence that such a cancellation indeed takes place~\cite{Christiansen:2015rva, Meibohm:2015twa, Dona:2015tnf, Denz:2016qks, Christiansen:2017cxa, Eichhorn:2017sok, Eichhorn:2018ydy, Eichhorn:2018akn, Eichhorn:2018nda}.

Let us illustrate this issue with an example. Consider a gravitational scattering of two scalar fields $\phi$ and $\chi$. We assume further that we can approximate the effective action by simply including the non-trivial momentum dependence of the graviton propagator in terms of form factors (as discussed in \cref{sect:ALESSIABENJAMIN_formfactors}),\footnote{This form of the effective action can indeed be arranged for by field redefinitions, if the scalar fields are massless and if no poles other than the massless ones appear in the gravitational sector. In that case, the only missing ingredient is the momentum-dependent scalar self-interaction.}
\begin{equation}
\begin{aligned}
    \Gamma \simeq \int \rmd^4x \, \sqrt{-g} \Bigg[ \frac{R}{16\pi \GN} + \frac{1}{6} R \, F_R(\Box) \, R &- \frac{1}{2} C^{\mu\nu\rho\sigma} \, F_C(\Box) \, C_{\mu\nu\rho\sigma} \\
    &- \frac{1}{2} (\covD_\mu \phi) (\covD^\mu \phi) - \frac{1}{2} (\covD_\mu \chi) (\covD^\mu \chi) \Bigg] \, .
\end{aligned}
\end{equation}
For such an action, the $s$-channel amplitude for the scattering $\phi\phi\to\chi\chi$ would read
\begin{equation}
    \scatteringamplitude_s(s,t) = \frac{4\pi\GN}{3} \, s^2 \left[ \frac{s^2 + 6t(s+t)}{s^2}\frac{1}{s(1+s \, F_C(s))} - \frac{1}{s(1+s \, F_R(s))} \right] \, .
\end{equation}
Here, we used the Mandelstam variables $s$ and $t$ as introduced in \cref{sec:ANNA}. If we now assume that the two form factors grow in a suitable manner and approach a constant value at high energies, the amplitude is asymptotically safe and unitary, see~\cite{Draper:2020bop} for a toy model.
However, this is not enough to claim success: more needs to be computed from first principles and studied. For instance, if we would naively consider the $t$-channel amplitude, whose expression can be obtained from the $s$-channel amplitude by crossing symmetry,
\begin{tcolorbox}
\begin{equation}
    \scatteringamplitude_t(s,t) = \scatteringamplitude_s(t,s) = \frac{4\pi\GN}{3} \, t^2 \left[ \frac{t^2 + 6s(s+t)}{t^2}\frac{1}{t(1+t \, F_C(t))} - \frac{1}{t(1+t \, F_R(t))} \right] \, ,
\end{equation}
\end{tcolorbox}
\noindent we would run into an issue: in the forward scattering limit at high energies, namely $s\to\infty$ with $t$ fixed, we would find a \emph{quadratic} divergence, independent of the precise form of the graviton propagator. Such a divergence would violate both the asymptotic safety condition and unitarity. This apparent issue stems from the fact that we only considered the effective action at quadratic order in the fields, disregarding corrections to the tree-level interaction vertices. Hence, this inconsistency may only be avoided if the momentum-dependent interaction vertices are computed and taken into account, along with the momentum-dependent propagator \eqref{eq:ALEBEN-propy} --- a task that will require substantial advancements in the field.

\subsection{Physical implications and open questions}\label{sec:ALEBEN-forefront}

In this section, we discuss some highlights from the forefront research in the field. First, we focus on the physical consequences of \ac{ASQG} stemming from three of its key features: gravity-matter dichotomy, quantum scale invariance, and gravitational anti-screening. Then, we provide a summary of the state-of-the-art of the field, starting from its milestones and arriving at open questions and challenges.

\subsubsection{Gravity-matter dichotomy}

Asymptotic safety in \ac{QG} requires that all gravitational couplings, including the Newton coupling, attain a non-trivial fixed point in the \ac{UV}: this ensures that the theory is renormalizable and \ac{UV} complete. We highlighted that the fixed point is stable under the inclusion of higher-derivative terms, but what about matter? Can matter, including any scalar, fermionic, and vectorial degree of freedom, destabilize the fixed point, thus compromising the \ac{UV} completion of the combined system? If this is the case, is the \ac{SM} compatible with \ac{ASQG}? These questions are very important not only in the context of \ac{ASQG}~\cite{Eichhorn:2022gku}, but in \ac{QG} in general --- as also emphasized by the idea of the swampland program~\cite{Vafa:2005ui, Brennan:2017rbf, Palti:2019pca, vanBeest:2021lhn, Agmon:2022thq} (cf. \cref{sec:swampland_stuff}). The most intuitive reason for this is that, even at a classical level, \ac{GR} teaches us that gravity influences the way matter moves, and matter decides how spacetime bends under its influence. At a quantum level, gravity is at least minimally coupled to everything, one reason being that the metric determinant multiplies everything else in the overall Lagrangian. As a consequence, gravity fluctuations enter matter loops and vice versa, resulting in gravity-matter beta functions that are always coupled: matter and gravity couplings influence the flow of all other couplings. The most direct consequence of this is that the combined \ac{UV} completion emerges from the interplay of gravitational and matter-field fluctuations. For instance, both collectively influence the running of \GNk. A critical balance between screening and anti-screening effects thus determines whether the combined fixed point exists and remains stable. In the following, we will explore the main consequences of this interplay, closely following the review~\cite{Eichhorn:2022gku}.

The \ac{RG} flow of the Newton coupling is encapsulated in its beta function $\beta_g$, which we derived in a simple setting in \cref{sect:ALESSIABENJAMIN_EH}. When including matter, and setting the cosmological constant to zero, this beta function can be expressed as~\cite{Eichhorn:2022gku}
\begin{equation}
\beta_g = 2 g - g^2 \left( a_{QG} + a_S N_S + a_F N_F + a_V N_V \right) + \orderneglected(g^3) \, ,
\end{equation}
where the first term, $2 g$, is the same that appears in \eqref{eq:betaskrunninggl} and arises from the canonical scaling dimension of $\GN$. The second term contains contributions from quantum fluctuations of both gravitational and matter fields:
\begin{itemize}
    \item $a_{QG}\equiv -\eta_N/g$: This contribution comes from the purely gravitational fluctuations. It is precisely the term we have derived and analyzed in the case of the Einstein-Hilbert truncation in \cref{sect:ALESSIABENJAMIN_EH}, where we found that $\eta_N$ is given by \eqref{eq:ALEBEN-anomalous}. Similarly to the simple example of \cref{sect:ALESSIABENJAMIN_EH}, computations show that this term can be positive, \ie{} anti-screening, and that the anomalous dimension can be $\eta_N\to-2$ at some $g_\ast>0$. If this is the case, the resulting compensation of screening and anti-screening ensures that $\beta_g$ vanishes at a non-trivial fixed point, and prevents $g$ from diverging at high energy. 
    
    \item $a_S, a_F, a_V$: These coefficients capture the effects of $N_S$ scalar, $N_F$ fermionic, and $N_V$ vector fields. If these fields are minimally coupled to gravity, then the coefficients $a_i$ are constants; otherwise, they become functions of the interaction couplings. Scalars and fermions typically screen gravity, $a_S < 0$ and $a_F < 0$, while vectors anti-screen it, $a_V > 0$~\cite{Eichhorn:2022gku}.
\end{itemize}
In the light of these corrections, the fixed-point value of the dimensionless Newton coupling is obtained by solving $\beta_{g} = 0$, and reads
\begin{equation}
g_\ast = \frac{2}{a_{QG} + a_S N_S + a_F N_F + a_V N_V}\,.
\end{equation}
This expression highlights the sensitivity of the fixed-point value to the balance of screening and anti-screening effects. If the screening contributions dominate, they can move the fixed point to unphysical regions, practically destroying asymptotic safety. Conversely, sufficiently anti-screening contributions can stabilize the system, allowing $g_\ast$ to remain finite and positive, but decreasing its value. Specifically, \cite{Dona:2013qba} and successive computations based on the background field approximation support the following picture:
\begin{itemize}
    \item \textbf{Scalars}: Minimally coupled scalar fields, such as the Higgs boson, introduce screening effects. Their contributions are universally negative ($a_S < 0$), weakening the gravitational interaction and increasing $g_\ast$.
    
    \item \textbf{Fermions}: Similarly, fermions screen gravity with $a_F < 0$, although their impact is typically weaker than that of scalar fields. The cumulative effect of a large number of fermions can significantly increase the fixed-point value $g^\ast$.
    
    \item \textbf{Gauge Fields}: By contrast, gauge bosons contribute positively ($a_V > 0$), introducing anti-screening effects that stabilize the fixed point. Their influence is particularly important in gauge theories with extended gauge groups, as the anti-screening effect grows with $N_V$. Indeed, the limit $N_V\to\infty$ makes the fixed point free with respect to $g$ (but still safe with respect to $\lambda$).
\end{itemize}
Computations beyond the background field approximation mostly confirm these features, modulo the case of fermions, whose subtleties make their screening character less established (see, \eg{}, \cite{Meibohm:2015twa}).

Beyond minimal coupling, matter fields can interact with curvature terms, modifying the coefficients $a_S$, $a_F$, and $a_V$. These non-minimal interactions can either enhance or reduce screening and anti-screening effects, leading to qualitative changes in the fixed-point structure.

The interplay between gravity and matter becomes particularly significant when considering the \ac{SM}. The \ac{SM} includes a specific set of matter fields:
\begin{itemize}
    \item Four scalar components (from the Higgs doublet),
    \item 45 Weyl fermions (accounting for quarks, leptons, and color multiplicities),
    \item 12 gauge bosons (the photon, gluons, and $W^\pm$, $Z^0$ bosons).
\end{itemize}
The contributions of these fields to the beta function for $g$ are quantified by their respective coefficients $a_S$, $a_F$, and $a_V$. Numerical studies demonstrate that the inclusion of \ac{SM} matter fields does not destabilize the gravitational fixed point. The contributions from gauge fields ($N_V = 12$) provide sufficient anti-screening to counterbalance the screening effects of scalars and possibly fermions, according to considerations and subtleties above. The resulting fixed-point value of $g$ remains finite and positive, indicating the compatibility of \ac{ASQG} with the observed matter content of the universe~\cite{Pastor-Gutierrez:2022nki}. Notably, the \ac{SM} requires non-zero Yukawa interactions for fermion mass generation via spontaneous symmetry breaking. Some conditions on the gravitational fixed-point values need to be fulfilled. These are only fulfilled if three generations of fermions are accounted for~\cite{Dona:2013qba}. Additional fields introduced by extensions of the \ac{SM}, such as right-handed neutrinos, axion-like particles, or scalar singlets for dark matter, further modify the \ac{RG} flow. While these extensions can increase screening contributions (\eg{}, from additional scalars), they remain consistent with asymptotic safety provided their number is constrained. Remarkably, grand unified theories introduce large gauge groups, increasing $N_V$ and enhancing anti-screening effects, which stabilize the fixed point.

So far we have only summarized how matter fluctuations affect the existence and location of the \ac{NGFP} of the combined system. Next, it is important to discuss how gravitational fluctuations affect the \ac{SM}. For a generic dimensionless matter coupling $c$, the \ac{RG} flow is governed by the beta function
\begin{equation}\label{eq:ALEBEN-mattergravity}
\beta_c = -f_c c + \beta_{c,1} c^n + \mathcal{O}(c^{n+1})\,,
\end{equation}
where $f_c $ is a \ac{QG}-induced correction that, when positive, may contribute to damping $c$ at high energies, $\beta_{c,1}$ is a pure-matter contribution at one loop, and $n$ depends on the specific coupling. This universal structure applies to all sectors of the matter couplings, including Yukawa, gauge, and scalar interactions. In the absence of gravity, \ie{}, if gravitational couplings are set to zero, $f_c\to0$. Matter couplings for which $\beta_{c,1}>0$ are thus affected by the \textit{triviality problem}: they vanish in the \ac{IR}, and diverge at a finite \ac{UV} scale --- leading to the infamous Landau poles, and the only way to remove these poles is to tune the coupling such that it vanishes at all scales, making the theory \textit{trivial}. On the one hand, this problem affects both the Higgs quartic coupling $\lambda_H$ and the Abelian hypercharge gauge coupling $g_Y$. On the other hand, the scale at which such Landau poles occur is trans-Planckian and thus it is expected that new physics --- perhaps \ac{QG} --- will resolve the problem. 

Thanks to the anti-screening contribution of \ac{QG} corrections in \ac{ASQG}, encoded in the term $-f_c$, there is growing evidence that Landau poles may be removed by \ac{ASQG}~\cite{Eichhorn:2017lry}, leading to an asymptotically safe \ac{SM}~\cite{Pastor-Gutierrez:2022nki}. For instance, gravitational corrections could address the Landau pole in the Abelian hypercharge gauge coupling $g_Y$. In the \ac{SM}, $g_Y$ grows logarithmically with energy, diverging at a finite energy scale and leading to an additional Landau pole. However, according to \eqref{eq:ALEBEN-mattergravity}, \ac{ASQG} introduces a linear damping term into the beta function
\begin{equation}
\beta_{g_Y} = b_Y g_Y^3 - f_Y g_Y\,,
\end{equation}
where $b_Y > 0$ represents the standard gauge contributions, and $-f_Y g_Y$ is the gravitational correction. For sufficiently large $f_Y=5g_\ast/18\pi$, the gravitational term dominates at high energies, causing $g_Y$ to asymptote to a finite value rather than a divergence~\cite{Eichhorn:2017lry}. While more work is necessary to test the stability of these results, \eg{} under different truncations and schemes, this mechanism may provide a way to remove Landau poles and ensure the consistency of the \ac{SM} at trans-Planckian energies.

Another consequence of the gravity-matter dichotomy is that some of the \ac{SM} couplings may come as a pre-/post-dictions of the combined asymptotically safe \ac{UV} completion. To give a concrete example, let us consider the gravitational corrections to the running of the Higgs quartic coupling $\lambda_H$,
\begin{equation}
\beta_{\lambda_H} = \frac{3}{2\pi^2}\lambda_H^2 +\tau_H(g_2,g_Y,y_t)+ \kappa_H(g_2,g_Y,y_t) \lambda_H - f_H \lambda_H + \mathcal{O}(\lambda_H^3)\,.
\end{equation}
Here the first term arises from the Higgs self-interaction, $\tau_H$ and $\kappa_H$ are functions of the Yukawa and gauge couplings (and vanish if they all vanish), and $-f_H \lambda_H$ is the gravitational correction. The term $-f_H \lambda_H$ provides an additional stabilization mechanism, counteracting the destabilizing effects of the large top Yukawa coupling. If gravity makes the gauge and the top Yukawa couplings free, and if $f_H<0$ (this is confirmed by computations to date~\cite{Oda:2015sma,Hamada:2017rvn,Eichhorn:2017ylw,Eichhorn:2016esv,Pawlowski:2018ixd}), then $\lambda_\ast=0$ is an \ac{IR} fixed point. In this case, \ac{QG} fluctuations would drive the running of $\lambda_H$ to zero below the Planck scale. This is crucial: starting from $\lambda_H(\MPl) \approx 0$, the flow is then driven by the Higgs quartic interaction, which in turn is regenerated by gauge and top-quark fluctuations, leading to a precise value for $\lambda_H$ in the \ac{IR}, which is related to the Higgs mass and vacuum expectation value $v = 246$ GeV via the well-known relationship 
\begin{equation}
    \lambda_{H}=\frac{1}{2}\left(\frac{m_H}{v}\right)^2\,.
\end{equation}
Specifically, starting from $\lambda_H ( \MPl )\approx 0$ --- a condition which is unexplained within the \ac{SM} --- leads precisely to a Higgs mass of $m_H \approx 126$ GeV.
This logic and the related calculations led to the prediction of the Higgs mass in \ac{ASQG}~\cite{Shaposhnikov:2009pv}. Similar arguments and computations also yield a post-diction of the top mass~\cite{Eichhorn:2017ylw}.

\subsubsection{Fixed points and approximate scale invariance of the power spectrum}\label{sec:ALESSIABENJAMIN-cosmology}

Primordial quantum fluctuations left lasting signatures, observable today in the sky as minuscule temperature anisotropies in the \ac{CMB}, with $\delta T / T \simeq 10^{-5}$. In the framework of the standard cosmological model, these temperature inhomogeneities are traced back to the quantum fluctuations from the pre-inflationary era. As the universe underwent exponential growth, these fluctuations were amplified and smoothened out, resulting in small variations in density at the last scattering surface. Consequently, the distribution of temperature anisotropies across the sky provides indirect insights into the physics of the universe's earliest stages.

The power spectra of scalar and tensor perturbations in momentum space are expressed as:
\begin{equation}
\mathcal{P}_s(k) \simeq A_s \left(\frac{k}{k_*}\right)^{n_s - 1}, \quad \mathcal{P}_t(k) \simeq A_t \left(\frac{k}{k_*}\right)^{n_t},
\end{equation}
where $k=|\vec{k}|$ denotes the magnitude of the three-momentum, and $k_* \sim 0.05$ Mpc$^{-1}$ serves as a reference scale. Observational data enable the determination of the spectral index $n_s$, and the tensor-to-scalar ratio $r \equiv A_t / A_s$. According to the most recent observations, $n_s$ is constrained to $n_s 0.9649 \pm 0.0042$ at a 68\% confidence level, while $r$ is restricted to values below $r<0.064$. This upper bound will hopefully be replaced by an approximate number in the near future~\cite{Matsumura:2013aja}. While the scalar power spectrum is nearly scale-invariant, exact scale invariance ($n_s = 1$) is ruled out.

The (approximate) scale invariance of the power spectrum should at this point ring a bell. Could it be related to the (approximate) scale invariance of \ac{RG} trajectories in the proximity of an \ac{NGFP}? This is not settled yet, but there are arguments that this might be the case. In the following, we will present a particularly simple argument, which is based on the scaling of the background graviton propagator and is based on \cite{Lauscher:2005qz}. Near the \ac{NGFP}, the background graviton propagator takes the form
\begin{equation}
\propG(p) \simeq \frac{1}{p^{2 - \eta_N}}\,.
\end{equation}
Having a non-trivial fixed point requires the anomalous dimension $\eta_N$ of the Newton coupling to approach the fixed-point value $\eta_{N\ast}=-2$ in the \ac{UV} limit. This is because the Newton coupling has mass dimension $-2$, cf. \cref{sect:ALESSIABENJAMIN_reno}. Under these conditions, in position space the fixed-point graviton propagator scales as \cite{Lauscher:2005qz}
\begin{equation}
\propG(x, y) \simeq \log|x - y|^2\,.
\end{equation}
If the temperature fluctuations are entirely driven by the quantum fluctuations of spacetime geometry during inflation, and these fluctuations originate in the Planck era, the corresponding density fluctuations $\delta \rho$ are characterized by a two-point correlation function \cite{Lauscher:2005qz}
\begin{equation}
\xi(\vec{x}) = \langle \delta \rho(\vec{x} + \vec{y}) \delta \rho(\vec{y}) \rangle \propto \langle \delta R(\vec{x} + \vec{y}, t) \delta R(\vec{y}, t) \rangle \simeq |\vec{x}|^{-4}\,,
\end{equation}
where $\delta \rho = \delta \rho / \bar{\rho}$ represents fractional density fluctuations, and $\delta R(\vec{y}, t)$ denotes fluctuations in the scalar curvature, induced by metric variations. The power spectrum in momentum space is the spatial Fourier transform of $\xi(\vec{x})$,
\begin{equation}
|\delta_{\vec{k}}|^2 = V \int \rmd^3\vec{x} \, \xi(\vec{x}) e^{-i \vec{k} \cdot \vec{x}}\,.
\end{equation}
The spectral index $n_s$, describing the power-law scaling of the spectrum, satisfies:
\begin{equation}
|\delta_{\vec{k}}|^2 \propto |\vec{k}|^{n_s} \, .
\end{equation}
For $\xi(\vec{x}) \simeq |\vec{x}|^{-4}$, this results in a perfectly scale-invariant power spectrum ($n_s = 1$). The nearly scale-invariant nature of $\mathcal{P}_s(k)$ can thus be attributed to the \ac{RG} trajectory's behavior near the \ac{NGFP}. This hypothesis has inspired studies exploring metastable \ac{dS} solutions in \ac{ASQG}, leading to a sufficiently extended period of ``\ac{NGFP}-driven inflation''~\cite{Bonanno:2008xp,Bonanno:2010mk,Bonanno:2015fga,Bonanno:2016rpx,Tronconi:2017wps,Bonanno:2018gck, Liu:2018hno,Platania:2019qvo,Wetterich:2019rsn}. The resulting scenario has been dubbed ``asymptotically safe inflation''~\cite{Weinberg:2009wa}.

\subsubsection{Gravitational anti-screening and singularity resolution}\label{sec:ALESSIABENJAMIN-BH}

\begin{figure}[t]
\centering\includegraphics[width=0.65\textwidth]{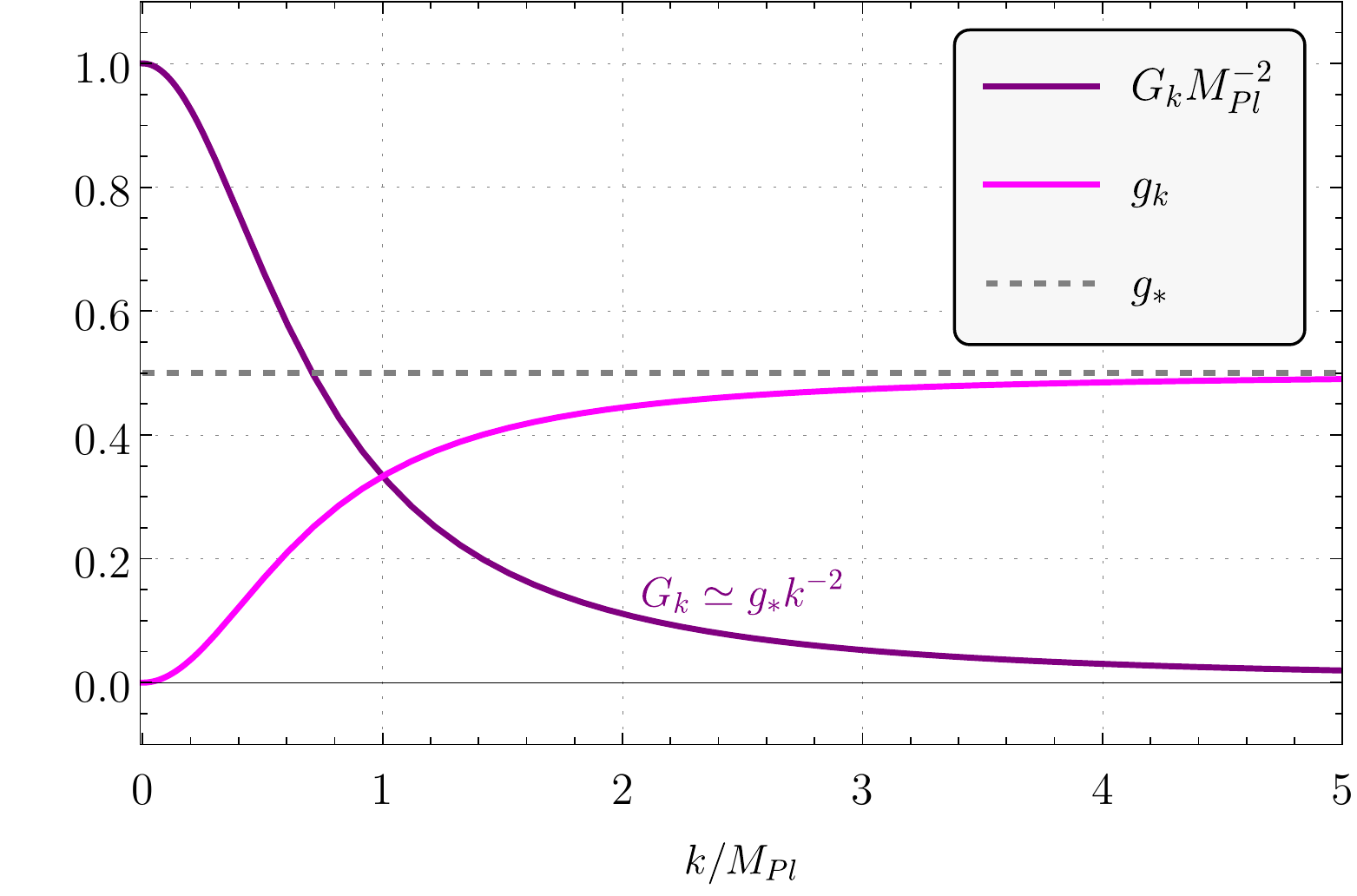}
\caption{Dependence of the dimensionless and dimensionful Newton coupling on the \ac{RG} scale $k$. As $k$ is raised above the Planck scale, the \ac{FRG} flow attains the fixed point regime, where $g_k\simeq g_\ast$ and $\GNk\simeq g_\ast k^{-2}$. The latter thus vanishes asymptotically, providing an intuitive way to think of gravitational anti-screening in \ac{ASQG}.\label{fig:ALESSIABENJAMIN-newton}}
\end{figure}

In \cref{sect:ALESSIABENJAMIN_EH} we have seen that the beta functions of the gravitational couplings in the Einstein-Hilbert truncation admit an \ac{NGFP} which is \ac{UV} attractive in the same sub-theory space. In general, when extending the calculation to higher-order truncations, one finds that the fixed point is stable and that it is \ac{UV} attractive for a subset of \ac{RG} trajectories in theory space --- those belonging to the basin of attraction of the fixed point. In a similar way as for \ac{QCD}, a \ac{UV} attractive fixed point is an indicator that anti-screening effects take over the screening ones. This is related to the physical mechanism underlying \ac{ASQG} \cite{Nink:2012vd}. An intuitive way to understand gravitational anti-screening is via the \textit{krunning} of the dimensionless Newton coupling. Neglecting the krunning of the cosmological constant, one can see that the beta function~\eqref{eq:betaskrunninggl} yields the following approximate krunning for the dimensionful Newton coupling~\cite{Bonanno:1998ye}
\begin{equation}
\GNk\simeq\frac{\GN}{1+g_\ast^{-1}\GN k^2}\,.
\end{equation}
This running is shown in \cref{fig:ALESSIABENJAMIN-newton}.

Close to the fixed point $\GNk \simeq g_\ast k^{-2}$: the existence of a non-trivial fixed point at $g_\ast \neq0$ thus implies that the dimensionful Newton coupling vanishes in the high-energy limit $k\to\infty$. Although this only happens with respect to the \ac{RG} scale $k$, it nicely encodes an essential feature of \ac{ASQG} that is at the core of gravitational anti-screening: gravity gets weaker at high energies. Assuming that gravitational anti-screening is qualitatively captured by an effective Newton coupling that vanishes at short distances --- which should hold if a decoupling mechanism is at work~\cite{Borissova:2022mgd} --- then gravitational singularities in \ac{GR} should get weaker or be resolved in \ac{ASQG}. To see this, recall that the lapse function of a Schwarzschild \ac{BH} reads
\begin{equation}
f(r)=1-\frac{2\GN{} M}{r}\,.
\end{equation}
If $\GN$ is replaced by an effective Newton coupling which depends on the radial coordinate $r$, and interpolates between the observed Newton constant $\GN$ asymptotically and zero at the would-be singularity, then $r=0$ ought to be replaced by a weaker singularity or perhaps even a regular \ac{BH} core. 

We cautioned in \cref{sect:ALEBEN-kruprun} that the krunning argument can give at best a qualitative idea of the consequences of gravitational anti-screening. In the simplest scenarios, \eg{}, in the case of a spherically symmetric static \ac{BH}, a similar picture may also arise from more formal arguments and computations: \ac{QG} is generally expected to correct \ac{GR} by (infinitely many) higher-derivative terms with specific Wilson coefficients whose values depend on the specific \ac{UV} completion. Once an effective action $\Gamma$ has been computed, \eg{}, from \ac{ASQG}, then the classical field equations are replaced by the effective ones
\begin{equation}
\frac{\delta\Gamma}{\delta g_{\mu\nu}}=0\,.
\end{equation}
The solutions to this will include spherically symmetric static \acp{BH}, and the higher-derivative corrections will result in a lapse function where the Newton coupling is replaced by an effective one. By consistency, this has to interpolate between zero and $\GN$ at large distances, which is the same picture presented before. Such a picture has been corroborated and refined by recent studies~\cite{Knorr:2022kqp,Pawlowski:2023dda,Daas:2023axu} based on first-principle calculations.

\subsubsection{Chronology of some milestones}\label{sect:ALESSIABENJAMIN_milestones}

In this section, we highlight some of the milestones achieved in the field:
\begin{itemize}
    \item \textbf{First computation in four dimensions}~\cite{Reuter:1996cp}: While the original idea of \ac{ASQG} goes back to the late seventies~\cite{Weinberg:1980gg}, it was only in the mid-nineties that the first non-per\-tur\-ba\-tive \ac{RG} computation in \ac{QG} was performed. Only then, modern functional methods had been developed to derive non-perturbative beta functions~\cite{Wetterich:1992yh, Morris:1993qb, Ellwanger:1993mw}. The original computation only differs in small details compared to our exposition of the derivation of the beta functions for Newton's constant and the cosmological constant.

    \item \textbf{Prediction of Higgs mass and postdiction of the Higgs to top mass ratio}~\cite{Shaposhnikov:2009pv, Eichhorn:2017ylw}: A theory of \ac{QG} does not only have to be internally consistent, it also has to be phenomenologically viable. One aspect of this is that it has to be compatible with measured values of \ac{SM} parameters like particle masses. This is relevant since if, \eg{}, an otherwise promising theory of \ac{QG} would predict that the electron is heavier than the Higgs boson, it is ruled out experimentally. The quantum scale invariance characterizing asymptotic safety, as any symmetry, can enhance its predictive power by adding constraints. Such constraints can turn into pre/post-dictions for some of the \ac{SM} couplings and masses. Two important examples are the prediction of the Higgs mass~\cite{Shaposhnikov:2009pv} and the postdiction of the Higgs to top mass ratio~\cite{Eichhorn:2017ylw}.
    
    \item \textbf{Matter matters in \ac{ASQG}}~\cite{Dona:2013qba}: While this is intuitively expected, since gravity couples (at least minimally) with everything, matter can influence gravity and gravity can affect matter. The article~\cite{Dona:2013qba} made this point clear by showing how the number of matter and gauge fields can impact the existence and properties of the Reuter fixed point. Since its appearance, increasingly sophisticated computations have been performed, highlighting how \ac{QG} can be constrained by requiring its compatibility with the \ac{SM} (see the recent review~\cite{Eichhorn:2022gku} and references therein).
    
    \item \textbf{The Goroff-Sagnotti term is asymptotically safe}~\cite{Gies:2016con, Baldazzi:2023pep}: The Goroff-Sagnotti term $C^3$ is the first non-subtractable divergence that appears at the two-loop order of the perturbative renormalization of \ac{GR} (see \cref{sec:LUCA_twoloopdivs}). This means that it cannot be removed by a local field redefinition. It hallmarks the breakdown of perturbative renormalizability, as it is the first of infinitely many such divergences. By contrast, in \ac{ASQG}, it was found that this term is irrelevant --- its coupling is a prediction of the theory. This is a significant result, as it shows that \ac{ASQG} performs strictly better than \ac{EFT} in terms of predictivity.

    \item \textbf{Towards a resolution of the triviality problem in the \ac{SM}}~\cite{Christiansen:2017gtg, Eichhorn:2017lry}: The triviality problem affects the $U(1)$ and Higgs sectors of the \ac{SM}: the couplings diverge at a finite \ac{RG} scale (the Landau pole) unless they are set to stay at the \ac{GFP}, in which case both couplings vanish at all scales, making the theory ``trivial''. The resolution of the Landau problem requires either new physics at sub-Planckian scales, \eg{}, supersymmetry or grand unification, or a non-trivial \ac{UV} completion via an interacting fixed point. Such a fixed point can likely be triggered if gravity is asymptotically safe, since in this case graviton loops would contribute with a non-zero constant in the matter beta functions~\cite{Christiansen:2017gtg, Eichhorn:2017lry}.

    \item \textbf{\ac{ASQG} can be unitary}: There is sometimes the misconception that due to truncations of the \ac{EAA}, the graviton propagator in \ac{ASQG} must have ghost degrees of freedom and hence the theory cannot be unitary, in analogy to quadratic gravity and as discussed in \cref{sec:lecture4}. However, since truncations are just approximations and do not reflect the full quantum dynamics, such ghosts are not only expected on general grounds but harmless: in toy models, the absolute value of their residues decreases with the truncation order and vanishes rapidly, hence decoupling such fictitious ghosts from amplitudes~\cite{Platania:2020knd, Platania:2022gtt}. First computations of the full graviton spectral function indicate that there is indeed no extra degree of freedom~\cite{Bonanno:2021squ, Fehre:2021eob}. Moreover, it has been shown that there exist form factors leading to unitary and asymptotically safe scattering amplitudes~\cite{Draper:2020bop}.
    
    \item \textbf{Lorentzian asymptotic safety}~\cite{Fehre:2021eob, Banerjee:2022xvi, DAngelo:2022vsh, DAngelo:2023tis}: A key challenge for \ac{ASQG} is to perform computations with the \ac{FRG} in Lorentzian signature. In the context of asymptotic safety, first steps have been taken in~\cite{Manrique:2011jc, Biemans:2016rvp, Knorr:2018fdu}. With a view to scattering amplitudes, Lorentzian correlation functions are a necessary ingredient. In~\cite{Fehre:2021eob}, the Lorentzian spectral function of the graviton was computed for the first time. Two key insights from this paper are first, that \ac{ASQG} only seems to feature a massless graviton (an important step towards establishing unitarity, see previous bullet point), and second, that a Wick rotation of Euclidean computations is qualitatively accurate~\cite{Bonanno:2021squ}. Lorentzian versions of the \ac{FRG} have been put forth in~\cite{Banerjee:2022xvi, DAngelo:2022vsh, DAngelo:2023tis}.
    
    \item \textbf{Asymptotically safe \ac{SM}}~\cite{Pastor-Gutierrez:2022nki}: On the path towards a complete description of our universe, we do not only have to quantize gravity, but also at least couple the \ac{SM} to it. A big first step was taken in~\cite{Pastor-Gutierrez:2022nki}, where multi-scale physics was resolved for the first time: \ac{ASQG} at high energies, electro-weak symmetry breaking at intermediate energies, and strongly coupled \ac{QCD} in the infrared. The analysis suggests that their are no new free parameters in the matter sector beyond the ones known in the \ac{SM}, emphasizing the high predictivity of the scenario.

     \item \textbf{Landscapes in \ac{ASQG}}~\cite{Basile:2021krr,Knorr:2024yiu}: In parallel to reinforcing the evidence for an asymptotically safe fixed point, determining the landscape of \acp{EFT} stemming from it, \ie{} determining its boundaries and geometry~\cite{Basile:2021krr,Knorr:2024yiu}, as well as its matter content~\cite{Eichhorn:2022gku}, is essential to compare the predictions of \ac{ASQG} --- encoded in its Wilson coefficients --- with theoretical constraints (such as those of \cref{sec:ANNA} and \cref{sec:FRANCESCO}) and observational bounds. Steps to compute the asymptotic safety landscape of photon-graviton systems have been taken in~\cite{Basile:2021krr,Knorr:2024yiu}. First comparisons with positivity bounds have been performed in~\cite{Knorr:2024yiu,Eichhorn:2024wba}.
\end{itemize}

\subsubsection{Outline of challenges and open questions}\label{sect:ALESSIABENJAMIN_challenges}

The achievements in the previous section come hand in hand with challenges and open questions: while the criticisms of~\cite{Donoghue:2019clr} have been either clarified~\cite{Bonanno:2020bil} or overcome~\cite{Fehre:2021eob}, several problems remain open. Below we list some of the most important open questions in \ac{ASQG}:

\begin{itemize}
        \item \textbf{Existence of fixed point in Lorentzian signature}: While there is some evidence that the results of Lorentzian computations are qualitatively and even quantitatively close to Euclidean results, the bulk of the evidence in favor of \ac{ASQG} still comes from Euclidean computations. It is necessary to understand this better, and Lorentzian computations are generally a strict necessity if we want to derive scattering amplitudes. First steps to make progress in this direction have been taken in~\cite{Manrique:2011jc, Biemans:2016rvp, Knorr:2018fdu, Eichhorn:2019ybe, Bonanno:2021squ, Fehre:2021eob}.

        \item \textbf{Unitarity and causality}: While some evidence exists that \ac{ASQG} only has the massless spin two pole and no ghost modes~\cite{Fehre:2021eob}, more solid computations have to be performed to check whether the theory is unitary. This also includes testing asymptotic safety against positivity bounds~\cite{Knorr:2024yiu, Eichhorn:2024wba}.
        
        \item \textbf{Form factors and amplitudes}: Computing even the simplest scattering process, two particles scattering into two particles, requires the full momentum dependence of a four-point function~\cite{Knorr:2022dsx, Pawlowski:2023gym}. This is extremely challenging --- the current state of the art is the full resolution of propagators, and selected channels of the three- and four-point functions~\cite{Christiansen:2015rva, Denz:2016qks, Knorr:2021niv}. This has to be systematically extended to get a reasonable estimate for scattering amplitudes.
        The computation of scattering amplitudes~\cite{Draper:2020knh} is also important conceptually, since the physical definition of \ac{ASQG} relies on the high-energy behavior of physical observables, which only depends on essential couplings~\cite{Weinberg:1980gg, Baldazzi:2021ydj}. This is also related to the problem of mapping out the asymptotic safety landscape~\cite{Basile:2021krr, Saueressig:2024ojx, Eichhorn:2024rkc}.

        \item \textbf{Number of free parameters}: How many of the free parameters of the \ac{SM} can be fixed by asymptotic safety? And how many does gravity add? There are several partial answers to this question, which apply to systems with a limited number of couplings. In such systems it appears that gravity brings in between two and four free parameters~\cite{Falls:2014tra, Denz:2016qks, Kluth:2020bdv, Knorr:2021slg, Kluth:2022vnq, Baldazzi:2023pep} whereas some couplings and masses can be computed from first principles~\cite{Shaposhnikov:2009pv, Eichhorn:2017ylw, Eichhorn:2022gku}. Yet, the full picture is to be understood.
        
        \item \textbf{Diffeomorphism invariance and background independence}: The necessity to use the background field method introduces all kinds of headaches due to the various symmetry breakings induced by both gauge fixing and especially the regulator~\cite{Pawlowski:2023gym}. A better conceptual understanding is necessary to obtain systematic schemes to restore full diffeomorphism invariance when $k\to0$.
        
        \item \textbf{State dependence of the \ac{FRG} flow}: A careful consideration of the \ac{FRG} on globally hyperbolic spacetimes indicates that beta functions might be state-dependent~\cite{Banerjee:2022xvi, DAngelo:2022vsh, DAngelo:2023tis}. This is a very recent and interesting development, and the consequences are still to be explored.
        
        \item \textbf{Reconstruction problem}: The \ac{FRG} is a powerful tool to test the existence of the Reuter fixed point and to extract predictions for the \ac{EAA}. However, even resolving the flow $\Gamma_k$ exactly, the limit $k\to\infty$ to the fixed-point action is not expected to recover the \textit{bare} action of the theory. This mismatch is known as the reconstruction problem~\cite{Manrique:2009tj}. Partial progress has been reported \eg{} in~\cite{Morris:2015oca, Fraaije:2022uhg}.
        
        \item \textbf{Truncation-independent statements}: Practically all results in \ac{ASQG} are based on computations in approximations. It is crucial to find ways to derive stronger statements that hold at the exact level, \ie{}, statements that are truncation-independent.
        
        \item \textbf{Quantum spacetimes}: A proper understanding of solutions to the quantum equations of motion is still lacking. On the one hand, this is related to the above points: more sophisticated computations in Lorentzian signature are needed. On the other hand, new solution strategies have to be developed to solve field equations in the presence of form factors, beyond asymptotic approximations or special cases~\cite{Holdom:2002xy,Anselmi:2013wha,Bonanno:2019rsq, deRham:2020ejn,Knorr:2022kqp, dePaulaNetto:2023cjw, Pawlowski:2023dda, Daas:2023axu, Koshelev:2024wfk}.
        
        \item \textbf{\ac{BH} thermodynamics in asymptotic safety}: \ac{BH} thermodynamics is observationally irrelevant, but highly constraining from a theoretical standpoint. Although the naive argument that a \ac{QFT} cannot be compatible with an area law is avoided by \eg{} giving up \ac{UV}-locality, \ac{ASQG} is left with the important task of explaining \ac{BH} thermodynamics in general and the area scaling in particular at a fundamental level~\cite{Basile:2025zjc}, \ie{}, beyond the Wald construction with higher derivatives~\cite{Conroy:2015wfa, Kaul:2000kf, Calmet:2021lny, Platania:2023uda}.
    
\end{itemize}
With both milestones and open questions in mind, we can proceed and wrap up this section.

\subsection{Conclusions}\label{sect:ALESSIABENJAMIN_conclusions}

In this section, we have introduced the \ac{FRG} --- a powerful framework with countless applications in condensed matter physics, particle physics, and gravity~\cite{Dupuis:2020fhh}. The \ac{FRG} is based on a formally exact equation describing the non-perturbative \ac{RG} flow of theories across the whole theory space. As such, it provides an ideal tool to investigate strongly interacting theories, including those that are non-perturbatively renormalizable. We then discussed one of the most conservative \ac{UV}-complete approaches to \ac{QG}: \ac{ASQG}. This approach is grounded on the framework of \ac{QFT} and on the idea that at a non-perturbative level, gravity is renormalizable, with a \ac{UV} completion defined at an interacting fixed point of the \ac{RG} flow. At this fixed point, the theory features ``quantum scale invariance'', \ie{}, dimensionless couplings reach constant values due to a compensation of screening and anti-screening effects at the level of the beta functions. In this way, in the \ac{UV}, \ac{ASQG} overcomes the renormalizability problem presented in~\cref{sec:LUCA}. In the \ac{IR}, \ac{ASQG} (as all other \ac{QG} theories) ought to give rise to Wilson coefficients in agreement with the positivity and causality bounds of \cref{sec:ANNA}. Renormalizability, causality, and unitarity are basic consistency tests that all theories should pass before even speaking about phenomenology and the topics in \cref{sec:FRANCESCO}.
While many questions in \ac{ASQG} remain open (cf.~\cref{sect:ALESSIABENJAMIN_challenges}), a unique selling point is its predictive power: quantum scale invariance at high energies induces non-trivial relationships between gravitational and \ac{SM} couplings, resulting in enhanced predictive power and non-trivial results, including the derivation of a top-to-Higgs mass ratio compatible with experiments~\cite{Eichhorn:2017ylw}. Assuming such non-trivial results (cf.~\cref{sect:ALESSIABENJAMIN_milestones}) are not mere coincidences, even if \ac{ASQG} would turn out not to be fundamental, it may still be realized at an effective level~\cite{deAlwis:2019aud,Basile:2021euh,Basile:2021krk}, and act as a bridge between gravitational \acp{EFT} (see \cref{sec:ANNA}) and a more fundamental description such as \ac{ST}, which is the topic of \cref{sec:IVANO}.


\section{\titleIvano}
\label{sec:IVANO}

\begin{tcolorbox}[colback=white,colframe=scipostblue]
{\bf Lecturer:} Ivano Basile, \briefaffiliationIvano

{\bf Email address:} \href{mailto:\emailIvano}{\emailIvano}
\tcblower
{\bf Lecture recordings:}
\begin{enumerate}[label= Lecture \arabic*:, leftmargin = 3.5cm, labelsep = 0.5cm, parsep = 0.0cm]
    \item \href{https://youtu.be/Bm7bhP8MUGY}{https://youtu.be/Bm7bhP8MUGY}
    \item \href{https://youtu.be/sCE2m96U0T8}{https://youtu.be/sCE2m96U0T8}
    \item \href{https://youtu.be/2qgbw_907h0}{https://youtu.be/2qgbw\_907h0}
    \item \href{https://youtu.be/gf-eAfKHPe0}{https://youtu.be/gf-eAfKHPe0}
\end{enumerate}

{\bf Abstract:}

This short introductory course on the basics of string theory is geared to students interested in quantum gravity in a broad sense. As such, I will provide extensive motivations from general quantum gravity considerations, placing string theory in this wider context. Concretely, the lectures will cover closed strings interacting weakly, their connection to gravitational effective field theory at low energy and their imprints at high energy.
\end{tcolorbox}

\subsection*{Preface}

These notes comprise my first attempt to give a short introductory course on \ac{ST} to an intended audience of theoretical physicists who care about the problem of \ac{QG} and fundamental physics, but are not necessarily interested in \ac{ST}. If hearing about mirror symmetry your reaction is ``so what?'' these lectures are for you. As such, I have two main goals. Firstly, I would like to present some aspects of the field in broad strokes with a modern approach, taking into account recent developments without focusing on excessive technicalities or special cases. Clearly, there must be some trade-off between breadth and depth; my strategy is to restrict the material to a particularly well-understood corner of \ac{ST}, namely closed strings which interact weakly. As we shall see, this is the relevant sector for gravity. This focus allows to communicate many of the most important conceptual lessons we have learned in this field since its inception, whilst keeping some level of detail. The second goal is to present the subject in a way that emphasizes what the intended audience actually wants to know. What are the take-home lessons? What does the theory say about \ac{QG}? What about our universe? This approach involves providing extensive motivations and addressing a number of misconceptions which are widespread outside of the field. This approach may be also useful for some practitioners in the field.

A background on \ac{GR} and (relativistic) \ac{QFT}, especially notions of \ac{EFT}, is ideal. Many advanced technical and conceptual subtleties are pointed out and deferred to the literature. Some useful references:

\begin{itemize}
    \item M. Green, J. Schwarz and E. Witten, \emph{Superstring Theory Vol. 1 \& 2} \cite{Green:2012oqa, Green:2012pqa}
    \item J. Polchinski, \emph{String Theory Vol. 1 \& 2} \cite{Polchinski:1998rq, Polchinski:1998rr}
    \item R. Blumenhagen, D. L\"{u}st and S. Theisen, \emph{Basic Concepts of String Theory} \cite{Blumenhagen:2013fgp}
    \item E. Kiritsis, \emph{String Theory in a Nutshell} \cite{Kiritsis:2019npv}
    \item D. Tong, \href{https://www.damtp.cam.ac.uk/user/tong/string.html}{\emph{String Theory, Part III Mathematical Tripos}} \cite{Tong:2009np}
    \item D. Isra\"{e}l, \href{https://www.lpthe.jussieu.fr/~israel/notes.pdf}{\emph{Lecture Notes on String Theory}}
    \item E. Witten, \href{https://www.youtube.com/watch?v=H0jLD0PphTY&t=1835s&pp=ygU0d2hhdCBldmVyeSBwaHlzaWNpc3Qgc2hvdWxkIGtub3cgYWJvdXQgc3RyaW5nIHRoZW9yeQ%3D%3D}{\emph{What Every Physicist Should Know About String Theory}}
    \item S. Cecotti, \emph{Introduction to String Theory} [recent addition to the roster!] \cite{Cecotti:2023dnp}
\end{itemize}

The presentation is intended to be somewhat informal and light-hearted to balance the novelty in the mathematical material and physical concepts that \ac{ST} introduces, as embodied by the writing style. Otherwise where's the fun? ;) I encourage any student to reach out for questions, comments, feedback, threats, cat pictures, typo and/or error corrections, and suggestions on good ice cream shops. \\

A good fraction of the conceptual aspects in these notes is the product of my personal \sout{de\-range\-ments} musings. As such, I am thankful for the many conversations I had over the years with several amazing colleagues, whose brilliant insights shaped my understanding and perspective. I am grateful to C. Aoufia, G. Contri, N. Cribiori, C. Markou, V. Larotonda, G. Leone, S. Raucci and N. Risso for their feedback and typo-spotting on these notes, and to the organizers of the PhD School on Quantum Gravity for the fantastic opportunity to present this beautiful subject from my point of view and with my unreasonable passion for footnotes. I also preemptively thank anyone who will inevitably spot some typos or errors, and my cat Melissa for obstrucig mu kybord. \\

These lectures are organized as follows.
\begin{description}

\item[Sec.~\ref{sec:part_i}:] We lay down the groundwork to motivate \ac{ST} from the bottom up. We begin from very general \ac{QG} considerations involving black holes, holography and S-matrix consistency, leading us to various motivations to consider \ac{ST}. We present a preliminary explanation of what \ac{ST} is and what we would like or expect from such a theory. We begin our journey from the worldline formulation of perturbative \ac{QFT}.

\item[Sec.~\ref{sec:part_ii}:] We build perturbative \ac{ST} from the ground up in its worldsheet formulation. Contrary to many accounts, we examine all possibilities at each step of the construction, treating bosonic strings and superstrings of the \ac{RNS}-\ac{RNS} and heterotic types all on the same footing. Moreover, we discuss general dimension and backgrounds from the outset, and only later specialize to the flat background when needed.

\item[Sec.~\ref{sec:part_iii}:] We discuss how \ac{ST} reduces to gravitational \ac{EFT} at low energies. We derive low-energy effective actions from anomaly cancellation and S-matrix matching. We examine the general structure of stringy \acp{EFT} and introduce the notion of string landscape.

\item[Sec.~\ref{sec:part_iv}:] We turn to the high-energy regime of \ac{ST}. We define the string S-matrix, providing some elementary examples. We discuss how some indications of black-hole formation can be glimpsed by estimating the resummed hard scattering of strings, matching the entropy computation at the black-hole threshold.

\item[Sec.~\ref{sec:IVANO_conclusions}:] We conclude by taking stock of what was presented in the lectures. We provide a summary of the various punchlines encountered along the way.

\end{description}


\subsection{The what, the why, and the how}\label{sec:part_i}

When introducing \ac{QG} \emph{aficionados} to \ac{ST} it is important to lay down a solid groundwork of motivations, and a big-picture summary of what the theory is about. What it says, what it doesn't say, and what we should expect such a theory to say in the first place. So let us begin from the reason why we are all here: the problem of \ac{QG}.

\subsubsection{Quantum fields and gravity}\label{sec:gravity_qft}

The spirit of this section is to start fresh and build the perspective that connects \ac{QG} to strings. Any seemingly arbitrary choice can be questioned; no stone shall be unturned. So what's all the fuss on \ac{QG} about? Let us look at it in natural units and emphatically mostly plus Lorentzian metric signature as in our common \hyperref[conventions]{conventions}, since doing otherwise would be irresponsible.

\subsubsection*{Renormalizability, but don't worry too much about it}

Some of you will probably think about renormalizability. \ac{GR} is not renormalizable in perturbation theory (see \eg{} \cref{sec:LUCA}).\footnote{The standard argument, besides power counting, is that although pure gravity is renormalizable at one loop, the Goroff-Sagnotti term~\cite{Goroff:1985th} arises at two loops. Including matter makes things worse. One could imagine a miraculous situation in which extra terms to include in the theory stop showing up at some order, analogously to the quest to find them in maximal supergravity (see, \eg{},~\cite{Bern:2018jmv}). However, it wouldn't solve the deeper issues we review below.} The history of theoretical physics teaches us that when something like this happens, it is a hint that the theory at stake is not unlike fluid dynamics: an incomplete macroscopic \ac{EFT} which works in the \ac{IR} regime of low energies,\footnote{Of course, in a relativistic setting one ought to speak of suitable invariant energies, such as center-of-mass energies or momentum transfer.} but requires a completion in the \ac{UV} regime. Thus, the Lagrangian (density) we know and love is the leading term in a Wilsonian \emph{effective action} $S_\text{eff}$, namely a (hopefully asymptotic to the full answer,\footnote{In the sense that its observables are formal series that are asymptotic to the corresponding observables in the full \ac{UV} completion. Recall that a formal series of functions $\sum_{k>0} f_k(x)$, with $f_k(x) \ll 1$ as $x \to 0$ (for example), is \emph{asymptotic} to $f(x)$ (as $x \to 0$) if and only if for any $N>0$ the distance $\abs{f(x) - \sum_{k<N} f_k(x)} = \orderneglected(f_N(x))$.} and probably divergent) series of terms of the schematic form
\begin{eqaed}\label{eq:generic_EFT}
    S_\text{eff} \sim \int \rmd^dx \, \sqrt{-g} \left(\lagrangian{}_\text{stuff} + \frac{\MPl^{d-2}}{2} \, R + \MPl^{d-2} \sum_k c_k \,  \frac{\mathcal{O}_k(\covD, \mathcal{R}, \dots)}{\UVcutoff^{k-2}} \right) ,
\end{eqaed}
where from now on (following our \hyperref[conventions]{conventions}) $d$ denotes the dimension of spacetime. The irrelevant operators $\{ \mathcal{O}_k \}$ of mass dimension $k$ depending on curvatures $\mathcal{R}$, covariant derivatives $\covD$, and the various fields are accompanied by the \emph{gravitational UV cutoff} $\UVcutoff$ and Wilson coefficients $\{c_k\}$, which generally depend on scalar fields (if any) on symmetry grounds. The non-gravitational sector $\lagrangian{}_\text{stuff}$ also contains irrelevant terms that are not linked to \ac{QG} effects, such as those arising by integrating out massive stuff. These can contain gravitational curvatures, as shown, \eg{}, by standard heat kernel computations~\cite{Vassilevich:2003xt} reviewed in \cref{sec:ALESSIABENJAMIN_HK}, but the corresponding suppressing scale does not depend on \MPl{}. The irrelevant terms we highlighted in \eqref{eq:generic_EFT}, instead, do depend on \MPl{} and are interesting for specifically quantum-gravitational new physics.

In this picture, attempting to use such a theory above the \ac{UV} cutoff \UVcutoff{} would be like discovering molecules by staring at a glass of water. At this level, a potential loophole is that strong-coupling (thus non-perturbative) effects could \emph{a priori} save the day, as discussed in the \ac{ASQG} lectures in \cref{sec:ALESSIABENJAMIN}. We shall briefly discuss this scenario in relation to some deeper and more impactful considerations on \ac{QG}. Of course, these considerations are still very much relevant: the problem is not to ``quantize gravity'' (many quantum theories do not even arise quantizing a classical counterpart), nor to ``make quantum mechanics and gravity compatible''; they are, at energy scales smaller than a threshold which is way up there~\cite{Donoghue:1994dn, Donoghue:2017pgk}. Quantum mechanics is a framework of physics, and \ac{GR} is a theory of one particular interaction. The two are not on the same conceptual footing. The problem is to come up with a quantum theory which never ceases to be reliable in any physical regime, and reduces to an \ac{EFT} which includes \ac{GR} as the leading gravitational sector. This is what we define as \ac{QG}. Oh, and before I forget --- $d > 3$ throughout unless otherwise stated. \ac{QG} in low dimensions is qualitatively different and not particularly interesting for what concerns us here. In particular, there are no gravitons! Also, you might recall that we seem to observe $d \geq 4$, and in particular four macroscopic dimensions. Amusingly, we will actually use low-dimensional \ac{QG} to begin our journey to \ac{ST}.

\subsubsection*{Fun with scales}

The \ac{UV} cutoff $\UVcutoff \lesssim \MPl$ cannot be parametrically larger\footnote{We define $f \lesssim g$ as the logical negation of $f \gg g$ (``$f$ is much larger than $g$''), which means that $\lim f/g$ is infinite. Any asymptotic statement requires fixing a limit in the parameter space of the theory (hence ``parametric''). Which is the relevant limit in a given setting is usually clear from the context.} than the Planck scale,\footnote{We define the $d$-dimensional Planck scale as $\MPl^{2-d} = 8\pi \GN$ in terms of Newton's constant in natural units, as per our \hyperref[conventions]{conventions}. The prefactor is irrelevant: by definition, scales are projective quantities independent of overall $\orderneglected(1)$ prefactors which do not scale in the asymptotic regime of interest.} which is a \emph{worst-case-scenario} scale. This is because gravitational interactions generate quantum contributions to irrelevant terms weighted by \MPl{}. In other words, since the effective quantum couplings of gravitons come in powers of $\frac{E}{\MPl{}}$ at some invariant energy scale $E$ probed by some experiment/observable, gravitons become strongly coupled by the time we reach the Planckian regime, and the gravitational \ac{EFT} ceases to be reliable. However, it could break down at much lower scales! One way this could happen is if gravitons become strongly coupled at some parametrically lower strong-coupling scale $\Lambda_\text{sc}$. We will estimate $\Lambda_\text{sc}$ for weakly coupled strings in \cref{sec:gross_mende_ooguri}.

Another way is if unknown massive species\footnote{In the context of \ac{EFT}, a species is a distinct quantum field, such as the electron field or the electromagnetic field, each of which has some physical polarizations.} of mass $m$ are present in whatever new physics is part of the \ac{UV} completion. This includes the case of $n$ compact extra dimensions, where Kaluza-Klein species arise in the effective $d$-dimensional description which ceases to be valid when the Kaluza-Klein mass gap $m_\text{KK}$ is reached. When such species are much lighter than \MPl{}, namely when the extra dimensions are super-Planckian in size, the \ac{EFT} breaks down much earlier when dialing up the energy scale; both the non-gravitational sector at $E = m_\text{KK}$ and the gravitational sector at $E = m_\text{KK}^{\frac{n}{n+d-2}} \MPl^{\frac{d-2}{n+d-2}}$~\cite{Basile:2023blg, Castellano:2023aum, Basile:2024dqq, Bedroya:2024ubj, Aoufia:2024awo}. As we shall see, in \ac{ST} there is a characteristic scale $M_s$ associated to the extended nature of strings relative to particles, and it also appears as a \ac{UV} cutoff suppressing irrelevant operators in the effective action. When strings are weakly coupled, $M_s \ll \MPl{}$, and new physics shows up much earlier than expected from the generic case.

These considerations are nice, but they seem to leave us with little hope of estimating when to expect new quantum-gravitational physics to show up in experiments unless we have some information on the \ac{UV} completion. Luckily, a simple calculation gives us an \emph{upper bound} to the above scales which can still be much smaller than \MPl{}. The relative one-loop correction to the graviton propagator in momentum space, which is well-defined in perturbation theory but not outside of it (see below), is roughly proportional to $N(E)\frac{E^{d-2}}{\MPl^{d-2}}$, where $N(E)$ is the number of species in the spectrum with masses up to $E$. Therefore, at least this effect becomes comparable to the tree-level contribution when $E$ reaches the \emph{species scale} $\Lambda_\text{sp}$ found by solving the (parametric) equation~\cite{Veneziano:2001ah, Dvali:2007hz, Dvali:2007wp, Dvali:2010vm, Calmet:2017omb}\footnote{As shown in the references, the same result can be derived considering the thermodynamics of semiclassical \acp{BH}.}
\begin{eqaed}\label{eq:species_scale}
    \Lambda_\text{sp} = \frac{\MPl{}}{N(\Lambda_\text{sp})^{\frac{1}{d-2}}} \, .
\end{eqaed}
Once more, computing the precise dependence of this scale on physical parameters of the theory hinges on some knowledge of its spectrum, but the above expression shows that including an incomplete spectrum still provides an upper bound. For example, $N_\text{SM}$ copies of the \ac{SM}~\cite{Dvali:2007wp} would provide an upper bound $\Lambda_\text{sp} \lesssim \MPl{}/\sqrt{N_\text{SM}}$ in $d=4$.

All in all, we have~\cite{Basile:2023blg, Aoufia:2024awo}
\begin{eqaed}\label{eq:species_hierarchy}
    \UVcutoff{} \lesssim \Lambda_\text{sc} \lesssim \Lambda_\text{sp} \lesssim \frac{\MPl{}}{N_\text{low-energy}^{\frac{1}{d-2}}} \leq \MPl{} \, .
\end{eqaed}
A scenario in which $\UVcutoff{} \ll \MPl{}$ means that the gravitational (sector of the) \ac{EFT} is not only weakly coupled at low energies, but in its entire range of validity and then some. This is the setting we will examine in the remainder of this section, unless otherwise stated. This scenario opens up a couple of intriguing prospects. Generically\footnote{Meaning up to a negligible (measure-zero) subset of \acp{EFT}. The relevant measure is usually clear from the specific context.} in the gravitational sector\footnote{Other sectors could have different and much smaller cutoffs, such as the electroweak scale for the Higgs mechanism.} $\UVcutoff{} = \MPl{}$, which indicates that direct detection of new quantum-gravitational physics is hopeless for humans. Planck-suppressed effects could however play a role in the context of inflation~\cite{Assassi:2013gxa, Price:2015xwa, Fumagalli:2019ohr, Tokareva:2023mrt}. If $\UVcutoff{} \ll \MPl{}$, perhaps there is a hope? Another consideration, which we will discuss shortly, is that this additional handle on \ac{QG} could be useful to restrict the options for viable \ac{UV} completions ``from the bottom up'', namely starting from physical principles rather than a specific ``top-down'' \ac{UV} completion. These considerations are enough motivation to consider weakly coupled \ac{QG}, but (\emph{at least} in $d=4$) it may be forced on us by the consistency of \ac{EFT}: in $d=4$, gravitational instantons of the Eguchi-Hanson type would be not parametrically controlled relative to the leading contribution, since their weight is schematically $\exp \left( - \, \text{const.} \times \frac{\MPl^2}{\UVcutoff^2}\right)$. If $d>4$ the same argument should go through by dimensional reduction. This was recently exploited in~\cite{Dvali:2024dlb}.

\subsubsection*{Black holes, entropy and holography}

The framework of gravitational \ac{EFT} is awesome. Sean Carroll calls it the \emph{core theory}.\footnote{You can even \href{https://www.preposterousuniverse.com/blog/2015/09/29/core-theory-t-shirts/}{get a T-shirt}. Perhaps the \ac{UV}-complete version will be the \emph{metalcore theory}?} In fact, it is so wise that it contains the seeds of its own demise, so to speak. Hints to what makes it break. In my opinion, the single most important piece of evidence that a field-theoretic description of \ac{QG} would be doomed, as for the metaphorical glass of water, lies in the \emph{mixing between \ac{UV} and \ac{IR} effects} that arises when dynamical gravity is involved. The standard non-gravitational \ac{QFT} lore teaches us that low-energy and long-distance physics is largely decoupled from high-energy and short-distance physics. But what does low and high even mean in the context of energy scales? We assign energy scales because the dynamical equations of motion of usual \acp{QFT} contain Laplacian operators such as $-\covD_\mu \covD^\mu$, whose spectrum gives meaning to energy scales via momenta, Fourier modes, and so on. In other words, energy scales make sense because we are looking at a fixed background spacetime.

In the presence of dynamical gravity spacetime is not fixed, and thus neither is the usual notion of low and high energy. Dimensionful quantities still exist, but there is no meaningful absolute ``ruler'' of energy scales. This observation is reinforced by another consequence of dynamical gravity, namely the existence of \acp{BH}. Here are a few important considerations:

\begin{itemize}
    \item \emph{Very massive \acp{BH} must be huge.} You cannot make a \ac{BH} of mass $M_\text{BH} \gg \MPl{}$ without its radius $R_\text{BH} = M_\text{BH}^{\frac{1}{d-3}} \MPl^{\frac{2-d}{d-3}} \gg \MPl^{-1}$ because of how event horizons work in \ac{GR}. By itself this is not necessarily a hint of \ac{UV}/\ac{IR} mixing, since cars are also large and massive, but this necessity together with the existence of a continuous spectrum for all masses $M_\text{BH} \gtrsim \UVcutoff^{3-d} \MPl^{d-2}$ is an omen for what's to come.
    
    \item \emph{Huge \acp{BH} are tame.} The typical curvature scale at the event horizon is $R_\text{BH}^{-2} \ll \MPl^2$. More precisely, \acp{BH} in the \ac{EFT} have $R_\text{BH} \gg \UVcutoff^{-1}$, and thus their typical curvature scale does not impact the \ac{EFT} for outside observers (as well as unfortunate infalling ones for long periods of time). Tidal forces in the \ac{EFT} are controlled by different powers of these scales~\cite{Horowitz:2023xyl, Horowitz:2024dch, Zigdon:2024ljl}, which can become dangerous in certain situations. Generally speaking, gravitational \ac{EFT} works well for large \acp{BH}, which means that any problematic conclusion derived within \ac{EFT} which holds for arbitrarily large \acp{BH} ought to be taken seriously.
    
    \item \emph{Smashing stuff hard makes huge \acp{BH}.} It is widely expected~\cite{Giddings:2007qq} that scattering particles at very high center-of-mass energies and sufficiently small impact parameter produces \acp{BH}, which can then act as intermediate resonances emitting outgoing particles as Hawking radiation~\cite{Hawking:1975vcx} (cf. \cref{sec:FRANCESCO}). This is where something starts to smell: the higher the energy, the \emph{larger} the resulting \acp{BH}, placing us back into \ac{IR} physics. This is a signature of \ac{UV}/\ac{IR} mixing.
    
    \item \emph{Huge \acp{BH} know about the \ac{UV}.} You could try to dismiss the above argument on the grounds that we do not understand \ac{BH} formation from microscopic scattering very well. However, the same signature still shows up. \acp{BH} have an entropy, which in the macroscopic limit is given by the remarkable Bekenstein-Hawking asymptotic formula~\cite{Bekenstein:1973ur, Hawking:1976de, Gibbons:1976ue, Wald:1999vt}
    \begin{eqaed}\label{eq:BH_entropy}
        S_\text{BH} = \frac{\text{Area}(R_\text{BH})}{4\GN} + \alpha \, \ln \frac{\text{Area}(R_\text{BH})}{\GN} + \dots \, ,
    \end{eqaed}
    with $\alpha$ \ac{EFT}-dependent, where $\text{Area}(R_\text{BH})$ denotes the area of the event horizon and the sub-leading corrections are weighted by inverse powers of $\UVcutoff^{d-2} \text{Area}(R_\text{BH})$. Among other things, this result implies that large \acp{BH} have microstates of very high energy, mixing \ac{UV} and \ac{IR} physics. These microstates cannot be ascribed to the same physics underlying a box of stuff or a lump of coal: if that were the case, the approximately local low-energy dynamics would yield an extensive entropy. The Bekenstein-Hawking asymptotics is instead \emph{holographic}, in the sense that it scales with the area of the \ac{BH}. In non-gravitational theories in a ``critical'' regime, certain solitons dubbed \emph{saturons} can exhibit similar properties~\cite{Dvali:2020wqi}, but do not comprise a continuous spectrum of arbitrarily large mass and size. In this sense, while the Bekenstein bound on entropy is not inherently gravitational~\cite{Casini:2008cr} since \MPl{} does not appear, the \emph{Bekenstein-Hawking} bound is. And since gravity couples to everything by the equivalence principle, these microstates will generically show up as intermediate resonances in high-energy scattering. In fact, they seem to dominate it, since the Bekenstein-Hawking scaling can be rewritten as $S_\text{BH} \propto E^{\frac{d-2}{d-3}}$ which is much larger than any field-theoretic degeneracy. This leads to the concepts of \emph{asymptotic darkness}~\cite{Banks:1999gd, Giddings:2007qq, Giddings:2009gj} and \emph{\ac{BH}/tower correspondence}~\cite{Basile:2023blg, Herraez:2024kux}, invalidating a field-theoretic representation of the density of states~\cite{Banks:1999az, Banks:2010tj} which would be inconsistent with local field operators even kinematically~\cite{Keltner:2015xda} (``non-localizable'', in Jaffe's terminology~\cite{Jaffe:1967nb}).
\end{itemize}

Let me delve a bit deeper into the idea of holography. The presence of an area law for \emph{large} \acp{BH}, together with the Bekenstein-Hawking bound on the maximum entropy in a region, suggests that in some sense the physics in a region of spacetime should be isomorphically encoded in its boundary, and that this encoding be \emph{local on the boundary}. Actually, there are additional hints pointing to this principle~\cite{tHooft:1993dmi, Susskind:1994vu}:

\begin{itemize}
    \item Canonical gravity in the Arnowitt–Deser–Misner formalism yields a Hamiltonian which is purely supported on the boundary. This is not surprising, since the bulk Hamiltonian is a constraint which generates certain diffeomorphisms. The physical dynamics appears to involve the boundary.
    
    \item Due to (\href{https://physics.stackexchange.com/a/346812}{active}) diffeomorphism invariance, gravity has no local observables. A work\-around would be to define local quantities relative to a gauge fixing, which would be a type of relational observable. However, no such gauge fixing is possible globally in the space of metrics~\cite{Singer:1978dk}. Locally it is possible, \eg{}, via the \ac{BRST} approach, which is why perturbation theory still makes sense. More intuitive relational observables, completing the perturbative construction of~\cite{Frob:2022ciq}, would not be sharply defined; a sufficiently precise measurement would involve a very heavy apparatus in the region at stake, in order to suppress fluctuations. Eventually, this leads again to \ac{BH} formation.\footnote{For a more in-depth discussion on this operational point, see \eg{} Nima Arkani-Hamed's interview in~\cite{Armas:2021yut}.} Finally, the notion of full background independence~\cite{Casadio:2022ozp} entails that no dependence on the spacetime manifold can show up in observables, except for an asymptotic boundary which does not fluctuate in the (semiclassical) path integral due to infinite action. Incorporating fluctuations of spacetime topology~\cite{Hawking:1978pog, Gibbons:1991tp, McNamara:2019rup, McNamaraThesis, McNamara:swamplandia} makes this explicit, and as a result the only sharply defined observables which can be in principle measured to arbitrary accuracy are boundary observables, such as scattering amplitudes.
    
    \item From the sum over spacetime topologies, in particular wormholes, the Bekenstein-Haw\-king entropy can be recovered from an overcomplete set of semiclassical states~\cite{Balasubramanian:2022gmo} in an independent way, suggesting a connection between the area law for \acp{BH} and topology fluctuations. Indeed, the latter follow from the holographic principle~\cite{McNamaraThesis}, and all these ideas are nicely connected.
\end{itemize}

The holographic principle is incompatible with a field-theoretic description of \ac{QG}. Given all the above considerations, we may try to seek an alternative which is compatible with them. From this markedly bottom-up point of view, how would one land on strings if not by accident, and only then realizing that it provides a theory of \ac{QG}~\cite{Schwarz:1982jn}?

\subsubsection{Why strings?}\label{sec:why}

As will hopefully become apparent when we will delve deeper in its construction, in \ac{ST} strings are sort of like phonons, but for everything (spacetime, matter, you name it). As such, it only really makes sense to talk about (fundamental\footnote{Solitonic strings can make sense also away from weak coupling.}) strings at weak coupling. In the \ac{EFT} language of the preceding section, the condition is $\UVcutoff{} \ll \MPl{}$. Is there a way to restrict the possible \ac{UV} completions of gravity and find some stringy hints from this?

The answer is yes. Let us consider asymptotically flat spacetimes (as we will do in the remainder of these notes unless otherwise states). Consider the (reduced, removing the mo\-men\-tum-conserving Dirac delta) amplitude to scatter two gravitons into two gravitons, expressed in terms of the standard Mandelstam variables $s, t, u$ (for gravitons $s+t+u=0$), defined as in \cref{sec:ANNA} by the external momenta $\{ p_i \}$ according to
\begin{eqaed}\label{eq:mandelstam_def}
    s \equiv - \, (p_1+p_2)^2 \, , \qquad t \equiv - \, (p_1 + p_3)^2 \, , \qquad u \equiv - \, (p_1+p_4)^2 \, .
\end{eqaed}
The amplitude then takes the form\footnote{In general, more kinematic factors can appear. This does not happen in the cases we shall consider. In any case, one can focus on \eg{} the term contributing to the maximally helicity-violating amplitude, which is already non-trivial at tree level in \ac{GR}.}
\begin{eqaed}\label{eq:graviton_amplitude}
    \scatteringamplitude = \mathbf{K} \, F(s,t,u) \, ,
\end{eqaed}
where $\mathbf{K}$ is a kinematic factor accounting for the polarizations of the gravitons with some momentum contractions, and $F$ is a crossing-symmetric function of $s,t,u$. In $d=4$ spacetime dimensions, the maximally helicity-violating amplitude has a \href{https://youtu.be/gd3g0rOG9w4?t=3021}{particularly simple kinematic prefactor} when expressed in terms of spinor-helicity variables~\cite{Elvang:2013cua}.

When gravitons are weakly coupled, only the tree-level contribution is relevant (when non-vanishing). Let us do some dimensional analysis: $\mathbf{K}$ and \GN{} have mass dimensions eight and $2-d$ respectively, while the amplitude has mass dimension $4-d$.\footnote{This can be easily derived from the expression of $\scatteringamplitude \, \delta^{(d)}(p_{\text{f}}-p_{\text{i}})$ as a matrix element of a dimensionless unitary evolution operator in momentum eigenstates. With these conventions, an $n$-point amplitude has mass dimension $d-\frac{d-2}{2}n$.} Hence, $F$ has dimension $-d-4$. At tree level there must be a factor of \GN{}, which means that $F_\text{tree}(s,t,u) = \GN{} \, \widetilde{F}(s,t,u)$ with $\widetilde{F}$ of dimension $-6$. Because of locality, $\widetilde{F}$ must have \emph{simple} poles. Crossing symmetry then implies $\widetilde{F} = \frac{1}{stu}$ up to a numerical prefactor which can be reabsorbed in $\mathbf{K}$. In \ac{GR} there is no additional scale, and couplings with matter cannot contribute at tree level, so there is no other option. All in all, in \ac{GR}~\cite{DeWitt:1967uc}
\begin{eqaed}\label{eq:graviton_amplitude_tree}
    \scatteringamplitude^\text{GR}_\text{tree} = \mathbf{K} \, \frac{\GN}{stu} \, .
\end{eqaed}
In a \emph{weakly coupled} \ac{UV} completion it makes sense to talk about tree-level amplitudes of gravitons. The general expression must take the form
\begin{eqaed}\label{eq:UV_amplitude}
    \scatteringamplitude^\text{UV}_\text{tree} = \mathbf{K} \, \frac{\GN{}}{stu} \, C\left(\frac{s}{\UVcutoff^2} \, , \, \frac{t}{\UVcutoff^2} \, , \, \frac{u}{\UVcutoff^2}\right)
\end{eqaed}
with some completion function $C$ encoding Wilson coefficients upon expanding it in powers of $s, t, u$. The kinematic factor can also receive corrections~\cite{Kawai:1985xq}. For example, the constant term $\alpha \, \MPl^{-6}$ in $\frac{C}{stu}$ in Planck units is a sharp observable related to the quartic Riemann term in the effective action. This is also true for the exact amplitude, which can be amenable to non-perturbative bootstrap methods~\cite{Guerrieri:2021ivu, Guerrieri:2022sod}. For instance, in 10-dimensional maximal supergravity the minimal value of $\alpha$ allowed by the purely bottom-up bootstrap computations of~\cite{Guerrieri:2021ivu} is $\alpha_\text{min} \approx 0.14$. A sector of \ac{ST} called 10-dimensional type II reduces to this class of \acp{EFT}, in which $\alpha^\text{string}$ is known exactly. It turns out that it spans all real values above $\alpha^\text{string}_\text{min} \approx 0.14$!

The idea here is that \emph{not anything goes} for a \ac{UV} completion. You cannot just choose any $C$ you like. The S-matrix bootstrap program provides us with a collection of methods to constrain $C$ imposing the physical requirements of unitarity, causality, crossing and so on that we have also encountered in \cref{sec:ANNA}. Without delving too much into the details, there is now a quite extensive body of literature~\cite{Camanho:2014apa, Caron-Huot:2016icg, Afkhami-Jeddi:2018apj, Alonso:2019ptb, Arkani-Hamed:2020blm, Huang:2022mdb, Caron-Huot:2022ugt, Geiser:2022icl, Geiser:2022exp,  Cheung:2022mkw, Cheung:2023adk, Cheung:2023uwn, Arkani-Hamed:2023jwn, Haring:2023zwu, Cheung:2024uhn, Cheung:2024obl} which indicates that for weakly coupled \ac{UV} completions of gravity $C$ must be stringy, and thus also its poles represent stringy resonances. The basic lessons are that \ac{UV}-completing graviton scattering perturbatively needs an infinite tower of higher-spin species~\cite{Camanho:2014apa, Afkhami-Jeddi:2018apj} to save causality --- which indeed string spectra do~\cite{DAppollonio:2015fly} --- and while non-stringy versions of non-gravitational amplitudes have been constructed, they fail to satisfy certain properties pertaining to the factorization of residues onto massive poles\footnote{This is a consistency requirement: tree-level scattering amplitudes are built gluing exchanged propagators and vertices. So there must exist a consistent set of couplings valid for all amplitudes and factorization channels.}~\cite{Arkani-Hamed:2023jwn}. Gravitational \ac{UV}-completions are even harder to find~\cite{Arkani-Hamed:2020blm, Geiser:2022exp, Geiser:2022icl, Cheung:2022mkw, Cheung:2023adk}, further supporting the rigidity of gravity with respect to the basic physical principles. The presence of a tower of new species gapped at the cutoff~\cite{Afkhami-Jeddi:2018apj} is also supported by information-theoretic arguments connected to the equivalence principle of gravity~\cite{Stout:2022phm}, and their extended nature follows applying this reasoning in small-radius limits of a compactification~\cite{Ooguri:2006in}.

Since \acp{BH} have played an important role in guiding our reasoning thus far, perhaps it could be worth examining their role in this discussion. Looking back at \eqref{eq:species_scale} and \eqref{eq:species_hierarchy}, it is apparent that the species scale and the \ac{UV} cutoff contain information on the nature of the new species needed to \ac{UV}-complete gravity at weak coupling. What would happen if we considered \emph{minimal} \acp{BH} of parametric size $R_\text{min} = \UVcutoff^{-1}$? It turns out that their thermodynamics is connected to that of the new species~\cite{Cribiori:2023ffn} in a calculable way~\cite{Basile:2023blg, Herraez:2024kux}, and similarly for their effect on the density of states extracted from graviton scattering~\cite{Bedroya:2024ubj}. The consistency of both pictures independently indicates that the only towers of new species which can perturbatively \ac{UV}-complete graviton scattering consistently are excitations of quantum strings. In particular, as we shall see in \cref{sec:spectra}, their entropy (logarithmic degeneracy) scales linearly in the mass in units of their mass gap $M_s$~\cite{Bedroya:2024ubj, Herraez:2024kux}.

This is a modern perspective which guides us to consider strings from a more solid footing. In particular, the rigidity of its structure, its connections to our beloved framework of \ac{QFT} (which go much deeper due to the \ac{AdSCFT} correspondence), and the indications garnered from but a handful of physical principles is what some researchers like David Gross refer to as ``radical conservatism'' --- to my understanding, it is the idea that given a solid theoretical framework it is much safer to, say, add particles here and change some interactions there rather than touch the foundational principles.

Intuitively, the reason why strings work better than particles to build \ac{UV}-complete dynamics is that their extended nature softens the \ac{UV} behavior. However, this intuition does not capture the several \emph{string}ent (get it?) consistency conditions we will encounter.

\subsubsection{What strings do and don't do --- some misconceptions}\label{sec:what}

Before outlining the road to string perturbation theory at the technical level, I would like to pause and discuss various aspects of the theory, its relation to phenomenology and some widespread misconceptions.

\subsubsection*{What (not) to expect}

To begin with, what should we even expect from a theory of \ac{QG} in the first place? Any such theory should produce physical quantities (observables) which are well-defined in all physical regimes and reduce to those of gravitational \ac{EFT} in the appropriate limit, as outlined above in the case of graviton scattering. From a low-energy perspective, deviations from the leading description are encoded in Wilson coefficients, as in \eqref{eq:generic_EFT}, and indeed they can be extracted\footnote{More precisely, only combinations of Wilson coefficients which are invariant under field redefinitions are meaningful, and only those can be extracted from physical observables.} from \eg{} S-matrix elements. In order to probe these effects with some statistical significance, one should somehow probe scales of the order of the \ac{UV} cutoff, which is generically Planckian. Even if smaller by various orders of magnitude,\footnote{For example, the ``dark dimension'' scenario of~\cite{Montero:2022prj} estimates $\UVcutoff \approx 10^{-10} \, \MPl{}$ for the cutoff of the \emph{gravitational} sector.} it is natural to expect that we should not see any new gravitational physics anytime soon, barring a lucky break such as unexpectedly large extra dimensions whose presence is somehow induced by quantum-gravitational consistency (as \eg{} in~\cite{Montero:2022prj}). For what concerns low-energy physics, in principle one could make do playing with a handful of \ac{EFT} parameters. Pragmatically speaking, there doesn't seem to be any need for a \ac{UV} completion in this sense.

Suppose you manage to measure one of the first Wilson coefficients in your lifetime. What does that tell you? How stringent a constraint on our favorite \ac{UV} completion this would entail depends on how much these coefficients can vary. The theory may have free parameters\footnote{Or it may have no free parameters~\cite{McNamara:2020uza}, such as in the case of \ac{ST}. There's a quote by Einstein along these lines in his autobiographical notes, but the only source I could find are some seminars by Hirosi Ooguri. \href{https://youtu.be/avPZnq48kqE?list=PLAMDKMENB2xAqm1dB7836_oOB1WYCzmk2&t=317}{Here's the most recent one.}} and/or multiple \acp{EFT} stemming from it. From the point of view of \ac{QG}, which naturally ``lives'' at Planckian scales, there is no reason to assume that the low-energy physics be simple and clear-cut to study and classify. The natural predictions of such a theory take place in the \ac{UV}, where the messy details of the \ac{IR} should be washed out and replaced by some universal dynamics. The fact that this regime is hard to experimentally probe for 21st century humans is due to the unfortunate truth that gravity in our universe is \emph{exceptionally weak}: even when probed with the heaviest known subatomic particle, the top quark, its quantum effects are weighted by powers of $\frac{m_\text{top}}{\MPl{}} \approx 10^{-17}$, which is staggeringly small. An alternative way to frame this problem, often pointed out by Nima Arkani-Hamed, is that the particles we observe are absurdly light relative to the fundamental Planck scale. What --- if anything --- is generating this hierarchy? Who knows.\footnote{Don't get me started on the cosmological hierarchy $\frac{\Lambda_\text{dark}}{\MPl^4} \approx 10^{-120}$\dots although according to the considerations in~\cite{Montero:2022prj} the two hierarchies could be connected, including also dark matter~\cite{Vafa:2024fpx}.}

Another issue is that any \ac{UV}-complete description of gravity must include everything else in the game. All matter, all interactions. This is because gravity couples to everything, and these effects cannot be selectively turned off. All these considerations boil down to one simple point: what one should expect from a quantum theory of gravity are sharp and hopefully universal predictions of what stuff does in the \ac{UV}, while the \ac{IR} can be messy but should also be calculable, at least in principle, similarly to how the mass of the proton should be calculable from \ac{QCD}.

There is however one more way that a theory of \ac{QG} can be useful to us puny humans even with a messy \ac{IR} with many vacua and \acp{EFT}. If the pool of possibilities that arises is much more restricted relative to the na\"{i}ve set of all \acp{EFT}, it means that in the \ac{UV} completion at stake \emph{not anything goes} when it comes to low-energy physics. In fact, \emph{almost nothing goes}! Sometimes there are ways to understand how the consistency of \ac{QG} with basic physical principles implies such \ac{IR} constraints via \ac{UV}/\ac{IR} mixing. We will explore this point in more detail in \cref{sec:landscape} and \cref{sec:swampland_stuff}. Ideas along these lines fall under the umbrella of the \emph{swampland program}~\cite{Vafa:2005ui, Brennan:2017rbf, Palti:2019pca, vanBeest:2021lhn, Agmon:2022thq}.

\subsubsection*{What string theory does}

\begin{itemize}
    \item It is a consistent quantum theory of gravity. Namely, its classical limits contain gravity (unavoidably!), matter and other interactions as described by a gravitational \ac{EFT} led by \ac{GR} and Yang-Mills theory coupled to matter, as befits the general structure of renormalizable weakly coupled relativistic \acp{QFT}~\cite{Anselmi:2019pdm}. In particular, closed strings lead to the gravitational sector, while open strings would necessarily produce closed strings by interaction. The theory avoids all possible gauge anomalies in a highly non-trivial fashion, since there is no room to change its rigid structure (more on that later).
    
    \item Depending on the vacuum configuration (which specifies the whole universe --- it's \ac{QG}, after all!), the \ac{IR} matter content, gauge interactions, couplings and masses etc. change, but they are completely determined and calculable, at least in principle.
    
    \item The set of \acp{EFT} it produces is numerous in absolute terms, but it appears to be negligible (countable, or even finite) compared to the continuum of naively consistent \acp{EFT} coupled to gravity. In this sense, one may view \ac{ST} as a framework or toolbox to produce consistent \acp{EFT} coupled to gravity (and also some that are not, with some decoupling limits). However, its uniqueness and rigid high-energy behavior tells us that the laws of physics it entails are those of a single theory.
    
    \item It has no free (continuous) parameters. Stringy effects are characterized by the string coupling constant $g_s$, which is the background value of a dynamical mode, and the string scale (inverse tension) $\alpha' \propto M_s^{-2}$, also historically known as Regge slope for reasons that will become clear in \cref{sec:examplitudes}. More generally, the theory is widely believed to be unique, since its apparently separated distinct limits are actually connected by dualities~\cite{Witten:1995ex}. Moreover, many consistent structures that have been investigated (special matrix models, supermembranes~\cite{Duff:1996zn} and more) turned out to be connected to it and became part of the framework. This suggests that, similarly to how a sphere cannot be covered by a single coordinate chart, there is no single complete formulation of the theory, rather a patchwork of charts.
    
    \item It produces scattering amplitudes which have a soft \ac{UV} behavior and match \ac{EFT} in the \ac{IR}. In particular, all Wilson coefficients are fixed by the vacuum configuration, while the \ac{UV} is universal~\cite{Gross:1987ar, Gross:1987kza, Mende:1989wt}. In some cases exact results are available, and they satisfy non-trivial consistency constraints coming the non-perturbative bootstrap~\cite{Guerrieri:2021ivu, Guerrieri:2022sod}.
    
    \item In all calculable cases thus far, it \emph{precisely} reproduces the Bekenstein-Hawking entropy from a microscopic counting of microstates, starting from the original work of~\cite{Strominger:1996sh}. More generally, the theory is holographic and is compatible with swampland conditions motivated independently from the bottom up. This holds beyond perturbation theory, which is good since swampland constraints do not rely on it. Some examples: the theory has no global symmetries. Whenever an Abelian gauge group is present there are massive particles whose mass in Planck units is lighter than their charges (``gravity is the weakest force''). Gauge groups are compact and the spectrum contains states carrying all representations (``completeness''). More on this stuff in \cref{sec:swampland_stuff}.
    
    \item It requires fermions to have a stable vacuum. This was recently proven for closed strings in~\cite{Angelantonj:2023egh}, as a corollary of a more general result. Open strings also get fermions for other reasons connected to anomaly cancellation. This could actually be a general requirement for \ac{QG}, see \eg{}~\cite{Dvali:2024dlb} for some arguments in this direction.
    
    \item As we shall see in \cref{sec:BH_transition}, at weak coupling its vacuum energy density is automatically small without (further) fine-tuning.
\end{itemize}

\subsubsection*{What string theory doesn't do}

\begin{itemize}
    \item It does not fix the number of spacetime dimensions, although it constrains it. To present knowledge, weakly coupled strings in a tame (namely weakly curved\footnote{More precisely, by \emph{tame spacetime} we mean a spacetime with all curvatures (gravitational, gauge, etc.) uniformly parametrically smaller than the \ac{UV} cutoff. It is a necessary condition to be able to talk about a \emph{bona fide} \ac{EFT} at low energies.}) and (at least classically) stable\footnote{Without requiring stability, the upper bound is $d \leq 26$.} spacetime predict the \emph{upper bound} $d\leq 10$, which is raised to $d \leq 11$ dropping the weak coupling requirement.\footnote{Literally, turned up to eleven.} Tame but not classically stable spacetimes are less understood. The maximal dimension provides the simplest settings, and lower-dimensional configurations can be attained by compactification. However, in many other configurations the required extra degrees of freedom do not have any direct connection to additional spatial dimensions. When the gravitational sector becomes weakly coupled, $\UVcutoff \ll \MPl$, due to \emph{something other} than a small string coupling $g_s \ll 1$, the extra degrees of freedom do appear to become additional dimensions which open up (``decompactify'' in the lingo)~\cite{Aoufia:2024awo, Ooguri:2024ofs}.
    
    \item It does not predict spacetime supersymmetry. More precisely, there exist sectors of the theory where spacetime supersymmetry is broken at the string scale or absent altogether, without any tachyons present. It turns out to be quite tricky to understand the resulting \ac{IR} physics in these situations, but perhaps they point to an ingredient for a realistic cosmology. Whether a \emph{non-perturbatively stable} vacuum requires spacetime supersymmetry is not settled, although it seems quite likely.\footnote{See also~\cite{Dvali:2024dlb} for some recent bottom-up arguments along similar lines.} When spacetime supersymmetry is actually present it provides powerful calculational tools, but there are no hints that it should somehow break (or ``super-Higgs'')  at energy scales accessible to humans. If anything, the natural scale for supersymmetry to break would be the string scale. On the other hand, \emph{worldsheet supersymmetry} is present, but it has nothing to do with the presence of superpartners in spacetime. Indeed, the worldline formulation of perturbative \ac{QFT} has the same feature to encode spinning (fermionic --- thanks, spin-statistic theorem!) particles in spacetime. In this sense, it is as much an ingredient as it is in standard \ac{QFT}.
    
    \item It does not uniquely fix the low-energy physics, since it has multiple vacuum configurations.\footnote{To be pedantic: in gravity the usual notion of vacuum is a bit shaky. To make it precise, one has to fix the asymptotic boundary of spacetime. But this is superselected anyway, since its fluctuations are infinitely suppressed in the semiclassical path integral. Within a given asymptotically flat (where the observables are S-matrix elements) or \ac{AdS} (where the observables are boundary correlators) sector, one can talk about vacua. For example, in asymptotically flat sectors, the positive energy theorem shows that flat spacetime is a \emph{bona fide} vacuum~\cite{cmp/1103919981}. As for \ac{dS}, it's weird and fraught with subtleties regarding holography, observables and stability~\cite{Banks:2001yp, Dyson:2002nt, Dvali:2017eba, Bedroya:2022tbh, Witten:2023qsv, Witten:2023xze, Banks:2024lvl}. See also~\cite{Banks:2010tj} for related considerations. We will not worry too much about this subtlety here, focusing on asymptotically flat sectors, which can also have cosmological histories (see~\cite{Dvali:2024dlb} for further recent comments on this point).} This is not surprising: the \ac{SM} \ac{EFT} (affectionately known as SMEFT, or GRSMEFT for committed folks) also has many such configurations~\cite{Arkani-Hamed:2007ryu, Gonzalo:2021zsp}. This is expected for any theory of gravity, simply because of the option to compactify a dynamical spacetime. Since these options are consistent on theoretical grounds, there must be some phenomenological input in order to narrow things down. In the case of the \ac{SM} it is pretty easy; for \ac{ST} not so much, although all things considered many admirable efforts got us pretty far (see \eg{}~\cite{Cvetic:2019gnh}, but the literature on string phenomenology is quite vast).
    
    \item Probably other things I can't think of at the moment.
\end{itemize}

Hopefully, we have convinced you that \ac{ST} is a worthwhile endeavor and that several hints point to it as an extension of the framework of quantum fields and gravity which complies with the basic principles of physics as we understand it. To conclude this first part of the section, we will set the stage for what comes next more concretely.

\subsubsection{How we'll proceed --- the worldline approach to QFT}\label{sec:how}

\begin{figure}[ht!]
    \centering
    \includegraphics[scale=0.6]{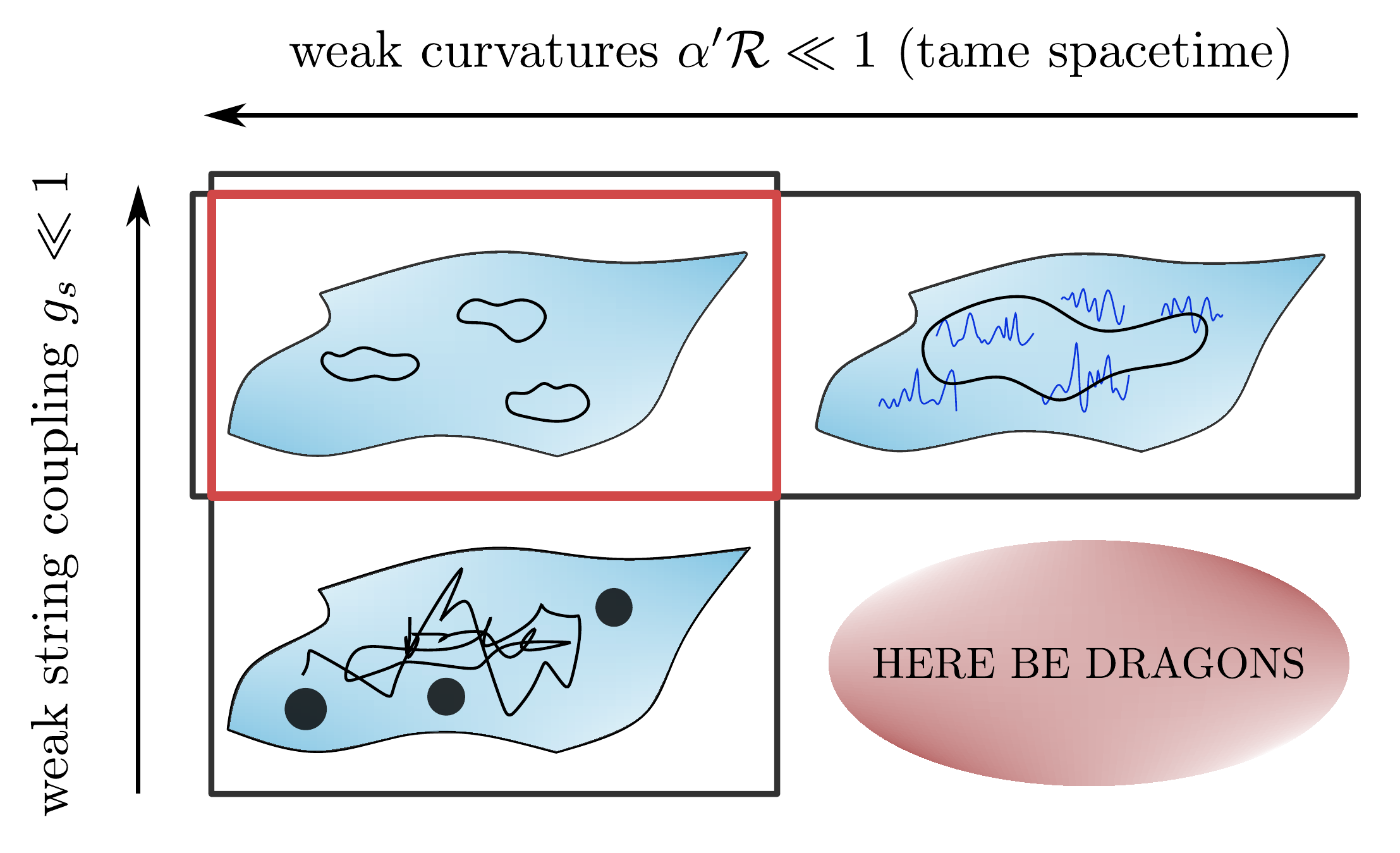}
    \caption{A depiction of various regimes of \ac{ST}. At weak string coupling $g_s \ll 1$ string perturbation theory is reliable, and it makes sense to distinguish strings from their background. The latter can be highly curved, in which case strings become big and wobbly or even tensionless~\cite{Maldacena:2000hw, Sundborg:2000wp, Gaberdiel:2017oqg, Eberhardt:2020bgq, Eberhardt:2021jvj, Demulder:2023bux}. Strongly coupled strings lose their distinction relative to the background, other solitonic extended objects and \acp{BH}. We focus on the highlighted upper-left corner, with closed oriented strings interacting weakly in a tame spacetime whose curvatures are parametrically smaller than the string scale $\alpha'$ (to be defined shortly). This is the regime in which strings are tough and rigid, and a weakly coupled \ac{EFT} including gravity arises at low energies.}
    \label{fig:classes}
\end{figure}

The basic idea should be more or less clear by now: we want to study the physics of quantum strings which are closed and interact weakly in a tame spacetime, as depicted in \cref{fig:classes}. The first restriction is not particularly important, since gravity comes from closed strings. Open strings are cool~\cite{Angelantonj:2002ct}, but they would take much more time than we have at our disposal. The second is mostly due to time constraints, but also to how much simpler weakly coupled systems are relative to their dastardly strongly interacting counterparts. In \ac{ST} in asymptotically flat sectors, it is always possible to take this limit. Finally, the third restriction is simply phenomenological (and a matter of convenience), although strings in highly curved backgrounds hold many intriguing lessons~\cite{Maldacena:2000hw, Sundborg:2000wp, Gaberdiel:2017oqg, Eberhardt:2020bgq, Eberhardt:2021jvj, Demulder:2023bux}.

\subsubsection*{One (st)ring to rule them all}

As a valiant warrior once said --- one does not simply study strings interacting in spacetime. There is no analog of a bare action principle, as should be expected from our preceding considerations. The formalism of open string field theory provides a notion of action principle, but attempting to extend it to the unavoidably present\footnote{Open strings can fuse into closed strings dynamically.} closed strings one runs into problems which likely reflect our considerations in the preceding sections. One can still build an effective action encoding perturbative string amplitudes, but it contains a very complicated structure with infinitely many vertices. Ultimately, it is more convenient and instructive for us to look elsewhere.

\begin{figure}[ht!]
    \centering
    \includegraphics[scale=0.6]{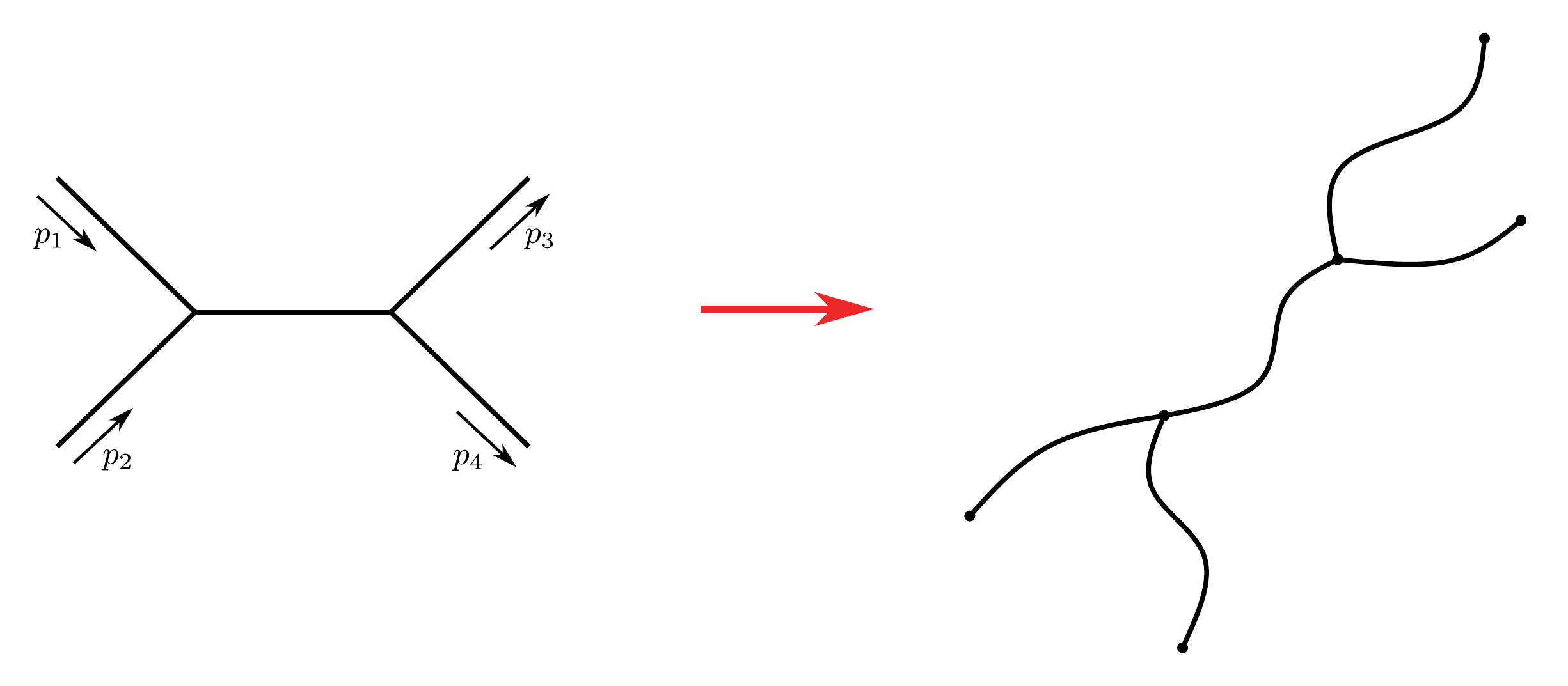}
    \caption{The worldline approach to perturbative \ac{QFT} builds Feynman diagrams from one-dimensional \ac{QG} on the worldline of particles in the theory. The interactions represented by vertices are specified in the prescription for the worldline path integral.}
    \label{fig:worldline}
\end{figure}

Luckily for us physics, like mathematics, often has redundancies in language. Some things can be recast in a different way using different structures, without changing the underlying substance. Since we are interested in weakly coupled strings, a natural starting point is to see if we can do this for weakly coupled particles.\footnote{Thinking along similar lines has led to a number of insights in recent work, \eg{}, in~\cite{Arkani-Hamed:2023lbd}.} Here, Feynman diagrams provide nice building blocks of perturbation theory, but they are often seen as mere computational tools. But if we take the pretty pictures we see in textbooks more seriously, we are led to consider each diagram as an actual process of particle propagation, decorated with some special spacetime events where some local interaction occurs and worldlines split or join. This can be made more precise by writing down a path integral encoding the amplitudes for particles to propagate along the diagram, as depicted in \cref{fig:worldline}. Conceptually, this means that we are dealing with a theory of a \emph{single particle}, and introduce interactions by hand. It turns out that the analogous (worldsheet) approach to strings has no such issue: the interactions are already included in the description of single-string propagation. One string to rule them all.

Let us try to do this for a particle first, to see how perturbative \ac{QFT} is recovered. The action of a particle propagating in the spacetime manifold $(M,g)$ is a functional of its worldline $W$, which is defined by an embedding $X : W \to M$, a curve in spacetime. In the following we use index-free notation whenever possible and unambiguous. This induces a metric $\widetilde{g} \equiv X^\ast g$ on $W$ given by pulling back the spacetime metric $g$ onto $W$. Using some local coordinate $\tau$ on $W$, the embedding map has components $X^\mu(\tau)$ and the pullback has components
\begin{eqaed}\label{eq:worldline_metric}
    \widetilde{g}_{\tau \tau} = \dot{X}^\mu \, \dot{X}^\nu \, g_{\mu \nu}(X(\tau)) \equiv \dot{X}^2 \, ,
\end{eqaed}
with $\dot{X}^\mu \equiv \frac{dX^\mu}{d\tau}$. The ``volume'' form is then $d\ell = \sqrt{-\widetilde{g}_{\tau \tau}} \, d\tau$, which integrates to the proper time of the particle. This is a natural action to use, and indeed the principles of \ac{EFT} applied to $W$ tell us that in general it is there but it can be also accompanied by other stuff, for instance terms containing higher derivatives of $X$. For now we simply ignore these as a matter of simplicity, but when doing the same for strings we will be forced to by renormalizability, since in that case we are looking for a weakly coupled \ac{UV} complete theory. We are thus led to consider the action of our particle as a \emph{one-dimensional \ac{QFT}} on $W$: the field $X$ is a \emph{worldline scalar} (the Lorentz indices pertain to spacetime!) expressing the spacetime position of the particle, but we know from representation theory (\eg{} of spacetime isometries) that particles can carry internal degrees of freedom, such as spin. These can be encoded in other (possibly fermionic) fields on $W$. Our tentative worldline (``wl'') action is then
\begin{eqaed}\label{eq:worldline_action}
    S^\text{tentative}_\text{wl}[X, \dots] = - \, m \int_W \rmd\ell + S_\text{other stuff} \, .
\end{eqaed}
The fact that the prefactor $m$ is the mass of the particle\footnote{For massless particles, whose proper time vanishes, one can simply start from the end result of the procedure we are outlining.} follows from Noether's theorem for momentum and energy. \eqref{eq:worldline_action} doesn't look much like a usual one-dimensional \ac{QFT}. Where are the kinetic terms? It looks like it would be difficult to build a path integral for this kind of theory and reproduce amplitudes given by Feynman diagrams. There is a way to massage the above action into a classically\footnote{At the quantum level things are more subtle, but a \href{https://physics.stackexchange.com/a/268810}{careful Hamiltonian analysis of constraints \`{a} la Dirac} suggests that it should work in the same way.} equivalent version, introducing a redundant degree of freedom, namely a \emph{dynamical metric on the worldline}. We will call it $\gamma$ to distinguish it from the spacetime metric $g$. The new action actually looks like a good ol' one-dimensional \ac{QFT} on $W$. It is given by
\begin{eqaed}\label{eq:worldline_action_g}
    S_\text{wl}[X,\gamma, \dots] = - \, \frac{1}{2}\int \rmd\tau \, \sqrt{-\gamma_{\tau \tau}} \left( \gamma^{\tau \tau} \, \dot{X}^\mu \, \dot{X}^\nu \, g_{\mu \nu}(X) + m^2\right) + S_\text{other stuff} .
\end{eqaed}
It is easy to see that placing $\gamma_{\tau \tau}$ on shell recovers the preceding action. The rest of the action is now also minimally coupled to $\gamma$: we are doing \emph{dynamical gravity on the worldline}! The attentive reader may notice a pressing issue in the above expression. Since the timelike direction $X^0$ appears with a different sign relative to the spatial components $X^i$ in the kinetic term, one may worry that ghosts might be present. Luckily, the coupling to $\gamma$ eliminates this spurious ghost because of gauge redundancy. There's no time to show it explicitly, but many of the quoted references present quantization in the so-called light-cone gauge in detail for strings, in particular~\cite{Angelantonj:2002ct}. At the level of canonical quantization (for example in flat spacetime), this can be seen as a consequence of (a representation of) the Virasoro algebra of constraints (more on this later). The upshot is that the field equation for $\gamma$ is a constraint because $\gamma$ itself was introduced as a spurious degree of freedom.

This is worth pausing for a moment. With a dynamical metric on the worldsheet, upon writing down a path integral we are doing \emph{\ac{QG} in one dimension}. There cannot be any Einstein-Hilbert term, but there is a cosmological one (the $m^2$ term) and minimal coupling to matter fields on $W$ (in particular, to the scalars $X$ describing the position of the particle in spacetime). Pretty much the most general thing we can write as a Lagrangian. There is a conceptual distinction between the worldline $W$ and spacetime $M$, also known as \emph{target space} in this context. For instance, spacetime isometries are encoded as \emph{internal symmetries} on the worldline. For those familiar with this notion, the first term in \eqref{eq:worldline_action_g} defines a \ac{NLSM} in one dimension with target space $M$, except that here, crucially, the metric on $W$ is also dynamical: it is an \ac{NLSM} \emph{coupled to gravity in one dimension}. In other words, we are doing one-dimensional \ac{QG} coupled to an \ac{NLSM} with target space $M$ in order to recover perturbative \ac{QFT} (without gravity) in spacetime. If you tap out for the rest of this section, this is the one lesson I would like you to take home.\footnote{As \href{https://www.youtube.com/watch?v=H0jLD0PphTY}{eloquently put by Edward Witten}, this connection we are about to unravel is one of the many cases in which ``physics rhymes''.}

What other stuff can appear in $S_\text{other stuff}$? Well, at the level of weakly coupled quantum fields on $W$ there is not a whole lot that can show up. Since gauge fields have no physical polarizations,\footnote{As for massive vectors, in the case of string worldsheets they are ruled out by consistency conditions.} the only truly distinct thing we can add are fermions: spinors or Rarita-Schwinger fields. The former can provide the spin degrees of freedom for the particle in spacetime. The latter are necessary to avoid problems when introducing the former. We will examine this in more detail in the case of (super)strings. Other than that, $S_\text{other stuff}$ could be a placeholder for a complicated strongly interacting theory which encodes some weird internal degrees of freedom of the particle. In the case of strings, we will encounter very strong consistency constraints on what these extra degrees of freedom can be; in particular, they must be there unless the spacetime dimension is fixed to a particular value. For the time being, we ignore them. As we shall see for superstrings, this road leads to \emph{worldline supergravity}. The type of supersymmetry which shows up here has \emph{nothing to do} with the one you may have heard of, which is \emph{spacetime supersymmetry}. The one on the worldline is an ingredient in perturbative \ac{QFT} with fermions! Similarly, it will show up as an essential ingredient in \ac{ST}. Unfortunately we don't have time to go into the gory details, but you can find more in many lecture notes, reviews and papers on the worldline formalism for (spinning) particles, for example \href{https://indico.cern.ch/event/206621/}{here}, \href{https://www-th.bo.infn.it/people/bastianelli/2-ch6-FT2-2018.pdf}{here}, and in~\cite{Schubert:2001he, Bonezzi:2018box, Bonezzi:2020jjq, Bonezzi:2024emt}.\footnote{An interesting intermediate approach between particles and strings in the worldline formalism, which is worth mentioning, was recently explored in~\cite{Abel:2019ufz}.}

A detail that is relevant for us is that the worldline fields are coupled to \emph{background fields} in spacetime such as the spacetime metric $g$ or an electromagnetic field $A$. This works by pulling these fields back onto the worldline as explained above. When computing scattering amplitudes \emph{for the spacetime \ac{QFT}}, the external states are described by operators on the worldline called \emph{vertex operators}, and they can be associated to these background fields as we shall see in more detail for strings where some extra magic happens. Remarkably, the consistency of this construction implies the well-known field equations for the background fields, such as Einstein and Yang-Mills equations. However, this formalism does not provide any \ac{UV} completion of the spacetime dynamics, rather it recovers it perturbatively in a different formalism.

In order to recover the simplest\footnote{S-matrix elements can also be computed using different Feynman rules, which in this formalism are captured by vertex operators.} Feynman diagrams, those associated to position space correlators in the spacetime $M$, we build the path integral for one-dimensional \ac{QG} on the worldline $W$. We path-integrate over the worldline fields (here we only display $X$ for notational simplicity) and the worldline metric $\gamma$. Because of diffeomorphism invariance, the (Euclidean, Wick-rotated\footnote{Here, Wick rotation acts on both the worldline and spacetime metrics. See \eg{} Polchinski's book \cite{Polchinski:1998rq, Polchinski:1998rr} for a more in-depth discussion.}) path integral over worldlines with proper time $T$ and embedded endpoints $X(0)=x_i$ and $X(T)=x_f$ looks like
\begin{eqaed}\label{eq:worldline_path_integral}
    \int_{X(0)=x_i}^{X(T)=x_f} \frac{\mathcal{D}X \mathcal{D}\gamma}{\text{Diff}(W)} \, e^{-S^E_\text{wl}[X,\gamma]} \, .
\end{eqaed}
The quotient notation indicates that we can gauge-fix diffeomorphisms on the worldline. There are many ways of doing this, using the metric $\gamma$ directly or an einbein $e$ instead. A quick 'n' dirty way using the latter is to notice that the only invariant quantity is the proper time $T = \int \, e \, \rmd\tau$, which is also the Fourier zero-mode of $e$. Since the path integral measure separates each Fourier mode by definition, gauge fixing say $e = T$ (assuming the integration domain $0 \leq \tau \leq 1$ for simplicity) removes everything except the integration $\int_0^\infty \rmd{}T$ over proper time, an imprint of one-dimensional \ac{QG}. In this context, $T$ is known as a \emph{Teichm\"{u}ller parameter}.

A cleaner way to derive the same result is using the standard Faddeev-Popov procedure, introducing the Faddeev-Popov factor $\Delta_\text{FP}$. Compared to, say, perturbative\footnote{Generally speaking, in non-Abelian gauge theory no gauge fixing exists globally~\cite{Singer:1978dk}, that is, non-perturbatively. The issue of Gribov copies has a long history but remains ultimately unsolved, although at least in the non-gravitational case lattice methods are an established alternative.} Yang-Mills theory, in this case there are \emph{moduli}, parameters which are left after all gauge redundancies are fixed. For a worldline with the topology of an interval, the proper time is a modulus. That is, the fiducial gauge-fixed metric $\hat{\gamma}(t)$ can be deformed by some moduli $t^i$ living in some finite-dimensional space. As we shall see in more detail in \cref{sec:BH_transition}, after quotienting by the full gauge group (not just the component connected to the identity) the Teichm\"{u}ller space reduces to the \emph{moduli space} $\mathcal{M}$, in this case $[0,\infty)$ as exemplified by the proper time $T$. Since there is a physically relevant parameter in the fiducial metric, care must be exercised when dealing with Faddeev-Popov ghosts: they are not worldline scalars! As more apparent in the stringy analog of this story, due to a two-dimensional worldsheet, the $b$ and $c$ ghosts produced by the straightforward Faddeev-Popov procedure are actually a covariant two-tensor and a vector respectively. As such, the correct path integral measure $\mathcal{D}b \mathcal{D}c$ secretly contains factors of $\sqrt{\text{det}\gamma}$, which in this case is simply $\sqrt{\gamma}$. This is entirely analogous to how the volume form on a Riemannian manifold contains a factor $\sqrt{g}$. This is usually dealt with by using Fourier modes which are orthogonal relative to an appropriate invariant inner product. In this case, this subtlety opens up the amusing option of simply changing variables, rescaling fields by the appropriate power of $\gamma$ or $e = \sqrt{\gamma}$ to express the Faddeev-Popov factor in terms of a path integral over worldline scalars.

Either way, the procedure is quite simple in one dimension: letting $t$ be our modulus (the result is coordinate-independent), define
\begin{eqaed}\label{eq:FP_det_wl}
    1 \equiv \Delta_\text{FP}[\gamma] \int \rmd{}t \int \mathcal{D}\xi \, \delta(\gamma - \hat{\gamma}(t)^\xi) \, ,
\end{eqaed}
where $\xi$ denotes a transformation parameter. As usual, inserting this factor of unity in the worldline path integral allows to integrate over $\gamma$, leaving an integral over moduli space. Thanks to the gauge invariance of $\Delta_\text{FP}$, which in turn follows from the Haar invariance of the group measure $\mathcal{D}\xi$ on $\text{Diff}(W)$, the resulting path integral looks like
\begin{eqaed}\label{eq:worldline_path_integral_fixed}
    \int \rmd{}t \, \Delta_\text{FP}[\hat{\gamma}(t)]\int_{X(0)=x_i}^{X(T)=x_f} \mathcal{D}X \, e^{-S^E_\text{wl}[X,\hat{\gamma}(t)]} \, .
\end{eqaed}
As a result, the Faddeev-Popov factor need only be evaluated at the fiducial metric, allowing for a convenient expression of the Dirac (functional) distribution in terms of a bosonic path integral which ``inverts'' to a Berezin integral over Grassmann variables for $\Delta_\text{FP}$. Indeed, from \eqref{eq:FP_det_wl} one can work near $\xi = 0$ and modulus $t$ writing the linearized variation $\delta \gamma|_{\gamma=\hat{\gamma}} = 2e \, \delta e|_{e=\hat{e}} = 2 \hat{e} \left(\delta t \, \partial_t \hat{e} - \, \partial_\tau (\hat{e} \, \xi) \right)$. The latter term is the usual gauge transformation leading to a functional determinant. In this case, taking into account the fact that the first-order kinetic operator $P$ which appears maps covariant rank-two tensors to vectors, the correct determinant is\footnote{The square root appears because the integral would morally give $\text{det} \, P$. In two dimensions, there is no square root.} $\sqrt{\text{det}(P^T P)} = \sqrt{\text{det}(-\covD^2)}$, where $\covD^2$ is the Laplace(-Beltrami) operator. The former term in the variation involves a variation in the modulus from its fiducial value $t$, and as explained \eg{} in Polchinski's book \cite{Polchinski:1998rq, Polchinski:1998rr} the resulting Berezin integral yields a factor of $(b, \partial_t \hat{\gamma})$, a $b$-ghost insertion within an inner product with the tangent vector $\partial_t \hat{\gamma}$ over the space of metrics. Finally, one must be careful about zero-modes of the Faddeev-Popov operator: on an interval with vanishing boundary conditions at the endpoints, $P$ has no non-trivial zero-modes, while $P^T$ does: it acts on a different space of functions, and constants are annihilated. Hence, the integral over $b$ splits into an integral over the zero-mode $b_0$ and the rest. The latter goes with the integral over $c$ to yield the reduced determinant $\sqrt{\text{det}'(-\covD^2)}$ devoid of zero-modes, defined by zeta regularization. Instead, the ghost insertion only survives at the zero-mode $(b_0, \partial_t \hat{\gamma})$. Since $\hat{e}$ is constant, the eigenvalues are simply $\frac{n^2 \pi^2}{\hat{e}^2}$ with $n>0$ integer. The result is
\begin{eqaed}\label{eq:PP_det}
    \text{det}'(-\covD^2) = \exp(- \, \frac{\rmd}{\rmd{}s} \left( \hat{e}^{2s} \, \zeta(2s) \right)  \bigg|_{s=0}) \propto \hat{e} \, .
\end{eqaed}
The correct normalization for the constant zero-mode (again taking $0 \leq \tau \leq 1$) is $\hat{e}^{\frac{3}{2}} \, b_0$ with measure $\rmd{}b_0$, so that using the Berezin integral $\int \rmd{}b_0 \, b_0 = 1$ and taking into account the index contractions in the inner product with two factors of $\hat{\gamma}^{\tau \tau} = \hat{e}^{-4}$ we have
\begin{eqaed}\label{eq:insertion_inner_product}
    \left(\hat{e}^{\frac{3}{2}} \, , \, \partial_t \hat{\gamma}\right) = \hat{e}^{-\frac{3}{2}} \, \partial_t \hat{\gamma} \, .
\end{eqaed}
Putting everything together, the moduli space measure finally reads
\begin{eqaed}\label{eq:worldline_final_moduli_measure}
    \int \rmd{}t \, \frac{\partial_t \hat{\gamma}}{\hat{e}} \propto \int \rmd\hat{e} = \int \rmd{}T \, ,
\end{eqaed}
using that $\hat{e} = T$ is the proper time when using integration extrema $0 \leq \tau \leq 1$ for convenience. See also \href{https://www.lpthe.jussieu.fr/~israel/notes.pdf}{Isra\"{e}l's lecture notes} for worldline computations with ghost path integrals.

The above derivation is useful because it extends to the only other allowed topology for a smooth worldline, namely a circle $W \simeq S^1$. In this case, in addition to the modulus $T$ (now proportional to the radius after Wick rotation) there is a zero-mode of $P$, usually dubbed a \ac{CKV} in the context of \ac{ST} and \ac{CFT}. This is due to the fact that periodic functions can be constant, annihilated by $\partial_\tau$ and thus by $P$. This changes a step in the derivation: evaluating \eqref{eq:FP_det_wl}, the integration over distinct zero-modes yields a factor of $T$ on the right-hand side, since on a circle $\tau \sim \tau + T$ are identified. As a result, the measure $\frac{\rmd{}T}{T}$ contains an extra factor of $T^{-1}$. This is crucial to reproduce the correct Schwinger representation of the relevant one-loop integrals.

As for the functional integral over $X$, it is a straightforward Gaussian integral with boundary conditions, of the same form of that of a non-relativistic free particle. The novelty is the integration over proper times $T$, with comes from one-dimensional \ac{QG}. For our purposes, it can be suggestively expressed introducing a spacetime momentum variable $p$, according to
\begin{eqaed}\label{eq:X_integral_wl}
    \int \frac{\rmd^dp}{(2\pi)^d} \, e^{ip \cdot(x_f - x_i)} \, e^{- T \, (p^2 + m^2)} \, .
\end{eqaed}
Integrating over proper times (in our language, over the physically distinct worldline metrics) one finally recovers the well-known Feynman propagator for a free scalar field in spacetime! For loops, using the above result one finds the correct logarithmic contribution. This structure generalizes to any kind of field and simplifies certain perturbative calculations, as extensively reported in the literature on the worldline formalism \cite{Schubert:2001he, Bonezzi:2018box, Bonezzi:2020jjq, Bonezzi:2024emt}. However, as anticipated, there is a catch: interaction vertices need to be added by hand, specifying their coupling factors. We can understand this undesirable feature from the point of view of one-dimensional \ac{QG}; it is simply the \emph{sum over topologies}. Smooth worldlines can only be intervals or circles, but allowing for singularities at vertices, and spelling out a rule whereby each vertex provides a factor of the corresponding coupling constant, we effectively recover perturbative \ac{QFT} in spacetime with its beloved Feynman rules. All in all, to reiterate once more, the main lesson is that \emph{one-dimensional \ac{QG} corresponds to perturbative \ac{QFT} in spacetime}.

With this background under our belts, we can move on to strings. As a final remark, we could have gotten here by tackling the problem of \ac{QG} from a different angle, namely starting in low dimension. We would have recovered perturbative \ac{QFT} in an auxiliary higher-dimensional space, but now in hindsight we identify it with physical spacetime, and the worldline as auxiliary instead. Going up, the next step is the worldsheet, which as we shall see shortly produces \emph{perturbative \ac{QG} in spacetime}. You may ask what would happen going further up: the resulting worldvolume theories are problematic to define in a \ac{UV}-complete fashion because membranes bend weirdly. Furthermore, the specter of gauge redundancy becomes an obstruction to gauge fixing~\cite{Singer:1978dk}, ruining any attempt to define the theory directly via a continuum path integral.


\subsection{Closed strings interacting weakly}\label{sec:part_ii}

\begin{figure}[ht!]
    \centering
    \includegraphics[scale=0.6]{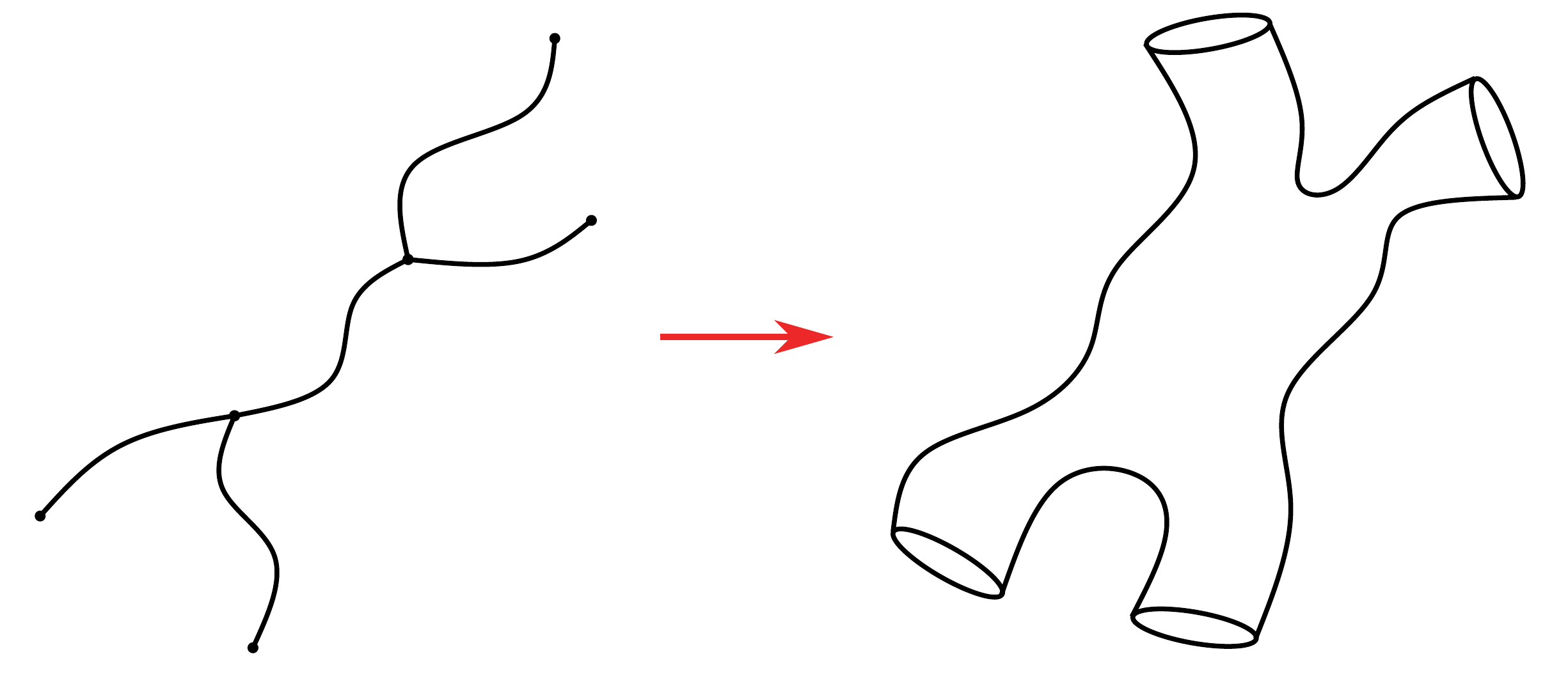}
    \caption{The worldsheet approach to perturbative \ac{ST} replaces particles and worldlines with strings and the worldsheets they sweep in spacetime. No interactions need external input, since the surface is smooth. We focus on closed, oriented strings.}
    \label{fig:worldsheet}
\end{figure}

All the work we put in the worldline formalism is now going to pay off. From here the construction of perturbative \ac{ST} from the worldsheet is mostly smooth sailing at the technical level, but there are several conceptual subtleties which ultimately make the story completely different in the resulting physics. A first observation is that with smooth worldsheets there is no need to introduce \emph{ad hoc} interaction vertices, as depicted in \cref{fig:worldsheet}. As mentioned above, we focus on closed oriented strings interacting weakly in a tame spacetime, meaning that the curvatures of spacetime fields are parametrically smaller than the string scale which we shall now define. In this regime, \ac{ST} reduces to gravitational \ac{EFT} at low energies.

\subsubsection{The view from the worldsheet}\label{sec:worldsheet}

Following the same procedure as before, the part of the two-dimensional theory on the worldsheet $\Sigma$ which describes strings moving in spacetime $(M,g)$ is given by the invariant (Lo\-rentz\-ian) area $A$ swept by the embedding $X : \Sigma \to M$. This is the \emph{Nambu-Goto action}
\begin{eqaed}\label{eq:nambu-goto}
    S_\text{NG} = - \, T \int_{\Sigma} \rmd{}A \, .
\end{eqaed}
The dimensionful constant $T \equiv \frac{1}{2\pi\alpha'}$ is \emph{the string tension}, and $M_s \equiv \sqrt{T}$ is the \emph{string scale}, the typical scale at which stringy stuff happens. Sometimes you will see terminology like ``string length'' $\ell_s \equiv \sqrt{\alpha'}$, but it does not actually mean that strings have a fixed length, just a typical length scale. The notation $\alpha'$ historically comes from the slope of ``Regge trajectories'' in hadron physics, the original home of quantum strings. After realizing that \ac{QCD} rocks and that quantum strings entail \ac{QG}, the physics developed accordingly, while the notation and terminology stuck. The units of measurement in which $\alpha' = 1$ are called string units.

The required procedure of introducing a dynamical worldsheet metric $\gamma$, thus doing two-dimensional \ac{QG} on $\Sigma$ coupled to an \ac{NLSM} with target space $M$, yields the \emph{Polyakov action}, the \ac{NLSM} with target space $M$, except defined on a two-dimensional worldsheet. Using local coordinates $\sigma = (\sigma^1, \sigma^2)$ on $\Sigma$, the total worldsheet (``ws'') theory we will work with is thus given by
\begin{eqaed}\label{eq:ws_action}
    S_\text{ws} = - \, \frac{1}{4\pi \alpha'} \int_{\Sigma} \rmd^2\sigma \, \sqrt{-\gamma} \left[ \gamma^{\alpha \beta} \, \partial_\alpha X^\mu \, \partial_\beta X^\nu \, g_{\mu \nu}(X)\right] + S_\text{other stuff} \, .
\end{eqaed}
This time, this is not just the leading action --- since the couplings of an \ac{NLSM} are encoded in the Riemann curvature of $g$ in string units, having a tame spacetime (weakly curved in string units) requires perturbative renormalizability of the worldsheet theory for \ac{UV} completeness. Also, the absence of the ``cosmological'' term is not coincidental; it is related to Weyl invariance. As before, $S_\text{other stuff}$ can also be a placeholder for a non-Lagrangian strongly coupled theory. It can also contain other fields, such as worldsheet fermions, as well as other couplings involving $X$. In fact, the most general Polyakov action we can use for our purposes involving only $X$ is
\begin{eqaed}\label{eq:polyakov_action}
    S_\text{P} = - \, \frac{1}{4\pi \alpha'} \int_{\Sigma} \rmd^2\sigma \, \sqrt{-\gamma} \left[ \left( \gamma^{\alpha \beta} \, g_{\mu \nu} + \epsilon^{\alpha \beta} \, B_{\mu \nu} \right) \partial_\alpha X^\mu \, \partial_\beta X^\nu + \alpha' \, \text{Ric}(\gamma) \, \phi \right]\, ,
\end{eqaed}
where the \emph{Kalb-Ramond} background field $B$, also known as \emph{$B$-field} for short, is a two-form (covariant anti-symmetric rank-two tensor field) in spacetime and $\phi$ is a scalar field called the \emph{dilaton}, which couples to the Ricci scalar of the worldsheet metric defined by the corresponding Levi-Civita connection. The arguments $X$ in the background fields are suppressed for ease of notation. Including worldsheet fermions leads to a supersymmetric version of \eqref{eq:polyakov_action}~\cite{Callan:1989nz}.

So we are now dealing with \ac{QG} on the worldsheet. Actually, it turns out that if we want classically stable vacuum configurations we explicitly need some worldsheet fermions in $S_\text{other stuff}$ to encode the spin degrees of freedom of our string (for this reason, Polyakov used the term ``fermionic string'' for what we now call superstring). Due to (local) Lorentz invariance in (tangent) spacetime, worldsheet spinor fields $\psi^\mu$ encoding spacetime spin degrees of freedom have a Lorentz index to match that of $X^\mu$. It would seem like there can be an analogous ghost issue for the temporal component $\psi^0$ as we encountered for $X^0$. It turns out that solving this issue requires coupling the worldsheet to two-dimensional \emph{supergravity}, introducing a gravitino $\chi_\alpha$ alongside the metric $\gamma_{\alpha \beta}$. We will not go into the details here, since they can be found in the references, in particular in the book by Green, Schwarz and Witten \cite{Green:2012oqa, Green:2012pqa}. The upshot is that there are now additional constraints similar to those arising from the field equations of $\gamma$, which remove the fermionic ghost modes from the spectrum. Gauge fixing via the Faddeev-Popov procedure now yields an additional set of bosonic \emph{superghosts}, commonly denoted $\beta$ and $\gamma$. They are Grassmann-even, since they ``Faddeev-Popovize'' Grassmann-odd fields. Here $\gamma$ is not the worldsheet metric, which can be gauge-fixed to $\hat{\gamma} = \delta$ (``conformal gauge'') unless anomalies appear; more on that later. For the time being, we will simply assume we can do this.

\subsubsection*{Conformal symmetry, Weyl invariance, and the (super-)Virasoro algebra}

Because the worldsheet theory is two-dimensional, something magical happens: after gauge-fixing the spurious degrees of freedom as outlined above, the resulting theory (including the $bc$ ghosts and the $\beta \gamma$ superghosts) must be \emph{conformally invariant}.\footnote{A complete description of the \acp{CFT} associated to the (super)ghosts has several subtleties, such as Q-vacua and the option of varying their scaling dimensions. We will not touch these issues here, but the interested reader can check some of them out in the vast literature on \ac{CFT}, or in the relevant chapters of Polchinski's book \cite{Polchinski:1998rq, Polchinski:1998rr}.} This follows from the fact that the classical theory has an additional gauge redundancy in two dimensions, namely \emph{Weyl invariance}, which rescales the worldsheet metric by a positive function, 
\begin{eqaed}\label{eq:weyl_rescaling}
    \gamma \mapsto \Omega^2(\sigma) \, \gamma \, .
\end{eqaed}
Although intimately connected, Weyl rescalings are not conformal transformations, which are diffeomorphisms: in conformal gauge it is convenient to use complex coordinates $z = \sigma^1 + i\sigma^2$, with $2 \, \rmd^2\sigma = \abs{\rmd{}z}^2$, so that conformal maps are \emph{holomorphic} maps $z \mapsto w(z)$ on $\Sigma$ viewed as a Riemann surface.\footnote{This is always possible because the structure group $SO(2) \simeq U(1)$ of the tangent bundle admits an almost complex structure given by $\frac{\pi}{2}$-rotation, and in two real dimensions it is integrable.} For superstrings, there is an analogous super-Weyl invariance. The quantum theory \emph{must be devoid of gauge anomalies} in order to consistently preserve unitarity and meaningful probabilities; we will examine this condition shortly. A proper introduction to this topic would take way too much time, so we will focus on the bare essentials for our purposes. The upshot of conformal invariance after gauge fixing is that the whole process of coupling the theory to two-dimensional (super)gravity can be understood in simpler terms as \emph{gauging a symmetry algebra} of the two-dimensional \ac{CFT}\footnote{From now on we will use this term here and there, but be afraid not: for our purposes, it simply refers to the two-dimensional \ac{QFT} on the worldsheet without coupling to (super)gravity.} on the worldsheet, namely keeping only invariant states, operators and stuff. In the purely bosonic case this is the Virasoro algebra of conformal transformations in two dimensions. For superstrings, it is a minimal \emph{superconformal algebra}. Looking at the story in this way is very useful, because it can be shown (see \eg{} table 11.1 in Polchinski's book \cite{Polchinski:1998rr} or section 4.5 of Green, Schwarz and Witten's book \cite{Green:2012oqa}) that \emph{no other gauging} is compatible with a tame Lorentzian spacetime. Its dimension would either have to be non-positive or with complex signature. This does not exclude extended \emph{global} supersymmetry on the worldsheet --- as a matter of fact, it arises \eg{} in Calabi-Yau compactifications when spacetime supersymmetry is present.

Fields living on a closed string can propagate waves in two directions, commonly referred to as \emph{left-movers} and \emph{right-movers}. At least at leading order in the spacetime curvature (namely in a flat spacetime), these two sectors of the Polyakov theory decouple: for example, the bosonic Lagrangian density $\delta_{\mu \nu} \, \partial X^\mu \overline{\partial} X^\nu$ leads to solutions $X = X_\text{L}(z) + X_\text{R}(\overline{z})$, the Euclidean analog of waves $X(\sigma^1 \pm \sigma^2)$ highlighted by light-cone coordinates $\sigma^{\pm} \propto \sigma^1 \pm \sigma^2$. This leaves three options to gauge a (super)conformal algebra:

\begin{itemize}
    \item \emph{Purely bosonic fields.} This leads to bosonic \ac{ST}, whose perturbative description is unstable due to a tachyon.
    
    \item \emph{\ac{RNS} superstrings.} If both left-moving and right-moving worldsheet spinors are present, one can separately gauge a minimal superconformal algebra on each. 
    
    \item \emph{Heterotic superstrings.} Since chiral spinors (in fact, Majorana-Weyl) can exist on the worldsheet, one can also gauge a bosonic Virasoro algebra in, say, the left-moving sector and a minimal superconformal algebra on the right-moving sector.
\end{itemize}

\ac{RNS} should really be called \ac{RNS}-\ac{RNS}, but we won't to save space. For us only the latter two are physically interesting due to classical stability, which rules out the former (at least perturbatively); however, bosonic strings are a very useful simplified setting in which many important features of superstrings are transparent. Before moving on, I'd like to remark that an alternative formalisms for superstrings exist: for instance the Green-Schwarz approach, in which supersymmetry \emph{in spacetime} is kept manifest, and the hybrid and pure spinor approaches, which are useful to discuss certain classes of backgrounds and perform certain computations. See \eg{}~\cite{Demulder:2023bux} for a recent review on some of these.

\subsubsection*{Central charges --- $c$ is for Cool}

One final ingredient we need to be able to talk about perturbative \ac{ST} also stems from conformal symmetry and the Virasoro algebra. At the classical level, the symmetry algebra generating conformal transformations is the \emph{Witt algebra} of vector fields on the worldsheet. At the quantum level, the Virasoro algebra appears: it differs by a central extension parametrized by a number called the \emph{central charge} $c$, which is positive for unitary theories. For free theories, it basically counts the number of degrees of freedom, with bosons contributing $c=1$ and fermions $c=\frac{1}{2}$. More precisely, there is a central charge $c_\text{L}$ for left-movers and a central charge $c_\text{R}$ for right-movers. Computing this quantity can be quite involved, and outside the scope of this section; however, it plays an absolutely paramount role in \ac{ST}, which is why I'm mentioning it, even if briefly. It is also very important in \ac{CFT} in general, since it controls a lot of interesting physical quantities. For instance, the thermal free energy is proportional to $c=c_\text{L} + c_\text{R}$, and similarly the \ac{vev} of the trace of the energy-momentum tensor $\langle T \rangle \propto c \, \text{Ric}(\gamma)$ on a curved worldsheet, and similarly the correlator $\langle T_{\alpha \beta} T_{\gamma \delta}\rangle$ on a flat worldsheet. The latter two are particularly useful to derive the anomaly under Weyl rescalings, as we shall discuss momentarily.

\subsubsection{Building string perturbation theory}\label{sec:string_perturbation_theory}

\begin{figure}[ht!]
    \centering
    \includegraphics[width=\textwidth]{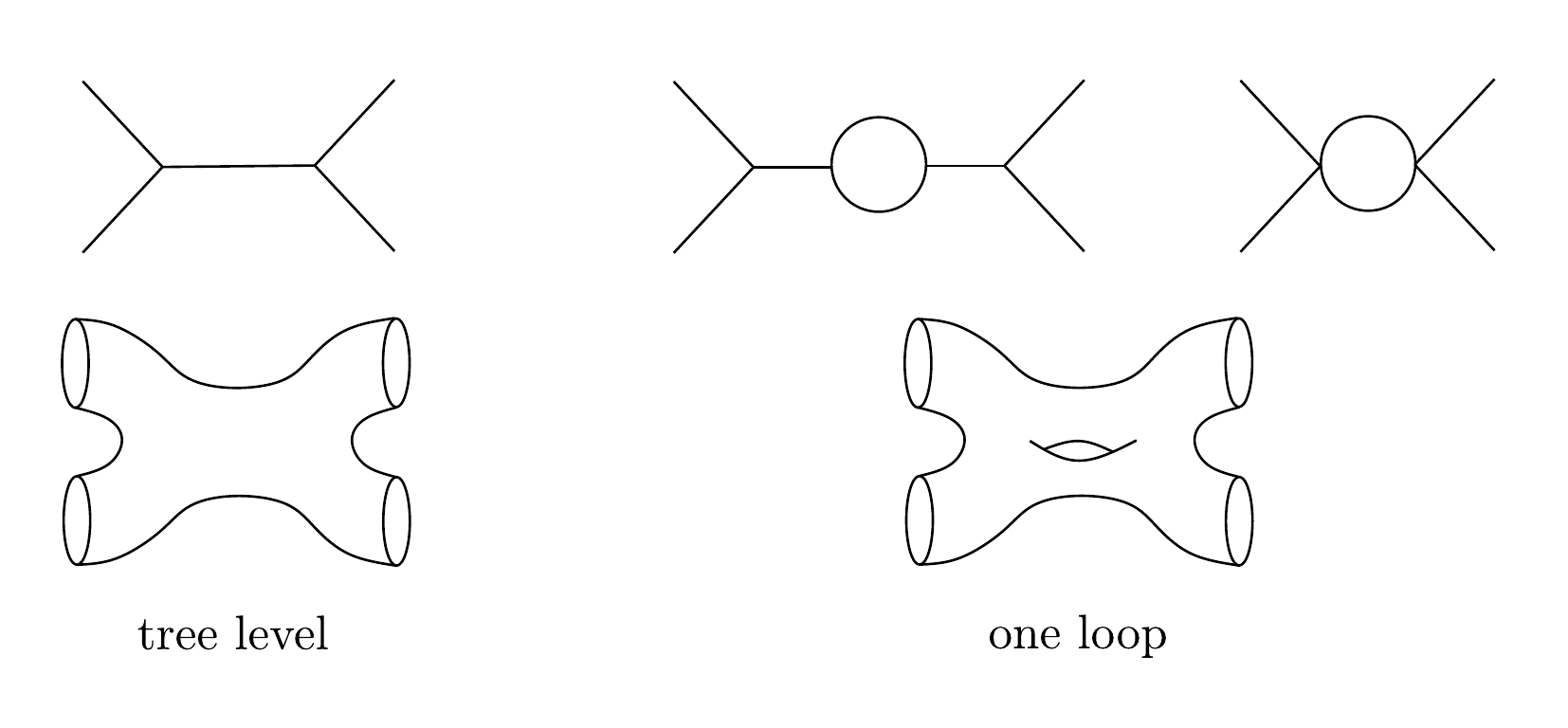}
    \caption{The loop expansion in string perturbation theory works analogously to the familiar organization of Feynman diagrams. Loops in the graph are replaced by any topology change of the worldsheet, which for closed oriented strings consists of handles increasing the genus. In this case, at each order in the loop expansion there is exactly one ``diagram'', as opposed to the very fast growth in \ac{QFT}. Even including open and unoriented strings, there are a handful of topologies to \emph{handle} (get it?) at each order, for a fixed number of external states.}
    \label{fig:loops}
\end{figure}

The construction of string perturbation theory as a theory in spacetime works similarly to the worldline story, with a few key differences. To begin with, the sum over topologies is now much simpler: it is a sum over \emph{genera} $g = 0, 1, 2, \dots$ where the genus $g$ is a topological invariant counting the number of handles in the surface. The Euler characteristic of the worldsheet is $\chi(\Sigma) = 2 - 2g$, and we will denote such worldsheets by $\Sigma_g$. For closed oriented strings this is the whole story, and it nicely simplifies the structure of the loop expansion of the spacetime interactions, as depicted in \cref{fig:loops}. 

\subsubsection*{Dangerous anomalies and where to find them}

Before looking at what observables we can compute or how perturbative \ac{QG} in spacetime pops up from this two-dimensional \ac{QG}, let us ponder the elephant in the room: the possibility of gauge anomalies. The worldsheet \ac{CFT} is coupled to worldsheet (super)gravity, and thus it has a $\text{Diff}(\Sigma) \times \text{Weyl}$ redundancy, or its supersymmetric counterpart for superstrings. What kinds of anomalies can arise? Generally speaking, a theory has \emph{local anomalies} when quantities such as partition functions are not gauge-invariant under transformations that are homotopic to the identity (``small''), whereas \emph{global anomalies} arise when the offending transformations are not homotopic to the identity (``large''). See~\cite{Alvarez-Gaume:2022aak} for a recent review on the topic, as well as references therein for more details.

\begin{itemize}
\item \emph{Diffeomorphisms.} Firstly, let us consider diffeomorphisms. They lead to \emph{gravitational anomalies}. In two dimensions there can be gravitational anomalies of the local type, due to small diffeomorphisms $\text{Diff}_0(\Sigma) \subset \text{Diff}(\Sigma)$, and it turns out that they are proportional to the difference $c_\text{L} - c_\text{R}$ between the central charges, which means that they must be equal to have a consistent theory. Then, depending on the global topology of the worldsheet $\Sigma$, there are large diffeomorphisms or various types. The quotient\footnote{More generally, one has a short exact sequence of groups which may not split.}
\begin{eqaed}\label{eq:MCG}
    \text{MCG}(\Sigma) \equiv \frac{\text{Diff}(\Sigma)}{\text{Diff}_0(\Sigma)}
\end{eqaed}
is the \emph{mapping class group}, which appears in the definition of the moduli space as in the case of the point particle (but cooler). Cancellation of global gravitational anomalies requires that stuff be invariant under this action. Remarkably, this all-order condition turns out to be equivalent to require that the worldsheet \ac{CFT} be \emph{modular-invariant}, a non-trivial condition on its spectrum and an important consequence of anomaly cancellation. From the spacetime viewpoint, it is merely a one-loop condition! We will study it in more detail in \cref{sec:BH_transition}.

\item \emph{Modular invariance.} At one-loop level in string perturbation theory, the up to conformal transformations the worldsheet $\Sigma \simeq T^2$ has the topology of a torus (possibly with punctures representing external states in a scattering process). As we shall see in more detail in \cref{sec:BH_transition}, the absence of global gravitational anomalies requires that the partition function of the worldsheet \ac{CFT} on this manifold be invariant under a rather special group, the \emph{modular group}, which we will discuss in \cref{sec:BH_transition}. This partition function captures aspects of the spectrum of states of the string, and modular invariance non-trivially relates their masses and degeneracies. At the level of the worldsheet path integral, this information comes from \emph{global data} of the \ac{CFT}. Namely, the worldsheet bosons $X$ are periodic, but fermionic fields can be either periodic or anti-periodic, and on a torus there are in total four independent choices,\footnote{You might wonder why we cannot choose different periodicity conditions for different components $\psi^\mu$. If we did, the fermionic Noether current generating worldsheet supersymmetry, schematically of the form $\psi^\mu \, \partial X_\mu$, would not be well-defined globally on the worldsheet.} encoding its four spin structures classified by the relevant cohomology group $H^1(T^2,\mathbb{Z}_2) \simeq \mathbb{Z}_2 \oplus \mathbb{Z}_2$. This means that the path integral contains a sum over these sectors. The integration measure weighting them fixes the so-called \emph{\ac{GSO} projection} of the Hilbert space.

For superstrings, a (gauge-fixed) path integral has the schematic form
\begin{eqaed}\label{eq:GSO_path_integral}
    \sum_{\text{genus } g} \int_{\mathcal{M}_g} \rmd\mu(t) \int_{\Sigma_g(t)} \sum_{\text{spin structure } s} C_s \, \mathcal{D}X \, \mathcal{D}\psi \, e^{-S_\text{ws}^E} \left( \text{insertions} \right) ,
\end{eqaed}
where the first integral is over the (super-)moduli space of closed (super-)Riemann surfaces of genus $g$, whose measure $\mu$ is encoded in the Faddeev-Popov factor, and the coefficients $C_s$ implement the \ac{GSO} projection. Their role is more apparent when computing the partition function of the worldsheet \ac{CFT} over the torus $\Sigma_1 \simeq T^2$, where they can be interpreted in terms of degeneracies. Generically, restricting the sum over spin structures would violate modular invariance, but its constraining power is much stronger --- there are usually but a few options for the \ac{GSO} coefficients. Including open and unoriented strings, further constraints arise from open-closed string duality~\cite{Angelantonj:2002ct}. This perturbation theory is mathematically quite intricate; see~\cite{Witten:2012bh} for a review. Some recent methods for more direct evaluations of loop-level amplitudes are discussed in~\cite{Eberhardt:2023xck, Eberhardt:2024twy}.

\item \emph{Weyl transformations.} Anomalies in Weyl transformations are trickier. In order to see them, let us observe that, locally, the (Wick-rotated) worldsheet metric can be gauge-fixed to a conformally flat form $\gamma = e^{2\omega(\sigma)} \, \delta$, removing two of its three independent components with a two-dimensional diffeomorphism. You can find out how in detail in \eg{} Tong's lecture notes \cite{Tong:2009np} or Polchinski's book \cite{Polchinski:1998rq, Polchinski:1998rr}. Choosing such a reference metric for gauge fixing, Weyl invariance requires that the path integrand be independent of the Weyl factor $\omega$. If so, we can go to conformal gauge $\omega = 0$. To begin with, this requires that the theory be scale-invariant, and thus conformally invariant since in the present settings they co-imply. But this is not enough; there are further constraints.

In order to see what's going on, we can look at the $\omega$ dependence of, say, a gauge-fixed partition function $\partitionfunction[\hat{\gamma} = e^{2\omega} \delta]$. Using the properties of the central charge and some clever path integral tricks (see \eg{} Polchinski's book \cite{Polchinski:1998rq, Polchinski:1998rr} or Tong's lecture notes \cite{Tong:2009np}), one can show that
\begin{eqaed}\label{eq:weyl_dependence}
    \ln \frac{\partitionfunction[e^{2\omega} \delta]}{\partitionfunction[\delta]} \propto \, c \int \rmd^2 \sigma \, (\partial \omega)^2 \, ,
\end{eqaed}
which requires the \emph{total} central charge vanish, $c = c_\text{ws} + c_\text{(super)ghosts} = 0$, including the contribution from the (super)ghosts. The type of $bc$ and $\beta \gamma$ (super)ghosts appearing in this construction have central charges of $-26$ and $+11$ respectively. Notice that this implies the absence of local gravitational anomalies. But wait, didn't we just say that unitarity requires $c > 0$? It turns out that this is not an issue: the \emph{spacetime \ac{QG} theory} is perfectly unitary, as shown by the celebrated \emph{no-ghost theorem}, analogously to the slogan that ``ghosts only run in loops'' in the context of perturbative gauge theory. What matters is that the spacetime S-matrix be unitary, which it is. As a final remark, you may have noticed that the dilaton coupling in \eqref{eq:polyakov_action} suspiciously breaks Weyl invariance at the \emph{classical level}. However, it is also at higher-order in the \ac{NLSM} perturbative expansion in ($\alpha' \times$ curvatures), and these two effects can compensate each other.
\end{itemize}

\emph{Summary of consistency conditions. ---} Specifying a consistent worldsheet theory defines a background around string perturbation theory can be performed, that is a \emph{string vacuum}. Having emphasized from the outset that there can be additional degrees of freedom in the theory, it is rather straightforward to piece the puzzle together. Since the ghosts form a \ac{CFT} by themselves, we can focus on the rest (also known as the ``matter \ac{CFT}''). We can state the following: a (classically stable) string vacuum is a two-dimensional unitary superconformal conformal field theory which satisfies
\begin{itemize}
    \item \emph{Modular invariance.} The partition function $\partitionfunction_{T^2}$ on the torus must be invariant under the action of the modular group on the moduli space of the torus.
    \item \emph{Criticality.} The central charges must be \emph{critical}. Namely, they must equal $(15,15)$ in the \ac{RNS} case or $(26,15)$ in the heterotic case.\footnote{In light-cone quantization one finds slightly different numbers due to how the longitudinal degrees of freedom of the string are handled.} This comes from the fact that the bosonic ghosts come with a central charge of -26, whereas a \ac{RNS} sector has both ghosts and superghosts, with a total central charge of $-26+11=-15$. The total worldsheet central charges must vanish by anomaly cancellation.
\end{itemize}
If we further require a tame $d$-dimensional spacetime, we can say more. The spacetime sector of the worldsheet \ac{CFT}, as discussed above, is simply a (supersymmetric) \ac{NLSM} with target space the spacetime $M$ itself, possibly decorated with additional background fields such as $B$ and $\phi$. Tameness means that the curvatures\footnote{I use the plural because not only the gravitational field has a curvature. Also other fields have meaningful curvatures, such as the dilaton and the Kalb-Ramond field, or the Yang-Mills fields appearing in the heterotic case. All of these appear in the low-energy \ac{EFT} with stringy effects suppressed by the string scale.} $\mathcal{R}$ satisfy $\alpha' \mathcal{R} \ll 1$, a necessary condition to get a low-energy \ac{EFT}. Serendipitously, this condition is equivalent to the statement that the spacetime sector of the worldsheet \ac{CFT} be weakly coupled! Indeed, a flat target space corresponds to a free theory, and deformations away from it must thus be invariantly encoded in operators weighted by curvatures. This can be seen explicitly using normal coordinates built from a geodesic expansion. Under these conditions, the central charges of the spacetime sector can be perturbatively expressed as asymptotic series dominated by
\begin{eqaed}\label{eq:central_charge_asymp}
    c_\text{spacetime} = \left(d_\text{bosonic} \text{ or } \frac{3}{2} \, d_\text{RNS}\right) + \orderneglected(\alpha' \mathcal{R}) \, ,
\end{eqaed}
depending on whether the sector is bosonic or \ac{RNS}. The spacetime bosons can never be chiral, so their contribution of $d$ to the left-moving and right-moving central charges is the same. By unitarity $c_\text{ws} = c_\text{spacetime} + c_\text{other stuff} > 0$, and therefore we can state the following property whose name I came up with:\footnote{Not to be confused with the notions of tameness in geometry and model theory, which have been recently applied to \ac{ST}~\cite{Grimm:2021vpn}.}

\begin{itemize}
    \item \emph{Tameness.} The worldsheet \ac{CFT} contains a tame spacetime sector, \ie{} a (supersymmetric, generalized) \ac{NLSM} on a weakly curved background. By unitarity and criticality, it follows that $d \leq 10$. A purely bosonic (and thus tachyonic) sector would have $d \leq 26$ instead. These conditions arise because curvature corrections to central charges are sub-leading, and thus must cancel on their own.
\end{itemize}

All in all, fixing the dimension $d$ of our tame spacetime, we can ignore that sector of the \ac{CFT} and focus on the ``internal'' sector (everything else prior to coupling to (super)gravity and fixing the gauge). These extra degrees of freedom have discrete spectra which must be present if $d < 10$, but they should be gapped, so that the their effect becomes suppressed at low energies. This condition is called \emph{compactness}, in close analogy with the discrete gapped Laplacian spectrum of a compact manifold. We can thus formulate a ``$d$-criticality'' condition for the internal sector: central charges must equal $(15-\frac{3}{2}d, 15-\frac{3}{2}d)$ (in the \ac{RNS} case) or $(26-d, 15-\frac{3}{2}d)$ (in the heterotic case). With this in mind, from now also dropping ``classically stable'' keeping it implicit, we can summarize the first important result in one cool-sounding, but accurate sentence:

\begin{tcolorbox}
\emph{String vacua with a tame $d$-dimensional spacetime are two-dimensional unitary compact superconformal field theories which are $d$-critical and modular-invariant.}
\end{tcolorbox}

\subsubsection*{String perturbation theory --- the loop expansion}

So far we managed to understand what a string background is. It is pretty wild, but we will build more intuition via examples. Each background brings along a spectrum of excitations, and the goal of string perturbation theory is to describe what observables can be defined and to compute them. Looking back at \eqref{eq:GSO_path_integral} and \cref{fig:loops}, you can recall that the loop order is given by the genus $g$. But what is the coupling constant? Going further back to \eqref{eq:polyakov_action}, the topologically inclined reader will spot something neat. In asymptotically flat spacetime, the boundary conditions include specifying the value of the dilaton $\phi(x) \to \phi_0$, which is constant because of the asymptotic isometries. Define $g_s \equiv e^{\phi_0}$ and $\widetilde{\phi} \equiv \phi - \phi_0$. Then, after Wick rotation, the dilaton coupling in the general Polyakov action of \eqref{eq:polyakov_action} evaluates to
\begin{eqaed}\label{eq:gs_euler}
    S_\text{P}^E|_{\text{dilaton}} & = \frac{\phi_0}{4\pi} \int_{\Sigma} \rmd^2\sigma \, \text{Ric}(\gamma) + \frac{1}{4\pi} \int_{\Sigma} \rmd^2\sigma \, \text{Ric}(\gamma) \, \widetilde{\phi} \\
    & = \chi(\Sigma) + \frac{1}{4\pi} \int_{\Sigma} \rmd^2\sigma \, \text{Ric}(\gamma) \, \widetilde{\phi} \, ,
\end{eqaed}
thanks to the two-dimensional Gauss-Bonnet theorem. Notice that the first contribution survives gauge fixing to a locally flat worldsheet, since it only cares about its global topology. As a result, the sum over worldsheet topologies entailed by the coupling to two-dimensional (super)gravity takes the schematic form
\begin{eqaed}\label{eq:sum_worldsheet_topologies}
    \sum_{\text{genus } g} g_s^{2g-2} \int_{\mathcal{M}_g} \rmd\mu(t) \left( \text{stuff} \right) .
\end{eqaed}
When including external states to compute a scattering amplitude, the (super-)moduli space $\mathcal{M}_g$ of closed genus-$g$ (super-)Riemann surfaces is modified by the presence of \emph{punctures}. Our main concern here is that $g_s$ plays the role of coupling constant for interactions, and $g_s \ll 1$ is the weak coupling limit. Contrary to what happens in field theory, $g_s$ is not like a knob external to the theory, such as, say, the (\ac{IR}) fine structure constant in \ac{QED}. Rather, it is given by the asymptotic value of a background field. At this point, this may not seem satisfactory enough for some of you to conclude that \ac{ST} has no free parameters; as we shall learn in \cref{sec:part_iii}, all background data is actually dynamical, and the one pertaining to the spacetime sector is described by a \emph{bona fide} \ac{EFT} at low energies. Speaking of which --- \emph{how} do we include external states? What do they look like?

\subsubsection{Spectra: gravitons, gauge bosons, matter and all that}\label{sec:spectra}

When discussing the worldline, we briefly mentioned that states in the corresponding perturbative \ac{QFT} in spacetime are encoded in certain operators called \emph{vertex operators}. For instance, starting from \eqref{eq:worldline_action_g} and \eqref{eq:worldline_path_integral_fixed}, the effect of deforming the spacetime metric $g \mapsto g + \delta g$ is encoded by insertions of
\begin{eqaed}\label{eq:graviton_vertex_wl}
    \mathcal{V}_{\delta g}^{\text{wl}} \equiv \gamma^{\tau \tau} \, \delta g_{\mu \nu}(X) \, \dot{X}^\mu \, \dot{X}^\nu \, .
\end{eqaed}
This is nice 'n' all, but in general operators have nothing to do with states on the worldline. This ``graviton vertex operator'' can be used to describe geometry fluctuations, but there is no correspondence with graviton \emph{states}, which can be present with sufficiently high worldline supersymmetry~\cite{Bonezzi:2018box}. Vertex operators are internal insertions, integrated over the worldline. In other words, the background geometry is not dynamical in the associated perturbative spacetime \ac{QFT}, although it is constrained to satisfy classical field equations by quantum consistency (\eg{} \ac{BRST} nilpotency~\cite{Bonezzi:2018box}).

\subsubsection*{Gravitons, and where to find them --- the state-operator correspondence}

In \ac{ST}, the story is quite different: the worldsheet theory is conformally invariant, and \acp{CFT} exhibit a correspondence between states and local operators! Roughly speaking, this comes from the fact that an operator insertion at some puncture on the worldsheet is equivalent to a stretching to infinity (where a state is specified) from where the puncture was. A cleaner derivation compares the theory on the cylinder $\Sigma \simeq S^1 \times \mathbb{R}$ to that on the plane $\Sigma \simeq \mathbb{R}^2$. The former represents a closed string propagating in time, while in the latter time becomes the \emph{radial direction} from the origin, which represents the infinite past. This is why a local insertion at the origin in the theory on the plane corresponds to specifying an initial state in the theory on the cylinder. This means that the worldsheet analog of \eqref{eq:graviton_vertex_wl},
\begin{eqaed}\label{eq:graviton_vertex_ws}
    \mathcal{V}_{\delta g}^{\text{ws}} \equiv \gamma^{\alpha \beta} \, \delta g_{\mu \nu}(X) \, \partial_\alpha X^\mu \, \partial_\beta X^\nu + \text{ possibly fermionic stuff} \, ,
\end{eqaed}
corresponds to a string state. It is a \emph{graviton} --- strings seen from afar look like particles, and this one turns out to be a massless helicity-two particle with the correct low-energy dynamics, as we shall see. The graviton is universally and \emph{unavoidably} present in \ac{ST}, which should be apparent from the origin of this vertex operator. To be more precise, \eqref{eq:graviton_vertex_wl} and \eqref{eq:graviton_vertex_ws} are modified by (super)ghost insertions, and the precise structure which survives the constraints of gauge invariance depends on the type of (super)string, namely bosonic, \ac{RNS} or heterotic in the terminology we introduced. Because of the operator-state correspondence, as opposed to field-theoretic gravitons, in \ac{ST} gravitons can actually deform the background, as can be explicitly seen using coherent states (more on that later). Since interactions are present and determined, general arguments by Feynman and Weinberg \cite{Feynman:1963ax, Feynman:1996kb} show that the leading low-energy dynamics of these gravitons is described by \ac{GR}, at least if the Planck mass is finite. In fact, the Planck mass can be expressed in terms of $g_s$ and the string scale $M_s$ (and possibly other stuff). Since this is the fundamental point of the section, we will explore it extensively both from the \ac{IR} and \ac{UV} points of view. In summary, in contrast to the case of the worldline, here \emph{two-dimensional \ac{QG} on the worldsheet recovers perturbative \ac{QG} in spacetime}. Doubling down on Witten's poetry metaphor quoted for the worldline, for the worldsheet we find a perfect rhyme.

Hopefully the above discussion sparked some interest on what the string spectrum looks like. Let us study it a bit more in depth.\footnote{Not as in-depth as in recent \emph{excavations}~\cite{Markou:2023ffh, Basile:2024uxn}, though.} As we just saw, states are isomorphic to vertex operators. But because string perturbation theory is not merely the worldsheet \ac{CFT}, rather it is coupled to two-dimensional (super)gravity, there is a (supersymmetric extension of) $\text{Diff}(\Sigma) \times \text{Weyl}$ gauge redundancy to worry about. Only gauge-invariant operators make sense. The standard way to deal with this is to perform a gauge fixing (say to the flat worldsheet metric) and use the \ac{BRST} construction to build vertex operators from representatives of \ac{BRST} cohomology classes.\footnote{This approach is useful: among other things, it addresses some subtleties involving physical polarizations encoded by some vertex operators and helps in proving gauge invariance and unitarity of observables. See \eg{} Polchinski's book \cite{Polchinski:1998rq, Polchinski:1998rr} for details on this stuff.} This produces operators like \eqref{eq:graviton_vertex_ws} multiplied by some ghost insertions, something like $c \, \overline{c} \, \mathcal{V}$. Another way to derive this is to realize that $c$-insertions are necessary to describe the correct vacuum in the ghost Hilbert space, and relatedly to obtain non-zero amplitudes upon integration over their zero-modes (due to the Riemann-Roch theorem, see \cite{Polchinski:1998rq, Polchinski:1998rr}). For superstrings, $\mathcal{V}$ also contains superghost insertions, which can be used to rewrite everything in terms of worldsheet superconformal fields and supergeometry. The presence of superghosts changes the structure of \eqref{eq:graviton_vertex_ws} for superstrings, schematically from $\partial X \partial X$ to $\psi \, \psi$ for \ac{RNS} superstrings or $\psi \, \partial X$ for heterotics. This is not a big deal, since these combinations can be written as superspace integrals of a supersymmetrization of $\partial X \, \partial X$~\cite{Callan:1989nz}. For what concerns us, this subtlety will not matter; hence, we will stick with \eqref{eq:graviton_vertex_ws} for illustrative purposes. Also, since we are ultimately interested in scattering amplitudes, it turns out that it is equivalent\footnote{As explained in~\cite{Witten:2012bh}, some subtleties can arise in some special cases. See also Polchinski's book \cite{Polchinski:1998rq, Polchinski:1998rr} for further comments.} to use \emph{integrated} vertex operators
\begin{eqaed}\label{eq:integrated_vertex_op}
    \int \rmd^2 \sigma \, \sqrt{\hat{\gamma}} \, \mathcal{V} \, ,
\end{eqaed}
now in Euclidean notation, since it is how we define the path integral. This also nicely clears up the air on how the integrated insertions arising from deformations of the background fields in the action relate to the local insertions dictated by the state-operator correspondence. These guys are manifestly diffeomorphism-invariant. What about Weyl rescalings? Since the volume form scales by a factor $e^{2\omega}$, which means that the integrand $\mathcal{V}$ must scale inversely.

Because we are working with the gauge-fixed \ac{CFT}, it makes more sense to recast this discussion in terms of scaling weights under conformal transformations. Recall that they act on the left-moving and right-moving sectors analogously to how conformal transformations on the complex plane are generated by holomorphic and anti-holomorphic maps, as we saw earlier. Hence, the scaling weight $\Delta$ of an irreducible representation of the conformal algebra splits into the sum of left-moving and right-moving conformal weights $h, \overline{h}$, with $\Delta = h + \overline{h}$. The difference $j = h - \overline{h}$ is the \emph{spin} of the state or operator. The remnant of Weyl invariance, which involves a generic function on the worldsheet, is full conformal invariance of these operators --- not just under rigid dilatons and rotations $w(z) \propto z$, rather any conformal map. In the \ac{CFT} lingo, such operators are called \emph{primary}, or \emph{conformal tensors}. In more detail, as we mentioned earlier, on a flat Euclidean worldsheet one can use complex coordinates $z = \sigma^1 + i \sigma^2 \in \mathbb{C}$, with $2 \, \rmd^2\sigma = \abs{\rmd{}z}^2$. We denote non-holomorphic functional dependence by the argument $(z, \overline{z})$. Then, a primary local operator (also called field in \ac{CFT}) $\mathcal{O}(z,\overline{z})$ has conformal weights $(h,\overline{h})$ if under a conformal transformation generated by a holomorphic map $z \mapsto w(z)$ the transformed operator $\widetilde{\mathcal{O}}$ is given by
\begin{eqaed}\label{eq:primary_def}
    \widetilde{\mathcal{O}}(w, \overline{w}) = \left(\frac{\rmd{}w}{\rmd{}z}\right)^{-h} \left(\frac{\rmd\overline{w}}{\rmd\overline{z}}\right)^{-\overline{h}} \mathcal{O}(z, \overline{z}) \, .
\end{eqaed}
All in all, due to diffeomorphism invariance $h = \overline{h}$, and due to Weyl invariance $h+\overline{h} = 2$. Therefore, vertex operators $\mathcal{V}$ \emph{corresponding to physical string states} are primary local operators of conformal weights $(h_\mathcal{V}, \overline{h}_\mathcal{V}) = (1,1)$. The field $X$ is \emph{not} primary, while $\partial X$ is. This makes geometric sense: coordinates do not have any intrinsic meaning, tangent vectors do.

Let us pause for a moment, since there is another little step to take to find an intelligible spectrum. The above result is still a little too abstract to put in a nice highlighted box. Backing up the logic that got us here, we are talking about states of a single string propagating in spacetime, which may have a bunch of other internal degrees of freedom. But in an asymptotically flat spacetime, asymptotic states have at least one universal degree of freedom due to the representation theory of isometries (or, you know, freshmen physics also does the trick). It is the momentum $p$! For a string, the relevant quantity is the momentum of its center of mass. In a flat spacetime background, the corresponding operator is the (quantum version of the) Noether charge generating spacetime translations, which the worldsheet sees as an internal symmetry. What does a vertex operator associated to a state of definite momentum look like? The answer can be guessed along the same lines as how the Schr\"{o}dinger equation is sometimes presented in class: using oscillating exponentials. The embedding $X : \Sigma \to M$ of the worldsheet into spacetime encodes how the string moves, so a good guess for the operator we seek is $e^{i p \cdot X}$, or more correctly its normally/radially ordered counterpart $: e^{i p \cdot X} :$, although we will suppress this annoying notation in the following. This guess can be shown to be correct computing the action of the momentum Noether charge on the vertex operator, which is given by a commutator. Alternatively, canonical quantization produces a Hilbert space where definite-momentum states can be matched to our vertex operator via standard creation and annihilation shenanigans.

So a general vertex operator with definite spacetime momentum looks like
\begin{eqaed}\label{eq:vertex_op_momentum}
    \mathcal{V}_p = V \, e^{ip \cdot X} \, ,
\end{eqaed}
where $V$ encodes polarizations or any other integral degree of freedom. A rather tedious but straightforward computation, which you can find \eg{} in Tong's notes \cite{Tong:2009np}, shows that the momentum piece contributes $\frac{\alpha'p^2}{4}$ to both conformal weights. Therefore, the dimensions $(h_V , \overline{h}_V)$ must be
\begin{eqaed}\label{eq:string_conformal_dims}
    h_V = \overline{h}_V = 1 - \, \frac{\alpha'p^2}{4} \, .
\end{eqaed}
Turning this around, since $p^2 = -m^2$ is the mass of the string state, we learn the following box-worthy lesson: excluding momentum contributions from conformal weights,

\begin{tcolorbox}
    \emph{If the spectrum of the worldsheet \ac{CFT} contains conformal weights $(h,\overline{h})$, the spectrum of physical string states has $h=\overline{h}$ and contains the squared masses}
    \begin{eqaed}\label{eq:string_spectrum_final}
        m^2 = \frac{4}{\alpha'} \left(h - 1\right) \, .
    \end{eqaed}
\end{tcolorbox}

The necessary equality of the weights is called \emph{level matching}, an important condition characterizing physical string states. These are of course the classical masses, which can receive quantum corrections lest forbidden by some protection mechanism. For example, gravitons are always exactly massless by diffeomorphism invariance. More generally, Weyl invariance \emph{forces momenta to be on-shell}, as befits \emph{bona fide} physical states. But wait, there's more --- because of Weyl invariance, there is no such thing as ``internal insertion''. The only well-defined physically meaningful quantity is the S-matrix, perfectly in line with our holographic arguments from \cref{sec:gravity_qft}. As a final technical remark, we derived \eqref{eq:string_spectrum} for vertex operators written as in \eqref{eq:vertex_op_momentum}, namely stripping away ghost and momentum contributions. For superstrings, the remaining operator $V$ contains other contributions from superghosts and ``spin fields'', as we shall now discuss. Stripping away these contributions as well and calling $(h, \overline{h})$ the remaining conformal weights for convenience, the $-1$ in \eqref{eq:string_spectrum} is replaced by $- \, \frac{1}{2}$ or $0$ depending on the boundary conditions for worldsheet fermions. This turns out to reflect whether the associated state is a boson or fermion \emph{in spacetime}! Of course \eqref{eq:string_spectrum} is still correct and general, but due to this consideration it is worthwhile discussing boundary conditions for worldsheet fermions. This also ties into a bit of an elephant-in-the-room situation in \eqref{eq:string_spectrum}; the attentive reader may notice that the expression in \eqref{eq:string_spectrum} may result in tachyons with $\alpha' m^2 < 0$, in particular due to the identity vertex operator. This is where the \ac{GSO} projection outlined in \eqref{eq:GSO_path_integral} comes to the rescue: classically stable vacua are such that tachyons are projected out by the combination of the coefficients $C_s$. Let us thus take this opportunity to go a bit deeper into this intriguing story.

\subsubsection*{Fermions, bosons and GSO projections}

Closed strings are topologically circles, which means that worldsheet spinors must be periodic or anti-periodic, as we mentioned earlier. When applied to the worldsheet spinors $\psi^\mu$ associated with the spacetime part of the \ac{CFT}, the former defines a \ac{R} sector, while the latter defines a \ac{NS} sector. Due to the split between left-movers and right-movers, \ac{RNS} superstrings have four options, creatively denoted \ac{R}-\ac{R}, \ac{NS}-\ac{NS}, \ac{NS}-\ac{R}, \ac{R}-\ac{NS}. Heterotic superstrings have \emph{chiral} worldsheet fermions $\psi^\mu_\text{R}$ with \ac{NS} and \ac{R} sectors. To achieve modular invariance, these sectors may combine in the \ac{GSO} projection with the ones arising from the internal degrees of freedom, which for heterotic superstrings are obligatory even for $d=10$ by criticality.

Each sector has different ground states, which show up explicitly in canonical quantization. In terms of vertex operators, one can define certain operators called \emph{spin fields} which create \ac{R} ground states. Their construction and behavior, alongside physical vertex operators is quite complicated,\footnote{They involve stuff like mutual locality, superghost ``bosonization'' and pictures, and Dirac deltas like $\delta(\gamma)$, the bosonic analog of $c$. On the flip side, they provide another perspective on the necessity of a \ac{GSO} projection. See \eg{}~\cite{Friedan:1985ge} for a classic presentation, and Polchinski's \cite{Polchinski:1998rq, Polchinski:1998rr} and Cecotti's \cite{Cecotti:2023dnp} books for a textbook account.} so we will draw some lessons from canonical quantization, schematically without worrying too much about gauge fixing, and talk directly about states. The two pictures are ultimately equivalent anyway. What happens is that a Hilbert space in an \ac{R} sector is built by creation operators from a vacuum, including periodic worldsheet spinors $\psi^\mu_\text{R}$. Their periodicity implies the existence of zero-modes $\widetilde{\psi}_0^\mu$ in their Fourier expansion, and canonical anti-commutation relations translate into a relation of the form
\begin{eqaed}\label{eq:clifford}
    \{\widetilde{\psi}_0^\mu \, , \, \widetilde{\psi}_0^\nu \} \propto \eta^{\mu \nu}
\end{eqaed}
in flat spacetime (the leading-order contribution of a tame background). Does \eqref{eq:clifford} look familiar? It is the defining property of (generators of) a Clifford algebra! Since zero-modes do not change the energy, the \ac{R} ground states form a \emph{spinor multiplet of spacetime isometries}. Depending on the dimension and the type of \ac{GSO} projection, these spacetime spinors can be chiral. Adding a center-of-mass momentum, one can show that the constraints on the Hilbert space imply the (massless) Dirac equation for these states, in this context also known as the \emph{Dirac-Ramond} equation. By a similar token, the \ac{NS} ground states are spacetime bosons without fermionic zero-modes. As we mentioned, the whole story can be in principle recast in terms of vertex operators, although it is technically involved. As we anticipated, the upshot of it is that stripping away superghosts and spin fields from the ``core'' vertex operators $V$ in \eqref{eq:vertex_op_momentum} is tantamount to replacing the $-1$ in \eqref{eq:string_spectrum} by $- \, \frac{1}{2}$ for \ac{NS} sectors and $0$ for \ac{R} sectors, for both left-movers and right-movers. 

The awesome mechanism we just described produces spacetime fermions\footnote{We intentionally interchange terms like fermions and spinors when not talking about (super)ghosts, since the spin-statistic theorem holds here.} from the worldsheet of a fermionic (``spinning'') string. Starting from \ac{R} ground states, acting with \emph{worldsheet} fermions \emph{does not make them into spacetime bosons}. This can be confusing. Repeat after me: \emph{the bosonic/fermionic character in spacetime is dictated by the sector}. Since \ac{R} (resp. \ac{NS}) sectors are fermionic (resp. bosonic), the resulting types of states for \ac{RNS} superstrings are summarized in the following table:

\begin{table}[h]
    \centering
    \begin{tcolorbox}[tab2,tabularx={Y|Y|Y},title=Spacetime bosons and fermions in \ac{RNS} superstrings,boxrule=0.5pt]
Sector of Hilbert space & Spacetime character & Example low-lying state(s) \\\hline\hline
\ac{R}-\ac{R}   & Bosons & $p$-forms \\
\ac{NS}-\ac{NS} & Bosons & Graviton, dilaton, $B$-field \\
\ac{NS}-\ac{R}   & Fermions & Gravitini, various fermions \\
\ac{R}-\ac{NS} & Fermions & Gravitini, various fermions \\
\end{tcolorbox}
    \caption{Closed-string spectra arranged by sector. The \ac{NS}-\ac{NS} sector contains the universal graviton, dilaton and B-field. The \ac{R}-\ac{R} sector contains spinor bilinears which decompose into p-form fields, while the \ac{NS}-\ac{R} and \ac{R}-\ac{NS} sectors contain spacetime fermions. In heterotic strings, the sectors of the Hilbert space are a bit different, and include gauge bosons.}
    \label{tab:RNS}
\end{table}

Some comments are in order. Covariant anti-symmetric tensors of various ranks ($p$-forms) arise because the \ac{R}-\ac{R} vacuum is a spinorial bilinear in spacetime, and thus it can be decomposed into $p$-forms via Fierz identities. The resulting physical states are creatively dubbed \ac{R}-\ac{R} forms. Gravitini arise when the \ac{GSO} projection brings along spacetime supersymmetry. In the heterotic case the situation is a bit more complicated, due to the obligatory internal degrees of freedom required by criticality. For instance, these can be free worldsheet chiral fermions with some internal global symmetry. As we will discuss in more generality in \cref{sec:swampland_stuff}, this gives rise to Yang-Mills quanta.

Still, the graviton, $B$-field and dilaton \emph{are always present} in both \ac{RNS} and heterotic superstrings. To see this, we observe that the \ac{GSO} projection can be thought of as gauging a $\mathbb{Z}_2$ symmetry on the worldsheet\footnote{Some recent developments in this direction were presented in~\cite{Kaidi:2019tyf}.} which contains the (exponentiated) right-moving \emph{worldsheet fermion number} $(-1)^{F_\text{R}}$. The latter is always a symmetry of the worldsheet superconformal field theory.\footnote{This allowed proving that heterotic superstrings with spacetime supersymmetry are devoid of \emph{any} kind of anomalies, regardless of the vacuum configuration~\cite{Tachikawa:2021mby}, extending the seminal work of~\cite{Schellekens:1986xh} on local anomalies.} In order to remove the tachyon, this projection must kill even numbers of \ac{NS} fermions, which means keeping odd numbers. By level matching, vertex operators of the schematic form $\psi_\text{L} \psi_\text{L}$ (resp. $\partial X_\text{L} \, \psi_\text{R}$) are allowed in \ac{RNS} (resp. heterotic) superstrings. As shown in \eqref{eq:string_spectrum}, vertex operators with more and more derivatives and field insertions create \emph{infinite towers} of states with unbounded mass. We will come back to this observation when computing their spectral density.

\subsubsection*{The \sout{donut} torus partition function}

\begin{figure}[ht!]
    \centering
    \includegraphics[width=\textwidth]{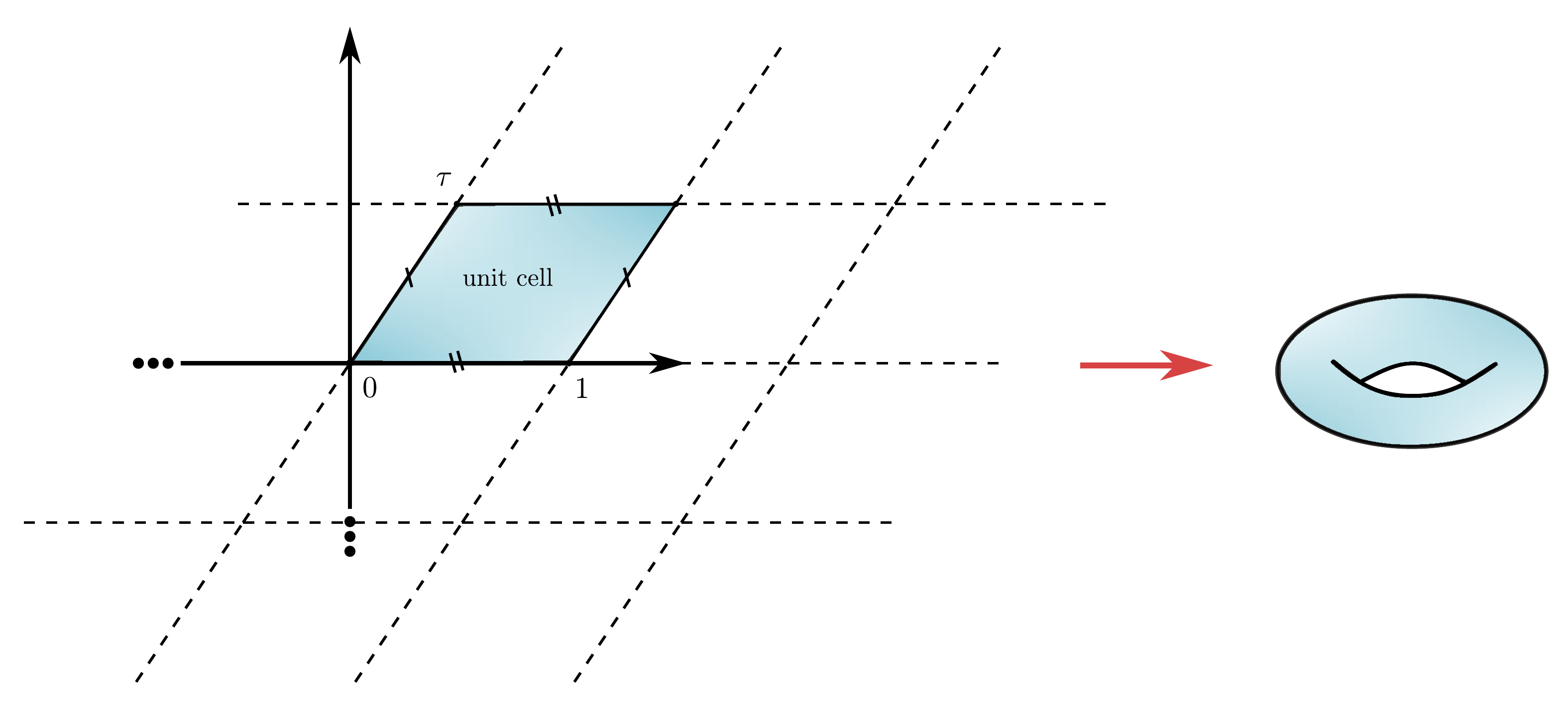}
    \caption{Defining a torus $T^2_\tau \equiv \mathbb{C}/\Lambda_\tau$ by quotienting the complex plane by the action of the lattice $\Lambda_\tau \equiv \mathbb{Z} \oplus \mathbb{Z} \tau$. The complex number $\tau$, up to modular equivalence, specifies the conformal (equivalently, complex) structure of the torus. The boundaries of the fundamental cell are identified accordingly.}
    \label{fig:torus}
\end{figure}

To see more directly how tachyons can be projected out, let us consider a more transparent quantity as a special case of \eqref{eq:GSO_path_integral}. The spectrum of a theory can be (at least partially) encoded by a thermal partition function, which for a theory on $\Sigma \simeq S^1_\text{space} \times \mathbb{R}_\text{time}$ translates to the partition function on the torus $\Sigma \simeq S^1_\text{space} \times S^1_\text{thermal}$. It is usually written in canonical language as $\partitionfunction(\beta) = \text{tr} \, e^{-\beta H}$ at inverse temperature $\beta$, but a more general quantity to consider is a grand-canonical partition function where a chemical potential $\xi$ for spatial momentum $P$ is added including $2\pi i \, \xi P$ in the exponent. Since $H$ generates time evolution on the cylinder, it generates dilations on the plane (recall radial quantization!). Thus, its eigenvalues are $h + \overline{h} - \, \frac{c}{24}$, the scaling dimensions, shifted by an extra ``Schwarzian'' or ``Casimir'' term due to the anomalous conformal properties of the Hamiltonian --- see \eg{} Tong's lecture notes \cite{Tong:2009np}. Long story short, one ends up with a trace over the spectrum of conformal weights $(h \, , \, \overline{h})$, conveniently packaged as the respective spectra of operators we'll call $L_0$ and $\overline{L_0}$. The notation comes from the fact that they are part of the generating set $\{L_n\}$ of the Virasoro algebra obtained by Fourier modes of the holomorphic and anti-holomorphic parts of the worldsheet energy-momentum tensor; these are their zero-modes. I guess the $L$ stands for\dots energy? Virasoro? Conformality? I'm too busy writing these notes to look it up.\footnote{I mean, the notation $\partitionfunction$ for partition functions comes the German word ``Zustandssumme'' for ``sum over states'', so I wouldn't be too surprised whatever the answer turns out to be.} Then $P = L_0 - \overline{L_0}$, since spatial translations on the cylinder are rotations on the plane. Thus, the torus partition function can be written
\begin{eqaed}\label{eq:torus_partition_function}
    \partitionfunction_{T^2} = \text{tr} \, q^{L_0-\frac{c_\text{L}}{24}} \, \overline{q}^{\overline{L_0}-\frac{c_\text{R}}{24}} \, ,
\end{eqaed}
where $q \equiv e^{2\pi i \tau}$, called the \emph{nome} of the torus, is defined by the parameter
\begin{eqaed}\label{eq:tau_def}
    \tau = \xi + i \, \frac{\beta}{2\pi} \equiv \tau_1 + i \, \tau_2
\end{eqaed}
whose real and imaginary parts contain the chemical potential and inverse temperature respectively. As we shall exploit in \cref{sec:BH_transition}, $\tau$ is actually the natural Teichm\"{u}ller parameter of the torus. This follows from the construction of $T^2_\tau$ as a quotient of the complex plane $\mathbb{C}$ by (the action of) a lattice, as shown in \cref{fig:torus}.

This partition function is particularly useful for a number of reasons. To begin with, it captures the spectrum of conformal weights, and it can be refined with further chemical potentials as usual in statistical mechanics. For the present discussion, it encodes the \ac{GSO} projection according to
\begin{eqaed}\label{eq:GSO_Z}
    \partitionfunction_{T^2} \; & = \sum_{\text{spin structures } s} C_s \, \partitionfunction_s = \sum_{s_\text{L} , s_\text{R}} C_{s_\text{L} , s_\text{R}} \, \partitionfunction_{s_\text{L} , s_\text{R}} \, .
\end{eqaed}
Here we see how tachyons can be projected out by the \ac{GSO} coefficients. Of course we refer to \emph{physical tachyons}, which are level-matched and are actual physical states. The torus partition function also includes states that are not level-matched, since at this point there is nothing telling it that the worldsheet theory is not just a two-dimensional \ac{CFT}, but rather it is coupled to (super)gravity to build \ac{QG} in spacetime. In \cref{sec:BH_transition} we will see how this story ends. For the time being, we can observe that \ac{GSO} coefficients must count degeneracies of states, and thus they must be integers whose signs must respect the spin-statistics theorem in spacetime. From the discussion on anomalies, we learned that $\partitionfunction_{T^2}$ must also be modular-invariant; all these constraints leave but a handful of consistent choices of \ac{GSO} projections. In fact, as we shall see around \cref{fig:hexagon}, in some cases all available options can be classified! Keep this excerpt in mind; it will come back \sout{haunt} help you later. You can find more details on this stuff in Polchinski's book \cite{Polchinski:1998rq, Polchinski:1998rr} and in~\cite{Angelantonj:2002ct}.

\subsubsection*{The lion, the witch, and the graviton --- generalities on string spectra}

Let's pause again to take stock of what we just discussed. We have a handle on the string spectrum, how tachyons are absent and fermions present. In fact, as already mentioned, the former turns out to non-trivially imply the latter due to the constraining power of modular invariance~\cite{Angelantonj:2023egh}. What we really care about, first and foremost, is the graviton, which is always present in the spectrum alongside the dilaton\footnote{More precisely, the dilaton is an effective field whose quanta are the low-energy description of these states. An analogous story goes for any other massless state/field, and thus often the terminology is a bit muddled. Hopefully, the conceptual distinction is clear.} whose background field's asymptotic value provides the string coupling constant $g_s$. For purely closed strings, (the quantum of the) the $B$-field is also there. These states are all \emph{massless}, and they arise from the decomposition of a rank-two tensor into irreducible representations of spacetime isometries: the general definite-momentum operator obtained deforming the background fields in \eqref{eq:polyakov_action} in complex coordinates has the schematic leading-order structure
\begin{eqaed}\label{eq:polyakov_vertex_op}
    V = \zeta_{\mu \nu}(p) \, \partial X^\mu \, \overline{\partial} X^\nu + \text{ possibly fermionic stuff} \, , 
\end{eqaed}
where the momentum contribution $e^{i p \cdot X}$ is left out for simplicity. As we explained above, this operator is not the actual unintegrated vertex operator for gravitons due to gauge invariance; there are (super)ghost insertions and not all terms in \eqref{eq:polyakov_vertex_op} survive superspace integration --- nevertheless, these states with these precise polarizations are always present in closed, oriented \ac{ST}. The polarization tensor $\zeta$ then decomposes into symmetric traceless, anti-symmetric and trace irreducible representations. The inquisitive reader may notice a puzzle here: on the one hand, the dilaton vertex operator is supposed to come from the Ricci scalar piece in \eqref{eq:polyakov_action}, but there is none in this gauge (``conformal gauge'') where the worldsheet is flat. On the other hand, the trace part of $\zeta$ would seem to correspond to a dilaton, but that's not where it came from. What's going on? Polchinski's book \cite{Polchinski:1998rq, Polchinski:1998rr} contains a detailed discussion of this point; see also~\cite{Polchinski:1988jq}. The upshot is that there is a relation between the coefficient $\zeta_\phi$ of the dilaton vertex operator (which is invisible in conformal gauge) and the trace part $\zeta_\mu^\mu$, which follows from Weyl invariance on a curved worldsheet. Solving the condition for Weyl invariance requires introducing an auxiliary unphysical momentum $\widetilde{p}$, which drops out of physical quantities. At the end of the day, solving these physical-state condition shows that $\zeta_\phi \propto \zeta_\mu^\mu$ as intuitively expected. This story has a counterpart the level of \ac{BRST} cohomology: the manifestly covariant ($\widetilde{p}$-independent) vertex operator contains the worldsheet Ricci curvature as expected, but it is not primary. Nevertheless, it lies in the same \ac{BRST} cohomology class as a primary operator which depends on $\widetilde{p}$. This auxiliary vector drops out of physical quantities thanks to the gauge redundancies associated to the graviton and $B$-field.

Within this universal sector of the spectrum, the massless states we just discussed are accompanied by string excitations, built from the spacetime sector, which arrange in higher-spin towers. Classically, they represent strings oscillating at different harmonic frequencies. Already in flat 10-dimensional spacetime, where no additional internal degrees of freedom are present and the conformal weights are known exactly, the spectrum contains infinite higher-spin towers arranged in Regge trajectories relating $m^2$ and spin, hence the name ``Regge slope'' for $\alpha'$. The structure of the spectrum as a whole is quite intricate~\cite{Markou:2023ffh, Basile:2024uxn}, but for our purposes these excited states are only (``only''!) relevant for \ac{UV} completeness as intermediate resonances, and we will not consider them as external states in scattering amplitudes. Ultimately, we care about finding a good behavior at high energies when scattering, say, gravitons, as well as recovering gravitational \ac{EFT} (coupled to other stuff) at low energies.

Other than these universal states, the spectrum can be messy depending on the string vacuum. It can comprise matter and gauge fields with various gauge groups. A way to see it goes like this: if the internal sector of the worldsheet \ac{CFT} contains some free fermions $\{\lambda^i\}$, such as in common heterotic constructions (see \eg{}~\cite{Florakis:2024ubz} for a recent review), their global internal symmetry acting on the index $i$ produces Yang-Mills fields in spacetime created by adjoint-valued combinations $\lambda^{[i}\lambda^{j]}$ due to anti-commutativity. We will outline a more general correspondence between (continuous) worldsheet internal symmetries and gauge redundancy in spacetime in \cref{sec:swampland_stuff}. Another commonly studied option is extra compact dimension in an internal \ac{NLSM} sector, whose spectrum contains Kaluza-Klein (and possibly winding) states. For instance, those built from operators $e^{ik \cdot Y}$, where $Y$ are the counterparts of $X$ for the internal compact dimensions and $k$ the counterparts of spacetime momenta $p$ encoding eigenvectors of the internal Laplace-Beltrami operator. We will not spend much time on these non-universal features.

\subsubsection*{Highly excited strings --- density of states and black-hole entropy}

In the spirit of drawing general lessons from \ac{ST}, independently of the particular vacuum/back\-ground configuration, there is one last thing we can look at regarding the spectrum. Namely, I wanna show you a universal high-mass behavior of the spectrum due to string excitations. A convenient quantity to package this information into is the (high-mass asymptotics of the) \emph{density of states} $\rho(m)$, which in the spacetime picture is a density of single-particle states in the sense of irreducible representations of spacetime isometries. A powerful tool to compute this quantity is --- you guessed it! --- the torus partition function of \eqref{eq:torus_partition_function}, alongside its modular invariance which we will define and extensively exploit in \cref{sec:BH_transition}.

Here is the story for \ac{RNS} superstrings in flat spacetime, which is the dominant contribution in tame backgrounds anyway.\footnote{See~\cite{Canneti:2024iyn} for a recent exploration of the density of states in curved backgrounds.} You can pick the simplest case with $d=10$ for concreteness, in order not to worry about the internal sector --- this will not actually affect the end result, since we care about the behavior of stringy excitations in spacetime, which is dominant at large mass. Similarly, the story for heterotic superstrings is analogous, albeit somewhat complicated by the obligatory internal degrees of freedom. We will, however, ask that $d$ be even to simplify some considerations on the \ac{R} sector. Let us compute some torus partition functions. In flat spacetime, since left-movers and right-movers decouple, traces factorize on the Hilbert spaces $\mathcal{H} = \mathcal{H}_\text{L} \otimes \mathcal{H}_\text{R}$ decomposed in \ac{R} and \ac{NS} sectors $\mathcal{H}_\text{R}, \mathcal{H}_\text{NS}$. This partially spells out the spin structure on the torus, the (anti-)periodicity of spinors along the spatial circle. The other piece of data is the one along the thermal circle, denoted $\pm$. It doesn't affect the Hilbert (sub)space; rather, it defines a different trace. As a result, spin structures $s_\text{L}, s_\text{R}$ for both left-movers and right-movers each take the four values $\{(\text{R},\pm), (\text{NS},\pm)\}$. We denote this by writing $s_\text{L} = (a_\text{L}, b_\text{L})$ and similarly for $s_\text{R}$. The final partition function in \eqref{eq:torus_partition_function} can then be recast in the form
\begin{eqaed}\label{eq:torus_partition_function_factorized}
    \partitionfunction_{T^2} \; & = \sum_{s_\text{L} , s_\text{R} \in \{(\text{R},\pm), (\text{NS},\pm)\}} C_{s_\text{L}, s_\text{R}} \, \text{tr}^{(b_\text{L})}_{\mathcal{H}_{a_\text{L}}} \, q^{L_0-\frac{c_\text{L}}{24}} \, \text{tr}^{(b_\text{R})}_{\mathcal{H}_{a_\text{R}}} \, \overline{q}^{\overline{L_0}-\frac{c_\text{R}}{24}} \\
    & \equiv \sum_{s_\text{L} , s_\text{R} \in \{(\text{R},\pm), (\text{NS},\pm)\}} C_{s_\text{L}, s_\text{R}}  \, \partitionfunction_{s_\text{L}} \, \overline{\partitionfunction_{s_\text{R}}} \, .
\end{eqaed}
Therefore, we can focus, say, on the left-movers (using holomorphic notation), where the four ``elementary traces'' appear. We momentarily drop the $\text{L}$ subscript, writing $\partitionfunction_s = \partitionfunction_{(a,b)}$ for the elementary traces. The structure of vertex operators, for example those in \eqref{eq:polyakov_vertex_op}, built from $X$ and $\psi$ fields shows that the conformal weights can be raised by including factors of $\partial^{n>0} X$ and $\partial^{n\geq 0} \psi$. For the \ac{R} sector, extra factors of $\psi$ and spin fields are present, in order to obtain the correct vertex operators creating \ac{R} states from the \ac{NS} vacuum. Moreover, superghosts insertions, whose precise form depends on the ``ghost picture'', are also present to ensure consistency with \eqref{eq:string_spectrum}. A thorough presentation of (and developments of) the general picture was recently given in~\cite{Markou:2023ffh, Basile:2024uxn}. The upshot is that worldsheet fermions contribute \emph{half-integer weights in the \ac{NS} sector and integer weights in the \ac{R} sector}. The same result can be derived much more easily in canonical quantization, where this shift arises from the appropriate Fourier expansion, while the effect of spin fields and superghosts is encoded in the zero-point energy of the \ac{R} ground states. Recall that the latter are degenerate, since they comprise a spacetime spinor.

We said \emph{lot word}, but \emph{few word do trick}: we can recycle standard techniques in statistical mechanics to compute the traces, as in~\cite{Angelantonj:2002ct}. The bosonic and fermionic traces factorize, since here the \acp{CFT} are decoupled (in general, they are only to leading order in the curvatures). We can drop the overall factors with exponents $\frac{c_\text{L,R}}{24}$, since they cancel, provided we take the (super)ghost contributions into account. 

\begin{itemize}
    \item \emph{Bosons.} The computation for bosons $X$ it is particularly straightforward: the trace over spacetime momenta contributes a factor
\begin{eqaed}\label{eq:momenta_contribution}
    \text{Vol}(M) \int \frac{\rmd^dp}{(2\pi)^d} \, e^{-\pi \alpha' p^2 \tau_2} = \frac{\text{Vol}(M)}{(4\pi^2 \alpha')^{\frac{d}{2}}} \, \tau_2^{- \frac{d}{2}} \, ,
\end{eqaed}
where the (formal) volume of spacetime appears due to the standard continuum measure in the functional trace, as familiar from statistical physics. It can also be derived from a path integral computation, isolating the zero-modes $X_0^\mu$. As for the stringy excitations, they arise including factors like $\prod_{k>0} (\partial^k X)^{n_k}$ in vertex operators, and thus contribute
\begin{eqaed}\label{eq:bosonic_oscillators}
    \sum_{\text{natural tuples } \{n_k\}} q^{\sum_{k > 0} k \, n_k} = \prod_{k>0} \sum_{n \geq 0} q^{kn} = \prod_{k>0} \frac{1}{1-q^k} \equiv \partitionfunction_\text{boson} \, .
\end{eqaed}
to the trace for each component $X^\mu$. The full result is thus simply $\partitionfunction_\text{boson}^d$. From a path integral perspective, the reduced functional determinant of the torus Laplacian is accompanied by the factor $\int \rmd^d X_0 = \text{Vol}(M)$, yielding the same combined result, as discussed \eg{} in \href{https://www.lpthe.jussieu.fr/~israel/notes.pdf}{Isra\"{e}l's lecture notes}. More precisely, the two expressions match including the neglected factors with the central charge. Upon doing so, the \emph{Dedekind $\eta$ function} shows up; similarly, for the fermions $\psi$ some pretty cool elliptic functions, the \emph{Jacobi $\vartheta$ functions}, appear. Unfortunately, we do not have time to go deeper in this direction, but you can look them up in~\cite{Angelantonj:2002ct}.

\item \emph{$bc$ ghosts.} From the structure of the Faddeev-Popov determinant, it should be apparent that they contribute a similar reduced Laplacian determinant as the bosons, except without spacetime momenta contributing. Since there are two such ghosts, the contribution is
\begin{eqaed}\label{eq:ghost_det}
    \partitionfunction_\text{ghosts} = \partitionfunction_\text{boson}^{-2} \, ,
\end{eqaed}
where once again we stripped away the central charge term. This result is important: the total power of $d-2$ that appears in the final expression is the number of \emph{transverse} spacetime dimensions to the worldsheet, which is the relevant number when doing canonical quantization in the light-cone gauge~\cite{Angelantonj:2002ct}.

\item \emph{Fermions.} Here is when the story gets interesting, since the trace depends on the spin structure $s=(a,b)$ on the torus. We already discussed the difference between the \ac{NS} and \ac{R} sectors in the weights of vertex operators. As for the thermal periodicity, it can be implemented via a suitable exponentiated operator in the trace,\footnote{Schematically, it acts as translations on the bosonized degrees of freedom. You can find the details in Polchinski's book \cite{Polchinski:1998rq, Polchinski:1998rr}.} defined in such a way as to conserve the various correlators of the fields. It turns out to be given by the exponentiated worldsheet fermion number $e^{i \pi F} = (-1)^F$, which shows up in \ac{GSO} projections and whose definition depends on the particular sector since \ac{R}-sector vertex operators are fermionic. Let us start from the bare traces in the two sectors: due to the difference between integer and half-integer weights, for a single component $\psi^\mu$ one finds
\begin{eqaed}\label{eq:NS_R_traces}
    \partitionfunction_{\text{NS},+}^{\text{single}} = \sum_{ \text{binary tuples } \{n_k \}} q^{\sum_{k>0} \left(k-\frac{1}{2}\right)n_k} & = \prod_{k>0} \left(1 + q^{k - \frac{1}{2}}\right) \, , \\
    \partitionfunction_{\text{R},+}^{\text{single}} = \text{dim(R)}\sum_{ \text{binary tuples } \{n_k \}} q^{\sum_{k>0} k \, n_k} & = \text{dim(R)} \prod_{k>0} \left(1 + q^k\right) \, .
\end{eqaed}
As for the bosons, the full results are the $d$-fold powers of the above expressions. For the \ac{R} sector there is a prefactor $\text{dim(R)}$, accounting for the Hilbert space of \ac{R} ground states. As explained above, it is the dimension of the spinorial representation of spacetime isometries.

The two remaining traces are slightly trickier: inserting $(-1)^F$ flips the sign to each fermionic contribution in each factor of the product, since they arise from single insertions of $\psi$. However, because of the spin fields (or, in the canonical approach, the definition of $(-1)^F$), the minus sign also affects the two chiral halves of the ground-state degeneracy, which now cancel each other in the trace. The resulting expressions are~\cite{Angelantonj:2002ct} (you can find more details in Kiritsis' book \cite{Kiritsis:2019npv}, for example)
\begin{eqaed}\label{eq:NS_R_traces_other}
    \partitionfunction_{\text{NS},-}^{\text{single}} = \prod_{k>0} \left(1 - q^{k - \frac{1}{2}}\right) \, , \qquad \partitionfunction_{\text{R},-}^{\text{single}} = 0 \, .
\end{eqaed}

\item \emph{Superghosts.} Similarly to the $bc$ ghosts, the $\beta \gamma$ superghosts contribute in such a way as to replace $d$-fold powers with $(d-2)$-fold powers, playing the same role as the light-cone gauge does in canonical quantization.
\end{itemize}

The complete partition function will contain these contributions weighted by \ac{GSO} coefficients, as well as factors accounting for internal degrees of freedom (if any). For concreteness, let's just pick the $(\text{R},+)$ option, the simplest to deal with without involving linear combinations of sectors. We also ignore the zero-mode contribution of \eqref{eq:momenta_contribution} and \eqref{eq:NS_R_traces}, since we care about the stringy excitations which dominate the high-energy regime anyway. The total partition function is
\begin{eqaed}\label{eq:total_R_plus}
    \partitionfunction^\text{total}_{\text{R},+} = \prod_{k>0} \left( \frac{1+q^k}{1-q^k} \right)^{d-2} .
\end{eqaed}
Since the partition function is a sum over states weighted by $q^k$, the degeneracy $d_k$ at ``mass level'' $k$ can be extracted by a contour integral over a loop $\mathcal{C}_0$ around $q=0$ in the complex plane,
\begin{eqaed}\label{eq:degeneracies}
    d_k = \frac{1}{2\pi i} \oint_{\mathcal{C}_0} \frac{\rmd{}q}{q^{k+1}} \, \partitionfunction^\text{total}_{\text{R},+} \, .
\end{eqaed}
At large $k$, the integral is amenable to a saddle-point expansion. The correct asymptotic expression contains an exponential term and a power-like prefactor; we will only derive the former, which is the universal term we mostly care about. In other words, we obtain an asymptotic expansion for $\ln \, d_k$. From \eqref{eq:total_R_plus} we can infer that the saddle point is located near $q = 1$. The useful manipulation
\begin{eqaed}\label{eq:log_trick}
    \sum_{k>0} \ln(1\pm q^k) = - \sum_{n,k>0} \frac{(\mp q^k)^n }{n} = - \sum_{n>0} \frac{(\mp)^n}{n} \, \frac{q^n}{1 - q^n}
\end{eqaed}
allows us to write
\begin{eqaed}\label{eq:log_expansion}
    \ln \partitionfunction^\text{total}_{\text{R},+} = 2(d-2) \sum_{n \text{ odd}} \frac{1}{n} \, \frac{q^n}{1-q^n} \overset{q \to 1}{\sim} \frac{2(d-2)}{1-q} \sum_{n \text{ odd}} \frac{1}{n^2} = \frac{\pi^2}{4} \, \frac{d-2}{1-q} \, .
\end{eqaed}
Including the $q^{-k-1} = -(k+1)\ln q$ factor, the saddle point $q_\ast$ is located at
\begin{eqaed}\label{eq:saddle_approx}
    1-q_\ast \overset{k \gg 1}{\sim} \sqrt{\frac{\pi^2(d-2)}{4k}} \, ,
\end{eqaed}
so that
\begin{eqaed}\label{eq:degeneracy_asymptotics}
    \ln d_k \overset{k \gg 1}{\sim} \ln \frac{\partitionfunction^\text{total}_{\text{R},+}}{q^{k+1}}\bigg|_{q=q_\ast} \overset{k \gg 1}{\sim} 2\pi\sqrt{\frac{(d-2)k}{4}} \, .
\end{eqaed}
This exponential behavior translates to closed strings as well as curved backgrounds (see \eg{}~\cite{Canneti:2024iyn} and references therein), and the internal degrees of freedom are no exception --- the reason is that their central charge universally governs a leading growth of the same type (``Cardy formula'').\footnote{For the curious, a general central charge $c>0$ yields $\ln d_k \overset{k \gg 1}{\sim} 2\pi\sqrt{\frac{c \, k}{6}}$. In this case, the transverse degrees of freedom of the superstring contribute $c=\frac{3}{2}(d-2)$, so it checks out.} This can be used to prove various cool stuff about string spectra, \eg{} the necessity of spacetime fermions from the absence of physical tachyons~\cite{Angelantonj:2023egh}. Since $m^2 \overset{k \gg 1}{\sim} \frac{4k}{\alpha'} = 8\pi \, M_s \, k$ for large $k$, finally
\begin{eqaed}\label{eq:density_mass}
    \ln \rho(m) \overset{m \gg M_s}{\sim} \sqrt{\frac{\pi (d-2)}{8}} \, \frac{m}{M_s}
\end{eqaed}
for this particular choice, although as we explained the scaling in $\frac{m}{M_s}$ is universal. The degeneracy of multi-string states can be be studied with similar methods~\cite{Bedroya:2024ubj}. As we shall see in \cref{sec:BH_transition}, this is crucial to yield \ac{BH} production. In a boxed summary:

\begin{tcolorbox}
    \emph{Unlike for point particles, the degeneracy of single-string states scales exponentially in their mass $m$ for $m \gg M_s$ with the scaling $\ln \rho(m) \overset{m \gg M_s}{\sim} \mathrm{const.} \times \frac{m}{M_s}$.}
\end{tcolorbox}

Deferring a more in-depth discussion to \cref{sec:BH_transition}, we can apply \eqref{eq:density_mass} to learn something enticing. As we argued in \cref{sec:gravity_qft}, a quantum theory of gravity ought to exhibit \ac{UV}/\ac{IR} mixing, and in particular \acp{BH} should dominate its high-energy behavior (``classicalization''~\cite{Dvali:2010jz}). The leading-order entropy of a Schwarzschild \ac{BH} of large mass $M_\text{BH} \gg \MPl{}$ (more precisely $\gg \UVcutoff^{3-d} \, \MPl^{d-2}$, where \UVcutoff{} for weakly coupled strings is the string scale $M_s$) is
\begin{eqaed}\label{eq:BH_leading}
    S_\text{BH}^\text{leading} \propto \left(\frac{M_\text{BH}}{\MPl}\right)^{\frac{d-2}{d-3}} = \left(g_s^2 \, \frac{M_\text{BH}^{d-2}}{M_s^{d-2}} \right)^{\frac{1}{d-3}} \, ,
\end{eqaed}
which intriguingly matches \eqref{eq:density_mass} for $m = M_\text{BH} = \frac{M_s}{g_s^2} \gg M_s$. But, as we shall see in \cref{sec:weyl_cancellation}, the Planck scale is given by\footnote{This is correct in the absence of other limits dominating over $g_s \ll 1$. This will play a role in \cref{sec:swampland_stuff}.} $\MPl^{d-2} = M_s^{d-2} \, g_s^{-2}$, which means that the matching scale can be recast according to
\begin{eqaed}\label{eq:matching_scale}
    M_\text{match} = M_s^{3-d} \, \MPl^{d-2} \, .
\end{eqaed}
But this is precisely the mass of a \ac{BH} of size $M_s^{-1} \gg \MPl^{-1}$! What matters for this scaling to be reliable is that the \ac{BH} be super-Planckian, whereas sub-leading terms are controlled by the \ac{EFT} cutoff. This parametric matching is very suggestive of some \ac{UV}/\ac{IR} mixing stuff going on connecting \acp{BH} and strings~\cite{Susskind:1993ws, Horowitz:1996nw}. Indeed, in much more sophisticated settings it is possible to perform an explicit counting of \ac{BH} microstates in \ac{ST}, and in all cases in which the comparison was possible the results match, including the precise prefactor~\cite{Strominger:1996sh}. Another related avenue to explore this deep connection is via high-energy string scattering, which we will discuss in \cref{sec:BH_transition}. For more details on both these intimately related story, see the recent review~\cite{Bedroya:2022twb}.

\subsubsection{The string landscape}\label{sec:landscape}

This journey has (hopefully) been pretty much forced on us so far, starting from our definition of \ac{QG} and looking at weakly coupled gravitons. Whenever there was a choice to make, we kept all the options. At least this is the spirit with which I undertook this pursuit. What have we learned? The tl;dr version is that string vacua have a sharp definition, and their spectra have fascinating universal features, chief among which are the presence of gravitons and the absence of free parameters. Where do we go from here? There are two avenues that I can think of. Having constructed the theory and looked at some of its generalities, it's time to do some physics! Our requirements for \ac{QG} were that its \ac{IR} be connected to, and dominated by, a gravitational \ac{EFT}, while its \ac{UV} ought to be finite and consistent with \ac{UV}/\ac{IR} mixing. We now set the stage to understand the former aspect in \cref{sec:part_iii}, leaving the latter stuff for \cref{sec:part_iv}. In order to understand how \ac{ST} reduces to gravitational \ac{EFT}, we need to talk about its vacua and their low-lying features with a broader scope.

\begin{figure}[ht!]
    \centering
    \includegraphics[scale=0.8]{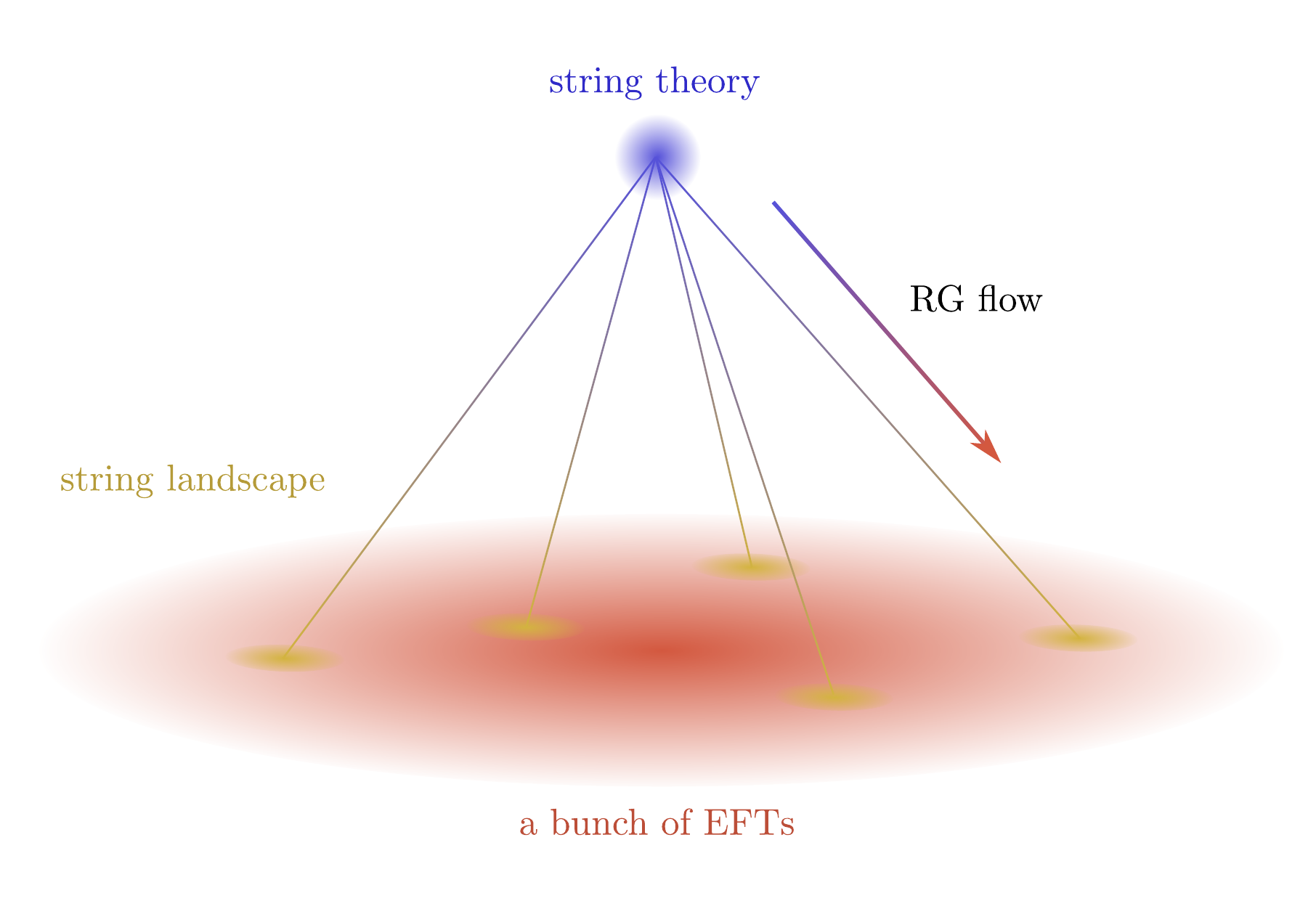}
    \caption{A depiction of the string landscape. Starting from \ac{ST}, which is believed to be unique due to (non-)perturbative dualities (and some swampland considerations), a number of \acp{EFT} arise, encoding the low-energy physics relative to a vacuum configuration. Although there seem to be many such inequivalent \acp{EFT}, there are several indications that they are \emph{finitely} many, or at least form a negligible subset of the set of all \acp{EFT}.}
    \label{fig:landscape}
\end{figure}

What we are hinting at here is the (in)famous concept of \emph{string landscape}. It is the set of vacua and their associated \acp{EFT} describing low-energy physics. For the corner of \ac{ST} we focus on here we have arrived at a definition of in \cref{sec:string_perturbation_theory} in terms of fairly abstract concepts, so it is instructive to go through some classes of examples. Bear in mind that having a definition is a far cry from having a complete and thorough understanding of the string landscape, even in this limited corner! The important thing for our purposes is that it looks discrete or even finite in a suitable sense,\footnote{Namely, if an \ac{EFT} has a moduli space of vacua, it still counts as one inequivalent point in the landscape.} as depicted in \cref{fig:landscape}. What's more, for some particularly simple classes of \acp{EFT}, a perfect match between the ones consistent with swampland conditions and the ones derived from \ac{ST} has been established~\cite{Adams:2010zy, Kumar:2010ru, Montero:2020icj, Bedroya:2021fbu, Tarazi:2021duw}. If you wanna check out this exciting research area, the keyword is ``string lamppost principle''.

\subsubsection*{The simplest sectors --- 10 dimensions}

To begin with, there is a distinguished class of vacua due to its simplicity. From the grueling construction we went through in the preceding sections, it should be clear that the maximal dimension $d=10$ for perturbative classically stable vacua is also the simplest, since there are no additional degrees of freedom to worry about. This is part of the reason why the slogan that \ac{ST} requires 10 dimensions is so widespread. Another reason is that many families of lower-dimensional vacua are connected to these by compactification, or some generalization thereof. Whether this is the full extent of the landscape is not clear, although some partial results are available~\cite{Dijkgraaf:1987vp, Seiberg:1988pf, Runkel:2002yb, Aoufia:2024awo}. Anywoosles, 10-dimensional vacua are extremely constrained due to the \ac{GSO} projections in \eqref{eq:GSO_path_integral} and \eqref{eq:GSO_Z}. As in \eqref{eq:torus_partition_function_factorized}, there are only four options at our disposal for the left-movers and right-movers of \ac{RNS} superstrings, and the \ac{GSO} coefficients arrange into a $4 \times 4$ matrix. The simplest heterotic superstrings are similar, replacing the left-moving \ac{R} and \ac{NS} sectors by periodic and anti-periodic worldsheet fermions pertaining to the internal (rather than spacetime) sector. Actually, in this case, it is more illuminating to use a different basis instead of the four elementary traces we have computed in the preceding section. Suitable linear combinations thereof encode representation-theoretic content\footnote{A more refined version of these affine characters are formal $q$-series with \emph{K-theory valued} coefficients rather than integer-valued. An analysis along these lines was performed in~\cite{Hanany:2010da}.} of spacetime isometries (and internal heterotic symmetries) via \emph{characters} of their affine realization on the worldsheet \ac{CFT}. This allows a more transparent interpretation of the \ac{GSO} coefficients and spectral degeneracies. We will not go into it, but you can find more details in~\cite{Angelantonj:2002ct} or in Kiritsis' book \cite{Kiritsis:2019npv}. At any rate, the \ac{GSO} coefficients are constrained by modular invariance (recall --- it is the remnant of global gravitational anomaly cancellation) and integrality conditions, to allow a sensible physical interpretation of expressions such as \eqref{eq:GSO_Z} and \eqref{eq:torus_partition_function_factorized}. Modular invariance acts on the four-component vectors of affine characters linearly, which means that the \ac{GSO} matrix must be invariant under conjugation by this action. This is basically a Diophantine linear algebra problem, and it can be fully solved in 10 dimensions. One finds a handful of non-tachyonic; for closed strings, there are five:

\begin{itemize}
    \item \emph{Type IIA and type IIB.} They arise from \ac{RNS} superstrings. They have no gauge group in spacetime, but they feature \ac{R}-\ac{R} forms. Their names come from the fact that they possess $\mathcal{N}=2$ spacetime supersymmetry, the maximal amount in 10 dimensions compatible with \ac{EFT} and Weinberg's soft theorem. The type IIA theory is non-chiral, whereas the type IIB theory is chiral and remarkably anomaly-free. They are the supersymmetric counterparts of the purely spacetime-bosonic (and hence tachyonic) type 0A and 0B superstrings, the two remaining consistent closed-string \ac{RNS} solutions to the \ac{GSO} constraints. The role of type II theories in string phenomenology usually brings along more sophisticated ingredients such as orientifolds. The low-energy limits of type II superstrings are the two type II supergravities in 10 dimensions.
    
    \item \emph{Exceptional and orthogonal heterotics.} As their names suggest, they arise from heterotic superstrings. They possess minimal spacetime supersymmetry, which implies that they are chiral and feature spacetime gauge algebras\footnote{If they did 
not, they would be inconsistent due to ``irreducible'' local anomalies~\cite{Alvarez-Gaume:2022aak}. Anomaly cancellation for heterotic superstrings is a fascinating subjects on its own, for which unfortunately we don't have time.} $\mathfrak{e}_8 \oplus \mathfrak{e}_8$ and $\mathfrak{so}(32)$ respectively, hence the names. The gauge groups are $(E_8 \times E_8) \rtimes \mathbb{Z}_2$ and $\text{Spin}(32)/\mathbb{Z}_2$. They immediately attracted interest in string phenomenology because of there features, although the challenges are complementary to the type II case. The low-energy limits of these heterotic superstrings are minimal supergravity coupled to super-Yang-Mills theory in 10 dimensions.
    
    \item \emph{The unique non-supersymmetric heterotic.} While there are a number of 10-dimensional heterotic constructions without spacetime supersymmetry, all but one are tachyonic. The unique option is again chiral, as all heterotic theories, and has a $\mathfrak{so}(16) \oplus \mathfrak{so}(16)$. The gauge group is thus a quotient of $\text{Spin}(16) \times \text{Spin}(16)$. The low-energy limit is not supergravity, rather some gravitational \ac{EFT} coupled to Yang-Mills theory, as well as some bosonic and fermionic matter fields. Basically your garden-variety, bread 'n' butter, run-of-the-mill \ac{EFT},\footnote{I ran out of synonyms.} with a scalar potential for the dilaton generated by quantum effects.
\end{itemize}

\begin{figure}[ht!]
    \centering
    \includegraphics[scale=0.7]{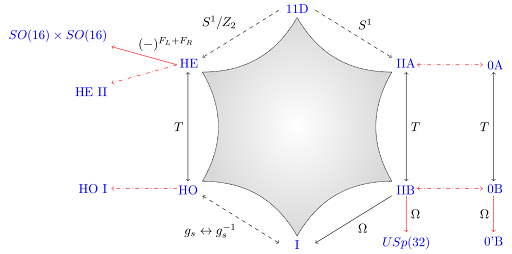}
    \caption{A depiction of the simplest classically stable string vacua, namely those in 10 dimensions. Amongst purely closed-string sectors there are only the type IIA, type IIB and three heterotic theories. The other theories involve open strings and/or non-perturbative ingredients (such as M-theory in 11 dimensions). Figure taken from \cite{Basile:2023knk}.}
    \label{fig:hexagon}
\end{figure}

The latter option is particularly attractive due to the absence of spacetime supersymmetry. Its phenomenological implications have been studied to some extent~\cite{Blaszczyk:2014qoa, Blaszczyk:2015zta}, but its ultimate fate hinges on quantum effects~\cite{Basile:2018irz, Antonelli:2019nar}. In addition to these, there are consistent solutions which involve open strings, and all these 10-dimensional settings are connected by string dualities~\cite{Witten:1995ex}, strongly hinting at an underlying uniqueness. The resulting web of limits and dualities for non-tachyonic vacua is depicted in \cref{fig:hexagon}, the celebrated ``duality hexagon'' of 10-dimensional superstrings, $11$-dimensional supergravity and their partially mysterious glue,\footnote{Sometimes by ``M-theory'' people refer exclusively to the corner without weakly coupled strings. I use the broader meaning of whatever description encompasses all connected limits. Its description goes beyond the mere low-energy limit~\cite{Banks:1996vh, Banks:1999az}, in some settings also allowing the use of the \ac{AdSCFT} correspondence.} ``M-theory''. Five of these vacua (six, including 11-dimensional supergravity) possess spacetime supersymmetry: type IIA, type IIB, type I and the exceptional and orthogonal heterotics. The rest has no spacetime supersymmetry and no tachyons in the perturbative spectrum, although they bring along several puzzling subtleties in their dynamics. The other options in 10 dimensions are tachyonic, although sometimes tachyons can disappear after compactification~\cite{Kaidi:2020jla}.

\subsubsection*{Compactifications --- geometry, non-geometry and stringy geometry}

A natural way to generate more vacua from the 10-dimensional constructions is by compactification. In the language we have developed in this section, one can either take the spacetime sector to be an \ac{NLSM} on a compactified 10-dimensional spacetime, or take the internal sector to be an \ac{NLSM} on a compact target space. Either way, the physical spacetime $M$ is
\begin{eqaed}\label{eq:compactification_manifold}
    M = M_\text{external} \times M_\text{internal} \, ,
\end{eqaed}
or more generally a fibration. In order for the worldsheet \ac{CFT} to make sense, the temporal component $X^0$ of the embedding fields should be kept in the external spacetime sector, as explained in Polchinski's book \cite{Polchinski:1998rq, Polchinski:1998rr}. It could be a technical limitations of our approach or, more suggestively, a stringy lesson about time. Other than that, the condition that the internal worldsheet \ac{CFT} be critical imposes non-trivial constraints on what $M_\text{internal}$ can be. At leading order in the curvatures, these are just the Einstein equations of the 10-dimensional \ac{EFT}, confirming that in \ac{ST} compactification is a dynamical gravitational notion. In \ac{QFT}, we can put a (consistent) theory on whatever manifold we want, insofar as it carries a compatible tangential structure (\eg{} an orientation, a (s)pin structure, and so on). Beyond leading order in the curvatures, there are some known vacua which are in fact $\alpha'$-exact.

\begin{itemize}
    \item \emph{Toroidal orbifolds.} When $M_\text{internal} \simeq T^n$ is an $n$-dimensional torus, the worldsheet \ac{CFT} is solvable. In fact, it remains solvable replacing $T^n$ by an \emph{orbifold} thereof, quotienting by the action of a discrete group~\cite{Dixon:1985jw}. These can sometimes be seen as limits of smooth compactifications, but they also admit ``asymmetric'' versions, acting differently on left-movers and right-movers, some of which\footnote{Some asymmetric orbifolds are perturbatively equivalent to geometric vacua~\cite{Blumenhagen:2000fp, Angelantonj:2000xf}.} are examples of \emph{non-geometric} string vacua: somewhat similar to actual compactifications, but not quite the same. Recent examples of this type without spacetime supersymmetry nor tachyons were found in~\cite{Baykara:2024tjr, Baykara:2023plc, Angelantonj:2024jtu}. Tori are simple examples where a stringy redundancy becomes manifest: it is \emph{T-duality}, an equivalence between large/small cycles (relative to the string scale) and Kaluza-Klein/winding modes. Particles can never exhibit it, since they cannot be wound around anything.
    
    \item \emph{Calabi-Yau compactifications.} Calabi-Yau manifolds are very cool spaces which afford a consistent worldsheet supersymmetric \ac{NLSM} \emph{to all orders in the curvatures}~\cite{Nemeschansky:1986yx, Jardine:2018sft}. They are often considered in string phenomenology because they lead to minimally supersymmetric \acp{EFT} in four dimensions after dimensional reduction. It would be nice if we could observe low-energy spacetime supersymmetry, but it is not clear what kinds of low energy scales would favor supersymmetry breaking (super-Higgsing) in this context. Calabi-Yau manifolds of real dimension six, which lead to four-dimensional reductions, correspond to superconformal worldsheet theories with $c=(9,9)$. The space of these \acp{CFT} has been examined thoroughly, both to find some universal consequences for four-dimensional physics~\cite{Feng:2010yx} and in order to understand whether Calabi-Yau compactifications are the only option~\cite{Seiberg:1988pf}. Unfortunately the situation remains to be settled, but in the simpler case of one internal dimension the analysis is much more comprehensive~\cite{Dijkgraaf:1987vp, Runkel:2002yb}.
\end{itemize}

Non-geometric vacua appear also beyond $\alpha'$-exact and/or perturbative settings, but we have no time to expand upon them. They exhibit pretty wild properties, such as non-commu\-ta\-tive (see also~\cite{Seiberg:1999vs}) and/or non-associative geometry.

\subsubsection*{More abstract superconformal field theories?}

One of the lessons I would like you to take home is that, even just at the perturbative level, \ac{ST} provides us with a potentially vast generalization of the notion of geometry (as in manifolds). Worldsheet \acp{CFT} can be in principle much more general and abstract, and geometry \emph{emerges} in some limits by moving in these spaces of theories. Non-perturbatively, mainly via the \ac{AdSCFT} correspondence, stringy matrix models and F-theory, there are different, but equally (if not more) interesting ways in which geometry emerges from something more fundamental. That being said, it's not like we have an embarrassment of riches --- modular invariance and criticality are extremely constraining, and besides two-dimensional \acp{CFT} are wild beasts which can hide secret equivalences.

That said, there are many well-studied classes of \acp{CFT} that do the job: minimal models, Gepner models~\cite{Gepner:1987qi}, Landau-Ginzburg models. Some can be deformed to geometric \acp{NLSM}, but sometimes ``rigid'' vacua can be found, \eg{} in~\cite{Baykara:2023plc} and most recently (at the time of writing, today!) in~\cite{Rajaguru:2024emw, Becker:2024ayh}. Otherwise, most of the well-understood examples have \emph{moduli}. Not to be confused with moduli of the worldsheet geometry, these are marginal couplings of the \ac{CFT} which correspond to massless scalar fields in spacetime, whose expectation values can be freely chosen in a ``field space'' (usually a Riemannian manifold). Moduli are problematic for phenomenology, since they give rise to additional long-range forces. A large fraction of the efforts in string phenomenology deals with solving this issue, and a smaller fraction with avoiding it from the outset. However, from the theoretical side of the coin, moduli are very useful! In a theory without free parameters, they are the only surrogate for a physical parameter. As dictated by (local) Lorentz invariance, couplings, masses etc. can in fact depend on moduli, and generally do. Their behavior at the boundary of moduli space often sheds light on a number of aspects of the string landscape~\cite{Ooguri:2006in}, and it can be connected to bottom-up considerations~\cite{Stout:2022phm}.

An alternative approach is to study the worldsheet \ac{CFT} in general, without specifying it explicitly. This allows proving some general facts about the landscape, or at least one of its corners~\cite{Banks:2010zn, Angelantonj:2023egh, Heidenreich:2024dmr, Aoufia:2024awo}. An example related to the emergence of geometry is that when the \ac{CFT} has a conformal manifold with points at infinity, it can be described by a compactification approaching them~\cite{Aoufia:2024awo, Ooguri:2024ofs}.

\subsection{Strings at low energies}\label{sec:part_iii}

Having discussed the construction of string perturbation theory and its vacua in detail, we can not turn to the first big question for \ac{QG} \emph{aficionados}: since \ac{ST} entails the presence of interacting gravitons, \emph{how does it reduce to gravitational \ac{EFT}}, as it must? How do background fields on the worldsheet connect to dynamical degrees of freedom at low energies? In order to answer this, let's piggyback on the discussion on vertex operators in \cref{sec:spectra}, which led us to gravitons from deformations of the background geometry. In order to see that background fields are affected, we can reverse the logic behind \eqref{eq:graviton_vertex_ws}. Namely, we can observe that computing physical quantities with some worldsheet path integral around a \emph{coherent state}, say of gravitons, is equivalent to computing that quantity in the vacuum of a \emph{deformed background}! Having developed the formalism and notation, it is a one-line argument. In canonical quantization, coherent states are exponentiated creation operators. Correspondingly, in the path integral description we exponentiate vertex operators. If a vertex operator $\mathcal{V}$ arises deforming the worldsheet action, $\delta S = \int \rmd^2 \sigma \, \sqrt{-\gamma} \, \mathcal{V}$, any path integral on a coherent state takes the form
\begin{eqaed}\label{eq:coherent_states}
    \int \mathcal{D}(\dots) \, \underbrace{ e^{-\int \rmd^2 \sigma \, \sqrt{-\gamma} \, \mathcal{V}} \, e^{-S^E_\text{ws}[\text{background}]} }_{e^{-S^E_\text{ws}[\text{deformed background}]}} \left( \text{other stuff} \right) .
\end{eqaed}
A small deformation of this type can be iterated, replacing the new vacuum with a coherent states of the deformed excitations. So we learn about the direct connection between (some) states and background data, which shows that the latter is \emph{always dynamical}. From the perspective of the spacetime \ac{QG} theory, they are configurations of the same theory! This means that the theory is indeed background independent, as expected from a theory of gravity. This reasoning can be pushed further in various directions, \eg{} when some quantities are protected from quantum corrections, or in some simple settings~\cite{Iqbal:2003ds, Eberhardt:2020bgq, Eberhardt:2021jvj} which show explicitly that this independence extends to spacetime topology, as expected from bottom-up considerations~\cite{McNamara:2019rup, McNamaraThesis, Casadio:2022ozp}. We can further corroborate this conclusion deriving a perfectly manifestly covariant \ac{EFT} with a well-defined Planck scale, which we do in the following. Precisely because all background fields are dynamical, we should be able to take two approaches to this end, and they ought to match. One is to match low-energy scattering amplitudes of strings to amplitudes computed in an \ac{EFT}; this approach uses the states directly. The other approach uses the background fields instead, showing that Weyl invariance requires that they satisfy field equations stemming from an effective action. Classical equations also arise from consistency in the worldline formalism for \ac{QFT}~\cite{Bonezzi:2018box}, but they do not contain stringy corrections (and thus no \ac{UV} completion either). Happily, the results of these two methods do match.

\subsubsection{Method I --- Weyl anomaly cancellation}\label{sec:weyl_cancellation}

Let us begin with the background-oriented approach. On a general background, the worldsheet action looks like \eqref{eq:polyakov_action} with the addition of internal degrees of freedom, if any. We focus on the bosonic metric terms for the type being, although the same conclusion holds for all the other terms, bosonic and fermionic~\cite{Callan:1989nz}. As we mentioned here and there in the preceding sections, in an \ac{NLSM} the notion of coupling constant is replaced by the background curvatures. There are various methods to concoct a perturbative expansion that makes this manifest, using the intrinsic geometric structures of the target space. For instance, applying the background field method \emph{to the worldsheet fields} around a constant solution $X(\sigma) = x + \sqrt{\alpha'} \, Y(\sigma)$, there are judicious parameterizations which involve Taylor expanding geodesic distances on target space (see \eg{}~\cite{Vassilevich:2003xt}). This leads to \emph{Riemann normal coordinates}, which simplify the expansion of the (Wick-rotated) spacetime metric $g$ around $x$ according to
\begin{eqaed}\label{eq:normal_coordinates}
    g_{\mu \nu}(X) \overset{Y \ll 1}{\sim} \delta_{\mu \nu} - \, \frac{\alpha'}{3} \, R_{\mu \rho \nu \sigma} \, Y^{\rho} Y^{\sigma} \, .
\end{eqaed}
The leading deviation from a free worldsheet theory is thus a quartic vertex of the schematic form $\alpha'\text{Riem} \times Y Y \partial Y \partial Y$. This is enough to derive the leading contribution to the renormalization of $g_{\mu \nu}(X)$ seen as a {coupling function} of the worldsheet scalars $X$. The resulting beta function(al)~\cite{Callan:1989nz}
\begin{eqaed}\label{eq:beta_g}
    \beta^{(g)}_{\mu \nu} \overset{\alpha' \text{Riem} \ll 1}{\sim} \alpha' \, R_{\mu \nu}
\end{eqaed}
is proportional to the \emph{Ricci tensor} of the metric. Indeed, we could have guessed it just by the tensorial structure of the various quantities and dimensional analysis. On a flat gauge-fixed worldsheet, this beta function(al) \emph{must vanish} in order to restore Weyl invariance. This recovers the Einstein equations in the vacuum, as in the worldline approach; in the case of strings, there is an asymptotic series of $\alpha'$ corrections encoding the extended nature of the string.

This approach can be extended to other background fields and higher orders~\cite{Callan:1989nz}, but it is perhaps more instructive to use another method to do so. The heat kernel expansion~\cite{Vassilevich:2003xt}, together with the geodesic coordinates, allow expanding the action around \emph{any background} $X \mapsto X + Y$, so that the renormalized background fields can be read off directly from the one-loop (in the $\alpha'$ sense) effective action~\cite{Vassilevich:2003xt} without computing Feynman diagrams. The approach based on the heat kernel expansion also handles curved worldsheet in a manifestly covariant way. Although we tried so hard\footnote{And got so far!} to gauge away the worldsheet metric, the dilaton coupling in \eqref{eq:polyakov_action} rears its head: it is impossible to see it directly on a flat worldsheet, but including it properly shows that the Weyl variation of the effective worldsheet action $\Gamma$ is \emph{not} controlled by the usual beta function(al)s $\beta$ governing the \ac{RG}. Rather, there are slightly different function(als) $\widetilde{\beta}$ in the functional Weyl variation of the effective action. In terms of the couplings in \eqref{eq:polyakov_action},
\begin{eqaed}\label{eq:eff_action_worldsheet_weyl}
    \delta_\omega \Gamma \propto \widetilde{\beta}_{\mu \nu}^{(g)} \, \partial X^\mu \cdot \partial X^\nu + \widetilde{\beta}_{\mu \nu}^{(B)} \, \epsilon^{\alpha \beta} \partial_\alpha X^\mu \, \partial_\beta X^\nu + \widetilde{\beta}^{(\phi)} \, \alpha' \text{Ric}(\gamma) \, \phi(X) \, ,
\end{eqaed}
where, at leading order in the curvatures $\mathcal{R}$, namely the gravitational $\text{Riem}$ and Kalb-Ramond $H \equiv \rmd{}B$ curvatures,
\begin{eqaed}\label{eq:beta_functionals_weyl}
    \widetilde{\beta}^{(g)}_{\mu \nu} & \overset{\alpha' \mathcal{R} \ll 1}{\sim} \alpha' R_{\mu \nu} - \frac{\alpha'}{4} \, H_{\mu \rho \sigma} \, {H_\nu}^{\rho \sigma} + 2 \, \alpha' \, \covD_\mu \covD_\nu \phi \, , \\
    \widetilde{\beta}^{(B)}_{\mu \nu} & \overset{\alpha' \mathcal{R} \ll 1}{\sim} - \, \frac{\alpha'}{2} \, \covD^\rho H_{\rho \mu \nu} + \alpha' \, \covD^\rho \phi \, H_{\rho \mu \nu} \, , \\
    \widetilde{\beta}^{(\phi)} & \overset{\alpha' \mathcal{R} \ll 1}{\sim} - \, \frac{\alpha'}{2} \, \covD^2 \phi - \, \frac{\alpha'}{24} \, H^2 + \alpha' \, (\covD \phi)^2 \, .
\end{eqaed}
These expressions differ from what would appear from an ordinary beta function(al) computation by the last term, which involves derivatives of dilaton --- precisely the structure that cannot be seen directly on a flat worldsheet. This structure actually holds to all orders in $\alpha'$~\cite{Tseytlin:1986tt, Tseytlin:1986ws,  Bonezzi:2021mub}, and allows computing the first two Weyl-anomaly coefficients in \eqref{eq:beta_functionals_weyl} to higher orders from ordinary beta function(al) on a flat worldsheet, as recently done in~\cite{Bonezzi:2021mub}. The condition that these quantities vanish\footnote{More precisely, when the first two vanish, the third is proportional to the central charge, due to the general structure of the Weyl anomaly on a curved worldsheet. The (super)ghosts and criticality take care of the rest.} gives the field equations for the background fields, and it turns out that they can be derived by a \emph{spacetime effective action} to all orders in $\alpha'$~\cite{Curci:1986hi, Tseytlin:2006ak, Papadopoulos:2024uvi}, a remarkable fact which is connected to the properties of $\widetilde{\beta}^{(\phi)}$. In terms of these background fields, one obtains the (now Lorentzian, for physical reasons) effective action for the massless fields in the \ac{NS}-\ac{NS} sector,
\begin{eqaed}\label{eq:NS-NS_action}
    S^{\text{NS-NS}}_\text{eff} \overset{\alpha' \mathcal{R} \ll 1}{\underset{e^\phi \ll 1}{\sim}} \frac{M_s^{d-2}}{2} \int \rmd^dx \, \sqrt{-g} \, e^{-2\phi} \left( R + 4 \left(\partial \phi \right)^2 - \, \frac{1}{12} \, H^2 \right) .
\end{eqaed}
Sticks and stones may break my bones, but this action is universal in this regime. In the above expression, it is crucial to involve all the worldsheet degrees of freedom in the central charge, in order to cancel some spurious terms which would ruin the hierarchical expansion in curvatures. Whenever an $\alpha'$-exact description is available, one need not worry about this subtlety, but those settings do not correspond to \emph{tame} spacetimes. The effective action in \eqref{eq:NS-NS_action} is expressed in terms of the natural spacetime metric probed by the string worldsheet, namely in the so-called \emph{string frame}. It is similar to the Jordan frame in scalar-tensor theories like Brans-Dicke. Here the scalar is the dilaton playing the role of local string coupling, since fields in the effective action vary slowly relative to the string scale. Subtracting the asymptotic value $\phi_0 = \ln g_s$, the fluctuation $\widetilde{\phi}$ can be used to pass to the Einstein frame. The Weyl rescaling of the spacetime metric (no problem here!) $g_\text{string} = e^{\frac{4}{d-2} \phi} \, g_\text{Einstein}$ does the trick, not only yielding a proper Einstein term but also showing that the dilaton has a canonical kinetic term with the correct sign. As a by-product, we learn that the Planck scale is given by
\begin{eqaed}\label{eq:planck_scale_string}
    \MPl{} = M_s \, g_s^{- \frac{2}{d-2}} \, .
\end{eqaed}
As promised, at weak coupling the \ac{EFT} cutoff $M_s \ll \MPl{}$ controls stringy corrections to \eqref{eq:NS-NS_action}.

As for other sectors and quantum effects, the second approach to computing spacetime effective actions is much more, well\dots effective.\footnote{In some cases, notably 10-dimensional type IIB superstrings, self-dual fields show up. There are various approaches to formulating their subtle dynamics in the language of Lagrangian field theory, such as in \cite{Hsieh:2020jpj}.} The resulting effective actions are more general than \eqref{eq:NS-NS_action}, but they exhibit a clear structure which we will outline in \cref{sec:stringy_EFTs}.

\subsubsection{Method II --- scattering amplitudes}\label{sec:scattering_amplitudes}

The S-matrix approach is more technical, as you may guess from the worldsheet path integral construction outlined by \eqref{eq:GSO_path_integral}. In \cref{sec:spectra} we discussed how to introduce external states and compute a scattering process. We will go into more details in \cref{sec:S-matrix}, since the high-energy behavior of string scattering is universal and instructive. The low-energy behavior, instead, is quite messy: given some perturbative string amplitude expressed in invariant independent Mandelstam variables\footnote{There are various convenient sets of such independent variables, see \eg{}~\cite{Arkani-Hamed:2023jwn, Arkani-Hamed:2023lbd}, follow-ups and references therein for thorough presentations of this formalism.} $\{s_i\}$, it must be expanded in powers of $\{ \alpha' s_i \}$ and matched with the corresponding computation at the same order from an \ac{EFT}. The latter will contain a bunch of Wilson coefficients, organized in field-redefinition-invariant combinations. Some technical subtleties with this approach are that Wilson coefficients are encoded in the ``analytic'' piece of the expansion, while stuff like effects of massless loops produce non-analytic functions like logarithms. The full expression for the \ac{UV}-complete string amplitude you would start from does not ``know'' how to split the two, giving rise to pesky, but ultimately unphysical, ambiguities in the computation.

\subsubsection*{A warm-up example --- gravitons at tree level}

Lemme give a concrete example. Since you care about \ac{QG} and we have introduced type II superstrings, let us do the simplest non-trivial\footnote{Three-point scattering is completely fixed by symmetries, locality and so on~\cite{Elvang:2013cua}.} one with gravitons. As we shall explore in more detail in \cref{sec:S-matrix}, the two-to-two graviton (reduced) amplitude \emph{at tree level} is given by a (gauge-fixed) integral over the moduli space of the four-punctured spherical worldsheet $\Sigma \simeq S^2 \simeq \mathbb{C}P^1$ seen for convenience as the Riemann sphere, where the external states are encoded inserting graviton vertex operators, schematically
\begin{eqaed}\label{eq:2-2_vertex_expression}
    \mathcal{S}^\text{tree}_{\lambda_1,\lambda_2, \lambda_3, \lambda_4}(p_1, p_2, p_3, p_4) \propto g_s^2 \int_\mathbb{C} \rmd^2z \, \langle \mathcal{V}_{\lambda_1, p_1}(0) \, \mathcal{V}_{\lambda_2, p_2}(1) \, \mathcal{V}_{\lambda_3, p_3}(\widetilde{\infty}) \, \mathcal{V}_{\lambda_4, p_4}(z) \rangle_{\mathbb{C}P^1} \, .
\end{eqaed}
After some work, the reduced amplitude $\mathcal{S} \equiv \text{identity} + i\scatteringamplitude \, (2\pi)^d \, \delta^{(d)}(p_{\text{f}}-p_{\text{i}})$ evaluates to~\cite{Schwarz:1982jn, Green:1999pv},\footnote{Up to a replacement of the kinematic prefactor $\mathbf{K}$ the result is the same for all theories~\cite{Kawai:1985xq}.} 
\begin{eqaed}\label{eq:2-2_graviton_type_II_tree}
    \scatteringamplitude^{\text{string}}_\text{tree} = \mathbf{K} \, \frac{\GN{}}{s t u} \, \frac{\Gamma(1 - \, \frac{\alpha's}{4})\Gamma(1 - \, \frac{\alpha't}{4})\Gamma(1 - \, \frac{\alpha'u}{4})}{\Gamma(1 + \, \frac{\alpha's}{4})\Gamma(1 + \, \frac{\alpha't}{4})\Gamma(1 + \, \frac{\alpha'u}{4})} \, ,
\end{eqaed}
which respects the general dimension-independent kinematic structure of \eqref{eq:UV_amplitude}, as expected. In the heterotic case the expression is the same, up to a modification of the kinematic factor $\mathbf{K}$, as you can check out \eg{} in Green, Schwarz and Witten's book \cite{Green:2012oqa, Green:2012pqa}. The fact that the \ac{UV}-completing function, which we called $C$ in \eqref{eq:UV_amplitude}, does not depend on the spacetime dimension either is due to the fact that it is a tree-level amplitude. This was all necessary for consistency: once we had determined that the Planck mass in \eqref{eq:planck_scale_string} is finite and gravitons exist in the spectrum, the structure had to be fixed by the various symmetries and redundancies in the game. Due to how constrained graviton scattering is, the above expression must reduce to \eqref{eq:graviton_amplitude}; one can immediately see by inspection that the massless poles in \eqref{eq:2-2_graviton_type_II_tree} pertain to graviton exchange, as they must in order to connect with an \ac{EFT}. As for the massive poles, since
\begin{eqaed}\label{eq:weierstrass_product}
    \Gamma(z) = \frac{e^{-\gamma_\text{E} \, z}}{z} \prod_{n>0} \left(1 + \frac{z}{n}\right)^{-1} e^{\frac{z}{n}}
\end{eqaed}
has an infinite tower of poles at non-positive integers, one finds an infinite tower of resonances. The \emph{residues} at the poles, say in $s$, are polynomials in $t$, which signify virtual exchanges of \emph{higher-spin} resonances up to a finite spin per pole. This is a hint of locality, albeit not the usual one arising at low energies: the higher-spin tower is infinite, as we learned studying the string spectrum, but somehow the interactions are ``just local enough'' to be consistent with unitarity and causality. More on that later.

The low-energy expansion is trickier, but doable~\cite{Green:1999pv}. Using the identity
\begin{eqaed}
    \ln \Gamma(1-z) = \gamma_\text{E} \, z + \sum_{n>1} \frac{\zeta(n)}{n} \, z^n \, ,
\end{eqaed}
the \ac{UV}-completing prefactor in \eqref{eq:2-2_graviton_type_II_tree} can be recast into the form
\begin{eqaed}\label{eq:exp_gamma}
    \exp\left(\sum_{n>0} \frac{2\zeta(2n+1)}{2n+1} \left(\frac{\alpha'}{4}\right)^{2n+1} \left(s^{2n+1} + t^{2n+1} + u^{2n+1} \right)\right) \, ,
\end{eqaed}
from which it is a straightforward, albeit tedious, matter to extract the low-energy expansion of the amplitude. The first correction comes from the constant $2\zeta(3)$, which corresponds to a $\text{Riem}^4$ term in the effective action. More generally, you can gleam from \eqref{eq:exp_gamma} that tree-level Wilson coefficients are Riemann zeta values, a mathematical connection which has been explored in the literature. As expected, \ac{ST} fixes the effective action to all orders in the curvatures; this approach can be used for all effective fields whose quanta describe massless states in the theory. Moreover, it can be extended in principle to all orders in the string coupling, capturing genuine \ac{QG} effects, which is probably what you're here for!

\subsubsection*{A cooler example --- one-loop and beyond}

The one-loop result is a bit more complicated, as one would expect. It is sensitive to more details of the theory, such as internal degrees of freedom (\eg{} extra compact dimensions) and contains non-analytic terms which are crucial for unitarity. In the lingo of the community studying these things, they ``unitarize'' scattering. Thus, hoping for a simple expression like \eqref{eq:2-2_graviton_type_II_tree} is misguided. It is more convenient to focus on the Wilson coefficient $\alpha$ of the quartic Riemann term, except now the low-energy expansion amplitude has the additional ``unitarizing'' non-analytic terms arising from massless loops~\cite{Green:1999pv, Guerrieri:2021ivu},
\begin{eqaed}
    \frac{1}{s t u} + \alpha \, \MPl^{-6} + \MPl^{2-d} \, f(s,t,u) + \dots \, ,
\end{eqaed}
which can be derived slightly more easily with a clever application of the optical theorem. Because of this, one must carefully separate the two terms.

For type II \ac{ST} \emph{in 10 dimensions} (some of which can be compactified on tori) this procedure has been carried out in detail in many papers, such as~\cite{Green:1999pv}. From the general structure of (closed-)string perturbation theory, we know that the one-loop term must have a factor of $g_s^2$ relative to the tree-level term. The prefactor can depend on a bunch of stuff: if there is an internal sector with some moduli or discrete values of some field (such as a flux of some ``electric'' field across some internal cycle), it can depend on those. Explicit results are known for toroidal internal spaces as the additional degrees of freedom~\cite{Green:1999pv, Green:2010wi, Blumenhagen:2024ydy}, as well as some of its general properties~\cite{Aoufia:2024awo}. The general story is beautifully connected to the mathematics of automorphic forms and group theory, but the actual expressions can become quite messy. Thus, let me just present the simplest expression; as always, it is the 10-dimensional case. Expressing the Wilson coefficient in Planck units,
\begin{eqaed}\label{eq:alpha_II_one-loop}
    \alpha^{\text{II}}_{10d} \overset{g_s \ll 1}{\sim} \frac{\sqrt{g_s}}{64} \left( \frac{2\zeta(3)}{g_s^2} + \frac{2\pi^2}{3} \right) \, .
\end{eqaed}
Actually, in 10-dimensional type IIA \ac{ST} this is the \emph{exact} value, because of the large amount of spacetime supersymmetry. In type IIB \ac{ST} the exact value is also known, but it is more complicated. Either way, this opens up the possibility of testing the consistency of this result at the non-perturbative level: regardless of the value of the string coupling, \eqref{eq:alpha_II_one-loop} is bounded below by $\alpha_\text{min} \approx 0.1403$, which is pretty much around the allowed value of $\alpha_\text{min}$ from \emph{non-perturbative} S-matrix bootstrap bounds~\cite{Guerrieri:2021ivu}! The story is analogous for type IIB \ac{ST}, as well as 11-dimensional M-theory and nine-dimensional type II superstrings~\cite{Guerrieri:2022sod}. This is a non-trivial consistency check of a genuine \ac{QG} effect.

\subsubsection{Structure of stringy EFTs}\label{sec:stringy_EFTs}

Let's take stock of what we learned. In \eqref{eq:NS-NS_action} we wrote down an effective spacetime action for the \ac{NS}-\ac{NS} sector at tree-level in string perturbation theory. Then we discussed how the low-energy expansion of string amplitudes produces everything else. So what does a general effective action look like? The \ac{UV}-completeness tells us that the various Wilson coefficients are fixed, while the structure of string perturbation theory tells us that, at weak coupling, they take the form of asymptotic series in (even, for closed strings) powers of $g_s$. Because of background independence, which in this context appears in the guise of the coherent states in \eqref{eq:coherent_states}, all occurrences of $g_s$ and dilaton fluctuations $\widetilde{\phi}$ will appear in the dynamical combination $\phi$, the dilaton itself. This means that all the curvature corrections $\mathcal{O}_k(\covD, \mathcal{R}, \dots)$ to the effective action are accompanied by asymptotic series in powers of $e^\phi$. The general pattern in the string frame is
\begin{eqaed}\label{eq:general_EFT_structure}
    M_s^{d-2} \sum_{k>2} \left(c_k^{0}(\varphi) \, e^{-2\phi} + c_k^{(1)}(\varphi) + c_k^{(2)}(\varphi) \, e^{2\phi} + \dots \right) \frac{\mathcal{O}_k(\covD, \mathcal{R}, \dots)}{\UVcutoff(\varphi)^{k-2}} \, ,
\end{eqaed}
where we allowed dependence on additional scalar fields $\varphi$, if any. Quantum fields of any type allowed by Weinberg's soft theorem can show up: some scalars $\{\varphi^i\}$, vectors $\{ A^a \}$ (and sometimes their higher-rank $p$-form cousins $\{ C^m_p \}$), spinors $\{ \psi^\alpha \}$, gravitini and, of course, the graviton. For example, in the more familiar Einstein frame, the general two-derivative structure for the bosonic terms looks like
\begin{eqaed}\label{eq:general_EFT_action}
    \frac{\MPl^{d-2}}{2} \int \rmd^dx \, \sqrt{-g} \left(R - \, \frac{1}{2} \, G_{ij}(\varphi) \, \gaugecovD_\mu \varphi^i \gaugecovD^\mu \varphi^j - V(\varphi) - \, \frac{1}{2} \, f_{ab}(\varphi) \, \text{tr} \, F^a_{\phantom{a}\mu \nu} \, F^{b \mu \nu} \right)
\end{eqaed}
$\gaugecovD{}$ is the gauge-covariant derivative, including gravity and Yang-Mills fields. $G_{ij}(\varphi)$ is some positive-definite matrix entering the kinetic terms plus, sometimes, higher-form terms like 
\begin{eqaed}\label{eq:higher-form_terms}
    - \sum_p \frac{w^{(p)}_{mn}(\varphi)}{2(p+1)!} \, \rmd{}C^m_p \cdot \rmd{}C^n_p
\end{eqaed}
and Chern-Simons terms. Basically, the \emph{pattern} one finds is the most general on \ac{EFT} and symmetry grounds. However, as we emphasized, the \emph{actual} precise field content and couplings are all but generic, and they are \emph{determined} by the chosen string vacuum. In principle, they can be computed with arbitrary accuracy. In the context of geometric compactifications, this is often stated as the fact that particles and interactions at low energies are encoded in the local geometry and global topology of the internal manifold. This is one of the indications that the landscape of allowed \acp{EFT} is discrete or even finite: topological data comes in discrete bits, like Hodge or Betti numbers. Moreover, amongst the classes of manifolds (or \acp{CFT}) that we control, there seems to be no way of dialing up these quantities without bound.\footnote{This is consistent with swampland considerations, which are very effective when combined with (extended) supersymmetry and anomaly cancellation~\cite{Adams:2010zy, Kumar:2010ru, Kim:2019vuc, Tarazi:2021duw}.} Continuous couplings are governed by scalars, but so is the \ac{UV} cutoff $\UVcutoff$. It turns out that, universally, $\frac{\UVcutoff}{\MPl} \to 0$ when dialing these fields to infinity. The \ac{EFT} ceases to be reliable! Not all couplings go in a sensible \ac{EFT}. An example in type IIA \ac{ST} is the \ac{UV} cutoff $\UVcutoff{} \propto \MPl{} \, \alpha^{- \frac{1}{6}}$ encoded in the exact 10-dimensional Wilson coefficient of \eqref{eq:alpha_II_one-loop}, with $g_s$ replaced by $e^\phi$. It vanishes in Planck units as $\phi \to \pm \infty$, and in particular it asymptotes to $M_s \overset{g_s \ll 1}{\sim} g_s^{\frac{1}{4}} \, \MPl{}$ as $\phi \to -\infty$ which is just the weak coupling limit. Once more, if you're tapping out at this point there's a nice boxy lesson to summarize what we learned and take home:

\begin{tcolorbox}
    \emph{The \acp{EFT} arising from string vacua are a tiny subset of the pool of all \acp{EFT}. The couplings and field content cannot be chosen willy-nilly: almost nothing goes!}
\end{tcolorbox}

\subsubsection{Aspects of low-energy physics, aka swampy stuff}\label{sec:swampland_stuff}

If I had more time, I would have given you many more details and examples in the preceding section. The issue is that they rely on several technical notions involved that do not directly relate to how \ac{ST} solves the problem of \ac{QG}, which is the point of this section. So the above lesson in the nice box may understandably leave you somewhat unsatisfied. Even if it is the case that almost nothing goes in the string landscape, numerically there are a lot of known vacua and \acp{EFT} in there. Finding our world, if it is there, is a scaled-up version of a needle-in-a-haystack problem, which likely requires a new conceptual leap in our understanding of the theory. Also, what do we learn, pragmatically, from the notion that almost nothing goes? Both of these considerations motivate seeking \emph{general low-energy properties} of the string landscape, which perhaps could match independently motivated bottom-up arguments which are outside the scope of this section. Here are some general facts about the corner of \ac{ST} we explored, although most considerations are expected to hold beyond its lamppost.

\subsubsection*{String theory has no global symmetries}

This one is easy to state. The way it works for \emph{continuous} global symmetries is that any such symmetry in the spacetime dynamics must be reflected by a global symmetry of the worldsheet \ac{CFT}. And, you know, continuous global symmetries come with Noether currents; say $J , \overline{J}$ in complex coordinates. Since they are locally conserved, they commute with the superconformal algebra; they are primary operators on the worldsheet with conformal weights $(1,0)$ and $(0,1)$ respectively. Then, the combinations
\begin{eqaed}\label{eq:gauge_vertex_ops}
    \epsilon_\mu(p) \, J \, \overline{\partial} X^\mu \, e^{ip \cdot X} \, , \qquad \epsilon_\mu(p) \, \overline{J} \, \partial X^\mu \, e^{ip \cdot X}
\end{eqaed}
give vertex operators for \emph{massless vectors}. The general structure of relativistic \ac{QFT} then implies that the corresponding states are gauge bosons, whose gauge redundancy is dictated by the symmetry group we started from. The bottom line is that \emph{continuous symmetries are gauged in \ac{ST}}. Discrete symmetries are also gauged in all known examples, and there are several compelling arguments for why it should be true. Unfortunately --- and certainly not for lack of trying! --- a proof like the above has not been found in perturbative \ac{ST}; perhaps because discrete redundancies are characteristically non-perturbative. For instance, they carry no local anomalies. On the flip side, a non-perturbative proof in the context of the \ac{AdSCFT} correspondence was presented in~\cite{Harlow:2018jwu, Harlow:2018tng}.

At the level of this discussion, this result is conceptually deep but not exceptionally constraining in practice: at low energies, \acp{EFT} can exhibit approximate/accidental global symmetries, which are broken by \ac{UV} effects. This seems to be the case in the \ac{SM}, for instance. The best we can say in general is that the breaking effects we expect from quantum-gravitational physics are at worst suppressed as instanton-like contributions
\begin{eqaed}
    \exp\left(- \, \text{const.} \times \left( \frac{\UVcutoff}{E}\right)^{k > 0} \right)
\end{eqaed}
at low (invariant) energies $E$ we probe.

\subsubsection*{Gravity is the weakest force}

As we just saw, continuous symmetries in \ac{ST} are gauged. The simplest example is an Abelian gauge redundancy. Since all gauge groups in \ac{ST} are compact (they better be, lest violating the Bekenstein-Hawking bound or charge completeness~\cite{Banks:2010zn}), the options are simply toroidal groups $G = U(1)^n$. Let's just pick $n=1$ for concreteness, without much loss of generality. Reversing the logic of the preceding argument, the worldsheet sees this as a $U(1)$ global symmetry in the \emph{internal sector} of the \ac{CFT}, and there are \emph{towers} of charged states which correspond to towers of charged particles in the spacetime theory. The existence of this tower can be proven using a torus partition function like in \eqref{eq:torus_partition_function}, but refined with a further chemical potential for the conserved charge. The resulting ``flavored'' partition function is not (and does not need to be) quite modular-\emph{invariant}, but its transformation properties allow proving various facts about the spectrum~\cite{Montero:2016tif, Heidenreich:2016aqi, Heidenreich:2024dmr}.

For example, it can be used to shown a form of \emph{spectral flow}, an automorphism on states leaving the spectrum invariant. In this case, the spectrum of the internal \ac{CFT} states with charges $(Q_\text{L} , Q_\text{R})$ is invariant under charge shifts $Q \mapsto Q + n$ according to
\begin{eqaed}\label{eq:spectral_flow}
    h_\text{int} & \mapsto h_\text{int} + \frac{1}{2} \left(Q_\text{L} + n_\text{L} \right)^2 - \, \frac{1}{2} \, Q_\text{L}^2 \, , \\
    \overline{h}_\text{int} & \mapsto \overline{h}_\text{int} + \frac{1}{2} \left(Q_\text{R} + n_\text{R} \right)^2 - \, \frac{1}{2} \, Q_\text{R}^2 \, .
\end{eqaed}
Therefore, starting from the universal graviton we get a tower of charged states with weights scaling as charges squared. The graviton has zero charges, which means that the spectrally flowed states have
\begin{eqaed}\label{eq:graviton_flow}
    h_\text{int} = h_\text{graviton} + \frac{1}{2} \, Q_\text{L}^2 \, , \qquad \overline{h}_\text{int} = \overline{h}_\text{graviton} + \frac{1}{2} \, Q_\text{R}^2 \, .
\end{eqaed}
Here the conformal weights $h_\text{graviton} , \overline{h}_\text{graviton}$ pertain to the \emph{internal \ac{CFT} contribution}, and thus vanish for the graviton. The combined conformal weights of these charged states need not be equal, because level matching applies to the \emph{total} conformal weight, not just the internal contribution. Still, $h_\text{int} - \overline{h}_\text{int} \in \mathbb{Z}$ by modular invariance, which means that any difference in total conformal weight can be saturated by including a suitable number of string excitations. Notice that this is only possible in $d>2$ where some physical modes of this type survive the gauge fixing --- it checks out since gravitons only exist in $d > 3$! As wisely put by Tong, this little graviton is seriously high-maintenance, but it is worth the effort for all the mileage we are able to squeeze out of it. All in all, including the contribution of the (super)ghosts to the structure of the vertex operators discussed in \cref{sec:spectra}, from \eqref{eq:string_spectrum} we find states with charges $(Q_\text{L}, Q_\text{R})$ and mass
\begin{eqaed}\label{eq:wgc_kinda}
    \frac{\alpha'}{4} \, m^2 \leq \frac{1}{2} \max \{ Q_\text{L}^2 , Q_\text{R}^2 \} .
\end{eqaed}
What do we learn from the bound in \eqref{eq:wgc_kinda}? It kinda looks like the self-gravitational force of these states is bounded by their self-repulsive electrostatic force. To actually see that this is the case takes a bit more work, in order to extract the actual self-force from string amplitudes~\cite{Heidenreich:2024dmr}. This idea sounds like gravity must be the weakest force, which is surely not something generic from the \ac{EFT} viewpoint. Why is this physically relevant for \ac{QG}? The reason is that the existence of such states is instrumental in allowing quasi-extremal charged \acp{BH} to decay without getting \href{https://physics.stackexchange.com/a/13800}{``constipated''}~\cite{Arkani-Hamed:2006emk}. Less prosaically, if a huge \ac{BH} of charge $Q \gg 1$ and mass $M \geq \kappa \, Q \, \MPl{} \gg \MPl{}$ respecting the extremality bound $\kappa = \orderneglected(1)$ starts radiating particles of mass $m$ and charge $q$, if all states in the spectrum are sufficiently massive relative to their charge the \ac{BH} would end up violating the extremality bound (thus producing naked singularities) or not being able to decay at all~\cite{Arkani-Hamed:2006emk} leaving thermodynamically problematic remnants~\cite{Susskind:1995da}. As explained in~\cite{Palti:2019pca}, the problems with \emph{charged} remnants are less severe with respect to the case of global symmetries discussed in the preceding section. 

Still, this ``weak gravity'' property holds for (super)strings, at least in the guise of \eqref{eq:wgc_kinda}. Although not apparent by simply staring at \eqref{eq:wgc_kinda}, the existence of such states is equivalent to solving the above issue with \ac{BH} physics~\cite{Heidenreich:2016aqi, Heidenreich:2024dmr}. In the broader context of \ac{QG}, this sharper avatar of the notion that gravity be the ``weakest force'' is known as the \emph{weak gravity conjecture}~\cite{Arkani-Hamed:2006emk, Harlow:2022ich}. Including the subtleties related to self-forces and extremality bound, an explicit proof was presented~\cite{Heidenreich:2024dmr} for bosonic strings in $d > 5$, and is being written down for (at least the \ac{NS}-\ac{NS} sector of) superstrings. You may also recall that, in our universe, the electron satisfies the bound with room to spare. From a purely \ac{EFT} standpoint, it didn't seem to \emph{have to} on consistency grounds --- the hallmark of a swampland condition.

\subsubsection*{There are no weakly interacting dS vacua}

This one looks intimidating, but it is going to be very short, since it involves several \ac{CFT} technicalities that we have no time to introduce. The main statement, presented in~\cite{Kutasov:2015eba}, is that closed superstrings have no \ac{dS} vacua at tree level in string perturbation theory, at least without \ac{R}-\ac{R} backgrounds. Translating this into a sharp statement about the worldsheet \ac{CFT} has to do with how (Wick-rotated) \ac{dS} isometries are implemented on the worldsheet, and how they play with unitarity. The proof involves a clever use of the properties of \emph{Kac-Moody algebras} and \href{https://conf.itp.phys.ethz.ch/esi-school/Lecture_notes/WZW%20models.pdf}{\emph{Wess-Zumino-Witten models}}, which we conveniently avoided talking about thus far. The remarkable aspect of this result is that it is exact in $\alpha'$, meaning that it holds no matter what the curvature is (as long as it does not scale with the string coupling $g_s$, which does not show up in this tree-level analysis). The limitations of not handling \ac{R}-\ac{R} backgrounds can be circumvented by other no-go theorems against the existence of \ac{dS} vacua~\cite{Maldacena:2000mw, Basile:2020mpt}, which however are complementarily limited to the leading-order \ac{EFT}.

Although it goes way beyond the scope of this section, the study of \ac{dS} constructions in \ac{ST}, involving a vast array of (non-)perturbative ingredients, is one of the most active fields in string phenomenology. It is subject of heated debate, with arguments~\cite{Kachru:2003aw}, counter-arguments~\cite{Lust:2022lfc, Bena:2023sks} and reviews~\cite{VanRiet:2023pnx}. Regardless of its outcome, it has led to many tour-de-force endeavors~\cite{McAllister:2024lnt} and the development of powerful new \href{https://cy.tools/}{computational machinery}. The whole thing about \ac{dS} is that it kinda looks like a good model for late-time (and inflationary) cosmology, but has several theoretical issues~\cite{Banks:2001yp, Dyson:2002nt, Banks:2010tj, Dvali:2017eba, Bedroya:2022tbh, Witten:2023qsv, Witten:2023xze, Banks:2024lvl} as we briefly mentioned in a long-winded footnote about vacua in \cref{sec:what}. Outside of the perturbative (or, more generally, parametrically controlled) corner of the string landscape, metastable \ac{dS} vacua could exist; the real theoretical issues deal with would-be eternal \ac{dS}. Whether we will need this for our cosmology in the long run remains to be seen; perhaps the recent spectroscopic measurements of DESI will shed some light on this issue. What we are saying here is that, at weak coupling, not even metastable \ac{dS} vacua arise.

\subsubsection*{Geometry emerges from conformal field theory}

Finally, let me present a short summary of some recent work I contributed to in~\cite{Aoufia:2024awo}. On the one hand, so far I tried emphasizing that extra dimensions, as far as we understand the theory, are \emph{not} a universal property of the string landscape. Even at the level of the worldsheet formalism, abstract \ac{CFT} extends and generalizes Riemannian geometry consistently with the expectations of an emergent spacetime in \ac{QG}~\cite{Banks:2010tj}. On the other hand, compactifications comprise the majority of constructions we understand. What gives? Of course, it could just be that we do not really know a lot about the landscape. But there are indications, both from the top down~\cite{Lee:2019xtm} and the bottom up~\cite{Basile:2023blg, Bedroya:2024ubj} that whatever additional degrees of freedom are present for $d < 10$ rearrange into extra dimensions \emph{in certain limits}. In the language of the worldsheet \ac{CFT}, limits in the space of \acp{CFT} in which the spectral gap of the internal sector vanishes. In plain English, from \eqref{eq:string_spectrum} this means that the scale $m_\text{gap} \ll \MPl$ at which the new physics effects \emph{coming from the internal sector} becomes small in Planck units; it drives the \ac{UV} cutoff $\UVcutoff{} \ll \MPl{}$ (both of the gravitational and non-gravitational sectors) to zero in Planck units. In fact, it is enough to assume the latter condition on the internal degrees of freedom~\cite{Aoufia:2024awo}. Assuming we're still talking about weakly coupled strings, the relationship between $m_\text{gap}$ and the gravitational \ac{UV} cutoff $\UVcutoff$ can be extracted from the same $\text{Riem}^4$ Wilson coefficient $\alpha$ we discussed in \cref{sec:scattering_amplitudes}. The corresponding cutoff scale is $\UVcutoff{} \propto \MPl{} \, \alpha^{- \frac{1}{6}}$, or at least a proxy thereof --- barring miraculous fine tunings~\cite{Heckman:2019bzm}, they should be parametrically identical~\cite{vandeHeisteeg:2023dlw}. In~\cite{Aoufia:2024awo} we were able to exploit modular invariance to find the universal scaling
\begin{eqaed}\label{eq:wilson_coeff_general}
    \alpha \overset{m_\text{gap} \ll \MPl{}}{\sim} \left( \frac{\MPl{}}{M_s} \right)^{8-d} \left(\frac{m_\text{gap}}{M_s}\right)^{-\hat{c}} \, ,
\end{eqaed}
where $\hat{c}$ is the ``reduced''~\cite{Afkhami-Jeddi:2020ezh, Abel:2024twz} central charge of the piece of the internal \ac{CFT} which undergoes the limit. (There could be another unaffected sector with string-sized gap). As a result, carefully taking into account how the Planck scale and string scale relate to each other, namely $M_s^{d+\hat{c}-2} = \MPl^{d-2} \, m_\text{gap}^{\hat{c}}$, when the dust settles
\begin{eqaed}\label{eq:UV_cutoff_general}
    \UVcutoff{} \overset{m_\text{gap} \ll \MPl{}}{\sim} \MPl{} \left(\frac{m_\text{gap}}{\MPl{}} \right)^{\frac{\hat{c}}{d+\hat{c}-2}} ,
\end{eqaed}
which is precisely the scaling arising from a geometric compactification in the large-volume limit~\cite{Basile:2023blg, Castellano:2023aum, Basile:2024dqq, Bedroya:2024ubj, Aoufia:2024awo}. What's more, this gap is accompanied by a light tower of states whenever $\UVcutoff \ll \MPl{}$~\cite{Ooguri:2024ofs, Aoufia:2024awo} (and thus gravitons are weakly coupled), which also implies $m_\text{gap} \ll \MPl{}$ if we had not started there, due to \eqref{eq:species_scale} and \eqref{eq:species_hierarchy}. The asymptotic tower of states looks like a Kaluza-Klein tower and the limiting worldsheet \ac{CFT} turns out to contain an \ac{NLSM} on $\mathbb{R}^{\hat{c}}$~\cite{Ooguri:2024ofs}, signifying the emergence of extra dimensions! This means that, although \emph{in general} \ac{ST} does not predict the number of dimensions, \emph{whenever the new physics pertaining to the internal degrees of freedom has a small \ac{UV} cutoff} they rearrange into extra dimensions.

Lemme summarize with a nice box what we learned to conclude this section:

\begin{tcolorbox}
    \emph{Some general properties of the string landscape:}

    \begin{itemize}
        \item \emph{There are no global symmetries. Just gauge redundancies.}
        \item \emph{Gravity is the weakest force. Abelian charges bring along light states.}
        \item \emph{There are no weakly interacting \ac{dS} vacua.}
        \item \emph{Limits of the internal worldsheet \ac{CFT} ``quack like geometry''.}
    \end{itemize}
\end{tcolorbox}

\subsection{Strings at high energies}\label{sec:part_iv}

In this final part of the section, we move on to study some aspects of the high-energy physics of \ac{ST}. As we outlined way back, it is the natural regime of stringy physics, where the messy details of \ac{IR} physics are washed out and the soft \ac{UV} behavior kicks in. A recurring theme in this section is that the S-matrix is the only well-defined observable in (asymptotically flat sectors of) \ac{QG}. String perturbation theory forces us to realize this due to Weyl invariance, placing vertex operators exclusively as external states in the path integral. Alternatively, moving closer to the standard definition of scattering amplitude, one can recast the description in terms of vertex operators in the language of states and inner products, as befits the state-operator correspondence. More details can be found in Chapter 6 of Polchinski's book \cite{Polchinski:1998rq}.

If you are not convinced by the necessity of the S-matrix at this point, you can try to poke the metaphorical bear. What happens is that it will show you its claws: if we tried defining some sort of local probe of the worldsheet of the form $\delta^{(d)}(X(\sigma) - x)$, or a bi-local worldsheet interaction of the form $\delta^{(d)}(X(\sigma)) - X(\sigma'))$, Weyl invariance would be lost: a Fourier representation of these expressions show that it includes all momenta in their would-be vertex operator, violating Weyl invariance. Once more, this is perfectly in line with what we would expect from a theory of gravity. Speaking of which, another recurring theme is that gravity is \ac{UV}/\ac{IR}-mixed; in the preceding section we saw some consequences of this, but we will present some other aspects related to \acp{BH}. Quite poetically, our journey will end where it first began.

\subsubsection{The string S-matrix}\label{sec:S-matrix}

Along the way, we collected all the required ingredients to construct the string S-matrix: gauge fixing, ghosts, zero-modes, Riemann surfaces and their moduli spaces, all of which has a worldsheet-supersymmetric counterpart for superstrings. We learned that external states, and only external states, are encoded in vertex operators inserted on the worldsheet. Finally, the prefactor for the correct S-matrix element is fixed by unitarity, as explained \eg{} in Polchinski's book \cite{Polchinski:1998rq, Polchinski:1998rr} --- we will not care about it, but it is fixed. Also, in order not to excessively overload some already pretty heavy notation, in the ensuing presentation we will not explicitly mention superghosts, odd moduli and so on, although they are implicitly there. The examples in \cref{sec:examplitudes} are not sensitive to the difference anyway.

The path-integral expression for the perturbative scattering amplitude involving $n$ definite-momentum states with polarizations $\{\lambda_i\}$ and momenta $\{ p_i \}$ then takes a form like \eqref{eq:GSO_path_integral}. For notational convenience we suppress the sum over spin structures, but we render explicit that the vertex operators add \emph{punctures} to the worldsheet. Its moduli space reflects this,
\begin{eqaed}\label{eq:moduli_space_gn}
    \mathcal{M}_{g,n} \equiv \frac{(\text{metrics on } \Sigma_g) \times \Sigma_g^n}{\text{Diff} \times \text{Weyl}} \, ,
\end{eqaed}
since we can insert vertex operators wherever on the worldsheet $\Sigma$. Indeed, it is equivalent to replace ghost insertions by an integral over the worldsheet positions, as in \eqref{eq:integrated_vertex_op}. In the quotient, the mapping class group of \eqref{eq:MCG} shows up, along small diffeomorphisms and Weyl rescalings, according to the schematic rearrangement $\text{Diff} \times \text{Weyl} \simeq \text{MCG} \times \text{Diff}_0 \times \text{Weyl}$. The important subgroup
\begin{eqaed}\label{eq:CKG}
    \text{CKG} \subset \text{Diff}_0 \times \text{Weyl} \, ,
\end{eqaed}
the \ac{CKG} generated by \acp{CKV} (as in \cref{sec:how}), preserves the metric, in the sense that the vector fields generate diffeomorphisms which are undone by a Weyl rescaling. Since they are zero-modes $c_0$ of the kinetic operator acting on $c$ ghosts, the corresponding path integral over $c_0$ cannot be treated with a functional determinant; instead, one can use this gauge freedom to \emph{fix some insertions to given positions on the worldsheet}. How many can we fix? Well, it turns out that the Riemann-Roch theorem (or, in physics jargon, the integrated anomaly of the ghost number) dictates that
\begin{eqaed}\label{eq:riemann-roch}
    \# \text{moduli} - \, \# \acp{CKV} = - \, 3 \, \chi(\Sigma_g) = 6g - 6 \, . 
\end{eqaed}
Furthermore, an ``energy functional'' argument (given \eg{} in Polchinski's book \cite{Polchinski:1998rq, Polchinski:1998rr}) shows that there are no moduli for $\chi(\Sigma_g) > 0$ and there are no \acp{CKV} for $\chi(\Sigma_g) < 0$. So if we have large enough $n$ we can fix all the gauge freedom provided by \acp{CKV}. We can either choose convenient numerical values of the coordinates, or use parameters which disappear from the amplitude once the Faddeev-Popov factor is included. Otherwise, it is often said that result would vanish; see, however, \cite{Erbin:2019uiz} for an alternative approach for the case $n=2$. With this proviso, letting $\mathcal{V}_\lambda(p)$ collectively denote both integrated and gauge-fixed vertex operators, the perturbative expression for the S-matrix amplitude reads
\begin{eqaed}\label{eq:S-matrix_formula}
    \mathcal{S}_{\lambda_i \, \dots \, \lambda_n}(p_1, \dots , p_n) \overset{g_s \ll 1}{\sim} \sum_{\text{genus } g} g_s^{2g-2} \int_{\mathcal{M}_{g,n}(t)} \rmd\mu(t) \, \langle \prod_{i=1}^n \mathcal{V}_{\lambda_i}(p_i) \rangle_{\Sigma_g(t)} \, ,
\end{eqaed}
where the (gauge-fixed) path integral over the worldsheet fields on $\Sigma_g(t)$ is expressed by the (connected) correlator. As premised at the beginning of this discussion, in the case of superstrings additional superghost contributions, and related odd moduli of super-Riemann surfaces, need to be taken into account.

\subsubsection{Examplitudes --- Veneziano and Virasoro-Shapiro}\label{sec:examplitudes}

\begin{figure}[ht!]
    \centering
    \includegraphics[scale=0.6]{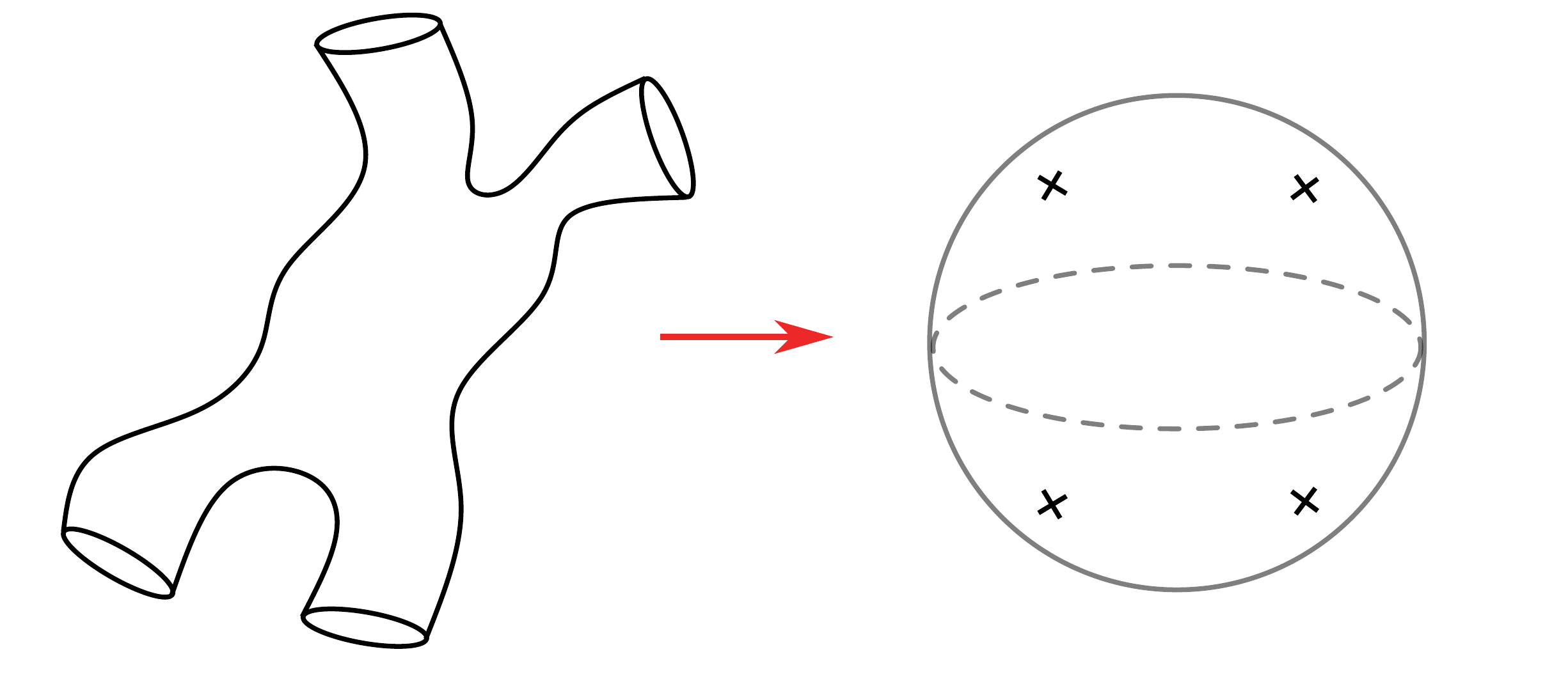}
    \caption{Weyl invariance allows treating a four-point scattering process in terms of a worldsheet with the topology of a sphere, where the four external tubes extended to infinity are replaced by vertex operators inserted at punctures due to the state-operator correspondence. Then, the local insertions can be replaced by integrated vertex operators as in \eqref{eq:integrated_vertex_op}~\cite{Witten:2012bh}.}
    \label{fig:four-point}
\end{figure}

As an example of closed-string scattering, the obvious starting point is the simplest non-trivial tree-level amplitude, namely that of two particles scattering into two particles. This is described by a worldsheet $\Sigma \simeq S^2$ with the topology of a sphere with four punctures representing the insertions as depicted in \cref{fig:four-point}. Since $\chi(S^2) = 2$, there are no moduli\footnote{You may think that the radius is a modulus, but it isn't because it changes the curvature. Indeed, a careful extension of the derivation in \cref{sec:how} to the worldsheet shows that Weyl invariance removes the trace part of metric deformations, namely the overall volume. Correspondingly, the $b$ ghost is a symmetric traceless tensor.} and we can fix six real coordinates of vertex operators, namely three out of the four positions. The proper normalization due to unitarity involves a factor $g_s^n$, so that the total prefactor is proportional to $g_s^2$ for $n=4$. This is precisely in line with what we expect from \eqref{eq:NS-NS_action} and \eqref{eq:planck_scale_string}, since $g_s^2 \propto \GN{}$. After gauge fixing to a flat worldsheet, the resulting correlator in \eqref{eq:S-matrix_formula} is doable, since the two-dimensional theory is non-interacting! This is an echo of \cref{fig:worldsheet}, in which interactions are completely determined by worldsheet topology.

Since Gaussian integrals are mostly straightforward, I'll tell you about a subtlety which gives most of the result, and then modify it accordingly at the end in order to avoid the annoying combinatorics of Wick-like contractions. Just as a teaser of how the correct form of the S-matrix elements shows up, let us consider only the pieces of vertex operators encoding the spacetime momentum of the states, $e^{ip \cdot X}$. The free correlator needs separating the zero-modes $X_0$ from $X$, since the kinetic term of the action vanishes when evaluated on the zero-modes. Then, the free correlator can be evaluated introducing a source $J(z) = \sum_k \, p_k \, \delta^{(2)}(z - z_k)$ according to
\begin{eqaed}\label{eq:exp_correlator}
    \langle \prod_{k=1}^n e^{ip_k \cdot X(\sigma_k)} \rangle & = \langle e^{i \sum_{k=1}^n p_k \cdot X_0} \rangle_\text{zero-modes} \, \langle e^{i \int \rmd^2z \, J \cdot X} \rangle_\text{no zero-modes} \\
    & = \int \rmd^dX_0 \, e^{i \sum_{k=1}^n p_k \cdot X_0} \, \langle e^{i \int \rmd^2z \, J \cdot X} \rangle_\text{no zero-modes} \\
    & = (2\pi)^d \, \delta^{(d)}\left(\sum_{k=1}^n p_k \right) \langle e^{i \int \rmd^2z \, J \cdot X} \rangle_\text{no zero-modes} \, .
\end{eqaed}
The non-exponential terms in vertex operators, such as $\partial X$, can also be addressed with the same technique, as detailed \eg{} in Polchinski's book \cite{Polchinski:1998rq, Polchinski:1998rr}. As for the remaining correlator, there is a prefactor given by the Laplacian determinant, as usual in Gaussian integrals. Completing the square in the exponent, one is left with a sourceless Gaussian integral (easily done) times a source-dependent factor
\begin{eqaed}\label{eq:source_term}
    \exp\left( \frac{\pi \alpha'}{2} \int \rmd^2z \, J \cdot (\propG \star J) \right) & = \exp\left( \frac{\alpha'}{4} \sum_{i,j=1}^n p_i \cdot p_j \, \ln\abs{z_i - z_j}^2 \right) \\
    & = \prod_{i < j} \abs{z_i - z_j}^{\alpha' p_i \cdot p_j} \, ,
\end{eqaed}
with $\propG(z, \overline{z}) \equiv \frac{1}{2\pi} \ln \abs{z}^2$ the scalar propagator convolved with $J$. This is the so-called \emph{Koba-Nielsen factor}.\footnote{This expression shows up everywhere in \ac{ST}, and as we shall see shortly it drives its high-energy behavior. In hindsight, it is perhaps not terribly surprising that it appears in modern approaches to \ac{QFT} amplitudes as well, through \emph{tropicalization}. See \eg{}~\cite{Arkani-Hamed:2023lbd} for a recent application of this idea, together with others inspired from string scattering.} Working out the contributions of non-exponential terms such as $\partial X$, they end up shifting the exponents. Because of their tensorial structure, they also bring along kinematic factors like the one in \eqref{eq:UV_amplitude}, encoding external polarizations. Let us consider the four-graviton amplitude, which takes the universal form in \eqref{eq:UV_amplitude} for bosonic, \ac{RNS} and heterotic strings up to replacing the kinematic factor. Hence, we can focus on the much simpler bosonic vertex operators.\footnote{Superstring amplitudes of this type are complicated by a subtlety due to superghosts, in this context often referred to as \emph{picture-changing operators}.} In this case, it is easy to guess that the exponents get shifted by $-2$ relative to the Koba-Nielsen factor in \eqref{eq:source_term}, since each factor $\partial X \overline{\partial} X$ contributes to the path integral by derivatives $\partial \propG \, \overline{\partial} \propG = \abs{z}^{-2}$. The actual computation is more complicated than that, since \eg{} polarizations also come into play, but this dirty trick gets us to the right answer.

For $n=4$, gauge-fixing three out of four positions, say to $z_{1,2,3}=0, 1, \widetilde{\infty}$ on the standard complex chart on the sphere viewed as the complex projective line $\mathbb{C}P^1$ (or Riemann sphere), the factors which do not depend on $z_4 \equiv z$ cancel due to the ghost functional integral, an avatar of Weyl invariance. Only powers of $\abs{z}$ and $\abs{1-z}$ remain, allowing an explicit evaluation of the integral. For gravitons, the shifted exponents rearrange into the Mandelstam combinations\footnote{The general expression includes factors of $z$ and $1-z$ whose effect can be absorbed in kinematic factors.}
\begin{eqaed}\label{eq:koba_nielsen}
    \abs{z}^{\alpha' p_1 \cdot p_4 - 2} \, \abs{1-z}^{\alpha' p_2 \cdot p_4 - 2} = \abs{z}^{-\frac{\alpha'u}{2} - 2} \, \abs{1-z}^{-\frac{\alpha't}{2} - 2} \, .
\end{eqaed}
Using the integral\footnote{As shown \eg{} in Green, Schwarz and Witten's book \cite{Green:2012oqa, Green:2012pqa} or Tong's notes \cite{Tong:2009np}, it can be evaluated using the Schwinger representation for each factor in the integral, evaluating the resulting Gaussian integral, and finally changing variables to reduce to the Euler beta function.}
\begin{eqaed}\label{eq:integral_trick}
    \int_\mathbb{C} \rmd^2z \, \abs{z}^{2a-2} \, \abs{1-z}^{2b-2} = 2\pi \, \frac{\Gamma(a)\Gamma(b)\Gamma(1-a-b)}{\Gamma(1-a)\Gamma(1-b)\Gamma(a+b)} \, ,
\end{eqaed}
the reduced amplitude takes precisely the form of \eqref{eq:2-2_graviton_type_II_tree} (upon taking the kinematic factor into account). The simpler case of tachyonic four-point scattering, yielding the \emph{Virasoro-Shapiro amplitude}, has the same ``$\frac{\Gamma\Gamma\Gamma}{\Gamma\Gamma\Gamma}$'' schematic structure with slightly different arguments. As such, \eqref{eq:2-2_graviton_type_II_tree} is also called with the same name in some of the literature~\cite{Arkani-Hamed:2020blm, Cheung:2023adk}.

There is a similar story (and nomenclature quirks) for the even simpler amplitude for \emph{open strings}, giving rise to gauge bosons instead of gravitons. The resulting (color-ordered) \emph{Ve\-ne\-zia\-no amplitude}, stripped of color and kinematic prefactors, reads
\begin{eqaed}\label{eq:veneziano}
    \scatteringamplitude_\text{tree}^\text{open} \propto \frac{\Gamma(- \, \frac{\alpha' s}{4})\Gamma(- \, \frac{\alpha' t}{4})}{\Gamma(1 - \, \frac{\alpha'(s+t)}{4})} \, .
\end{eqaed}
This amplitude and its cousins are actually the precursors of \ac{ST} as we know it. They were found by attempting to bootstrap hadron scattering before the advent of \ac{QCD}. It is thus particularly amusing to note that, although unitarity was indirectly proven via the no-ghost theorem, a direct understanding has only begun to surface in recent years~\cite{Arkani-Hamed:2022gsa}. The simpler structure of \eqref{eq:veneziano}, despite not containing gravity directly, secretly contains the information in \eqref{eq:2-2_graviton_type_II_tree} due to a concrete incarnation of open-closed string duality, namely the \emph{Kawai-Levellen-Tye relations} (``KLT'')~\cite{Kawai:1985xq}. These remarkable identities are the inspiration for the field-theoretic ideas of ``double copy''~\cite{Adamo:2022dcm}, and are being developed beyond tree level~\cite{Stieberger:2022lss}. Without delving into these fascinating topics, already from \eqref{eq:veneziano} and \eqref{eq:2-2_graviton_type_II_tree} one can see by inspection that there is an infinite tower of resonances reproducing the string spectrum. The residues at $s = \text{pole}$ are polynomials in $t$ (recall $s+t+u=0$ here), a smoking gun of higher-spin exchanges. The poles exhibit \emph{dual resonance}: the amplitude can be equally well expanded in terms of the $s$-channel or the $t$-channel. Pictorially it makes perfect sense --- \cref{fig:worldsheet} shows how different Feynman diagrams correspond to the same ``string diagram''.

\subsubsection{Scattering strings very hard}\label{sec:gross_mende_ooguri}

With the above background, we are finally ready to step away from \ac{EFT} and take a high-energy limit. The ideal setting is four-point scattering with high center-of-mass energy at fixed angle,
\begin{eqaed}\label{eq:regime}
    \alpha' s \gg 1 \, , \qquad s/t \text{ fixed} \, .
\end{eqaed}
This is the hard scattering regime which will turn out to reflect the expected \ac{UV}/\ac{IR} properties of \ac{BH} dominance which we have described in the preceding sections.

This limit can be applied to \eqref{eq:2-2_graviton_type_II_tree} using Stirling's asymptotic approximation. As for the density of states in \eqref{eq:density_mass}, we focus on the exponential scaling; to this end, let us define $\scatteringamplitude \equiv \text{kinematic factors} \times F(s,t,u)$, as in \eqref{eq:graviton_amplitude}. It is convenient to express everything in string units $\alpha' = 1$ (mind that much literature used $\alpha' = \frac{1}{2}$ instead). Then,
\begin{eqaed}\label{eq:virasoro-shapiro_hard}
    \ln F_\text{tree} \overset{\text{hard}}{\sim} - \, \frac{1}{2} \left(s \ln s + t \ln t + u \ln u \right) ,
\end{eqaed}
which is an extremely soft \ac{UV} behavior, unlike anything expected from \ac{QFT}. Indeed, the Martin-Cerulus bound $F \gtrsim e^{- \text{const.} \times \sqrt{s} \ln s}$ appears to be violated, possibly due to non-locality \cite{Gross:1987kza}. If that were the case, it would be a very subtle type of non-locality, qualitatively different from a garden-variety non-local deformation of a relativistic \ac{QFT} which is very likely to induce causality violations (if not more). This violation is not actually problematic, since \ac{ST} is not a \ac{QFT} and it is consistent with unitarity and causality; furthermore, its spectrum is not gapped, although there is a gap in single-particle masses, as there should be. Deep as this consideration may be, the violation of the Martin-Cerulus bound may also be an artifact of the tree-level approximation.

In order to assess the situation, we should look at higher-order contribution to $F$. Luckily, there is no need to deal with the intricate mathematics of Riemann surfaces and moduli spaces, nor with the complications of (super)ghosts, fermions and internal degrees of freedom~\cite{Gross:1987ar, Gross:1987kza}. The MVP is the representation of the amplitude as an integral over the moduli space, as we saw earlier. The universal contribution arising from spacetime momenta is the Koba-Nielsen factor, where invariant energies appear exponentially --- this is precisely the situation where a bell should ring in your head saying ``saddle-point asymptotics''! Since the propagator is some sort of ``electrostatic'' potential in two dimensions, solving the Poisson equation, saddles of the Koba-Nielsen exponent are equilibria of an auxiliary two-dimensional electrostatics problem on the worldsheet~\cite{Gross:1987ar, Gross:1987kza}. This analogy is often useful in \ac{CFT}, and it simplifies some computations via symmetry arguments. The upshot of some very clever manipulations on branched coverings of the sphere is that~\cite{Gross:1987ar, Gross:1987kza}
\begin{eqaed}\label{eq:branched_trick}
    \ln F_{g} \overset{\text{hard}}{\sim} \frac{1}{g+1} \ln F_\text{tree} \, , 
\end{eqaed}
which means that the tree-level and one-loop terms become comparable at a scale $s$ such that, parametrically (no $\orderneglected(1)$ prefactor matters),
\begin{eqaed}\label{eq:tree_loop_scale}
    g_s^2 \, \exp\left(- \, A \, s \ln s  \right) = \exp\left(- \, \frac{A}{2} \, s \ln s  \right)
\end{eqaed}
where the constant $A$ depends on the fixed scattering angle. The same result follows taking any two genera, due to \eqref{eq:branched_trick}. The resulting strong-coupling scale is roughly (ignoring terms whose logarithm is sub-leading) $\Lambda_\text{sc} = M_s \, \sqrt{\ln g_s^{-2}}$, which is formally reminiscent to the species scale $\Lambda_\text{sp} = M_s \, \ln g_s^{-2}$ found by solving \eqref{eq:species_scale} at $g_s \ll 1$~\cite{Castellano:2022bvr, Basile:2023blg, Aoufia:2024awo}. In this regime $\Lambda_\text{sc} \ll \Lambda_\text{sp}$, a non-trivial consistency check of \eqref{eq:species_hierarchy}. It makes sense that the strong-coupling scale be related to the coupling constant; in the weak coupling limit, asymptotically perturbation theory never breaks down, as befits a \ac{UV}-complete description. The existence of a strong-coupling scale much higher than the \ac{UV} cutoff and controlled by the coupling is exactly what we would have expected from a weakly coupled theory, since at some point non-perturbative effects must come into play.\footnote{Although we argued for this ubiquitous phenomenon on general grounds, for string perturbation theory this has been discussed in~\cite{Baker:1988tw, Gross:1988ib, Gross:1988tx}.} More importantly, the general pattern in \eqref{eq:branched_trick} requires going to all orders in string perturbation theory~\cite{Gross:1987ar, Gross:1987kza}. A (Borel) resummation of the the saddle-point contributions of~\cite{Gross:1987ar, Gross:1987kza} was studied in~\cite{Mende:1989wt}, but already in~\cite{Gross:1987kza} an upper estimate for the result was provided assuming that all the phases be aligned. Since the prefactor has a power-like dependence on the genus $(g_s^2 \, s)^g$, a saddle-point analysis in the genus for $g_s^2 s \leq 1$ reveals that the largest contribution in the asymptotic series arises at $g \sim \sqrt{\frac{A s}{\abs{\ln g_s^2 s}}}$, which \emph{restores the Martin-Cerulus behavior}, as pointed out in~\cite{Mende:1989wt}! Perhaps, \ac{ST} is somehow just local enough to satisfy a field-theoretic bound even without being a \ac{QFT} itself. This estimate is qualitatively confirmed by the more refined analysis of~\cite{Mende:1989wt}, including the phase estimates of~\cite{Gross:1987ar}. This more sophisticated asymptotic analysis allows to go beyond the limit of the above estimation, obtaining the overall exponential scaling $\exp\left( - \, \sqrt{\frac{s f(\phi)}{\ln s}}\right)$ above the strong-coupling scale, with some function of the scattering angle $\phi$.

\subsubsection{Black holes and UV/IR mixing}\label{sec:BH_transition}

\begin{figure}[ht!]
    \centering
    \includegraphics[width=\textwidth]{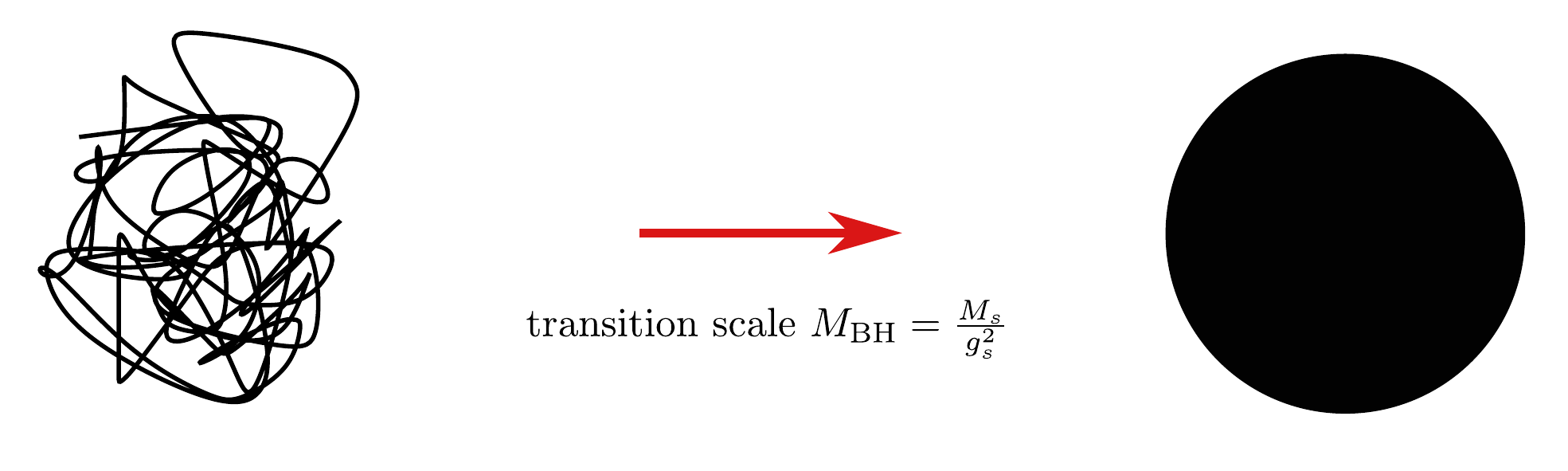}
    \caption{A depiction of the transition between strings and \acp{BH}~\cite{Susskind:1993ws, Horowitz:1996nw, Bedroya:2022twb} at the mass scale $M_\text{BH} = \frac{M_s}{g_s^2}$, where the parametric dependence of entropies and scattering amplitudes matches in the two pictures.}
    \label{fig:BHs}
\end{figure}

From the above results we learn that \ac{ST} seems to be \emph{juuust local enough} to be consistent, with a characteristic universal behavior of high-energy cross sections. Moreover, we can repeat an analysis along the lines of \eqref{eq:BH_leading} and \eqref{eq:matching_scale}: the schematic $\sqrt{s}$ scaling in the exponent can be compared with the expected ``asymptotically dark'' behavior due to (virtual) \ac{BH} production~\cite{Giddings:2007qq}. As review in (Appendix A of)~\cite{Bedroya:2022twb}, the expected amplitude for two-to-two scattering starting at the \ac{BH} threshold $s \gtrsim M_\text{th}$ is $\ln F \overset{s \gg M_\text{th}}{\sim} - \, S_\text{BH}(M_\text{BH} = \sqrt{s})$, with the leading-order entropy given in \eqref{eq:BH_leading}. Here the \ac{BH} threshold is the (parametric) mass of the smallest \ac{BH} which makes sense in the \ac{EFT}, in this case
\begin{eqaed}
    M_\text{th} = M_s^{3-d} \, \MPl^{d-2} = \frac{M_s}{g_s^2}
\end{eqaed}
as in \eqref{eq:matching_scale}. The fact that the Borel-resummed estimate of hard string scattering has precisely a $\sqrt{\frac{s}{M_s^2}}$ factor, derived \emph{independently} of the exponential growth of \eqref{eq:density_mass}, means that once more the two amplitudes \emph{match at the \ac{BH} threshold}! Together with the earlier result in \eqref{eq:matching_scale}, other investigations on this classicalization effect~\cite{Dvali:2014ila} and the growing number of microstate countings, starting from the seminal work of~\cite{Strominger:1996sh}, there is a compelling picture that \ac{ST} ``knows'' about \acp{BH}, and confirms all the expectations about asymptotic darkness and \ac{BH} dominance discussed in \cref{sec:gravity_qft}, as depicted in \cref{fig:BHs}. This is worth boxing up in a summary:

\begin{tcolorbox}
    \emph{Hard string scattering has a universal $\approx \exp(- \sqrt{s})$ behavior, barely surviving the Martin-Cerulus bound and transitioning to \ac{BH} production at $\sqrt{s} = M_\mathrm{th}$.}
\end{tcolorbox}

\noindent If you want to be pedantic, the $\sqrt{s}$ piece of an amplitude or cross section can be extracted taking a double logarithm with some absolute values for good measure. As promised, \emph{the extreme \ac{UV} ``quacks'' like the extreme \ac{IR}} due to \ac{QG} effects. Our journey indeed ended where it once began long ago in a section far far away.

\subsubsection*{One last lesson about UV/IR mixing}

\begin{figure}[ht!]
    \centering
    \includegraphics[width=0.6\textwidth]{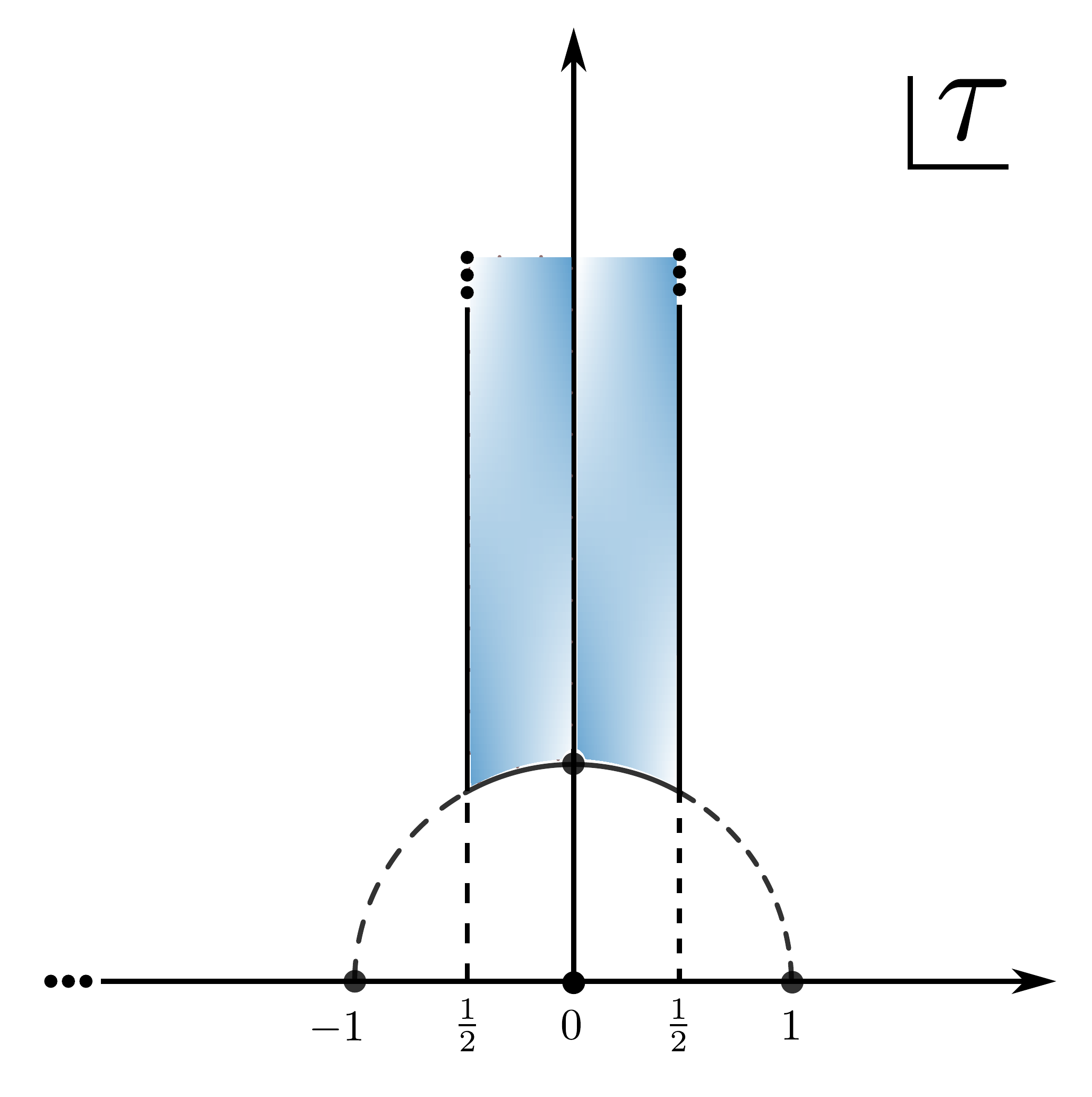}
    \caption{The most popular choice of fundamental domain $\mathcal{F}$ for the inequivalent complex structures $\tau$ of the torus, a representative of the moduli space $\mathcal{M} = \mathbb{H}/PSL(2,\mathbb{Z})$. It arises quotienting the upper-half plane $\mathbb{H}$ by the action of the modular group $PSL(2,\mathbb{Z})$, the mapping class group of the torus. Notably, associating $\tau$ to a sort of Schwinger proper time $t$ of particles, the \ac{UV} region of small $t$ is absent from the one-loop integration domain, reflecting the good \ac{UV} behavior of strings. Modular invariance maps this \ac{UV} region to the \ac{IR} region of large $\Im{\tau}$, showing an explicit instance of \ac{UV}/\ac{IR} mixing.} 
    \label{fig:modular}
\end{figure}

I'd like to conclude this section with one more cool thing about \ac{UV}/\ac{IR} mixing, namely modular invariance. We hinted at it many times and it plays a crucial role in many aspects of \ac{ST}; now we can discuss it in more detail. Recall from \cref{sec:spectra} that the torus partition function of the worldsheet \ac{CFT} contains data about the string spectrum, and from \cref{sec:string_perturbation_theory} we know that it must be modular-invariant due to cancellation of global gravitational anomalies. However, as we emphasized along the way, string (perturbation) theory is not simply the worldsheet \ac{CFT}; rather, it is the spacetime theory obtained from the analogous recipe as in the case of a worldline studied in \cref{fig:worldline}. The relevant physical quantity for the spacetime theory is then the integral over geometries on a toroidal worldsheet $\Sigma \simeq T^2$ then reduces to an integral over the Teichm\"{u}ller parameter $\tau$, due to the construction depicted in \cref{fig:torus}. This is nothing but the \emph{one-loop vacuum amplitude} of the spacetime \ac{QG} theory; the ``stringy bubble diagram'', if you will. Looking back at \eqref{eq:MCG}, there is a residual mapping class group, generated by the discrete transformations
\begin{eqaed}\label{eq:S_T_modular}
    \tau \mapsto \tau + 1 \, , \qquad \tau \mapsto - \, \frac{1}{\tau} \, .
\end{eqaed}
As it can be checked from \cref{fig:torus}, these are large diffeomorphisms without a small component, and thus belong to the mapping class group
\begin{eqaed}\label{eq:modular_group}
    \text{MCG}(T^2) & = \left\{ \tau \mapsto \frac{a \tau + b}{c \tau + d} \, : \, ad-bc = 1 \right\}/\left\{(a,b,c,d) \sim (-a,-b,-c,-d) \right\} \\
    & \simeq PSL(2,\mathbb{Z}) \equiv SL(2,\mathbb{Z})/\mathbb{Z}_2 \, ,
\end{eqaed}
dubbed the \emph{modular group}. The Teichm\"{u}ller space is the upper-half plane $\mathbb{H}$, so that the moduli space $\mathcal{M}_1 = \mathbb{H}/PSL(2,\mathbb{Z})$ can be represented by any domain $\mathcal{F} \subset \mathbb{H}$ of unique representatives for equivalence classes $[\tau]$ of Teichm\"{u}ller parameters. These parametrize the inequivalent conformal structures (or equivalently complex structures, as apparent from our construction) of the torus. The standard example of such a \emph{fundamental domain} $\mathcal{F}$ is depicted in \cref{fig:modular}.

From \cref{fig:modular} you can immediately gleam something cool: interpreting $\tau$ as \href{https://youtu.be/lCcHAUW1D2E?t=527}{a stringy extension of the Schwinger proper time} (see \eg{}~\cite{Angelantonj:2002ct}), the \ac{UV} region is \emph{completely absent}. It has been removed by the gauge redundancy necessary for the spacetime picture. This is \emph{not} what happened in \cref{sec:how} when discussing one-loop diagrams; it is a hallmark of the \ac{UV}-finiteness of stringy dynamics, with similar avatars in other physical quantities. Similarly to that discussion, however, the most convenient way to get the correct measure for integration over $\tau$ is to simply integrate over the ghost zero-modes associated to \acp{CKV}. This time there are two (real) moduli, $\tau_1 = \Re{\tau}$ and $\tau_2 = \Im{\tau}$, and thus two \acp{CKV} since $\chi(T^2) = \chi(S^1)^2 = 0$ (cf. \eqref{eq:riemann-roch}). These \acp{CKV} are simply translations on the torus, so the volume of the corresponding \ac{CKG} is simply $\text{Vol}(T^2_\tau) = \tau_2$. As for the case of a worldline loop, this factor ends up in the denominator of the measure, leaving us with $\int_{\mathcal{F}} \frac{\rmd^2\tau}{\tau_2}$. You may worry that this measure is not actually modular-invariant, but fear not --- it is simply because we have not arranged the full amplitude with modular-invariant building blocks. Indeed, taking the ever-present contribution in \eqref{eq:momenta_contribution} from the functional trace over spacetime momenta, we see that the integral rearranges into something like
\begin{eqaed}\label{eq:modular_measure}
    \int_{\mathcal{F}} \frac{\rmd^2\tau}{\tau_2^2} \, \frac{1}{\tau_2^{\frac{d-2}{2}}} \left( \text{excitations} \right) \equiv \int_{\mathcal{F}} \rmd\mu_{T^2}(\tau) \, \partitionfunction(\tau) \equiv \mathcal{T} \, .
\end{eqaed}
The measure is now modular-invariant, and separately also the new partition function $\tilde\partitionfunction$, where the spacetime momenta contribute according to the same power of $d-2$ arising from the physical transverse mode of the string as we saw in \cref{sec:spectra}. In light-cone quantization in flat spacetime, this comes out more or less automatically, at the price of sacrificing manifest covariance.

Due to the absence of a \ac{UV} region, the amplitude in \eqref{eq:modular_measure} is \emph{finite} in the absence of tadpoles or physical tachyons. As well-known from \ac{QFT}, it contributes to the vacuum energy density (with the appropriate power $M_s^d$ of the string scale in front), except this time the vacuum energy density is physical due to coupling to gravity. But then what's the tree-level contribution? Well, as we discussed in \cref{sec:S-matrix}, the zero-point sphere amplitude vanishes due to gauge redundancy! This is a very deep fact about string (perturbation) theory: \emph{the tree-level contribution to the vacuum energy density vanishes.} Therefore, at weak coupling, the leading Einstein-frame contribution to the vacuum energy density (which depends on scalar fields, possibly also through $\mathcal{T}$) is
\begin{eqaed}
    \CC \overset{e^\phi \ll 1}{\sim} - \, e^{\frac{2d}{d-2} \phi} \, M_s^d \, \mathcal{T} \, ,
\end{eqaed}
which is both \emph{naturally small} (in Planck units) and \emph{controlled by the mass gap of the spectrum}. The former would not be the case in a gravitational \ac{EFT}, where the vacuum energy density is still physical but generically (in a Bayesian sense with respect to the appropriate measure on theory space, akin to Weinberg's anthropic estimate) dominated by the heavy states, or at least a much larger (generically Planckian) \ac{UV} cutoff. The latter would also not work in a generic \ac{EFT} for the same reason. When non-zero, the above expression is generically of the order of the Einstein-frame string scale, but it can be much smaller according to the results of \cref{sec:swampland_stuff}. Thus, we learn once again that stringy \acp{EFT} look highly fine-tuned from the perspective of a generic \ac{EFT}; their \ac{UV} features feed back into their \ac{IR} features in a highly non-trivial fashion. These properties may be instrumental in relating a small dark energy in our universe to \ac{UV} features of \ac{QG}~\cite{Montero:2022prj}. As should be abundantly clear by now, the MVP here is modular invariance, which also constrains $\tilde\partitionfunction(\tau)$ so much that a number of general results can be derived purely on these grounds~\cite{Abel:2021tyt, Abel:2023hkk, Aoufia:2024awo, Abel:2024twz}. In particular, one can discuss a kind of \ac{RG} flow which takes modular invariance and \ac{UV}/\ac{IR} mixing into account, painting a picture in which the ``true \ac{UV}'' is actually in the middle of the energy ruler. Let us write a final boxed summary before closing our journey for good:

\begin{tcolorbox}
    \emph{Unlike in a generic gravitational \ac{EFT}, the vacuum energy density in perturbative \ac{ST} is driven by the mass gap of the \ac{UV} completion, due to the \ac{UV}/\ac{IR} mixing induced by modular invariance. It encodes properties of the \ac{UV} spectrum.}
\end{tcolorbox}

\subsection{Conclusions}\label{sec:IVANO_conclusions}

I hope the journey we undertook together was worthwhile and insightful. I attempted not to leave any subtlety unmentioned, while also not delving into the technicalities to solve them; this way, the interested reader can explore the referenced literature. I also attempted to structure the topics targeting what's mostly relevant for \ac{QG}, at least in the restricted context of weakly coupled closed strings. The idea was to present nothing as arbitrary or chosen without some physical motivation; the road (perhaps a \ac{UV}/\ac{IR}-mixed loop?) began from the very basics of what \ac{QG} should be and what \acp{EFT} do when dynamical gravity is involved. In the rest of these lectures, we learned a lot about perturbative gravity in \cref{sec:LUCA} and \ac{EFT} in \cref{sec:ANNA}. Here we explored a way to keep both in a \ac{UV}-completion, since strings can be weakly coupled way above the \ac{EFT} cutoff. The price to pay is exiting the framework of quantum fields in spacetime, which forced us to introduce a number of concepts which can be somewhat abstract and technical. Hopefully, it wasn't too bad! With some experience, you can build physical intuition about the physics of strings, much like you had to do for fields and particles.

\ac{ST} is a vast topic, much broader than I could have possibly covered in this section. But I hope I could convey at least some of the key lessons for someone who genuinely wants to know what strings can teach us about \ac{QG}, and possibly --- hopefully --- the world we live in.

\subsubsection*{A summary of summaries (summary$^2$)}

\begin{tcolorbox}
\emph{String vacua with a tame $d$-dimensional spacetime are two-dimensional unitary compact superconformal field theories which are $d$-critical and modular-invariant.}
\end{tcolorbox}

\begin{tcolorbox}
    \emph{If the spectrum of the worldsheet \ac{CFT} contains conformal weights $(h,\overline{h})$, the spectrum of physical string states has $h=\overline{h}$ and contains the squared masses}
    \begin{eqaed}\label{eq:string_spectrum}
        m^2 = \frac{4}{\alpha'} \left(h - 1\right) \, . \nonumber
    \end{eqaed}
\end{tcolorbox}

\begin{tcolorbox}
    \emph{Unlike for point particles, the degeneracy of single-string states scales exponentially in their mass $m$ for $m \gg M_s$ with the scaling $\ln \rho(m) \overset{m \gg M_s}{\sim} \mathrm{const.} \times \frac{m}{M_s}$.}
\end{tcolorbox}

\begin{tcolorbox}
    \emph{The \acp{EFT} arising from string vacua are a tiny subset of the pool of all \acp{EFT}. The couplings and field content cannot be chosen willy-nilly: almost nothing goes!}
\end{tcolorbox}

\begin{tcolorbox}
    \emph{Some general properties of the string landscape:}

    \begin{itemize}
        \item \emph{There are no global symmetries. Just gauge redundancies.}
        \item \emph{Gravity is the weakest force. Abelian charges bring along light states.}
        \item \emph{There are no weakly interacting \ac{dS} vacua.}
        \item \emph{Limits of the internal worldsheet \ac{CFT} ``quack like geometry''.}
    \end{itemize}
\end{tcolorbox}

\begin{tcolorbox}
    \emph{Hard string scattering has a universal $\approx \exp(- \sqrt{s})$ behavior, barely surviving the Martin-Cerulus bound and transitioning to \ac{BH} production at $\sqrt{s} = M_\mathrm{th}$.}
\end{tcolorbox}

\begin{tcolorbox}
    \emph{Unlike in a generic gravitational \ac{EFT}, the vacuum energy density in perturbative \ac{ST} is driven by the mass gap of the \ac{UV} completion, due to the \ac{UV}/\ac{IR} mixing induced by modular invariance. It encodes properties of the \ac{UV} spectrum.}
\end{tcolorbox}


\section{\titleFrancesco}
\label{sec:FRANCESCO}

\begin{tcolorbox}[colback=white,colframe=scipostblue]
{\bf Lecturer:} Francesco Di Filippo, \briefaffiliationFrancesco

{\bf Email address:} \href{mailto:\emailFrancesco}{\emailFrancesco}
\tcblower
{\bf Lecture recordings:}
\begin{enumerate}[label= Lecture \arabic*:, leftmargin = 3.5cm, labelsep = 0.5cm, parsep = 0.0cm]
    \item \href{https://youtu.be/AwSabAeuX44}{https://youtu.be/AwSabAeuX44}
    \item \href{https://youtu.be/EOAWAFtmIe8}{https://youtu.be/EOAWAFtmIe8}
    \item \href{https://youtu.be/y56Azm8kef0}{https://youtu.be/y56Azm8kef0}
    \item \href{https://youtu.be/MdtaL0itUc4}{https://youtu.be/MdtaL0itUc4}
\end{enumerate}

{\bf Abstract:}

Black holes are one of the primary motivations for seeking a theory of quantum gravity, as both quantum and strong gravitational effects are expected to be significant in these spacetimes. While we still lack a full theory, this course focuses on some quantum effects in black holes spacetimes that can be derived without the knowledge of a specific theory of quantum gravity (\ie{}, Hawking radiation) and will examine their consequences.
\end{tcolorbox}

\subsection*{Preface}

This section covers some basics aspects regarding the interplay between \ac{QFT} and \ac{GR} in the context of \ac{BH} spacetimes. Classical \acp{BH} hide spacetime singularities in their interior that are often regarded as one of the main motivations to look for a theory of \ac{QG}. In fact, singularities mark the breakdown of the classical theory of \ac{GR}, and we expect that a full theory of \ac{QG} will make sense of singularities. However, we still lack a complete understanding of the \ac{UV} behavior of gravity. As seen in the previous sections (cf. \cref{sec:lecture4}, \cref{sec:ALESSIABENJAMIN}, and \cref{sec:IVANO}), we do have a few promising approaches to \ac{QG}, and the study of the predictions of these approaches in relation to the singularity problem of \ac{BH} spacetimes is a very exciting and active area of research. However, these investigations go well beyond the scope of this section. Instead, we are going to discuss some effects that we expect to be universal, and for which we can argue should be obtained independently of the specific theory of \ac{QG}. In fact, we will treat gravity in a classical way, and we will just consider the effect of quantum matter propagating on the classical \ac{BH} spacetime. 
We will see that this is enough to obtain some non-trivial effects. 

The lectures are organized as follows.
\begin{description}
\item[Sec.~\ref{Sec:Francesco-Preliminaries}:] We introduce some preliminary information that will be crucial in the rest of the course. In particular, we review some basic notions regarding the physics of \ac{BH} in classical \ac{GR} and some elements of \ac{QFT} in curved spacetime.
\item[Sec.~\ref{Sec:Francesco-Haw-rad}:] We discuss the phenomenon of Hawking radiation. Probably every physics student knows that classical \acp{BH} cannot emit anything, while Hawking radiation is a quantum effect that allows the emission of particles~\cite{Hawking:1974rv,Hawking:1975vcx}.
However, this result is also quite misunderstood. There is a heuristic argument to explain the effect that focuses on quantum fluctuations of the vacuum near the \ac{BH} horizon. According to this argument, particle-antiparticle pairs near the horizon can be separated by tidal forces, leading to particle production. This argument was introduced by Hawking himself, who was, however, well aware of its limitations. Unfortunately, due to its simplicity, this explanation is now widespread and its limitations are often overlooked. We will see that the true root of Hawking radiation is in the time dependence of the geometry.
\item[Sec.~\ref{Sec:Francesco-SET}:] We introduce the renormalized quantum stress-energy tensor and study some of its basic properties. We will study the main vacuum states that can be chosen for a Schwarzschild \ac{BH}. From this, we will also better understand some crucial questions regarding energy conservation and the limitations of the heuristic arguments for Hawking radiation.
\item[Sec.~\ref{sec:Francesco_Info_loss}:] We  discuss the information loss problem. As for Hawking radiation, the information loss problem is also an extremely well-known subject. However, once again, the precise formulation is often overshadowed by simple arguments whose limitations and assumptions are often overlooked. Therefore, the focus of the section will be devoted to a precise formulation of the problem.
\item[Sec.~\ref{sec:FRANCESCO_conclusions}:] We discuss the main conclusions and highlight the most important lessons from these lectures.
\end{description}

I will try to give my personal viewpoint on the field, focusing on a few aspects that are sometimes misinterpreted. However, the topics discussed here are very well-known and presented in textbooks that are reference points in the literature. In particular, the content of \cref{Sec:Francesco-Preliminaries} regarding classical \acp{BH} can be found in most \ac{GR} textbooks. In writing the content of the \ac{QFT} in curved spacetime part of \cref{Sec:Francesco-Preliminaries} and for \cref{Sec:Francesco-Haw-rad} and \cref{Sec:Francesco-SET} I particularly used the books
\begin{itemize}
    \item  N. D. Birrell and P. C. W. Davies, Quantum Fields in Curved Space, Cambridge Monographs on Mathematical Physics. Cambridge University Press, Cambridge, UK, (1982) \cite{Birrell:1982ix},
    \item A. Fabbri and J. Navarro-Salas, Modeling black hole evaporation, Imperial College Press (2005)~\cite{Fabbri:2005mw}.
\end{itemize}
The content of \cref{sec:Francesco_Info_loss} follows the logic of the paper
\begin{itemize}
    \item L. Buoninfante, F. Di Filippo and S. Mukohyama, On the assumptions leading to the information loss paradox, JHEP 10, 081 (2021)~\cite{Buoninfante:2021ijy}.
\end{itemize}
Finally, I will make use of other references that will be cited throughout the text.

\subsection{Preliminaries}\label{Sec:Francesco-Preliminaries}

In this first subsection, we introduce the main concepts we will require for our discussions. We will first review some aspects of classical \acp{BH} spacetimes, and then we will introduce some basic notions of \ac{QFT} in curved spacetimes.
\subsubsection{Penrose-Carter diagrams}

\begin{figure}[ht]
    \centering
   \includegraphics[width=0.25\linewidth]{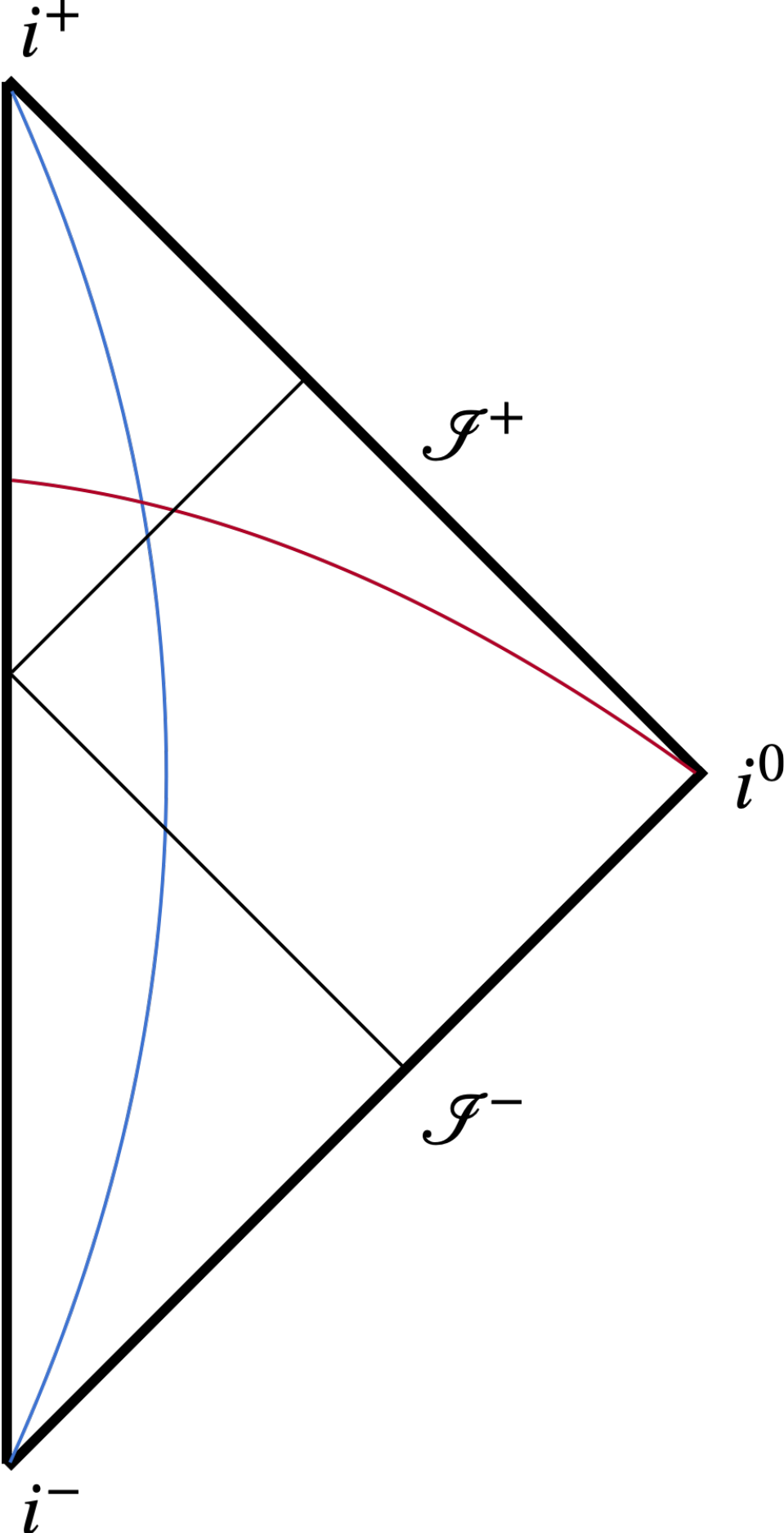} 
    \caption{\ac{PC} diagram of flat spacetime. The black $45^\circ$ lines that originate at the past null infinity are radial null trajectories. Examples of timelike and spacelike trajectories are depicted as blue and red lines, respectively.}
    \label{fig:Francesco_flat}
\end{figure}

\ac{PC} diagrams give us a way to visualize the causal relations between points of spacetime. The idea is to rewrite the line element using conformal transformations to obtain a metric that is conformally equivalent to the starting one but in which the coordinates have a finite range. Each point of a \ac{PC} diagram is equivalent to a 2-sphere. Therefore the visualization provided by these diagrams works best for spherically symmetric spacetimes. As a simple example, \cref{fig:Francesco_flat} shows the \ac{PC} diagram of flat spacetime. The relevant properties and elements of a diagram are the following:
\begin{itemize}
    \item \textit{Radial null trajectories} are $45^\circ$ lines that originates at the past null infinity $\mathscr{I}^-$ and end at the future null infinity $\mathscr{I}^+$.
    \item \textit{Timelike trajectories} form an angle of less than $45^\circ$  with the vertical axis and originate at the past infinity $i^-$ and end at the future infinity $i^+$.
    \item \textit{Spacelike trajectories} form an angle of less than $45^\circ$ with the vertical axis and have one hand at the asymptotic infinity $i^0$.
\end{itemize}
While the causal structure of Minkowski spacetime is quite simple, \ac{PC} diagrams constitute a very powerful tool to visualize more complex spacetimes.

\subsubsection{Schwarzschild black hole}

\begin{figure}[ht]
    \centering
   \includegraphics[width=0.6\linewidth]{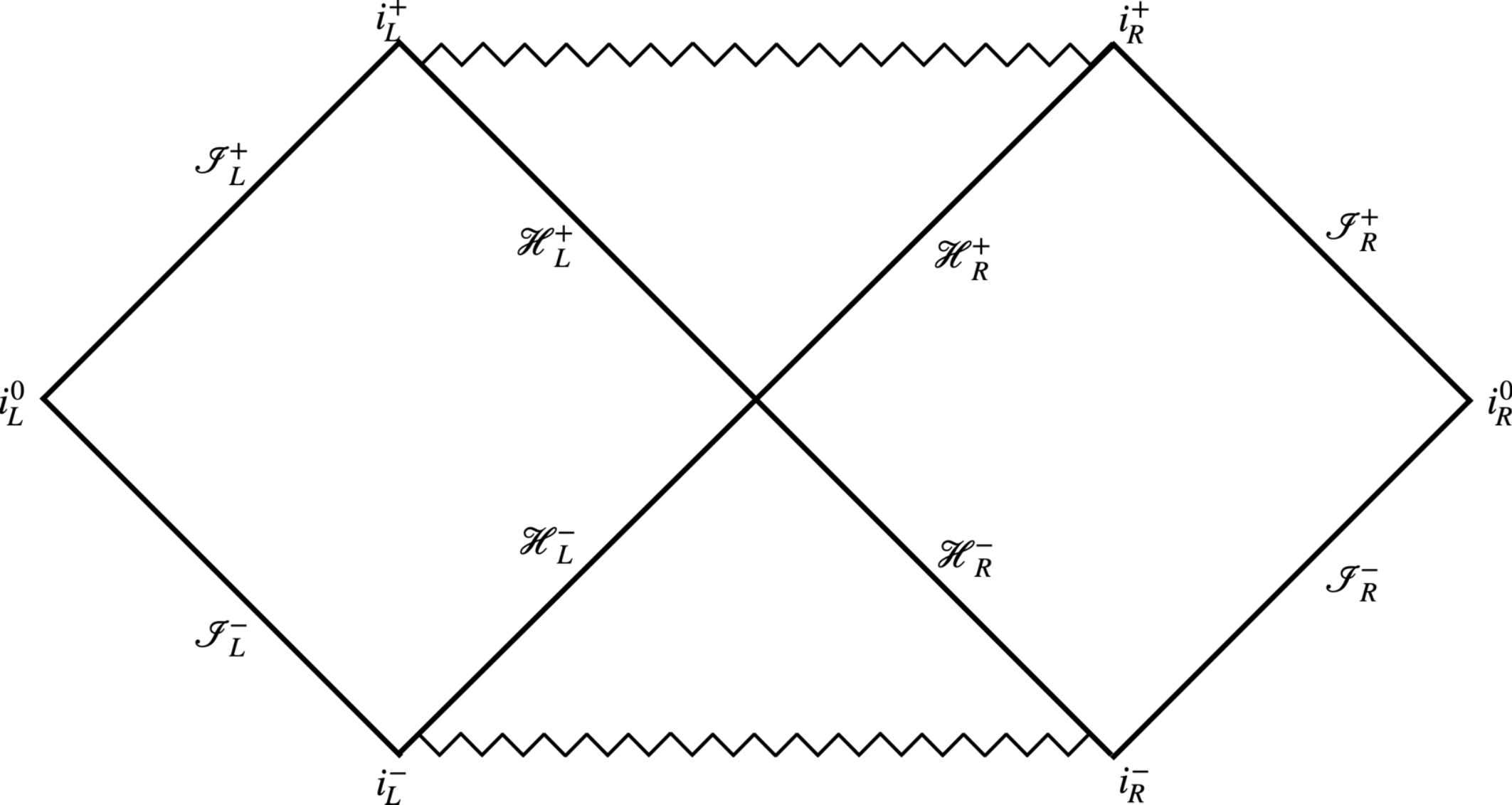} 
   \hspace{2cm}
    \includegraphics[width=0.2\linewidth]{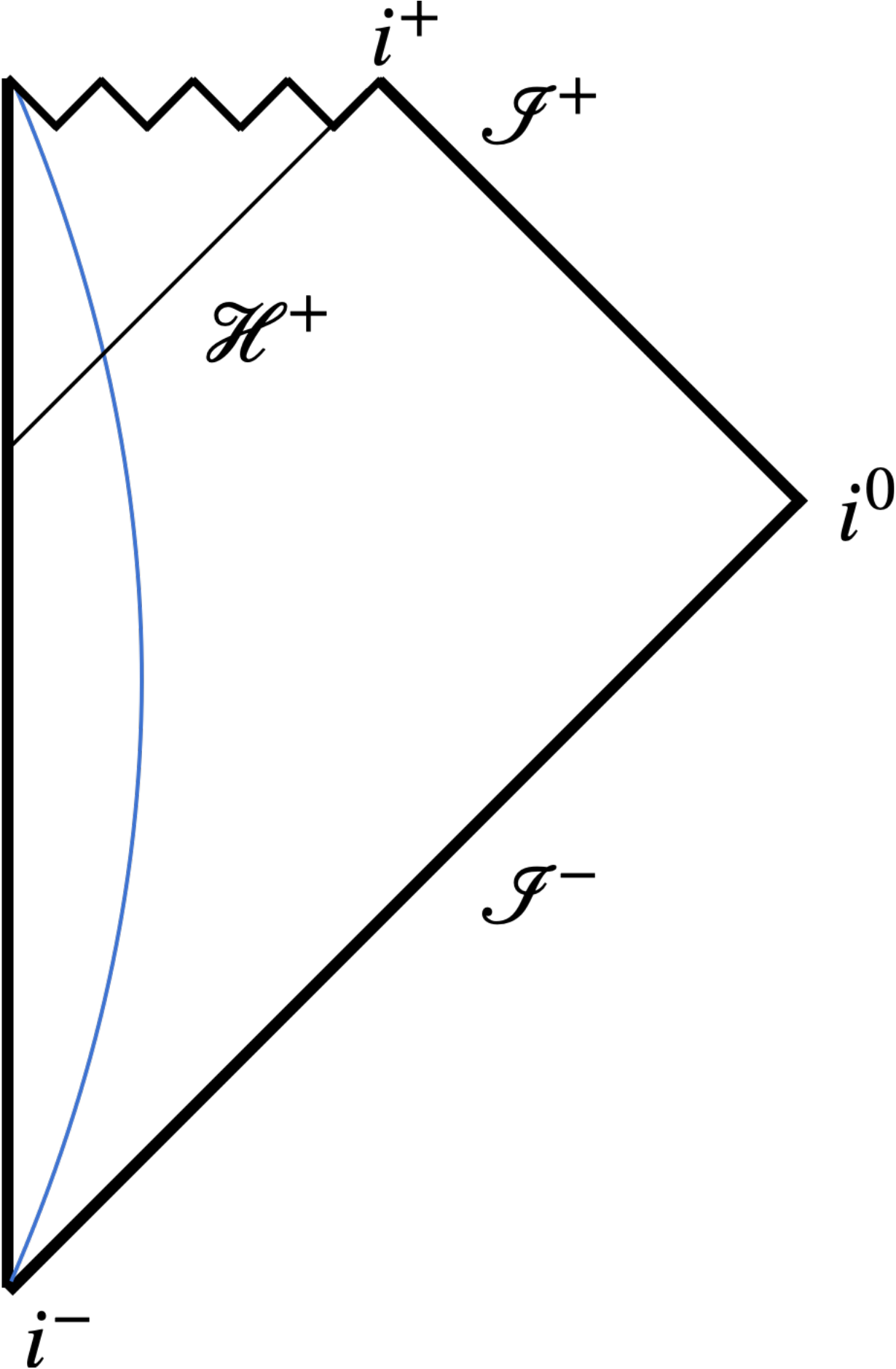} 
    \caption{\ac{PC} diagram of an eternal \ac{BH} (left panel) and a \ac{BH} formed via gravitational collapse. In the eternal case, there are two asymptotic regions labeled by the indexes $L,R$. Besides the future event horizon $\mathscr{H}^+$, the eternal geometry has also a past white hole horizon $\mathscr{H}^-$.}
    \label{fig:Francesco_Schwarz}
\end{figure}

Let us start by reminding some crucial properties of Schwarzschild \acp{BH}. 
The line element of a Schwarzschild \ac{BH} is a vacuum solution of the Einstein equations and can be written as
\begin{equation}
   \rmd{}s^2=-\left(1-\frac{2\GN M}{r}\right) \rmd{}t^2+\left(1-\frac{2\GN M}{r}\right)^{-1} \rmd{}r^2+r^2\rmd{}\theta^2+r^2\sin^2\theta \rmd{}\phi^2\,.
\end{equation}
This metric is singular at $r=0$ and at $r=2\GN M$. The singularity at $r=0$ is physical as we can confirm by computing some curvature invariants and noticing that they diverge at $r=0$. For instance
\begin{equation}
    R_{\mu\nu\rho\sigma}R^{\mu\nu\rho\sigma}=\frac{48\GN^2M^2}{r^6}\,.
\end{equation}
On the other hand, the singularity at $r=2\GN M$ is due to the choice of coordinates. We can define a set of coordinates that covers the horizon by adapting the coordinates to radial null geodesics. To this end, we introduce the advanced ($v$) and retarded ($u$) time coordinates as
\begin{equation}
    v=t+r^\ast\,,\qquad u=t-r^\ast\,.
\end{equation}
where the tortoise coordinate $r^\ast$ is defined such that
\begin{equation}\label{eq:Francesco_rtor}
    \frac{\rmd{}r^\ast}{\rmd{}r}=\sqrt{\frac{g_{rr}}{g_{tt}}}\,\implies r^\ast=r+2\GN M\ln{\left|\frac{r}{2\GN M}-1\right|}\,.
\end{equation}
Right-going and left-going null observers move on $u=\text{const.}$ and $v=\text{const.}$ trajectories.
The metric, using the $(v,r)$, $(u,r)$ or $(u,v)$ coordinates reads
\begin{equation}\label{eq:Francesco_metric_vr}
       \rmd{}s^2=-\left(1-\frac{2\GN M}{r}\right) \rmd{}v^2+2\rmd{}v\rmd{}r+r^2\rmd{}\theta^2+r^2\sin^2\theta \rmd{}\phi^2\,,
\end{equation}
\begin{equation}\label{eq:Francesco_metric_ur}
       \rmd{}s^2=-\left(1-\frac{2\GN M}{r}\right) \rmd{}u^2-2\rmd{}u\rmd{}r+r^2\rmd{}\theta^2+r^2\sin^2\theta \rmd{}\phi^2\,,
\end{equation}
\begin{equation}\label{eq:Francesco_metric_uv}
       \rmd{}s^2=-\left(1-\frac{2\GN M}{r}\right) \rmd{}v\rmd{}u+r^2\rmd{}\theta^2+r^2\sin^2\theta \rmd{}\phi^2\,.
\end{equation}
From the near horizon limit of \eqref{eq:Francesco_rtor} we note that the region $r=2\GN M$ is mapped into $v-u=+\infty$, and 
\begin{equation}
    1-\frac{2\GN M}{r}\approx  e^{(v-u)/4\GN M}\,.
\end{equation}
We can now define some new coordinates as
\begin{equation}\label{eq:Francesco_coord_Kru}
    U=\mp e^{-u/4\GN M}\,,\qquad V=e^{v/4\GN M}\,.
\end{equation}
These coordinates go by the name of \textit{Kruskal} coordinates. In terms of Kruskal coordinates, the metric is manifestly regular at the horizon
\begin{equation}
    \rmd{}s^2=-\frac{32\GN^3M^3}{r}e^{-r/2\GN M}\rmd{}U\rmd{}V+r^2\rmd\Omega^2\,.
\end{equation}
The horizon is mapped into $U=0$ and $V=0$. From  \eqref{eq:Francesco_coord_Kru} the range of the coordinates is
\begin{equation}
    U\in (-\infty,0)\,,\qquad V\in(0,+\infty)\,.
\end{equation}
However, both coordinates can be extended to negative values obtaining the maximally extended spacetime. To visualize it, we can perform a compactification of the coordinates (for instance via $U'=\arctan U$  and $V'=\arctan V$) to obtain the \ac{PC} diagram depicted in the left panel of \cref{fig:Francesco_Schwarz}.

Finally, note that this description is only true for eternal \acp{BH}. The \ac{PC} diagram of a \ac{BH} that is formed by gravitational collapse is depicted in the right panel of \cref{fig:Francesco_Schwarz}. It is obtained by gluing the causal structure in the asymptotic past (before the collapse begins and the spacetime is approximately flat) with the causal structure after the collapse that is given by the static Schwarzschild metric.

\subsubsection{Different notions of horizons}
There are several different notions of horizons, sometimes confused with each other. At least three of them are very relevant for these lectures section: event horizon, trapping horizon, and Cauchy horizon. We will not need the precise mathematical formulation of each definition. The interested reader can check standard \ac{GR} books or review articles on the topic for more details and the definitions of other types of horizons that are not relevant for our discussion (see \eg{}~\cite{Wald:1984rg,Ashtekar:2004cn}). However, it is important to keep in mind the qualitative differences among these different types of horizons:
\begin{itemize}
    \item \textbf{Event Horizon.} The event horizon is the boundary of the spacetime region that cannot reach infinity. This is a non-local definition. You must know the full manifold in order to define the event horizon.
    \item \textbf{Trapping horizon.} The trapping horizon is the boundary of the locally trapped region. In spherical symmetry, this means that causal observers must move in the direction of decreasing area radius. For static geometries, the event horizon and the trapping horizon coincide.
    \item \textbf{Cauchy horizon.} A Cauchy horizon denotes the end of the predictability of the theory. We say that a theory is predictable when we can give initial conditions on a spacelike slice and evolve these initial conditions with the field equations. Beyond the Cauchy horizon, the evolution is not unique. Examples of Cauchy horizons include the inner horizons of Kerr and Reissner-Nordstr\"om \acp{BH}~\cite{Wald:1984rg}.
\end{itemize}
In the last section, we will discuss the spacetime of a \ac{BH} formed by gravitational collapse that evaporates in finite time. As we will see in \cref{fig:Francesco_Complete_evaporation}, this is an example of a spacetime which has all three types of the horizons just described, and we will see that all the horizons are different.

\subsubsection{QFT on flat spacetime}
Having introduced the main basic concepts on classical \acp{BH} we will use, we now have to introduce some aspects of \ac{QFT} that were not covered in \cref{sec:LUCA}. We start from the simplest possible case, \ie{}, a free massless scalar field on flat spacetime, with equation of motion
\begin{equation}\label{eq:Francesco_EOMphi}
    \partial_\mu\partial^\mu\phi=0 \, .
\end{equation}
To quantize the field, we decompose a generic solution of the Klein--Gordon equation into positive and negative frequency modes,
\begin{equation}
    \phi=\sum_{i}\left(  a_{i}f_{i}+a^\dagger_{i}f^\ast_{i}\right) \, .
\end{equation}
where the $f_i$ modes are the positive frequency ones. For instance, we can consider plane waves for which
\begin{equation}\label{eq:FRANCESC0-mode_flat}
    f_{\vec k}=\frac{1}{\sqrt{16\pi\omega_k}} e^{-i\omega_{k}t\pm i{\vec k}\cdot \vec x}\,,\qquad\omega_k=|\vec k|>0\,.
\end{equation}
The normalization factor is chosen to make the base orthonormal with respect to the scalar product
\begin{equation}
    \left(f_i,f_j\right)\equiv-i\int \rmd^3\vec x \left(f_i\partial_tf_j^\ast-f_j^\ast\partial_tf_i\right)=\delta_{ij}\,.
\end{equation}
The standard commutation relations apply
\begin{equation}
    \left[a_{\vec k},a_{\vec k'}\right]=
    \left[a^\dagger_{\vec k},a^\dagger_{\vec k'}\right]=0\,,\quad    \left[a_{\vec k},a^\dagger_{\vec k'}\right]=\delta^3(\vec{k}-\vec{k'}) \, .
\end{equation}
We can define the vacuum state as the state which is annihilated by all the operators $a_{\vec{k}}$
\begin{equation}
    a_{\vec{k}}\left|0\right>=0\,,
\end{equation}
and the states containing particles as those obtained acting with the creation operators $a^\dagger_{\vec k}$. 
The choice of the vacuum state depends on the time coordinates we choose as the splitting of time and spatial coordinates in \eqref{eq:FRANCESC0-mode_flat} is not unique. That is, it depends on the observer. In flat spacetime this is not a big problem as the state is invariant under Poincar\'e transformations. Therefore, while the choice of vacuum is not unique, all inertial observers agree on the definition of the vacuum. We will now see that this is no longer the case in curved spacetime. This effect is arguably the main difference between \ac{QFT} in flat and curved spacetime. 
\subsubsection{QFT on curved spacetime}
We now have to generalize the construction of the previous section to curved spacetime. 
The first uncertainty consists of the generalization of the equation of motion \eqref{eq:Francesco_EOMphi}. We consider the simplest possibility, namely, promoting partial to covariant derivatives:
\begin{equation}
    g^{\mu\nu}\covD_\mu\covD_\nu\phi=0\,.
\end{equation}
A more crucial uncertainty concerns the split into positive and negative modes, as in general there is no privileged set of observers. 
If the spacetime is stationary, \ie{}, there is an everywhere-defined timeline killing vector $\xi$, we can choose the decomposition according to such notion of time,
\begin{equation}
    \phi=\sum_i a_if_i+a^\dagger_if^\ast_i\,,
\end{equation}
where $f_i$ is a basis chosen such that
\begin{equation}
    \xi^\mu\covD_\mu f_i=-i\omega_i f_i\,,\qquad\text{with}\qquad\omega_i>0\,.
\end{equation}
We can go on and define the vacuum state, the multi-particle states, and the scalar product among states in the same way it was done for flat spacetimes.
To deal with non-stationary spacetimes, we make use of the facts that \textit{(i)} most physically relevant scenarios are approximately stationary in the far past and far future, and \textit{(ii)} the dynamics is localized in some intermediate transient. 
If that is the case, we can define $in$ and $out$ states for the two asymptotic regions.
For the $in$ region
\begin{equation}\label{eq:Francesco_phi_in}
    \phi=\sum_i a_i^{in}f^{in}_i+a^{\dagger in}_if^{in\ast}_i\,,
\end{equation}
while for the $out$ region
\begin{equation}\label{eq:Francesco_phi_out}
    \phi=\sum_i a_i^{out}f^{out}_i+a^{\dagger out}_if^{out\ast}_i\,,
\end{equation}
These two bases can be related to each other introducing the Bogolyubov coefficients.
\begin{equation}
f^{out}_i = \sum_j \left( \alpha_{ij} f^{in}_j + \beta_{ij} f^{in\ast}_j \right).
\end{equation}
Given the orthonormality of the modes, we can extract the Bogolyubov coefficients by performing the appropriate scalar products
\begin{equation}\label{eq:Francesco_bog}
\alpha_{ij}=\left(f_i^{out},f_j^{in}\right)\,,\qquad\beta_{ij}=-\left(f_i^{out},f_j^{in\ast}\right)
\end{equation}
The orthonormality of $f_i^{out}$ can be used to prove the following relations among the Bogolyubov coefficients:
\begin{equation}\label{eq:Francesco_ort_discrete}
\sum_k (\alpha_{ik} \alpha_{jk}^\ast - \beta_{ik} \beta_{jk}^\ast) = \delta_{ij}, \quad \sum_k (\alpha_{ik} \beta_{jk} - \beta_{ik} \alpha_{jk}) = 0 \, ,
\end{equation}
or, if the indexes can vary continuously,
\begin{equation}\label{eq:Francesco_ort_cont}
    \int \rmd\omega \, \alpha_{\omega_1\omega'}\alpha_{\omega_2\omega'}-\beta_{\omega_1\omega'}\beta_{\omega_2\omega'^\ast}=\delta(\omega_1-\omega_2)\,.
\end{equation}
The $\left|in\right>$ vacuum state is the state annihilated by all the destruction operators $a_i^{in}$ while the $\left|out\right>$ is annihilated by the operators $a_i^{out}$
\begin{equation}
    a_i^{in}\left|in\right>=0\,\qquad    a_i^{out}\left|out\right>=0\,.
\end{equation}
We can now ask what is the number of particles in the vacuum $\left|in\right>$ state as seen by an observer in the $out$ region. To this end, we just need to evaluate the expectation value of the particle number operator in the $\left|in\right>$ state
\begin{equation}\label{eq:Francesco_numb_opr} \left<in\right|N_i^{out}\left|in\right>=\left<in\right|a_i^{out\dagger} a_i^{out}\left|in\right>\,.
\end{equation}
Next, we have to express the $out$ operators in terms of the $in$ operators. To this end, we can simply perform the scalar product $\left(\phi,f^{out}_i\right)$ using \eqref{eq:Francesco_phi_in}, \eqref{eq:Francesco_phi_out} and \eqref{eq:Francesco_bog}. We obtain
\begin{equation}
    a_i^{out}=\sum_j\alpha_{ij} a_j^{in}-\beta_{ij}^\ast a_j^{in\dagger}\,.
\end{equation}
It is now straightforward to evaluate \eqref{eq:Francesco_numb_opr} to get
\begin{tcolorbox}
\begin{equation}
 \left<in\right|a_i^{out\dagger} a_i^{out}\left|in\right>=\sum_j\left|\beta_{ij}\right|^2  \,.
\end{equation}
\end{tcolorbox}
\noindent Therefore, if $\beta\neq 0$, \ie{}, if $out$ and $in$ observers do not agree on the definition of positive and negative frequency modes, the vacua for the $in$ and $out$ regions are different.

This concludes the content of this section. Before moving to the following section, we invite the readers to ponder on the question below.
\begin{tcolorbox}
\begin{center}
\textbf{Question for the readers.}
\end{center}We have seen that there is no particle production for stationary spacetimes. However, you are probably aware that stationary and static \acp{BH} emit particles due to Hawking radiation (which will be the topic of the next section). Is this in contradiction with what we have discussed so far? 
\end{tcolorbox}

\subsection{Hawking radiation}\label{Sec:Francesco-Haw-rad}

We now have the necessary tools to compute the radiation emitted by a Schwarzschild \ac{BH} via the Hawking effect. We will start by providing a heuristic argument that is widespread due to its simplicity. We will then discuss a more rigorous derivation of the phenomenon which will highlight the limitations of the heuristic argument.

\subsubsection{Heuristic derivation}
Hawking evaporation can be heuristically understood by looking at the vacuum fluctuations of quantum fields near the \ac{BH} horizon. In \ac{QFT}, the vacuum is not empty, but exhibits temporary particle-antiparticle pair production that can arise anywhere, including in the proximity of the horizon. 

Near the horizon, intense tidal forces can separate the pairs with one constituent falling into the \ac{BH}, while the other escapes to infinity. Particles escaping to infinity become real observable particles with positive energy, while their partner falling into the \ac{BH} will have negative energy (as measured from an external observer) and fall into the singularity. The negative energy of the infalling particle implies that the mass of the \ac{BH} shrinks as a consequence of this process. Furthermore, the particle pairs are maximally entangled. Thus the radiation reaching infinity will have large entropy. An asymptotic observer would see this process as particle emission from the \ac{BH} horizon. 

This heuristic picture captures some interesting results that we will confirm later on. However, it falls short in a number of ways. The most clear indication of the failure of this description is the fact that it seems to be applicable to static spacetimes. However, as seen in the previous section, static spacetime cannot have any particle production. 

In the remainder of this section, we will provide a proper derivation of Hawking radiation. In \cref{Sec:Francesco-SET} we will discuss the limitations of this heuristic argument more concisely.

\subsubsection{Gravitational collapse}
We are now going to answer the question asked at the end of the previous section. If stationary spacetimes cannot produce particles, how do we explain the phenomenon of Hawking radiation? 
The answer is that, despite the Schwarzschild geometry being static, \acp{BH} are not static (nor stationary for the rotating case) spacetimes. In fact, \ac{BH} are produced dynamically via gravitational collapse and are described by the Schwarzschild solution only at late times. Hawking radiation is due to the time-dependence of the geometry, and it is obtained considering a gravitational collapse in which the geometry evolves dynamically from a spacetime that closely resembles Minkowski to Schwarzschild. The analysis of eternal \acp{BH} is non-trivial and requires a dedicated analysis that will be the subject of the next section.

\begin{figure}[ht]
\centering
 \hspace*{1cm}  \includegraphics[width=0.35\linewidth]{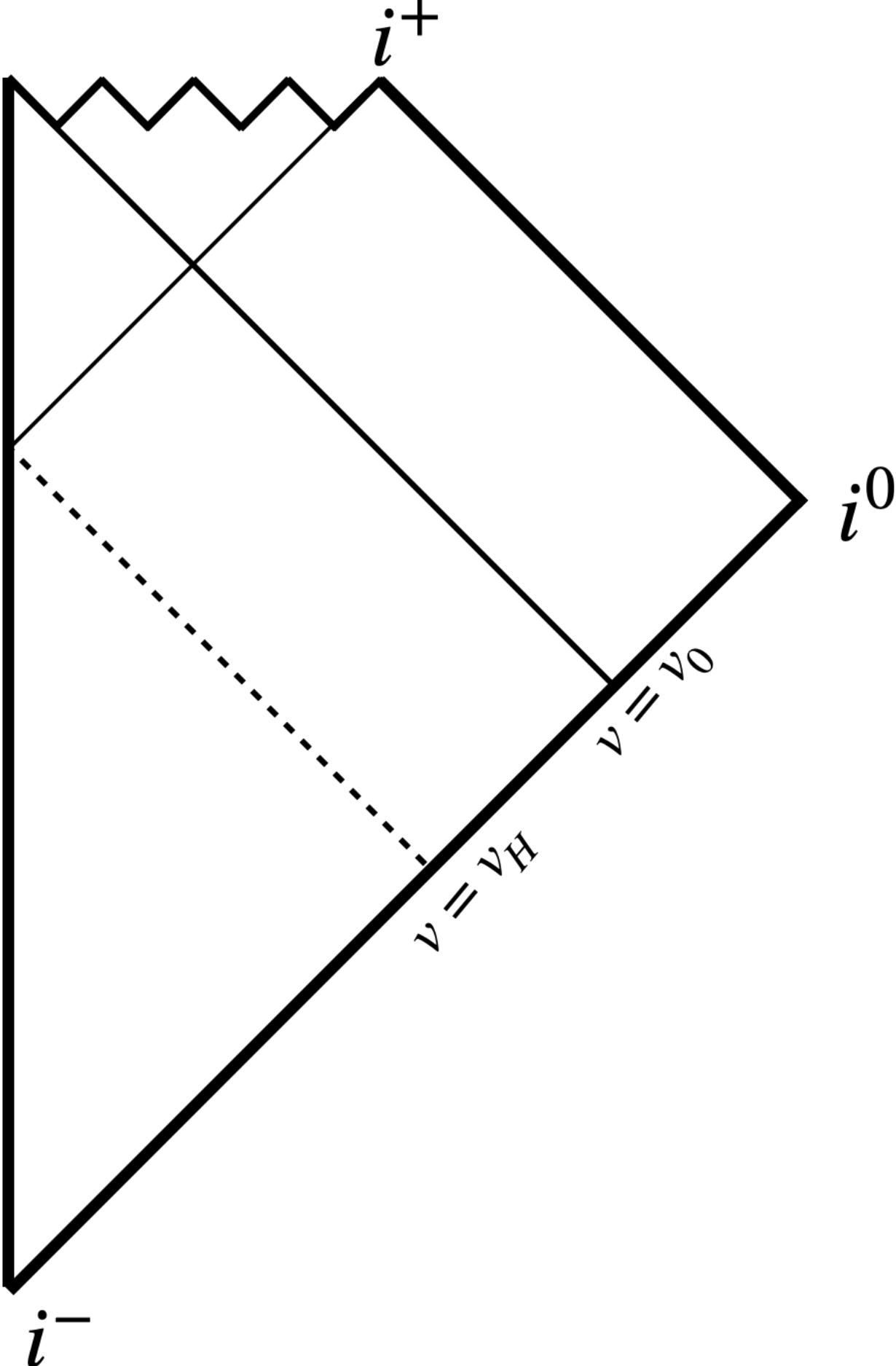} 
    \caption{Null shell collapsing into a \ac{BH} at $v=v_0$. Note that the event horizon forms at $v=v_H$ with $v_H<v_0$. In other words, it forms in the Minkowski region showing that there is no local observable that can determine the presence of an event horizon. Note that for the clarity of the picture the labels $\mathscr{I}^\pm$ are omitted.}
    \label{fig:Francesco_Schwarz_2}
\end{figure}

The details of the collapse can complicate the details of the discussion. However, we will argue that the late time flux of radiation (which is what we want to discuss) has a universal behavior. Therefore, we will consider a very idealized setting in which the collapsing matter is modeled by a single null shell. The \ac{PC} diagram of the resulting spacetime is depicted in \cref{fig:Francesco_Schwarz_2}. 

As customary, we denote by $v$ the null coordinate on $\mathscr{I}^-$. The shell is at 
\begin{equation}
    v=v_0=\text{const.}\,.
\end{equation}
We decompose the scalar field into spherical harmonics
\begin{equation}
    \phi=\sum_{l,m}\frac{\phi_{lm}(t,r)}{r}Y^l_m(\theta,\varphi) \, .
\end{equation}
For $v<v_0$ the spacetime is flat
\begin{equation}
    \rmd{}s^2=-\rmd{}u^{in}\rmd{}v+r^2\rmd\Omega^2\,,
\end{equation}
and the wave equation reads (dropping the $\{l,m\}$ subscript)
\begin{equation}
 \frac{\partial\phi}{\partial t^2}+ \frac{\partial\phi}{\partial r^{\ast2}}+  \frac{l(l+1)}{r^2}\phi=0\,.
\end{equation}
Let us now consider the portion of spacetime with $v>v_0$. The metric is given by the Schwarz\-schild \ac{BH}
\begin{equation}
    \rmd{}s^2=-\left(1-\frac{2\GN M}{r}\right)\rmd{}u^{out}\rmd{}v+r^2\rmd\Omega^2 \, ,
\end{equation}
the wave equation reads (dropping the $\{l,m\}$ subscript)
\begin{equation}
 \frac{\partial\phi}{\partial t^2}+ \frac{\partial\phi}{\partial r^{\ast2}}+   \left(1-\frac{2\GN M}{r}\right)\left(\frac{l(l+1)}{r^2}+\frac{2\GN M}{r^3}\right)\phi=0\,.
\end{equation}
We will consider a simple approximation in which the potential term is set to zero. This approximation can appear very crude. However, most of the physics will be based on effects very close to the horizon or at asymptotic distances. In both regions the potential is approximately zero, thus it is reasonable to expect that this approximation captures the main physical effects. We are going to quickly discuss the effects of this term later on.

In the Minkowski region, the solution of the wave equation in this region can be decomposed into left-going and right-going modes. 
On $\mathscr{I}^-$ a base of modes is given by the left-going modes associated with constant $v$ observers
\begin{equation}
    f_\omega^{in}=\frac{1}{4\pi\sqrt{\omega}} e^{-i\omega v} \, . 
\end{equation}
There are also right going modes that are obtained by reflecting the modes across $r=0$, with the regularity boundary condition $\phi(r=0)=0$.

In the Schwarzschild region, we again use plane wave as a base for the state. At $\mathscr{I}^+$ we are interested in the right-going modes\footnote{Note that, contrary to $\mathscr{I}^-$, $\mathscr{I}^+$ is not a Cauchy hypersurface. To be complete, we should include the left-going modes at the horizon as well. However, these modes are not visible for an asymptotic observer, which is what we are interested in.}
\begin{equation}
    f_\omega^{out}=\frac{1}{4\pi\sqrt{\omega}} e^{-i\omega u^{out}} \,.
\end{equation}
From the discussion of the next section, it follows that observers at $\mathscr{I}^-$ and $\mathscr{I}^+$ see two different vacuum states, because of the time dependence of the geometry. It should also be clear that we can relate the two vacua by computing the Bogolyubov coefficients $\alpha$ and $\beta$ defined as
\begin{align}
    \alpha_{\omega\omega'} &= (f_{\omega'}^{out},f_\omega^{in})=-i\int_{\mathscr{I}^-}\rmd{}r \, r^2\left(f_\omega^{out}\partial_vf^{in\ast}_{\omega'}-f_\omega^{in\ast}\partial_vf^{out}_{\omega'}\right)\, , \\
    \beta_{\omega\omega'} &= -(f_\omega^{out},f_{\omega'}^{in\ast})=i\int_{\mathscr{I}^-}\rmd{}r \, r^2\left(f_\omega^{out}\partial_vf^{in}_{\omega'}-f_\omega^{out}\partial_vf^{in}_{\omega'}\right)\,.
\end{align}
After integrating by parts, we get
\begin{equation}\label{eq:Francesco_Bogol}
\alpha_{\omega\omega'}=-2i\int_{\mathscr{I}^-}\rmd{}r \, r^2f_\omega^{in}\partial_vf^{out}_{\omega'}\,,\qquad
    \beta_{\omega\omega'}=2i\int_{\mathscr{I}^-}\rmd{} r \, r^2f_\omega^{in}\partial_vf^{out}_{\omega'}\,.
\end{equation}
In the next subsections, we will discuss how to evaluate these integrals. These will give us all the ingredients that we need to obtain the result of particle production by \acp{BH}.
\subsubsection{Tracing the out mode on \texorpdfstring{$\mathscr{I}^-$}{scri minus}}
\begin{figure}[ht]
    \centering
   \includegraphics[width=0.35\linewidth]{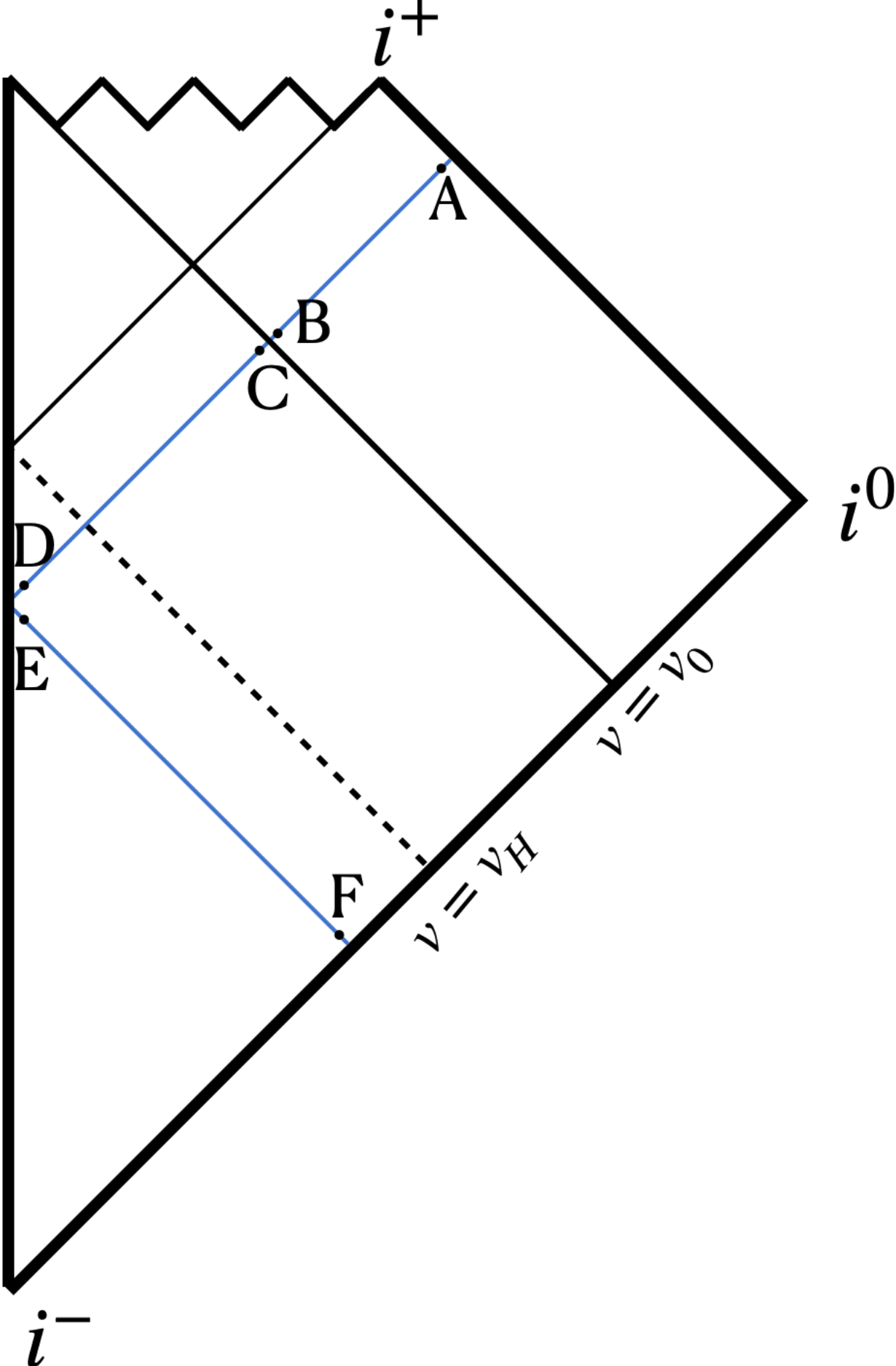} 
    \caption{Schematic representation of the tracing of a mode from $\mathscr{I}^+$ to $\mathscr{I}^-$. We start from a mode near $\mathscr{I}^+$ (A). The mode propagates freely to the past up to the point the position of the shell (B). The junction conditions are used to obtain the mode at the other side of the shell (C). The mode propagates freely up to $r=0$ (D) where it is reflected (E). Finally, the modes freely propagate once again until $\mathscr{I}^-$ (F). }
    \label{fig:Francesco_mode-tracing}
\end{figure}

To evaluate the integrals \eqref{eq:Francesco_Bogol}, we need to determine the behavior of the  $f^{out}_\omega$ states on $\mathscr{I}^-$.
To this end, we need to propagate the $out$ modes back in time starting from $\mathscr{I}^+$ to $\mathscr{I}^-$. This is schematically explained in \cref{fig:Francesco_mode-tracing} and it is quite easy to do for the setup under consideration of a \ac{BH} formed by the collapse of a single null shell. In fact, the modes propagate freely both in the Schwarzschild and in the Minkowski region. We only need to study the matching condition along the $v=v_0$ hypersurface and the reflection at $r=0$.

\paragraph{Matching at the shell.}
Let us start by matching the modes at the null shell. The form of the mode remains unchanged while we trace it back to the shell. Just before the shell, the $out$ mode is given by
\begin{equation}
        f_\omega^{out}=\frac{1}{4\pi\sqrt{\omega}} e^{-i\omega u^{out}\left(u^{in}\right)} \,.
\end{equation}
We can obtain the mode by expressing the null coordinate $u^{out}$ in terms of $u^{in}$. To this end, we remind that the radial coordinate $r$ is continuous at the shell~\cite{Poisson:2009pwt}. In the Schwarzschild region, we have
\begin{equation}
    r^\ast=r+2\GN M\ln\left|\frac{r}{2\GN M}-1\right|=\frac{v_0-u^{out}}{2}\, ,
\end{equation}
while in the Minkowski region,
\begin{equation}\label{eq:Francesco_match_Mink}
   r =\frac{v_0-u^{in}}{2}\,.
\end{equation}
Combining these two relations, we get
\begin{equation}\label{eq:Francesco_out-in}
    u^{out}=u^{in}-4\GN M\ln\left|\frac{v_0-4\GN M-u^{in}}{4\GN M}\right| \, .
\end{equation}
From this relation, we can also find the position of the event horizon. In fact, the event horizon is at $u^{out}=+\infty$. From \eqref{eq:Francesco_out-in} we get
\begin{equation}
    u^{in}_H=v_0-4\GN M\,.
\end{equation}
We can also determine the value of $v_H$ previously defined using the relation analogous of \eqref{eq:Francesco_match_Mink} for $u_H^{in}$ at $r=0$, obtaining
\begin{equation}
    v_H=v_0-4\GN M\,.
\end{equation}

\paragraph{Reflection at $r=0$.}
We now need to determine the reflection at $r=0$ in order to propagate the modes until $\mathscr{I}^-$. We start by noticing that at $r=0$ a left-going mode becomes a right-going mode. Furthermore, we must impose the regularity condition at $r=0$
\begin{equation}
    \phi(r=0)=0\,.
\end{equation}
Therefore, the left-going mode is obtained from the right-going mode by replacing $u_{in}$ with $v$ and changing the overall sign.
\begin{equation}
            f_\omega^{out}=-\frac{1}{4\pi\sqrt{\omega}} e^{-i\omega u^{out}\left(v\right)} \theta(v-v_H)\,,
\end{equation}
where $u^{out}\left(v\right)$ is obtained from \eqref{eq:Francesco_out-in} by replacing $u^{in}\to v$,
\begin{equation}\label{eq:Francesco_out-v}
    u^{out}=v-4\GN M\ln\left|\frac{v_H-v}{4\GN M}\right|\,,
\end{equation}
and the Kronecker delta encapsulates the fact that the reflection at $r=0$ is only possible for $v<v_H$. In fact, the left-going modes for $v>v_H$ enter the \ac{BH} and after the reflection at $r=0$ reach the singularity, not $\mathscr{I}^+$. 

At early times, $u^{out}\to-\infty$, and \eqref{eq:Francesco_out-v} implies $u^{out}\approx v$. Therefore, there is no particle production as the out modes match the in modes $f_\omega^{out}\approx f_\omega^{in}$. At late times,  $u^{out}\to-\infty$, we have $v\to v_H$ and 
\begin{equation}
    u^{out}\approx v_H-4\GN M\ln\left|\frac{v_H-v}{4\GN M}\right|\,.
\end{equation}
The out modes are infinitely blueshifted and there is non-trivial particle production.
\subsubsection{Particle number on \texorpdfstring{$\mathscr{I}^+$}{scri plus}}
We now have all the ingredients to compute the integrals in \eqref{eq:Francesco_Bogol}. The full computation is straightforward and can be checked, \eg{}, in~\cite{Fabbri:2005mw}. Here, we just provide a relation among the coefficients that we need:
\begin{equation}
    \left|\alpha_{\omega\omega'}\right|=e^{4\GN M\omega}    \left|\beta_{\omega\omega'}\right| \, .
\end{equation}
If we were to compute the expectation number using a continuous varying $\omega$ we would get an infinite result. This is because we are considering a \ac{BH} that is emitting for infinite time. Therefore, the total emitted number of particles is infinite. To avoid the infinite result, we can replace the plane waves with wave packets sharply peaked at $\omega$, so that $\omega$ can assume a discrete set of values. 
We can now use the property of the Bogolyubov coefficients in \eqref{eq:Francesco_bog}
\begin{equation}  \sum_{\omega'}\alpha_{\omega\omega'}\alpha^\ast_{\omega''\omega'}-\beta_{\omega\omega'}\beta^\ast_{\omega''\omega'}=\delta_{\omega\omega''} \, ,
\end{equation}
to write
\begin{equation}  \sum_{\omega'}\left|\alpha_{\omega\omega'}\right|^2-\left|\beta_{\omega\omega'}\right|^2=
    \left(e^{8\GN M\omega}-1\right)\sum_{\omega'}\left|\beta_{\omega\omega'}\right|^2=1 \, ,
\end{equation}
so we get 
\begin{tcolorbox}
\begin{equation}
     N_\omega^{out}=\frac{1}{e^{8\pi \GN M\omega}-1}\,.
\end{equation}
\end{tcolorbox}
\noindent Comparing with the Planck distribution of thermal radiation for bosons
\begin{equation}
     \frac{1}{e^{\omega/k_BT}-1} \, ,
\end{equation}
we obtain that \acp{BH} emit particles following a thermal spectrum with temperature (reinstating for the moment $\hbar$ and $c$)
\begin{tcolorbox}
\begin{equation}\label{eq:FRANCESCO-Hawk_temp}
     T=\frac{\hbar c^3}{8\pi k_B \GN M }\,.
\end{equation}
\end{tcolorbox}
\noindent Readers are encouraged to appreciate the beauty of this equation that contains natural constants that usually arise in different sector of physics (the Planck constant, the speed of light, the Boltzmann constant and Newton's constant), showing that this equation truly connects fundamental principles of nature across diverse physical domains. It is instructive to insert the values of the constants to obtain the temperature in Kelvin. We get
\begin{equation}
     T\approx 10^{-7}\left(\frac{M_\odot}{M}\right)\,\text{K},
\end{equation}
where $M_\odot$ is the mass of the sun. We can see that solar mass \acp{BH} have an extremely low temperature that is even much lower than that of the \ac{CMB} radiation (recall that $T_\text{CMB}\approx 2.7$ K.

\subsubsection{Thermal state on \texorpdfstring{$\mathscr{I}^+$}{scri plus}}
The fact that the spectrum follows the thermal distribution for photons does not imply that the radiation is thermal. To prove that the radiation is truly thermal, we need to prove that there is no correlation among modes:
\begin{equation}
    \left<in\right|N^{out}_\omega N^{out}_{\omega'}
    \left|in\right>=\left<in\right|N^{out}_\omega \left|in\right> \left<in\right|N^{out}_{\omega'}
    \left|in\right>\,,\qquad\text{for}\,\omega\neq\omega'\,.
\end{equation}
This relation, as well as similar relations for higher-order modes can be explicitly checked with calculations very similar to the one just performed (see~\eg{}~\cite{Fabbri:2005mw}). The state is, therefore, truly thermal and must be described in terms of a density matrix.In other words, it cannot be described by a pure state. 
\subsubsection{Role of the potential}
Let us now discuss the role of the potential that we have ignored. This term will change the tracing of the modes from $\mathscr{I}^+$ to $\mathscr{I}^-$. 
We will not perform a precise computation, but we can address this problem with some simple considerations that lead to the correct result. A formal calculation can be found in several books on the topic (see, \eg{}, \cite{Birrell:1982ix, Fabbri:2005mw}).

We start by noticing that a mode approaching the \ac{BH} is blueshifted while moving from $\mathscr{I}^-$ to $r\sim 2\GN M$, and then redshifted moving from $r\sim 2\GN M$ to $\mathscr{I}^-$. In a static configuration, these effects would cancel. For a collapsing geometry, the redshift exceeds the blueshift. In particular, if an event horizon forms, the redshift is arbitrarily large, while the blueshift is finite. The frequency of the modes on $\mathscr{I}^+$ is known because the radiation follows a thermal distribution with temperature given by \eqref{eq:FRANCESCO-Hawk_temp}. In particular, this frequency at $\mathscr{I}^+$ is not arbitrarily small. Tracing the modes back on $\mathscr{I}^-$, we realize that the initial frequency must have been extremely large to compensate for the arbitrary large redshift. This is referred to as the \textit{trans-Planckian problem}, and constitutes an issue as the computation is performed in a semiclassical description while Planckian scale physics might be required. However, if we ignore this issue, the very large frequency simplifies the computation. In fact, it tells us that we can use the geodesic approximation. This implies that the final result does not strongly depend on the geometry before the collapse and it is universal. Furthermore, the reflection coefficient must be zero. So the effect of the potential term does not influence the result in this region. 
Now consider the propagation of the modes close to the event horizon. In the Minkowski region, the geodesic approximation holds. In the Schwarzschild region, the potential vanishes. 
Therefore, the only non-trivial contribution of the potential is due to the fact that when the modes reach the maximum of the potential at $r\sim 3\GN M$, a fraction of them will be backscattered and fall into the \ac{BH}. 
This is the effect of depleting the asymptotic region of a fraction of the modes $1-\Gamma_{\omega l}$, where $\Gamma_{\omega l}$ is the transmission coefficient. We get
\begin{equation}\label{eq:FRANCESCO_gray_body}
    N_\omega^{out}=\frac{\Gamma_{\omega l}}{e^{8\pi \GN M\omega}-1}\,.
\end{equation}
Therefore the spectrum is not fully thermal. The coefficient $\Gamma_{\omega l}$ goes by the name of gray-body factor.
We can still consider the radiation thermal in the sense that it is the same spectrum we would see if the \ac{BH} was replaced by a thermal source without changing the potential. Note, however, that the radiation is \textbf{not} emitted from the horizon.

\begin{tcolorbox}
\begin{center}
\textbf{Question for the readers.}
\end{center}
We have just shown that the pure $\left|in\right>$ state on $\mathscr{I}^-$ evolves into a thermal state on $\mathscr{I}^+$. However, a unitary evolution should evolve pure states into pure states. Is this result paradoxical/problematic?
\end{tcolorbox}
\noindent We will answer this question in \cref{sec:Francesco_Info_loss}. While the answer might appear very simple (the title of \cref{sec:Francesco_Info_loss} is ``Information loss problem''), readers are encouraged to actually think about this question as the answer is not as trivial as it might seem.


\subsection{Quantum stress-energy tensor}\label{Sec:Francesco-SET}
In the study of quantum effects in generic spacetimes, it is extremely useful to define a notion of a quantum stress-energy tensor. Given a field $\phi$ in a generic state $\left|\psi\right>$ and a stress-energy tensor $T_{\mu\nu}$ defined as
\begin{equation}
    T_{\mu\nu}=\frac{1}{\sqrt{-g}}\frac{\delta S_\phi}{\delta g^{\mu\nu}}\,,
\end{equation}
we consider the expectation value of the stress-energy tensor on the state $\left|\psi\right>$,
\begin{equation}
    \left<\psi\right|\hat{T}_{\mu\nu}\left|\psi\right>\,.
\end{equation}
This quantity is crucial to studying the gravitational dynamics in regimes where the quantum nature of the gravitational field can be ignored, but the quantum nature of the matter fields plays an important role. In fact, we can consider the semiclassical Einstein equations
\begin{equation}
    G_{\mu\nu}=8\pi \GN T_{\mu\nu}^\text{cl}+ 8\pi \GN \left<\psi\right|\hat{T}_{\mu\nu}\left|\psi\right>\,.
\end{equation}
which are expected to provide a first approximation for the dynamics of a theory of \ac{QG}. 
As we will see, even in the test field approximation we can learn valuable lessons from the expression of the stress-energy tensor. In particular, we will discuss the answer to two crucial questions. 
\begin{enumerate}
    \item We have seen that Hawking radiation is a finite flux of energy at infinity that continues forever (or for a very long time if we include backreaction). However, at the horizon all this energy piles up in a region that is crossed by an infalling observer in a very short timescale.  Does it mean that an incoming observer will measure an infinite number of particles and infinite energy?
    \item The \ac{BH} emits energy at infinity. Can energy be conserved?
\end{enumerate}

\subsubsection{Two-dimensional black holes}

Let us now show that the two-dimensional stress-energy tensor corresponds to the $s$-wave approximation for a spherically symmetric spacetime.
The classical action for a free scalar field in four dimensions reads
\begin{equation}
    S=\int \rmd^4x\sqrt{-g^{(4)}} \, \left[ -\frac{1}{2} \left( \covD \phi \right)^2 \right] \,.
\end{equation}
The line element is
\begin{equation}
    \rmd{}s^2=g_{ab}\rmd{}x^a\rmd{}x^b+r^2\rmd\Omega^2\,.
\end{equation}
The field $\phi$ can be expanded in spherical harmonics. The $s-$wave approximation consists of considering only the term related to $l=m=0$. In turn, this implies that the scalar field does not have any angular dependence. We get
\begin{equation}\label{eq:Francesco_S_swave}
        S^{(4)}=4\pi\int \rmd^2x\sqrt{-g^{(2)}} \, r^2 \, \left[ -\frac{1}{2} \, \left( \covD \phi \right)^2 \right] \,.
\end{equation}
Simply comparing the functional variation of the action in \eqref{eq:Francesco_S_swave} with the functional variation of a two-dimensional action
\begin{equation}
            S^{(2)}=\int \rmd^2x\sqrt{-g^{(2)}} \, \left[ -\frac{1}{2} \, \left( \covD \phi \right)^2 \right] \,,
\end{equation}
we get that, for the non-angular components
\begin{equation}\label{eq:Francesco_Polya}
        T_{ab}^{(4)}=\frac{1}{4\pi r^2 }  T_{ab}^{(2)} \,.
\end{equation}
This means that the results of the two-dimensional analysis can be transferred to a spherically symmetric four-dimensional system. Of course, this approximation breaks down near $r=0$ where the prefactor in \eqref{eq:Francesco_Polya} diverges. However, near $r=0$ we already know that we cannot trust the semiclassical approximation.

\subsubsection{Generic results}
We are able to compute the expectation value of the stress-energy tensor for a two-dimensional spacetime because we can relate it to the so-called conformal anomaly.
If we consider an action which is invariant under conformal symmetry
\begin{equation}\label{eq:Francesco_Weyl_trans}
    g_{\mu\nu}\to\Omega^2(x)g_{\mu\nu}\,,
\end{equation}
we can easily show that the trace of the classical stress-energy tensor is zero.
In fact, considering an infinitesimal version of \eqref{eq:Francesco_Weyl_trans}, with $\Omega^2=1+\omega$, we get
\begin{equation}
    \delta g_{\mu\nu}=\omega g_{\mu\nu}\,,\qquad0=\delta S=\int \rmd^nx\sqrt{-g} \, \omega T^{\mu\nu}g_{\mu\nu}\,\Leftrightarrow T^{\mu\nu}g_{\mu\nu}=0\,.
\end{equation}
However, at the quantum level, the expectation value of the trace picks up a non-zero value during the renormalization procedure. The trace can be computed exactly. In two dimensions we obtain~\cite{Birrell:1982ix}
\begin{equation}
    \left<T^{(2)}\right>=a R\,,
\end{equation}
where $R$ is the Ricci scalar and $a$ is a constant that can be evaluated. 

The trace anomaly can be computed in four dimensions as well. What is special about the two-dimensional case is that the knowledge of the trace can be used to compute the full stress-energy tensor. The full computation goes beyond the scope of this section as it is quite long and requires some notions on the normalization of the stress-energy tensor. However, the main reason why this computation is possible is that any two-dimensional metric is conformally flat, \ie{}, it is always possible to choose $\Omega$ such that
\begin{equation}\label{eq:Francesco_conf_flat}
    g_{\mu\nu}=\Omega^2(x)\eta_{\mu\nu} \, .
\end{equation}
Given that the action is conformally invariant, the problem is said to be ``conformally trivial" as, modulo a conformal transformation, we can reduce the problem to the corresponding problem in flat spacetime. The trace anomaly can provide information regarding the effect of the conformal transformation.

We will not discuss the full computation, but we can get an idea of the steps involved. The full computation is discussed in~\cite{Birrell:1982ix}.
\begin{itemize}
    \item First, we need to define an effective action $\Gamma$  which is related to the expectation value of the stress-energy tensor in the same way the classical action is related to the classical stress-energy tensor, \ie{},
\begin{equation}
    \left<\hat{T}_{\mu\nu}\right>=\frac{-2}{\sqrt{-g}}\frac{\delta \Gamma}{\delta g^{\mu\nu}}\,.
\end{equation}
\item We now need to get rid of the divergences by a renormalization procedure.
\item Using the transformation of \eqref{eq:Francesco_conf_flat} we get
\begin{equation}
    \Gamma\left[g\right]=\Gamma\left[\eta\right]+\int \rmd^2x\sqrt{-g}   \left<\hat{T}_{\phantom{\mu}\mu}^\mu\right>\delta \Omega^2 \, .
\end{equation}
\end{itemize}
When performing the functional variation, the first term will give a contribution that depends on the vacuum state and that does not directly depend on the spacetime curvature. The second term will give a vacuum ``polarization'' contribution that depends on the spacetime curvature, and it can be obtained via the information on the trace anomaly.

\subsubsection{Regularity conditions}
Before showing the results of the expectation values of the stress energy tensor for some particularly interesting vacuum states, let us briefly discuss one regularity requirement for a well-posed quantum stress energy tensor~\cite{Christensen:1977jc, Balbinot:2023vcm}.

We require that all the components are finite in a coordinate basis that is well-behaved everywhere. Such coordinates, for instance, are the Kruskal coordinates introduced in \cref{Sec:Francesco-Preliminaries}.
The results are often discussed in the $(u,v)$ coordinates which are not well-behaved at the horizon. We can easily change coordinates and obtain
\begin{align}
     \left<\hat{T}_{UU}\right> &= \left<\hat{T}_{uu}\right>\left(\frac{\partial U}{\partial u}\right)^{-2}\propto U{^{-2}}\left<\hat{T}_{uu}\right> \, , \\
    \left<\hat{T}_{UV}\right> &= \left<\hat{T}_{uv}\right>\left(\frac{\partial U}{\partial u}\right)\left(\frac{\partial V}{\partial v}\right)\propto U^{-1}V^{-1}\left<\hat{T}_{uv}\right> \, , \\
    \left<\hat{T}_{VV}\right> &= \left<\hat{T}_{vv}\right>\left(\frac{\partial V}{\partial v}\right)^{-2}\propto V{^{-2}}\left<\hat{T}_{vv}\right> \, .
\end{align}
With reference to \cref{fig:Francesco_Schwarz} at the future horizon $\mathscr{H}^+_R$, $V$ is finite while $U\propto1-2\GN M/r=0$. Therefore, the stress-energy tensor is regular on $\mathscr{H}^+_R$ if 
\begin{tcolorbox}
\begin{equation}
\begin{aligned}
    \left(1-\frac{2\GN M}{r}\right)^{-2}\left<\hat{T}_{uu}\right> &< \infty \, ,\\
    \left(1-\frac{2\GN M}{r}\right)\left<\hat{T}_{uv}\right> &< \infty \, , \\
    \left<\hat{T}_{vv}\right> &< \infty \, .
\end{aligned}
\end{equation}
\end{tcolorbox}
\noindent On the other hand, at the past horizon $\mathscr{H}^-_R$, $U$ is finite while $V\propto1-2\GN M/r=0$. Therefore, the stress-energy tensor is regular on $\mathscr{H}^-_R$ if 
\begin{tcolorbox}
\begin{equation}
\begin{aligned}
    \left<\hat{T}_{uu}\right> &< \infty \, , \\
    \left(1-\frac{2\GN M}{r}\right)\left<\hat{T}_{uv}\right> &< \infty \, , \\
    \left(1-\frac{2\GN M}{r}\right)^{-2}\left<\hat{T}_{vv}\right> &< \infty \, .
\end{aligned}
\end{equation}
\end{tcolorbox}
We will now study a few important vacuum states and check if they satisfy these regularity conditions.

\subsubsection{Different choices of the vacuum state}
We will now show and discuss the expectation value of the stress-energy tensor in a few particularly interesting quantum states. We consider a static two-dimensional Schwarzschild \ac{BH}. The \ac{PC} diagram is the one in the left panel of \cref{fig:Francesco_Schwarz}.

\subsubsubsection*{Boulware state}
The first noteworthy state is the Boulware state $\left|B\right>$, which is a state that is vacuum both for observers at $\mathscr{I}^-$ and $\mathscr{I}^+$. To obtain this state, we consider the left-going modes
\begin{equation}
    f_L\propto e^{-i\omega v}\,.
\end{equation}
This choice fixes the splitting into positive and energy modes on $\mathscr{I}^-$ but they do not represent a basis as we also need the right-going modes. To guarantee a vacuum state on $\mathscr{I}^+$ we choose as right-going modes
\begin{equation}
    f_R\propto e^{-i\omega u}\,,
\end{equation}
so we can write the field as
\begin{equation}\label{eq:Francesco_B_state}
    \phi=\sum_\omega \left[ \frac{1}{4\pi\sqrt{\omega}} e^{-i\omega v} a_\omega+\frac{1}{4\pi\sqrt{\omega}} e^{-i\omega u} a_\omega \right] + h.c.\,.
\end{equation}
There is no Hawking radiation in this state as the splitting of positive and negative frequency is the same in the $in$ and $out$ region. However, the expectation value of the stress-energy tensor is non-zero due to what we can call ``vacuum polarization". This is an effect of the spacetime curvature. A detailed computation shows~\cite{Birrell:1982ix}
\begin{equation}
\begin{aligned}
    \left<B\right|\hat{T}_{uu}\left|B\right>=\left<B\right|\hat{T}_{vv}\left|B\right> &= \frac{1}{24\pi}\left(-\frac{\GN M}{r^3}+\frac{3}{2}\frac{\GN^2M^2}{r^4}\right) \, , \\
    \left<B\right|\hat{T}_{uv}\left|B\right> &= -\frac{1}{24\pi}\left(1-\frac{2\GN M}{r}\right)\frac{\GN M}{r^3} \, .
\end{aligned}
\end{equation}
We can see that all the components are finite and vanish at infinity, and that the flux vanishes as well, \ie{}, asymptotic observers do not detect any particles neither on $\mathscr{I}^-$ nor on $\mathscr{I}^+$. However, the $(v,v)$ and $(u,u)$ components do not vanish at the horizon. For what is explained in the previous section, this implies that the state is not regular neither at the future horizon $\mathscr{H}^+_R$, nor at the past horizon $\mathscr{H}^-_R$. This implies that an observer crossing the horizon would measure an infinite flux of energy.

Different states can be obtained by changing the splitting into positive and negative energy modes. One way to do that would be to introduce new coordinates 
\begin{equation}\label{eq:Francesco_new_state}
    \Tilde{u}(u)\,,\qquad  \text{and}\qquad \Tilde{v}(v)\,.
\end{equation}
In this way \eqref{eq:Francesco_B_state} changes to
\begin{equation}\label{eq:Francesco_gen_state}
    \phi=\sum_\omega \left[ \frac{1}{4\pi\sqrt{\omega}} e^{-i\omega \Tilde{v}} a_\omega+\frac{1}{4\pi\sqrt{\omega}} e^{-i\omega \Tilde{u}} a_\omega \right] + h.c.\,.
\end{equation}
The stress energy tensor becomes~\cite{Birrell:1982ix}
\begin{equation} 
    \begin{aligned}
    \left<\Tilde{0}\right|\hat{T}_{uu}\left|\Tilde{0}\right> &= \left<B\right|\hat{T}_{uu}\left|B\right>-\frac{1}{24\pi}\left\{\Tilde{u},u\right\} \, , \\
    \left<\Tilde{0}\right|\hat{T}_{vv}\left|\Tilde{0}\right> &= \left<B\right|\hat{T}_{vv}\left|B\right>-\frac{1}{24\pi}\left\{\Tilde{v},v\right\} \, , \\
    \left<\Tilde{0}\right|\hat{T}_{uv}\left|\Tilde{0}\right> &= \left<B\right|\hat{T}_{uv}\left|B\right> \, ,
\end{aligned}
\end{equation}
where the curly brackets denote the Schwarzian derivative
\begin{equation}
    \left\{f(x),x\right\}\equiv\frac{f'''(x)}{f'(x)}-\frac{3}{2}\left(\frac{f''(x)}{f'(x)}\right)\,.
\end{equation}

\subsubsubsection*{Unruh state}
Next, we want to consider the Unruh state $\left|U\right>$. This state is a vacuum on $\mathscr{I}^-$ and it is chosen to be well-behaved on the future horizon. Therefore, the left-going modes are the same as the Boulware state, while the right-going mode are replaced with modes relative to the Kruskal coordinate $U$ which is well-behaved at the horizon:
\begin{equation}\label{eq:Francesco_U_state}
    \phi=\sum_\omega \left[ \frac{1}{4\pi\sqrt{\omega}} e^{-i\omega v} a_\omega+\frac{1}{4\pi\sqrt{\omega}} e^{-i\omega U} a_\omega \right] + h.c.\,.
\end{equation}
Using the definition of the Kruskal coordinate, we can easily derive
\begin{equation}
    -\frac{1}{24\pi}\left\{\Tilde{u},u\right\}=\frac{k^2}{48\pi} \, ,
\end{equation}
and the stress energy tensor
\begin{equation}\label{eq:Francesco_T_U}
\begin{aligned}
\left<U\right|\hat{T}_{uu}\left|U\right> &= \left<B\right|\hat{T}_{uu}\left|B\right>-\frac{1}{24\pi}\left\{U,u\right\} \\
&= \frac{1}{32\GN^2M^2}\left(1-\frac{2\GN M}{r}\right)^2\left(1 + 4 \frac{\GN M}{r} + 12 \left(\frac{\GN M}{r} \right)^2 \right) \, , \\
\left<U\right|\hat{T}_{vv}\left|U\right> &= \left<B\right|\hat{T}_{vv}\left|B\right> \, , \\
    \left<U\right|\hat{T}_{uv}\left|U\right> &= \left<B\right|\hat{T}_{uv}\left|B\right> \, .
\end{aligned}
\end{equation}
Therefore, the stress energy tensor is well-behaved at the future horizon, but divergent at the past horizon.
Note that $\left<U\right|\hat{T}_{uu}\left|U\right>$ does not vanish asymptotically. This means that there is an out-going flux that is extending all the way to $\mathscr{I}^+$. This is the Hawking radiation.
\subsubsubsection*{Hartle--Hawking state}
Another interesting state is the Hartle-Hawking state $\left|HH\right>$ which is regular everywhere. 
To obtain it, we replace $v$ with the Kruskal coordinate $V$: 
\begin{equation}\label{eq:Francesco_HH_state}
    \phi=\sum_\omega \left[ \frac{1}{4\pi\sqrt{\omega}} e^{-i\omega V} a_\omega+\frac{1}{4\pi\sqrt{\omega}} e^{-i\omega U} a_\omega \right] + h.c. \, .
\end{equation}
It is straightforward to check that the resulting expectation value is everywhere regular. However, neither $\left<HH\right|\hat{T}_{uu}\left|HH\right>$ nor $\left<HH\right|\hat{T}_{vv}\left|HH\right>$ vanish asymptotically. This implies that there is an ingoing flux of radiation which is originating on $\mathscr{I}^-$ and an outgoing flux that extends all the way to $\mathscr{I}^+$. As this is the only state that is everywhere regular, this is the correct state to use to describe a quantum particle on an eternal \ac{BH}. This implies that eternal \acp{BH} do not create particles (that would be impossible without any time dependence), but they are in a thermal bath.

\subsubsubsection*{\texorpdfstring{$\left|in\right>$}{in} state}
The final state we want to study is the $\left|in\right>$ discussed in the previous class. The outgoing modes are the ones obtained by reflecting the ingoing modes across $r=0$, 
\begin{equation}\label{eq:Francesco_Uin_state}
    \phi=\sum_\omega \left[ \frac{1}{4\pi\sqrt{\omega}} e^{-i\omega v} a_\omega+\frac{1}{4\pi\sqrt{\omega}} e^{-i\omega u^{in}} a_\omega \right] + h.c. \, .
\end{equation}
We can compute the Schwarzian derivative of $u^{in}$ with respect to $u=u^{out}$. A shortcut is to note that at late times, \eqref{eq:Francesco_out-in} shows that $u^{in}\approx U$. Therefore, at late times the $\left|in\right>$ state is approximated by the Unruh state. The divergence at the past horizon is not particularly worrisome as the $\left|in\right>$ state describes the quantum field on a background of a \ac{BH} formed by gravitational collapse. The past horizon is not part of the spacetime.

We can now answer the two questions given at the beginning of the section:
\begin{enumerate}
    \item \textit{The Hawking radiation is a finite flux of energy at infinity. At the horizon, we have infinite blueshift. Does it mean that an incoming observer will measure an infinite number of particles and infinite energy?}
\end{enumerate}
The contribution of the stress energy tensor due to the Hawking flux is given by the term proportional to the Schwarzian derivative in \eqref{eq:Francesco_T_U}. This term is indeed infinite. However, this term is canceled by the divergent vacuum polarization (the first term in \eqref{eq:Francesco_T_U}). Therefore, an observer would only observe a finite energy.
\begin{enumerate}
\setcounter{enumi}{1}
    \item \textit{The \ac{BH} emits energy at infinity. Can energy be conserved?}
\end{enumerate}
At the future event horizon (where the state is perfectly approximated by the Unruh state)
\begin{equation}\label{eq:Francesco_inco_rad_hor}
     \left.\left<in\right|\hat{T}_{vv}\left|in\right>\right|_{r=2\GN M}=-\frac{1}{648\GN^2M^2\pi}\,.
\end{equation}
On the other hand, at late times on $\mathscr{I}^+$,
\begin{equation}
     \left.\left<in\right|\hat{T}_{uu}\left|in\right>\right|_{r\to \infty}=\frac{1}{648\GN^2M^2\pi}\,.
\end{equation}
Therefore, besides the positive flux of energy at infinity due to Hawking radiation, at the horizon there is an in-going flux of negative energy. This is a first step towards energy conservation. We have discarded the backreaction on the geometry, therefore the mass of the \ac{BH} is considered constant.

However, this is just an approximation. If the backreaction was included, the mass of the \ac{BH} should decrease, not because anything is escaping the horizon, but because negative energy is coming in.

\subsubsubsection*{Other states}

Of course it is possible to consider other choices for the vacuum states. We might wonder if it is possible to construct a state for which the expectation value of the stress energy tensor vanishes on $\mathscr{I}^-$ and it is regular everywhere. In other words, we would like to construct a state that approximates the Boulware state on $\mathscr{I}^-$ and the Hartle--Hawking state on  $\mathscr{H}^-$ and  $\mathscr{H}^+$. In principle this is possible, but it would lead to an expectation value that is time-dependent. This can be very easily seen if we restrict to the states obtained via the transformations \eqref{eq:Francesco_new_state}. In fact, to have a time-independent expectation value, the Schwarzian derivative term must be a function only of $r$. However, $\left\{\Tilde{u},u\right\}$ is a function of $u=t-r^\ast(r)$. Therefore, the only possibility to cancel the time dependence is for the Schwarzian derivative to be a constant. So, it cannot remove the divergences at the horizon and still vanish at infinity. A similar consideration applies to $\Tilde{v}$.

\subsection{Information loss problem}\label{sec:Francesco_Info_loss}

In \cref{Sec:Francesco-Haw-rad}, we have learned that \acp{BH} emit Hawking radiation. We have also discussed that the emitted radiation is thermal, meaning that it has a Planckian spectrum and that the emitted modes have no correlation on $\mathscr{I}^+$. We have closed the section with a question for the readers, asking if this result is paradoxical/problematic. This is going to be the topic of this section, in which we will discuss the information loss paradox/problem~\cite{Hawking:1976ra}. 
\subsubsection{Pure states and mixed states}
Let us start by reminding the readers about the notions and some properties of pure and mixed states. A pure quantum state is a state which can be described by a single state vector among the eigenstate defining a complete set. On the other hand, a mixed quantum state is a statistical ensemble of pure states and must be described by a density matrix
\begin{equation}
    \rho=\sum_i\rho_i\left|\psi_i\right>\left<\psi_i\right| \, .
\end{equation}
Simple example: Consider a system for which the waveform can exist in two configurations $\psi_1$ and $\psi_2$. A density matrix corresponding to an ensemble half in the first state and half in the second state is
\begin{equation}
    \rho=\frac{1}{2}\left(\left|\psi_1\right>\left<\psi_1\right|+\left|\psi_2\right>\left<\psi_2\right|\right)\,.
\end{equation}
On the other hand, a superposition of the two configurations would be described by a state
\begin{equation}
    \left|\psi\right>=\frac{1}{\sqrt{2}}\left(\left|\psi_1\right>+\left|\psi_2\right>\right)\,,\quad\rho=\left|\psi\right>\left<\psi\right|=\frac{1}{2}\left(\left|\psi_1\right>\left<\psi_1\right|+\left|\psi_1\right>\left<\psi_2\right|+\left|\psi_2\right>\left<\psi_1\right|+\left|\psi_2\right>\left<\psi_2\right|\right)\,.
\end{equation}
We also need to define the von Neumann and thermodynamic entropies. The von Neumann entropy is defined as
\begin{equation}
    S_\text{vN}=-\text{Tr}\left(\rho\ln\rho\right)\,.
\end{equation}
The thermodynamic entropy is obtained by considering only a few macroscopic quantities and maximizing the von Neumann entropy among all the microstates reproducing those macroscopic variables (\eg{}, energy, pressure, \dots). Some relevant properties are:
\begin{itemize}
    \item From the definition, it immediately follows that the von Neumann entropy cannot exceed the thermodynamic entropy,
    \begin{equation}\label{eq:Francesco_Entr_comp}
        S_\text{Th}\geq S_\text{vN}\,.
    \end{equation}
    
    \item The von Neumann entropy vanishes if and only if the state is pure.
    
    \item The von Neumann entropy is conserved for unitary evolution of the system.
    
    \item Consider a quantum system in a pure state made of two parts $A$ and $B$. We can define the von Neumann entropy for the subsystem $A$ ($B$) by considering the density matrix $\rho_A$ ($\rho_B$) obtained by tracing out the remaining degrees of freedom. The resulting entropies are non-zero and equal to each other
    \begin{equation}\label{eq:Francesco_Ent_states}
        S_\text{vN}(A \cup  B)=0\,,\qquad S_\text{vN}(A)=S_\text{vN}(B)\neq 0\,.
    \end{equation}
\end{itemize}

\subsubsection{State of Hawking radiation}\label{sec:Francesco-state_Haw_rad}
Let us now discuss the answer to the question with which we closed \cref{Sec:Francesco-Haw-rad}. Is starting from a pure state on $\mathscr{I}^-$ and obtaining a mixed state on $\mathscr{I}^+$ a problematic result? Indeed, this result might appear paradoxical. However, it is not (yet)! In fact, we have ignored the modes that enter the horizon. Contrary to $\mathscr{I}^-$, $\mathscr{I}^+$ is not a Cauchy hypersurface. We need to add the horizon to not ignore a portion of spacetime. Basically, considering only the asymptotic portion of the Cauchy hypersurface we are tracing out the degrees of freedom inside the \ac{BH} event horizon. We know that tracing out a subset of the degrees of freedom turns a pure state into a mixed state. No paradox here. 

In fact, it is possible to explicitly compute the correlations among modes on $\mathscr{I}^+$ and at the horizon to show that there are correlations among late time modes on $\mathscr{I}^+$ and modes that enter the horizon at early times. It is important to stress that even without doing the explicit computation we know that result. The theory and the field equations we have used lead to a well-defined initial value problem and, in particular, to a unitary evolution of the quantum states. 

However, there are issues concerning the evolution of quantum states regarding the evaporation of \acp{BH}. These issues arise once extra elements are added to the analysis. 
In particular, the derivation of Hawking radiation is only valid in the test field approximation, so it did not consider the backreaction of the field to the metric. This is only a reasonable approximation if the total energy of the radiation is much smaller than the \ac{BH} mass. Otherwise, we would have violations of energy conservation. For stellar mass \acp{BH}, this is definitely a good approximation for a very long time. However, if we do not include the backreaction, the evaporation will continue for an infinite time, emitting infinite energy. Eventually, the approximation must break down. To avoid this problem, we artificially include the fact that the mass of the \ac{BH} must decrease at the specific rate needed to compensate for the energy loss by Hawking radiation. Since, for large \acp{BH}, the radiation is emitted at a very low rate, we can confidently assume the adiabatic or quasi-static approximation and assume that Hawking radiation proceeds with temperature given by  \eqref{eq:FRANCESCO-Hawk_temp} with a time-dependent mass. This is a quite reasonable assumption that can also be formalized~\cite{Barcelo:2010pj}.

The information loss paradox/problem arises when we try to incorporate the mass loss of the \ac{BH} and add some other assumptions that cannot be derived within the model. We will give a precise formulation of the problem in the next sections.

\subsubsection{Information loss problem: complete evaporation}\label{subsec:Francesco_par-A}
The simplest way to formulate the problem is by assuming that the evaporation continues in the way predicted by semiclassical gravity until the \ac{BH} completely vanishes. We also need to assume that the end point of the evaporation is a regular spacetime.

\begin{figure}[ht]
    \centering
   \includegraphics[width=0.34\linewidth]{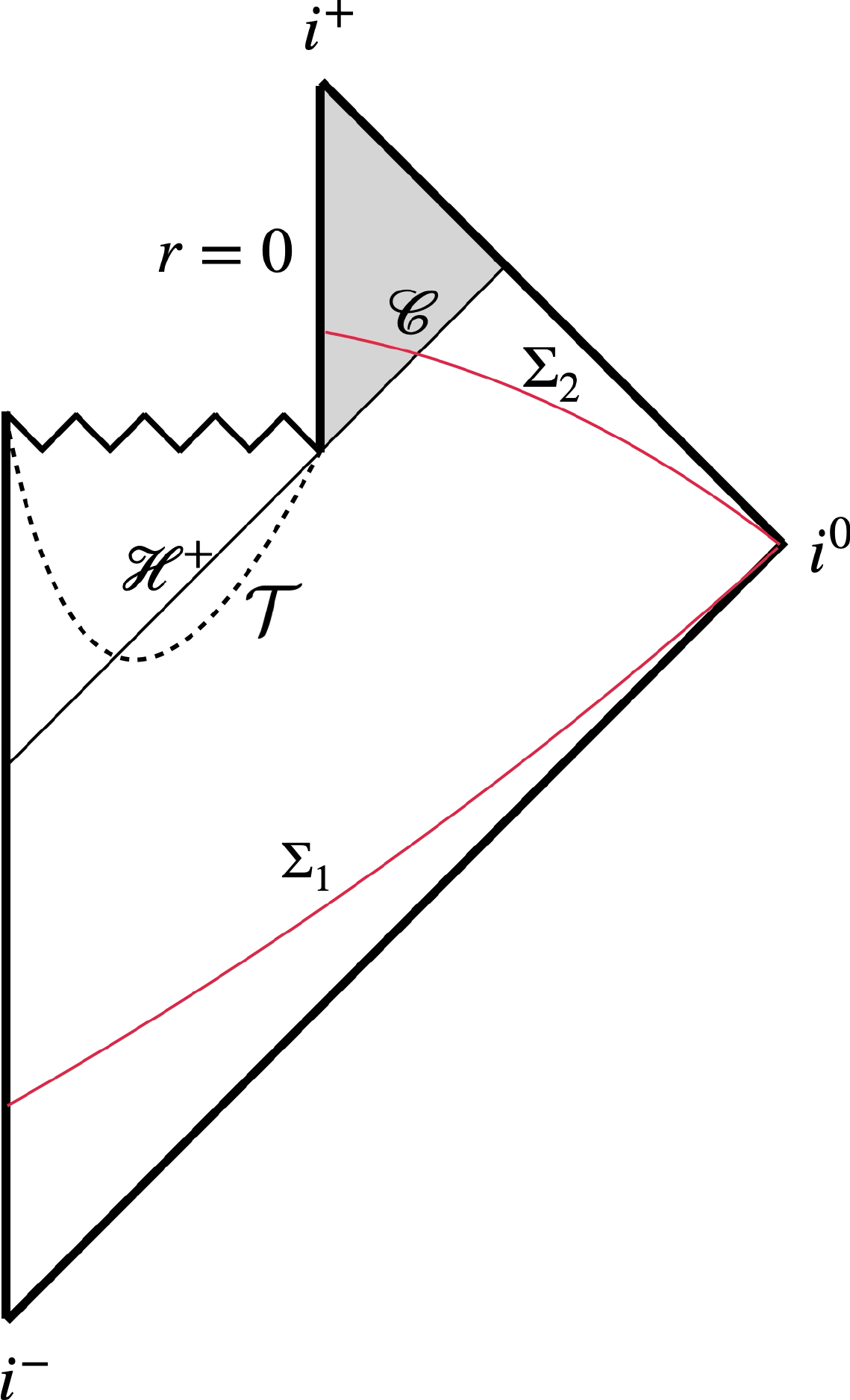} 
    \caption{\ac{PC} diagram of a \ac{BH} formed by gravitational collapse that evaporates completely in finite time. The evaporation process is assumed to leave a spacetime without singularity, where $r=0$ is regular and timelike after the complete evaporation of the \ac{BH}. The spacetime has an event horizon ($\mathscr{H^+}$), a Cauchy horizon ($\mathscr{C}$) and a trapping horizon ($\mathcal{T}$). The red lines $\Sigma_{1,2}$ are two Cauchy hypersurfaces. We can see that $\Sigma_2$ extends beyond the Cauchy horizon, thus we cannot trust the predictions of semiclassical gravity to determine the state on $\Sigma_2$.}
    \label{fig:Francesco_Complete_evaporation}
\end{figure}

This configuration is represented in \cref{fig:Francesco_Complete_evaporation}, in which the \ac{BH} evaporates completely and the line $r=0$ is regular and timelike. If we assume that the evaporation continues in the way described in \cref{Sec:Francesco-Haw-rad} , we arrive at the conclusion that a pure state on $\Sigma_1$ evolves into a mixed state on $\Sigma_2$ (red lines in \cref{fig:Francesco_Complete_evaporation}). 
This clearly contradicts our assumptions. However, there is a lot that can go wrong. First of all, the semiclassical computation is expected to break down at some point. Assuming its validity until the end of the evaporation is not well-justified. Furthermore, even if we believe to this picture, would it be reasonable to expect that a pure state on $\Sigma_1$ evolves into a pure state on $\Sigma_2$? In fact, according to this picture, the spacetime has an event horizon. By definition, anything that enters the event horizon is lost for an outside observer. We actually expect the state to be mixed if we only focus on the region outside the event horizon. Furthermore, in \cref{fig:Francesco_Complete_evaporation} we can see that the spacetime has a Cauchy horizon $\mathscr{C}$ as at the end of the evaporation, we have a region of the spacetime that is in the causal future of the singularity. We can see that the Cauchy hypersurface $\Sigma_2$ crosses the Cauchy horizon. From the point of view of the semiclassical theory, we cannot trust any prediction beyond the Cauchy horizon, so it is a conceptual error to draw conclusions on the quantum state on $\Sigma_2$. For completeness, \cref{fig:Francesco_Complete_evaporation} also shows the trapping horizon (dashed line), which does not agree with the event horizon as the spacetime is not stationary. In particular, as the \ac{BH} is losing mass via Hawking radiation, the trapping horizon extends beyond the event horizon.

Therefore, this formulation of the information loss paradox is not particularly worrisome. In the next section, we are going to discuss a second formulation of the problem which is, arguably, more relevant.

\subsubsection{Information loss problem: entropy problem}\label{subsec:Francesco_par-B}
In this section we present a second formulation of the problem. This formulation follows the so-called Page argument and we will refer to it as the ``entropy problem'' \cite{Page:1993wv}.

For this formulation of the problem, we only need to assume the validity of the semiclassical picture as a low-energy effective theory to describe \ac{BH} physics far from the Planckian regime. In particular, we do not assume anything regarding the final state of \ac{BH} evaporation. However, the price to pay is that we need to add the following
\begin{center}
    \textit{\textbf{Assumption:} As seen from the outside, a \ac{BH} behaves like a quantum system whose number of degrees of freedom is bounded by $A/4\GN$, where $A$ is the horizon area.\footnote{It is usually not specified if the word ``horizon'' refers to the event horizon or other types of horizon. This is not relevant for the formulation of the problem.}}
\end{center}

Let us discuss why the inclusion of this hypothesis leads to a contradiction without the need of assuming anything on the end point of gravitational collapse~\cite{Page:1993wv, Page:1993up, Almheiri:2020cfm}. After that, we will explain the reasoning behind this assumption.
Let us denote with $\mathcal{H}_{\rm bh}$ and $\mathcal{H}_{\rm rad}$ the Hilbert spaces of the \ac{BH} and of the radiation, respectively. As customary, we consider a configuration in which the initial state before \ac{BH} formation is a pure state, but similar considerations can be done with a different initial state. The joint state of \ac{BH} plus radiation, \ie{}, $\left|\psi \right\rangle \in \mathcal{H}_{\rm bh}\otimes \mathcal{H}_{\rm rad},$ must remain pure if the dynamical evolution is unitary.

\begin{figure}[hb]
    \centering
   \includegraphics[width=0.5\textwidth]{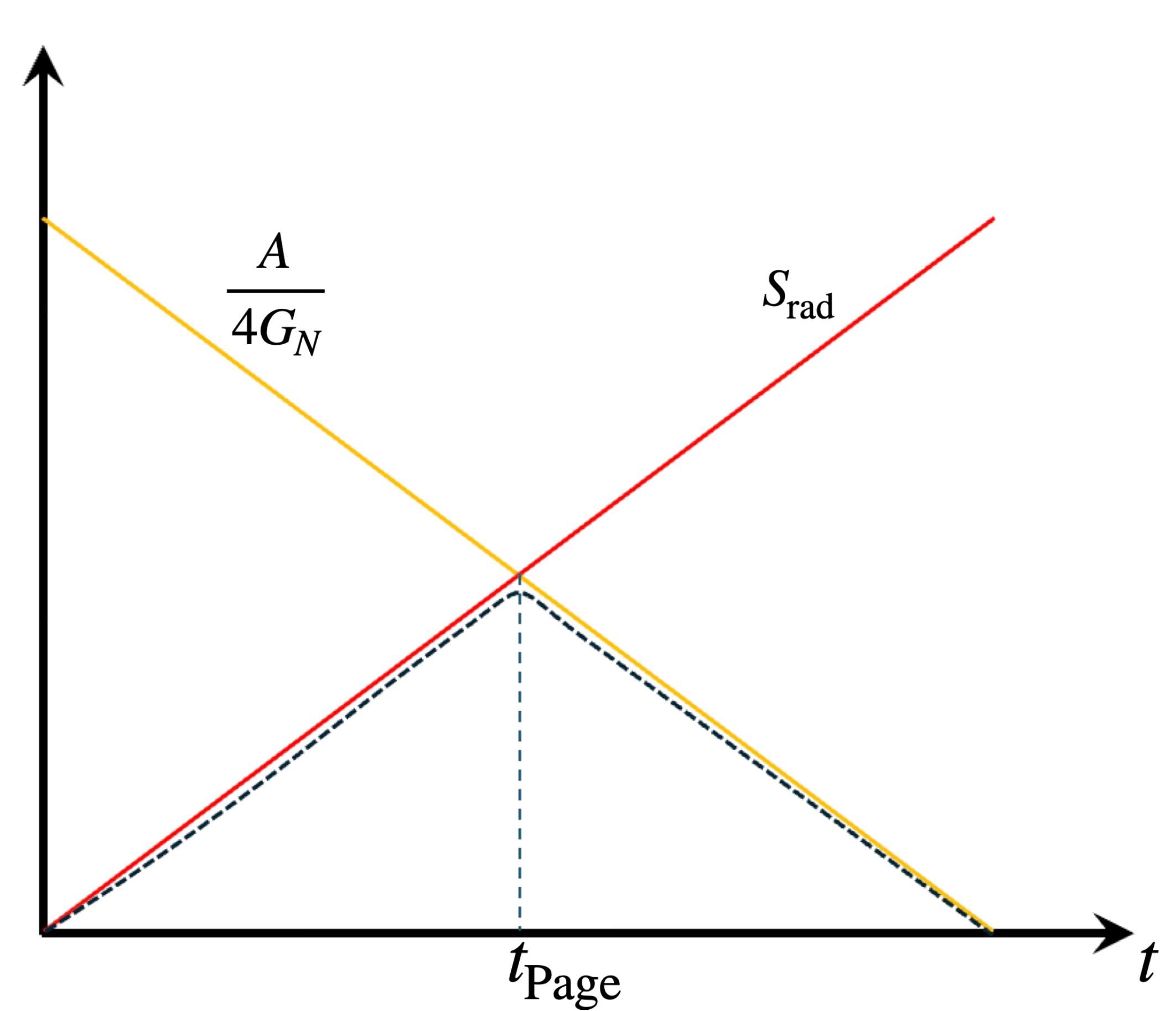} 
    \caption{Behavior of different relevant entropies as the \ac{BH} evaporates. The von Neumann entropy $S_{\rm rad}$ of the radiation (red line) increases monotonically, while the area of the \ac{BH} (yellow line) decreases. In order to respect the area limit, the entropy of the radiation should start decreasing latest at the Page time, and follow a curve qualitatively similar to the black dashed line. Figure adapted from \cite{Buoninfante:2021ijy}.}
    \label{fig:Francesco_Page-curve}
\end{figure}

If we now trace over the \ac{BH} degrees of freedom, we obtain a mixed state with a von Neumann entropy $S_{\rm rad}$. 
As explained in \eqref{eq:Francesco_Ent_states}, this entropy must be equal to the von Neumann entropy $S_{\rm bh}$ of the \ac{BH}. 
The semiclassical computation tells us that the von Neumann entropy of the emitted radiation increases with time which implies that $S_{\rm bh}$ must also increase with time. However, during the evaporation process the area of the \ac{BH} shrinks. The area limit implies that the maximum value of the degrees of freedom inside the \ac{BH}, and hence the maximum possible value of the von Neumann entropy $S_{\rm bh}$ is also decreasing. 
As a consequence, there must be a timescale\footnote{Sometimes it is underappreciated that the notion of time is not very well-defined in this context. We can have notions of time defined by the foliation, meaning that we foliate the spacetime with Cauchy hypersurfaces that must be spacelike. The direction orthogonal to the hypersurfaces is the time direction. However, changing the foliation would also change the time behavior of the entropy of the \ac{BH}.} $t_{\rm Page}$ --- known as the Page time --- after which the entropy of the radiation exceeds the maximum possible entropy of the \ac{BH} (see \cref{fig:Francesco_Page-curve}).  Therefore, the total state cannot be pure after the Page time. It is straightforward to check that at the Page time, the \ac{BH} is far from the Planckian regime~\cite{Page:1993wv, Page:1993up}.

\subsubsubsection*{Why the area limit}
We have just seen that the area limit hypothesis leads to a problem for the preservation of unitary evolution well before Planckian physics become relevant. It is therefore very important to justify the assumption and quantify the evidence that supports it. 

\textbf{\ac{BH} thermodynamics:} Arguably one of the main motivations comes from the thermodynamical properties of \acp{BH}. In fact, \acp{BH} obey physical laws analogue to the standard thermodynamic laws~\cite{Bardeen:1973gs, Wald:1999vt} with an entropy given by the Bekenstein-Hawking formula\footnote{Note that the notation might be confusing as $S_{\rm BH}$ indicates the Bekenstein-Hawking entropy while $S_{\rm bh}$ indicates the von Neumann
entropy of the \ac{BH}.}~\cite{Bekenstein:1972tm, Bekenstein:1973ur}, 
\begin{equation}\label{eq:Francesco_BH_ent}
S_{\rm BH}=\frac{A}{4\GN}.
\end{equation} 
Furthermore, \eqref{eq:Francesco_Entr_comp} tells us that the thermodynamic entropy gives an upper bound for the von Neumann entropy.
Therefore, the thermodynamical entropy \eqref{eq:Francesco_BH_ent} is expected to limit the maximum number of internal degrees of freedom of the \ac{BH}.

\textbf{Bekenstein bound:} Another argument in favor of the area limit is a bound discussed by Bekenstein~\cite{Bekenstein:1980jp}, according to which the entropy that any quantum system can contain is bounded by its radius $R$ and the total energy $E$ of the system:
\begin{equation}\label{eq:Francesco_Bek}
    S\leq 2\pi ER\,.
\end{equation}
In particular, for a spherically symmetric \ac{BH}, $R=2\GN M$ and $E=M$. The Bekenstein bound \eqref{eq:Francesco_Bek} reads
\begin{equation}
    S\leq\frac{A}{4\GN}=S_\text{BH}\,.
\end{equation}
However, we must stress that a rigorous proof of this bound can only be obtained for a quantum system in flat spacetime~\cite{Casini:2008cr}.

\textbf{Holography and \ac{AdSCFT} correspondence:} More recently, with the use of the Ryu-Takayanagi entropy formula~\cite{Ryu:2006bv, Lewkowycz:2013nqa}, obtained in the context of the \ac{AdSCFT} correspondence~\cite{Maldacena:1997re}, it was shown that it is possible to reproduce a behavior for the von Neumann entropy compatible with the Page curve. This can be taken as an indication of the validity of the area limit~\cite{Almheiri:2020cfm}. 

\subsubsection{Towards resolution of the problem?}

The resolution of the problem is definitely beyond the scope of this section, and is indeed part of a very active research area (see, \eg{}, \cite{Almheiri:2020cfm,Penington:2019npb,Penington:2019kki,Almheiri:2019psf,Almheiri:2019hni,Almheiri:2019qdq,Marolf:2020rpm,Dvali:2011aa,Dvali:2012en,Susskind:1993if,Visser:2014ypa,tHooft:1996rdg,Mathur:2005zp,Hayward:2005oet,Giddings:2012gc,Hawking:2014tga,Hawking:2016msc,Frolov:2014jva,Frolov:2017rjz,Bardeen:2014uaa,DAmbrosio:2020mut} and references therein for a very incomplete list). The main message of this section is that, while the information loss paradox is often introduced as an incompatibility between the unitary evolution, one of the pillars of quantum mechanics, and semiclassical gravity, the situation is not so simple. Obtaining the contradiction either requires pushing the semiclassical models well beyond their regimes of validity, as discussed in \cref{subsec:Francesco_par-A}, or the addition of an extra assumption in the form of the area limit, as discussed in \cref{subsec:Francesco_par-B}.  
Therefore, the resolution requires to abandon either the validity of semiclassical gravity as a good effective description far from the Planckian regime, or the area limit on the numbers of degrees of freedom~\cite{Buoninfante:2021ijy}. 
There are several approaches discussed in the literature that obtain a mathematically consistent picture getting rid of either one of these assumptions . However, to solve a physical problem, it is not enough to find a mathematically consistent description, but we should understand which (if any) of these descriptions is realized in nature. Due to the technical impossibility of observing \ac{BH} evaporation, this problem is expected to remain open for a very long time.

\subsection{Conclusions}\label{sec:FRANCESCO_conclusions}

In this section, we have discussed some consequences of considering quantum physics in \ac{BH} spacetimes. Despite a complete analysis of quantum effects in \ac{BH} spacetimes would require the knowledge of a full theory of \ac{QG}, we have seen that the study of quantum matter on classical spacetimes is enough to obtain some very interesting results.

This section was not intended to cover the topic in a comprehensive way, which is much wider than what can be presented in a short course. Instead, we focused on some specific aspects that are often misunderstood but that are crucial to properly understand the physics at play. 
Let us here summarize the most important take home messages. 

\begin{description}

\item[The cause of Hawking radiation:]
The cause of Hawking radiation is the time dependence of the spacetime. This can seem counterintuitive as we are used to think about \acp{BH} as stationary spacetimes. However, \acp{BH} are formed via gravitational collapse. The time dependence of the geometry, which transitions from something close to Minkowski in the asymptotic past to Schwarzschild in the asymptotic future, is the reason why the vacuum states for observers on $\mathscr{I}^-$ and on $\mathscr{I}^+$ do not agree.

\item[Eternal \acp{BH} do not evaporate:]
Related to the previous point, eternal \acp{BH} do not evaporate. Instead, we have seen that eternal \acp{BH} are in thermal equilibrium with a thermal bath. This might appear a minor difference. However, it is a crucial point as there cannot be particle creation in static geometries.

\item[Difference between the heuristic and the rigorous picture:]
Hawking radiation is often explained in a simplified way. This description has some merits. It correctly predicts that an asymptotic observer on $\mathscr{I}^+$ would see a flux of energy coming from the \ac{BH}. It also explains that the flux is entangled with an ingoing flux of negative energy. However, the oversimplification of the heuristic explanation fails in several regards. 
\begin{itemize}
    \item  First and foremost, reasoning in terms of particle-antiparticle production near the horizon does not capture the real root of Hawking radiation. Particles are created because \acp{BH} form via gravitational collapse and so the geometry is time-dependent. 
    
    \item According to the heuristic argument, Hawking quanta are produced near the horizon. However, it is not correct to identify a region where particles are produced, as we can formally talk about particles only in the asymptotic regions. 
    
    \item Finally, in the heuristic picture, the negative energy particle are entangled with the outgoing Hawking particles. However, we have discussed in \cref{sec:Francesco-state_Haw_rad} that the Hawking particles are entangled with quanta that enter the horizon at early times. On the other hand, \cref{eq:Francesco_inco_rad_hor} shows that the negative energy flux enters the horizon at a fixed rate. Therefore, it is clear that the flux of negative energy and the flux of entropy cannot be carried by the same virtual particles.
\end{itemize}

\item[Information loss problem:]
The information loss problem is also often misunderstood. It is usually described as an incompatibility between \ac{QFT} and \ac{GR}. However, we have discussed that there are extra ingredients necessary to formulate the problem.
\end{description}
Taken together, these results highlight how \ac{QFT} on classical curved spacetimes can offer some partial yet highly non-trivial insights into the interplay between quantum mechanics and gravity even in the absence of a fundamental theory of \ac{QG}.

\section{FAQ in Quantum Gravity}
\label{sec:FAQ}

\subsubsection*{QG in general}

\begin{itemize}
    \item \textbf{What is \ac{QG}?}

    Roughly speaking, a physical theory which consistently incorporates quantum mechanics and gravitation. A more precise definition is a quantum-mechanical theory which reduces to \ac{GR} (possibly coupled to other fields) in some low-energy or semiclassical regime. This meaning is the most widespread, but it is not universally shared in some research communities. Other researchers propose for instance that quantum mechanics be modified (see below).
    
    \item \textbf{Is it true that \ac{GR} is incompatible with quantum mechanics?}

    Not quite: quantum \ac{GR} makes sense as a low-energy \ac{EFT}. The relevant energy scales are determined by curvatures of the background, combinations of momenta of asymptotic states being scattered, or other invariant quantities. The issue is finding a completion of this \ac{EFT} which does not break down in some regime.
    
    \item \textbf{Should \ac{QG} be ``quantized gravity'', ``gravitized quantum mechanics'' or neither?}

    All these options have proponents in the wider research community. Some propose that the gravitational field should undergo quantization, in a similar manner as other classical systems can be mapped onto a quantum system by applying one of various quantization procedures. Then the consistency of the classical limit, if any, would need to be assessed. Others propose that the principles of quantum mechanics should be modified in some way to take gravitation into account. Others yet propose neither: in other words, the principles of quantum mechanics should remain untouched, and some quantum theory should produce gravity and/or spacetime as an emergent phenomenon at large scales.
    
    \item \textbf{What are some \ac{QG} phenomena that could be tested?}

    Presumably, scattering particles at very high energies would produce some quantum-gravitational effects, such as certain signatures in cross sections. The leading corrections to the gravitational \ac{EFT} may also be detectable in gravitational waveforms from \ac{BH} mergers, corrections to \ac{BH} geometries (Love numbers), or any other quantity. The issue is that typically, these corrections are extremely small relative to our capabilities. We may hope to see some amplifications in cosmological quantities such as the tensor-to-scalar ratio or multi-point correlators of cosmological perturbations. \ac{QG} effects can also generate a signal in the primordial \ac{GW} spectrum at LISA frequencies. Effects like a slight enhancement of couplings between gravitons and matter in \ac{EFT} easily lead to high-frequency \ac{GW} signals from reheating in the early universe which may be detectable in the future. Furthermore, we could be very lucky, and some quantum-gravitational effects may imply some low-energy consequences that could be tested. For instance, the very small observed dark energy $\CC \approx 10^{-120} \, \MPl^4$ is sensitive to new physics, and could be linked to other detectable phenomena like mesoscopic extra dimensions~\cite{Montero:2022prj}. While these features may not appear to be directly linked to \ac{QG}, they could follow from its consistency requirements, namely they could be swampland conditions for our universe. Additionally, extremal \acp{BH} may act as amplifiers of new physics, enhancing the role of \ac{EFT} corrections~\cite{Horowitz:2023xyl}. Another set of observables could be related to Lorentz invariance violations and modified dispersion relations~\cite{ Addazi:2021xuf, AlvesBatista:2023wqm}. Finally, regardless of the specific \ac{QG} approach, it might be possible that in the not-too-distant future, we will be able to detect some quantum features of gravity in a laboratory, in the non-relativistic regime. Indeed, recent proposals for tabletop experiments aim at detecting the quantum nature of gravity by looking for gravity-induced entanglement correlations between two superposed massive particles~\cite{Bose:2017nin,Marletto:2017kzi}. These correlations would presumably be captured by a quantum-gravitational \ac{EFT}.
    
    \item \textbf{While we wait for future experiments that will detect some \ac{QG} signatures, what are we doing?} 

    Even if we were to give up on testing \ac{QG} effects in our lifetime, it is still worthwhile to study \ac{QG} from a theoretical standpoint. In the past few decades, we have learned that \ac{QFT} is an extremely rich and rigid framework, and that \ac{QG} is vastly more rigid than that. In other words, the theoretical consistency of \ac{QG} with basic physical principles (unitarity, causality) may be enough to draw new conclusions and make progress. However, it would be extremely interesting if a future experiment were to indicate that one of these principles must be given up. For example, detecting some effects induced by the violation of some form of causality could be useful not only to learn about new physics, but also to discriminate between \ac{QG} approaches.
    
    \item \textbf{Can't we just do \ac{GR}-\ac{SM}-\ac{EFT} with 50-whatever parameters and call it a day for the next 500 years?}

    From a pragmatic point of view, yes. But physics is not just about fitting data to sufficiently many parameters. Such a procedure would completely miss the deeper physical understanding that we are seeking when studying \ac{QG}. Moreover, there could be new physics that cannot be captured by Wilson coefficients: if new particle species show up, the \ac{EFT} would only be able to encode the effects induced on the known degrees of freedom.
    
    \item \textbf{Does \ac{QG} only affect tiny distances?}

    With the premise that distances should be given an invariant meaning in a (general) relativistic theory, not necessarily. Some consequences of \ac{QG} may be amplified and affect detectable regimes. One option is for cosmological observations to pick up on such amplified effects, for example by looking at quantum correlations in the primordial \ac{GW} spectrum. Another option is for the consistency conditions of \ac{QG} to trickle down to observable physics. For instance, some new anomalies due to gravity can constrain some symmetries of models beyond the \ac{SM}, or they could require the existence of light particles such as axions.
    
    \item \textbf{What are the most important open problems in \ac{QG}?}

    From a theoretical standpoint, the basic problem is understanding what the underlying physical principles are. Is it strong coupling physics? \acp{BH} and holography? The emergence of spacetime? What are the degrees of freedom and what do they do? From a phenomenological standpoint, we would like \ac{QG} to produce a realistic model of our universe, hopefully bringing along some explanation for puzzles in cosmology and particle physics.

    \item \textbf{Is there a phase transition in gravity when it enters a strong coupling regime?}

    This is not obvious. One such a transition could be \textit{classicalization}~\cite{Dvali:2014ila}, namely, the transition to \ac{BH} dominance when probing physics beyond the \ac{BH} production threshold $E_\text{threshold} = \UVcutoff^{3-d} \MPl^{d-2}$, with $\UVcutoff$ being the \ac{UV} cutoff and $d$ the number of spacetime dimensions. Another independent possibility is related to the existence of a non-trivial fixed point of the \ac{RG} flow (as in \ac{ASQG}), and the possibility that this fixed point describes a second-order phase transition.

    \item \textbf{Can gravity and spacetime emerge from other degrees of freedom?}

    As mentioned above, the emergence of gravity and spacetime is a scenario to realize \ac{QG} where the fundamental degrees of freedom are not those of a quantized gravitational field. In fact, due to the possibility of dualities, it may not even be possible to ascribe all physical regimes to a single set of fundamental degrees of freedom. Examples in \ac{QFT} such as Montonen-Olive electromagnetic duality~\cite{Montonen:1977sn} also appear in \ac{ST} via S-duality. The \ac{AdSCFT} correspondence and matrix theories~\cite{Banks:1996vh, Ishibashi:1996xs, Dijkgraaf:1997vv} are models of this type, where spacetime and gravity emerge from something else. In perturbative \ac{ST}, spacetime arises as a special case of the more abstract \ac{CFT} on the worldsheet. 

\end{itemize}

\subsubsection*{Perturbative QG}

\begin{itemize}

\item \textbf{What does the adjective ``perturbative'' mean in the expression ``perturbative \ac{QG}''?}

It means that interactions are weak, and we can identify one or more small dimensionless quantities (such as dimensionless interaction couplings or ratios between energy scales and dimensionful interaction couplings) in terms of which we can define a perturbative expansion. Roughly speaking, the validity of the ``perturbative'' \ac{QFT} framework requires that interaction terms are smaller than the kinetic terms. 

\item \textbf{What is the entity that needs to be quantized in perturbative \ac{QG}?}

In the perturbative \ac{QFT} framework, one typically quantizes field fluctuations. In a gravitational context, these are usually chosen to be metric fluctuations around some (arbitrary) background which is kept fixed. While the metric fluctuation field is quantized, the background is kept classical. The quantum counterpart of the metric perturbation defines the \textit{graviton field} (see also the next question). It is important to remark that background independence ensures that physical quantities (\eg{}, scattering amplitudes) are independent of how one splits background and fluctuations. In particular, one could equivalently quantize the vielbein instead of the metric. Furthermore, for gravitational theories treating the metric and the connection as two independent objects, it is also necessary to quantize the latter. The physical consequences of quantizing the connection together with the metric are however less explored, particularly because at low energies standard \ac{GR} is solid, and there is no hint that an independent connection plays a role. Similar arguments hold for other versions of \ac{GR} with torsion or non-metricity~\cite{BeltranJimenez:2019esp}.

\item \textbf{What is a graviton?}

In the \ac{QFT} framework, quantum particles are described as excited quantum states on the top of the vacuum. These particle states can be excited by acting with some field operator on the vacuum state. In perturbative \ac{QG}, the quantum state populated by a single energy excitation is called the \textit{graviton}. Gravitons are massless, and in four spacetime dimensions they propagate helicities $\pm 2$ on-shell. A classical \ac{GW} can also be described as a collection of gravitons in a certain quantum state and with a frequency distribution peaked around the frequency of the classical wave. In particular, the larger the amplitude of the \ac{GW}, the more gravitons populate the (semiclassical) state.

\item \textbf{Does the failure of perturbative renormalizability of \ac{GR} imply that the latter is not compatible with quantum mechanics?}

No. It is important to clarify that \ac{QFT} and \ac{GR} can indeed be compatible, at least in the low-energy regime. Indeed, one can formulate a consistent \ac{EFT} of \ac{GR} that is valid up to some cutoff energy scale, \eg{} the Planck mass $\MPl\sim 10^{18}$ GeV in pure gravity when the massless graviton is the only active degree of freedom at energies below $\MPl$. This means that there exists a \ac{QFT} framework where consistent computations can be performed and \ac{QG} predictions can be trusted up to finite errors proportional to inverse powers of the cutoff. In the \ac{EFT} of \ac{GR} the local part of the gravitational Lagrangian contains all possible counterterms that are compatible with the symmetries of \ac{GR} (which is an infinite number). 

\item \textbf{Does the expression ``perturbative \ac{QG}'' only refer to \ac{GR} as \ac{QFT}?}

Although it is often used for that, the term may be extended to denote any possible consistent perturbative \ac{QFT} of the gravitational interaction. There are in principle several candidates, each with different advantages and shortcomings. One example is super-renormalizable \ac{QG}~\cite{Anselmi:2017ygm}: it makes gravity super-renormalizable thanks to the addition of higher derivative terms up to the sixth (or higher) order, but there are still open questions about unitarity and causality. Additionally, there are at least three examples of theories that are strictly renormalizable: quadratic gravity~\cite{Salvio:2018crh}, Ho\v{r}ava–Lifshitz gravity~\cite{Herrero-Valea:2023zex}, and metric-affine gravity~\cite{Percacci:2020bzf}. The first achieves renormalizability by the addition of higher-derivative terms up to quadratic order, at the price of introducing a massive spin-two ghost. The second is renormalizable and seemingly ghost-free, but at the cost of breaking local Lorentz invariance and diffeomorphism invariance down to foliation-preserving diffeomorphisms. The third one introduces non-metricity and torsion as additional degrees of freedom, in particular the metric and the connection are two independent objects. 

\item \textbf{Is the requirement of ``strict'' renormalizability still a valid criterion to select \acp{QFT} when applied to gravity?}

All \ac{SM} interactions (electromagnetic, weak, and strong) are described by strictly renormalizable \acp{QFT}, where the word ``strictly'' means that the relevant couplings in the \ac{UV} are dimensionless. Therefore, the most natural attempt would be to look for a strictly renormalizable gravitational \ac{QFT} of the metric field preserving the symmetries of \ac{GR}. In four spacetime dimensions, this \ac{QFT} is quadratic gravity. This theory is considered interesting because it has a built-in explanation for inflation: a spin-zero field that can play the role of the inflaton and dynamics resembling that of Starobinsky inflation. However, there are still important open questions about quadratic gravity, including unitarity and \ac{UV} completeness (see below).

\item \textbf{Is unitarity violated in quadratic gravity due to the presence of the spin-two ghost?}

The answer is yes if the standard quantization procedure is used. However, the answer could be no if alternative quantization prescriptions are implemented.
Let us remember that unitarity means that quantum probabilities are conserved. In standard two-derivative \acp{QFT}, the notion of unitarity is compatible with the existence of one single arrow of causality (Feynman causal prescription) and with physical quantum states having positive norms. In quadratic gravity, the presence of the spin-two ghost introduces unusual minus signs in the amplitudes. In this case, unitarity can still be satisfied, but it cannot be compatible simultaneously with a single arrow of causality and with positive norms. Indeed, a unitary quantization of quadratic gravity may require giving up some concept of causality (at energies of the order of the mass of the spin-two ghost), or the existence of physical states with negative norms. The viability of two coexisting arrows of causality at the microscopic level or of negative-norm states is the subject of active investigations~\cite{Salvio:2018crh, Donoghue:2021cza}. Another type of quantization that could reconcile perturbative renormalizability and unitarity is the so-called \textit{fakeon} prescription~\cite{Anselmi:2017ygm, Piva:2023bcf}. This converts the massive spin-two ghost into a purely virtual particle that can only appear off-shell through internal lines in Feynman diagrams but never as an on-shell propagating degree of freedom.

\item \textbf{Is quadratic gravity a \ac{UV}-complete 
theory of \ac{QG}?}

It is still not clear. For example, the \ac{UV} behavior of scattering amplitudes in quadratic gravity is not understood. Despite the property of renormalizability and the possibility to have asymptotic freedom (see \cref{sec:renorm-quad-grav}), the tree-level $2\rightarrow 2$ scattering amplitude between gravitons and the corresponding inclusive cross section grow with the energy as in \ac{GR}~\cite{Dona:2015tra, Holdom:2021hlo}. This fact seems to suggest that perturbativity breaks down at the Planck scale. Nonetheless, it has also been argued that the only relevant physical quantities are the totally-inclusive cross sections, and these can be shown to be suppressed in the \ac{UV} regime~\cite{Holdom:2021hlo}. Furthermore, when coupling quadratic gravity (or any other \ac{QG} theory) to matter, one should ensure the \ac{UV} completeness of the combined interacting system. The matter content of the \ac{SM} introduces Landau poles, so either \ac{QG} or matter beyond the \ac{SM} should remove them. Quadratic gravity coupled to \ac{SM}, with an additional non-minimal coupling between the Higgs field and the Ricci scalar, gives rise to a perturbatively (strictly) renormalizable \ac{QFT} of gravity and matter. However, it can be shown that in this theory the \ac{SM} gauge couplings do not receive any gravitational correction~\cite{Salvio:2014soa, Anselmi:2018ibi}. This means that the hypercharge gauge coupling will still hit a Landau pole at some trans-Planckian energy scale. It is unclear whether any (perhaps non-perturbative) solution to these problems exists.

\end{itemize}

\subsubsection*{EFT of gravity and positivity bounds}

\begin{itemize}
\item \textbf{What are the limitations of the \ac{EFT} description? How to determine the scale where it breaks down?}

It depends on the particular \ac{EFT}: the cutoff is the lowest scale where some principle (\eg{}, unitarity, ordering of operators in the action, etc.) breaks down. A rough estimate can be obtained by looking at the suppression scale of all \ac{EFT} operators and picking the lowest scale. For instance, if the cutoff scale is determined by $2\to 2$ scattering, this scale can be defined as the lowest value of $s$ when \ac{PWU} breaks down. However, one should have in mind that the theory may have interactions which do not contribute to $2\to 2$ scattering but still dominate unitarity breaking. This is a typical feature of scalar field \acp{EFT} with exponential potentials~\cite{Bezrukov:2010jz} which are commonly used for early universe inflation. In such kinds of theories, the cutoff would be determined by the scattering of a large number of particles.

\item \textbf{Does the absence of ghosts imply \ac{PWU}?}

No. The presence of ghosts with cubic interactions makes it impossible to avoid negative probabilities (assuming a standard quantization and unmodified rules of quantum mechanics and \ac{QFT}) and thus a violation of \ac{PWU} conditions. However, the absence of ghost states is not enough to meet all unitarity requirements. For example, in pure ghost-free \ac{GR} the scattering amplitudes violate \ac{PWU} above the Planck scale. Let us mention here that \ac{UV}-completing a theory by introducing a ghost field would violate positivity bounds, as it was discussed in~\cite{Caron-Huot:2020cmc} for a scalar toy model.

\item \textbf{The propagator of an unstable particle may have poles located outside the real axis of the complex $s$-plane. Does this contradict analyticity properties required in the derivation of positivity bounds?}

No. Analyticity is required in the first sheet of the complex plane while the poles caused by unstable particles emerge in the second sheet.

\item \textbf{Causality implies analyticity of the amplitudes, but do the right analyticity properties of the full non-perturbative amplitude guarantee a causal propagation?}

Most likely not. Causal propagation of the signal is an extra requirement which has to be checked separately. For asymptotic causality, it is only required that the real parts of the partial wave amplitudes grow with $s$. This is certainly a different requirement, not connected to analyticity.

\item \textbf{Why are four dimensions so special for graviton-mediated scattering?}

Mathematically, the eigenfunctions of angular momentum in four dimensions are Legendre polynomials. The expansion of the graviton pole part of the amplitude $s^2/t$ in Legendre polynomials is not well-defined because the partial wave amplitudes are expressed through logarithmically divergent integrals. The elimination of these divergences leaves the dependence of the partial waves (or amplitude in impact parameter space) on a certain infrared scale, which is often just set to be the size of the universe. The physical reason for all these problems is connected to the fact that in four dimensions, the emission of soft gravitons cannot be neglected, and it is hard to rigorously define extremely soft gravitons as asymptotic states. In higher dimensions, soft emission is more suppressed.

\item \textbf{How can we use dispersion relations if there are different degrees of freedom in the \ac{IR} and \ac{UV} theories (like in \ac{QCD}: pions in the \ac{IR} and quarks and gluons in the \ac{UV})?}

This question is related to the definition of low-energy asymptotic states which are scattered. In the case of \ac{QCD}, these states are pions (maybe boosted for the case of hard scattering). For this reason, the amplitude is still well-defined, even though in the high-energy domain, non-perturbative physics plays a major role. The technique of dispersion relations requires just a set of assumptions about the \ac{UV}, not a detailed description. 

\item \textbf{Can early universe inflation be completely described within the regime of applicability of some \acp{EFT}?}

From one side, the energy scale of inflation, or energy density of the inflaton field is about $10^{-10}~\MPl^4$  which naively makes it safe from \ac{QG} corrections. Inflation usually requires a large excursion of the inflaton field. For example, in the Starobinsky model, the theory should be well-defined up to background field values of the order of $6\MPl$ in the Einstein frame. Although this sounds like a breakdown of the \ac{EFT}, it is usually argued that the asymptotic shift symmetry of the potential at large field values protects the stability of this potential. However, strictly speaking, this interplay between large backgrounds and large energies in \acp{EFT} still requires a more rigorous study.

\item \textbf{Imagine we found (or bootstrapped) a full non-perturbative S-matrix for the scattering of gravitons and all other states. Does it fully define the theory in all situations, including large classical backgrounds, \acp{BH}, etc.?}

Certainly, the $2\rightarrow 2$ scattering amplitude is not enough to define the theory non-perturbatively. Even if we have non-perturbative data for all $n\rightarrow m$ scatterings, it does not cover all situations. For example, the theory may have different vacua, while perturbative scattering represents an expansion around only one selected vacuum. The S-matrix of the particle-like states represents a solid ground for a (semi)perturbative definition of the theory around asymptotically flat spacetime (or, more generally, a spacetime with an \ac{AdS}-like boundary also allows to define the theory through the boundary correlators). However, it may not be enough for a complete description of all non-perturbative phenomena. The full non-perturbative S-matrix does not depend on a background, and could contain more asymptotic states, including solitons and coherent states. Even in this case, to the best of our knowledge, it is not fully understood whether it is enough and what is enough to define the theory. 

\end{itemize}

\subsubsection*{Asymptotic safety}

\begin{itemize}
    \item \textbf{Is asymptotic safety fundamental?}
    
    We do not know. In its original incarnation, \ac{ASQG} has been proposed to be a fundamental \ac{QG} theory, with the fixed point describing its \ac{UV} completion. Yet, as the fixed point is consistently found in \ac{RG} computation, even if \ac{ASQG} is not fundamentally realized, it could still be a low-energy approximation of a more fundamental theory. A scenario in which the Reuter fixed point acts as a pivot to a more fundamental description has been dubbed ``effective asymptotic safety''~\cite{Percacci:2010af, deAlwis:2019aud, Basile:2021euh, Basile:2021euh}. 
    
    \item \textbf{Is there any known example of asymptotic safety?}
    
    The Gross-Neveu model~\cite{Kivel:1993wq} and gravity in $(2+\epsilon)$-dimensions~\cite{Christensen:1978sc, Gastmans:1977ad} are both proven to be asymptotically safe, using perturbation theory. Scalar field theories in three dimensions also have a non-trivial fixed point --- the well-known Wilson-Fisher fixed point. Its properties have been tested in the context of critical phenomena. More recently, a set of four-dimensional theories has been constructed that is asymptotically safe in the perturbative regime~\cite{Litim:2014uca, Bond:2017lnq, Bond:2022xvr, Litim:2023tym}.
    
    \item \textbf{Can we see asymptotic safety in \ac{QG} via perturbation theory?}

    It is not clear. Concrete computations point towards a near-perturbativity of \ac{ASQG}~\cite{Eichhorn:2018ydy}. On the one hand, it is known that perturbation theory can see asymptotic safety in certain \emph{ad hoc} non-gravitational models~\cite{Litim:2014uca, Bond:2017lnq, Bond:2022xvr}. On the other hand, in \ac{ASQG}, there is so far no parametric control as in the above examples. 
    
    \item \textbf{Does asymptotic safety break unitary because truncated actions like the quadratic one have ghosts?}
    
    No, this is a common misconception. Truncations are approximations of the full effective action, and such truncations can generate \emph{fictitious ghosts}, which however decouple in a controlled way as the truncation order is increased~\cite{Platania:2020knd, Platania:2022gtt}. The full effective action of asymptotic safety will include infinitely many operators and derivatives (yet a finite number of free parameters, dictated by the dimension of the \ac{UV} critical surface), hence the contribution to the propagator \emph{could} re-sum to yield ghost-free form factors~\cite{Draper:2020bop}. In particular, there is growing evidence~\cite{Knorr:2021niv,Bonanno:2021squ}, even based on fully-Lorentzian \emph{computations}~\cite{Fehre:2021eob}, that asymptotic safety is unitary.

    \item \textbf{Does \ac{ASQG} predict extra gravitational degrees of freedom in addition to the massless spin-two graviton?}

    The question is not settled yet. More precisely, derivations of the transverse-traceless part of the non-perturbative graviton propagator~\cite{Fehre:2021eob} have shown no pole other than the massless one corresponding to the graviton. Investigations on the scalar part of the propagator have not been performed yet; these are certainly necessary, as they could in principle show the existence of a scalaron degree of freedom, and this could drive inflation in the early universe.

    \clearpage
    
    \item \textbf{Does the Newton coupling run?}
    
    No. The interaction coupling multiplying the Ricci scalar can only have a dependence on the unphysical \ac{RG} scale $k$, but not on the physical momentum $p$~\cite{Draper:2020bop} (see \cref{sect:ALEBEN-kruprun}). While the so-called cutoff identification may be used to build \ac{QG}-inspired models, one has to be extremely careful in how to apply it (it may only be justified in single-scale systems or multi-scale systems with decoupling~\cite{Borissova:2022mgd, Platania:2023srt}). One should not draw any definite conclusions from such identifications and, in any case, since the Newton coupling does not run, its coordinate-dependent extension could at best be dubbed effective Newton coupling.
    
    \item \textbf{How is asymptotic safety in \ac{QG} physically realized?}
    
    The mechanism is similar as in Yang-Mills theories: it is about the ``paramagnetic'' dominance of curvature operators over the ``diamagnetic'' ones encoded in the Laplacian operators. This leads to gravitational anti-screening, allowing for the formation of a fixed point~\cite{Nink:2012vd}.

    \item \textbf{Is gravity plus matter asymptotically safe?}

     The answer to this question generally depends on the number and type of fields considered. As a key example, when considering the matter content of the \ac{SM}, there is evidence that the combined gravity-matter theory is asymptotically safe~\cite{Dona:2013qba,Biemans:2017zca, Pastor-Gutierrez:2022nki}. Small modifications to the \ac{SM} seem also to be asymptotically safe (see~\cite{Eichhorn:2022gku} and reference therein). 
    
    \item \textbf{Is there any way to theoretically test asymptotic safety beyond the \ac{FRG}?}
    
    Yes. Similarly to the case of \ac{QCD}, one can use lattice methods in \ac{QG}. The programs aiming at assessing the asymptotic safety conjecture in gravity are the so-called Euclidean and Causal Dynamical Triangulations~\cite{Laiho:2016nlp, Loll:2019rdj, Brunekreef:2023ljt}. In that context, looking for asymptotic safety is tantamount to searching for a second-order phase transition. In principle, one could also use an alternative version of the \ac{FRG} using 2PI effective actions~\cite{Blaizot:2021ikl}, but these techniques are still under development in the context of non-gravitational \acp{QFT}. Finally, since \ac{ASQG} should feature quantum scale symmetry at high energies, some form of non-perturbative S-matrix bootstrap may be able to test whether this property is compatible with basic axioms such as unitarity and causality.
\end{itemize}

\subsubsection*{String theory}

\begin{itemize}
    \item \textbf{Does \ac{ST} postulate that tiny strings are the fundamental constituents of everything?}

    Sort of. Despite its name, we now understand that \ac{ST} is not a theory of strings, rather it contains extended objects of various dimensions (including particles!). The special role of strings shows up within perturbative regimes, where there always appears a unique one-dimensional object playing the role of fundamental degrees of freedom. In these limits, the other extended objects manifest as heavy solitons. However, much like phonons in solids, generally this description is only valid at weak coupling. In the '90s it was discovered that the strong-coupling limit of one stringy description is the weak-coupling limit of another. The emerging picture is thus that \ac{ST} is a unique theory with many connected perturbative limits. Its generic regime, where couplings are of order one and all objects are equally important, is not completely understood.

    \item \textbf{Is \ac{ST} \ac{UV}-complete?}

    As far as we can tell, it is. Its observables are well-behaved at high energies, where they appear to match the expected scaling governed by \ac{BH} formation. Dualities express strong-coupling limits in terms of different weakly coupled degrees of freedom. In several cases, exact computations interpolating between these limits are also available.

    \item \textbf{Since \ac{ST} is not fully understood, does it make sense to attempt string phenomenology?}

    While understanding and developing the theory is important, we know enough about it to attempt constructing progressively more realistic configurations with it. These efforts also teach us about features of the string landscape, guiding further theoretical development and giving indications for low-energy predictions. Finally, there are reasons to believe that, if our universe lies in the string landscape, it is in some weakly coupled corner where we can hope to make progress without the whole theoretical picture~\cite{Grana:2021zvf, Castellano:2021mmx, Montero:2022prj}.

    \item \textbf{Does the existence of the string landscape mean that the theory predicts a multiverse?}

    No. Landscapes are a rather generic feature of any theory including gravity, as exemplified by the \ac{SM} itself. It simply means that the theory has many (meta)stable vacua as possible states, hopefully with our universe in one of them.

    \item \textbf{Is \ac{ST} predictive/falsifiable?}

    Yes, but it is hard to obtain predictions that are both sharply quantitative and relevant for (comparatively) low-energy experiments. Much like any physical theory, once boundary conditions are fixed, the observables are also predicted as functions of the free parameters in the theory. For instance, one can pick a vacuum state and compute scattering amplitudes. Relative to \acp{QFT} in particle physics, \ac{ST} is much less flexible since it has no dimensionless free parameters. However, it is much harder to find a vacuum state or configuration whose low-energy excitations are realistic. If this can be done, the rest would also be predicted by the theory. On the flip side, high-energy scattering amplitudes have a universal profile which predicts the regime between particle scattering and \ac{BH} formation. This high-energy stringy regime appears to be much simpler to nail down theoretically, but much harder to access experimentally. As for currently experimentally accessible low-energy physics, finding bounds on Wilson coefficients or universal features of the string landscape could provide concrete predictions which can be tested in the foreseeable future. An example could the presence of mesoscopic extra dimensions~\cite{Montero:2022prj}, if it turns out that they are required in \ac{ST} by the smallness of the observed dark energy~\cite{Basile:2024lcz}.

    \item \textbf{Does \ac{ST} recover the \ac{SM} with its particle content, interactions, and gauge groups in the low-energy regime?}

    We do not know. Over several decades, many attempts at model building have made progress in various directions, such as recovering cosmological features~\cite{Antoniadis:2020stf, McAllister:2024lnt} or particle spectra resembling those of the \ac{SM}. It is very difficult to achieve everything simultaneously. For instance, minimally supersymmetric brothers of the \ac{SM} can be constructed~\cite{Cvetic:2019gnh}. Braneworld scenarios~\cite{Danielsson:2023alz} offer a realistic cosmology and non-supersymmetric \ac{SM}-like spectra, but it is not clear whether Yang-Mills fields behave as in the \ac{SM}.

    \item \textbf{Does \ac{ST} require extra dimensions and supersymmetry? What about their lack of detection at particle accelerators?}

    According to our current understanding, the short answer is no. The longer answer is a bit more subtle! In \ac{ST}, compact dimensions can blend with more exotic, non-geometric things. In some configurations there may be no extra dimensions at all, and then they can emerge by varying some fields. In other words, non-geometric degrees of freedom can geometrify and vice versa, for lack of a better word. As for supersymmetry, there exist non-supersymmetric configurations all the way to (at least) the string scale, thus from \ac{ST} alone there is no phenomenological hint whatsoever on whether supersymmetry should be realized and at what scale. Indeed, the expectations and excitement for its detection at particle accelerators were motivated partly by the electroweak hierarchy problem and partly by the simplicity of the configurations of \ac{ST} which do feature spacetime supersymmetry. It is however worth mentioning that, despite the existence of non-supersymmetric configurations, it seems likely that the only exactly stable vacua are in fact supersymmetric. Our universe is not in a stable vacuum, but there may be interesting implications.

    \item \textbf{What do we know about non-perturbative \ac{ST}?}
    
    Most of our current understanding of non-perturbative \ac{ST} is somehow tied to supersymmetry, which can provide a great deal of control in strongly coupled settings. Whether it is ultimately a fundamental ingredient for exact vacuum stability or dualities is still unclear. In sufficiently supersymmetric settings, much has been learned about the network of string dualities since the second superstring revolution~\cite{Witten:1995ex}. Dualities relate different perturbative descriptions to each other, and some physical quantities such as BPS masses, which are exact in the couplings, can thus be fully tracked between two perturbative limits across the strongly coupled regime. Many instances of this type, and their mutual consistency (see \eg{}~\cite{Lee:2019wij, Lee:2019xtm}), support the existence of a strongly coupled regime that glues together all perturbative limits. Moreover, some sectors of string theory can be described non-perturbatively via the \ac{AdSCFT} correspondence~\cite{Maldacena:1997re} or matrix models such as BFSS~\cite{Banks:1996vh} for eleven-dimensional M-theory, IKKT~\cite{Ishibashi:1996xs} for ten-dimensional type IIB and~\cite{Dijkgraaf:1997vv} for ten-dimensional type IIA. Lower-dimensional descriptions of this type become increasingly more difficult to handle~\cite{Seiberg:1997ad}. Another corner in which recent progress has been made non-perturbatively is the topological sector of string theory~\cite{Marino:2024tbx, Hattab:2024ssg}. Tensionless string limits or hard scattering limits also allow some resummations or estimates~\cite{Gross:1987ar, Gross:1987kza, Mende:1989wt, Eberhardt:2020bgq, Eberhardt:2021jvj}.

    \item \textbf{Is \ac{ST} effectively local? Can the dispersion relations based on polynomial boundedness be used if \ac{ST} (especially when strongly coupled) is a \ac{UV}-completion of gravity?}

    \ac{ST} behaves effectively locally at low energies, where it reduces to \ac{EFT}. However, in order to match \ac{BH} formation at high energies, some degree of non-locality kicks in, and it is not clear to which extent bounds related to locality can be trusted. However, as discussed in~\cite{Mende:1989wt}, estimating the resummed high-energy behavior of string scattering leads to behavior closer to \ac{QFT} than the naive tree-level analysis would suggest (\eg{} the Martin-Cerulus bound is restored).

    \item \textbf{Is \ac{ST} background-independent?}

    As far as we can tell, it is. Already at the perturbative level, backgrounds can be deformed into neighboring ones via coherent states including gravitons. At low energies, the \ac{EFT} description of the physics is manifestly covariant in the usual way, although the sum over topologies cannot be seen in this limit. Sometimes this can be shown to hold beyond weak coupling, for example via supersymmetry which can completely fix the low-energy effective action. Beyond perturbation theory or low energies, some quantities can be shown to be background-independent directly or indirectly. For instance, tensionless strings in asymptotically \ac{AdS}$_3$ backgrounds allow resumming the partition function showing that it only depends on boundary data and not on the background. In fact, it can be written as a sum over bulk geometries~\cite{Eberhardt:2020bgq, Eberhardt:2021jvj}. The topological sector of \ac{ST} also allows an explicit computation of similar effects. In settings where a dual \ac{CFT} description is available via the \ac{AdSCFT} correspondence, bulk quantities clearly only depend on boundary data and are thus background-independent, as they should. In stringy matrix models, spacetime itself is absent from the mathematical description from the outset.

    \item \textbf{What is the deal with dark energy in \ac{ST}?}

    The short answer is that the question is open. To date, there are no fully controlled \ac{dS} constructions in \ac{ST}, and any such construction would be at best metastable. In some regimes there are no-go theorems against the existence of \ac{dS} vacua. Whether metastable \ac{dS} vacua do not exist at all in the string landscape is an open question, but there are bottom-up arguments indicating that their lifetime cannot be arbitrarily parametrically larger than the Hubble time. Another possibility is to realize an accelerated cosmological expansion without a metastable vacuum with positive vacuum energy~\cite{Shiu:2023nph, Andriot:2023wvg}.
\end{itemize}

\subsubsection*{Quantum effects in BH spacetimes}

\begin{itemize}
    \item \textbf{Where are Hawking particles created?}

    According to the heuristic picture of Hawking radiation, the phenomenon is generated by the formation of particle-antiparticle pairs in the vicinity of the event horizon. Tidal forces separate the pairs with the positive energy particle escaping to infinity, while the negative energy partner falls into the \ac{BH}. However, we have seen that this heuristic picture is too simplistic. We can only talk about particles in the asymptotic regions, thus the question of where such particles are formed is not well-posed. The heuristic picture does capture some aspects of the physics at play, as there is a negative energy flux going into the horizon. However, the information only enters the horizon at very early times (meaning that particles in the asymptotic region are entangled with modes that enter the horizon right after the collapse), while the negative energy flux is always present. This shows that these two physical quantities are not carried by the same physical entity, contrary to what the heuristic picture would suggest.
    
    \item \textbf{Can we have evaporation without a horizon?}
    
    Yes, it is possible to have evaporation even without the formation of any horizon. The evaporation is present every time we have a time-dependent geometry. However, in the absence of a horizon, the radiation must eventually switch off after the geometry relaxes into a static configuration~\cite{Barcelo:2010xk}.  

    \item \textbf{What is the role of the horizon in \ac{BH} evaporation?}
    
    While, as explained in the previous question, horizons are not fundamental to particle creation, they do play an important role in \ac{BH} evaporation.
    To start, as it was discussed in \cref{Sec:Francesco-SET}, the vacuum states that are regular at the event horizon are not empty at asymptotic distances. Therefore, a horizon necessarily leads to the presence of a thermal flux, either because of particle creation (like for \acp{BH} formed by gravitational collapse) or because they are in a thermal bath (like for eternal \acp{BH}). 
    Furthermore, while without a horizon the evaporation must eventually switch off, the horizon's formation makes the evaporation continue until the horizon evaporates (or at least until the breakdown of semiclassical physics).
    Finally, the presence of a horizon leads to universal radiation which is independent of the details of the gravitational collapse. \clearpage

    \item \textbf{Is the Hawking radiation thermal?}
    
    There are two points to be discussed here. 
    First of all, Hawking evaporation is non-thermal because of the gray body factor (see \eqref{eq:FRANCESCO_gray_body}).
    Regardless, we can consider the radiation thermal as it has the same spectrum it would have if the \ac{BH} was replaced by a thermal source without changing the potential. 
    Furthermore, Hawking evaporation is only thermal (up to the gray body factor) within the approximation in which it is derived. In particular, the derivation assumes no backreaction, \ie{} the mass of the \ac{BH} does not change. Modifications due to the change of the mass of the \ac{BH} lead to deviations from thermality. However, such deviations are very small for \acp{BH} with masses much larger than the Planck mass.
    
    \item \textbf{Do eternal \acp{BH} emit Hawking radiation?}   
    
    No, eternal \acp{BH} do not emit Hawking radiation, as the latter can only be emitted by dynamical geometries. 
    However, eternal \acp{BH} are in equilibrium with a thermal bath, as the only regular vacuum state has a non-zero flux of radiation both on $\mathscr{I}^-$ and on $\mathscr{I}^+$. At sufficiently late times, there is no detectable difference between the vacuum state of a \ac{BH} formed by a gravitational collapse and an eternal \ac{BH}. However, the difference between evaporating and being in thermal equilibrium with a thermal bath is conceptually very relevant. 

    \item \textbf{Do extremal \acp{BH} emit Hawking radiation?}
    
    The temperature at which \acp{BH} evaporate is proportional to the surface gravity. Extremal \acp{BH} have zero surface gravity and hence also zero temperature and they do not evaporate.
    
    \item \textbf{Is the information loss problem a paradox?}

    A paradox is something that is self-contradictory or logically inconsistent. Therefore, the information loss problem is better described as an open issue or a conundrum rather than a paradox. In particular, the picture described by semiclassical gravity is consistent although definitely incomplete. Semiclassical gravity predicts the formation of a singularity shielded by an event horizon. The theory is incomplete as it stops being predictive at the singularity. However, it is perfectly consistent regarding the fate of information. Everything that crosses the event horizon could never be retrieved by external observers, and will inevitably be destroyed in the singularity. The information loss arises when we try to push semiclassical gravity beyond its regime of validity, or when we add extra (reasonable) assumptions that cannot be predicted by the model. In this sense, we still have a lot to understand, but we are not facing a logical impossibility.

    \item \textbf{Can \acp{BH} evaporate completely?} 
    
    \acp{BH} gradually lose mass and energy through Hawking radiation. In theory, they could evaporate completely, but the final stages of evaporation involve energies and curvatures requiring a full theory of \ac{QG} for a proper description. Hence, the answer to this question may not be universal.

    \item \textbf{For an external observer, any object falling into a static \ac{BH} needs an infinite time to reach the horizon. How does the picture change when considering the evaporation of the \ac{BH}, given that its lifetime is finite for an external observer?}
    
    First of all, we need to specify what horizon we are talking about. A classical asymptotic observer cannot see an object falling into the event horizon. This means that it either takes infinite time (see \cref{fig:Francesco_Schwarz}), or classical observers cross a Cauchy horizon in finite time (see \cref{fig:Francesco_Complete_evaporation}). Which of the two possibilities is realized depends on the final stages of \ac{BH} evaporation that cannot be predicted without the knowledge of the full theory of \ac{QG}. However, this is so almost by construction. In fact, the event horizon is the boundary of the region that can reach asymptotic infinity, so observers crossing the event horizon cannot be causally connected with asymptotic observers. The question is more interesting if we refer to trapping horizons (which for any practical purposes is the correct definition for the boundary of the \ac{BH}). If we assume some suitable energy conditions, there are theorems stating that trapping horizons are always inside event horizons (see \eg{}~\cite{Wald:1984rg}). This implies that no classical asymptotic observer can see an object entering the trapping horizon (as it would first enter the event horizon, and we have discussed that this is not possible). On the other hand, Hawking radiation violates such energy conditions, and the trapping horizon of an evaporating \ac{BH} is outside the event horizon. Therefore, asymptotic observers will see infalling objects cross the trapping horizon in finite time (see \cref{fig:Francesco_Complete_evaporation}) and without crossing any Cauchy horizon. 
\end{itemize}

\clearpage

\section{Conclusions}
\label{sec:conclusions}

The Nordita Scientific Program \href{https://indico.fysik.su.se/event/8133/}{\textit{``Quantum Gravity: from gravitational effective field theories to ultraviolet complete approaches''}} brought together communities working on different aspects of \ac{QG}, with a focus on extensively discussing questions from different viewpoints. The ultimate scope was to start building solid common grounds and boosting progress in \ac{QG}. The PhD school \textit{``Towards Quantum Gravity''} at the beginning of the program prepared students for the coming weeks of discussions across approaches. The idea was to provide a pedagogical overview of selected topics in \ac{QG}, including some that are typically discussed separately but that are deeply intertwined.

In this spirit, these lecture notes have been prepared cohesively across the mini-courses taught at the PhD school. The scope is to provide a coherent (yet inevitably partial) picture of basic knowledge in \ac{QG}. As a whole, they are intended as a bridge between standard university courses and forefront research.

The first two sets of lecture notes (\cref{sec:LUCA} and \cref{sec:ANNA}) discuss common grounds that all approaches ought to reproduce at low energies: the perturbative framework for \ac{QG} and \ac{EFT}, including the consistency bounds stemming from the latter. Moreover, the last part of \cref{sec:LUCA} gives a first glimpse into \ac{QG} by presenting quadratic gravity --- historically the first attempt to build a renormalizable quantum theory of gravity --- and discussing its features, modern developments, and open questions related to unitarity and the high-energy behavior of the theory. Departing from the common grounds of \cref{sec:LUCA} and \cref{sec:ANNA} requires finding a \ac{UV} completion, and several proposals have been put forth to this end. 
In particular, in \cref{sec:ALESSIABENJAMIN} the notion of non-perturbative renormalizability is introduced. This is key to explain one of the approaches to \ac{QG} based on \ac{QFT}, namely, \ac{ASQG}. Both advantages and shortcomings are discussed. While \ac{ASQG} was initially formulated as a fundamental approach to \ac{QG}, it may also be an intermediate framework bridging \ac{EFT} and a more fundamental description beyond \ac{QFT}. The lecture notes thus continue with one of such possible \ac{UV} completions, \ie{}, \ac{ST} (\cref{sec:IVANO}). Due to the breadth of this field, the lectures focused on its perturbative formulation, while keeping an eye on general lessons for \ac{QG}. In particular, the emergence of gravitational \ac{EFT} at low energies and its behavior at high energies were stressed. The latter is simpler and more universal, whereas low-energy physics is not uniquely determined. Nevertheless, several features of \ac{QG} at low energies are universal, \ie{}, independent of any particular \ac{UV} completion. The last set of lectures (\cref{sec:FRANCESCO}) focuses on some of these aspects. Specifically, it discusses quantum aspects in \acp{BH} spacetimes, that ought to be universal and should be recovered by all \ac{QG} approaches.

In summary, these lecture notes aim at connecting together different patches of the \ac{QG} puzzle, providing the readers with an overall picture that is as coherent as possible. Even in the literature of forefront research, these patches correspond to different areas that are often kept apart. We hope that this contribution, together with the \textit{``Visions in Quantum Gravity''}~\cite{Buoninfante:2024yth}, will encourage further research at the interface between \ac{EFT} and different \ac{QG} approaches.

\clearpage

\subsection*{Acknowledgements}

Luca Buoninfante, Benjamin Knorr, and Alessia Platania would like to thank Nordita for sponsoring the Nordita program. The lecturers are grateful to all the participants of the PhD school for helping to create a very pleasant and stimulating atmosphere.

\paragraph{Author contributions}

Introduction, conclusions, and FAQs were jointly written by all authors. Individual sections have been written by the lecturers of the corresponding mini-courses, as reported at the beginning of each set of lectures.
The style and formatting of TeX, as well as the tracking and uniformization of notation across lecture notes, have been handled by Benjamin Knorr. Additionally, to maximize the coherence of the whole manuscript on several levels, each lecturer provided detailed feedback on the other sets of lectures.

\paragraph{Funding information}

Ivano Basile acknowledges financial support from the Origins Excellence Cluster. Luca Buoninfante acknowledges financial support from the European Union’s Horizon 2020 research and innovation programme under the Marie Sk\l{}odowska-Curie Actions (grant agreement ID: 101106345-NLQG). Francesco Di Filippo acknowledges financial support from the PRIMUS/23/SCI/005 and UNCE24/SCI/016 grants by Charles University, and the GAR 23-07457S grant from the Czech Science Foundation. Benjamin Knorr was partially supported by Nordita. Nordita is supported in part by NordForsk.
The research of Alessia Platania is supported by a research grant (VIL60819) from VILLUM FONDEN. The work of Anna Tokareva is supported by the National Natural Science Foundation of China (NSFC) under Grant No. 1234710.

{\setstretch{0.9}
\section*{List of acronyms}
\addcontentsline{toc}{section}{List of acronyms}

\if\acroswithtooltips1
\begin{acronym}[AdS/CFT]
 \tooltipacro{AdS}{anti-de Sitter}
 \acro{AdSCFT}[\pdftooltip{AdS/CFT}{anti-de Sitter/conformal field theory}]{anti-de Sitter/conformal field theory}
 \tooltipacro{ASQG}{asymp\-to\-ti\-cally safe quantum gravity}
 \tooltipacro{BRST}{Becchi-Rouet-Stora-Tyutin}
 \tooltipacro{BKL}{Belinski–Khalatnikov–Lifshitz}
 \tooltipacro{BH}{black hole}
 \tooltipacro{CDT}{causal dynamical triangulations}
 \tooltipacro{CFT}{conformal field theory}
 \acrodefplural{CFT}{conformal field theories}
 \tooltipacro{CKG}{conformal Killing group}
 \tooltipacro{CKV}{conformal Killing vector}
 \tooltipacro{CMB}{cosmic microwave background}
 \tooltipacro{dS}{de Sitter}
 \tooltipacro{EAA}{effective average action}
 \tooltipacro{EFT}{effective field theory}
 \acrodefplural{EFT}{effective field theories}
 \tooltipacro{FLRW}{Friedmann-Lema\^itre-Robertson-Walker}
 \tooltipacro{FU}{full unitarity}
 \tooltipacro{FRG}{functional renormalization group}
 \tooltipacro{GFP}{Gaussian fixed point}
 \tooltipacro{GR}{General Relativity}
 \tooltipacro{GSO}{Gliozzi-Scherk-Olive}
 \tooltipacro{GW}{gravitational wave}
 \tooltipacro{HZ}{Harrison-Zeldovich}
 \tooltipacro{IR}{infrared}
 \tooltipacro{LSS}{large-scale structure}
 \tooltipacro{NS}{Neveu–Schwarz}
 \tooltipacro{NGFP}{non-Gaussian fixed point}
 \acro{NLSM}[\pdftooltip{NL$\sigma$M}{non-linear sigma model}]{non-linear sigma model}
 \tooltipacro{PWU}{partial wave unitarity}
 \tooltipacro{PC}{Penrose-Carter}
 \tooltipacro{QCD}{quantum chromodynamics}
 \tooltipacro{QED}{quantum electrodynamics}
 \tooltipacro{QFT}{quantum field theory}
 \acrodefplural{QFT}{quantum field theories}
 \tooltipacro{QG}{quantum gravity}
 \tooltipacro{R}{Ramond}
 \tooltipacro{RNS}{Ramond–Neveu–Schwarz}
 \tooltipacro{RG}{renormalization group}
 \tooltipacro{SM}{Standard Model of Particle Physics}
 \tooltipacro{ST}{string theory}
 \tooltipacro{UV}{ultraviolet}
 \tooltipacro{vev}{vacuum expectation value}
\end{acronym}
\else
\begin{acronym}[AdS/CFT]
 \acro{AdS}[AdS]{anti-de Sitter}
 \acro{AdSCFT}[AdS/CFT]{anti-de Sitter/conformal field theory}
 \acro{ASQG}[ASQG]{asymp\-to\-ti\-cally safe quantum gravity}
 \acro{BRST}[BRST]{Becchi-Rouet-Stora-Tyutin}
 \acro{BKL}[BKL]{Belinski–Khalatnikov–Lifshitz}
 \acro{BH}[BH]{black hole}
 \acro{CDT}[CDT]{causal dynamical triangulations}
 \acro{CFT}[CFT]{conformal field theory}
 \acrodefplural{CFT}{conformal field theories}
 \acro{CKG}[CKG]{conformal Killing group}
 \acro{CKV}[CKV]{conformal Killing vector}
 \acro{CMB}[CMB]{cosmic microwave background}
 \acro{dS}[dS]{de Sitter}
 \acro{EAA}[EAA]{effective average action}
 \acro{EFT}[EFT]{effective field theory}
 \acrodefplural{EFT}{effective field theories}
 \acro{FLRW}[FLRW]{Friedmann-Lema\^itre-Robertson-Walker}
 \acro{FU}[FU]{full unitarity}
 \acro{FRG}[FRG]{functional renormalization group}
 \acro{GFP}[GFP]{Gaussian fixed point}
 \acro{GR}[GR]{General Relativity}
 \acro{GSO}[GSO]{Gliozzi-Scherk-Olive}
 \acro{GW}[GW]{gravitational wave}
 \acro{HZ}[HZ]{Harrison-Zeldovich}
 \acro{IR}[IR]{infrared}
 \acro{LSS}[LSS]{large-scale structure}
 \acro{NS}[NS]{Neveu–Schwarz}
 \acro{NGFP}[NGFP]{non-Gaussian fixed point}
 \acro{NLSM}[NL$\sigma$M]{non-linear sigma model}
 \acro{PWU}[PWU]{partial wave unitarity}
 \acro{PC}[PC]{Penrose-Carter}
 \acro{QCD}[QCD]{quantum chromodynamics}
 \acro{QED}[QED]{quantum electrodynamics}
 \acro{QFT}[QFT]{quantum field theory}
 \acrodefplural{QFT}{quantum field theories}
 \acro{QG}[QG]{quantum gravity}
 \acro{R}[R]{Ramond}
 \acro{RNS}[RNS]{Ramond–Neveu–Schwarz}
 \acro{RG}[RG]{renormalization group}
 \acro{SM}[SM]{Standard Model of Particle Physics}
 \acro{ST}[ST]{string theory}
 \acro{UV}[UV]{ultraviolet}
 \acro{vev}[vev]{vacuum expectation value}
\end{acronym}
\fi

}

\bibliography{combined_bib.bib}

\end{document}